\documentclass[twoside,openright,titlepage,numbers=noenddot,headinclude, lineheaders footinclude=true,cleardoublepage=empty,
                                BCOR=5mm,paper=a4,fontsize=12pt ]{scrbook} 
\usepackage[usenames,dvipsnames,svgnames,table,x11names]{xcolor}
\usepackage[unicode,colorlinks,hyperindex]{hyperref}

\usepackage[utf8]{inputenc}
\usepackage[eulerchapternumbers,beramono]{classicthesis}
\usepackage[english]{babel}
\usepackage[T1]{fontenc}
\usepackage{estilo}
\usepackage{stmaryrd}
\usepackage{soul}
\usepackage{upgreek}
\usepackage{graphicx}
\usepackage{amsmath}
\usepackage{titlesec}
\usepackage{tocloft}
\usepackage{breqn}   

\usepackage{amsfonts}
\usepackage{amssymb}
\usepackage{amsthm}
\usepackage{mathrsfs}
\usepackage{physics}
\usepackage{array}
\usepackage{multirow}
\usepackage{enumerate}
\usepackage{bbm}
\usepackage{cite}
\usepackage{lastpage}
\usepackage{blindtext}
\usepackage{geometry}
\usepackage{cancel}
\usepackage{pdfpages}
\usepackage{makeidx}
\usepackage{mathtools,tikz-cd}
\usepackage{systeme}
\usetikzlibrary{arrows}
\usepackage[colorinlistoftodos]{todonotes}
\usepackage{color,graphicx}
\usepackage{pst-node}
\usepackage{fancyvrb} 
\usepackage{bm}
\usepackage{tikz}
\usetikzlibrary{shapes,arrows,intersections}
\usetikzlibrary{matrix,fit,calc,trees,positioning,arrows,chains,shapes.geometric,shapes}
\usepackage{wrapfig}
\usepackage{relsize}
\usepackage{float}
\usepackage{appendix}
\usepackage{framed}
\usepackage{scrlayer-scrpage}

\usepackage{tikz,siunitx}
\usetikzlibrary{shapes.geometric,shapes.symbols}

\newcommand{\tikzsymbol}[2][circle]{\tikz[baseline=-0.5ex]\node[inner
sep=1.2pt,shape=#1,draw,#2]{};}%

\def\bomdia{\tikzsymbol[diamond]{fill=black} }

\hypersetup{
  colorlinks = true,
  linkcolor  = gray,
  citecolor = darkgray,
}

\newcommand{\sysdots}{%
  \hbox{\kern-0em$\cdotp\cdotp\cdotp\cdotp\cdotp\cdotp$}%
}
\definecolor{formalshade}{rgb}{0.95,0.95,1}

\usepackage{mdframed}

\makeatletter
\newcommand\bigDiamond{\mathop{\mathpalette\bigDi@mond\relax}}
\newcommand\bigDi@mond[2]{%
  \vcenter{\hbox{\m@th
    \scalebox{\ifx#1\displaystyle 2\else1.2\fi}{$#1\Diamond$}%
  }}%
}
\newcommand\bigLozenge{\mathop{\mathpalette\bigL@zenge\relax}}
\newcommand\bigL@zenge[2]{%
  \vcenter{\hbox{\m@th
    \scalebox{\ifx#1\displaystyle 2\else1.2\fi}{$#1\blacklozenge$}%
  }}%
}

\makeatother

\newtheorem{theorem}{Theorem}[chapter]
\newtheorem{lemma}{Lemma}[chapter]
\newtheorem{propos}{Proposition}[chapter]
\newtheorem{cor}{Corollary}[chapter]
\newtheorem{defin}{Definition}[chapter]
\newtheorem{ex}{Example}[chapter]
\newtheorem{rem}{Remark}[chapter]

\newcommand{\BigW}{\mathord{\adjustbox{valign=B,totalheight=.7\baselineskip}{$\bigwedge$}}}
\def\'#1{\mathbf{#1}}
\def\rt{\mathbb{R}^3}
\def\upb{\upbeta}
\def\upa{\upalpha}
\def\clt{\mathcal{C}\ell_{3}}
\def\cl{\mathcal{C}\ell}
\def\Cl{\mathcal{C}\ell}
\def\clm{\mathcal{C}\ell_{1,3}}
\def\clpq{\mathcal{C}\ell_{p,q}}
\def\clvg{\mathcal{C}\ell(V,g)}

\def\proof{\noindent \textit{Proof.}\;\;}
\def\rr{\mathbb{R}}
\def\rrpq{\mathbb{R}^{p,q}}
\def\cos{\text{cos}}
\def\sen{\text{sen}}
\def\exp{\text{exp}}

\def \eskvetr3{\BigW^k(\mathbb{R}^3)}

\def\matcomp{\text{Mat}(2,\mathbb{C})}
\def\matcompq{\text{Mat}(4,\mathbb{C})}

\def\matreald{\text{Mat}(2,\mathbb{R})}
\def\matrealq{\text{Mat}(4,\mathbb{R})}

\def\cc{\mathbb{C}}
\def\kk{\mathbb{K}}
\def\zz{\mathbb{Z}}
\def\id{\text{id}}
\def\hh{\mathbb{H}}
\def\hsh{\mathbb{H} \oplus \mathbb{H}}
\def\rsr{\mathbb{R} \oplus \mathbb{R}}
\def\endo{\text{End}}
\def\A{\mathcal{A}}
\def\B{\mathcal{B}}
\def\M{\mathcal{M}}
\def\E{\mathcal{E}}
\def\O{\text{O}}
\def\SO{\text{SO}}
\def\Pin{\text{Pin}}
\def\Spin{\text{Spin}}

\def\a{\alpha}
\def\b{\beta}
\def\tensor{\otimes}

\def\dim{\text{dim\,}}
\def\sp{\text{span\,}}
\def\Gpq{\Gamma_{p,q}}

\def\inter{\cap}

\def\alt{\text{Alt}}
\def\Alt{\text{Alt}}
\def\Ap{A_{[p]}}
\def\Bq{B_{[q]}}
\def\Cr{C_{[r]}}

\def\Cinf{\mathcal{C}^{\infty}}

\def\Tmd{T^{*}M}

\def\Autpq{\text{Aut}(\mathcal{C}\ell_{p,q})}

\def\*#1{\mathbf{#1}}

\numberwithin{equation}{chapter}
\definecolor{mycolor}{rgb}{0.88,0.82,0.11}
\newenvironment{tftheorem}{\begin{mdframed}\vspace{-0.2cm}\begin{theorem}}
  {\end{theorem}\end{mdframed}}

  \newenvironment{tfcor}{\begin{mdframed}\vspace{-0.2cm}\begin{cor}}
  {\end{cor}\end{mdframed}}

\newenvironment{tfpropos}{\begin{mdframed}\vspace{-0.2cm}\begin{propos}}{\end{propos}\end{mdframed}}

 \newenvironment{tflemma}{\begin{mdframed}\vspace{-0.2cm}\begin{lemma}}
  {\end{lemma}\end{mdframed}}

\newenvironment{definvermelho}
   {\colorlet{shadecolor}{Red!15}\begin{shaded}\begin{defin}}
   {\end{defin}\end{shaded}}

\newenvironment{definlaranja}
   {\colorlet{shadecolor}{Orange!15}\begin{shaded}\begin{defin}}
   {\end{defin}\end{shaded}}

\newenvironment{definamarelo}
   {\colorlet{shadecolor}{yellow!15}\begin{shaded}\begin{defin}}
   {\end{defin}\end{shaded}}

\newenvironment{definverde}
   {\colorlet{shadecolor}{Green!15}\begin{shaded}\begin{defin}}
   {\end{defin}\end{shaded}}

\newenvironment{definazul}
   {\colorlet{shadecolor}{cyan!15}\begin{shaded}\begin{defin}}
   {\end{defin}\end{shaded}}
\newenvironment{definanil}
   {\colorlet{shadecolor}{blue!15}\begin{shaded}\begin{defin}}
   {\end{defin}\end{shaded}}
\newenvironment{definvioleta}
   {\colorlet{shadecolor}{DarkOrchid!15}\begin{shaded}\begin{defin}}
   {\end{defin}\end{shaded}}
   \newenvironment{definrosa}
   {\colorlet{shadecolor}{Magenta!12}\begin{shaded}\begin{defin}}
   {\end{defin}\end{shaded}}

\newenvironment{defincinza}
   {\colorlet{shadecolor}{gray!15}\begin{shaded}\begin{defin}}
   {\end{defin}\end{shaded}}

\newcommand{\autor}{Deborah Gonçalves Fabri}

\newcommand{\titulo}{From Clifford Bundles to Spinor Classification:  Algebraic and Geometric Approaches,\newline and  
New Spinor Fields in Flux Compactifications}
\def\mestrado{}
\newcommand{\orientador}{Roldão da Rocha}
\newcommand{\fomento}{Coordenação de Aperfeiçoamento de Pessoal de Nível Superior - Brasil (CAPES) - Finance Code 001.}
\def\versaofinal{}

\newcommand{\centro}{Centro de Matemática, Computação e Cognição \xspace}
\newcommand{\titulacao}{Mestre em Matemática \xspace}
\newcommand{\palavraschaves}{álgebra de Clifford, fibrado de Clifford, fibrado de Kähler-Atiyah, fibrado espinorial, espinores, bilineares covariantes,
identidades de Fierz, classificação de espinores.}
\newcommand{\keywords}{Clifford algebras, Clifford bundle, Kähler-Atiyah bundle, bundle of spinors, spinors, bilinear covariants, Fierz identities, spinor classification.} 
\makeindex

\NewDocumentCommand\myred{m}{\textcolor{red}{#1}}
\NewDocumentCommand\myorange{m}{\textcolor{orange}{#1}}
\NewDocumentCommand\myyellow{m}{\textcolor{yellow}{#1}}
\NewDocumentCommand\mygreen{m}{\textcolor{green}{#1}}
\NewDocumentCommand\mycyan{m}{\textcolor{cyan}{#1}}
\NewDocumentCommand\myblue{m}{\textcolor{blue}{#1}}
\NewDocumentCommand\myviolet{m}{\textcolor{DarkOrchid}{#1}}
\NewDocumentCommand\mymagenta{m}{\textcolor{Fuchsia}{#1}}

\DeclareSymbolFont{yhlargesymbols}{OMX}{yhex}{m}{n} 
\DeclareMathAccent{\yhwidehat}{\mathord}{yhlargesymbols}{"62}
\DeclareSymbolFont{yhlargesymbols}{OMX}{yhex}{m}{n} 
\DeclareMathAccent{\yhwidetilde}{\mathord}{yhlargesymbols}{"65}
\begin{document}

\begin{center}\thispagestyle{empty}
\includegraphics[height=2cm, keepaspectratio=true]{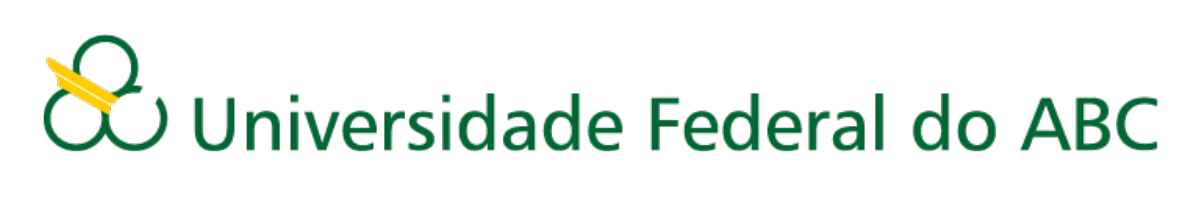} \end{center}\hfill

\begin{center}
  {\Large \scshape Programa De Pós-Graduação
em Matemática}
\end{center}
\vspace*{2cm}
\begin{center}
  {\Large \scshape \autor}
\end{center}
\vspace{1cm}
\begin{center}
    \includegraphics[scale=0.2]{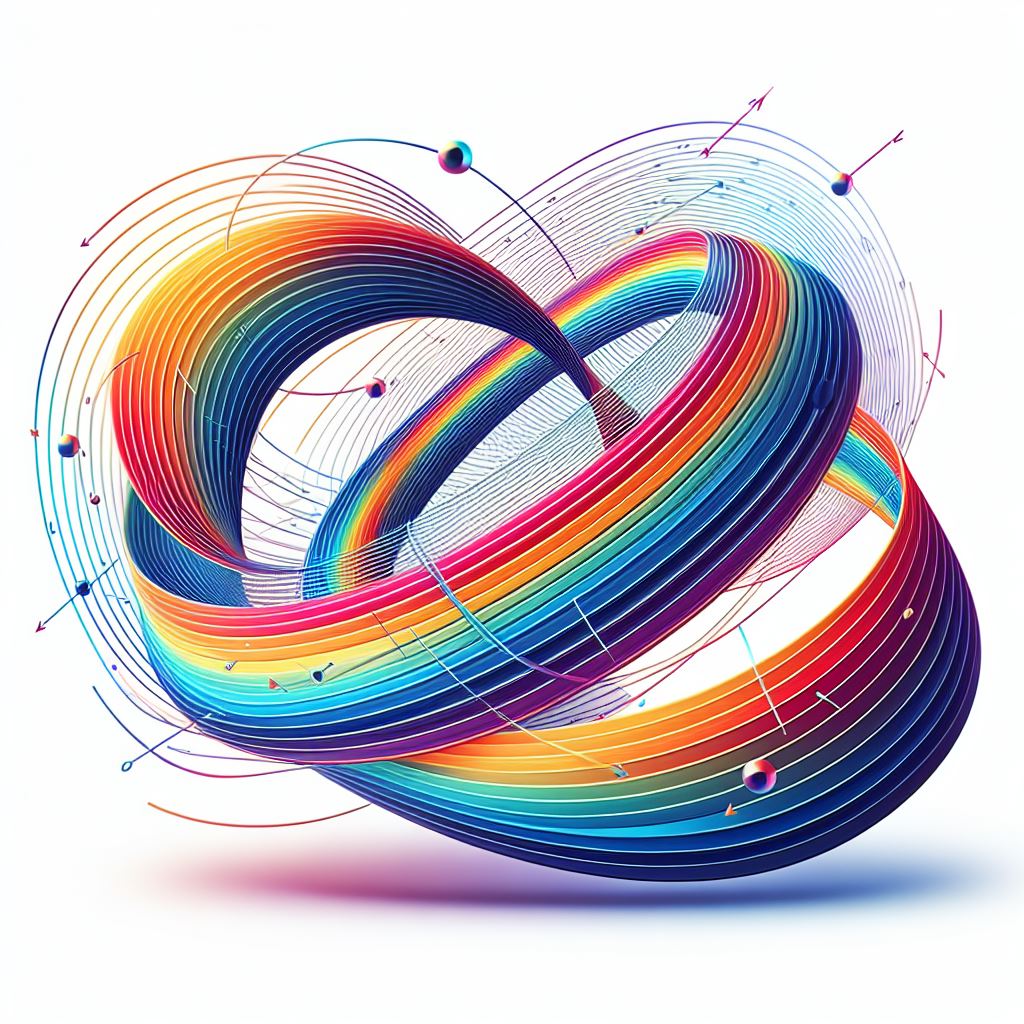}
\end{center}
\vspace{1cm}
\begin{center}
  {\LARGE \scshape \bfseries \titulo}
\end{center}
\vfill
\vspace{1cm}
\begin{center}
  { This study was financed by \fomento \\} 
  \vspace{1cm}
  {\bfseries Santo  André, 2024}
\end{center}
\frontmatter

\thispagestyle{empty}
\begin{center}
    \includegraphics[height=2.7cm, keepaspectratio=true]{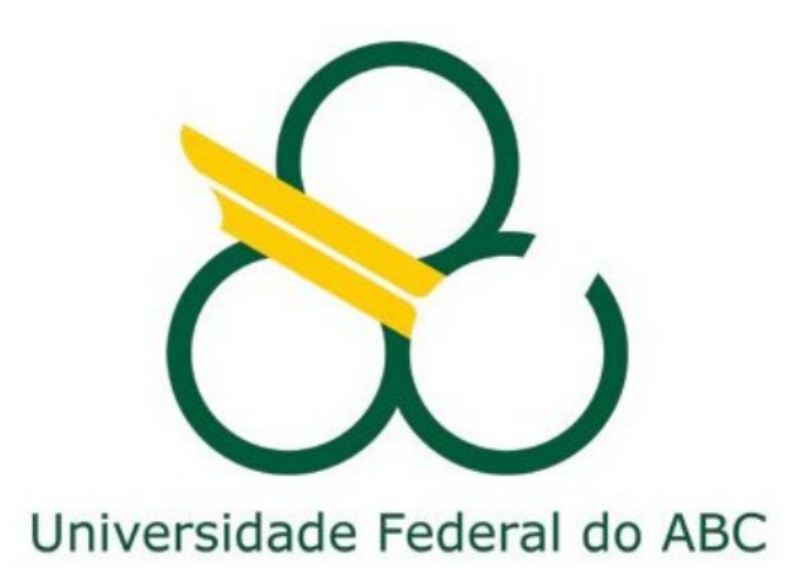} 
\end{center}\hfill

\vspace{1.37cm}
\begin{center}
  {\large \scshape \bfseries Universidade Federal do ABC
  \vspace{.5cm}

\centro}
\end{center}
\vspace{.7cm}
\begin{center}
  {\large \scshape \bfseries \autor}
\end{center}
\vspace{.7cm}
\begin{center}
  {\LARGE \scshape \bfseries \titulo}
\end{center}
\vspace{1cm}
{\bfseries
\noindent
Orientador\ifx\femaleOrientador\undefined
\else
a\fi: Prof\ifx\femaleOrientador\undefined
\else
a\fi. Dr\ifx\femaleOrientador\undefined
\else
a\fi. \orientador
\vspace{.25cm}

\ifx\coorientador\undefined
\else
\noindent
Coorientador\ifx\femaleCoorientador\undefined
\else
a\fi: Prof\ifx\femaleCoorientador\undefined
\else
a\fi. Dr\ifx\femaleCoorientador\undefined
\else
a\fi. \coorientador
\fi
}

\vspace{1.4cm}
\begin{flushright}
  \begin{minipage}[c]{.6\textwidth}
    \begin{flushleft}
      \ifx\mestrado\undefined
      \noindent Tese de doutorado
      \else
    \noindent   Dissertação de mestrado
      \fi
      apresentada ao \centro para \\ \noindent  obtenção do título de
      \titulacao
    \end{flushleft}
  \end{minipage}
\end{flushright}
\vspace{1.3cm}
  \ifx\versaofinal\undefined
\noindent 
{\footnotesize \scshape
Esta é a versão original da tese, tal como\\
submetida à Comissão Julgadora.\\
}
\else
\noindent 
{\footnotesize \scshape
Este exemplar foi revisado e alterado em relação à versão original, de acordo com as observações levantadas pela banca examinadora no dia da defesa, sob responsabilidade única da autora e com a anuência do orientador.}
\fi

\vspace{1.2cm}

\vfill
\begin{center}
  {\small \scshape \bfseries Santo André, 2024}
\end{center}

\includepdf{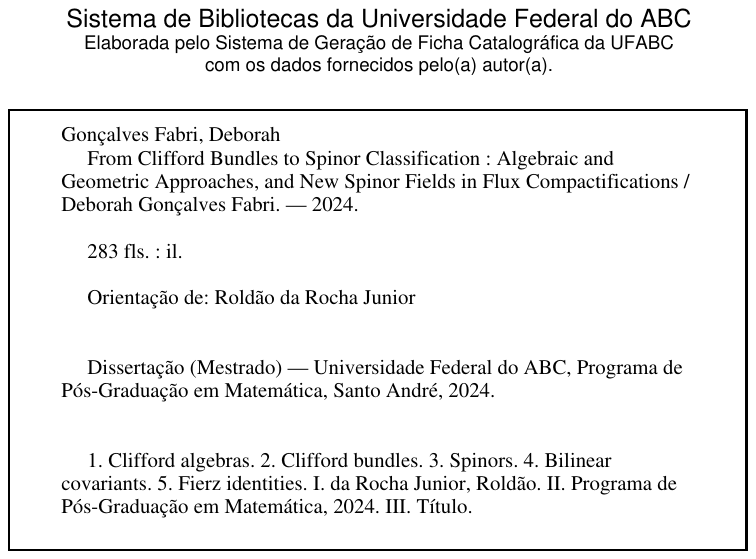}

\chapter*{Agradecimentos}
\paragraph{ } Agradeço imensamente ao meu orientador, Roldão, pela orientação brilhante, excepcional e impecável, por ter sido o melhor orientador que alguém poderia ter.

\paragraph{ } O presente trabalho foi realizado com apoio da Coordenação de Aperfeiçoamento de Pessoal de Nível Superior - Brasil (CAPES) - Código de Financiamento 001.

%

%

%

%
%
\chapter*{Resumo}

Ao conectar conceitos geométricos e algébricos, esta dissertação estabelece as bases para um estudo abrangente das estruturas de Clifford em fibrados e espinores. Exploramos o fibrado de Kähler-Atiyah, que captura a essência das álgebras de Clifford e oferece insights profundos sobre as estruturas algébricas subjacentes aos contextos geométricos. São abordadas as definições algébricas e clássicas de espinores dentro das álgebras de Clifford, bem como sua definição como seções do fibrado de espinores, construído dentro de uma estrutura de espinores em uma variedade, com base no fibrado de Kähler-Atiyah, que serve como uma estrutura eficaz para aplicações envolvendo espinores, como a sua classificação baseada em seus bilineares covariantes e identidades de Fierz. Formas diferenciais homogêneas desempenhando o papel de bilineares covariantes podem se anular devido a obstruções algébricas em uma compactificação com fluxo deformado AdS$_3 \times \M_8$, resultando na identificação de novas classes de campos de espinores.

\vspace{.5cm}
\textbf{Palavras-chave}: \palavraschaves

\chapter*{ Abstract}
\selectlanguage{english}

By bridging geometric and algebraic concepts, this dissertation lays the groundwork for a comprehensive study of the Clifford structures on bundles and spinor fields. We delve into the Kähler-Atiyah bundle, which encapsulates the essence of Clifford algebras and provides profound insights into the algebraic structures underlying geometric frameworks. The algebraic and classical definitions of spinors within Clifford algebras are concerned, as well as their global realisation as sections of the bundle of spinors constructed within a spin structure on a manifold from the Kähler-Atiyah bundle background which serves as an effective framework for applications involving spinors such as their classification based on their bilinear covariants and Fierz identities. Homogeneous differential forms playing the role of bilinear
covariants can vanish due to algebraic obstructions in a warped flux compactification AdS$_3 \times \M_8$ yielding the identification of new spinor field classes.

\vspace{.5cm}
\textbf{Keywords}: \keywords

\selectlanguage{english}

\setcounter{tocdepth}{1}
\tableofcontents
\mainmatter

\newpage
\phantomsection
\addcontentsline{toc}{chapter}
{\;\;\;\;\;\textsc{introduction}}
\chapter*{\textsc{introduction}}
\paragraph{ } Matter is made of fermions which are described by spinors. Some fermions are elementary particles, such as electrons. In the beginning of the 20th century there was a great interest in developing a theory that describes the electron. By virtue of the electron's nature, this theory must have a quantum and a relativistic approach. The Schrödinger equation

\begin{equation}
    i\hbar\frac{\partial\psi}{\partial t} = \frac{-\hbar^2}{2m} \nabla^2\psi + W\psi,
\end{equation}

\noindent describes all atomic phenomena except those involving magnetism and relativity. Due to the distinct orders of time and space derivatives on that equation the Schrödinger equation is not suitable to deal with relativistic phenomena. The different orders of derivatives do not make the equation invariant under space-time transformations and this contradicts the principle that the laws of physics must be the same for all inertial observers. Therefore, since the Schrödinger equation is incompatible with the theory of relativity another path was to start from the relativistic energy equation,

\begin{equation}\label{eq_espd_eqrelativist}
   E^2 = \'p^2c^2 + m^2c^2
\end{equation}

\noindent By the correspondence principle, inserting energy and momentum operators

\begin{equation}
    E = i\hbar \frac{\partial}{\partial t}, \;\;\;\;\;\;\;\; \'p = -i\hbar\nabla,
\end{equation}

\noindent into that equation, results in the Klein-Gordon equation

\begin{equation}
\hbar^2 \left (- \frac{1}{c^2}\frac{\partial^2}{\partial t^2} + \frac{\partial^2}{\partial x_{1}^2} + \frac{\partial^2}{\partial x_{2}^2}+ \frac{\partial^2}{\partial x_{3}^2}  \right)\psi = m^2c^2 \psi.
\end{equation}

The Klein-Gordon equation solves the asymmetry problem in the derivatives, however, only for spinless particles. Also, this equation has a second derivative problem that can give rise to what would be a probability density that is not always positive. Therefore, what the physicist Paul Dirac did was linearise the Klein-Gordon equation, or replaced it with a first-order equation,

\begin{equation}\label{eq_espdireq000}
    i\hbar \left (\gamma_{0} \frac{1}{c^2}\frac{\partial}{\partial t} + \gamma_{1}\frac{\partial}{\partial x_{1}} + \gamma_{2}\frac{\partial}{\partial x_{2}}+ \gamma_{3}\frac{\partial}{\partial x_{3}}  \right)\psi = mc \psi,
\end{equation}

\noindent this way the derivative in time would become linear, therefore time and space are treated on an equal footing which solves the problem that the Klein-Gordon equation was facing. In addition, Dirac brought very important interpretations such as the existence of antiparticles as an explanation for the existence of negative energies that the Klein-Gordon equation presents. Hence, Dirac's theory takes into account the invariance under space-time isometries and also takes into account spins' interaction, in particular, the Dirac equation considers relativistic effects as well as spin-$\frac{1}{2}$ particles achieving the main goal of establishing a theory that describes the electron \cite{Dir28}. However, the development of this theory was only possible because Dirac introduced a set of interesting $4 \times 4$ complex matrices $\gamma_\mu$, $\mu = 0,1,2,3$:

\begin{align}
\begin{aligned}\label{eq_matdiracquatro}
&\gamma_{0} = \begin{pmatrix} 
    1 & 0 & 0 & 0 \\
    0 & 1 & 0 & 0\\
    0 & 0 & -1 & 0\\
    0 & 0 & 0 & -1
    \end{pmatrix}, &&&\gamma_{1} = \begin{pmatrix} 
    0 & 0 & 0 & -1 \\
    0 & 0 & -1 & 0\\
    0 & 1 & 0 & 0\\
    1 & 0 & 0 & 0
    \end{pmatrix}, \\
    &\gamma_{2} = \begin{pmatrix} 
    0 & 0 & 0 & i \\
    0 & 0 & -i & 0\\
    0 & -i & 0 & 0\\
    i & 0 & 0 & 0
    \end{pmatrix}, &&&\gamma_{3} = \begin{pmatrix} 
    0 & 0 & -1 & 0 \\
    0 & 0 & 0 & 1\\
    1 & 0 & 0 & 0\\
    0 & -1 & 0 & 0
    \end{pmatrix}.
\end{aligned}
\end{align}

\noindent satisfying a very specific relation, that is:

\begin{align}
\begin{aligned}
& \gamma_{0}^{2} = I, \;\;\; \gamma_{1}^{2} = \gamma_{2}^{2} = \gamma_{3}^{2} = -I, \\
&\gamma_{\mu}\gamma_{\nu} = -\gamma_{\nu}\gamma_{\mu} \;\;\;\;\text{for} \;\;\;\;\mu \neq \nu.
\end{aligned}
\end{align}

\noindent or in terms of the metric $\eta_{\mu \nu}$ of the Minkowski space-time $\rr^{1,3}$

\begin{equation}
  \gamma_{\mu}\gamma_{\nu} + \gamma_{\nu}\gamma_{\mu} = 2\eta_{\mu \nu}I
\end{equation}

\noindent This seemingly innocent condition is in fact quite deep. It first appeared 1878, in an attempt to unify the structures introduced by Grassmann and Hamilton, namely, the exterior algebra and the quaternions respectively.  William Kingdom Clifford in his work \textit{“Applications Of Grassmann's extensive algebra”}, introduced the analogue of the quaternion product on the multivector structure of exterior algebra in which he established a new algebraic structure called Geometric Algebra or Clifford algebra. In contrast to the quaternion algebra or the Gibbs algebra, this new structure is not limited to be defined on a particular space, but in any quadratic space, dimensions and signatures, which makes Clifford algebras rich in structure and its applications. In particular, since the Dirac equation now lies in an matrix Clifford algebra the solution $\psi$ of that equation is no longer a simply wave function but a new object that rely on the space of the column vectors, namely, a \textit{spinor}.

\begin{equation}\label{eq_diracspinor}
     \psi = \begin{pmatrix} 
    \psi_{1} \\
    \psi_{2} \\
    \psi_{3} \\
    \psi_{4}
    \end{pmatrix} \in \mathbb{C}^{4}
\end{equation}

However, the concept of spinors is not limited to four-dimensional spacetime; they can be defined in any dimension and signature, with Clifford algebras providing the necessary algebraic structure for their definition and manipulation. The notion of spinors first appeared in the early 20th century, introduced by the mathematician Élie Cartan in the 1910s. Cartan developed the mathematical framework for spinors within the context of differential geometry and group theory \cite{ECart}. The link between Clifford algebras and spinors became increasingly explicit and evolved significantly throughout the 20th century, particularly with the advancements in quantum mechanics and quantum field theory, including the contributions of Paul Dirac. Clifford algebras lay the groundwork for both classical and algebraic definitions of spinors, which are related through minimal left ideals and irreducible modules/representations. A major reason for the substantial interest in spinors, Clifford algebras, and related topics within the physics community is the remarkable success of Dirac fields in describing spin-$\frac{1}{2}$ particles with fermionic statistics in the Standard Model. Spinors can be realised in the context of differential geometry, the concept of spin structures on manifolds allow the definition of spinor fields in the geometric background. In this sense, the Kähler-Atiyah Bundle plays a fundamental role in this geometric approach. This bundle encapsulates the essence of geometric algebra, offering a profound insight into the algebraic structures underlying a geometric framework. The typical approach to spinors is viewing a bundle of spinors over a pseudo-Riemannian manifold as a bundle of modules over the real Kähler-Atiyah bundle with module structure specified by a particular morphism of bundles of algebras.  

\paragraph{ } This dissertation embarks on a journey to unify algebraic and geometric perspectives in the study of Clifford algebras, spinors, and their classification. We aim to offer a thorough examination of Clifford structures on bundles and spinor fields, culminating in a detailed classification of spinors across various dimensions and metric signatures. The classification of spinors is important both mathematically and physically. Mathematically, it provides a systematic framework for understanding and categorising these fundamental objects, which in turn reveals the structure of Clifford algebras and their representations.  Physically, the introduction of the Dirac spinor has already demonstrated its significance in describing fundamental particles and their interactions, indicating how the classification of spinors is crucial for constructing models in quantum field theory. However, it is important to clarify that this dissertation does not delve into any physical aspects of spinors. Instead, the focus is strictly on the mathematical description and answering questions related to the structure and classification of spinors. Any references to physical implications serve solely as context and motivation.

\paragraph{ } This dissertation is divided into two main parts, each addressing the mathematical framework necessary for the definition and classification of spinors. Both parts aim to dissect the underlying mathematical structures, though they approach the topic from different perspectives: algebraic and geometric. The first part focuses purely on algebraic methods. We begin by thoroughly examining the structure of Clifford algebras, which forms the foundation for defining spinors. This section culminates in the classification of spinors based on bilinear covariants, which act like an inner product in the spinor space but are confined to Minkowski space-time. The second part introduces a geometric flavour to our study. Here, we build upon the algebraic groundwork by integrating geometric concepts, we take the Clifford algebra and promote it to spin structures and spinor bundles over manifolds and explore how the Clifford algebra formalism is imported to the bundle setting. This involves defining and exploring Clifford structures within fibre bundles, particularly the Kähler-Atiyah bundle. With this algebraic-geometric framework, we extend the notion of bilinear covariants beyond Minkowski space-time, allowing for a generalisation applicable to various dimensions and metric signatures.

\paragraph{ } In the \textcolor[rgb]{0.82,0.01,0.11}{Chapter 1} we start with the fundamental concepts of Clifford algebras, including their definition, construction, structure, representation and classification. We demonstrate their periodicity and natural relation to matrix algebras that make possible their classification. This chapter lays the groundwork for understanding the algebraic properties that underpin all the other topics of this dissertation. In 
\textcolor[rgb]{0.96,0.65,0.14}{Chapter 2} we delve into groups and symmetries that can be defined within Clifford algebras. We examine the study of rotations and reflections in various dimensions, leading to the introduction of the Pin and Spin groups. This exploration is crucial for understanding the symmetries associated with spinors. \textcolor[rgb]{0.88,0.82,0.11}{Chapter 3} proceed with the definition and classification of algebraic and classical spinors. We define algebraic spinors as elements carrying irreducible representation spaces of the reduced Spin group, constructed within Clifford algebras. Moving forward, \textcolor[rgb]{0.47,0.81,0.1}{Chapter 4} introduces bilinear covariants, which serve as a form of inner product in the space of spinors. We focus on bilinear covariants in Minkowski space-time, discussing their categorisation and the geometric and physical information they encode. This chapter concludes with Lounesto’s spinor classification, providing a comprehensive view of spinor properties in this context.

\begin{center}

\tikzset{every picture/.style={line width=0.75pt}} 

\begin{tikzpicture}[x=0.75pt,y=0.75pt,yscale=-1,xscale=1]

\draw    (187.67,137.93) -- (187.67,172.33) ;
\draw    (141.67,198.93) -- (141.67,233.33) ;
\draw    (75.67,91.93) -- (75.67,170.93) ;
\draw    (135.67,84.93) -- (166.67,115.93) ;

\draw (33.33,62) node [anchor=north west][inner sep=0.75pt]  [font=\small] [align=left] {{\fontfamily{pcr}\selectfont \textcolor[rgb]{0.82,0.01,0.11}{\textbf{Clifford Algebras}}}};
\draw (168.67,118.93) node [anchor=north west][inner sep=0.75pt]  [font=\small] [align=left] {{\fontfamily{pcr}\selectfont \textcolor[rgb]{0.96,0.65,0.14}{\textbf{Groups and Symmetries}}}};
\draw (8,177.67) node [anchor=north west][inner sep=0.75pt]  [font=\small] [align=left] {{\fontfamily{pcr}\selectfont \textbf{\textcolor[rgb]{0.88,0.82,0.11}{Algebraic and Classical Spinors}}}};
\draw (12,237) node [anchor=north west][inner sep=0.75pt]  [font=\footnotesize] [align=left] {{\fontfamily{pcr}\selectfont \textcolor[rgb]{0.47,0.81,0.1}{\textbf{Bilinear Covariants in Minkowski Space-Time}}}};

\end{tikzpicture}
\end{center}

The second part of the dissertation shifts to a geometric perspective, building upon the algebraic groundwork laid in the first part. \textcolor[rgb]{0.41,0.67,0.97}{Chapter 5} introduces the foundational concepts of tangent and cotangent bundles, vector bundles, and principal bundles. These preliminaries are essential for understanding spin structures and defining spinors in a geometric context, setting the stage for the geometric approach. In \textcolor[rgb]{0.08,0.18,0.88}{Chapter 6}, we delve into the Kähler-Atiyah bundle, exploring its intricate properties and its role in unifying algebraic and geometric concepts. We discuss the Clifford algebra on the exterior bundle, where non-commutative multiplication induces a non-commutative associative multiplication on inhomogeneous differential forms. This chapter also examines the volume form and key isomorphic subalgebras, emphasising the richness of the Kähler-Atiyah bundle. \textcolor[rgb]{0.74,0.06,0.88}{Chapter 7} explores the profound interplay between algebra and geometry in the context of spinors and presents spinors within the context of differential geometry by defining spin structures and their associated bundles. The final \textcolor[rgb]{0.98,0.4,0.91}{Chapter 8} extends the concept of bilinear covariants to any dimension and metric signature using the Kähler-Atiyah framework. We explore how algebraic constraints can cause some spinor bilinears to vanish, and we identify new classes of spinor fields within an eight-dimensional manifold derived from an eleven-dimensional warped flux compactification, AdS$_3 \times \M_8$.  These findings extend Lounesto's spinor classification and demonstrate the extensive applicability of the Kähler-Atiyah bundle.

\begin{center}

\tikzset{every picture/.style={line width=0.75pt}} 

\begin{tikzpicture}[x=0.75pt,y=0.75pt,yscale=-1,xscale=1]

\draw    (195.67,70.93) -- (314,70.93) ;
\draw    (187.67,137.93) -- (187.67,172.33) ;
\draw    (551.67,31.93) -- (517.67,64.93) ;
\draw    (473.67,84.93) -- (473.67,141.93) ;
\draw    (556.67,31.93) -- (556.67,141.93) ;
\draw    (513.67,167.93) -- (513.67,227.93) ;
\draw    (141.67,198.93) -- (141.67,233.33) ;
\draw    (75.67,91.93) -- (75.67,170.93) ;
\draw    (135.67,84.93) -- (166.67,115.93) ;
\draw    (366,242.93) -- (398,242.93) ;

\draw (33.33,62) node [anchor=north west][inner sep=0.75pt]  [font=\small] [align=left] {{\fontfamily{pcr}\selectfont \textcolor[rgb]{0.82,0.01,0.11}{\textbf{Clifford Algebras}}}};
\draw (527,15.33) node [anchor=north west][inner sep=0.75pt]  [font=\small] [align=left] {{\fontfamily{pcr}\selectfont \textcolor[rgb]{0.41,0.67,0.97}{\textbf{Bundles}}}};
\draw (332.67,64.93) node [anchor=north west][inner sep=0.75pt]  [font=\small] [align=left] {{\fontfamily{pcr}\selectfont \textcolor[rgb]{0.08,0.18,0.88}{\textbf{The Kähler-Atiyah Bundle}}}};
\draw (168.67,118.93) node [anchor=north west][inner sep=0.75pt]  [font=\small] [align=left] {{\fontfamily{pcr}\selectfont \textcolor[rgb]{0.96,0.65,0.14}{\textbf{Groups and Symmetries}}}};
\draw (8,177.67) node [anchor=north west][inner sep=0.75pt]  [font=\small] [align=left] {{\fontfamily{pcr}\selectfont \textbf{\textcolor[rgb]{0.88,0.82,0.11}{Algebraic and Classical Spinors}}}};
\draw (12,237) node [anchor=north west][inner sep=0.75pt]  [font=\footnotesize] [align=left] {{\fontfamily{pcr}\selectfont \textcolor[rgb]{0.47,0.81,0.1}{\textbf{Bilinear Covariants in Minkowski Space-Time}}}};
\draw (457,148) node [anchor=north west][inner sep=0.75pt]  [font=\small] [align=left] {\textcolor[rgb]{0.74,0.06,0.88}{\textbf{{\fontfamily{pcr}\selectfont Spin Geometry}}}};
\draw (402.67,236) node [anchor=north west][inner sep=0.75pt]  [font=\small,color={rgb, 255:red, 251; green, 81; blue, 216 }  ,opacity=1 ] [align=left] {{\fontfamily{pcr}\selectfont \textcolor[rgb]{0.98,0.4,0.91}{\textbf{Generalised Bilinear Covariants}}}};

\end{tikzpicture}
\end{center}

\part{\myred{A}\;\myorange{L}\;\myyellow{G}\;\mygreen{E}\;\mycyan{B}\;\myblue{R}\;\myviolet{A}\;\mymagenta{I}\;\myred{C}\;\;\;\myorange{A}\;\myyellow{P}\;\mygreen{P}\;\mycyan{R}\;\myblue{O}\;\myviolet{A}\;\mymagenta{C}\;\myred{H}\;\myorange{E}\;\myyellow{S}}
\colorlet{chapter}{red!50}
\chapter{Clifford algebras}\label{chap1}

\hypersetup{
  colorlinks = true,
  linkcolor  = BrickRed,
  citecolor = Red,
}

\paragraph{ } In the first part of this work we aim to move from \textcolor[rgb]{0.82,0.01,0.11}{Clifford algebras} to the \textcolor[rgb]{0.47,0.81,0.1}{classification of spinors in the Minkowski space-time}, while the second part adopts a geometric approach with the goal to define and study a groundwork in which it is possible generalising the classification of spinors in any dimension and signature. For both the first and second parts, this first chapter is fundamental, as we will introduce in detail the theory of Clifford algebras, which is crucial for defining the underlying structures within this algebra in the first part and for the second part, where we will work globally with Clifford structures in the exterior cotangent bundle, specifically the \textcolor[rgb]{0.08,0.18,0.88}{Kähler-Atiyah bundle}. This involves defining a geometric framework within \textcolor[rgb]{0.41,0.67,0.97}{bundle} structures that adhere to the fundamental algebraic relations of the Clifford product. More specifically, in the Kähler-Atiyah bundle, the Clifford product is defined on the underlying sections of the exterior bundle, transforming the fibres into Clifford algebras. Thus, it is important to begin this work by presenting with details the fundamental aspects of Clifford algebras that will be essential for any subsequent part of this dissertation.

 The theory of the Clifford algebras is exposed in this Chapter together with examples. We present its definition, we show the fact that in any quadratic space is possible to define a Clifford algebra, we present operations that can be defined on those algebras based on their multivectorial structure, some important subspaces that can be defined from them and present he the main theorems concerning the Clifford algebras structure that unravel an important behaviour of these algebras: its periodicity. The periodicity is an essential property within Clifford algebras because makes it possible their classification based on representation theory. We conclude this chapter then by presenting the way the classification of Clifford algebras is conducted, taking into account their periodicity and its matrix representation.
 
It is worth to emphasise that Clifford algebra is naturally related to matrix algebra and this representation is quite important for at least the following reasons: a priori, Clifford algebras may seem very abstract, but through their representation by matrix algebra, relevant, concrete properties and behaviours of these algebraic structures can be explicitly observed. Calculations involving matrix algebra can be simpler and more straightforward compared to those on the abstract structure of Clifford algebra in depending on the context. Moreover, the interplay between matrix algebra and various scientific fields in mathematics and physics coupled with their natural relation to Clifford algebras, enhances the richness of applications for Clifford algebras. One of these applications pertains specifically to the study of \textcolor[rgb]{0.88,0.82,0.11}{spinors} as objects that can be defined within the Clifford structure.

To introduce Clifford algebras, the preliminary topics include multilinear algebra, tensor algebra, and exterior algebra. This is because, to define a Clifford algebra, we need a quadratic form on our vector space, a topic covered in Appendix \ref{app1}. Clifford algebras are essentially a tensor algebra quotient by a specific ideal, therefore the tensor algebra is addressed in Appendix \ref{app2}. Additionally, the tensor product will be treated multiple times in various constructions in this Chapter, and it is also defined in the same appendix. Exterior algebra, in turn, plays a crucial role in Clifford algebras, which utilise the exterior product and contraction to define the Clifford product. These operations are defined in Appendix \ref{app3}, along with the detailed development of exterior algebra which is itself is a Clifford algebra and is isomorphic to Clifford algebras for the same quadratic space, possessing the same multivector structure.

\section{Definition}\label{secdefacs}

\paragraph{ } To define a (real) Clifford  algebra, a
{\color{red} \underline{\color{black} vector space}} $V$ over $\rr$ and a {\color{red} \underline{\color{black} quadratic form}} $Q$ defined in $V$ are the basic ingredients. A Clifford algebra is an algebra generated from the vectors of $V$ in such a way that the square of an element of $V$ is related to the quadratic form. Quadratic forms are defined in Def. \ref{def_5t4r8uemc9dwzasex}. In order to define a Clifford algebra we must first characterise what we mean by an algebra (over a field).

\begin{definvermelho}
An \textbf{algebra} over a field $\mathbb{K}$ is a vector space $\mathcal{A}$ over $\mathbb{K}$ endowed with a bilinear product $\mathcal{A} \times \mathcal{A} \to \mathcal{A} $ ,\,$(a,b) \to ab$.
\end{definvermelho}

\noindent In other words, let $V$ be a $\mathbb{K}$-vector space, the pair $\mathcal{A} = (V, *)$ such that $* : \mathcal{A} \times \mathcal{A} \to \mathcal{A} $ is called an algebra over $\mathbb{K}$ if given $a,b,c \in V$ and $\lambda \in \mathbb{K}$ the product $*$ satisfies the following conditions:

\begin{align*}
\begin{aligned}
&\textcolor{BrickRed}{\textbf{(i)}}  &&a * (b + c) = a * b = a * c, \\
&\textcolor{BrickRed}{\textbf{(ii)}}  &&(a + b) * c = a * c + b * c, \\
&\textcolor{BrickRed}{\textbf{(iii)}} &&\lambda (a * b) = (\lambda a) * b = a * (\lambda b).\\
\end{aligned}
\end{align*}
\noindent  These conditions are the bilinearity of the multiplication $*$. Furthermore, if there exists an element $1 \in V$ such that $1*a = a*1 = a$ for all $a \in V$, the algebra is said to be an algebra with unity. Moreover, if $a * b = b * a$ for all $a,b \in V$ one has a commutative algebra and $(a*b) * c = a * (b *c)$ for all $ a,b,c \in V$, defines an associative algebra. The tensor algebra $(T(V), \tensor)$ defined in the Appendix \ref{appb2_tensoralg} and the exterior algebra $(\bigwedge(V), \wedge)$ in the Appendix \ref{sec4_app3} are important examples of algebras. An example of algebra that is studied in introductory modules of a wide range of science courses is the Gibbs-Heaviside algebra within the three-dimensional Euclidean space. This algebra can be used as motivation to present Clifford algebras. From a orthonormal basis $\{\mathbf{e}_{1}, \mathbf{e}_{2},\mathbf{e}_{3}\}$ of the euclidean space $\rr^{3}$ and a bilinear form $B$ defined on it, the square of a vector $\mathbf{v} = v_1\mathbf{e}_{1}+v_2\mathbf{e}_{2}+v_3\mathbf{e}_{3} \in \rr^{3}$ leads to:

\begin{align}
    \begin{aligned}
         \mathbf{v}^{2} &:= B(v_1\mathbf{e}_{1}+v_2\mathbf{e}_{2}+v_3\mathbf{e}_{3},v_1\mathbf{e}_{1}+v_2\mathbf{e}_{2}+v_3\mathbf{e}_{3})\\
        &=
         v_1^{2}
B(\mathbf{e}_1, \mathbf{e}_1) + v_2^{2}
B(\mathbf{e}_2, \mathbf{e}_2) + v_3^{2}
B(\mathbf{e}_3, \mathbf{e}_3) \\&+v_1v_2[B (\mathbf{e}_1, \mathbf{e}_2) + B (\mathbf{e}_2, \mathbf{e}_1)] +v_1v_3[B (\mathbf{e}_1, \mathbf{e}_3) + B (\mathbf{e}_3, \mathbf{e}_1)]\\
&+v_2v_3[B (\mathbf{e}_2, \mathbf{e}_3) + B (\mathbf{e}_3, \mathbf{e}_2)].
    \end{aligned}
\end{align}

\noindent Within an orthonormal basis the result is required to be  $\mathbf{v}^{2} =v_1^{2} + v_2^{2} + v_3^{2}$. Choosing 

\begin{equation}\label{eq_yutrjviuvn48vurm}
    \color{BrickRed} \begin{cases} \color{black}
        B\'(\'e_i, \mathbf{e}_i) = 1 \text{ for } i = 1,2,3\\ \color{black}
        \color{black} B (\mathbf{e}_i, \mathbf{e}_j) = B (\mathbf{e}_j, \mathbf{e}_i) = 0 \text{ for } i \neq j \color{black}
    \end{cases}\color{black},
\end{equation}

\noindent leads to  $\mathbf{v}^{2} = v_1^{2} + v_2^{2} + v_3^{2}$ as required. The product of two vectors $\mathbf{v}, \'u \in \rr^{3}$ is such that

\begin{align}\label{eq_54tu93jewmcx}
    \begin{aligned}
        B(\mathbf{v},\'u) &= B(v_1\mathbf{e}_{1}+v_2\mathbf{e}_{2}+v_3\mathbf{e}_{3},u_1\mathbf{e}_{1}+u_2\mathbf{e}_{2}+u_3\mathbf{e}_{3})\\    
         &= v_1 u_1 B(\mathbf{e}_1, \mathbf{e}_1) + v_1 u_2 B(\mathbf{e}_1, \mathbf{e}_2) + v_1 u_3 B(\mathbf{e}_1, \mathbf{e}_3) \\
&\quad + v_2 u_1 B(\mathbf{e}_2, \mathbf{e}_1) + v_2 u_2 B(\mathbf{e}_2, \mathbf{e}_2) + v_2 u_3 B(\mathbf{e}_2, \mathbf{e}_3) \\
&\quad + v_3 u_1 B(\mathbf{e}_3, \mathbf{e}_1) + v_3 u_2 B(\mathbf{e}_3, \mathbf{e}_2) + v_3 u_3 B(\mathbf{e}_3, \mathbf{e}_3) \\
    \end{aligned}
\end{align}

\noindent The choice in Eq. \eqref{eq_yutrjviuvn48vurm} yields $B(\mathbf{v},\'u) = v_1u_1 + v_2u_2 + v_3u_3.$ which is precisely the standard inner product $\mathbf{v} \cdot \'u$ on $\rr^{3}$. This choice of orthogonality assumes that $B(\mathbf{e}_i, \mathbf{e}_j)$ is a scalar and gives rise to Gibbs-Heaviside algebra. However, this is not the only assumption that can be made. There is no reason that $B(\mathbf{e}_i, \mathbf{e}_j)$ needs to be a scalar, it can be a \textit{bivector} $\'e_i\'e_j$ which is different from a scalar and a vector. This leads to a multivectorial structure of the algebra. The bivector arises from defining $B (\mathbf{e}_i, \mathbf{e}_j) = \'e_i\'e_j$ as juxtaposition and establishing the choice:
\begin{equation}\label{eq_yut85mkjnhbgwqe}
   \color{BrickRed}  \begin{cases}
       \color{black} \'e_i^{2} = 1 \text{ for } i = 1,2,3\\\color{black}
     \color{black}  \mathbf{e}_i\mathbf{e}_j + \mathbf{e}_j\mathbf{e}_i = 0 \text{ for } i \neq j
    \end{cases}\color{black}.\color{black}
\end{equation}

\noindent The geometric intuition behind a vector is its realisation as a oriented line segment, similarly this new multivector object called bivector can be realised as oriented plane fragment as shown bellow

\begin{center}

\tikzset{every picture/.style={line width=0.75pt}} 

\begin{tikzpicture}[x=0.75pt,y=0.75pt,yscale=-1,xscale=1]

\draw  [fill={rgb, 255:red, 208; green, 2; blue, 27 }  ,fill opacity=0.32 ] (139.03,96.74) -- (205.26,136.03) -- (165.97,202.26) -- (99.74,162.97) -- cycle ;
\draw    (99.74,162.97) -- (164.25,201.24) ;
\draw [shift={(165.97,202.26)}, rotate = 210.68] [color={rgb, 255:red, 0; green, 0; blue, 0 }  ][line width=0.75]    (10.93,-3.29) .. controls (6.95,-1.4) and (3.31,-0.3) .. (0,0) .. controls (3.31,0.3) and (6.95,1.4) .. (10.93,3.29)   ;
\draw    (165.97,202.26) -- (204.24,137.75) ;
\draw [shift={(205.26,136.03)}, rotate = 120.68] [color={rgb, 255:red, 0; green, 0; blue, 0 }  ][line width=0.75]    (10.93,-3.29) .. controls (6.95,-1.4) and (3.31,-0.3) .. (0,0) .. controls (3.31,0.3) and (6.95,1.4) .. (10.93,3.29)   ;
\draw    (138.2,156.6) .. controls (129.7,128.1) and (163.5,132.1) .. (170.1,141.1) .. controls (176.4,149.69) and (172.58,159.2) .. (153.41,163.52) ;
\draw [shift={(150.6,164.1)}, rotate = 349.51] [fill={rgb, 255:red, 0; green, 0; blue, 0 }  ][line width=0.08]  [draw opacity=0] (8.93,-4.29) -- (0,0) -- (8.93,4.29) -- cycle    ;
\draw  [fill={rgb, 255:red, 208; green, 2; blue, 27 }  ,fill opacity=0.32 ] (299.53,104.74) -- (365.76,144.03) -- (326.47,210.26) -- (260.24,170.97) -- cycle ;
\draw    (260.24,170.97) -- (324.75,209.24) ;
\draw [shift={(326.47,210.26)}, rotate = 210.68] [color={rgb, 255:red, 0; green, 0; blue, 0 }  ][line width=0.75]    (10.93,-3.29) .. controls (6.95,-1.4) and (3.31,-0.3) .. (0,0) .. controls (3.31,0.3) and (6.95,1.4) .. (10.93,3.29)   ;
\draw    (326.47,210.26) -- (364.74,145.75) ;
\draw [shift={(365.76,144.03)}, rotate = 120.68] [color={rgb, 255:red, 0; green, 0; blue, 0 }  ][line width=0.75]    (10.93,-3.29) .. controls (6.95,-1.4) and (3.31,-0.3) .. (0,0) .. controls (3.31,0.3) and (6.95,1.4) .. (10.93,3.29)   ;
\draw    (297.88,161.33) .. controls (292.86,136.43) and (324.26,140.46) .. (330.6,149.1) .. controls (337.2,158.1) and (332.7,168.1) .. (311.1,172.1) ;
\draw [shift={(298.7,164.6)}, rotate = 253.39] [fill={rgb, 255:red, 0; green, 0; blue, 0 }  ][line width=0.08]  [draw opacity=0] (8.93,-4.29) -- (0,0) -- (8.93,4.29) -- cycle    ;

\draw (150.22,203) node [anchor=north west][inner sep=0.75pt]  [rotate=-24.89]  {$\mathbf{e}_{i}$};
\draw (209.54,139.62) node [anchor=north west][inner sep=0.75pt]  [rotate=-28.32]  {$\mathbf{e}_{j}$};
\draw (139.06,100.69) node [anchor=north west][inner sep=0.75pt]  [rotate=-30.04]  {$\mathbf{e}_{i} \mathbf{e}_{j}$};
\draw (311.29,210) node [anchor=north west][inner sep=0.75pt]  [rotate=-27.93]  {$\mathbf{e}_{i}$};
\draw (369.39,148.98) node [anchor=north west][inner sep=0.75pt]  [rotate=-28.49]  {$\mathbf{e}_{j}$};
\draw (299.56,108.69) node [anchor=north west][inner sep=0.75pt]  [rotate=-30.04]  {$\mathbf{e}_{j} \mathbf{e}_{i}$};

\end{tikzpicture}
\end{center}

Likewise, this new choice leads to $\mathbf{v}^{2} = v_1^{2} + v_2^{2} + v_3^{2}$ as required. However, the Eq. \eqref{eq_54tu93jewmcx} of product of two vectors $\mathbf{v}, \'u \in \rr^{3}$ turns out to be significantly different:
\begin{align}\label{eq_098jnhybgtvfrcd}
    \begin{aligned}
         B(\mathbf{u},\mathbf{v}) = \mathbf{u}\mathbf{v} &= u_1v_1 + u_2v_2 + u_3v_3 + 
(u_2v_3 - u_3v_2)\mathbf{e}_{2}\'e_3 \\&\;\;\;\;+ (u_3v_1 - u_1v_3)\mathbf{e}_{3}\'e_1 + (u_1v_2 - u_2v_1)\mathbf{e}_{1}\'e_2.
    \end{aligned}
\end{align}

\noindent The Euclidean space $\rr^{3}$ with this new product satisfying Eq. \eqref{eq_yut85mkjnhbgwqe} defines a Clifford algebra for the Euclidean space. The advantage of Clifford algebras over the Gibbs-Heaviside are vast, for instance, they are not limited to just the Euclidean space and can be defined in spaces of any dimension and signature. We proceed to the general definition and construction of Clifford algebras.

\begin{definvermelho}\label{def_a15d1s}
Let $V$ be a vector space over $\rr$ equipped with a symmetric bilinear form $g$. Let $\mathcal{A}$ be an associative algebra with unity $1_{\mathcal{A}}$ and let $\gamma$ be a linear mapping $\gamma: V \to \mathcal{A}$. The pair $(\mathcal{A},\gamma)$ is a \textbf{real Clifford algebra} for the quadratic space $(V,g)$ if $\mathcal{A}$ is generated as an algebra by $\{\gamma(\mathbf{v}) \;|\; \mathbf{v} \in V \}$ and $\{x1_{\mathcal{A}} \;|\; x \in \rr  \}$, and $\gamma$ 
satisfies for all $\mathbf{v},\mathbf{u} \in V$ the relation: 

\begin{equation}\label{eq_clif1}
    \gamma(\mathbf{v}) \gamma(\mathbf{u}) + \gamma (\mathbf{u}) \gamma (\mathbf{v}) = 2g(\mathbf{u},\mathbf{v})1_{\mathcal{A}}.
\end{equation}

\end{definvermelho}

The Eq. \eqref{eq_clif1} can be written as

{\colorlet{shadecolor}{Red!10}\begin{shaded} \begin{align}\label{eq_clifrelideal}
    \begin{aligned}
    \gamma(\mathbf{v})^{2} = Q(\mathbf{v})1_{\mathcal{A}} = g(\mathbf{v},\mathbf{v})1_{\mathcal{A}},
    \end{aligned}
\end{align}\end{shaded}}

\noindent for all $\mathbf{v} \in V$.  The mapping $\gamma$ is called \textit{Clifford mapping} and as Eq. \eqref{eq_clifrelideal} shows, it is a kind of square root of the quadratic form  $Q(\mathbf{v}) = g (\mathbf{v},\mathbf{v})$. 

\hspace{0.5cm} Fundamental properties of Clifford algebras can be derived by exploring how the relation in Eq. \eqref{eq_clif1} behaves with respect to an orthonormal basis $\{\mathbf{e}_{1},\ldots, \mathbf{e}_{n}\}$ of $V$. For a Clifford algebra $(\mathcal{A},\gamma)$ for $(V,g)$ one has

\begin{align}
\begin{aligned}\label{eq_apclborto}
   \color{BrickRed} \begin{cases} \color{black}\gamma(\mathbf{e}_{i})\gamma(\mathbf{e}_{j}) + \gamma(\mathbf{e}_{j})\gamma(\mathbf{e}_{i}) =0,\color{black} \text{ for } i \neq j \\\color{black}
\color{black}\gamma(\mathbf{e}_{i})^{2} \color{black} = \color{black}g(\mathbf{e}_{i},\mathbf{e}_{i})1_{\mathcal{A}} \text{ for } i=1,\ldots,n\color{black}
\end{cases}\color{black}.
\end{aligned}
\end{align}

\noindent This is then the general version of the \textit{choice} in the Eq. \eqref{eq_yut85mkjnhbgwqe}. By using the relations \eqref{eq_apclborto}, any product involving $\gamma(\mathbf{e}_i)$ and their powers can be reordered to yield \cite{Roc16}

\begin{equation}
   \gamma(\mathbf{e}_{1})^{\mu_{1}}\gamma(\mathbf{e}_{2})^{\mu_{2}}\cdots\gamma(\mathbf{e}_{n})^{\mu_{n}}, \;\;\;\; \mu_{i} = 0,1,\;\; (i=1,\ldots,n)
\end{equation}

\noindent and since $\mathcal{A}$ is generated by $\{\gamma(\mathbf{v}) \;|\; \mathbf{v} \in V \}$ and $\{x1_{\mathcal{A}} \;|\; x \in \rr  \}$, it follows that it is generated by the products

\begin{equation}
    \mathcal{A} = \text{span} \{\gamma(\mathbf{e}_{1})^{\mu_{1}}\gamma(\mathbf{e}_{2})^{\mu_{2}}\cdots\gamma(\mathbf{e}_{n})^{\mu_{n}} \;\;|\;\; \mu_{i} = 0,1 \},
\end{equation}

\noindent such that the identity is denoted by $\gamma(\mathbf{e}_{1})^{0}\gamma(\mathbf{e}_{2})^{0}\cdots\gamma(\mathbf{e}_{n})^{0} = 1_{\mathcal{A}}$. One can notice that the maximum number of elements of type $\gamma(\mathbf{e}_{1})^{\mu_{1}}\gamma(\mathbf{e}_{2})^{\mu_{2}}\cdots\gamma(\mathbf{e}_{n})^{\mu_{n}}$ with $\mu_{i} = 0,1 $ is  $2^n$. As a result, the maximal dimension of a Clifford algebra is $2^n$. Such algebras are said to be a \textit{universal Clifford algebra} \cite{Roc16}.

\begin{definvermelho}\label{def_clifuniv}
A Clifford algebra $(\mathcal{A},\gamma)$ for the quadratic space $(V,g)$ is said to be an \textbf{universal Clifford algebra} if for each Clifford algebra $(\mathcal{B},\rho)$ for $(V,g)$ there exists a unique homomorphism $\phi: \mathcal{A} \to \mathcal{B}$ such that $\rho = \phi \circ \gamma$. In these conditions, a universal Clifford algebra for $(V,g)$ is denoted by $\cl(V,g)$.
\end{definvermelho}

\begin{tftheorem}\label{teo2cliffordalgdi}
Let  $(\mathcal{A},\gamma)$ be a Clifford algebra for the quadratic space $(V,g)$ with $\dim V = n$. If dim $\mathcal{A} = 2^n$ then $(\mathcal{A},\gamma)$ is a universal Clifford algebra for $(V,g)$.
\end{tftheorem}
\noindent \textcolor{BrickRed}{\textit{Proof.}} Consider $\{\mathbf{e}_{1},\ldots, \mathbf{e}_{n}\}$x an orthonormal basis of $V$. Since  $(\mathcal{A},\gamma)$ is a Clifford algebra the relation \eqref{eq_apclborto} holds. Moreover, from the hypothesis we have that $[$dim $\mathcal{A} = 2^n]$ therefore the set $\{\gamma(\mathbf{e}_{1})^{\mu_{1}}\gamma(\mathbf{e}_{2})^{\mu_{2}}\cdots\gamma(\mathbf{e}_{n})^{\mu_{n}}$ $|\;\mu_{i} = 0,1 \}$ generates and is a basis for $\mathcal{A}$. Now, let $(\mathcal{B},\rho)$ be an arbitrary Clifford algebra for  $(V,g)$. It holds the same relation \eqref{eq_apclborto} for $\mathcal{B}$ thence the set $\{\rho(\mathbf{e}_{1})^{\mu_{1}}\rho(\mathbf{e}_{2})^{\mu_{2}}\cdots\rho(\mathbf{e}_{n})^{\mu_{n}}$ $|\;\mu_{i} = 0,1 \}$ generates $\mathcal{B}$. Define the linear mapping $\phi: \mathcal{A} \to \mathcal{B}$ such that

\begin{equation}
    \phi (\gamma(\mathbf{e}_{1})^{\mu_{1}}\gamma(\mathbf{e}_{2})^{\mu_{2}}\cdots\gamma(\mathbf{e}_{n})^{\mu_{n}}) = \rho(\mathbf{e}_{1})^{\mu_{1}}\rho(\mathbf{e}_{2})^{\mu_{2}}\cdots\rho(\mathbf{e}_{n})^{\mu_{n}}.
\end{equation}

\noindent That way defined, $\phi$ is an algebra isomorphism satisfying

\begin{equation}
    \phi (\gamma(\mathbf{e}_{j}))  \phi (\gamma(\mathbf{e}_{i})) +   \phi (\gamma(\mathbf{e}_{i}))  \phi (\gamma(\mathbf{e}_{j})) = 2g(e_i,e_j)1_{\mathcal{B}}.
\end{equation}

\noindent Thus, by the Definition \ref{def_clifuniv}, $(\mathcal{A},\gamma)$ is a universal Clifford algebra $\cl(V,g)$. \textcolor{BrickRed}{$\Box$}

 \paragraph{ } The universal Clifford algebra $\cl(V,g)$, if exists, it is unique up to a unique homomorphism. In other words, the above Definition \ref{def_clifuniv} state that for each linear application $\rho: V \to \mathcal{B}$ there is a unique algebra homomorphism $\phi: \mathcal{A} \to \mathcal{B}$ making the following diagram commute

\begin{equation*}
\begin{tikzcd}
V \arrow[black]{d}[black,swap]{\rho} \rar{\gamma} & [15pt] \mathcal{A} \arrow[BrickRed, dashed]{dl}[black]{\phi} \\ [20pt] \mathcal{B}
\end{tikzcd} 
\end{equation*}{}

Thereby, if a universal Clifford algebra $\cl(V,g)$ exists, by the uniqueness of the homomorphism, $\cl(V,g)$ will be unique. An important question rises: \textcolor{BrickRed}{\textit{for any quadratic space $(V,g)$ given, is there a universal Clifford algebra $\cl(V,g)$ associated to this space?}} The answer is affirmative and is given by the following theorem.

\begin{tftheorem}\label{teocliff1}
For all quadratic space $(V,g)$ there exists a universal Clifford algebra. 
\end{tftheorem}
\noindent \textcolor{BrickRed}{\textit{Proof.}} Let $(V,g)$ be a quadratic space and $T(V)$ the algebra of the contravariant tensors. In order to construct a Clifford algebra as a quotient of the tensor algebra by a two-sided ideal, is required that the elements of our Clifford algebra obey the fundamental relation \eqref{eq_clifrelideal}, let us consider then the ideal $\mathcal{I}$ of $T(V)$ generated by elements of type

\begin{equation}
    \mathbf{v}  \otimes \mathbf{v} - Q(\mathbf{v})1_{T}
\end{equation}

\noindent where $ Q(\mathbf{v}) = g(\mathbf{v},\mathbf{v})$ and $1_{T(V)}$ is the identity of the tensor algebra. That way, the two-sided ideal $\mathcal{I}$ consists of all sums

\begin{equation}
    \sum_{i} A_i \otimes ( \mathbf{v} \otimes \mathbf{v} - Q(\mathbf{v})1_{T}) \otimes B_i
\end{equation}

\noindent where $A_i, B_i \in T(V)$. However, one can also realise that the ideal $\mathcal{I}$ is generated equivalently by the elements 

\begin{equation}
    \mathbf{v}  \otimes \'u  + \'u \otimes \mathbf{v} - 2g(\mathbf{v},\'u)1_{T}.
\end{equation}

\noindent Consider now the quotient map $\pi : T(V) \to T(V)/ \mathcal{I}$ and the inclusion $\iota : V \hookrightarrow T(V)$ then one can define $\gamma : V \to T(V)/ \mathcal{I}$ such that the diagram below commutes.

\begin{equation*}
\begin{tikzcd}
V \arrow[BrickRed]{d}[black,swap]{\gamma}   \rar[hookrightarrow]{\iota} & T(V) \arrow[black]{dl}[black]{\pi}  \\ [20pt] T(V)/ \mathcal{I}
\end{tikzcd} 
\end{equation*}{}

\noindent In fact, let us prove that $\gamma = \pi \circ \iota$ is a Clifford mapping. Consider the equivalence relation

\begin{equation}
    A \sim B \iff A = B + x, \;\;\;\; x \in \mathcal{I}.
\end{equation}

\noindent Let $\'u,\mathbf{v} \in V$ and notice that

\begin{align}
    \begin{aligned}
    \gamma(\mathbf{v})\gamma(\'u) &= \pi(\iota(\mathbf{v})\pi(\iota(\'u))
    = \pi(\iota(\mathbf{v})\otimes(\iota(\'u)) = [ \mathbf{v} \otimes \'u].
    \end{aligned}
\end{align}

\noindent Such that $[ \mathbf{v} \otimes \'u]$ is the equivalence class of $\mathbf{v} \otimes \'u$. In addition, we have that $ \mathbf{v} \otimes \'u$ can be expressed as

\begin{align}
    \begin{aligned}
 \mathbf{v} \otimes \'u &= \frac{1}{2}(  \mathbf{v} \otimes \'u -  \'u \otimes \mathbf{v}) + g(\mathbf{v},\'u)1_{T} + \frac{1}{2}(  \mathbf{v} \otimes \'u + \'u \otimes \mathbf{v}) - g(\mathbf{v},\'u)1_{T} \\
&=  \frac{1}{2}(  \mathbf{v} \otimes \'u -  \'u \otimes \mathbf{v}) + g(\mathbf{v},\'u)1_{T} + \frac{1}{2}[  (\mathbf{v} + \'u) \otimes (\mathbf{v}+ \'u) - g(\mathbf{v} + \mathbf{v},\mathbf{v} + \'u)1_{T} \\ &\;\;\;\;- \mathbf{v} \otimes \mathbf{v} + g(\mathbf{v},\mathbf{v})1_{T} - \'u \otimes \'u + g(\'u,\'u)1_{T}].
    \end{aligned}
\end{align}

\noindent As we can see, the term in the brackets is an element of the ideal $\mathcal{I}$, therefore it yields

\begin{equation}
    \mathbf{v} \otimes \'u \sim \frac{1}{2}(  \mathbf{v} \otimes \'u -  \'u \otimes \mathbf{v}) + g(\mathbf{v},\'u)1_{T} 
\end{equation}

\noindent or equivalently

\begin{equation}\label{eq_tensclaseqcl}
    \mathbf{v} \otimes \'u \sim \mathbf{v} \wedge \'u + g(\mathbf{v},\'u)1_{T}. 
\end{equation}

\noindent Hence,
\begin{align}
    \begin{aligned}
    \gamma(\mathbf{v})\gamma(\'u) + \gamma(\'u)\gamma(\mathbf{v}) &=  [\mathbf{v} \otimes \'u] + [\'u \otimes \mathbf{v}]  \\ &= [\mathbf{v} \wedge \'u + g(\mathbf{v},\'u)1_{T}] + [\'u \wedge \mathbf{v} + g(\'u,\mathbf{v})1_{T}]  \\
    &= 2 g(\mathbf{v},\'u)1_{T},
    \end{aligned}
\end{align}

\noindent which means that $\gamma: V \to T(V)/ \mathcal{I}$ is a Clifford mapping. Therefore, $T(V)/ \mathcal{I}$ is a Clifford algebra for $(V,g)$. Let us prove now the universality condition establish by the Definition \ref{def_clifuniv} for $T(V)/ \mathcal{I}$. Suppose that there is another Clifford algebra $(\mathcal{B},\rho)$ for the same quadratic space $(V,g)$. For $(\mathcal{B}, \rho)$ the function $\rho: V \to \mathcal{B}$ such that $\rho(\mathbf{v})^2 = Q(\mathbf{v})$ is considered. Since the vector space of the multilinear mappings $\mathcal{L}(T_k(V), \mathcal{B}) \simeq \mathcal{L}_{(k)}(V, \ldots , V; \mathcal{B})$ this mapping $\rho$ can be extended to $T(V)$ as the linear map $\rho' : T(V) \to \mathcal{B}$ given by $\rho = \rho' \circ \iota$

\begin{equation*}
\begin{tikzcd}
V \arrow[black]{d}[black,swap]{\rho}   \rar[hookrightarrow]{\iota} & T(V) \arrow[BrickRed, dashed]{dl}[black]{\rho'}  \\ [20pt] \mathcal{B}
\end{tikzcd} 
\end{equation*}{} 

\noindent such that 

\begin{equation}
    \rho' (\mathbf{v}_1 \tensor \cdots \tensor \mathbf{v}_k) =  (\rho' \circ \iota) (\mathbf{v}_1) \cdots (\rho' \circ \iota) (\mathbf{v}_k) = \rho (\mathbf{v}_1) \cdots \rho (\mathbf{v}_k).
\end{equation}

\noindent Since $\rho(\mathbf{v})^2 = Q(\mathbf{v})$, we have 

\begin{align}
    \begin{aligned}
        \rho'(\mathbf{v}  \otimes \mathbf{v} - Q(\mathbf{v})1_{T}) = 0.
    \end{aligned}
\end{align}

\noindent Moreover, we also have that since elements of that type $\mathbf{v}  \otimes \mathbf{v} - Q(\mathbf{v})1_{T}$ are the generators of the two-sided ideal $\mathcal{I}$ it follows that $\mathcal{I} \subset \ker\rho'$. Let us consider then $T(V)/\ker\rho'$. In this case there exists a map $\phi : T(V)/\ker\rho' \to \mathcal{B}$ such that for all $ x \in T(V)$

\begin{equation}
    \phi ([x]) = \rho' (x),
\end{equation}

\noindent  which is an homomorphism

\begin{equation}
    \phi ([x] [y]) = \phi ([x \tensor y] = \rho'(x \tensor y) = \rho' (x) \rho' (y) =  \phi ([x]) \phi ( [y]). 
\end{equation}

\noindent On the other hand, we have that if $U_i$ is a subspace of a vector space $W_i \, (i =1,2)$, and if $f: W_1 \to W_2$ is a linear map such that $f(U_1) \subset U_2$, then $f$ induces a linear map $f' : W_1/U_1 \to W_2/U_2$. Moreover, when $f$ is surjective, then $f'$ is surjective too. Therefore, the surjective homomorphism $T(V)/\mathcal{I} \to T(V)/\ker \rho'$ shows that dim $T(V)/\mathcal{I} \geq$ dim $T(V)/\ker\rho'$. The homomorphism $T(V)/\mathcal{I} \to \mathcal{B}$ follows from $T(V)/\mathcal{I} \to T(V)/\ker \rho' \to \mathcal{B}$, say $\phi'$, then

\begin{equation*}
\begin{tikzcd}
V \arrow[black]{d}[black,swap]{\rho}  \arrow[bend left]{rr}[black]{\gamma} \rar[hookrightarrow]{\iota} & T(V) \arrow[BrickRed, dashed,swap]{dl}[black]{\rho'} \rar{\pi} & {T(V)/\mathcal{I}} \arrow[Red, dashed]{dll}[black]{\phi'} \\ [20pt] \mathcal{B}
\end{tikzcd} 
\end{equation*}{}

\noindent Hence,

\begin{align}
    \begin{aligned}
    \phi' \circ \pi  &= \rho' \\
    \phi' \circ (\pi \circ \iota)  &= \rho' \circ \iota \\
    \phi' \circ \gamma  &= \rho
    \end{aligned}
\end{align}

\noindent as desired. It is straightforward to show the uniqueness of $\phi'$ since if there were $\phi'_1 \circ \gamma  = \rho$ and $\phi'_2 \circ \gamma  = \rho$ they would coincide on the vector space $V$ the generator of $T(V)$ and thereby on $T(V)/\mathcal{I}$ then $\phi_1,\phi_2$ would be the same. Finally, we conclude that $T(V)/\mathcal{I}$ is the universal Clifford algebra for the quadratic space $(V,g)$, i.e.,  $T(V)/\mathcal{I} = \cl(V,g)$. \textcolor{BrickRed}{$\Box$}

\begin{rem}
    The homomorphism between the Clifford algebras $\phi':  \cl(V,g) = T(V)/\mathcal{I} \to \mathcal{B}$ defined at the end of the proof of Theorem \ref{teocliff1} is injective and even bijective if $\mathcal{B}$ is generated by $\rho(V)$, then in this case, they are isomorphic.
\end{rem}

\paragraph{ } The explicit construction of the universal Clifford algebra is given as the quotient of the tensor algebra of $V$ by a specific two-sided ideal. This is shown with details in the Appendix \ref{app4}. This construction is made in a similar way as it is done for the exterior algebra in the Appendix \ref{app3}. For simplicity the word \textit{universal} will be suppressed henceforward. Now since for any quadratic space is possible to define a Clifford algebra, let $g$ be a symmetric bilinear form in $\rr^{n}$ of signature $(p,q)$ where $p + q = n$ and let $\{\mathbf{e}_1, \ldots, \mathbf{e}_n \}$ be the orthonormal basis of such vector space. The symmetric bilinear form can be evaluated on a vector $\mathbf{v} = v^{i}\mathbf{e}_i \in \rr^{n}$ and yields 

\begin{equation}
    g(\mathbf{v},\mathbf{v}) = (v^{1})^{2} + \cdots +(v^{p})^{2} - (v^{p + 1})^{2} - \cdots -(v^{p+q})^{2}.
\end{equation}

\noindent This quadratic space is denoted by $\rr^{p,q}$. As a consequence of the Theorem \ref{teocliff1}, the corresponding Clifford algebra of $\rrpq$ exists and it is denoted by $\cl_{p,q}$:

{\colorlet{shadecolor}{red!15}\begin{shaded} $$\cl_{p,q} = \cl (\rr^{p,q}).$$ \end{shaded}}

There is a result that follows from the previous Theorem \ref{teocliff1} that gives us the expression for the Clifford product on Clifford algebras \cite{Roc16}.

\begin{tfcor}\label{cor_235qwsad}
For $\mathbf{v} \in V$, $A_{[p]} \in \bigwedge_p(V)$, if $[\mathbf{v}], [A_{[p]}] \in T(V)/\mathcal{I} = \cl(V,g)$ (where $T(V)/\mathcal{I}$ is defined in the proof of Theorem \ref{teocliff1}) then the Clifford product, denoted by juxtaposition, is given by

\begin{equation}\label{eq_345esaghnb}
    \mathbf{v} A_{[p]} = \mathbf{v}\wedge A_{[p]} + \mathbf{v}_{\flat} \rfloor A_{[p]}
\end{equation}

\noindent where $\wedge$ is the exterior product (Def. \ref{definprod_ex}), $\rfloor$ is the left contraction (Def. \ref{def_fd65g415sad}) and $\mathbf{v}_{\flat}$ is such that $\flat$ is the musical isomorphism (Def. \ref{defin_w23485rte}).
\end{tfcor}

\noindent \textcolor{BrickRed}{\textit{Proof.}} It yields from Eq. \eqref{eq_tensclaseqcl} that the product of vectors $\mathbf{v}, \'u$ in the quotient algebra $T(V)/\mathcal{I} = \cl(V,g)$, without the bracket notation, is given by

\begin{equation}\label{eqq_tensclr3434}
    \mathbf{v} \'u = \mathbf{v} \wedge \'u + g (\mathbf{v},\'u).
\end{equation}

\noindent The next step is to generalise the above expression \eqref{eqq_tensclr3434} considering $\mathbf{v}, \'u,\'w \in V$. The exterior product $\mathbf{v} \wedge \'u \wedge \'w$ can be written as 

\begin{equation}\label{eqq_clifwedge3}
    \mathbf{v} \wedge \'u \wedge \'w = \frac{1}{3}(\mathbf{v} \tensor (\'u \wedge \'w) - \'u \tensor (\mathbf{v} \wedge \'w) + \'w \tensor (\mathbf{v} \wedge \'u)).
\end{equation}

\noindent By Eq. \eqref{eq_tensclaseqcl} we have that

\begin{align}
    \begin{aligned}
     \mathbf{v} \otimes \'u &\sim \mathbf{v} \wedge \'u + g(\mathbf{v},\'u)\\
     \'w \otimes (\mathbf{v} \otimes \'u ) &\sim \'w \otimes (\mathbf{v} \wedge \'u + g(\mathbf{v},\'u)) \\
     \'w \otimes \mathbf{v} \otimes \'u  &\sim \'w \otimes (\mathbf{v} \wedge \'u) + g(\mathbf{v},\'u)\'w
    \\ \implies \'w \otimes (\mathbf{v} \wedge \'u)  &\sim  \'w \otimes \mathbf{v} \otimes \'u  -  g(\mathbf{v},\'u)\'w.
    \end{aligned}
\end{align}

\noindent Hence, we have the following relations

\begin{align}
    \begin{aligned}\label{eq_tensclas00}
    \'u \otimes (\mathbf{v} \wedge \'w)  &\sim  \'u \otimes \mathbf{v} \otimes \'w  -  g(\mathbf{v},\'w)\'u, \\
    \'w \otimes (\mathbf{v} \wedge \'u)  &\sim  \'w \otimes \mathbf{v} \otimes \'u  -  g(\mathbf{v},\'u)\'w.
    \end{aligned}
\end{align}

\noindent Since $\wedge$ is antisymmetric it follows from Eq. \eqref{eq_tensclaseqcl}
that 
\begin{align}\label{eqq_tensclas1234}
    \begin{aligned}
   \'u \tensor \mathbf{v} + \mathbf{v} \tensor \'u &\sim 2g(\'u,\mathbf{v}) \implies (\'u \tensor \mathbf{v}) \tensor \'w + (\mathbf{v} \tensor \'u) \tensor \'w \sim 2g(\'u,\mathbf{v})\'w, \\
   \'w \tensor \mathbf{v} + \mathbf{v} \tensor \'w &\sim 2g(\'w,\mathbf{v}) \implies (\'w \tensor \mathbf{v}) \tensor \'u  + (\mathbf{v} \tensor \'w) \tensor \'u  \sim 2g(\'w,\mathbf{v})\'u.
    \end{aligned}
\end{align}

\noindent By using the above relations \eqref{eqq_tensclas1234} in Eq. \eqref{eq_tensclas00} it yields

\begin{align}\label{eqq_tensclas1234000}
    \begin{aligned}
\'u \tensor (\mathbf{v} \wedge \'w) &\sim - \mathbf{v} \tensor \'u \tensor \'w + 2g(\'u,\mathbf{v})\'w - g(\mathbf{v}, \'w)\'u, \\
\'w \tensor (\mathbf{v} \wedge \'u) &\sim - \mathbf{v} \tensor \'w \tensor \'u  + 2g(\'w,\mathbf{v})\'u - g(\mathbf{v}, \'u)\'w.
    \end{aligned}
\end{align}

\noindent Using the relations written in Eq. \eqref{eq_tensclas00} we get

\begin{align}\label{eq_tensclass99}
    \begin{aligned}
        \mathbf{v} \tensor \'u \tensor \'w &\sim \mathbf{v} \tensor (\'u \wedge \'w) + g(\'u,\'w)\'w, \\
        -\mathbf{v} \tensor \'w \tensor \'u &\sim - \mathbf{v} \tensor (\'w \wedge \'u) - g(\'w, \'u)\mathbf{v} =  \mathbf{v} \tensor (\'u \wedge \'w) - g(\'w, \'u)\mathbf{v}.
    \end{aligned}
\end{align}

\noindent Substituting the equations \eqref{eqq_tensclas1234000} and \eqref{eq_tensclass99} in Eq. \eqref{eqq_clifwedge3} we obtain  

\begin{align}
    \begin{aligned}
    \mathbf{v} \wedge \'u \wedge \'w &= \frac{1}{3}(\mathbf{v} \tensor (\'u \wedge \'w) - \'u \tensor (\mathbf{v} \wedge \'w) + \'w \tensor (\mathbf{v} \wedge \'u))\\
    &\sim \frac{1}{3}(\mathbf{v} \tensor (\'u \wedge \'w)
    + \mathbf{v} \tensor \'u \tensor \'w - 2g(\'u,\mathbf{v})\'w + g(\mathbf{v}, \'w)\'u \\
    &\;\;\;\; - \mathbf{v} \tensor \'w \tensor \'u  + 2g(\'w,\mathbf{v})\'u - g(\mathbf{v}, \'u)\'w \\
    &\sim \frac{1}{3}(\mathbf{v} \tensor (\'u \wedge \'w)  
    + \mathbf{v} \tensor (\'u \wedge \'w) + g(\'u,\'w)\'w + \mathbf{v} \tensor (\'u \wedge \'w) - g(\'w, \'u)\mathbf{v} \\
    &\;\;\;\;+ 3g(\mathbf{v}, \'w)\'u -3g(\'u,\mathbf{v})\'w  \\
    &\sim  \mathbf{v} \tensor (\'u \wedge \'w)  + g(\mathbf{v}, \'w)\'u  -g(\mathbf{v},\'u)\'w.
    \end{aligned}
\end{align}

Recalling the musical isomorphism flat $\flat$ and the contraction expression for bivectors stated in Eq. \eqref{eq_bivectextcontr32}, we have that

\begin{align}
\begin{aligned}
  g(\mathbf{v},\'u)\'w - g(\mathbf{v}, \'w)\'u  = - \mathbf{v}_{\flat}(\'u)\'w - \mathbf{v}_{\flat}(\'w)\'u 
    =\mathbf{v}_{\flat} \rfloor (\'u \wedge \'w),
    \end{aligned}
\end{align}

\noindent which implies that the product of a vector and a bivector on $\cl(V,g)$ can be written as

\begin{equation}
    \mathbf{v} (\'u \wedge \'w) = \mathbf{v} \wedge \'u \wedge \'w + \mathbf{v}_{\flat}\rfloor (\'u \wedge \'w).
\end{equation}

\noindent By induction, one may generalise the Clifford product of a vector $\mathbf{v}$ and a p-vector $A_{[p]}$ as \cite{Roc16}

\begin{equation}\label{eqq_prodclifprofteo}
    \mathbf{v} A_{[p]} = \mathbf{v} \wedge A_{[p]} + \mathbf{v}_{\flat} \rfloor A_{[p]},
\end{equation}

\noindent which proves the corollary. \textcolor{BrickRed}{$\Box$}

\subsection*{EXAMPLES}

\begin{ex} \label{ex_acs1comp} {{\textcolor{BrickRed}{{$\blacktriangleright$\;}}}}  $\mathbb{C}$. \normalfont Let $V$ be a 1-dimensional vector space with basis $\{\mathbf{e}\}$, i.e., for all $ \mathbf{v} \in V, \mathbf{v} = y\mathbf{e}$. Consider $g$  the symmetric bilinear functional such that $g(\mathbf{e},\mathbf{e}) = -1$, then, for all $\mathbf{v} \in V, g(\mathbf{v},\mathbf{v}) = -y^2$.  Consider now the subalgebra $\mathcal{A}$ of the matrix algebra $\text{Mat}(2, \mathbb{R})$ defined by
 
 \begin{equation}
    \mathcal{A} = \left\{ \begin{pmatrix} 
    x & y \\
    -y & x 
    \end{pmatrix} \, ; \, x,y \in \rr \right\}.
 \end{equation}

\noindent  $\mathcal{A}$ is generated by
\begin{align}
\begin{aligned}
 \left\{ y\begin{pmatrix} 
    0 & 1 \\
    -1 & 0 
    \end{pmatrix} \, ; \, y \in \rr \right\} &&\text{and}&&  \left\{x \begin{pmatrix} 
    1 & 0 \\
    0 & 1
    \end{pmatrix} = x1_{\mathcal{A}} \, ; \, x \in \rr \right\}.
\end{aligned}
\end{align}

\noindent Define  $\gamma : V \to \mathcal{A}$ as

\begin{equation}
    \gamma(\mathbf{e}) = \begin{pmatrix} 
    0 & 1 \\
    -1 & 0 
    \end{pmatrix} \implies \gamma(\mathbf{v}) = \begin{pmatrix} 
    0 & y \\
    -y & 0 
    \end{pmatrix},
\end{equation}

\noindent thus,

\begin{equation}
\gamma(\mathbf{v})^{2} = \begin{pmatrix} 
    0 & y \\
    -y & 0 
    \end{pmatrix}\begin{pmatrix} 
    0 & y \\
    -y & 0 
    \end{pmatrix} = -y^{2}\begin{pmatrix} 
    1 & 0 \\
    0 & 1
    \end{pmatrix} = g(\mathbf{v},\mathbf{v})1_{\mathcal{A}}.
\end{equation}

\noindent That is, $\gamma$ is a Clifford mapping. Considering the following isomorphism

\begin{align}
\begin{aligned}
\mathbb{C} \leftrightarrow  \text{Mat}(2, \mathbb{R}), && x + iy \leftrightarrow \begin{pmatrix} 
    x & y \\
    -y & x 
    \end{pmatrix}.
\end{aligned}
\end{align}

\noindent We conclude that $\mathcal{A} \simeq \mathbb{C}$, therefore $\mathbb{C}$ is an example of a Clifford algebra.\;$\textcolor{BrickRed}{{\blacktriangleleft}}$ \end{ex}

\begin{ex}\label{ex1matripauli}  \normalfont {{\textcolor{BrickRed}{{$\blacktriangleright$\;}}}} $\text{Mat}(2,\cc)$. Take $V = \rt$, the canonical basis $\{\mathbf{e}_1,\mathbf{e}_2,\mathbf{e}_3\}$ and $g(\mathbf{v},\mathbf{u}) = \left \langle \mathbf{v},\mathbf{u} \right \rangle$ the usual scalar product in $\rt$. Consider $\mathcal{A}= \matcomp$ the algebra of $2 \times 2$ complex matrices

\begin{equation}
    \mathcal{A} =  \left\{ \begin{pmatrix} 
    z_1 & z_3 \\
    z_2 & z_4 
    \end{pmatrix} \, ; \, z_{i} \in \cc \right\}.
\end{equation}

\noindent Consider now the Pauli matrices $\sigma_{1}, \sigma_{2}, \sigma_{3}$

\begin{align}
\begin{aligned}\label{eq_mpauli1}
\sigma_{1} = \begin{pmatrix} 
    0 & 1 \\
    1 & 0 
    \end{pmatrix}, \;\;\;\;\;\; \sigma_{2} = \begin{pmatrix} 
    0 & -i \\
    i & 0 
    \end{pmatrix}, \;\;\;\;\;\; \sigma_{3} = \begin{pmatrix} 
    1 & 0 \\
    0 & -1 
    \end{pmatrix},
\end{aligned}
\end{align}

\noindent The algebra $\mathcal{A}$ is generated by $\{x_1\sigma_1, x_2\sigma_2, x_3\sigma_3 \; | \; x_{i} \in \cc \}$ and $\{x_0 I \;|\; x_0 \in \cc \}$. Define the mapping $\gamma : V \to \mathcal{A}$ as 

\begin{equation}
\gamma(\mathbf{e}_1) = \sigma_1,  \gamma(\mathbf{e}_2) = \sigma_2, \gamma(\mathbf{e}_3) = \sigma_3.\end{equation}
\noindent As a result, for the vectors $\mathbf{v} = v_1\mathbf{e}_1 +  v_2\mathbf{e}_2 + v_3\mathbf{e}_3$ and $ \mathbf{u} = u_1\mathbf{e}_1 + u_2\mathbf{e}_2 + u_3\mathbf{e}_3  \in \mathbb{R}^3$,

\begin{align}
\begin{aligned}
\gamma(\mathbf{v}) &= \gamma(v_1\mathbf{e}_1 +  v_2\mathbf{e}_2 + v_3\mathbf{e}_3) = v_1\sigma_1 +  v_2\sigma_2 + v_3\sigma_3 \\&=   \begin{pmatrix}
    v_3 & v_1 - iv_2 \\
    v_1 + iv_2 & -v_3 
    \end{pmatrix} \\
\implies &\gamma(\mathbf{v}) \gamma(\mathbf{u})  + \gamma(\mathbf{u}) \gamma(\mathbf{v}) = 2 ( v_1u_1 + v_2u_2 + v_3u_3) \begin{pmatrix}
    1 & 0 \\
    0 & 1
    \end{pmatrix} = 2g(\mathbf{v},\mathbf{u})1_{\mathcal{A}}.
\end{aligned}
\end{align}

\noindent Therefore, $\gamma$ is a Clifford mapping and $\mathcal{A} = \matcomp$ is a Clifford algebra.\;$\textcolor{BrickRed}{{\blacktriangleleft}}$ \end{ex}

\begin{ex}\label{ex2cl3}  {{\textcolor{BrickRed}{{$\blacktriangleright$\;}}}} $\cl_{3}$. \normalfont Consider $V = \rt$ endowed with the usual scalar product again. The universal Clifford algebra of such quadratic space is denoted by $\cl_{3}$ and is generated by $\{1, \mathbf{e}_1,\mathbf{e}_2,\mathbf{e}_3\}$, with dimension $2^{3} = 8$ such that

 \begin{align}
     \begin{aligned}
     \clt &= \text{span} \{\gamma(\mathbf{e}_{1})^{\mu_{1}}\gamma(\mathbf{e}_{2})^{\mu_{2}}\gamma(\mathbf{e}_{3})^{\mu_{3}} \;\;|\;\; \mu_{i} = 0,1 \}\\&=  \text{span}\{1, \mathbf{e}_{1}, \mathbf{e}_{2},\mathbf{e}_{3},\mathbf{e}_{1}\mathbf{e}_{2},\mathbf{e}_{1}\mathbf{e}_{3},\mathbf{e}_{2}\mathbf{e}_{3},\mathbf{e}_{1}\mathbf{e}_{2}\mathbf{e}_{3}\}
     \end{aligned}
 \end{align}
 
\noindent and satisfies $
 \mathbf{e_i}\mathbf{e_j} + \mathbf{e_j}\mathbf{e_i} = 2\delta_{ij}1,$ such that $\delta_{ij}$ is the Kronecker delta. The Clifford product between two vectors $\mathbf{u} = u_1\mathbf{e}_1 + u_2\mathbf{e}_2 + u_3\mathbf{e}_3$ and $\mathbf{v} = v_1\mathbf{e}_1 +  v_2\mathbf{e}_2 + v_3\mathbf{e}_3$ is given by

\begin{align}\label{eq_produtodecl3}
\begin{aligned}
\mathbf{u}\mathbf{v} &= u_1v_1 + u_2v_2 + u_3v_3 + 
(u_2v_3 - u_3v_2)\mathbf{e}_{23} \\ &\;\;\;\;+ (u_3v_1 - u_1v_3)\mathbf{e}_{31} + (u_1v_2 - u_2v_1)\mathbf{e}_{12} \\ &= \'u \cdot \mathbf{v} + \'u \wedge \mathbf{v}.
\end{aligned}
\end{align} 
 
\noindent 
This is precisely the product in the Eq. \eqref{eq_098jnhybgtvfrcd} where the Clifford algebra of the Euclidean space was used to motivate and introduce Clifford algebras in general. One may also denote a multivector $\mathbf{e}_i\mathbf{e}_j$ as $\mathbf{e}_{ij}$ where convenient for simplicity. The multivector structure of $\cl_{3}$ is represented below.

\begin{center}

\tikzset{every picture/.style={line width=0.75pt}} 

\begin{tikzpicture}[x=0.75pt,y=0.75pt,yscale=-1,xscale=1]

\draw   (157.8,22.7) -- (275.1,129.75) -- (157.8,236.79) -- (40.5,129.75) -- cycle ;
\draw  [fill={rgb, 255:red, 208; green, 2; blue, 27 }  ,fill opacity=0.5 ] (92.6,82.1) -- (222.6,82.1) -- (263.6,119.6) -- (51.1,119.6) -- cycle ;
\draw  [fill={rgb, 255:red, 208; green, 2; blue, 27 }  ,fill opacity=0.5 ] (62.1,150.1) -- (252.6,150.1) -- (215.1,185.1) -- (100.6,185.1) -- cycle ;

\draw (134,48.9) node [anchor=north west][inner sep=0.75pt]    {$\mathbf{e}_{1}\mathbf{e}_{2}\mathbf{e}_{3}$};
\draw (106,157.9) node [anchor=north west][inner sep=0.75pt]    {$\mathbf{e}_{1}$};
\draw (152,157.9) node [anchor=north west][inner sep=0.75pt]    {$\mathbf{e}_{2}$};
\draw (194.5,157.9) node [anchor=north west][inner sep=0.75pt]    {$\mathbf{e}_{3}$};
\draw (152,202.9) node [anchor=north west][inner sep=0.75pt]    {$1$};
\draw (80.5,91.9) node [anchor=north west][inner sep=0.75pt]    {$\mathbf{e}_{1}\mathbf{e}_{2}$};
\draw (139,92.4) node [anchor=north west][inner sep=0.75pt]    {$\mathbf{e}_{1}\mathbf{e}_{3}$};
\draw (198,92.4) node [anchor=north west][inner sep=0.75pt]    {$\mathbf{e}_{2}\mathbf{e}_{3}$};
\draw (210.5,49) node [anchor=north west][inner sep=0.75pt]  [font=\scriptsize] [align=left] {\textit{trivector}};
\draw (270,93.5) node [anchor=north west][inner sep=0.75pt]  [font=\scriptsize] [align=left] {\textit{bivector}};
\draw (253,167.5) node [anchor=north west][inner sep=0.75pt]  [font=\scriptsize] [align=left] {\textit{vector}};
\draw (199.5,216.5) node [anchor=north west][inner sep=0.75pt]  [font=\scriptsize] [align=left] {\textit{scalar}};

\end{tikzpicture}
\end{center}

 As we expect from the Definition \ref{def_clifuniv}, both Clifford algebras $\matcomp$ and the universal Clifford algebra $\cl_3$ are isomorphic and an element of $\cl_3$ can be represented as a matrix on $\matcomp$ through the identification $\mathbf{e}_{1} \leftrightarrow  \sigma_{1},\;
\mathbf{e}_{2} \leftrightarrow  \sigma_{2}, \;
\mathbf{e}_{3} \leftrightarrow  \sigma_{3}.$ The discussion about Clifford algebra and its representations will be presented later on. \;$\textcolor{BrickRed}{{\blacktriangleleft}}$\end{ex}

\begin{ex} {{\textcolor{BrickRed}{{$\blacktriangleright$\;}}}} $\cl_{1,3}$. \normalfont Let $V = \rr^{1,3}$ be the Minkowski space-time. Considering an orthonormal basis of such space as $\{\mathbf{e}_0, \mathbf{e}_1,\mathbf{e}_2,\mathbf{e}_3\}$ and $\mathbf{v} = (v_{0},v_{1},v_{2},v_{3})$ an arbitrary vector on this four dimensional real space endowed with the metric $g$, 

\begin{align*}
\begin{aligned}
g(\mathbf{v},\mathbf{v}) &= (v_{0}\mathbf{e}_0 + v_{1}\mathbf{e}_1 + v_{2}\mathbf{e}_2 + v_{3}\mathbf{e}_3)(v_{0}\mathbf{e}_0 + v_{1}\mathbf{e}_1 + v_{2}\mathbf{e}_2 + v_{3}\mathbf{e}_3) \\
&= v_{o}^{2}(\mathbf{e}_0)^{2} + v_{1}^{2}(\mathbf{e}_1)^{2} + v_{2}^{2}(\mathbf{e}_2)^{2} + v_{3}^{2}(\mathbf{e}_3)^{2}  \\
&\;\;\;\;+ v_{0}v_{1}(\mathbf{e}_0\mathbf{e}_1 + \mathbf{e}_1\mathbf{e}_0) 
+ v_{0}v_{2}(\mathbf{e}_0\mathbf{e}_2 + \mathbf{e}_2\mathbf{e}_0)
+ v_{0}v_{3}(\mathbf{e}_0\mathbf{e}_3 + \mathbf{e}_3\mathbf{e}_0) \\
&\;\;\;\;+ v_{1}v_{2}(\mathbf{e}_1\mathbf{e}_2 + \mathbf{e}_2\mathbf{e}_1)
+ v_{1}v_{3}(\mathbf{e}_1\mathbf{e}_3 + \mathbf{e}_3\mathbf{e}_1)
+ v_{2}v_{3}(\mathbf{e}_2\mathbf{e}_3 + \mathbf{e}_3\mathbf{e}_2)\\
&= v_{0}^{2} - v_{1}^{2} - v_{2}^{2} - v_{3}^{2}.
\end{aligned}
\end{align*}

\noindent The Clifford algebra of such space is denoted by $\cl_{1,3}$ and is generated by  \begin{align}
\begin{aligned}
      \clm &= \text{span} \{\gamma(\mathbf{e}_{0})^{\mu_{0}}\gamma(\mathbf{e}_{1})^{\mu_{1}}\gamma(\mathbf{e}_{2})^{\mu_{2}}\gamma(\mathbf{e}_{3})^{\mu_{3}} \;\;|\;\; \mu_{i} = 0,1 \} \\
      &=  \text{span}\{1, \mathbf{e}_{0}, \mathbf{e}_{1}, \mathbf{e}_{2},\mathbf{e}_{3},\mathbf{e}_{0}\mathbf{e}_{1},\mathbf{e}_{0}\mathbf{e}_{2}, \mathbf{e}_{0}\mathbf{e}_{3}, \mathbf{e}_{1}\mathbf{e}_{2}, \\ &\mathbf{e}_{1}\mathbf{e}_{3}, \mathbf{e}_{2}\mathbf{e}_{3},\mathbf{e}_{0}\mathbf{e}_{1}\mathbf{e}_{2},
      \mathbf{e}_{0}\mathbf{e}_{1}\mathbf{e}_{3}, \mathbf{e}_{0}\mathbf{e}_{2}\mathbf{e}_{3},\mathbf{e}_{1}\mathbf{e}_{2}\mathbf{e}_{3},\mathbf{e}_{0}\mathbf{e}_{1}\mathbf{e}_{2}\mathbf{e}_{3}\}.
\end{aligned}
\end{align}

\noindent The basis elements satisfy the following relations

\begin{align}
\begin{aligned}
&(\mathbf{e}_0)^{2} = 1,\\
&(\mathbf{e}_i)^{2} = -1, &  &&i = 1,2,3.\\
&(\mathbf{e}_{\mu}\mathbf{e}_{\nu} + \mathbf{e}_{\nu}\mathbf{e}_{\mu}) = 0,  &\mu \neq \nu  &&\mu, \nu = 0,1,2,3.
\end{aligned}
\end{align}

\begin{center}

\tikzset{every picture/.style={line width=0.75pt}} 

\begin{tikzpicture}[x=0.75pt,y=0.75pt,yscale=-1,xscale=1]

\draw   (157.8,22.7) -- (275.1,129.75) -- (157.8,236.79) -- (40.5,129.75) -- cycle ;
\draw  [fill={rgb, 255:red, 208; green, 2; blue, 27 }  ,fill opacity=0.5 ] (103.1,71.6) -- (210.6,71.1) -- (246.1,103.6) -- (69.6,104.1) -- cycle ;
\draw  [fill={rgb, 255:red, 208; green, 2; blue, 27 }  ,fill opacity=0.5 ] (70.6,157.1) -- (245.1,157.1) -- (205.1,193.6) -- (110.6,193.6) -- cycle ;

\draw (143.5,44.9) node [anchor=north west][inner sep=0.75pt]  [font=\small]  {$\mathbf{e}_{0123}$};
\draw (161.85,166) node [anchor=north west][inner sep=0.75pt]  [font=\small]  {$\mathbf{e}_{2}$};
\draw (196.5,166) node [anchor=north west][inner sep=0.75pt]  [font=\small]  {$\mathbf{e}_{3}$};
\draw (153.5,205.4) node [anchor=north west][inner sep=0.75pt]  [font=\small]  {$1$};
\draw (156,125) node [anchor=north west][inner sep=0.75pt]  [font=\small]  {$\mathbf{e}_{12}$};
\draw (252.5,79) node [anchor=north west][inner sep=0.75pt]  [font=\scriptsize] [align=left] {\textit{trivector}};
\draw (292,124) node [anchor=north west][inner sep=0.75pt]  [font=\scriptsize] [align=left] {\textit{bivector}};
\draw (241.5,175.5) node [anchor=north west][inner sep=0.75pt]  [font=\scriptsize] [align=left] {\textit{vector}};
\draw (196.5,213) node [anchor=north west][inner sep=0.75pt]  [font=\scriptsize] [align=left] {\textit{scalar}};
\draw (130.85,166) node [anchor=north west][inner sep=0.75pt]  [font=\small]  {$\mathbf{e}_{1}$};
\draw (99.35,166) node [anchor=north west][inner sep=0.75pt]  [font=\small]  {$\mathbf{e}_{0}$};
\draw (192.5,125) node [anchor=north west][inner sep=0.75pt]  [font=\small]  {$\mathbf{e}_{13}$};
\draw (227,125) node [anchor=north west][inner sep=0.75pt]  [font=\small]  {$\mathbf{e}_{32}$};
\draw (95.5,125) node [anchor=north west][inner sep=0.75pt]  [font=\small]  {$\mathbf{e}_{02}$};
\draw (126.5,125) node [anchor=north west][inner sep=0.75pt]  [font=\small]  {$\mathbf{e}_{03}$};
\draw (64.5,125) node [anchor=north west][inner sep=0.75pt]  [font=\small]  {$\mathbf{e}_{01}$};
\draw (210.5,39) node [anchor=north west][inner sep=0.75pt]  [font=\scriptsize] [align=left] {\textit{quadrivector}};
\draw (93,80.4) node [anchor=north west][inner sep=0.75pt]  [font=\small]  {$\mathbf{e}_{012}$};
\draw (162,80.4) node [anchor=north west][inner sep=0.75pt]  [font=\small]  {$\mathbf{e}_{123}$};
\draw (125,80.4) node [anchor=north west][inner sep=0.75pt]  [font=\small]  {$\mathbf{e}_{023}$};
\draw (198,80.9) node [anchor=north west][inner sep=0.75pt]  [font=\small]  {$\mathbf{e}_{013}$};

\end{tikzpicture}
\end{center}

The multivector structure of $\cl_{1,3}$ is represented above. Moreover, an arbitrary element $A$ of $\cl_{1,3}$ is written as 

\begin{align}
\begin{aligned}
A &=  a + a_{0}\mathbf{e}_{0} + a_{1}\mathbf{e}_{1} + a_{2}\mathbf{e}_{2} + a_{3}\mathbf{e}_{3} + a_{01}\mathbf{e}_{0}\mathbf{e}_{1} + a_{02}\mathbf{e}_{0}\mathbf{e}_{2} \\&\;\;\;\;+ a_{03}\mathbf{e}_{0}\mathbf{e}_{3} +
a_{12}\mathbf{e}_{1}\mathbf{e}_{2} +  a_{13}\mathbf{e}_{1}\mathbf{e}_{3} + a_{23}\mathbf{e}_{2}\mathbf{e}_{3} + a_{012}\mathbf{e}_{0}\mathbf{e}_{1}\mathbf{e}_{2} \\&\;\;\;\;+ a_{013}\mathbf{e}_{0}\mathbf{e}_{1}\mathbf{e}_{3} +
a_{023}\mathbf{e}_{0}\mathbf{e}_{2}\mathbf{e}_{3} +
a_{123}\mathbf{e}_{1}\mathbf{e}_{2}\mathbf{e}_{3} +  a_{0123}\mathbf{e}_{0}\mathbf{e}_{1}\mathbf{e}_{2}\mathbf{e}_{3}. \;\;\;\textcolor{BrickRed}{{\blacktriangleleft}}
\end{aligned}
\end{align}\end{ex}

\begin{ex}{{\textcolor{BrickRed}{{$\blacktriangleright$\;}}}} 
    The exterior algebra. \normalfont Consider the exterior algebra defined in the Appendix \ref{app3}.
The exterior algebra is a Clifford algebra. Indeed, let $V$ be a vector space and $\left ( \bigwedge(V), \wedge \right )$ be the exterior algebra of $V$. Take the mapping $\gamma: V \to \bigwedge^{1} (V)$. For arbitrary elements $\mathbf{u},\mathbf{v} \in V$ it reads that $\gamma(\mathbf{v})\gamma(\mathbf{u}) = \mathbf{v} \wedge \mathbf{u}$. Each element of $\bigwedge^{k} (V)$ is given by $\mathbf{v}_{1} \wedge  \cdots \wedge \mathbf{v}_{k} = \gamma(\mathbf{v}_1)\cdots\gamma(\mathbf{v}_k)$. Therefore, $\bigwedge(V)$ is generated by  $\{\gamma(\mathbf{v}) \;|\; \mathbf{v} \in V \}$ e $\{x1_{\mathcal{A}} \;|\; x \in \rr  \}$. For the symmetric bilinear form $g = 0$:
\begin{equation}
     \gamma(\mathbf{v}) \gamma(\mathbf{u}) + \gamma (\mathbf{u}) \gamma (\mathbf{v}) = \mathbf{v} \wedge \mathbf{u} + \mathbf{u} \wedge \mathbf{v} = 0 = 2g(\mathbf{u},\mathbf{v})1_{\mathcal{A}}.
\end{equation}

\noindent As a result, the exterior algebra is a Clifford algebra with respect to the null bilinear form $g = 0$.  $\textcolor{BrickRed}{{\blacktriangleleft}}$
 \end{ex}

\section{Structure of Clifford Algebras}\label{sec_strc}

\paragraph{ } This section provides an exposition of the structure of Clifford algebras. We present several operations that can be defined on these algebras based on their multivector structure and introduce some of their important subspaces. These operations, in turn, are very important and are used to define several underlying structures in the other chapters. The main goal of this section is also to present the key theorems concerning the structure of Clifford algebras, which reveal important properties such as their periodicity, the crucial property for the classification of Clifford algebras, which will be discussed in the following section.

\hspace{0.5cm} Let $V$ be a vector space and $\{\mathbf{e}_1, \ldots, \mathbf{e}_n\}$ be an orthogonal basis for $V$. It yields from Eq. \eqref{eqq_prodclifprofteo} that

\begin{align}
\mathbf{e}_i \mathbf{e}_j = \mathbf{e}_i \wedge \mathbf{e}_j + (\mathbf{e}_{i})_{\flat} \rfloor \mathbf{e}_j = \mathbf{e}_i \wedge \mathbf{e}_j + (\mathbf{e}_{i})_{\flat} (\mathbf{e}_j) = \mathbf{e}_i \wedge \mathbf{e}_j + g(\mathbf{e}_i, \mathbf{e}_j) = \mathbf{e}_i \wedge \mathbf{e}_j  
\end{align}

\noindent for every $i \neq j$. One may generalise the above result and obtain the following relation

\begin{equation}\label{eq_839wreifjodms}
    \mathbf{e}_{\mu_1} \cdots \mathbf{e}_{\mu_p} = \mathbf{e}_{\mu_1} \wedge \cdots \wedge \mathbf{e}_{\mu_p} \;\;\;\;\;\;\; ({\mu_1} \neq \cdots \neq \mu_p).
\end{equation}

\noindent As it is derived in the Eq. \eqref{eq_app3dimeext}, the dimension of $\cl(V,g)$ is given by

\begin{equation}
    \text{dim} \,\cl(V,g) = \sum_{p \, = \, 0}^{n} \binom{n}{p} = 2^{n} = 2^{ \text{\, dim}\,V}.
\end{equation}

\noindent Therefore, the converse of Theorem \ref{teo2cliffordalgdi} is established: 

{\colorlet{shadecolor}{Red!10}\begin{shaded} Every universal Clifford algebra has dimension $2^n$ ($n = $ dim $V$). \end{shaded}}

Once the exterior algebra $\bigwedge (V)$ (defined in the Appendix \ref{sec4_app3}) and the Clifford algebra $\cl(V,g)$ have the same dimension, a \textit{vector space isomorphism} exists between them. The Eq. \eqref{eq_839wreifjodms} establishes the isomorphism itself which is canonical and does not depend on any particular choice of basis. Therefore, one important point to note here is the fact that Clifford algebra inherits the multivector structure from the exterior algebra. 

  {\colorlet{shadecolor}{Red!10}\begin{shaded}$\cl(V,g)$ can be written as:

\begin{equation}
    \cl(V,g) = \bigoplus_{p\,=\,0}^{n} \bigwedge_{p} (V).
\end{equation}\end{shaded}}

This means that the study of the exterior algebra is fundamental to understand the multivector structure of the Clifford algebras. A complete discussion of the exterior algebra can be found in the Appendix \ref{app3}.

\subsection*{OPERATIONS ON CLIFFORD ALGEBRAS}

\paragraph{ } Some operations can be defined in a multivector algebra based on its own multivector structure. They can be used to unravel some properties in the algebraic structure leading to important applications. Those operations can be defined equivalently with respect to the exterior algebra $\bigwedge(V)$ or in terms of Clifford algebras $\mathcal{C}\ell(V,g)$.

\begin{definvermelho}\label{def_asd4aa5ngf43} Let $A \in \mathcal{C}\ell(V,g)$ be a multivector, the \textbf{projector operator} is defined as

\begin{align}
\begin{aligned}\label{eq_opproj}
\langle \; \cdot \;\rangle_p :\;    &\mathcal{C}\ell(V,g)  &\longrightarrow \;\;\;\;\; &\bigwedge_{p}(V) \\
&\;\;\;\;\;\;\left \langle A \right \rangle_{p}  &\longmapsto \;\;\;\;\; &A_{[p]}, \;\;\;\;\;\;\;\;\;\;\;
\end{aligned}
\end{align}
\noindent such that $A_{[p]} \in \bigwedge _{p}(V)$ is the $p$-vector component of $A$.
\end{definvermelho}

\begin{definvermelho}\label{def15op} Let $A \in \mathcal{C}\ell(V,g)$ be a multivector, the \textbf{grade involution} is defined as
\begin{align}
\begin{aligned}\label{eq_graduacao}
\yhwidehat{\;\cdot\;}\;:\;    &\mathcal{C}\ell(V,g)  &\longrightarrow \;\;\;\;\; &\mathcal{C}\ell(V,g) \\
&\;\;\;\;\;\;\yhwidehat{A_{[p]}}  &\longmapsto \;\;\;\;\; &(-1)^{p}A_{[p]}.
\end{aligned}
\end{align}
\end{definvermelho}

\begin{definvermelho}\label{def16op}  Let $A \in \mathcal{C}\ell(V,g)$ be a multivector, the \textbf{reversion} is defined as follows
\begin{align}
\begin{aligned}\label{eq_reversao1}
\yhwidetilde{\;\cdot \;}:\;    &\mathcal{C}\ell(V,g)  &\longrightarrow \;\;\;\;\; &\mathcal{C}\ell(V,g) \\
&\;\;\;\;\;\;\yhwidetilde{A_{[p]}}  &\longmapsto \;\;\;\;\; & (-1)^{\frac{p(p-1)}{2}} A_{[p]}. 
\end{aligned}
\end{align}

\end{definvermelho}

\begin{definvermelho}\label{def17op} Let $A \in \mathcal{C}\ell(V,g)$ be a multivector,  the \textbf{ conjugation}, denoted by $\overline{A}$, is defined as the composition of the grade involution and the grade involution:
\begin{align}
\begin{aligned}\label{eq_conjugacao1}
\yhwidetilde{\yhwidehat{A_{[p]}}} = \yhwidehat{\yhwidetilde{A_{[p]}}} = \overline{A_{[p]}}.
\end{aligned}
\end{align}
\end{definvermelho}

  {\colorlet{shadecolor}{Red!15}\begin{shaded}
  In summary, for $A \in \cl(V,g)$:
    \begin{itemize}
 \item [\textcolor{BrickRed}{\textbf{(i)}}]  $\left \langle A \right \rangle_{p}  = A_{[p]}$, \;\;\;\;\;\;\;\;\;\;\;\;\;\;\;\;\;\;\;\;\;\;\;\;\;\;\;\;\; (projection);
    \item [\textcolor{BrickRed}{\textbf{(ii)}}] $\yhwidehat{A_{[p]}} = (-1)^p A_{[p]}$, \;\;\;\;\;\;\;\;\;\;\;\;\;\;\;\;\;\;\;\;\;\; (grade involution);
    \item[\textcolor{BrickRed}{\textbf{(iii)}}] $\yhwidetilde{A_{[p]}} = (-1)^{\frac{p(p-1)}{2}} A_{[p]}$, \;\;\;\;\;\;\;\;\;\;\;\,\;\;\;\;\; (reversion);
    \item[\textcolor{BrickRed}{\textbf{(iv)}}] $  \overline{A_{[p]}} = \yhwidetilde{\yhwidehat{A_{[p]}}} =  \yhwidehat{\yhwidetilde{A_{[p]}}}$, \;\;\;\;\;\;\;\;\;\;\;\;\;\;\;\;\;\;\;\;\;(conjugation).
\end{itemize}\end{shaded}} 

\begin{ex}{{\textcolor{BrickRed}{{$\blacktriangleright$\;}}}} Operations in $\clm$. \normalfont Consider the Clifford algebra $\clm$ and an arbitrary $A \in \clm$ given by
\begin{align}
\begin{aligned}
A &=  a + a_{0}\mathbf{e}_{0} + a_{1}\mathbf{e}_{1} + a_{2}\mathbf{e}_{2} + a_{3}\mathbf{e}_{3} + a_{01}\mathbf{e}_{0}\mathbf{e}_{1} + a_{02}\mathbf{e}_{0}\mathbf{e}_{2} \\&\;\;\;\;+ a_{03}\mathbf{e}_{0}\mathbf{e}_{3} +
a_{12}\mathbf{e}_{1}\mathbf{e}_{2} +  a_{13}\mathbf{e}_{1}\mathbf{e}_{3} + a_{23}\mathbf{e}_{2}\mathbf{e}_{3} + a_{012}\mathbf{e}_{0}\mathbf{e}_{1}\mathbf{e}_{2} \\&\;\;\;\;+ a_{013}\mathbf{e}_{0}\mathbf{e}_{1}\mathbf{e}_{3} +
a_{023}\mathbf{e}_{0}\mathbf{e}_{2}\mathbf{e}_{3} +
a_{123}\mathbf{e}_{1}\mathbf{e}_{2}\mathbf{e}_{3} +  a_{0123}\mathbf{e}_{0}\mathbf{e}_{1}\mathbf{e}_{2}\mathbf{e}_{3}.
\end{aligned}
\end{align}

\noindent The multivector $A$ can be written in a compact form as follows

\begin{equation}
    A =  A_{[0]} + A_{[1]} + A_{[2]} + A_{[3]} + A_{[4]}.
\end{equation}

\noindent The projector operator yields:
\begin{align}
\begin{aligned}
&\left \langle A \right \rangle_{0} = a = A_{[0]}\in \bigwedge ^{0}(\mathbb{R}^{1,3}),  \\
&\left \langle A \right \rangle_{1} = a_{0}\mathbf{e}_{0} + a_{1}\mathbf{e}_{1} + a_{2}\mathbf{e}_{2} + a_{3}\mathbf{e}_{3} = A_{[1]} \in \bigwedge ^{1}(\mathbb{R}^{1,3}),\\
&\left \langle A \right \rangle_{2} = a_{01}\mathbf{e}_{0}\mathbf{e}_{1} + a_{02}\mathbf{e}_{0}\mathbf{e}_{2} + a_{03}\mathbf{e}_{0}\mathbf{e}_{3} +
a_{12}\mathbf{e}_{1}\mathbf{e}_{2} +  a_{13}\mathbf{e}_{1}\mathbf{e}_{3} + a_{23}\mathbf{e}_{2}\mathbf{e}_{3} = A_{[2]} \in \bigwedge ^{2}(\mathbb{R}^{1,3}),  \\
&\left \langle A \right \rangle_{3} = a_{012}\mathbf{e}_{0}\mathbf{e}_{1}\mathbf{e}_{2} + a_{013}\mathbf{e}_{0}\mathbf{e}_{1}\mathbf{e}_{3} +
a_{023}\mathbf{e}_{0}\mathbf{e}_{2}\mathbf{e}_{3} +
a_{123}\mathbf{e}_{1}\mathbf{e}_{2}\mathbf{e}_{3}   = A_{[3]} \in \bigwedge ^{3}(\mathbb{R}^{1,3}). \\
&\left \langle A \right \rangle_{4} = a_{0123}\mathbf{e}_{0}\mathbf{e}_{1}\mathbf{e}_{2}\mathbf{e}_{3}  = A_{[4]} \in \bigwedge ^{4}(\mathbb{R}^{1,3}). 
\end{aligned}
\end{align}

\noindent With respect to the other operations, one has:

\begin{align}
\begin{aligned}
\yhwidehat{A} &=  A_{[0]}- A_{[1]} + A_{[2]} - A_{[3]} + A_{[4]}, &&\text{grade involution},\\
\yhwidetilde{A} &=  A_{[0]}+A_{[1]} - A_{[2]} - A_{[3]} + A_{[4]}, &&\text{reversion},\\
\overline{A} &=  A_{[0]} - A_{[1]} - A_{[2]} + A_{[3]} + A_{[4]}, &&\text{conjugation}.\;  \textcolor{BrickRed}{{\blacktriangleleft}} 
\end{aligned}
\end{align}\end{ex}

\paragraph{ } The following Proposition \ref{propos_2341rta} presents a characterisation for the operations on multivectors $A_{[p]}$ based on the number $p$ of their component.

\begin{tfpropos}\label{propos_2341rta}
Let $A_{[p]} \in \bigwedge_{p} (V)$, then $\yhwidehat{A_{[p]}} = \pm A_{[p]} $, $\yhwidetilde{A_{[p]}} = \pm A_{[p]} $ and $\overline{A_{[p]}} = \pm  A_{[p]} $ with the plus or minus sign depending only on $p \mod 4$ as indicated in the following table:

\begin{table}[H]\centering
\begin{tabular}{c|cccc}
\rowcolor{red!15} $p \mod 4$       & $0$ & $1$ & $2$ & $3$ \\ \hline
$\text{grade involution }(\yhwidehat{\;\cdot\;})$   & $+$ & $-$ & $+$ & $-$ \\
$\text{reversion } (\yhwidetilde{\;\cdot\;})$ & $+$ & $+$ & $-$ & $-$ \\
$\text{conjugation }(\overline{\;\cdot\;})$  & $+$ & $-$ & $-$ & $+$
\end{tabular}
\end{table}\end{tfpropos}
\textcolor{BrickRed}{\textit{Proof.}} For the grade involution it is straightforward since $\yhwidehat{A_{[p]}} = (-1)^p A_{[p]}$. For the reversion, one has that $\yhwidetilde{A_{[p]}} = (-1)^{\frac{p(p-1)}{2}} A_{[p]}$. Let us consider the four cases for $p \in \mathbb{N}$.
Suppose that $p = 0 \mod 4$. In this case, $p = 4x$ for some $x \in \mathbb{N}$, hence $ {\frac{1}{2}}p(p-1) = \frac{1}{2} 4x (4x-1)$ which in turn is an even number. Suppose that $p = 1 \mod 4$. In this case, $p = 4x + 1$, thus $ {\frac{1}{2}}p(p-1) = \frac{1}{2} (4x +1) (4x)$ which is an even number. Now let $p = 2 \mod 4$, therefore ${\frac{1}{2}}p(p-1) = \frac{1}{2} (4x+2) (4x-1)$ which is an odd number. For $p = 3 \mod 4$ it follows that ${\frac{1}{2}}p(p-1) = \frac{1}{2} (4x+3) (4x-2)$ which is an odd number. Finally, once the conjugation is the composition of the grade involution and the reversion, so is its signal. \textcolor{BrickRed}{$\Box$}

\subsection*{SUBSPACES OF CLIFFORD ALGEBRAS}
\hspace{0.5cm} Let us now introduce some important subspaces of $\mathcal{C}\ell(V,g)$.

\begin{definvermelho}
The \textbf{centre} of a Clifford algebra $ \cl_{p,q}$ is defined as being the set of elements on $\cl_{p,q}$ that commutes with all elements of $\cl_{p,q}$\normalfont

\begin{equation}
    \text{Cen}(\cl_{p,q}) = \{a \in \cl_{p,q} \;|\; ax = xa, \forall\, x \in \cl_{p,q} \}.
\end{equation}
\end{definvermelho}

\noindent The $\text{Cen}(\cl_{p,q})$, $n = p +q$, can be also characterised as\cite{Roc16}

\begin{equation}\label{eq_centrws65g1}
      \text{Cen}(\cl_{p,q}) = 
    \color{BrickRed} \begin{cases} \color{black}
     \bigwedge_{0}(\rr^{p,q}), \text{\;if\;} n \text{\, is even,}\\
    \color{black} \bigwedge\,\!\!_{0}(\rr^{p,q}) \oplus  \bigwedge\,\!\!_{n}(\rr^{p,q}), \text{\;if\;} n \text{\, is odd.}
    \color{black} \end{cases}\color{black}
\end{equation}

The next definition allows the decomposition of Clifford algebras as a direct sum of two subspaces.

\begin{definvermelho}\label{subsp_defq5asw}
The following subspaces of $\mathcal{C}\ell(V,g)$\normalfont

\begin{align}
\begin{aligned}
\cl^{\text{even}}(V,g) &= \{ A \in \mathcal{C}\ell(V,g) \;|\; \yhwidehat{A} = A \}, \\
\cl^{\text{odd}}(V,g) &= \{ A \in \mathcal{C}\ell(V,g) \;|\; \yhwidehat{A} = -A \},
\end{aligned}
\end{align}

\noindent \textit{are said to be respectively the \textbf{even
part} and the \textbf{odd part} of $\mathcal{C}\ell(V,g)$.} \end{definvermelho}

Therefore, it is possible to write

\begin{equation}
    \mathcal{C}\ell(V,g) = \cl^{\text{even}}(V,g) \oplus \cl^{\text{odd}}(V,g).
\end{equation}

Those subspaces unravel one important property of the Clifford algebra structure concerning its grading.

\begin{definvermelho}
Let $G$ be an abelian group. An algebra $\mathcal{A}$ is said to be \textbf{G-graded} if there exists subspaces $\mathcal{A}_k \,(k \in G)$ such that $\mathcal{A} = \bigoplus \mathcal{A}_k $ and, if given $x_k \in \mathcal{A}_k,\, y_l \in \mathcal{A}_l$ it follows that $x_k y_l \in \mathcal{A}_{k+l}$·
\end{definvermelho}

\noindent The elements of $\mathcal{A}_k$ are said to be\textit{ homogeneous of degree }$k$. In general, the following notation is used 

\begin{equation}
    k = \text{deg}(x_k), \;\;\; x_k \in \mathcal{A}_k.
\end{equation}

\noindent Since $G$ is abelian it implies that

\begin{equation}
    \text{deg}(x_k y_l) = \text{deg}(x_k) + \text{deg}(y_l).
\end{equation}

\noindent It is assumed that for the scalars $a \in \mathbb{K} = \mathbb{C}, \mathbb{R}$ that deg$(a)=0$ and that the null vector must be considered homogeneous for all degrees since every subspace $\mathcal{A}_k$ contains it. If the unique element which is negative-graded is the null vector, the algebra is said to be positive-graded.\\

Since the grade involution is an automorphism \cite{Roc16}

\begin{equation}
   \yhwidehat{A_{[p]}B_{[q]}} = \yhwidehat{A_{[p]}} \,\yhwidehat{B_{[q]}} = (-1)^{p+q}A_{[p]}B_{[q]} \in \bigwedge \,\!\!_{p+q}(V)
\end{equation}

\noindent the following relations for the Clifford algebra $\mathcal{C}\ell(V,g)$ can be derived 

   {\colorlet{shadecolor}{Red!15}\begin{shaded}\begin{align}
\begin{aligned}\label{eq_acz22222grad}
\cl^{\text{even}}(V,g) \,\cl^{\text{even}}(V,g) \subset \cl^{\text{even}}(V,g), \\
\cl^{\text{even}}(V,g) \,\cl^{\text{odd}}(V,g) \subset  \cl^{\text{odd}}(V,g), \\
\cl^{\text{odd}}(V,g) \,\cl^{\text{even}}(V,g)\subset \cl^{\text{odd}}(V,g), \\
\cl^{\text{odd}}(V,g) \,\cl^{\text{odd}}(V,g) \subset  \cl^{\text{even}}(V,g).
\end{aligned}
\end{align}\end{shaded}}

Thus, $\mathcal{C}\ell(V,g)$ is a $\mathbb{Z}_2$-graded algebra and $\cl^{\text{even}}(V,g)$ is a subalgebra of $\mathcal{C}\ell(V,g)$ called \textit{even subalgebra}. 

It is worth to emphasise that in contrast to the grade involution, the reversion is an anti-automorphism, that is \cite{Roc16}

\begin{equation}
    \yhwidetilde{A_{[p]}B_{[q]}} =\yhwidetilde{B_{[q]}}\, \yhwidetilde{A_{[p]}}.
\end{equation}

\subsection*{THEOREMS ON CLIFFORD ALGEBRAS STRUCTURE}

\paragraph{ } In this section, some important theorems concerning the structure of Clifford algebras is presented. Aside from the inherent importance of such theorems, they are used to construct the classification of the Clifford algebras. Although, before presenting those theorems, we start by defining an important concept regarding the \textit{complexification} of a Clifford algebra. Since an algebra is a vector space, we recall the complexification of a vector space.

\begin{definvermelho}\label{def_complexificacao}
Let $V$ a vector space over $\rr$ with dimension $\dim V = n$. The \textbf{complexification} $V_{\mathbb{C}}$ is the space of elements in the form $v+ iu$, such that $v,u \in V$ and $i$ is the imaginary unit.
\end{definvermelho}

\noindent The space $V_{\mathbb{C}}$ is a vector space with sum and multiplication by a complex scalar $(a +ib)$ defined by

\begin{align}
\begin{aligned}
(\mathbf{v}_{1} + i\'u_{1}) + (\mathbf{v}_{2} + i\'u_{2}) &= (\mathbf{v}_{1} + \mathbf{v}_{2}) + i(\'u_{1} + \'u_{2}), \\
(a + ib)(\mathbf{v} + i\'u) &= (a\mathbf{v} - b\'u) + i(b\mathbf{v} + a\'u).
\end{aligned}
\end{align}

\noindent The dimension of $V_{\cc}$ is $\dim_{\cc}\, V_{\cc} = n$ over $\mathbb{C}$ and $\dim_{\rr}\, V_{\cc} = 2n$ over $\rr$. It can be noticed that

\begin{equation}
    V_{\cc} = \cc \tensor V.
\end{equation}

\noindent By the nature of the tensor product, any vector $\mathbf{v} \in V_{\mathbb{C}}$ is written in a unique way as \cite{Roc17}

\begin{equation}
    \mathbf{v} = \mathbf{v}_{1} \otimes 1 + \mathbf{v}_{2} \otimes i,
\end{equation}

\noindent such that $\mathbf{v}_{1}, \mathbf{v}_{2} \in V$. However, for simplicity, one can just write  $\mathbf{v} = \mathbf{v}_{1} + i\mathbf{v}_{2}$ as shown in the Definition \ref{def_complexificacao}.

Given a symmetric bilinear form $g$ endowing $V$, its extension $g_{\cc} : V_{\cc} \cross V_{\cc} \to \cc$ may be defined as

\begin{equation}
g_{\mathbb{C}}(\mathbf{v},\'u) = g_{\mathbb{C}}(\mathbf{v}_{1} + i \mathbf{v}_{2}, \'u_{1} + i \'u_{2}) = g(\mathbf{v}_1,\'u_1) - g(\mathbf{v}_2,\'u_2) + i(g(\mathbf{v}_{1},\'u_{2}) + g(\mathbf{v}_2,\'u_1)).
\end{equation}

In the first Section \ref{secdefacs}, the Clifford algebras have been introduced by the Definition \ref{def_a15d1s} regarding \textbf{real} Clifford algebras. We are now in a position to present the \textit{complex Clifford algebras} through the next theorem that unravel its structure.

\begin{tftheorem}\label{teo_acsestrcomplex}
Let $(V,g)$ a quadratic space over $\rr$ and $\cl(V,g)$ its associated Clifford real algebra.  Consider the complex Clifford algebra  $\cl(V_{\mathbb{C}},g_{\mathbb{C}})$ for the complexified quadratic space $(V_{\mathbb{C}},g_{\mathbb{C}})$. Hence

\begin{equation}\label{eq_complexalgclc}
\cl(V_{\mathbb{C}},g_{\mathbb{C}}) \simeq  \cl_{\mathbb{C}}(V,g), 
\end{equation}

\noindent where $\cl_{\mathbb{C}}(V,g) = \mathbb{C} \otimes \cl(V,g)$ denotes the complexification of  $\cl(V,g)$.
\end{tftheorem}
\noindent \textcolor{BrickRed}{\textit{Proof.}} Let us first notice that $\mathbb{C} \otimes \cl(V,g)$ is an algebra such that for all $a,b \in \cc$ and $A, B \in \cl(V,g)$ the product is given by

\begin{equation}
    (a \otimes A)(b \otimes B) = ab \otimes AB.
\end{equation}

\noindent Since $\dim_{\rr}\,\cl(V,g) = $ $2^{\,\text{dim} \,V} $, the dimension of $\cl_{\mathbb{C}}(V,g)$ over $\rr$ is given by dim$_{\,\rr}\, \cl_{\mathbb{C}}(V,g)$ $= 2 \, \text{dim}_{\,\rr}\,\cl(V,g) $ $= 2 \cdot 2^{\,\text{dim}\,V}$. If $\gamma: V \to \cl (V,g)$ denotes the Clifford map, one may define a map $\Gamma: V_{\cc} \to \cl_{\mathbb{C}}(V,g)$ as a linear map on $\cc$ such that

\begin{equation}
    \Gamma = 1 \otimes \gamma,
\end{equation}

\noindent where $1$ denotes the identity of $V_{\cc}$. Hence, for $a \otimes \mathbf{v} \in \cc \otimes V$ we have that

\begin{equation}
    \Gamma (a \otimes \mathbf{v}) = a \otimes \gamma(\mathbf{v}).
\end{equation}

\noindent We claim that the map $\Gamma$ is a Clifford map. Indeed,

\begin{equation}
    (\Gamma (1 \otimes \mathbf{v}))^2 = (1 \otimes \gamma(\mathbf{v}) ) (1 \otimes \gamma(\mathbf{v}) ) = 1 \otimes g(\mathbf{v},\mathbf{v}).
\end{equation}

\noindent On the other hand, since $\cl(V_{\mathbb{C}},g_{\mathbb{C}})$ with a Clifford map $\gamma_{\cc}$ is a universal Clifford algebra for $(V_{\mathbb{C}},g_{\mathbb{C}})$, by the universal property there exists a homomorphism $\phi: \cl(V_{\mathbb{C}},g_{\mathbb{C}}) \to \cl_{\mathbb{C}}(V,g)$ for which the following diagram commutes

\begin{equation*}
\begin{tikzcd}
V_{\cc} \arrow[black]{d}[black,swap]{\Gamma} \rar{\gamma_{\cc}} & [15pt] \cl(V_{\mathbb{C}},g_{\mathbb{C}}) \arrow[BrickRed, dashed]{dl}[black]{\phi} \\ [20pt] \cl_{\mathbb{C}}(V,g) = \mathbb{C} \otimes \cl(V,g)
\end{tikzcd} 
\end{equation*}{}

\noindent With respect to the dimension, we have that: dim$_{\,\cc}\,\cl(V_{\mathbb{C}},g_{\mathbb{C}}) = 2^{\,\text{dim}\,V}$ or equivalently dim$_{\rr}\,\cl(V_{\mathbb{C}},g_{\mathbb{C}}) = 2 \cdot 2^{\,\text{dim}\,V}$. Since dim$\,\cl(V_{\mathbb{C}},g_{\mathbb{C}})$ = dim$\,\cl_{\mathbb{C}}(V,g)$ we conclude that $\phi$ is an isomorphism.\textcolor{BrickRed}{$\Box$}
\paragraph{ } The above result shows us that \textcolor{BrickRed}{\textit{complex Clifford algebras} are isomorphic to the \textit{real Clifford algebras of the complexified vector space}}. This means that the tensor product of the Clifford algebra with respect to $\mathbb{C}$ can come out of the parentheses, that is,

   {\colorlet{shadecolor}{Red!15}\begin{shaded}\begin{equation}
\cl(\mathbb{C} \tensor V,g_{\mathbb{C}}) \simeq  \mathbb{C} \otimes \cl(V,g). 
\end{equation}\end{shaded}}

\noindent Once the real Clifford algebras structure is known, the complex Clifford algebras are immediately obtained via the complexification.
It alludes that in order to describe the complex Clifford algebras structure it suffices to understand the real ones.

\paragraph{ }  Clifford algebras exhibit a periodic structure, which repeats every eight dimensions for the real case and every two dimensions for the complex case. This periodicity is one of their most significant features, allowing their classification and deeply influencing their applications across mathematics and physics. The following theorem is a start for the study of the periodicity of Clifford algebras and it is important to the construction of their classification.

\begin{tftheorem}   \label{teo_casperiod1}
Let $\clpq$ be the Clifford algebra associated with the quadratic space $\rr^{p,q}$. Then the following isomorphisms hold:\normalfont

\begin{align}\label{eq_isowesdkjhgf}
    \begin{aligned}
    &&\textcolor{BrickRed}{\textbf{(i)}}\;\;\, \; \cl_{p+1, q+1} &\simeq\; \cl_{1,1} \tensor \clpq;\\
    &&\textcolor{BrickRed}{\textbf{(ii)}} \;\;\;\;\;\;    \cl_{q+2,p} &\simeq\; \cl_{2,0} \tensor \clpq;\\
   &&\textcolor{BrickRed}{\textbf{(iii)}}\;\;\;  \;\;  \cl_{q,p+2} &\simeq\; \cl_{0,2} \tensor \clpq.
    \end{aligned}
\end{align}

\noindent \textit{where either $p,q>0$ and $\tensor$ denotes the usual tensor product.}\end{tftheorem}
\noindent \textcolor{BrickRed}{\textit{Proof.}} Let $U$ be a $2$-dimensional space, with orthonormal basis $\{\'f_1, \'f_2\}$ endowed with a symmetric bilinear form $g_{U}$ such that for $\'u = u^{1}\'f_1 + u^{2}\'f_2 \in U$, one may write

\begin{equation}
    g_{U} = \lambda_1 (u^{1})^2 + \lambda_2 (u^{2})^2, \;\;\;\; \lambda_{1}, \lambda_2 = \pm 1.
\end{equation}

\noindent That is,

\begin{equation}
  g_{U} =   \color{BrickRed}\begin{cases}
    \color{black} (u^{1})^2 +  (u^{2})^2, \;\;\;\;\;\; \text{if}\; U = \rr^{2,0},\\
      \color{black}(u^{1})^2 -  (u^{2})^2, \;\;\;\;\;\; \text{if}\; U = \rr^{1,1},\\
      \color{black} -(u^{1})^2 -  (u^{2})^2, \;\;\; \text{if}\; U = \rr^{0,2}.
    \end{cases}\color{black}
\end{equation}

\noindent With respect to the $n$-dimensional quadratic space $\rr^{p,q}$, $n= p+q$, with an orthonormal basis $\{\'e_1,\ldots,\'e_n\}$, for $\mathbf{v} = v^{i}\'e_{i} \in \rrpq$ we have that

\begin{equation}
    g(\mathbf{v},\mathbf{v}) = (v^{1})^{2} + \cdots +(v^{p})^{2} - (v^{p + 1})^{2} - \cdots -(v^{p+q})^{2}.
\end{equation}

\noindent A linear map $\Gamma: \rrpq \oplus U \to \cl(U,g_{U}) \otimes \clpq$ is defined for all $\'u \in U$ and  $\mathbf{v} \in \rrpq$ as

\begin{equation}
    \Gamma (\mathbf{v} + \'u) = \'f_1\'f_2 \otimes \mathbf{v} + \'u \tensor 1
\end{equation}

\noindent where $\'f_1 = \rho (\'f_1),\; \'f_2 = \rho (\'f_2),\; \'u = \rho (\'u),\; \mathbf{v} = \gamma (\mathbf{v})$ and $\rho: U \to \cl(U,g_{U}), \gamma: \rrpq \to \clpq$ are the Clifford mappings. We claim that $\Gamma$ is also a Clifford map. In fact,

\begin{align}\label{eq_gammavucmaps}
    \begin{aligned}
      (\Gamma (\mathbf{v} + \'u))^2 &= (\'f_1\'f_2 \otimes \mathbf{v} + \'u \tensor 1)(\'f_1\'f_2 \otimes \mathbf{v} + \'u \tensor 1) \\
      &= (\'f_1\'f_2)^2 \otimes \mathbf{v}^2 + (\'u\'f_1\'f_2 + \'f_1\'f_2\'u) \otimes \mathbf{v} + \'u^2 \tensor 1.
    \end{aligned}
\end{align}

\noindent Recalling that $\{\'f_1,\'f_2 \}$ is orthonormal, it follows that \begin{equation}
    \'f_1\'f_2 = \'f_1 \wedge \'f_2 = -\'f_2 \'f_1.
\end{equation}

\noindent It then yields

\begin{align}
    \begin{aligned}
    \'u\'f_1\'f_2 + \'f_1\'f_2\'u &= \'u(\'f_1 \wedge\'f_2)+ (\'f_1\wedge\'f_2)\'u \\
    &= 2\'u \wedge \'f_1\wedge\'f_2\\
    &= 2( u^{1}\'f_1 + u^{2}\'f_2 ) \wedge \'f_1\wedge\'f_2 = 0.
    \end{aligned}
\end{align}
 
\noindent In addition,

\begin{equation}
    (\'f_1\'f_2)^2 = \'f_1\'f_2 \'f_1\'f_2  = - (\'f_1)^2(\'f_2)^2 = -\lambda_1 \lambda_2.
\end{equation}

\noindent Therefore, for $  (\Gamma (\mathbf{v} + \'u))^2$ in Eq. \eqref{eq_gammavucmaps}, it reads

\begin{align}
    \begin{aligned}
      (\Gamma (\mathbf{v} + \'u))^2 &=  -\lambda_1 \lambda_2 \tensor g(\mathbf{v},\mathbf{v}) + g_{U}(\'u,\'u) \tensor 1 \\
      &=  [\lambda_1 (u^{1})^2 + \lambda_2 (u^{2})^2 -\lambda_1\lambda_2((v^{1})^{2}\\ &\;\;\;+ \cdots +(v^{p})^{2} - (v^{p + 1})^{2} - \cdots -(v^{p+q})^{2})]\, 1 \tensor 1.
    \end{aligned}
\end{align}

\noindent Then, the map $\Gamma$ is a Clifford map with $\Gamma: \rrpq \oplus U \to \cl(W,g_{W})$ where $W$ is a $(n+2)$-dimensional space endowed with a symmetric bilinear functional $g_{W}$ given by

\begin{equation}
g_{W} =  \lambda_1 (u^{1})^2 + \lambda_2 (u^{2})^2 -\lambda_1\lambda_2((v^{1})^{2} + \cdots +(v^{p})^{2} - (v^{p + 1})^{2} - \cdots -(v^{p+q})^{2})\end{equation}

\noindent where $\'w = \'u + \mathbf{v} = u^{1}\'f_1 + u^{2}\'f_2 + v^{i}\'e_i.$ Hence, three possibilities arise:

\begin{itemize}
    \item[\textcolor{BrickRed}{\textbf{(i)}}] If $U = \rr^{1,1}$ then $W = \rr^{p+1,q+1}$, since
\begin{equation}
g_{W} =   (u^{1})^2 -  (u^{2})^2 + ((v^{1})^{2} + \cdots +(v^{p})^{2} - (v^{p + 1})^{2} - \cdots -(v^{p+q})^{2}).\end{equation}
    \item[\textcolor{BrickRed}{\textbf{(ii)}}] If $U = \rr^{2,0}$ then $W = \rr^{q+2,p}$, since
\begin{equation}
g_{W} =   (u^{1})^2 +  (u^{2})^2 - ((v^{1})^{2} + \cdots +(v^{p})^{2} - (v^{p + 1})^{2} - \cdots -(v^{p+q})^{2}).\end{equation}
    \item[\textcolor{BrickRed}{\textbf{(iii)}}] If $U = \rr^{0,2}$ then $W = \rr^{q,p+2}$, since
\begin{equation}
g_{W} =   -(u^{1})^2 -  (u^{2})^2 - ((v^{1})^{2} + \cdots +(v^{p})^{2} - (v^{p + 1})^{2} - \cdots -(v^{p+q})^{2}).\end{equation}
\end{itemize}

In addition, the isomorphism follows from the Clifford algebra universality

\begin{equation*}
\begin{tikzcd}
\rrpq \oplus U \arrow[black]{d}[black,swap]{\,} \rar{\,} & [15pt] \cl(W,g_{W}) \arrow[BrickRed, dashed]{dl}[black]{\simeq} \\ [20pt] \cl(U,g_{U}) \otimes \clpq
\end{tikzcd} 
\end{equation*}

\noindent That is,

\begin{equation*}
\begin{tikzcd}
\rr^{p+1,q+1} \arrow[black]{d}[black,swap]{\,} \rar{\,} & [15pt] \cl_{p+1, q+1} \arrow[BrickRed, dashed]{dl}[black]{\simeq} \\ [20pt] \cl_{1,1} \tensor \clpq
\end{tikzcd} \;\;
\begin{tikzcd}
\rr^{q+2,p} \arrow[black]{d}[black,swap]{\,} \rar{\,} & [15pt] \cl_{q+2,p} \arrow[BrickRed, dashed]{dl}[black]{\simeq} \\ [20pt] \cl_{2,0} \tensor \clpq
\end{tikzcd} \;\;
\begin{tikzcd}
\rr^{q,p+2} \arrow[black]{d}[black,swap]{\,} \rar{\,} & [15pt] \cl_{q,p+2} \arrow[BrickRed, dashed]{dl}[black]{\simeq} \\ [20pt] \cl_{0,2} \tensor \clpq
\end{tikzcd} 
\end{equation*}{}

\noindent which proves the Theorem. \textcolor{BrickRed}{$\Box$}

\begin{tfcor}\label{cor_casperiod1}
    Let $\clpq$ a Clifford algebra associated with the quadratic space $\rr^{p,q}$ and let $k>0$. Then the following isomorphisms hold:\normalfont
    \begin{align}
    \begin{aligned}
    &&\textcolor{BrickRed}{\textbf{(i)}} \;\;\;\;\;\cl_{p+k, q+k} &\simeq\; \cl_{k,k} \tensor \clpq;\\
    &&\textcolor{BrickRed}{\textbf{(ii)}} \;\;\;\;\;\;    \cl_{q+2k,p} &\simeq\; \cl_{2k,0} \tensor \clpq;\\
   &&\textcolor{BrickRed}{\textbf{(iii)}}\;\;\;  \;\;  \cl_{q,p+2k} &\simeq\; \cl_{0,2k} \tensor \clpq.
    \end{aligned}
\end{align}

\noindent \textit{where either $p,q>0$ and $\tensor$ denotes the usual tensor product.}
\end{tfcor}
\noindent \textcolor{BrickRed}{\textit{Proof.}} Let $U$ be a $2k$-dimensional space such that $U = \rr^{k,k}, U= \rr^{0,2k}, U= \rr^{2k,0}$. Therefore, the same steps shown in the proof of the Theorem \ref{teo_casperiod1} provide the desired isomorphisms. \textcolor{BrickRed}{$\Box$}

\paragraph{ } By combining the above isomorphisms in Eq. \eqref{eq_isowesdkjhgf} several others can be obtained. For instance, notice that the repeated use of the relation \textcolor{BrickRed}{\textbf{(i)}} in the Theorem \ref{teo_casperiod1} yields 
\begin{equation}\label{eq_acisoclppcl11}
    \cl_{p,p} \simeq     \color{BrickRed}\overbrace{\color{black}\; \cl_{1,1} \tensor \cdots \tensor \cl_{1,1}\;}^{\text{\color{black}$p$ factors}}  \color{black}= {\tensor^{p}}\, \cl_{1,1}.
\end{equation}

\noindent By the same relation \textcolor{BrickRed}{\textbf{(i)}} in the Theorem \ref{teo_casperiod1}, for $p<q$ it follows that

\begin{align}
    \begin{aligned}
    \clpq &\simeq \cl_{1,1} \tensor \cl_{p-1,q-1}\\
    &\simeq \cl_{1,1} \tensor \cl_{1,1}  \tensor \cl_{p-2,q-2}\\
    &\vdots\\
    &\simeq {\tensor^{p}}\, \cl_{1,1} \tensor \cl_{0,q-p}\\
    &\simeq  \cl_{p,p} \tensor \cl_{0,q-p}.
    \end{aligned}
\end{align}

\noindent As a result, the following relations are derived:

\begin{align}\label{eq_acsclpqppqq}
    \begin{aligned}
    \clpq &\simeq \cl_{p,p} \tensor \cl_{0,q-p} \;\;\;\; (p < q),\\
    \clpq &\simeq \cl_{q,q} \tensor \cl_{p-q,0} \;\;\;\; (p > q).
    \end{aligned}
\end{align}

\noindent In particular, by the relation \textcolor{BrickRed}{\textbf{(iii)}} in Theorem \ref{teo_casperiod1} it is straightforward that

\begin{equation}\label{eq_cl040220}
    \cl_{0,4} \simeq \cl_{0,2} \tensor \cl_{2,0}.
\end{equation}

\noindent and that

\begin{equation}\label{eq_4p9}
    \cl_{2,2} \simeq \cl_{0,2} \tensor \cl_{2,0} \simeq \cl_{1,1} \tensor \cl_{1,1}.
\end{equation}

\noindent Therefore, by the relations \textcolor{BrickRed}{\textbf{(ii)}} and \textcolor{BrickRed}{\textbf{(iii)}} in the Theorem \ref{teo_casperiod1} and the Eq. \eqref{eq_cl040220} it follows that

\begin{align}
    \begin{aligned}
    \cl_{p,q+4} =  \cl_{p,(q+2)+2} &\simeq \cl_{0,2} \tensor \cl_{q+2,p} \\
    &\simeq \cl_{0,2} \tensor \cl_{2,0} \tensor \clpq \\
    &\simeq \cl_{0,4} \tensor \clpq.
    \end{aligned}
\end{align}

\noindent Consequently,

\begin{equation}\label{eq_4p88}
    \cl_{0,8} \simeq \cl_{0,4} \tensor \cl_{0,4}.
\end{equation}

\noindent One also has

\begin{align}\label{eq_r4p10}
    \begin{aligned}
    \cl_{p,q+8} = \cl_{p,(q+4)+4} &\simeq \cl_{0,4} \tensor \cl_{p,q+4} \\
    &\simeq \cl_{0,4} \tensor \cl_{0,4} \tensor \clpq\\
    &\simeq \cl_{0,8} \tensor \clpq.
    \end{aligned}
\end{align}

Another isomorphism (not following from the Theorem  \ref{teo_casperiod1}) that has outstanding importance is given in the next lemma.

\begin{tflemma}\label{lemma_cl2cl11}
$\cl_{2,0} \simeq \cl_{1,1}.$
\end{tflemma}
\noindent \textcolor{BrickRed}{\textit{Proof.}} Each $A \in \cl_{2,0}$ and $B \in \cl_{1,1}$ has the form

\begin{align}
    \begin{aligned}
    A &= a_0 + a_1\'e_1 + a_2\'e_2 + a_{12}\'e_{1}\'e_2, \\
    B &= b_0 + b_1\'f_1 + b_2\'f_2 + b_{12}\'f_{1}\'f_2,
    \end{aligned}
\end{align}

\noindent such that those basis elements satisfy the following relations

\begin{align}
    \begin{aligned}
    (\'e_1)^2 = 1, \;\; (\'e_2)^2 = 1;\\
    (\'f_1)^2 = 1, \;\; (\'f_1)^2 = -1.
    \end{aligned}
\end{align}

\noindent Therefore, one can define a linear map $\phi: \cl_{2,0} \to \cl_{1,1}$ by setting

\begin{equation}
    \phi(1) = 1, \;\;\;\; \phi(\'e_1) = \'f_1, \;\;\;\; \phi(\'e_2) = \'f_1, \;\;\;\; \phi(\'e_1\'e_2) = \'f_2,
\end{equation}

\noindent which is an algebra isomorphism. \textcolor{BrickRed}{$\Box$}

\paragraph{ } From the relations \textcolor{BrickRed}{\textbf{(i)}} and \textcolor{BrickRed}{\textbf{(ii)}} in the Theorem \ref{teo_casperiod1} and the above Lemma \ref{lemma_cl2cl11}, another isomorphism is derived:

\begin{align}\label{eq_casisomorsubalg}
    \begin{aligned}
    \cl_{p+1,q} &\simeq \cl_{1,1} \tensor \cl_{p,q-1} \\
    &\simeq \cl_{2,0} \tensor \cl_{p,q-1} \\
    &\simeq \cl_{q+1,p}.
    \end{aligned}
\end{align}

Such results indicate that by knowing the following low-dimensional Clifford algebras
   {\colorlet{shadecolor}{Red!15}\begin{shaded}\begin{equation}\label{eq_s5d1fgh}
    \cl_{1,0}, \;\;\;\; \cl_{0,1}, \;\;\;\; \cl_{0,2}, \;\;\;\; \cl_{1,1} \simeq \cl_{2,0}
\end{equation}\end{shaded}}

\noindent then we know all of them in arbitrary finite dimensions, since any other Clifford algebras can be constructed using the isomorphisms that have been presented. Those isomorphisms introduce a method for the classification of Clifford algebras that shall be regarded in the next section. Before proceeding, we present the next theorem which will be very important in the classification of spinors in the Chapter \ref{chap_spinors}.

\begin{tftheorem}
    \label{teo_seven}
Let $\clpq$ a Clifford algebra associated with the quadratic space $\rr^{p,q}$ and $\cl_{p,q}^{\text{even}}$ its even subalgebra. Then the following isomorphisms hold:

\begin{equation}
    \cl_{p,q}^{\text{even}} \simeq \cl_{q,p-1} \simeq \cl_{p,q-1} \simeq \cl_{q,p}^{\text{even}}.
\end{equation}
\end{tftheorem}
\noindent \textcolor{BrickRed}{\textit{Proof.}} Let $\{\'e_i, \'f_k\}$ with $i \in \{1,\ldots,p\}$ and $k \in \{1,\ldots,q\}$ an orthonormal basis of the vector space $V$ such that $\cl_{p,q}$ is generated by $1$ and $\{\'e_i, \'f_k\}$, for $i,j \in \{1,\ldots,p\}$ and $k,l \in \{1,\ldots,q\}$ such that 

\begin{align}
\begin{aligned}
       &(\'e_i)^2 = 1, \;\;\;\; (\'f_k)^2 = -1, \\
       &\'e_i\'e_j + \'e_j\'e_i = 0 \;\; (i \neq j), \\
       &\'f_k\'f_l + \'f_l\'f_k = 0 \;\; (k \neq l) ,\\
       &\'e_i\'f_k + \'f_k\'e_i = 0. \;\; 
\end{aligned}
\end{align}

\noindent Then the vector space $\bigwedge^2(\mathbb{R}^{p,q})$ consists of the elements $\{\'e_i\'e_j \;(i \neq j), \'f_k\'f_l \; (k \neq l), \'e_i\'f_l\}$. However, not all such quantities generate the even subalgebra  $\cl_{p,q}^{\text{even}}$. Actually, there is a redundancy. For example, all the bivectors $\{\'f_k\'f_l \; (k \neq l)\}$ may be written in terms of the bivectors of type $\{\'e_i\'f_l\}$, since

\begin{equation}
    (\'e_i\'f_k)(\'e_i\'f_l) = -(\'e_i)^2 \'f_k\'f_l = - \'f_k\'f_l \;\; (k \neq l).
\end{equation}

\noindent Choosing an arbitrary vector, for instance $\'e_1$,  it follows that the set $\{\'e_1,\'e_m, \'e_1,\'f_k\}$ with $m \in \{2,\ldots,p\}$ and $k \in \{1,\ldots,q\}$ generates the space $\bigwedge^2(\mathbb{R}^{p,q})$; consequently it generates the even subalgebra $\clm^{\text{even}}$. Writing such generators of $\clm^{\text{even}}$ as $\xi_{a} = \'e_1 \'e_{a+1}$ for $a \in \{1,\ldots, p-1\}$ and $\zeta_{b} = \'e_1 \'f_{b}$ for $b \in \{1,\ldots, q\}$  it yields

\begin{align}
\begin{aligned}
&(\xi_{a})^2 = -(\'e_1)^2(\'e_{a+1})^{2} = -1, \\
&(\zeta_{b})^2 = -(\'e_1)^2 (\'f_{b})^2 = 1, \\
&\xi_{a}\zeta_{b} + \zeta_{b}\xi_{a} = 0, \\
&\xi_{a}\xi_{c} + \xi_{c}\xi_{a} = 0 \;\; (a \neq c), \\
&\zeta_{b}\zeta_{d} + \zeta_{d}\zeta_{b} = 0 \;\;  (b \neq d).
\end{aligned}
\end{align}

\noindent Therefore, the quantities $\{\zeta_b, \xi_a\}$ for $b \in \{1,\ldots, q\}$  and   $a \in \{1,\ldots, p-1\}$ are the generators of a Clifford algebra associated with a quadratic space $\rr^{q, p-1}$, that is,  $  \cl_{p,q}^{\text{even}} \simeq \cl_{q,p-1} $. The other above isomorphisms naturally follow from the isomorphism written in Eq. \eqref{eq_casisomorsubalg}. \textcolor{BrickRed}{$\Box$}

\section{Classification of Clifford Algebras}\label{section_acsclass}

\paragraph{ } In this section the periodicity of finite dimensional Clifford algebras is explored to finally classify them. The prominence of representation theory is here concerned, since Clifford algebras are naturally related to matrix algebra and their classification is based on it. The representation theory in the algebra field is very important by at least the following reasons: a priori, Clifford algebras seem to be very abstract, however, from their representation by matrix algebra some relevant concrete properties and behaviour of those algebraic structures are explicit seen. Calculations involving matrix algebra is oftentimes easier and simpler rather than the calculations on the Clifford algebras abstract structure. In addition, the matrix algebra dialogue with an abundance of areas and the natural relation of them with Clifford algebras makes Clifford algebras rich in applications. Moreover, based on the periodicity of the Clifford algebras, we work on the representation of the low-dimensional ones to develop the classification of all Clifford algebras. We start by defining a representation of an algebra.

\begin{definvermelho}

Let $\mathcal{A}$ be a real algebra and $V$ a vector space over $\mathbb{K} = \rr, \cc, \hh$. A linear map $\rho : \mathcal{A} \to \endo_{\kk}(V)$ satisfying $\rho(1_{\mathcal{A}}) = 1_{V}$ and $\rho(ab)=\rho(a)\rho(b)$, for all $a,b \in \A$, is called a \textbf{$\kk$-representation} of $\A$.
\end{definvermelho}

\begin{itemize}
    \item[\textcolor{BrickRed}{{$\triangleright$}}] Such vector space $V$ is called the \textit{representation space} or \textit{carrier space} of $\A$. 
    \item[\textcolor{BrickRed}{{$\triangleright$}}]Two representations $\rho_1: \A \to \endo_{\kk}(V_1)$ and $\rho_2: \A \to \endo_{\kk}(V_2)$ are \textit{equivalent} if there exists a $\kk$-isomorphism $\phi: V_1 \to V_2$ satisfying $\rho_2(a) = \phi \circ \rho_1 (a) \circ \phi^{-1}$, for all $ a \in \A.$
     \item[\textcolor{BrickRed}{{$\triangleright$}}] A representation is said to be \textit{faithful} if $\ker \rho = \{0\}$.
     \item[\textcolor{BrickRed}{{$\triangleright$}}] A representation is \textit{irreducible} or \textit{simple} if the only invariant subspaces of $\rho(a)$, for all $a \in \A$ are $V$ and $\{0\}$. It is said to be \textit{reducible} or \textit{semisimple} if $V = V_1 \oplus V_2$ where $V_1$ and $V_2$ are invariant subspaces under the action of $\rho(a)$, for all $a \in \A$.
\end{itemize}

\begin{ex}\label{ex_reprec1}
{{\textcolor{BrickRed}{{$\blacktriangleright$\;}}}} Representations for $\mathbb{C}$. \normalfont Consider the algebra of complex numbers $\cc$. As an algebra over $\cc$ it has two representations $\rho(a+ib) = a + ib$ and $\bar{\rho}(a+ib) = a- ib$. Since there does not exists any linear map $\phi_z: \cc \to \cc; (a+ib) \mapsto \phi_z(a+ib) = z(a+ib$), where $z = x + iy$, s.t., $\bar{\rho}(a+ib) = z\rho(a+ib)z^{-1}$, these two $\cc$-representations are not equivalent. Although, every $\cc$-representation (and also every $\hh$-representation) is a $\rr$-representation. One can define two real representations $\sigma : \cc \to \matreald$ and $\bar{\sigma}: \cc \to \matreald$ as

\begin{align}
    \begin{aligned}
    \sigma (a+ib) = \begin{pmatrix} 
    a & b \\
    -b & a
    \end{pmatrix}, 
    &&&\bar{\sigma}(a + ib) = \begin{pmatrix} 
    a & -b \\
    b & a
    \end{pmatrix}
    \end{aligned}.
\end{align}

\noindent Such representations are equivalent, namely, there exists an isomorphism $\phi: \rr^2 \to \rr^2$ such that $\bar{\sigma}(a + ib) = \phi \sigma(a+ib)\phi^{-1}.$ For instance,

\begin{equation}
    \phi = \frac{1}{\sqrt{2}}\begin{pmatrix} 
    1 & 1 \\
    1 & -1
    \end{pmatrix} = \phi^{-1}.
\end{equation}

\noindent Those $\rr$-representations are irreducible. An example of a reducible $\rr$-representation is provided by

\begin{equation}
    \xi (a+ib) = \begin{pmatrix} 
    a & b & 0 & 0 \\
    -b & a & 0 & 0\\
    0 & 0 & a & -b\\
    0 & 0 & b & a
    \end{pmatrix}. \;\;\textcolor{BrickRed}{{\blacktriangleleft}}
\end{equation}
\end{ex}

\begin{ex}\label{ex_reprec22}
{{\textcolor{BrickRed}{{$\blacktriangleright$\;}}}}  The representation of $\clt$ by \normalfont $\text{Mat}(2,\mathbb{C})$. We recall the Clifford algebras $\text{Mat}(2,\mathbb{C})$ generated by the Pauli matrices $\{1,\sigma_{1},\sigma_{2},\sigma_{3}\}$ described in the Example \ref{ex1matripauli} and the universal Clifford algebra $\clt$ 
described in the Example \ref{ex2cl3}. By setting the following identification in the generators:

\begin{align}
\begin{aligned}\label{eq_mpauli4}
\mathbf{e}_{1} \leftrightarrow  \sigma_{1}, \;\;\;\;\;\;
\mathbf{e}_{2} \leftrightarrow  \sigma_{2}, \;\;\;\;\;\;
\mathbf{e}_{3} \leftrightarrow  \sigma_{3}, \;\;\;\;\;\;
\end{aligned}
\end{align}

\noindent one has the following correspondence of the basis elements of those Clifford algebras:

\begin{center}
\begin{tabular}{cc|cc}
\rowcolor{red!15} & $\text{Mat}(2,\mathbb{C})$  & $\clt$ & \tabularnewline
\cline{1-4} 
 & $I$ & $1$ \tabularnewline
 & $\sigma_{1}$,\;\;$\sigma_{2}$,\;\;$\sigma_{3}$ &  $\mathbf{e}_{1}$,\;\;$\mathbf{e}_{2}$,\;\;$\mathbf{e}_{3}$  \tabularnewline

 & $\sigma_{1}\sigma_{2}$,\;\;$\sigma_{1}\sigma_{3}$,\;\; $\sigma_{2}\sigma_{3}$ & $\mathbf{e}_{1} \mathbf{e}_{2}$,\;\;$\mathbf{e}_{1} \mathbf{e}_{3}$,\;\; $\mathbf{e}_{2} \mathbf{e}_{3}$  \tabularnewline

 & $\sigma_{1}\sigma_{2}\sigma_{3}$ & $\mathbf{e}_{1} \mathbf{e}_{2} \mathbf{e}_{3}$  \tabularnewline
\end{tabular}
\par \end{center}

Therefore, by that identification, an arbitrary element $A \in $ $\clt$

\begin{align}
\begin{aligned}\label{eq_mpauli5A}
A &= a + a_{1}\mathbf{e}_{1} + a_{2}\mathbf{e}_{2} + a_{3}\mathbf{e}_{3} + a_{12}\mathbf{e}_{1}\mathbf{e}_{2} + a_{13}\mathbf{e}_{1}\mathbf{e}_{3} + a_{23}\mathbf{e}_{2}\mathbf{e}_{3} + a_{123}\mathbf{e}_{1}\mathbf{e}_{2}\mathbf{e}_{3}
\end{aligned}
\end{align}

\noindent can be written in terms of $\text{Mat}(2,\mathbb{C})$  as:

\begin{align}
\begin{aligned}\label{eq_mpauli67A}
A &\leftrightarrow \begin{pmatrix} 
    (a + a_{3}) + i (a_{12} + a_{123}) & (a_{1} - a_{13}) - i(a_{2} - a_{23})\\
    (a_{1} + a_{13}) + i(a_{2} + a_{23}) & (a - a_{3}) - i (a_{12} - a_{123})
    \end{pmatrix}=\begin{pmatrix} 
    z_{1} & z_{3} \\
    z_{2} & z_{4} 
    \end{pmatrix}.\\
\end{aligned}
\end{align}
The operations of grade involution, reversion and conjugation defined in the Section \ref{sec_strc} in $\clt$ are given respectively by:
\begin{align}
\begin{aligned}
\hat{A} &= A_{0} -A_{1} + A_{2} - A_{3},\\
\yhwidetilde{A} &= A_{0} + A_{1} - A_{2} - A_{3},\\   
\overline{A} &= A_{0} - A_{1} - A_{2} + A_{3}.
\end{aligned}
\end{align}

\noindent In terms of  $\text{Mat}(2,\mathbb{C})$ they are represented, respectively, by the following matrices:

\begin{align}
\begin{aligned}\label{eq_mpauli7}
\hat{A} &\leftrightarrow \begin{pmatrix} 
    (a  -  a_{3})  -  i(a_{123}  -  a_{12}) & ( -a_{13}  -  a_{1})  +  i(a_{2}  +  a_{23})\\
    (a_{13}  -  a_{1})  +  i(a_{23}  -  a_{2}) & (a  +  a_{3})  -  i (a_{12}  +  a_{123})
    \end{pmatrix}  =  \begin{pmatrix} 
    z_{4}^{*} &  -z_{2}^{*} \\
     -z_{3}^{*} & z_{1}^{*} 
    \end{pmatrix},\\
\yhwidetilde{A} & \leftrightarrow   \begin{pmatrix} 
    (a  +  a_{3})  -  i (a_{12}  +  a_{123}) & (a_{1}  +  a_{13})  -  i(a_{2}  +  a_{23}) \\
    (a_{1}  -  a_{13})  +   i(a_{2}  -  a_{23})\ & (a  -  a_{3}) + i (a_{12}  -   a_{123})
    \end{pmatrix} = \begin{pmatrix} 
    z_{1}^{*} & z_{3}^{*} \\
    z_{2}^{*} & z_{4}^{*} 
    \end{pmatrix},\\     
\overline{A} & \leftrightarrow  \begin{pmatrix} 
     (a  -  a_{3})  -  i (a_{12}  -  a_{123}) & (a_{13}  -  a_{1})  +  i(a_{2}  -  a_{23})\\
    ( -a_{1}  -  a_{13})  -  i(a_{2}  +  a_{23}) & (a  +  a_{3})  +  i (a_{12}  +  a_{123})
    \end{pmatrix}  =  \begin{pmatrix} 
    z_{4} &  -z_{3}  \\
     -z_{2}  & z_{1} 
    \end{pmatrix}.
\end{aligned}
\end{align}

In addition, the elements of the odd part  $A_{\text{odd}} \in \mathcal{C}\ell_{3}^{\text{odd}}$, the even subalgebra $A_{\text{even}} \in \mathcal{C}\ell_{3}^{\text{even}}$, and the centre $A_{\text{cen}} \in \text{Cen}(\mathcal{C}\ell_{3})$ in terms of the algebra $\text{Mat}(2,\mathbb{C})$ are given by:

\begin{align}
\begin{aligned}\label{eq_mpaulicenip}
A_{\text{odd}} = A_{1} + A_{3} &\leftrightarrow 
    \begin{pmatrix} 
    a_{3} + ia_{123} & a_{1} - ia_{2} \\
    a_{1} + ia_{2} & -a_{3} + ia_{123}
    \end{pmatrix}=\begin{pmatrix} 
    w_{1} & w_{2}^{*} \\
    w_{2} & -w_{1}^{*} 
    \end{pmatrix},\\
A_{\text{even}} = A_{0} + A_{2}  &\leftrightarrow \begin{pmatrix} 
    a + ia_{12} & -a_{13} + ia_{23} \\
    a_{13} + ia_{23} & a -  ia_{12}
    \end{pmatrix} =\begin{pmatrix} 
    x_{1} & -x_{2}^{*} \\
    x_{2} & x_{1}^{*} 
    \end{pmatrix},\\
A_{\text{cen}} = A_{0} + A_{3} &\leftrightarrow \begin{pmatrix} 
    a \, + \, ia_{123} & \;0\; \\
    \;0\; & a \, + \, ia_{123} 
    \end{pmatrix}=\begin{pmatrix} 
    y_{1} & 0 \\
    0 & y_{1}
    \end{pmatrix}.\;\;\textcolor{BrickRed}{{\blacktriangleleft}}
\end{aligned}
\end{align}
\end{ex}

\begin{ex}\label{ex_reprec}
{{\textcolor{BrickRed}{{$\blacktriangleright$\;}}}}  The representation of $\cc \otimes \clm$ \normalfont by $\matcompq$.  Consider now the complex Clifford algebra $\cl(\mathbb{C} \otimes V ,g_{\mathbb{C}}) = \mathbb{C} \otimes \cl(V,g)$, this algebra is isomorphic to the matrix algebra $\text{Mat}(4,\mathbb{C})$ generated by the Dirac matrices $\{\gamma_0, \gamma_1, \gamma_2, \gamma_3\}$ presented in the Introduction in the Eq. \eqref{eq_matdiracquatro}. The  generators of $\clm$ are identified with the Dirac matrices 

\begin{equation}
    \mathbf{e}_{0} \leftrightarrow \gamma_{0}, \;\;\;\;\; \mathbf{e}_{1} \leftrightarrow \gamma_{1},\;\;\;\;\; \mathbf{e}_{2} \leftrightarrow \gamma_{2},\;\;\;\;\; \mathbf{e}_{3} \leftrightarrow \gamma_{3}.
\end{equation}

Therefore, we can write an arbitrary element $C \in  \cc \otimes \clm$ in terms of $\matcompq$. This element has the form:

\begin{align}
\begin{aligned}
C &= A + iB \;\;\;\; A,B \in \clm \\
C &=  a + a_{0}\mathbf{e}_{0} + a_{1}\mathbf{e}_{1} + a_{2}\mathbf{e}_{2} + a_{3}\mathbf{e}_{3} + a_{01}\mathbf{e}_{0}\mathbf{e}_{1} + a_{02}\mathbf{e}_{0}\mathbf{e}_{2} + a_{03}\mathbf{e}_{0}\mathbf{e}_{3} +
a_{12}\mathbf{e}_{1}\mathbf{e}_{2} \\&+  a_{13}\mathbf{e}_{1}\mathbf{e}_{3}+ a_{23}\mathbf{e}_{2}\mathbf{e}_{3} + a_{012}\mathbf{e}_{0}\mathbf{e}_{1}\mathbf{e}_{2} + a_{013}\mathbf{e}_{0}\mathbf{e}_{1}\mathbf{e}_{3} +
a_{023}\mathbf{e}_{0}\mathbf{e}_{2}\mathbf{e}_{3} +
a_{123}\mathbf{e}_{1}\mathbf{e}_{2}\mathbf{e}_{3} \\&+  a_{0123}\mathbf{e}_{0}\mathbf{e}_{1}\mathbf{e}_{2}\mathbf{e}_{3} +  i(b + b_{0}\mathbf{e}_{0} + b_{1}\mathbf{e}_{1} + b_{2}\mathbf{e}_{2} + b_{3}\mathbf{e}_{3} + b_{01}\mathbf{e}_{0}\mathbf{e}_{1} + b_{02}\mathbf{e}_{0}\mathbf{e}_{2} \\&+ b_{03}\mathbf{e}_{0}\mathbf{e}_{3} +
b_{12}\mathbf{e}_{1}\mathbf{e}_{2} +  b_{13}\mathbf{e}_{1}\mathbf{e}_{3} + b_{23}\mathbf{e}_{2}\mathbf{e}_{3} + b_{012}\mathbf{e}_{0}\mathbf{e}_{1}\mathbf{e}_{2} + b_{013}\mathbf{e}_{0}\mathbf{e}_{1}\mathbf{e}_{3} \\&+
b_{023}\mathbf{e}_{0}\mathbf{e}_{2}\mathbf{e}_{3} +
b_{123}\mathbf{e}_{1}\mathbf{e}_{2}\mathbf{e}_{3} +  b_{0123}\mathbf{e}_{0}\mathbf{e}_{1}\mathbf{e}_{2}\mathbf{e}_{3}).
\end{aligned}
\end{align} 

In terms of $\matcompq$, one has

\begin{align*}
\begin{aligned}
    &I \!=\! \begin{pmatrix} 
   \!1 & \!0 & \!0 & \!0 \\
    \!0 & \!1 & \!0 & \!0\\
    \!0 & \!0 & \!1 & \!0\\
    \!0 & \!0 & \!0 & \!1
    \end{pmatrix}, \!&\!\gamma_{0} \!=\! \begin{pmatrix} 
   \!1\! & \!0\! & \!0 & \!0 \!\\
    \!0\! & \!1\! & \!0 & \!0\!\\
    \!0\! & \!0\! & \!\!-1\!\! & \!0\!\\
    \!0\! & \!0\! & \!0\! & \!\!-1\!\!
    \end{pmatrix}, \;\;\!&\!\gamma_{1} \!=\! \begin{pmatrix} 
    \!0\! & \!0\! & \!0\! & \!\!-1\! \\
    \!0 \!& \!0\! & \!\!-1\!\! & \!0\!\\
    \!0\! & \!1\! & \!0 & \!0\!\\
    \!1\! & \!0\! & \!0\! & \!0\!
    \end{pmatrix}, \!&\!\gamma_{2} \!=\! \begin{pmatrix} 
   \!0\! & \!0\! & \!0 & \!i\! \\
    \!0 \!& \!0\! & \!\!-i\! & \!0\!\\
    \!0\! & \!\!-i\! & \!0\! & \!0\!\\
    \!i\! & \!0 \!& \!0\! & \!0\!
    \end{pmatrix}, \\
     &\gamma_{3} \!=\! \begin{pmatrix} 
   \!0\! & \!0\! & \!\!-1\!\! & \!0\! \\
    \!0\! & \!0\! & \!0\! & \!1\\
    \!1\! & \!0\! & \!0\! & \!0\!\\
    \!0\! & \!\!-1\!\! & \!0\! & \!0\!
    \end{pmatrix}, \!&\!\gamma_{01} \!=\! \begin{pmatrix} 
   \!0 \! & \! 0 \! & \! 0 \! & \!\!\! -\!1\!\!\\
\!0 \! & \! 0 \! & \!\! -\!1\! & \! 0 \!\\
\!0 \! & \!\!\! -\!1\!\! & \! 0 \! & \! 0 \!\\
\!\!\! -\!1\!\!& \! 0 \! & \! 0 \! & \! 0 \!
    \end{pmatrix}, \;\;\!&\!\gamma_{02} \!=\! \begin{pmatrix} 
    \!0 \! & \! 0 \! & \! 0 \! & \! i \!\\
\!0 \! & \! 0 \! & \! \!-\!i \! & \! 0 \!\\
\!0 \! & \! i \! & \! 0 \! & \! 0 \!\\
\!-\!i \! & \! 0 \! & \! 0 \! & \! 0 \!
    \end{pmatrix},\!&\!\gamma_{03} \!=\! \begin{pmatrix} 
0 \! & \! 0 \! & \! \!-\!1 \! & \! 0 \!\\
\!0 \! & \! 0 \! & \! 0 \! & \! 1 \!\\
\!-\!1 \! & \! 0 \! & \! 0 \! & \! 0 \!\\
\!0 \! & \! 1 \! & \! 0 \! & \! 0\!
    \end{pmatrix}, \\
     &\gamma_{12} \!=\! \begin{pmatrix} 
\!-\!i \! & \! 0 \! & \! 0 \! & \! 0\!\\
\!0 \! & \! i \! & \! 0 \! & \! 0 \!\\
\!0 \! & \! 0 \! & \! \!-\!i \! & \! 0 \!\\
\!0 \! & \! 0 \! & \! 0 \! & \! i \!
    \end{pmatrix}, \!&\!\gamma_{13} \!=\! \begin{pmatrix} 
\!0 \! & \! 1 \! & \! 0 \! & \! 0 \!\\
\!-\!1 \! & \! 0 \! & \! 0 \! & \! 0 \!\\
\!0 \! & \! 0 \! & \! 0 \! & \! 1 \!\\
\!0 \! & \! 0 \! & \! \!-\!1 \! & \! 0 \!
    \end{pmatrix}, \;\;\!&\!\gamma_{23} \!=\! \begin{pmatrix} 
\! 0 \! & \! \!-\!i \! & \! 0 \! & \! 0 \!\\
\!-\!i \! & \! 0 \! & \! 0 \! & \! 0 \!\\
\!0 \! & \! 0 \! & \! 0 \! & \! \!-\!i \!\\
\!0 \! & \! 0 \! & \! \!-\!i \! & \! 0 \!
    \end{pmatrix}, \!&\!\gamma_{012} \!=\! \begin{pmatrix} 
\!-\!i \! & \! 0 \! & \! 0 \! & \! 0 \!\\
\!0 \! & \! i \! & \! 0 \! & \! 0 \!\\
\!0 \! & \! 0 \! & \! i \! & \! 0 \!\\
\!0 \! & \! 0 \! & \! 0 \! & \! \!-\!i \!
    \end{pmatrix}, \\ \!&\!\gamma_{013} \!=\! \begin{pmatrix} 
\!0 \! & \! 1 \! & \! 0 \! & \! 0 \!\\
\!\!-\!1 \!\! & \! 0 \! & \! 0 \! & \! 0 \!\\
\!0 \! & \! 0 \! & \! 0 \! & \! \!\!-\!1\! \!\\
\!0 \! & \! 0 \! & \! 1 \! & \! 0  \!
    \end{pmatrix}, \!&\!\gamma_{023} \!=\! \begin{pmatrix} 
\!0 \! & \! \!-\!i \! & \! 0 \! & \! 0 \!\\
\!-\!i \! & \! 0 \! & \! 0 \! & \! 0 \!\\
\!0 \! & \! 0 \! & \! 0 \! & \! i \!\\
\!0 \! & \! 0 \! & \! i \! & \! 0 \!
    \end{pmatrix}, \;\;\!&\!\gamma_{123} \!\!=\!\! \begin{pmatrix} 
\!0 \! & \! 0 \! & \! 1 \! & \! 0 \!\\
\!0 \! & \! 0 \! & \! 0 \! & \! \!\!-\!1\! \!\\
\!\!-\!1 \!\! & \! 0 \! & \! 0 \! & \! 0 \!\\
\!0 \! & \! 1 \! & \! 0 \! & \! 0 \!
    \end{pmatrix}, \!&\!\gamma_{0123} \!=\! \begin{pmatrix} \!0 \! & \! 0 \! & \! 1 \! & \! 0 \!\\ 
\!0 \! & \! 0 \! & \! 0 \! & \!\! \!-\!1 \!\\
\!1 \! & \! 0 \! & \! 0 \! & \! 0 \!\\
\!0 \! & \!\! \!-\!1 \! & \! 0 \! & \! 0 \!
    \end{pmatrix}.
\end{aligned}
\end{align*}

\noindent Applying the correspondence, it follows that: 

\begin{align*}
\begin{aligned}
& C_{0}= aI + ibI,  \\
&C_{1} = a_{0}\gamma_{0} + 
a_{1}\gamma_{1} + 
a_{2}\gamma_{2} + 
a_{3}\gamma_{3} + i(b_{0}\gamma_{0} + 
b_{1}\gamma_{1} + 
b_{2}\gamma_{2} + 
b_{3}\gamma_{3}), \\
&C_2 =
a_{01}\gamma_{01} + 
a_{02}\gamma_{02} + 
a_{03}\gamma_{03} +
a_{12}\gamma_{12} +
a_{13}\gamma_{13} +
a_{23}\gamma_{23} + i(b_{01}\gamma_{01} + 
b_{02}\gamma_{02} + 
b_{03}\gamma_{03} \\
&+
b_{12}\gamma_{12} +
b_{13}\gamma_{13} +
b_{23}\gamma_{23}),  \\
&C_{3} =
a_{012}\gamma_{012}  + 
a_{013}\gamma_{013} +
a_{023}\gamma_{023} +
a_{123}\gamma_{123} + i(b_{012}\gamma_{012}  + 
b_{013}\gamma_{013} +
b_{023}\gamma_{023} +
b_{123}\gamma_{123}), \\
&C_{4} = a_{0123}\gamma_{0123} + ib_{0123}\gamma_{0123}.
\end{aligned}
\end{align*}

\noindent Therefore $C = C_0 + C_1 + C_2 + C_3 + C_4 \in \cc \otimes \clm$ is represented by the matrix:
\begin{equation}
C =  \begin{pmatrix} 
   m_{11} & m_{12} & m_{13} &  m_{14} \\
   m_{21} & m_{22} & m_{23} &  m_{24} \\
   m_{31} & m_{32} & m_{33} &  m_{34} \\
   m_{41} & m_{42} & m_{43} &  m_{44} \\
    \end{pmatrix} \in \matcompq,
  \end{equation}
  
\noindent such that each term is given by

\begin{align}
\begin{aligned} 
&m_{11} = (a + a_0 + b_{12} + b_{012}) + i( -a_{12} -a_{012} 
+ b + b_0), \\
&m_{12} =  (a_{13} + a_{013} +  b_{23}  + b_{023} ) + i(-a_{23} - a_{023} + b_{13} + b_{013}), \\
&m_{13} = (-a_3 -a_{03} + a_{123} +  a_{0123}) + i(- b_3 - b_{03} + b_{123} + b_{0123}), \\
&m_{14} = (-a_1 -a_{01} - b_2  - b_{02})
+ i(a_2 + a_{02} -b_1- b_{01})  \\
&m_{21} = (-a_{13}  -a_{013}+ b_{23} + b_{023}) +i(- a_{23}  -a_{023} - b_{13} - b_{013} ) ,\\
&m_{22} =( a + a_0 - b_{12} - b_{012}) + i(a_{12} + a_{012} + b + b_0 ) , \\ 
&m_{23} = (-a_1 -a_{01} + b_2 + b_{02}) + i( - a_2  - a_{02} + b_1  - b_{01} ),  \\
&m_{24} = (a_3 + a_{03} -a_{123} -a_{0123}) + i(+ b_3 + b_{03} - b_{123} - b_{0123}) ,\\
&m_{31} = (a_3 -a_{03} -a_{123} + a_{0123})  
+ i(b_3 - b_{03} - b_{123} + b_{0123}),  \\
&m_{32} = (a_1  -a_{01} + b_2 - b_{02}) + i( - a_2  + a_{02} + b_1 - b_{01} ), \\
&m_{33} = (a -a_0  + b_{12} - b_{012} ) 
+ i(-a_{12} + a_{012}  + b - b_0),\\
&m_{34} =  (a_{13}-a_{013} + b_{23} - b_{023})
+i(-a_{23}  + a_{023} + b_{13} - b_{013}), \\
&m_{41} = (a_1  -a_{01} - b_2 + b_{02}) + i( a_2- a_{02} + b_1 - b_{01} ), \\
&m_{42} = (-a_3 + a_{03}  +  a_{123} -a_{0123}  + b_{23})  + i(-a_{23}-b_3 + b_{03} + b_{123} - b_{0123}),\\
&m_{43} =  (-a_{13} + a_{013} - b_{023}) + i(a_{023} -b_{13} + b_{013}),  \\
&m_{44} = (a -a_0 - b_{12} + b_{012}) + i(a_{12} -a_{012} + b -b_0 ). \\
\end{aligned}
\end{align}

\noindent This example explicitly provides the representation of an arbitrary element in the complex Clifford algebra $\cc \otimes \clm$. \; $\textcolor{BrickRed}{{\blacktriangleleft}}$ \end{ex} 

\paragraph{ } We recall the important result developed in the previous Section \ref{sec_strc} in the Eq. \eqref{eq_s5d1fgh} which states that by knowing four low-dimensional Clifford algebras, namely, $\cl_{1,0},$  $\cl_{0,1},$ $\cl_{0,2}$ and $ \cl_{1,1} \simeq \cl_{2,0}$, then we know the next ones by applying the Theorem \ref{teo_casperiod1}. Our next goal, aiming the classification, is therefore to establish the representation of such low-dimensional Clifford algebras.

{\colorlet{shadecolor}{red!10}\begin{shaded}  \textcolor{BrickRed}{\textbf{(I)}} \textbf{The Clifford Algebra }$\cl_{0,1}$\centering \end{shaded}}


\paragraph{ } A Clifford algebra associated to the quadratic space $\rr^{0,1}$ was introduced in the Example \ref{ex_acs1comp}. If $\mathbf{e}$ is an unit vector such that $g(\mathbf{e},\mathbf{e}) = -1$, an arbitrary element $\psi \in \cl_{0,1}$ is written as

\begin{equation}
    \psi = a+b\mathbf{e} \in \cl_{0,1}
\end{equation}

\noindent where $\mathbf{e}^2 = -1$. This algebra is isomorphic to the complex algebra $\cc$, namely, the set of pairs $(a,b) \in \rr^2$ endowed with the multiplication given by

\begin{equation}
    (a,b)(c,d) = (ac - bd, ad + bc).
\end{equation}

\noindent The isomorphism is establish by $\rho: \cl_{0,1} \to \cc$ such that $\rho(1) = (1,0)$ and $\rho(\mathbf{e}) = (0,1) = i$. Therefore

\begin{equation}
    \textcolor{BrickRed}{\textbf{(I)}}\;\;\;   \cl_{0,1} \simeq \cc.
\end{equation}

{\colorlet{shadecolor}{red!10}\begin{shaded}  \textcolor{BrickRed}{\textbf{(II)}} \textbf{The Clifford Algebra }$\cl_{1,0}$ \centering \end{shaded}}


\paragraph{ } Consider the quadratic space $\rr^{1,0}$. Taking the unit vector $\mathbf{e}$ such that $g(\mathbf{e},\mathbf{e}) = 1$, an arbitrary element $\psi$ of $\cl_{1,0}$ reads

\begin{equation}
    \psi = a + b\mathbf{e} \in \cl_{1,0}
\end{equation}

\noindent with $\mathbf{e}^2 = 1$. The difference between this case and the previous one is that in this case $\mathbf{e}^2 = 1$ instead of $\mathbf{e}^2 = -1$. This leads some drastic distinct consequences. To be more precise, 
consider the set of the pairs of numbers $(a,b) \in \rr^2$ with multiplication defined by

\begin{equation}
    (a,b)(c,d) = (ac + bd, ad +bc).
\end{equation}

\noindent One denotes this set by $\mathbb{D}$, whose elements have different denominations: double numbers, perplex numbers, duplex or Lorentz numbers \cite{Roc16}. In particular, such set is not a field, but a ring. It is neither a division ring, since $(1,1)(1,-1) = (0,0)$. Despite the above defined multiplication being appropriate to compare $\mathbb{D}$ to $\cc$, it is not so suitable for the Clifford algebras classification. Therefore, let us consider the set of the pairs of numbers $(a,b) \in \rr^2$ endowed with a product defined by

\begin{equation}
    (a,b) * (c,d) = (ac,bd)
\end{equation}

\noindent This algebra is the direct sum of real algebras, namely, $\rsr$. It is isomorphic to  the $2 \cross 2$ diagonal matrices. Such isomorphism is given by

\begin{equation}
    \phi(a,b) = \begin{pmatrix} 
    a & 0 \\
    0 & b
    \end{pmatrix}
\end{equation}

\noindent The algebra $\rsr$ is also isomorphic to $\mathbb{D}$. The isomorphism $\phi: \mathbb{D} \to \rsr$ reads

\begin{equation}
    \varphi (a,b) = (a+b, a-b).
\end{equation}

\noindent In fact, let $(a,b), (c,d) \in \mathbb{D}$

\begin{align}
\begin{aligned}
\varphi ((a,b)(c,d)) &= \varphi (ac + bd, ad + bc) \\
&= (ac + bd + ad + bc, ac + bd - ad -bc) \\
&= ((a+b)(c+d),(a-b)(c-d)) \\
&= \varphi(a,b) * \varphi(c,d).
\end{aligned}
\end{align}

\noindent On the other hand, $\cl_{1,0}$ is isomorphic to  $\mathbb{D}$ by the identification $1 \leftrightarrow (1,0)$ and $\mathbf{e}  \leftrightarrow (0,1)$. Hence,

\begin{equation}
    \textcolor{BrickRed}{\textbf{(II)}}\;\;\;   \cl_{1,0} \simeq \rsr.
\end{equation}

{\colorlet{shadecolor}{red!10}\begin{shaded}  \textcolor{BrickRed}{\textbf{(III)}} \textbf{The Clifford Algebra } $\cl_{0,2}$\centering \end{shaded}} 
 \paragraph{ } Consider now the quadratic space $\rr^{0,2}$ and an orthonormal basis $\{\mathbf{e}_1, \mathbf{e}_2\}$, it follows that

\begin{equation}
    g(\mathbf{e}_1,\mathbf{e}_1) = g(\mathbf{e}_2,\mathbf{e}_2) = -1, \;\;\;\;\;\; g(\mathbf{e}_1,\mathbf{e}_2) = g(\mathbf{e}_2,\mathbf{e}_1) = 0.
\end{equation}

\noindent An arbitrary element $\psi$ of $\cl_{0,2}$ is written as

\begin{equation}
    \psi = a + b\mathbf{e}_1 + c\mathbf{e}_2 + d\mathbf{e}_1\mathbf{e}_2 \in \cl_{0,2},
\end{equation}

\noindent such that $a,b,c,d \in \rr$ and

\begin{equation}
    (\mathbf{e}_1)^2 = (\mathbf{e}_2)^2  = -1, \;\;\;\; \mathbf{e}_1\mathbf{e}_2 + \mathbf{e}_2\mathbf{e}_1 = 0, \;\;\;\; (\mathbf{e}_1\mathbf{e}_2)^2 = -1.
\end{equation}

\noindent The Clifford algebra $\cl_{0,2}$ is isomorphic to the quaternion algebra $\hh$ through the identification \cite{Roc16}
\begin{equation}
    1 \leftrightarrow 1, \;\;\;\; \mathbf{e}_1 \leftrightarrow i, \;\;\;\; \mathbf{e}_2 \leftrightarrow j, \;\;\;\; \mathbf{e}_1\mathbf{e}_2 \leftrightarrow k,
\end{equation}
\begin{center}
\begin{tabular}{cc|cc}
\rowcolor{red!15} &  $\cl_{0,2}$  & $\hh$ &\; \; \tabularnewline
\cline{1-4} 
 & $\mathbf{e}_{1}$ &  $i$ \tabularnewline
 & $\mathbf{e}_{2}$ &  $j$  \tabularnewline
 & $\mathbf{e}_{1}\mathbf{e}_{2}$ &  $k$  \tabularnewline
\end{tabular}
\par \end{center}

\noindent where $i,j,k$ are the quaternion units

\begin{equation}
\begin{gathered}
i^2 = j^2 = k ^2 = ijk = -1, \\
jk = -kj = i, \\
ki = -ik = j, \\
ij = -ji = k. \\
\end{gathered}\label{eq_quaternionrelfund}   
\end{equation}

\noindent Therefore,

\begin{equation}
  \textcolor{BrickRed}{\textbf{(III)}}\;\;\;  \cl_{0,2} \simeq \hh.
\end{equation}

{\colorlet{shadecolor}{red!10}\begin{shaded}\textcolor{BrickRed}{\textbf{(IV)}} \textbf{The Clifford Algebra } $\cl_{2,0} \simeq \cl_{1,1}$\centering\end{shaded}} 


The Clifford algebras associated to the quadratic spaces $\rr^{2,0}$ and $\rr^{1,1}$ were shown to be isomorphic in Lemma \ref{lemma_cl2cl11}. Hence, it suffices to consider just one of those spaces. For instance, let us consider $\rr^{2,0}$, it holds for an orthonormal basis $\{ \mathbf{e}_1, \mathbf{e}_2\}$  that

\begin{equation}
    g(\mathbf{e}_1,\mathbf{e}_1) = g(\mathbf{e}_2, \mathbf{e}_2) = 1, \;\;\;\; g(\mathbf{e}_1,\mathbf{e}_2) = g(\mathbf{e}_2,\mathbf{e}_1) = 0.
\end{equation}

\noindent One may write an arbitrary element $\psi$ of $\cl_{2,0}$ as

\begin{equation}
       \psi = a + b\mathbf{e}_1 + c\mathbf{e}_2 + d\mathbf{e}_1\mathbf{e}_2 \in \cl_{2,0},
\end{equation}

\noindent where $a,b,c,d \in \rr$ and

\begin{equation}
    (\mathbf{e}_1)^2 = (\mathbf{e}_2)^2  = 1, \;\;\;\; \mathbf{e}_1\mathbf{e}_2 + \mathbf{e}_2\mathbf{e}_1 = 0, \;\;\;\; (\mathbf{e}_1\mathbf{e}_2)^2 = -1.
\end{equation}

Let $\matreald$ be the set of the real $2 \cross 2$ matrices. The generator of such algebra is the following set

\begin{equation}\label{eq_acl2100}
    \left\{  \begin{pmatrix} 
    1 & 0 \\
    0 & 1
    \end{pmatrix}, \begin{pmatrix} 
    1 & 0 \\
    0 & -1
    \end{pmatrix},\begin{pmatrix} 
    0 & 1 \\
    1 & 0
    \end{pmatrix},\begin{pmatrix} 
    0 & 1 \\
    -1 & 0
    \end{pmatrix}\right\}.
\end{equation}

\noindent Furthermore,

\begin{equation}\label{eq_acl211112}
    \begin{pmatrix} 
    1 & 0 \\
    0 & -1
    \end{pmatrix}^2 = \begin{pmatrix} 
    1 & 0 \\
    0 & 1
    \end{pmatrix} = \begin{pmatrix} 
    0 & 1 \\
    1 & 0
    \end{pmatrix}^2, \;\;\;\;\;\; \begin{pmatrix} 
    0 & 1 \\
    -1 & 0
    \end{pmatrix}^2 = -\begin{pmatrix} 
    1 & 0 \\
    0 & 1
    \end{pmatrix}.
\end{equation}

\noindent Comparing the relations in the Eqs. \eqref{eq_acl2100} and \eqref{eq_acl211112}, an isomorphism between those algebras can be constructed by the identification

\begin{align}\label{eq_ac_cl2isomorhicmat2r}
    \begin{aligned}
    1 \leftrightarrow \begin{pmatrix} 
    1 & 0 \\
    0 & 1
    \end{pmatrix}, \;\;\;\;\; \mathbf{e}_1 \leftrightarrow \begin{pmatrix} 
    1 & 0 \\
    0 & -1
    \end{pmatrix}, \;\;\;\;\; \mathbf{e}_2 \leftrightarrow \begin{pmatrix} 
    0 & 1 \\
    1 & 0
    \end{pmatrix}, \;\;\;\;\; \mathbf{e}_1\mathbf{e}_2 \leftrightarrow \begin{pmatrix} 
    0 & 1 \\
    -1 & 0
    \end{pmatrix}
    \end{aligned}.
\end{align}

\noindent Hence,

\begin{equation}
    \textcolor{BrickRed}{\textbf{(IV)}}\;\;\; \cl_{2,0} \simeq \cl_{1,1} \simeq \matreald.
\end{equation}

\subsection*{Clifford Algebras Classification}\label{subsec245}

 Once the isomorphisms 

{\colorlet{shadecolor}{red!10}\begin{shaded}\vspace{-0.4cm}\begin{align}\label{eq_isos}
    \begin{aligned}
    &\textcolor{BrickRed}{\textbf{(I)}}  &&\cl_{0,1} \simeq \cc, \\
   &\textcolor{BrickRed}{\textbf{(II)}}   &&\cl_{1,0} \simeq \rsr, \\
    &\textcolor{BrickRed}{\textbf{(III)}}  &&\cl_{0,2} \simeq \hh,\\
    &\textcolor{BrickRed}{\textbf{(IV)}}   &&\cl_{2,0} \simeq \cl_{1,1} \simeq \matreald
    \end{aligned}
\end{align} \vspace{-0.5cm}\end{shaded}}

\noindent has been established, by using the isomorphisms presented in the previous Chapter, the classification of arbitrary Clifford algebras can proceed. For instance, by using Eq. \eqref{eq_acisoclppcl11} and the following relation

\begin{equation}
    \text{Mat}(m,\rr) \tensor \text{Mat}(n,\rr) \simeq \text{Mat}(mn,\rr)
\end{equation}

\noindent it follows that

\begin{equation}
  \cl_{p,p} \simeq  {\tensor^{p}}\, \cl_{1,1} \simeq {\tensor^{p}}\, \matreald \simeq \text{Mat}(2^{p}, \rr).
\end{equation}

\noindent This result together with the relation \textcolor{BrickRed}{\textbf{(III)}} in Eq. \eqref{eq_isos}  and the Eq. \eqref{eq_4p9}, allow us to conclude that

\begin{equation}
    \hh \tensor \hh \simeq \cl_{0,2} \tensor \cl_{0,2} \simeq \cl_{1,1} \tensor \cl_{1,1} \simeq  \text{Mat}(2^2,\rr) \simeq \matrealq.
\end{equation}

\noindent  Eq. \eqref{eq_cl040220} and the relations \textcolor{BrickRed}{\textbf{(III)}} and \textcolor{BrickRed}{\textbf{(IV)}} in Eq. \eqref{eq_isos} yield

\begin{equation}
    \cl_{0,4} \simeq \cl_{0,2} \tensor \cl_{2,0} \simeq \hh \tensor \matreald \simeq \text{Mat}(2,\hh) \simeq \matreald \tensor \hh \simeq \cl_{4,0}.
\end{equation}

\noindent Together with those previous results, it follows from Eq. \eqref{eq_4p88} that

\begin{align}\label{eq_cl08mat16}
    \begin{aligned}
       \cl_{0,8} &\simeq \cl_{0,4} \tensor \cl_{0,4} \simeq \text{Mat}(2,\hh) \tensor \text{Mat}(2,\hh) \simeq \matreald \tensor \hh \tensor \hh \tensor \matreald \\
       &\simeq \matreald \tensor \matrealq \tensor \matreald \simeq \text{Mat}(16,\rr).
    \end{aligned}
\end{align}

\noindent Now, using the above result \eqref{eq_cl08mat16} and Eq. \eqref{eq_r4p10} the following theorem has been proved.

\begin{tftheorem}\label{teo_att_shapiro}
\textbf{(Atiyah-Bott-Shapiro Periodicity Theorem \cite{Atiyah})}. For every quadratic space $\rr^{p,q}$ it follows that \normalfont

\begin{equation}
    \cl_{p,q+8} \simeq \clpq \tensor \text{Mat}(16,\rr).
\end{equation}
\end{tftheorem}

\noindent The Atiyah-Bott-Shapiro Periodicity Theorem \ref{teo_att_shapiro} has a very important consequence: it tells us that we only need to explicitly obtain the classification of the Clifford algebras up to $\dim V = p +q =8$, since for higher dimensions one can use the isomorphism $\cl_{p,q+8} \simeq \clpq \tensor \text{Mat}(16,\rr).$ By using all the previous results in this Chapter, we obtain:

\begin{align*}
    \begin{aligned}
    \cl_{2,0} &\simeq \cl_{1,1} \simeq \matreald \\
    \cl_{3,0} &\simeq \cl_{1,2} \simeq \text{Mat}(2,\cc)\\
    \cl_{0,4} &\simeq \cl_{4,0} \simeq \cl_{1,3} \simeq \text{Mat}(2,\hh) \\
    \cl_{3,1} &\simeq \cl_{2,2} \simeq \matrealq \\
    \cl_{5,0} &\simeq \cl_{1,4} \simeq \text{Mat}(2, \hsh) \\
    \cl_{0,5} &\simeq \cl_{4,1} \simeq  \cl_{2,3} \simeq  \matcompq \\
    \cl_{6,0} &\simeq \cl_{5,1} \simeq \cl_{1,5} \simeq \cl_{2,4} \simeq \text{Mat}(4,\hh) \\
    \cl_{7,0} &\simeq \cl_{1,6} \simeq \cl_{5,2} \simeq \cl_{3,4} \simeq \text{Mat}(8,\cc) \\
    \cl_{0,7} &\simeq  \cl_{4,3} \simeq \text{Mat}(8, \rsr)\\
    \cl_{6,1} &\simeq \cl_{2,5} \simeq \text{Mat}(4, \hsh)
    \end{aligned}
\end{align*}

One can organise the algebras of dimension $n<8$ according to $p-q$, obtaining

\begin{align*}
    \begin{aligned}
    &p - q = 0:\;\; \cl_{0,0},\;\; \cl_{1,1},\;\; \cl_{2,2},\;\; \cl_{3,3} &&\text{Mat}(2^{\left[\frac{n}{2} \right]}, \rr) \\
    &p - q = 1:\;\; \cl_{1,0},\;\; \cl_{2,1},\;\; \cl_{3,2},\;\; \cl_{4,3},\;\; \cl_{0,7}\;\;\;\; &&\text{Mat}(2^{\left[\frac{n}{2} \right]}, \rr) \oplus \text{Mat}(2^{\left[\frac{n}{2} \right]}, \rr)\\
    &p - q = 2:\;\; \cl_{2,0},\;\; \cl_{3,1},\;\; \cl_{4,2},\;\; \cl_{0,6} &&\text{Mat}(2^{\left[\frac{n}{2} \right]}, \rr) \\
    &p - q = 3:\;\; \cl_{3,0},\;\; \cl_{4,1},\;\; \cl_{5,2},\;\; \cl_{0,5},\;\; \cl_{1,6} \;\;\;\; &&\text{Mat}(2^{\left[\frac{n}{2} \right]}, \cc) \\
    &p - q = 4:\;\; \cl_{4,0},\;\; \cl_{5,1},\;\; \cl_{0,4},\;\; \cl_{1,5} &&\text{Mat}(2^{\left[\frac{n}{2} \right]-1}, \hh) \\
    &p - q = 5:\;\; \cl_{5,0},\;\; \cl_{6,1},\;\; \cl_{0,3},\;\; \cl_{1,4},\;\; \cl_{2,5} &&\text{Mat}(2^{\left[\frac{n}{2} \right]-1}, \hh) \oplus \text{Mat}(2^{\left[\frac{n}{2} \right]-1}, \hh) \\
    &p - q = 6:\;\; \cl_{6,0},\;\; \cl_{0,2},\;\; \cl_{1,3},\;\; \cl_{2,4} &&\text{Mat}(2^{\left[\frac{n}{2} \right] -1}, \hh)\\
    &p - q = 7:\;\; \cl_{7,0},\;\; \cl_{0,1},\;\; \cl_{1,2},\;\; \cl_{2,3},\;\; \cl_{3,4}\;\;\;\; &&\text{Mat}(2^{\left[\frac{n}{2} \right]}, \cc)
    \end{aligned}
\end{align*}

\noindent The Appendix \ref{clifclassapp} contains the detailed computations. Consequently, the real Clifford algebras over $\rr^{p,q}$ are classified as

\begin{table}[H]
\begin{center}
\scalebox{1.33}{\begin{tabular}{c|c}
\rowcolor{red!15} 
\rule{0pt}{3ex} $p - q \mod 8 $ & $\cl_{p,q}$                   \\ \hline \rule{0pt}{3ex}$0$ & $\text{Mat}(2^{\left[\frac{n}{2} \right]}, \rr)$ \\
\rowcolor{red!15} \rule{0pt}{3ex} $1$ & $\text{Mat}(2^{\left[\frac{n}{2} \right]}, \rr) \oplus \text{Mat}(2^{\left[\frac{n}{2} \right]}, \rr)$ \\
\rule{0pt}{3ex} $2$& $\text{Mat}(2^{\left[\frac{n}{2} \right]}, \rr)$\\
\rowcolor{red!15}\rule{0pt}{3ex} $3$ & $\text{Mat}(2^{\left[\frac{n}{2} \right]}, \cc) $ \\
\rule{0pt}{3ex}$4$ & $\text{Mat}(2^{\left[\frac{n}{2} \right]-1}, \hh)$ \\
\rowcolor{red!15} \rule{0pt}{3ex}$5$  & $\text{Mat}(2^{\left[\frac{n}{2} \right]-1}, \hh) \oplus \text{Mat}(2^{\left[\frac{n}{2} \right]-1}, \hh)$ \\
\rule{0pt}{3ex}$6$  & $\text{Mat}(2^{\left[\frac{n}{2} \right] -1}, \hh)$ \\
\rowcolor{red!15} \rule{0pt}{3ex}$7$ & $\text{Mat}(2^{\left[\frac{n}{2} \right]}, \cc)$
\end{tabular}}\caption{Real Clifford Algebras Classification.}\label{table1}
\end{center}
\end{table}

\begin{ex}
{{\textcolor{BrickRed}{{$\blacktriangleright$\;}}}} Real Clifford algebras classification table usage. \normalfont Let us illustrate the usage of the classification Table \ref{table1}. Consider the Clifford algebra $\cl_{3,0}$, where $p - q  = 3$, by looking at the above table it can be seen that it is isomorphic to $\text{Mat}(2^{\left[\frac{n}{2} \right]}, \cc)$. Since $n = p + q = 3$ and $\left [\frac{n}{2} \right]= \left [\frac{3}{2} \right] = 1$, we can conclude that $\cl_{3,0} \simeq \text{Mat}(2^1, \cc) = \matcomp$. Consider now the algebra $\cl_{0,2}$. In this case $p - q = -2 = 6 \mod 8$ and the corresponding algebra is $\text{Mat}(2^{\left[\frac{n}{2} \right] -1}, \hh)$. Since $n = p + q = 2$  and $\left[\frac{n}{2} \right] = \left[\frac{2}{2} \right] = 1$ it follows that $\cl_{0,2} \simeq \text{Mat}(2^{1-1}, \hh) = \text{Mat}(1,\hh) = \hh$ as already seen previously. \;$\textcolor{BrickRed}{{\blacktriangleleft}}$\end{ex}

\paragraph{ } With respect to the complex case the classification can be obtained by the relation $\cl(V_{\mathbb{C}},g_{\mathbb{C}}) \simeq  \mathbb{C} \otimes \cl(V,g) $ shown in Theorem \ref{teo_acsestrcomplex}. We denote $\cc \tensor \clvg$ as $\cl_{\cc}(n)$. It follows that\cite{Roc16}

\begin{align*}
    \begin{aligned}
    &p - q = 0:\;\; &&\cc \otimes \text{Mat}(2^n,\rr) \simeq \text{Mat}(2^n,\cc)\\
     &p - q = 1:\;\; &&\cc \otimes \text{Mat}(2^{n}, \rr) \oplus \text{Mat}(2^n, \rr) \simeq \text{Mat}(2^{n}, \cc) \oplus \text{Mat}(2^n, \cc)\\
    &p - q = 2:\;\; &&\cc \tensor \text{Mat}(2^n, \rr) \simeq \text{Mat}(2^n,\cc)\\
    &p - q = 3:\;\; &&\cc \tensor \text{Mat}(2^n, \cc) \simeq \text{Mat}(2^n, \cc) \oplus \text{Mat}(2^n, \cc) \\
    &p - q = 4:\;\; &&\cc \tensor \text{Mat}(2^{n-1}, \hh) \simeq \text{Mat}(2^n, \cc) \\
    &p - q = 5:\;\; &&\cc \tensor \text{Mat}(2^{n-1}, \hh) \oplus \text{Mat}(2^{n-1}, \hh) \simeq \text{Mat}(2^n, \cc) \oplus \text{Mat}(2^n, \cc) \\
    &p - q = 6:\;\; &&\cc \tensor \text{Mat}(2^{n -1}, \hh)  \simeq \text{Mat}(2^n,\cc) \\
    &p - q = 7:\;\; &&\cc \tensor \text{Mat}(2^n, \cc) \simeq \text{Mat}(2^n, \cc) \oplus \text{Mat}(2^n, \cc)
    \end{aligned}
\end{align*}

\noindent As we have seen, the complex Clifford algebra $\cl_{\cc}(n)$ depends only on the parity of $n = p + q$. Hence, the complex Clifford algebra classification is given by the following table

\begin{table}[H]
\begin{center}
\scalebox{1.2}{\begin{tabular}{c|c}
\rowcolor{red!15}  & \\ \rowcolor{red!15} 
$n$ even & $\cl_{\cc}(2k) \simeq \text{Mat}(2^{k}, \cc)$                                                                               \\ 
\rowcolor{red!15} & \\ \hline \\
$n$ odd             & $\cl_{\cc}(2k+1) \simeq \text{Mat}(2^{k}, \cc) \oplus \text{Mat}(2^{k}, \cc)$ \\                   \\          
\end{tabular}}  \caption{Complex Clifford Algebras Classification.}\label{table2}
\end{center}
\end{table}

\paragraph{ } Despite the classification provides for all Clifford algebras the isomorphic matrix algebra associated, it does not explicitly assert how to develop such isomorphisms. For instance, the table tells us that $\cl_{2,0} \simeq \matreald$ but it does not tell us a way to explicitly obtain such isomorphism as given by Eq. \eqref{eq_ac_cl2isomorhicmat2r}. However, in cases involving lower dimensional algebras like $\cl_{2,0}$, it is not so complicated to find a isomorphism, but once the dimension of the space increases, it is not trivial explicitly assert how to write such isomorphisms. Along with that, the Appendix \ref{app7} contains an exhibition of the method to obtain a matrix representation for Clifford algebras.

\colorlet{chapter}{orange!50}
\chapter{Groups and Symmetries}\label{chap_groups}

\hypersetup{
  colorlinks = true,
  linkcolor  = RedOrange,
  citecolor = Orange,
}
\paragraph{ } Orthogonal transformations are linear transformations that preserve the inner product of a vector space. Orthogonal transformations and Clifford algebras are interconnected in both mathematical and physical contexts.  \textcolor[rgb]{0.82,0.01,0.11}{Clifford algebras} provide a natural framework for representing orthogonal groups and enable the unification of rotations and reflections under a single algebraic structure. In this chapter, we explore the essential concepts of groups and symmetries, focusing on the groups that can be defined within Clifford algebras. We want to understand rotations and reflections with the Clifford algebra framework in various dimensions, leading to the study of \textcolor[rgb]{0.88,0.82,0.11}{spinors}. This way, the final goal of this chapter is to introduce \textcolor[rgb]{0.96,0.65,0.14}{the Pin and Spin groups}. The Spin group, which is a double cover of the special orthogonal group $\SO(n)$, can be constructed within the Clifford algebras and is crucial for defining spinors — objects that carry representations of this group. The Appendix \ref{app5} has a discussion about orthogonal transformations and the group $\O(p,q)$ and its components. The main reference of this Chapter is Ref. \cite{Roc16}.

\section{Orthogonal Symmetries and Reflections}\label{sec_23w4ertygfd}

\paragraph{ } We begin by studying orthogonal symmetries and reflections, the goal of this section is to describe that a reflection of a vector $\mathbf{v}$ on the hyperplane orthogonal to the vector $\mathbf{u}$ can be expressed in the form '$\mathbf{u}\mathbf{v}\mathbf{u}^{-1}$'. Using this formulation, we introduce the double covering homomorphism from the Spin group to the special orthogonal group in the next section. Additionally, we present the Cartan-Dieudonné Theorem, which provides another way to express any orthogonal transformation in a finite-dimensional vector space. This theorem is particularly significant for representing the Spin group, as it offers an alternative and insightful characterisation of the group's structure.

\paragraph{ }Let $V$ be a quadratic vector space and $U$ be a non-isotropic subspace of $V$, that is, for all $\mathbf{u} \in U$ one has $g(\mathbf{u},\mathbf{u}) \neq 0$. Consider now the orthogonal complement of $U$, 

\begin{equation}
    U^{\perp} = \{ \mathbf{v} \in V : g(\mathbf{v},\mathbf{u}) = 0 \;\; \text{for all} \;\; \mathbf{u} \in U\}.
\end{equation}

The vector space $V$ can be written as $V = U \oplus U^{\perp}$. Given a vector $v = \mathbf{v}_{\parallel} + \mathbf{v}_{\perp} \in V$ one defines the \textit{orthogonal symmetry} $S_{U}$ with respect to $U$ as \cite{Roc16}

\begin{equation}
    S_{U}(\mathbf{v}_{\parallel} + \mathbf{v}_{\perp}) = -\mathbf{v}_{\parallel} + \mathbf{v}_{\perp}
\end{equation}

\noindent such that $\mathbf{v}_{\parallel} \in U$ and  $\mathbf{v}_{\perp} \in  U^{\perp}$. That way, using the identity operator $\id$, $S_{U} $ can be written as $S_{U} = -\id_{U} + \id_{U^{\perp}}$, consequently:

\begin{equation}\label{eq_acgdet004}
    \det S_{U} = (-1)^{\dim U}
\end{equation}

Let us choose a vector $\mathbf{u} \in V$ such that $g(\mathbf{u},\mathbf{u}) \neq 0$ and take $U = \{a\mathbf{u} : a \in \rr \}$, the subspace generated by $\mathbf{u}$. For $\mathbf{v} \in V$, by defining, 

\begin{equation}
    \'w = \mathbf{v} - \frac{g(\mathbf{v},\mathbf{u})}{g(\mathbf{u},\mathbf{u})}\mathbf{u}
\end{equation}

\noindent it follows that:

\begin{align}
    \begin{aligned}
    g(\mathbf{u},\'w) = g\left (\mathbf{u}, \mathbf{v} - \frac{g(\mathbf{v},\mathbf{u})}{g(\mathbf{u},\mathbf{u})}\mathbf{u} \right ) =  g (\mathbf{u},\mathbf{v}) - \frac{g(\mathbf{v},\mathbf{u})}{g(\mathbf{u},\mathbf{u})} g(\mathbf{u},\mathbf{u} ) =0.
    \end{aligned}
\end{align}

\noindent Therefore $\'w \in U^{\perp}$. In that manner, one can take $\mathbf{v} = \mathbf{v}_{\parallel} + \mathbf{v}_{\perp} \in V$ such that:

\begin{align}
    \begin{aligned}
    \mathbf{v}_{\perp} &= \mathbf{v} - \frac{g(\mathbf{v},\mathbf{u})}{g(\mathbf{u},\mathbf{u})}\mathbf{u} \in U^{\perp} ,\\
    \mathbf{v}_{\parallel} &=\frac{g(\mathbf{v},\mathbf{u})}{g(\mathbf{u},\mathbf{u})}\mathbf{u} \in U.
    \end{aligned}
\end{align}

The symmetry $S_{\mathbf{u}}$ is thus given by

\begin{align}\label{eq_acg517}
    \begin{aligned}
    S_{\mathbf{u}}(\mathbf{v}) &= S_{\mathbf{u}}(\mathbf{v}_{\parallel} + \mathbf{v}_{\perp})\\ &= -\mathbf{v}_{\parallel} + \mathbf{v}_{\perp}\\
    &= - \frac{g(\mathbf{v},\mathbf{u})}{g(\mathbf{u},\mathbf{u})}\mathbf{u} + \mathbf{v} - \frac{g(\mathbf{v},\mathbf{u})}{g(\mathbf{u},\mathbf{u})}\mathbf{u} \\
    &= \mathbf{v} - 2\frac{g(\mathbf{v},\mathbf{u})}{g(\mathbf{u},\mathbf{u})}\mathbf{u}.
    \end{aligned}
\end{align}

Since $\det S_{\mathbf{u}} = (-1)^{1} = -1$ by the Eq. \eqref{eq_acgdet004}, the symmetry $S_{\mathbf{u}}$ is a reflection. Therefore, $S_{\mathbf{u}}(\mathbf{v})$ is a reflection of the vector \textcolor{rgb, 255:red, 245; green, 113; blue, 35 }{\textbf{v}} that occurs on the \textcolor{rgb, 255:red, 255; green, 184; blue, 36 }{hyperplane} orthogonal to the vector $\mathbf{u}$.

\begin{center}

\tikzset{every picture/.style={line width=0.75pt}} 

\begin{tikzpicture}[x=0.75pt,y=0.75pt,yscale=-1,xscale=1]

\draw    (331.3,48.22) -- (331.3,276.22) ;
\draw    (196.6,166) -- (491.1,166) ;
\draw  [draw opacity=0][fill={rgb, 255:red, 255; green, 184; blue, 36 }  ,fill opacity=0.8 ] (456.23,122.09) -- (555.59,122.09) -- (366.91,221.49) -- (267.55,221.49) -- cycle ;
\draw [color={rgb, 255:red, 0; green, 0; blue, 0 }  ,draw opacity=1 ]   (401.69,161.92) -- (344.65,104.88) ;
\draw [shift={(343.24,103.47)}, rotate = 45] [color={rgb, 255:red, 0; green, 0; blue, 0 }  ,draw opacity=1 ][line width=0.75]    (10.93,-3.29) .. controls (6.95,-1.4) and (3.31,-0.3) .. (0,0) .. controls (3.31,0.3) and (6.95,1.4) .. (10.93,3.29)   ;
\draw  [dash pattern={on 4.5pt off 4.5pt}]  (401.69,161.92) -- (421.44,181.67) ;
\draw    (466.24,226.47) -- (421.44,181.67) ;
\draw  [dash pattern={on 4.5pt off 4.5pt}]  (331.19,188.12) -- (331.19,222.12) ;
\draw [color={rgb, 255:red, 245; green, 113; blue, 35 }  ,draw opacity=1 ] [dash pattern={on 0.84pt off 2.51pt}]  (371,72.47) -- (512.63,214.1) ;
\draw [color={rgb, 255:red, 245; green, 113; blue, 35 }  ,draw opacity=1 ]   (411.57,171.79) -- (389.69,118.14) -- (370.86,71.95) ;
\draw [shift={(370.1,70.1)}, rotate = 67.82] [color={rgb, 255:red, 245; green, 113; blue, 35 }  ,draw opacity=1 ][line width=0.75]    (10.93,-3.29) .. controls (6.95,-1.4) and (3.31,-0.3) .. (0,0) .. controls (3.31,0.3) and (6.95,1.4) .. (10.93,3.29)   ;
\draw [color={rgb, 255:red, 245; green, 113; blue, 35 }  ,draw opacity=1 ]   (439.6,183.6) -- (510.79,213.33) ;
\draw [shift={(512.63,214.1)}, rotate = 202.67] [color={rgb, 255:red, 245; green, 113; blue, 35 }  ,draw opacity=1 ][line width=0.75]    (10.93,-3.29) .. controls (6.95,-1.4) and (3.31,-0.3) .. (0,0) .. controls (3.31,0.3) and (6.95,1.4) .. (10.93,3.29)   ;
\draw [color={rgb, 255:red, 245; green, 113; blue, 35 }  ,draw opacity=1 ] [dash pattern={on 4.5pt off 4.5pt}]  (411.57,171.79) -- (439.6,183.6) ;

\draw (345.9,120.6) node [anchor=north west][inner sep=0.75pt]    {$\mathbf{u}$};
\draw (352.4,67.1) node [anchor=north west][inner sep=0.75pt]    {$\mathbf{\textcolor[rgb]{0.96,0.44,0.14}{v}}$};

\end{tikzpicture}\end{center}

Two arbitrary non-isotropic vectors $\mathbf{v}, \mathbf{u} \in V$ with the same norm, $g(\mathbf{v},\mathbf{v}) = g(\mathbf{u},\mathbf{u}) \neq 0$, can be related by at most two reflections. In fact, consider the case when $g(\mathbf{v} - \mathbf{u}, \mathbf{v} - \mathbf{u}) \neq 0$. Hence

\begin{align}\label{eq_acg518}
    \begin{aligned}
    S_{\mathbf{v} - \mathbf{u}}(\mathbf{v}) &= \mathbf{v} - 2\frac{g(\mathbf{v}, \mathbf{v} - \mathbf{u})}{g(\mathbf{v} - \mathbf{u}, \mathbf{v} - \mathbf{u})}(\mathbf{v} - \mathbf{u})\\
    &=\mathbf{v} - 2\frac{g(\mathbf{v}, \mathbf{v}) - g(\mathbf{v},\mathbf{u})}{g(\mathbf{v},\mathbf{v}) + g(\mathbf{u},\mathbf{u}) -2 g(\mathbf{v},\mathbf{u})}(\mathbf{v} - \mathbf{u})\\
    &=\mathbf{v} - \frac{g(\mathbf{v}, \mathbf{v}) - g(\mathbf{v},\mathbf{u})}{g(\mathbf{v},\mathbf{v}) - g(\mathbf{v},\mathbf{u})}(\mathbf{v} - \mathbf{u})\\
    &= \mathbf{v} - (\mathbf{v} - \mathbf{u}) = \mathbf{u}.
    \end{aligned}
\end{align}

\noindent Now, supposing $g(\mathbf{v} - \mathbf{u}, \mathbf{v} - \mathbf{u}) = g(\mathbf{v},\mathbf{v}) + g(\mathbf{u},\mathbf{u}) -2 g(\mathbf{v},\mathbf{u}) = 0$, since $g(\mathbf{v},\mathbf{v}) = g(\mathbf{u},\mathbf{u})$ it follows that $g(\mathbf{v},\mathbf{v}) = g(\mathbf{u},\mathbf{u})  = g(\mathbf{v},\mathbf{u}) \neq 0$ and $g(\mathbf{v} + \mathbf{u}, \mathbf{v} + \mathbf{u}) \neq 0$. Therefore,

\begin{align}
    \begin{aligned}
    S_{\mathbf{v} + \mathbf{u}}(\mathbf{v}) &= \mathbf{v} - 2\frac{g(\mathbf{v}, \mathbf{v} + \mathbf{u})}{g(\mathbf{v} + \mathbf{u}, \mathbf{v} + \mathbf{u})}(\mathbf{v} + \mathbf{u})\\
    &=\mathbf{v} - 2\frac{g(\mathbf{v}, \mathbf{v}) + g(\mathbf{v},\mathbf{u})}{g(\mathbf{v},\mathbf{v}) + g(\mathbf{u},\mathbf{u}) + 2 g(\mathbf{v},\mathbf{u})}(\mathbf{v} + \mathbf{u})\\
    &=\mathbf{v} - \frac{g(\mathbf{v}, \mathbf{v}) + g(\mathbf{v},\mathbf{u})}{g(\mathbf{v},\mathbf{v}) + g(\mathbf{v},\mathbf{u})}(\mathbf{v} + \mathbf{u})\\
    &= \mathbf{v} - (\mathbf{v} + \mathbf{u}) = -\mathbf{u}.
    \end{aligned}
\end{align}

\noindent Finally,

\begin{equation}\label{eq_acg520}
    S_{\mathbf{u}}S_{\mathbf{u} + \mathbf{v}}(\mathbf{v}) =  S_{\mathbf{u}}(-\mathbf{u}) = -\mathbf{u} - 2\frac{g(-\mathbf{u}, \mathbf{u})}{g(\mathbf{u},\mathbf{u})}\mathbf{u} = \mathbf{u}.
\end{equation}

In order to generalise this result, one has the following theorem:

\begin{tftheorem}\label{theo_a6s5d1fg3}
\textbf{(Cartan-Dieudonné) }Any orthogonal transformation $T$ in a finite dimensional vector space $V$ can be expressed as the product of symmetries (reflections) with respect to non-isotropic hyperplanes.
\end{tftheorem}
\noindent \textcolor{RedOrange}{\textit{Proof.}} We proceed by induction on $\dim V = n$. It is clear that it holds for $n = 1$. Now suppose that it holds for $\dim V = n$, we shall prove the assertion for $\dim V = n + 1$. Let $\mathbf{v} \in V$ be such that $g(\mathbf{v},\mathbf{v}) \neq 0$ and consider the vector space $U = \sp \{\mathbf{v} \}$. Therefore, the vector space $U^{\perp}$ is $n$-dimensional. Consider now an orthogonal transformation $T : V \to V$, by Definition \ref{def_as54dfd3245} $g(T(\mathbf{v}),T(\mathbf{v})) = g(\mathbf{v},\mathbf{v})$, hence by the previously construction shown in Eq. \eqref{eq_acg518} and \eqref{eq_acg520},  the vectors $T(\mathbf{v})$ and $\mathbf{v}$ are related by at most two reflections $S$, that is, $S(T(\mathbf{v})) = \mathbf{v}$. Since $U$ is $S \circ T$-invariant, so is the space $U^{\perp}$, implying that $S \circ T$ is an orthogonal transformation. Therefore, since $\dim U^{\perp} = n$, by the induction hypothesis it follows that $S \circ T$ is a finite product of reflections, let us call, $S \circ T = \Sigma$. That way, $T = S^{-1} \circ \Sigma$ and the proof is concluded. \textcolor{RedOrange}{$\Box$}

\paragraph{ } Let us recall the Eq. \eqref{eq_acg517} which establish the symmetry $S_{\mathbf{u}}$. One can ask how this equation can be written in terms of a Clifford algebra $\clpq$. It is known that in terms of $\clpq$, for $\mathbf{u} = \gamma(\mathbf{u}), \mathbf{v}  = \gamma(\mathbf{v})$, one has $\mathbf{v} \mathbf{u} + \mathbf{u} \mathbf{v} = 2g(\mathbf{v},\mathbf{u})$ and $\mathbf{u}^2 = g(\mathbf{u},\mathbf{u})$. That is,

\begin{equation}
    \mathbf{u} \frac{\mathbf{u}}{g(\mathbf{u},\mathbf{u})} = 1.
\end{equation}

\noindent In this manner, the element $\frac{\mathbf{u}}{g(\mathbf{u},\mathbf{u})}$ can be interpreted as the inverse of the element $\mathbf{u}$, namely, $\mathbf{u}^{-1} = \frac{\mathbf{u}}{g(\mathbf{u},\mathbf{u})}$. Therefore, one can rewrite the Eq.\eqref{eq_acg517} as

\begin{equation}
     S_{\mathbf{u}}(\mathbf{v}) = \mathbf{v} - 2\frac{g(\mathbf{v},\mathbf{u})}{g(\mathbf{u},\mathbf{u})}\mathbf{u} = \mathbf{v} - \frac{\mathbf{v}\mathbf{u} + \mathbf{u} \mathbf{v}}{g(\mathbf{u},\mathbf{u})}\mathbf{u} = \mathbf{v} - (\mathbf{v}\mathbf{u} + \mathbf{u} \mathbf{v})\mathbf{u}^{-1} = -\mathbf{u}\mathbf{v}\mathbf{u}^{-1}.
\end{equation}

  {\colorlet{shadecolor}{Peach!20}\begin{shaded} The symmetry $S_{\mathbf{u}}$ is thus written in terms of $\clpq$ as

\begin{equation}\label{eq_acgr523}
    S_{\mathbf{u}}(\mathbf{v}) =  -\mathbf{u}\mathbf{v}\mathbf{u}^{-1} = \yhwidehat{\mathbf{u}}\mathbf{v}\mathbf{u}^{-1}.
\end{equation} \end{shaded}}

\section{First Groups on Clifford Algebras} 
\paragraph{ } We are now in a position to define the first groups within a Clifford algebra. From these groups, we  present the construction of an important homomorphism that in turn will be settled as double covering homomorphism from the Spin group to the orthogonal group. We start with the largest and simplest group that can be defined in a Clifford algebra $\clpq$, the group of the invertible elements:

\begin{equation}
    \clpq^{*} = \{ a \in \clpq : \exists b \in \clpq : ab = ba = 1\}.
\end{equation}

  {\colorlet{shadecolor}{Peach!20}\begin{shaded} 
The Clifford-Lipschitz group $\Sigma_{p,q}$ is a subgroup of $\clpq^{*}$ of great interest:

\begin{equation}
    \Sigma_{p,q} = \{ a \in \clpq^{*} : a\mathbf{v}a^{-1} \in V, \; \text{for all} \;\; \mathbf{v} \in V = \rrpq\}.
\end{equation} \end{shaded}}

The Clifford-Lipschitz group may be defined for
any Clifford algebra $\cl(V,g)$ but we will consider hereinafter only the case $V = \rrpq$.
This term is not used randomly, it is intrinsically linked with symmetries and orthogonal transformations that were discussed in the previous section \ref{sec_23w4ertygfd}. Our goal here is to define orthogonal transformations within a Clifford algebra. This is crucial for working with spinors, which inherently carry the notion of symmetry. Everything that happens in the spinor space (in terms of symmetries and orthogonal transformations) also happens in the physical space $V = \rrpq$. The two are interconnected. The adjoint representation that will be defined in the sequel plays a fundamental role to connect those spaces.

\paragraph{ } The set of all automorphism on $\clpq$ form a group denoted by $\Autpq$.

\begin{definlaranja}
    The \textbf{adjoint representation}  \normalfont $\text{Ad}: \Sigma_{p,q} \to \Autpq$ \textit{is defined by sending}  $a \mapsto \text{Ad}_{a} \in \Autpq$  \textit{given by }  $\text{Ad}_{a}(x) = axa^{-1}$. 
\end{definlaranja}

\noindent The importance of defining this automorphism is because the induced group action will provide a correspondence between the Clifford-Lipschitz group and the space of orthogonal transformations in the associated vector space $V = \rrpq$. Take $\mathbf{v}, \mathbf{u} \in V = \rrpq$, such that $\mathbf{v} \mathbf{u} + \mathbf{u} \mathbf{v} = 2g(\mathbf{v},\mathbf{u}).$ Therefore, for the vectors $\text{Ad}_{a}(\mathbf{v}), \text{Ad}_{a}(\mathbf{u}) \in V$ it follows that:

\begin{align}
    \begin{aligned}
    2g(\text{Ad}_{a}(\mathbf{v}), \text{Ad}_{a}(\mathbf{u})) &=  \text{Ad}_{a}(\mathbf{v})\text{Ad}_{a}(\mathbf{u}) + \text{Ad}_{a}(\mathbf{u})\text{Ad}_{a}(\mathbf{v})\\
    &= a\mathbf{v}a^{-1} a\mathbf{u} a^{-1} + a\mathbf{u}a^{-1}a\mathbf{v}a^{-1}\\
    &= a\mathbf{v}\mathbf{u}a^{-1} + a\mathbf{u}\mathbf{v}a^{-1}\\
    &= a (\mathbf{v}\mathbf{u} + \mathbf{u}\mathbf{v})a^{-1}\\
    &= a 2g(\mathbf{v},\mathbf{u}) a^{-1}\\
    &=2g(\mathbf{v},\mathbf{u}).
    \end{aligned}
\end{align}

\noindent We conclude that $\text{Ad}_{a} \in \O(p,q)$, since it satisfies the orthogonality condition (Def. \ref{def_as54dfd3245}): $g(\text{Ad}_{a}(\mathbf{v}), \text{Ad}_{a}(\mathbf{u})) = g(\mathbf{v},\mathbf{u})$ for $\mathbf{u}, \mathbf{v} \in \rrpq$. 

\begin{tfpropos}
The mapping \normalfont$\text{Ad} : \Sigma_{p,q} \to \O(p,q)$ \textit{is a group homomorphism.}
\end{tfpropos}
\proof In fact, for any $\mathbf{v} \in \rrpq, \,a,b \in \Sigma_{p,q}$:

\begin{equation}
    \text{Ad}_{ab}(\mathbf{v}) = ab\mathbf{v}(ab)^{-1} = ab\mathbf{v}b^{-1}a^{-1} = \text{Ad}_{a}\text{Ad}_{b}(\mathbf{v}) \implies \text{Ad}_{ab} = \text{Ad}_{a} \text{Ad}_{b}. \;\; \Box
\end{equation}

\paragraph{ } Let $\{\'e_1, \ldots, \'e_n\}$ be an orthonormal basis of $\rrpq$. We proceed to obtain the image of the mapping $\text{Ad}$. In this sense, it is necessary to study its determinant. The determinant of $\text{Ad}_{a}$ can be expressed in terms of the exterior product in what follows:

\begin{equation}
    \left(\det \text{Ad}_{a}\right) \'e_1 \wedge \cdots \wedge \'e_n = \text{Ad}_{a}(\'e_1) \wedge \cdots \wedge \text{Ad}_{a}(\'e_n). 
\end{equation}

\noindent Besides that, for vectors $\mathbf{v}, \mathbf{u} \in V = \rrpq$ it follows that:

\begin{align}\begin{aligned}\label{eq_acgr534}
    \text{Ad}_{a}(\mathbf{v}) \wedge \text{Ad}_{a}(\mathbf{u}) &= (a \mathbf{v} a^{-1}) \wedge (a \mathbf{u} a^{-1})\\
    &= \frac{1}{2}(a\mathbf{v}a^{-1}a\mathbf{u}a^{-1} - a\mathbf{u}a^{-1}a\mathbf{v}a^{-1})\\
    &= a\frac{1}{2}(\mathbf{v}\mathbf{u} - \mathbf{u}\mathbf{v}) a^{-1}\\
    &= a(\mathbf{v} \wedge \mathbf{u})a^{-1}\\
    &= \text{Ad}_{a}(\mathbf{v} \wedge \mathbf{u}).\end{aligned}
\end{align}

\noindent More generally,

\begin{equation}\label{eq_acg54987}
    \left(\det \text{Ad}_{a}\right) \'e_1 \wedge \cdots \wedge \'e_n = a(\'e_1 \wedge \cdots \wedge \'e_n)a^{-1}.
\end{equation}

\noindent The task is now to compute the right hand side of Eq.\eqref{eq_acg54987}. Take $a$ as the homogeneous multivector $a = a_{[k]} \in \bigwedge_{k}(V)$, since the general result for an arbitrary multivector follows by linearity. Therefore,

\begin{align}\label{eq_acg498794}
    \begin{aligned}
    a_{[k]}(\'e_1 \wedge \cdots \wedge \'e_n) &= a_{[k]\flat} \rfloor(\'e_1 \wedge \cdots \wedge \'e_n)\\
    &= (-1)^{k(n-1)}(\'e_1 \wedge \cdots \wedge \'e_n)\lfloor a_{[k]\flat}\\
    &= (-1)^{k(n-1)}(\'e_1 \wedge \cdots \wedge \'e_n)a_{[k]}.
    \end{aligned}
\end{align}

\noindent Notice that the result of the above equation depends on the parity of the $\dim V = n$. If $n$ is odd, then $n - 1$ is even and if $n$ is even, then $n - 1$ is odd. Such conditions in Eq. \eqref{eq_acg498794} give us the following result:

\begin{align}
    \begin{aligned}
     a_{[k]}(\'e_1 \wedge \cdots \wedge \'e_n) &=  (\'e_1 \wedge \cdots \wedge \'e_n)a_{[k]} &&&&(n \text{ odd}),\\
      a_{[k]}(\'e_1 \wedge \cdots \wedge \'e_n) &=  (-1)^{k}(\'e_1 \wedge \cdots \wedge \'e_n)a_{[k]} &&&&(n \text{ even}).\\
    \end{aligned}
\end{align}

Consequently, by Eq.\eqref{eq_acg54987} we conclude that when $n$ is odd, $\det \text{Ad}_{a} = 1$ whereas for $n$ even, the expression $\det \text{Ad}_{a} = \pm 1$ holds. More specifically:

\begin{align}
    \begin{aligned}
    \det \text{Ad}_{a} &= 1, &&&&\text{if } a \in \clpq^{\text{even}}, \\
       \det \text{Ad}_{a} &= -1, &&&&\text{if } a \in \clpq^{\text{odd}}.
    \end{aligned}
\end{align}

  {\colorlet{shadecolor}{Peach!20}\begin{shaded} 
  We conclude that, \begin{align}\label{eq_acg2640057}
    \begin{aligned}
    \text{Ad}(\Sigma_{p,q}) &= \O(p,q), &&&&( n \text{ even}),\\
       \text{Ad}(\Sigma_{p,q})  &= \SO(p,q),&&&&( n \text{ odd}).
    \end{aligned}
\end{align}
  \end{shaded}}

\begin{definlaranja}\label{def_twofold2345}
     The \textbf{twisted adjoint representation} \normalfont $\yhwidehat{\text{Ad}}$ \textit{is defined as follows:}

\begin{equation}
    \yhwidehat{\text{Ad}}_{a}(\mathbf{v}) = \yhwidehat{a}\mathbf{v}a^{-1}.
\end{equation}
\end{definlaranja}

Where $\yhwidehat{\;\cdot\;}$ is the grade involution operator (Def. \ref{def15op}). One important consequence of the twisted adjoint representation is to remove the condition on the dimension of the vector space $V = \rrpq$ as we have seen in the Eq.\eqref{eq_acg2640057}. Therefore, we consider the \textbf{twisted Clifford-Lipschitz} group $\yhwidehat{\Sigma}_{p,q}$:

\begin{equation}
    \yhwidehat{\Sigma}^{p,q} = \{a \in \clpq^{*} : \yhwidehat{a}\mathbf{v}{a}^{-1} \in V, \text{ for all } \mathbf{v} \in V = \rrpq \}
\end{equation}

It is straightforward to notice that $\Sigma_{p,q} \simeq \yhwidehat{\Sigma}_{p,q}$, however, in terms of a representation there is a huge difference due to the presence of the grade involution. The results obtained for the adjoint representation can be handled in much the same way for the twisted representation. Therefore, for $a = a_{[k]} \in \bigwedge_{k}(V)$ in terms of $\yhwidehat{\text{Ad}}$ it follows that\cite{Roc16}:

\begin{align}
    \begin{aligned}
    \left(\det \yhwidehat{\text{Ad}}\left(a_{[k]}\right)\right) \'e_1 \wedge \cdots \wedge \'e_n  &=
    \yhwidehat{\text{Ad}}  \left(a_{[k]}\right)(\'e_1) \wedge \cdots \wedge \yhwidehat{\text{Ad}}\left(a_{[k]}\right)(\'e_n)\\ &= (-1)^{k(n-1)}\yhwidehat{\text{Ad}}  \left(a_{[k]}\right)(\'e_1 \wedge \cdots \wedge \'e_n)\\
    &= (-1)^{k(n-1)}(-1)^{k}\text{Ad}  \left(a_{[k]}\right)(\'e_1 \wedge \cdots \wedge \'e_n) \\
    &=  (-1)^{kn}\text{Ad}  \left(a_{[k]}\right)(\'e_1 \wedge \cdots \wedge \'e_n) \\
    &= (-1)^{kn}a_{[k]}(\'e_1 \wedge \cdots \wedge \'e_n)a_{[k]}^{-1}.
    \end{aligned}
\end{align}

\noindent Now, by Eq. \eqref{eq_acg498794} one has

\begin{align}
    \begin{aligned}
      \left(\det \yhwidehat{\text{Ad}}\left(a_{[k]}\right)\right) \'e_1 \wedge \cdots \wedge \'e_n &= (-1)^{kn}(-1)^{k(n-1)}(\'e_1 \wedge \cdots \wedge \'e_n) \\
      &= (-1)^{2kn}(-1)^{-k}(\'e_1 \wedge \cdots \wedge \'e_n) \\
      &= (-1)^{k}(\'e_1 \wedge \cdots \wedge \'e_n).
    \end{aligned}
\end{align}

\noindent Consequently,

\begin{equation}
\det \yhwidehat{\text{Ad}}\left(a_{[k]}\right) = (-1)^{k} = \pm 1,
\end{equation}

\noindent which does not depend on the dimension $n$ as required. Consider $\Sigma_{p,q}^{\text{even}} = \Sigma_{p,q} \inter \clpq^{\text{even}}$, that is, $\Sigma_{p,q}^{\text{even}}$ consists of the even elements of the Clifford-Lipschitz group. Now, let us remind the Eq. \eqref{eq_acgr523} that establish a reflection to the orthogonal hyperplane to a given vector $\mathbf{u}$ in terms of $\clpq$, namely,  $S_{u}(\mathbf{v}) = \yhwidehat{\mathbf{u}}\mathbf{v}\mathbf{u}^{-1}$. In that manner, one can notice that

  {\colorlet{shadecolor}{Peach!20}\begin{shaded} 
\begin{equation}
   \yhwidehat{\text{Ad}}_{\mathbf{u}} = S_{\mathbf{u}}.
\end{equation}\end{shaded}

\noindent That is, the Clifford-Lipschitz group can be considered as the set of invertible elements within the Clifford algebra that perform reflections of vectors across their orthogonal hyperplanes. According to the Cartan-Dieudonné Theorem \ref{theo_a6s5d1fg3}, any orthogonal transformation can be written as a finite product of reflections of type $S_{\mathbf{u}} =\yhwidehat{\text{Ad}}_{\mathbf{u}}$.  Therefore, for any $T \in \O(p,q)$ it follows that there exists some vectors $\{\mathbf{u}_{1}, \ldots, \mathbf{u}_{k}\}$ such that

\begin{equation}
    T = \yhwidehat{\text{Ad}}_{\mathbf{u}_{1}} \cdots \yhwidehat{\text{Ad}}_{\mathbf{u}_{k}} = \yhwidehat{\text{Ad}}_{\mathbf{u}_{1} \cdots \mathbf{u}_{k}}.
\end{equation}

\noindent Hence, $\mathbf{u}_{1} \cdots \mathbf{u}_{k} \in \Sigma_{p,q}$ and $\yhwidehat{\text{Ad}}$ is surjective. The same conclusion can be drawn for $\yhwidehat{\text{Ad}} : \Sigma_{p,q}^{\text{even}} \to \SO(p,q)$. This way, we have proved the following proposition:

\begin{tfpropos}\label{propos_5rtdgfijvknm}
    The mappings \normalfont $\yhwidehat{\text{Ad}}: \Sigma_{p,q} \to \O(p,q)$  \textit{and} $\yhwidehat{\text{Ad}}: \Sigma_{p,q}^{\text{even}} \to \SO(p,q)$ \textit{are surjective homomorphisms.}
\end{tfpropos}

  {\colorlet{shadecolor}{Peach!20}\begin{shaded}  That is,  \begin{align}
    \begin{aligned}
    \yhwidehat{\text{Ad}}(\Sigma_{p,q}) &= \O(p,q),\\
    \yhwidehat{\text{Ad}}(\Sigma_{p,q}^{\text{even}}) &= \SO(p,q).
    \end{aligned}
\end{align}  
  \end{shaded}}

\paragraph{ } This result can be used to characterise the Clifford-Lipschitz group as the group consisting of the product of all non-null vectors of $\clpq$:

   {\colorlet{shadecolor}{Peach!20}\begin{shaded}\begin{equation}
    \Sigma_{p,q} = \{a \in \clpq^{*} : a = \mathbf{u}_{1} \cdots \mathbf{u}_{k}, \text{ such that } \mathbf{u}_{i} \in \rrpq \text{ and } g(\mathbf{u}_i, \mathbf{u}_i) \neq 0 \}.
\end{equation}\end{shaded}}

It is worth to emphasise that the decomposition $a = \mathbf{u}_{1} \cdots \mathbf{u}_{k}$ is not unique. In addition, one has \cite{Roc16}:

\begin{align}
    \begin{aligned}
    k = \text{ odd }&\color{RedOrange} \longleftrightarrow \color{black}\text{ reflection, } \\
     k = \text{ even }& \color{RedOrange} \longleftrightarrow \color{black}\text{ rotation. } 
    \end{aligned}
\end{align}

Furthermore, consider the set

\begin{equation}
    \ker \yhwidehat{\text{Ad}} = \{a \in \Sigma_{p,q} : \yhwidehat{\text{Ad}}_{a} = 1 \}
\end{equation}

\noindent such that $1$ is the unity in $\O(p,q)$. 

\begin{tfpropos}
  \normalfont  $\ker \yhwidehat{\text{Ad}} = \rr \setminus \{0\}.$
\end{tfpropos}

\noindent \textcolor{RedOrange}{\textit{Proof.}} In fact, take $a \in \ker \yhwidehat{\text{Ad}}$, hence

\begin{align}\label{eq_agcr5361}
    \begin{aligned}
    \yhwidehat{\text{Ad}}_{a}(\mathbf{v}) &= \mathbf{v}\\
    \yhwidehat{a}\mathbf{v} a^{-1} &=\mathbf{v}\\ 
    \yhwidehat{a}\mathbf{v} &= \mathbf{v} a.    \end{aligned}
\end{align}

\noindent By writing $a \in \clpq = \clpq^{\text{even}} + \clpq^{\text{odd}}$ in terms of its even and odd part, namely, $ a= (a)_{\text{even}} + (a)_{\text{odd}}$, it follows from the above Eq. \eqref{eq_agcr5361} that

\begin{align}
    \begin{aligned}\label{eq_acg79415645771}
    (a)_{\text{even}}\mathbf{v} &= \mathbf{v} (a)_{\text{even}},\\
    ( a)_{\text{odd}}\mathbf{v} &= -\mathbf{v} (a)_{\text{odd}}.
    \end{aligned}
\end{align}

\noindent The second condition holds only if $(a)_{\text{odd}} = 0$. Indeed, let us suppose that there exists $x \in \clpq$ such that $x\mathbf{v} = -\mathbf{v}x$ for all $\mathbf{v} \in \rrpq \subset \clpq$. Therefore, since $\mathbf{v}$ is arbitrary it must hold for all generators:

\begin{align}\label{eq_acgr87wq9e87wq}
    \begin{aligned}
    x\'e_{1} &= -\'e_{1}x,\\
       x\'e_{2} &= -\'e_{2}x,\\
       &\vdots \\
          x\'e_{n} &= -\'e_{n}x.\\
    \end{aligned}
\end{align}

\noindent However, the first two lines of Eq.\eqref{eq_acgr87wq9e87wq} yield:

\begin{align}
    \begin{aligned}
     x\'e_{1} = -\'e_{1}x \implies x\'e_{1}\'e_{2} = -\'e_{1}x\'e_{2} \implies x\'e_{1}\'e_{2}  = \'e_{1}\'e_{2} x \\ \implies -\'e_{2}x\'e_{1} = \'e_{1}\'e_{2} x 
     \implies \'e_{2}\'e_{1}x = \'e_{1}\'e_{2} x \implies -\'e_{1}\'e_{2}x = \'e_{1}\'e_{2} x.
    \end{aligned}
\end{align}

\noindent Hence, $-\'e_{1}\'e_{2}x = \'e_{1}\'e_{2} x$ just holds for $x = 0$. That is, there is no element of $\clpq$ that anti-commutes with all generators of the algebra. On the other hand, the first condition in the Eq. \eqref{eq_acg79415645771} holds for $(a)_{\text{even}} \in \text{Cen}(\clpq)$. We recall
the characterisation of the centre presented in the Eq. \eqref{eq_centrws65g1} in the Chapter \ref{chap1}:

\begin{equation}
     \text{Cen}(\cl_{p,q}) = 
    \color{RedOrange} \begin{cases} \color{black}
     \bigwedge\,\!\!_{0}(\rr^{p,q}), \text{\;if\;} n=p+q \text{\, is even,}\\
    \color{black} \bigwedge\,\!\!_{0}(\rr^{p,q}) \oplus  \bigwedge\,\!\!_{n}(\rr^{p,q}), \text{\;if\;} n=p+q \text{\, is odd.}
    \color{black} \end{cases}\color{black}
\end{equation}

\noindent That way, if $n$ is even then $(a)_{\text{even}} \in \text{Cen}(\cl_{p,q}) = \bigwedge\,\!\!_{0}(\rr^{p,q}) = \rr$ and if $n$ is odd, an element of
the space $\bigwedge\,\!\!_{n}(\rr^{p,q})$ is odd
then the second condition of Eq. \eqref{eq_acg79415645771} must hold, instead of the first one, leading to $(a)_{\text{odd}} = 0$. Therefore, the two conditions are only
satisfied when $a \in \text{Cen}(\cl_{p,q}) = \bigwedge\,\!\!_{0}(\rr^{p,q}) = \rr$. Moreover, since $a$ is invertible it follows that $a \neq 0$ and $\ker \yhwidehat{\text{Ad}} = \rr \setminus \{0\}$ as claimed. \textcolor{RedOrange}{$\Box$}

\begin{ex}
{{\textcolor{RedOrange}{{$\blacktriangleright$\;}}}} Groups in the Clifford algebra $\cl_{2,0}$. \normalfont An arbitrary element $\psi \in \cl_{2,0}$ is given by

\begin{equation}
    \psi = a_0 + a_1\'e_1 + a_2e_2 + a_{12}\'e_{1}\'e_{2}.
\end{equation}

\noindent In addition 
\begin{align}
    \begin{aligned}
\overline{\psi} &= a_0 - a_1\'e_1 - a_2\'e_2 - a_{12}\'e_1\'e_2  \implies \overline{\psi}\psi &=  a_0^2 - a_1^2 - a_2^2 + a_{12}^2  .
    \end{aligned}
\end{align}


\noindent If $a_0^2 - a_1^2 - a_2^2 + a_{12}^2  \neq 0$ then $\psi^{-1}$ can be defined as

\begin{align}\label{eq_acg321507}
    \begin{aligned}
    \psi^{-1} =\frac{\overline{\psi}}{\overline{\psi}\psi} =  \frac{a_0 - a_1\'e_1 - a_2\'e_2 - a_{12}\'e_1\'e_2}{a_0^2 - a_1^2 - a_2^2 + a_{12}^2}.
    \end{aligned}
\end{align}

\noindent Therefore, one can establish the group of invertible elements of $\cl_{2,0}$ as follows:

\begin{equation}
    \cl_{2,0}^{*} = \{\psi = a_0 + a_1\'e_1 + a_2e_2 + a_{12}\'e_{1}\'e_{2} \in   \cl_{2,0} : a_0^2 - a_1^2 - a_2^2 + a_{12}^2  \neq 0  \}.
\end{equation}

\noindent We proceed to determine the Clifford-Lipschitz group. On that group, the relation $\psi\mathbf{v}\psi^{-1} \in \rr^{2,0}$ must hold for every $\mathbf{v} = v_1 \'e_1 + v_2 \'e_2 \in \rr^{2,0}$. Since $\psi^{-1} = \frac{\overline{\psi}}{\overline{\psi}\psi}$ one may consider the similar condition $\psi\mathbf{v}\overline{\psi} \in \rr^{2,0}$ that reads:

\begin{align}
    \begin{aligned}
    \psi\mathbf{v}\overline{\psi} &= \left[ v_1(a_0^{2} - a_1^{2} + a_2^2 - a_{12}^2) + v_2(2a_0a_{12} - 2a_1a_2) \right ]\'e_1\\
    &\;\;\;\;+ \left[  v_1(-2a_0a_{12} - 2a_1a_2) + v_2(a_0^{2} + a_1^{2} - a_2^2 - a_{12}^2) \right ]\'e_2\\
    &\;\;\;\;+ \left[  v_1(-2a_1a_{12} - 2a_0a_2) + v_2(2a_0a_{1} - 2a_2a_{12}) \right ]\'e_1\'e_2.
    \end{aligned}
\end{align}

\noindent Hence, for $\psi\mathbf{v}\overline{\psi} \in \rr^{2,0}$ the following equalities must hold:

\begin{align}
    \begin{aligned}
    \color{RedOrange}\begin{cases}
    \color{black} a_1a_{12} + a_0a_2 = 0,\\
    \color{black} a_0a_1 - a_2a_{12} = 0.
    \end{cases}
    \end{aligned}
\end{align}

\noindent That is

\begin{align}
    \begin{aligned}
    \color{RedOrange}\begin{cases}
    \color{black} a_1a_{12} + a_0a_2 = 0,\\
    \color{black} a_0a_1 - a_2a_{12} = 0. 
    \end{cases} \color{black}\implies\;\;\;   \color{RedOrange}\begin{cases}
    \color{black} a_0a_1a_{12} + a_0^2a_2 = 0,\\
    \color{black} -a_0a_1a_{12} + a_2a_{12}^2 = 0. 
    \end{cases}  \color{black}\implies\;\;\; (a_0^2 + a_{12}^2)a_2 = 0.  
    \end{aligned}
\end{align}

\noindent The last result leads us to two conditions that allow us to describe the Clifford-Lipschitz group, that is: $a_2 = 0$ or $a_0 = a_{12} = 0$. Hence, if $a_2 = 0$ and $a_0^2 + a_{12}^2 \neq 0$ then $a_1 = 0$, therefore in this case an element $\psi$ of $\Sigma_{2,0}$ is written as $\psi = a_0 + a_{12}\'e_1\'e_2$. In the another case, $\psi \in \Sigma_{2,0}$ is written as $\psi = a_1\'e_1 + a_2\'e_2 = (a_2 +a_1\'e_1\'e_2)\'e_2$.\\

 In addition, to better realise what is being accomplished in this example, we proceed to present another way to obtain the inverse element $\psi^{-1}$ by using matrix representation. The Table \ref{table1} in the Chapter \ref{chap1}, shows the fact that each Clifford algebra is actually isomorphic to a matrix algebra or sum of two matrix algebras. Therefore, once one identifies the isomorphism, the procedure to obtain the inverse element in a Clifford algebra is just to obtain an inverse of a matrix. It is worth to emphasise that the  matrix representation of $\cl_{2,0}$ was studied the Chapter \ref{chap1}, by Eq. \eqref{eq_ac_cl2isomorhicmat2r} and obtained in the Appendix \ref{app7} in the Eq. \eqref{eq_clirep282319}. An element $\psi \in \cl_{2,0}$ can be represented by the matrix $\Psi = \rho(\psi) \in \text{Mat}(2,\rr)$ given by
 
 \begin{equation}
     \Psi = \begin{pmatrix}
     a_0 + a_1 & a_2 + a_{12} \\
     a_2 - a_{12} & a_0 - a_1
     \end{pmatrix}.
 \end{equation}
 
\noindent The inverse matrix $\Psi^{-1}$ corresponds to

\begin{equation}
    \Psi^{-1} = \frac{1}{\det \Psi}  \begin{pmatrix}
     a_0 - a_1 & -a_2 - a_{12} \\
     -a_2 + a_{12} & a_0 + a_1
     \end{pmatrix} = \frac{1}{a_0^2 - a_1^2 - a_2^2 + a_{12}^2} \begin{pmatrix}
     a_0 - a_1 & -a_2 - a_{12} \\
     -a_2 + a_{12} & a_0 + a_1
     \end{pmatrix}
\end{equation}

\noindent One can see that the condition $\det \Psi \neq 0$ necessary to the existence of $\Psi^{-1}$ is exactly the condition $\overline{\psi}\psi \neq 0$ concerning about the existence of $\psi^{-1}$ presented in the Eq. \eqref{eq_acg321507}. To summarise:

\begin{align}
    \begin{aligned}
    \rho(\psi) &= \Psi =  \begin{pmatrix}
     a_0 + a_1 & a_2 + a_{12} \\
     a_2 - a_{12} & a_0 - a_1
     \end{pmatrix},\\
     \rho(\psi^{-1}) &= \Psi^{-1} =  \frac{1}{a_0^2 - a_1^2 - a_2^2 + a_{12}^2} \begin{pmatrix}
     a_0 - a_1 & -a_2 - a_{12} \\
     -a_2 + a_{12} & a_0 + a_1
     \end{pmatrix},\\
     \rho(\overline{\psi}) &=  \begin{pmatrix}
     a_0 - a_1 & -a_2 - a_{12} \\
     -a_2 + a_{12} & a_0 + a_1
     \end{pmatrix}. \;\;\;\; \textcolor{RedOrange}{{\blacktriangleleft}}
    \end{aligned}
\end{align}
\end{ex}

\section{The Pin and Spin Groups}\label{sec_pinspin}
\paragraph{ } Now we can finally have a description of the Pin and Spin groups from the previous constructions, they are subgroups of the Clifford-Lipschitz group. Since $\ker \yhwidehat{\text{Ad}} = \rr \setminus \{0\}$, the Clifford-Lipschitz group $\Sigma_{p,q}$ is bigger than necessary to describe the orthogonal transformations.  Therefore, the objective is to find a subgroup of $\Sigma_{p,q}$ such that is sufficient to describe them, in a manner that is as \textit{close} to a one-to-one correspondence as possible. However, if $ a \in \ker \yhwidehat{\text{Ad}}$ the element $-a$ is in the kernel as well. This implies that the smallest kernel that could possibly be  restricted turns out to be $\{+1,-1\} \simeq \zz_{2}$, which can be obtained via a suitable normalisation process of the elements of $\Sigma_{p,q}$.

  {\colorlet{shadecolor}{Peach!20}\begin{shaded} A \textbf{norm} of a multivector $a \in \clpq$ can be written as

\begin{equation}
    N(a) =  \langle \yhwidetilde{a} a \rangle_{0}.
\end{equation}\end{shaded}}

\noindent Where $\yhwidetilde{\;\cdot \;}$ holds for the reversion (Def. \ref{def16op}) and $ \langle \,\cdot \, \rangle_{0}$ is the projector operator (Def. \ref{def_asd4aa5ngf43}), this way, the resulting scalar part of the product $\yhwidetilde{a} a $ are taken. Another equivalent definition for the norm is given by the other anti-automorphism
\begin{equation}
    N'(a) =  \langle \overline{a} a \rangle_{0},
\end{equation}

\noindent where $\overline{a}$ is the conjugation of the element $a \in \clpq$ (Def. \ref{def17op}). A relation between these norms for $a_{[k]} \in \bigwedge_{k}(\rr^{p,q})$ is given by the following equation\cite{Roc16}

\begin{equation}\label{eqgcr564}
    N'\left(a_{[k]}\right) = (-1)^{k}N\left(a_{[k]}\right).
\end{equation}

To restrict the Clifford-Lipschitz group with respect with the norm function, it is required to prove that the multiplication is preserved.

\begin{tfpropos} The mappings $N: \Sigma_{p,q} \to \rr \setminus \{0\}$, $N': \Sigma_{p,q} \to \rr \setminus \{0\}$ are homomorphisms.
\end{tfpropos}
\textcolor{RedOrange}{\textit{Proof.}} Consider $a = \mathbf{u}_{1} \cdots \mathbf{u}_{k} \in \Sigma_{p,q}$ hence

\begin{align}
\begin{aligned}
    N(a) &= \langle \mathbf{u}_{k} \cdots \mathbf{u}_{1}\mathbf{u}_{1} \cdots \mathbf{u}_{k} \rangle_{0}\\ &= \mathbf{u}_{k} \cdots \mathbf{u}_{1}\mathbf{u}_{1} \cdots \mathbf{u}_{k}\\ &= N(\mathbf{u}_1) \cdots N(\mathbf{u}_{k})
\end{aligned}
\end{align}

\noindent It follows that

\begin{align}
    \begin{aligned}
    N(ab) &= \langle \yhwidetilde{(ab)}ab \rangle_0
    = \langle \yhwidetilde{b} \yhwidetilde{a} ab \rangle_0
    =  \langle  \yhwidetilde{a} ab\yhwidetilde{b} \rangle_0\\
    &=  \langle \mathbf{u}_{k} \cdots \mathbf{u}_{1}\mathbf{u}_{1} \cdots \mathbf{u}_{k}\mathbf{v}_{1} \cdots \mathbf{v}_{l}\mathbf{v}_{l} \cdots \mathbf{v}_{1} \rangle_{0}\\
    &=N(\mathbf{u}_{1}) \cdots N(\mathbf{u}_{k})N(\mathbf{v}_{1}) \cdots N(\mathbf{v}_{l})\\
    &= N(a)N(b).
    \end{aligned}
\end{align}

\noindent The same argument can be drawn for $N'$. \textcolor{RedOrange}{$\Box$}

\paragraph{ }  Since $\ker \yhwidehat{\text{Ad}} = \rr \setminus \{0\}$ the only scalars in $\rr \setminus \{0\}$ with norm $+1$ or  $-1$ are precisely $+1$ and $-1$, therefore, the elements of $\Sigma_{p,q}$ can be normalised, obtaining subgroups whose respective kernels are smaller, namely, $ \{+1,-1\} \simeq \mathbb{Z}_2$. The candidate for a subgroup of $\Sigma_{p,q}$ that maps onto the orthogonal group with kernel $\zz_{2}$ is the \textit{Pin Group}.

\subsection*{Pin Group}

  {\colorlet{shadecolor}{Peach!20}\begin{shaded} The group $\Pin(p,q)$ is defined as \cite{Roc16}
\begin{equation}
    \Pin(p,q) = \{a \in \Sigma_{p,q} : N(a) = \pm 1\}.
\end{equation}  
  \end{shaded}}

\noindent For $\yhwidehat{\text{Ad}} : \Pin(p,q) \to \O(p,q)$, it follows that

\begin{equation}
    \ker \yhwidehat{\text{Ad}} |_{\Pin(p,q)} = \{+1,-1\} = \mathbb{Z}_2.
\end{equation}

\noindent Similarly, the group $\Pin'(p,q)$ with respect to the norm $N'$ is defined as

\begin{equation}
    \Pin'(p,q)  =  \{a \in \Sigma_{p,q} : N'(a) = \pm 1\}.
\end{equation}

\noindent In addition, from Eq. \eqref{eqgcr564} it can be seen that for $a \in \clpq^{\text{even}}$, $N'(a) = N(a)$, therefore,

\begin{equation}\label{eqgcr571}
    \Pin(p,q) \inter \clpq^{\text{even}} = \Pin'(p,q) \inter \clpq^{\text{even}}.
\end{equation}

 If the focus is solely on rotations, excluding reflections, one would need to identify a specific subgroup that is mapped to the special orthogonal group. The Cartan-Dieudonné Theorem \ref{theo_a6s5d1fg3} states that any orthogonal transformation can be broken down into a finite number of reflections. However, reflections have a determinant of $-1$, thus every rotation can be expressed as a product of an even number of reflections. Proposition \ref{propos_5rtdgfijvknm} indicates that the elements mapped to rotations by the twisted adjoint representation are exactly those in the even subalgebra. This leads to a definition of the \textit{Spin Group} by restricting the Pin group to the even subalgebra.

\subsection*{Spin Group}

  {\colorlet{shadecolor}{Peach!20}\begin{shaded}  The group $\Spin(p,q)$ is defined as\cite{Roc16}
  \begin{equation}
    \Spin(p,q) = \{a \in \Sigma_{p,q}^{\text{even}} : N(a) = \pm 1\}.
\end{equation}
  \end{shaded}}

\noindent Similarly, for $\yhwidehat{\text{Ad}} : \Spin(p,q) \to \SO(p,q)$:

\begin{equation}
    \ker \yhwidehat{\text{Ad}} |_{\Spin(p,q)} = \{+1,-1\} = \mathbb{Z}_2
\end{equation}

\noindent which implies that

\begin{equation}
    \Spin(p,q) = \Pin(p,q) \inter \clpq^{\text{even}}.
\end{equation}

\noindent We can certainly conclude that the definition of the group $\Spin'(p,q)$ is redundant, since by Eq \eqref{eqgcr571}:

\begin{align}
    \begin{aligned}
    \Spin'(p,q) &= \{a \in \Sigma_{p,q}^{\text{even}} : N'(a) = \pm 1\}\\
    &=\Pin'(p,q) \inter \clpq^{\text{even}}\\
    &= \Pin(p,q) \inter \clpq^{\text{even}}\\
    &= \Spin(p,q).
    \end{aligned}
\end{align}

A sequence of groups and group homomorphisms is said to be \textit{exact} if the kernel of each homomorphism is the image of the preceding one.

  {\colorlet{shadecolor}{Peach!20}\begin{shaded} The following sequences are exact:
  \begin{equation}
\begin{tikzcd}
 1 \rar{} & \{\pm 1\} \rar{}  & \Pin(p,q) \rar{\yhwidehat{\text{Ad}}} & \O(p,q) \rar{} &1,\\
  1 \rar{} & \{\pm 1\} \rar{}  & \Spin(p,q) \rar{\yhwidehat{\text{Ad}}} & \SO(p,q) \rar{} &1.
\end{tikzcd} 
\end{equation}{}
  \end{shaded}}

As the fundamental group of $\text{SO}(n)$ is $\zz_2$ when $n>2$ it follows that in this case $\Spin(n)$ is a double cover of $\text{SO}(n)$. If $p,q$ are both $>1$ then $\Spin(p,q)$ is characterised as the unique double cover that induces a double cover of both $\text{SO}(p)$ and $\text{SO}(q)$.

\subsection*{The Reduced Pin and Spin Groups}

\paragraph{ } The components of the orthogonal group $\O(p,q)$ :

\begin{equation}
    \O^{\uparrow}_{+}(p,q),\;\; \O^{\uparrow}_{-}(p,q),\;\; \O^{\downarrow}_{+}(p,q), \;\; \O^{\downarrow}_{-}(p,q)
\end{equation}

\noindent together with its subgroups

\begin{equation}
    \O^{\uparrow}_{+}(p,q),\;\;  \O^{\uparrow}(p,q),\;\;  \O_{+}(p,q),\;\; \O^{\uparrow}_{+}(p,q) \cup \O^{\downarrow}_{-}(p,q).
\end{equation}

\noindent are defined in the Appendix \ref{sec_sd54fd}, respectively in the Eqs. \eqref{eq_lfhjknb654} and \eqref{eq_as564d1f}. In this subsection we present the reduced Pin and Spin groups that are related to those components.

  {\colorlet{shadecolor}{Peach!20}\begin{shaded}  The reduced Pin and Spin groups: $\Pin_{+}(p,q),\, \Pin_{+}'(p,q)$ and $\Spin_{+}(p,q)$ are respectively defined as follows \begin{align}
    \begin{aligned}
    \Pin_{+}(p,q) &= \{a \in \Sigma_{p,q} : N(a) = 1\},\\
    \Pin_{+}'(p,q) &=\{a \in \Sigma_{p,q} : N'(a) = 1\},\\
    \Spin_{+}(p,q) &= \{a \in \Sigma_{p,q}^{\text{even}} : N(a) = 1\}.
    \end{aligned}
\end{align}  
  \end{shaded}}

\noindent From the above definitions it is straightforward to notice that 

\begin{equation}\label{eqgcr821}
     \Spin_{+}(p,q) =  \Pin_{+}(p,q)  \inter \Pin_{+}'(p,q).
\end{equation}

\paragraph{ } Consider now $a \in \Pin_{+}(p,q)$ therefore $a \in \Sigma_{p,q}$ with $N(a) = 1$ that is

\begin{equation}
    N(a) = N(\mathbf{u}_1 \cdots \mathbf{u}_k) = N(\mathbf{u}_1) \cdots N(\mathbf{u}_k) = 1.
\end{equation}

\noindent If $N(\mathbf{u}_j) = \mathbf{u}_j^2 = g(\mathbf{u}_j,\mathbf{u}_j) = 1$ for all $j = 1,\ldots,k$ then $a \in \Pin_{+}(p,q)$ is such that $\yhwidehat{\text{Ad}}_{a}$ consists of the product of reflections on hyperplanes orthogonal to vectors of type $g(\mathbf{u}_j,\mathbf{u}_j) = 1$. In fact,

\begin{equation}
   \yhwidehat{\text{Ad}}_{a} = \yhwidehat{\text{Ad}}(\mathbf{u}_1 \cdots \mathbf{u}_k) = \yhwidehat{\text{Ad}}(\mathbf{u}_1) \cdots \yhwidehat{\text{Ad}}(\mathbf{u}_k) \text{ such that } g(\mathbf{u}_j,\mathbf{u}_j)  = 1,\;\; j = 1,\ldots,k.
\end{equation}

\noindent Consequently, if T(a) is the matrix 

\begin{equation}
 T(a) =  \begin{pmatrix} A_p & B_{p,q} \\ C_{q,p} & D_q \end{pmatrix}
\end{equation}

\noindent which represents $\yhwidehat{\text{Ad}}_{a}$, then $\det D_q > 0$. That yields:

\begin{equation}
    \yhwidehat{\text{Ad}}\big(\Pin_{+}(p,q)\big) = \O_{+}(p,q).
\end{equation}

\noindent Similarly, for $\mathbf{u} \in \rrpq$ such that $N'(\mathbf{u}) = -\mathbf{u}^2 = -g(\mathbf{u},\mathbf{u})$ and $N(\mathbf{u})$ implies that $g(\mathbf{u},\mathbf{u}) = 1$. For this case $a \in \Pin'_{+}(p,q)$ is such that $\yhwidehat{\text{Ad}}_{a}$ consists of the product of reflections on hyperplanes orthogonal to vectors of type $g(\mathbf{u}_j,\mathbf{u}_j) = -1$. Therefore, the matrix $T(a)$ representing such orthogonal transformation is given by $\det A_p$, which does not change the sign under this transformation. Hence

\begin{equation}
    \yhwidehat{\text{Ad}} (\Pin'_{+}(p,q) ) = O^{\uparrow}(p,q).
\end{equation}

\noindent In addition, by Eqs. \eqref{eqgcr821} and \eqref{eqagcr511}:

\begin{equation}
    \yhwidehat{\text{Ad}}(\Spin_{+}(p,q)) = \SO_{+}(p,q).
\end{equation}

\noindent Since the kernel of that mapping is $\zz_2$, the group $\Spin_{+}(p,q)$ is the \textit{2-fold covering} of the group $\SO_{+}(p,q)$.

  {\colorlet{shadecolor}{Peach!20}\begin{shaded}  More generally, 
  \begin{align}
    \begin{aligned}
    \Pin_{+}(p,q)/\zz_2 &\simeq \O_{+}(p,q),\\
    \Pin_{+}'(p,q)/\zz_2 &\simeq \O^{\uparrow}(p,q),\\
    \Spin_{+}(p,q)/\zz_2 &\simeq \SO_{+}(p,q).
    \end{aligned}
\end{align}
  \end{shaded}}

\begin{ex}\label{ex_spintres}
{{\textcolor{RedOrange}{{$\blacktriangleright$\;}}}}    The group \normalfont $ \Spin(3)$.  To introduce the $\Spin(3)$ group, we are going to work with rotations within the Clifford algebra $\mathcal{C}\ell_{3}$ \cite{Vaz97}. The lack of need for additional structure when performing a rotation in this space is a significant advantage compared to the case of vector algebra, where it needs to be supplemented by the structure of matrix algebra to apply a rotation through linear transformations. Consider a vector $\mathbf{v} \in \mathbb{R}^3$ upon which we will perform an arbitrary rotation such that $\mathbf{v} \to \mathbf{v}'$ after the rotation. The Clifford product $\mathbf{v}'\mathbf{v}$ is     $    \mathbf{v}'\mathbf{v} = \mathbf{v}' \cdot \mathbf{v} + \mathbf{v}' \wedge \mathbf{v} $. The first part $\'v' \cdot \'v$ is precisely the dot product, and that $\mathbf{v}' \wedge \mathbf{v}$ is related to the cross product through $\mathbf{v}' \times \mathbf{v} = \star(\mathbf{v}' \wedge \mathbf{v}) = -\'I(\mathbf{v}' \wedge \mathbf{v})$ where $\'I = \'e_1\'e_2\'e_3$. Therefore,  $\mathbf{v}'\mathbf{v} = \mathbf{v}' \cdot \mathbf{v} + \'I(\mathbf{v}' \times \mathbf{v}).$ The angle $\theta$ between $\mathbf{v}'$ and $\mathbf{v}$ is defined by the expressions

\begin{align}
\begin{aligned}\label{eq_espang}
\mathbf{v}' \cdot \mathbf{v} &=  |\mathbf{v}'||\mathbf{v}|\,\cos\,\theta, \\
|\mathbf{v}' \times \mathbf{v}|  &= |\mathbf{v}'||\mathbf{v}|\,\sin\,\theta.
\end{aligned}
\end{align}

One has $\mathbf{v}' \times \mathbf{v} \perp \mathbf{v}' \wedge \mathbf{v}$. Take $\'w$ a unitary normal vector in which $\mathbf{v}' \times \mathbf{v}  = |\mathbf{v}' \times \mathbf{v} |{\'w} = |\mathbf{v}'||\mathbf{v}|\,\sin\,\theta\,{\'w}$. Therefore,
\begin{align}
\begin{aligned}\label{eq_esp5}
\mathbf{v}'\mathbf{v} = \mathbf{v}' \cdot \mathbf{v} + \'I(\mathbf{v}' \times \mathbf{v}) &= |\mathbf{v}'||\mathbf{v}|\,\cos\,\theta + 
 |\mathbf{v}'||\mathbf{v}|\,\sin\,\theta\,\'I{\'w}= |\mathbf{v}'||\mathbf{v}|(\cos\,\theta + \sin\,\theta\,\'I{\'w}).
\end{aligned} 
\end{align}

On the other hand, one has

\begin{equation}
    \exp (\'I{\'w}\theta) = \sum_{n=0}^{\infty}\frac{(\'I{\'w}\theta)^n}{n!} =  \cos\,\theta + \'I{\'w}\,\sin\,\theta.
\end{equation} 

Hence the Clifford product is $ \mathbf{v}'\mathbf{v} =  |\mathbf{v}'||\mathbf{v}|\,\exp (\'I{\'w}\theta)$. Since we are working with rotations $|\mathbf{v}'| = |\mathbf{v}|$ and it follows that

\begin{equation}\label{eq_derot1}
    \mathbf{v}' =  \exp (\'I{\'w}\theta) \mathbf{v}.
\end{equation}

The equation \eqref{eq_derot1} above describes the rotation of the vector $v$ through a plane whose normal vector is $w$ by an angle $\theta$. However, by using appropriate trigonometric identities it can be written in a more suitable way as follows

\begin{equation}\label{eq_derotof}
    \mathbf{v}' = \left( \cos\frac{\theta}{2} + \'I\hat{\'w}\sin\frac{\theta}{2}\right)\mathbf{v} \left( \cos\frac{\theta}{2} - \'I\hat{\'w}\sin\frac{\theta}{2}\right) = \exp \left(\'I\hat{\'w}\frac{\theta}{2} \right) \mathbf{v}\,\exp \left( -\'I\hat{\'w}\frac{\theta}{2}\right)
\end{equation}

The above expression in the Eq. \eqref{eq_derotof} becomes more suitable for the following reason: take the scalar $\alpha \in \mathbb{R}$, by Eq. \eqref{eq_derot1} one has

\begin{align}
\begin{aligned}\label{eq_escarot1}
\alpha' = \exp(\'I{\'w}\theta) \alpha = \alpha (\cos\,\theta + \'I{\'w}\,\sin\,\theta).
\end{aligned} 
\end{align}

Whereas from Eq. \eqref{eq_derotof}, one has 
\begin{align}
\begin{aligned}\label{eq_escarotof}
\alpha' 
&= \alpha \left( \cos\frac{\theta}{2} + \'I\hat{\'w}\sin\frac{\theta}{2}\right)\left( \cos\frac{\theta}{2} - \'I\hat{\'w}\sin\frac{\theta}{2}\right)\\
&= \alpha \left(\cos^2\frac{\theta}{2} + \sin^2\frac{\theta}{2} + \'I\hat{\'w}\cos\frac{\theta}{2}\sin\frac{\theta}{2} -  \'I\hat{\'w}\cos\frac{\theta}{2}\sin\frac{\theta}{2} \right)\\
&= \alpha,
\end{aligned} 
\end{align}

That is, by Eq. \eqref{eq_derotof}, the scalars remain unchanged under a rotation, which is indeed how a scalar should be understood in the context of a rotation. The same result applies to bivectors $\mathbf{u} \wedge \mathbf{v}$ by using Eq.\eqref{eq_derotof}. Therefore, we conclude that the rotation of any multivector $A \in \cl_3$ by an angle $\theta$ through the plane whose normal is $\mathbf{w}$, such that $A \to A'$, is given by the expression:
\begin{equation}\label{eq_defeqrotspin3}
A' = \exp \left(\'I\hat{\mathbf{w}}\frac{\theta}{2} \right) A \exp \left( -\'I\hat{\mathbf{w}}\frac{\theta}{2}\right) .
\end{equation}

\noindent In addition, if we define $R = \exp\,{\'B}$ such that $\'B = \'I\hat{\'w}\theta/2 \in \bigwedge^2(\rr^3)$ one has $R \in \mathcal{C}\ell_{3}^{\text{even}}$. We can also observe that with $\theta = 0$, $\exp\,{0} = 1.$

Applying the reversion operation to 
$R$ yields

\begin{align}
\begin{aligned}\label{eq_spin3relrev1}
\yhwidetilde{R} = \yhwidetilde{\exp{\mathbf{B}}} = \exp\yhwidetilde{{\mathbf{B}}} = \exp({-\mathbf{B}}).
\end{aligned}
\end{align}

\noindent An important relation is derived when multiplying $\yhwidetilde{R}R$:

\begin{align}
\begin{aligned}\label{eq_spin3relrev2}
\yhwidetilde{R}R &= \exp({-B})\exp{B} = 1\
R\yhwidetilde{R} &= \exp{B}\,\exp({-B}) = 1 \
\Rightarrow \yhwidetilde{R} &= R^{-1}.
\end{aligned}
\end{align}

Notice that if $R_1, R_2 \in \mathcal{C}\ell_{3}^{\text{even}}$ such that $\yhwidetilde{R}_i = R_i^{-1}$, since $\mathcal{C}\ell_{3}^{\text{even}}$ is a subalgebra it follows that $R_{1}R_{2} = R_{3} \in \mathcal{C}\ell_{3}^{\text{even}}$, moreover, $\yhwidetilde{R_{3}}R_{3} = \yhwidetilde{R_{1}R_{2}}R_{1}R_{2} = \yhwidetilde{R_{2}}\yhwidetilde{R_{1}}R_{1}R_{2} = 1$. Therefore, the set of elements  $R \in \mathcal{C}\ell_{3}^{\text{even}}$ that satisfy $\yhwidetilde{R} =R^{-1}$ forms a group and this group is precisely the  $\Spin(3) = \{R \in \mathcal{C}\ell_{3}^{\text{even}} \,|\, \yhwidetilde{R} = R^{-1}\}$ group. From the equation \eqref{eq_defeqrotspin3}, we can see how $R$ appears paired to the right and left of $A$. Both $R \in \Spin(3)$ and $-R \in \Spin(3)$ describe the same rotation. In fact, let $R = \exp (I{\'w}\theta)/2 \in \Spin(3)$, which describes a rotation of angle $\theta$ through a plane with normal ${\'w}$. Note that:

\begin{equation}
\exp \left (\frac{I(-\'w)(2\pi - \theta )}{2}\right) = \exp \left(\frac{I\'w\theta}{2}\right)\exp\left(\frac{-I\'w\pi}{2}\right) = -\exp\left(\frac{I\'w\theta}{2}\right) = -R,
\end{equation}

\noindent which means that $-R$ describes a rotation of $(2\pi - \theta)$ around the plane with normal ${-\'w}$. In other words, the rotation of angle $\theta$ that $R$ describes on the plane with normal $\hat{\mathbf{w}}$ is equivalent to the rotation $(2\pi - \theta)$ that $-R$ describes in the same plane but in the opposite direction (normal ${-\'w}$). That is, two opposite elements in $\Spin(3)$ represent the same rotation in $\SO(3)$ which shows us that $\Spin(3)$ is the double cover of $\SO(3)$ as we mentioned before.

\begin{equation}
\frac{\Spin(3)}{\{\pm 1\}} \simeq \SO(3).
\end{equation}

Finally, the algebra $\clt$ is represented by the matrix algebra $ \text{Mat}(2,\mathbb{C})$, through the representation it follows that,

\begin{align}
\begin{aligned}\label{eq_spin3matrizpauli}
R\yhwidetilde{R} = 1\leftrightarrow \begin{pmatrix} 
    x_{1} & -x_{2}^{*} \\
    x_{2} & x_{1}^{*} 
    \end{pmatrix}\begin{pmatrix} 
    x_{1}^{*} & x_{2}^{*} \\
    -x_{2} & x_{1} 
    \end{pmatrix} = \begin{pmatrix} 
    1 & 0 \\
    0 & 1
    \end{pmatrix},
\end{aligned} 
\end{align}

\noindent that is, 
\begin{equation}
|x_{1}|^2 + |x_{2}|^2 = \text{det} \begin{pmatrix} 
    x_{1} & -x_{2}^{*} \\
    x_{2} & x_{1}^{*}
    \end{pmatrix} = 1.
\end{equation}

\noindent Thus, it establishes the isomorphism  $ \Spin\text{(3)} \simeq \text{SU}(2).$ $\; \textcolor{RedOrange}{{\blacktriangleleft}}$ 

\begin{ex}
{{\textcolor{RedOrange}{{$\blacktriangleright$\;}}}}    The Group \normalfont $\Spin(1,3)$.   In this case we consider the Lorentz group within the Clifford algebra $\mathcal{C}\ell_{1,3}$. The Lorentz group is better explored in the Appendix \ref{app62}. The Lorentz groups  $\O(1,3), \SO_{+}(1,3), \SO^{\uparrow}_{+}(1,3)$ are such that, respectively, their double coverings are \cite{Roc16}:
\end{ex}

\begin{align}
\begin{aligned}
&\text{Pin}(1,3) = \{S \in \clm^{\text{even}} \cup \clm^{-} \;|\; S\yhwidetilde{S} = \pm 1 \},\\
&\Spin(1,3) = \{S \in \clm^{\text{even}} \;|\;  S\yhwidetilde{S} = \pm 1 \}, \\
&\text{Spin}_{+}(1,3) = \{S \in \clm^{\text{even}} \;|\;  S\yhwidetilde{S} = 1 \} \simeq \text{SL}(2, \mathbb{C}).
\end{aligned} 
\end{align}

 It is also possible to perform rotations within $\mathcal{C}\ell_{1,3}$ \cite{Vaz00}. However, it needs a more carefully look because $\mathcal{C}\ell_{1,3}$ is a richer structure than $\mathcal{C}\ell_{3}$. For instance, when considering the space of bivectors, $\bigwedge^{2}(\mathbb{R}^{1,3})$, generated by the elements $\{e_{0}e_{1}, e_{0}e_{2}, e_{0}e_{3}, e_{1}e_{2}, e_{1}e_{3}, e_{2}e_{3}\}$, one has bivectors $B$ satisfying $B^2 > 0$, $B^2 < 0$, as well as $B^2 = 0$. In particular, there exist $B \in \bigwedge^{2}(\mathbb{R}^{1,3})$ with $B^2 = 1$, such as $e_{0}e_{1}, e_{0}e_{2}, e_{0}e_{3}$, as well as convenient linear combinations of these, and there are bivectors with $B^2 = -1$, such as $e_{1}e_{2}, e_{1}e_{3}, e_{2}e_{3}$, as well as convenient linear combinations of these. One can treat each case separately. Starting with the case of bivectors such that $B^2 = -1$, as in the Euclidean plane, such bivectors generate rotations in the plane defined by them. As examples of proper rotations, consider a bivector $B = \'e_{1}\'e_{3}$, a vector $\'x$ and the operation $\'x \mapsto S\'xS^{-1}$ such that $S = \text{exp}((\phi/2)\'e_{1}\'e_{3})$. Similarly to Eq. \eqref{eq_defeqrotspin3}, one has

\begin{align}
\begin{aligned}
&\exp \left(\frac{\phi}{2} \'e_{1}\'e_{3}\right) \'e_0 \,\exp \left(\frac{-\phi}{2} \'e_{1}\'e_{3}\right) = \'e_0 ,\\
&\exp \left(\frac{\phi}{2} \'e_{1}\'e_{3}\right) \'e_1 \,\exp \left(\frac{-\phi}{2} \'e_{1}\'e_{3}\right) = \'e_1\cos \phi + \'e_3\sen \phi, \\
&\exp \left(\frac{\phi}{2} \'e_{1}\'e_{3}\right) \'e_2 \,\exp \left(\frac{-\phi}{2} \'e_{1}\'e_{3}\right) = \'e_2 ,\\
&\exp \left(\frac{\phi}{2} \'e_{1}\'e_{3}\right) \'e_3 \,\exp \left(\frac{-\phi}{2} \'e_{1}\'e_{3}\right) = \'e_3\cos \phi - \'e_1\sen \phi.
\end{aligned} 
\end{align}

\noindent These equations show us that is there a rotation in the plane of the vectors $e_1, e_3$. As for the case of bivectors where $B^2 = 1$, such bivectors generate hyperbolic rotations in the plane they define. Take the bivector  $B = \'e_0 \'e_3$ with $(\'e_0\'e_3)^2 = 1 $, one has

\begin{align}
\begin{aligned}
&\exp \left(\frac{\phi}{2} \'e_{0}\'e_{3}\right) \'e_0 \,\exp \left(\frac{-\phi}{2} \'e_{0}\'e_{3}\right) = \'e_0 \cosh \phi - \'e_3 \sinh \phi, \\
&\exp \left(\frac{\phi}{2} \'e_{0}\'e_{3}\right) \'e_1 \,\exp \left(\frac{-\phi}{2} \'e_{0}\'e_{3}\right) = \'e_1, \\
&\exp \left(\frac{\phi}{2} \'e_{0}\'e_{3}\right) \'e_2 \,\exp \left(\frac{-\phi}{2} \'e_{0}\'e_{3}\right) = \'e_2 , \\
&\exp \left(\frac{\phi}{2} \'e_{0}\'e_{3}\right) \'e_3 \,\exp \left(\frac{-\phi}{2} \'e_{0}\'e_{3}\right) = \'e_3 \cosh \phi - \'e_0 \sinh \phi,
\end{aligned} 
\end{align}
\end{ex}

\noindent which is indeed a hyperbolic rotation in the plane of the vectors $e_0$ and $e_3$. The operation $x \mapsto SxS^{-1}$ with $S = \exp (\phi B/2)$ describes a spatial rotation if $B^2 = -1$ and describes a hyperbolic rotation if $B^2 = 1$. Furthermore, if $x \mapsto SxS^{-1}$ is an arbitrary rotation in spacetime, then we can uniquely write $S$ as $S = LU$ with $L$ being a hyperbolic rotation and $U$ a spatial rotation \cite{Vaz00}. The group that describes such rotations is the \text{Spin}$_{+}(1,3)$. Therefore, in general, a Lorentz transformation $L \in \O(1,3)$ is given by $L(x) = Sx\yhwidetilde{S}^{-1}$, however, a special Lorentz transformation given by $L \in \SO_{+}(1,3)$ corresponds to an $S$ from the even subalgebra so it can be written as $L(x) = SxS^{-1}$. $\; \textcolor{RedOrange}{{\blacktriangleleft}}$

\colorlet{chapter}{yellow!70}
\chapter{Algebraic and Classical Spinors}\label{chap_spinors}

\hypersetup{
  colorlinks = true,
  linkcolor  = Goldenrod,
  citecolor = [rgb]{0.88,0.8,0.01},
}

\paragraph{ } We introduced the dissertation
showing that \textcolor[rgb]{0.88,0.82,0.11}{spinors} are essential in quantum mechanics and in the study of fermions in particle physics. In addition we also presented the fact that they are naturally described by \textcolor[rgb]{0.82,0.01,0.11}{Clifford algebras}. However, the concept of spinors is not exclusive in a four-dimensional spacetime, in fact they can be defined in any dimension and signature with Clifford algebras providing the algebraic structure necessary to define and manipulate them. In this Chapter we present the definition and the classification of algebraic and classical spinors whereas in the seventh, we present spinors in the context of differential geometry, the concept of \textcolor[rgb]{0.74,0.06,0.88}{spin structures} on manifolds allow the definition of spinor fields in this context. The main reference of this chapter is Ref. \cite{Roc16}. The concept of algebraic spinors is connected to the idea of algebraic ideals in simple and semisimple Clifford algebras, as detailed in the Appendix \ref{app7} while the notion of classical spinors pertains to the representation space of the \textcolor[rgb]{0.96,0.65,0.14}{spin group}, as defined in the previous Chapter \ref{chap_groups}. However, we start by visiting some concepts in representation theory, since the classical and algebraic definition of spinors, in turn, are equivalent and are linked by the relation between minimal left ideals and irreducible modules/representations. 

\begin{definamarelo}
    A \textbf{left $\mathcal{A}$-module} for a $\kk$-algebra
$\mathcal{A}$ is a vector space $V$ over $\kk$ with a left bilinear multiplication $\mathcal{A} \cross V \to V$, $(a,v) \mapsto av$ such that $(a_1a_2)v = a_1 (a_2 v)$ for all $a_1,a_2 \in \mathcal{A}$ and $v \in V$. 
\end{definamarelo}

Modules and representations of an algebra or a group are two sides of the same coin. Given a representation $\rho: \mathcal{A} \to \text{End}_{\kk}(V)$ of  $\mathcal{A}$, the vector space $V$ becomes an $\mathcal{A}$-module if the required map $\mathcal{A} \cross V \to V$ is defined by the rule $(a,v) \mapsto av = \rho(a)v$ for all $a \in \mathcal{A}$ and $v \in V$. Conversely if $V$ is an  $\mathcal{A}$-module, for each $a \in \mathcal{A}$ a representation $\rho$ can be defined by setting for $\rho(a): V \to V$, $\rho(a)v = av$ for all $v \in V$. In this sense, a representation is irreducible if its corresponding module is irreducible, that is, if it is non-zero nad it has no submodule apart from $\{0\}$ and itself. The \textbf{left regular $\mathcal{A}$-module} is defined by setting $V=\mathcal{A}$ with the module operations being those of $\mathcal{A}$ (Def. \ref{defin_lkjhgcve}).  Let $\mathcal{A}$ be a $\kk$-algebra. A \textbf{left ideal} in $\mathcal{A}$ is a submodule of the left regular module. That is,

\begin{definamarelo}
    A \textbf{left ideal} of $\mathcal{A}$  is a nonempty subset $I$ of $\mathcal{A}$ satisfying:

    \begin{itemize}
        \item [1.] $x + y \in I$ for $x,y \in I$,
        \item [2.] $cx \in I$ for $x \in I$ and $c \in \kk$,
        \item [3.] $ax \in I$ for $x \in I$ and $a \in \mathcal{A}$
    \end{itemize}
\end{definamarelo} 

A left ideal is said to be \textit{minimal} if $I \neq \{0\}$ and there are no nonzero left ideals $J \subset I$ with $J \neq I$.

   {\colorlet{shadecolor}{Yellow!17}\begin{shaded}   The left ideal $I$ is minimal if and only if it is an irreducible left $\mathcal{A}$-module.\end{shaded}}

\noindent The Appendix \ref{app7} contains more details about this fact.  Moreover, there is a particularly convenient method to construct minimal ideals in terms of idempotent elements which are objects of fundamental importance in representation theory. This discussion is also shown in the Appendix \ref{app7}.

\begin{definamarelo}
An algebra $\A$ is said to be \textbf{simple} if the unique two-sided ideals of $\A$ are $\A$ and $\{0\}$ and  \textbf{semisimple} if is the sum of simple algebras.
\end{definamarelo}

Now we are in a position to proceed to the definition of algebraic spinors.

\section{Algebraic Spinors}

\begin{definamarelo}
An element of a minimal left ideal of a Clifford algebra $\clvg$ is said to be an \textbf{algebraic spinor} if $\clvg$ is a simple algebra and an \textbf{algebraic semispinor} if $\clvg$ is semisimple.
\end{definamarelo}

\paragraph{ } As discussed in the Appendix \ref{app7}, every algebra of type $ \text{Mat}(N,\kk)$ is simple and every simple algebra is isomorphic to $ \text{Mat}(N,\kk)$, where $\kk$ in this case is a
division algebra. In this sense, from the Clifford algebra classification -- real (Table \ref{table1}) or complex (Table \ref{table2}) -- we could see that every Clifford algebra is either a simple algebra or a sum of simple algebras. Once the classification of the Clifford algebras has been presented, we are now in a position to identify the algebraic spinors according to that classification. For instance, for a simple Clifford algebra, the isomorphism $\clpq \simeq  \text{Mat}(N,\kk)$ holds, and a minimal left ideal of $\clpq$ is isomorphic to $\kk^N$, that is used for the algebraic spinors classification. In the case of a semisimple Clifford algebra, the isomorphism $\clpq \simeq  \text{Mat}(N,\kk) \oplus  \text{Mat}(N,\kk)$ holds, a minimal left ideal of $\clpq$ is, therefore, isomorphic to $\kk^N$ and the algebraic semispinors classification follows from this isomorphism. The sum of algebraic semispinors is called an algebraic spinor, thus for a semisimple Clifford algebra, an algebraic spinor can be classified as $\kk^N \oplus \kk^N$. \\

According to the real Clifford algebras classification Table \ref{table1}, presented in the Subsection \ref{subsec245}, we proceed with the classification of the algebraic spinors \cite{Roc16}.

\begin{itemize}
    \item[\textcolor{Goldenrod}{{$\diamond$}}]\textcolor{Goldenrod}{\textbf{$p + q = 0,2 \mod 8:$}}$\;\; \clpq \simeq  \text{Mat}(2^{\left[\frac{n}{2} \right]}, \rr).$
    \item[ ] An algebraic spinor is thus an element of a minimal left ideal isomorphic to $\rr^{2^{\left[n/2 \right]}}$.
\end{itemize}

\begin{itemize}
    \item[\textcolor{Goldenrod}{{$\diamond$}}]\textcolor{Goldenrod}{\textbf{$p + q = 4,6 \mod 8:$}}$\;\; \clpq \simeq  \text{Mat}(2^{\left[\frac{n}{2} \right]-1}, \hh).$
    \item[ ] An algebraic spinor is thus an element of a minimal left ideal isomorphic to $\hh^{2^{\left[n/2 \right]-1}}$.
\end{itemize}

\begin{itemize}
    \item[\textcolor{Goldenrod}{{$\diamond$}}]\textcolor{Goldenrod}{\textbf{$p + q = 3,7 \mod 8:$}}$\;\; \clpq \simeq  \text{Mat}(2^{\left[\frac{n}{2} \right]}, \cc)$
    \item[ ] An algebraic spinor is thus an element of a minimal left ideal isomorphic to $\cc^{2^{\left[n/2 \right]}}$.
\end{itemize}

\begin{itemize}
    \item[\textcolor{Goldenrod}{{$\diamond$}}]\textcolor{Goldenrod}{\textbf{$p + q = 5 \mod 8:$}}$\;\; \clpq \simeq  \text{Mat}(2^{\left[\frac{n}{2} \right]-1}, \hh) \oplus  \text{Mat}(2^{\left[\frac{n}{2} \right]-1}, \hh) $
    \item[ ] The space of the algebraic semispinors is thus isomorphic to $\hh^{2^{\left[n/2 \right]-1}}$, whereas the space of algebraic spinors is isomorphic to $\hh^{2^{\left[n/2 \right]-1}} \oplus \hh^{2^{\left[n/2 \right]-1}}$.
\end{itemize}

\begin{itemize}
    \item[\textcolor{Goldenrod}{{$\diamond$}}]\textcolor{Goldenrod}{\textbf{$p + q = 1 \mod 8:$}}$\;\; \clpq \simeq  \text{Mat}(2^{\left[\frac{n}{2} \right]}, \rr) \oplus  \text{Mat}(2^{\left[\frac{n}{2} \right]}, \rr) $
    \item[ ] The space of the algebraic semispinors is thus isomorphic to $\rr^{2^{\left[n/2 \right]}}$, whereas the space of algebraic spinors is isomorphic to $\rr^{2^{\left[n/2 \right]}} \oplus \rr^{2^{\left[n/2 \right]}}$.
\end{itemize}

We summarise the above classification for algebraic spinors $S^{A}_{p,q}$ in the following Table, where $p+q = n$ and $\left[\frac{n}{2} \right]$ denotes the integer part of $\frac{n}{2}$

\begin{table}[H]
\centering
\scalebox{1.2}{
    \begin{tabular}{c|c}
        \rowcolor{Yellow!17} 
        \rule{0pt}{3ex} $p - q \mod 8 $ & $S^{A}_{p,q}$                                                 \\ \hline 
        \rule{0pt}{3ex} $0$             & $\rr^{2^{\left[n/2 \right]}}$                                                   \\
        \rowcolor{Yellow!17} 
        \rule{0pt}{3ex} $1$             & $\rr^{2^{\left[n/2 \right]}} \oplus \rr^{2^{\left[n/2 \right]}}$     \\
        \rule{0pt}{3ex} $2$             & $\rr^{2^{\left[n/2 \right]}}$                                                   \\
        \rowcolor{Yellow!17} 
        \rule{0pt}{3ex} $3$             & $\cc^{2^{\left[n/2 \right]}}$                                                  \\
        \rule{0pt}{3ex} $4$             & $\hh^{2^{\left[n/2 \right]-1}}$                                                 \\
        \rowcolor{Yellow!17} 
        \rule{0pt}{3ex} $5$             & $\hh^{2^{\left[n/2 \right]-1}} \oplus \hh^{2^{\left[n/2 \right]-1}}$ \\
        \rule{0pt}{3ex} $6$             & $\hh^{2^{\left[n/2 \right]-1}}$                                               \\
        \rowcolor{Yellow!17} 
        \rule{0pt}{3ex} $7$             & $\cc^{2^{\left[n/2 \right]}}$                           \\
    \end{tabular}
}
\caption{Algebraic Spinors Classification (Real Case).}
\label{table3}
\end{table}

As before, the case involving complex Clifford algebras is simpler since the complex Clifford algebras $\cl_{\cc}(n)$ depends only on the parity of $\dim V = n = p + q$.

\begin{itemize}
    \item[\textcolor{Goldenrod}{{$\diamond$}}]\textcolor{Goldenrod}{\textbf{$n = \dim V$ even}}
    \item[ ]  When $n$ is an even number, the isomorphism $\cl_{\cc}(2k) \simeq  \text{Mat}(2^{k}, \cc)$ holds, hence the space of algebraic spinors is isomorphic to $\cc^{2^{k}}$.
\end{itemize}

\begin{itemize}
    \item[\textcolor{Goldenrod}{{$\diamond$}}]\textcolor{Goldenrod}{\textbf{$n = \dim V$ odd}}
    \item[ ]  If $n$ is odd, the following isomorphism $\cl_{\cc}(2k+1) \simeq  \text{Mat}(2^{k}, \cc) \oplus  \text{Mat}(2^{k}, \cc)$ holds. The space of the algebraic semispinors is, therefore, isomorphic to $\cc^{2^{k}}$ and the space of algebraic spinors is isomorphic to $\cc^{2^{k}} \oplus \cc^{2^{k}}$.
\end{itemize}

\begin{table}[H]
\begin{center}
\scalebox{1.2}{\begin{tabular}{c|c}
\rowcolor{Yellow!17}  & \\ \rowcolor{Yellow!17} 
$n$ even & $\cc^{2^{k}}$                                                                               \\ 
\rowcolor{Yellow!17} & \\ \hline \\
$n$ odd             & $\cc^{2^{k}} \oplus \cc^{2^{k}}$ \\                   \\          
\end{tabular}}  \caption{Algebraic Spinors Classification (Complex Case).}\label{table4}
\end{center}
\end{table}

\section{Classical Spinors}\label{sec_tr5dfijsk}

\paragraph{ } Now we proceed to the general definition of classical spinors. The previous Chapter \ref{chap_groups} becomes very important not only to constitute the classic definition of spinor but also to highlight that spinors are objects with an intrinsic geometric nature.

\begin{definamarelo}
    Let $\rr^{p,q}$ be a quadratic space, $\clpq$ its associated Clifford algebra and $\Spin_{+}(p,q)$ the associated reduced Spin group. An element of the irreducible representation space of $\Spin_{+}(p,q)$ is said to be a \textbf{classical spinor}.
\end{definamarelo}

That is, by considering the mapping

\begin{equation}
    \yhwidehat{\text{Ad}}|_{\Spin_{+}(p,q)}:  \Spin_{+}(p,q) \to \text{GL}(S)
\end{equation}

\noindent the classical spinors are the elements in $S$.

Algebraic spinors are defined as elements of a minimal left ideal of Clifford algebras and since minimal left ideals may be viewed as representation spaces for irreducible representations, this algebraic definition turns out to be equivalent to the classical one.

Since the group $\Spin_{+}(p,q) = \{a \in \Gpq^{\text{even}} : N(a) = 1 \}$ is the set of even elements of the Clifford-Lipschitz group, an irreducible representation of $\Spin_{+}(p,q)$ descends from an irreducible representation of the even subalgebra $\clpq^{\text{even}}$. Let us recall the Theorem \ref{teo_seven} in the Chapter \ref{chap1} which establishes the following isomorphisms

\begin{equation}
    \clpq^{\text{even}} \simeq \cl_{q,p-1} \simeq \cl_{p,q-1} \simeq \cl_{q,p}^{\text{even}}.
\end{equation}

\noindent From this result, an irreducible representation of $\clpq^{\text{even}} \simeq \cl_{q,p}^{\text{even}}$ is thus obtained from an irreducible representation of $\cl_{q,p-1} \simeq \cl_{p,q-1}$ which has been previously classified. Therefore, a  classical spinor in a quadratic space $\rrpq$ or $\rr^{q,p}$ is an algebraic spinor (or an algebraic semispinor) in a quadratic space that is either $\rr^{q,p-1}$ or $\rr^{p,q-1}$. Moreover, a table for the real even subalgebra $\clpq^{\text{even}}$ classification can be constructed regarding such isomorphism $\clpq^{\text{even}} \simeq \cl_{q,p-1} \simeq \cl_{p,q-1}$. That is, for $p + q = n$ and $\kappa = (n-1)/2$ one has the following table:

\begin{table}[H]
\centering
\scalebox{1.2}{%
    \begin{tabular}{c|c}
        \rowcolor{Yellow!17} 
        \rule{0pt}{3ex} $p - q \mod 8 $ & $\cl_{p,q}^{\text{even}}$ \\ \hline 
        \rule{0pt}{3ex} 
        $0$ & $ \text{Mat}(2^{\left[\kappa \right]}, \rr) \oplus  \text{Mat}(2^{\left[\kappa \right]}, \rr)$ \\ 
        \rowcolor{Yellow!17} 
        \rule{0pt}{3ex} 
        $1$ & $ \text{Mat}(2^{\left[\kappa \right]}, \rr)$ \\ 
        \rule{0pt}{3ex} 
        $2$ & $ \text{Mat}(2^{\left[\kappa \right]}, \cc) $ \\ 
        \rowcolor{Yellow!17} 
        \rule{0pt}{3ex} 
        $3$ & $ \text{Mat}(2^{\left[\kappa \right]-1}, \hh)$ \\ 
        \rule{0pt}{3ex} 
        $4$ & $ \text{Mat}(2^{\left[\kappa \right]-1}, \hh) \oplus  \text{Mat}(2^{\left[\kappa \right]-1}, \hh)$ \\ 
        \rowcolor{Yellow!17} 
        \rule{0pt}{3ex} 
        $5$ & $ \text{Mat}(2^{\left[\kappa \right]-1}, \hh) $ \\ 
        \rule{0pt}{3ex} 
        $6$ & $ \text{Mat}(2^{\left[\kappa \right]}, \cc)$ \\ 
        \rowcolor{Yellow!17} 
        \rule{0pt}{3ex} 
        $7$ & $ \text{Mat}(2^{\left[\kappa \right]}, \rr)$ \\ 
    \end{tabular}%
}
\caption{Real Even Subalgebras Classification.}
\label{table5}
\end{table}

We are now in a position to describe the classification of classical spinors \cite{Roc17}. Let us introduce $p', q'$ such that $\clpq^{\text{even}} \simeq \cl_{p,q-1} = \cl_{p',q'}$.

\begin{itemize}
    \item[\textcolor{Goldenrod}{{$\diamond$}}]\textcolor{Goldenrod}{\textbf{$p + q = 1,7 \mod 8:$}}$\;\; \clpq^{\text{even}}  \simeq  \text{Mat}(2^{\left[\frac{n - 1}{2} \right]}, \rr).$
    \item[ ] In this case, $p' - q' = p - q + 1 = 0,2 \mod 8$.  A classical spinor is thus an element the representation space $\rr^{2^{\left[n - 1/2 \right]}}$, such that $n = p + q$.
\end{itemize}

\begin{itemize}
    \item[\textcolor{Goldenrod}{{$\diamond$}}]\textcolor{Goldenrod}{\textbf{$p + q = 2,6 \mod 8:$}}$\;\; \clpq^{\text{even}} \simeq  \text{Mat}(2^{\left[\frac{n - 1}{2} \right]}, \cc)$
    \item[ ] In this case, $p' - q' = p - q + 1 = 3,7 \mod 8$.  A classical spinor is thus an element the representation space $\cc^{2^{\left[n-1/2 \right]}}$.
\end{itemize}

\begin{itemize}
    \item[\textcolor{Goldenrod}{{$\diamond$}}]\textcolor{Goldenrod}{\textbf{$p + q = 4,6 \mod 8:$}}$\;\; \clpq^{\text{even}} \simeq  \text{Mat}(2^{\left[\frac{n-1}{2} \right]-1}, \hh).$
    \item[ ]  In this case, $p' - q' = p - q + 1 = 4,6 \mod 8$.  A classical spinor is thus an element the representation space $\hh^{2^{\left[n-1/2 \right]-1}}$.
\end{itemize}

\begin{itemize}
    \item[\textcolor{Goldenrod}{{$\diamond$}}]\textcolor{Goldenrod}{\textbf{$p + q = 4  \mod 8:$}}$\;\; \clpq^{\text{even}} \simeq  \text{Mat}(2^{\left[\frac{n-1}{2} \right]-1}, \hh) \oplus  \text{Mat}(2^{\left[\frac{n-1}{2} \right]-1}, \hh) $
    \item[ ]  In this case, $p' - q' = p - q + 1 = 5 \mod 8$ and the even subalgebra is semisimple. There are two non equivalent representations of $\Spin_{+}(p,q)$. The even subalgebra $\clpq^{\text{even}}$ can be decomposed into positive and negative parts by writing $\clpq^{\text{even}} = _{+}\clpq^{\text{even}} \oplus _{-}\clpq^{\text{even}}$ in such way the elements of the representation space of $_{+}\clpq^{\text{even}}$ are denominated positive classical spinors and the elements of the representation space of $_{-}\clpq^{\text{even}}$ are called negative classical spinors.
    A classical spinor positive or negative is an element the representation space $\hh^{2^{\left[n - 1/2 \right]-1}}$. 
\end{itemize}

\begin{itemize}
    \item[\textcolor{Goldenrod}{{$\diamond$}}]\textcolor{Goldenrod}{\textbf{$p + q = 0 \mod 8:$}}$\;\; \clpq^{\text{even}} \simeq  \text{Mat}(2^{\left[\frac{n-1}{2} \right]}, \rr) \oplus  \text{Mat}(2^{\left[\frac{n-1}{2} \right]}, \rr) $
    \item[ ]  In this case, $p' - q' = p - q + 1 = 1 \mod 8$ and the even subalgebra is semisimple. There are two non equivalent representations of $\Spin_{+}(p,q)$. The even subalgebra $\clpq^{\text{even}}$ can be decomposed into positive and negative parts by writing $\clpq^{\text{even}} = _{+}\clpq^{\text{even}} \oplus _{-}\clpq^{\text{even}}$ in such way the elements of the representation space of $_{+}\clpq^{\text{even}}$ are denominated positive classical spinors and the elements of the representation space of $_{-}\clpq^{\text{even}}$ are called negative classical spinors.
    A classical spinor positive or negative is an element the representation space $\rr^{2^{\left[n-1/2 \right]}}$.
\end{itemize}

We summarise the above classification for classical spinors $S^{C}_{p,q}$ in the following Table, where $p+q = n$ and $\left[\frac{n-1}{2} \right]$ denotes the integer part of $\frac{n-1}{2}$.

\begin{table}[H]
\centering
\scalebox{1.2}{%
    \begin{tabular}{c|c}
        \rowcolor{Yellow!17} 
        \rule{0pt}{3ex} $p - q \mod 8 $ & $S^{C}_{p,q}$ \\ \hline 
        \rule{0pt}{3ex} 
        $0$ & $\rr^{2^{\left[n-1/2 \right]}} \oplus \rr^{2^{\left[n-1/2 \right]}}$ \\ 
        \rowcolor{Yellow!17} 
        \rule{0pt}{3ex} 
        $1$ & $\rr^{2^{\left[n-1/2 \right]}}$ \\ 
        \rule{0pt}{3ex} 
        $2$ & $\cc^{2^{\left[n-1/2 \right]}}$ \\ 
        \rowcolor{Yellow!17} 
        \rule{0pt}{3ex} 
        $3$ & $\hh^{2^{\left[n-1/2 \right]-1}}$ \\ 
        \rule{0pt}{3ex} 
        $4$ & $\hh^{2^{\left[n-1/2 \right]-1}} \oplus \hh^{2^{\left[n-1/2 \right]-1}}$ \\ 
        \rowcolor{Yellow!17} 
        \rule{0pt}{3ex} 
        $5$ & $\hh^{2^{\left[n-1/2 \right]-1}}$ \\ 
        \rule{0pt}{3ex} 
        $6$ & $\cc^{2^{\left[n-1/2 \right]}}$ \\ 
        \rowcolor{Yellow!17} 
        \rule{0pt}{3ex} 
        $7$ & $\rr^{2^{\left[n-1/2 \right]}}$ \\ 
    \end{tabular}%
}
\caption{Classical Spinors Classification (Real Case).}
\label{table6}
\end{table}

\noindent These spaces arises from \textit{spin representations} which are given by\cite{spincali}:
\begin{table}[H]
    \centering
    \begin{tabular}{>{$}c<{$} | >{$}l<{$}}
        \toprule
        \rowcolor{Yellow!17} p-q \mod 8 & \text{Representation Details} \\
        \midrule
        0 & \varrho^{(+)} \oplus \varrho^{(-)} : \text{Spin}(p, q) \subset \cl_{p,q}^{\text{even}} \cong \text{End}_{\mathbb{R}}(S^{(+)}) \oplus \text{End}_{\mathbb{R}}(S^{(-)}) \\
          & _\pm\cl_{p, q}^{\text{even}} \cong \text{End}_{\mathbb{R}}(S^{(\pm)}) \\
        4 &  \varrho^{(+)} \oplus \varrho^{(-)} :  \text{Spin}(p, q) \subset \cl_{p,q}^{\text{even}} \cong \text{End}_{\mathbb{H}}(S^{(+)}) \oplus \text{End}_{\mathbb{H}}(S^{(-)}) \\
          & _\pm\cl_{p, q}^{\text{even}} \cong \text{End}_{\mathbb{H}}(S^{(\pm)}) \\
        2, 6 & \varrho : \text{Spin}(p, q) \subset \cl_{p,q}^{\text{even}} \cong \text{End}_{\mathbb{C}}(S) \\
        1, 7 & \varrho :\text{Spin}(p, q) \subset \cl(p, q)^{\text{even}} \cong \text{End}_{\mathbb{R}}(S)  \\
        3, 5 & \varrho  : \text{Spin}(p, q) \subset \cl(p, q)^{\text{even}} \cong \text{End}_{\mathbb{H}}(S) \\
        \bottomrule
    \end{tabular}
    \caption{The spin representations according to the real classical spinor classification.}\label{table7_spin}
\end{table}

In the next section, spin representations shall be inspected closely. Concerning the complex Clifford algebras, the situation is again straightforward. By Theorem \ref{teo_acsestrcomplex}, $\cl_{\cc}^{\text{even}}(n) \simeq \cc \tensor \clpq^{\text{even}}$, consequently, $\cl_{\cc}^{\text{even}}(n) \simeq \cc \tensor \cl_{p,q-1} \simeq \cl_{\cc}^{\text{even}}(n-1)$.

\begin{itemize}
    \item[\textcolor{Goldenrod}{{$\diamond$}}]\textcolor{Goldenrod}{\textbf{$n = \dim V$ even}}
    \item[ ]  If $\dim V = n$ is even, the isomorphism $\cl_{\cc}^{\text{even}}(2k) \simeq \cl_{\cc}(2k-1) \simeq  \text{Mat}(2^{k-1},\cc) \oplus  \text{Mat}(2^{k-1},\cc)$. There are two nonequivalent irreducible representations. The classical positive or negative spinors are elements of $\cc^{2^{k-1}}$.
\end{itemize}

\begin{itemize}
    \item[\textcolor{Goldenrod}{{$\diamond$}}]\textcolor{Goldenrod}{\textbf{$n = \dim V$ odd}}
    \item[ ]  If $\dim V = n$ is odd, then the following isomorphism 
   $\cl_{\cc}^{\text{even}}(2k+1) \simeq \cl_{\cc}(2k) \simeq  \text{Mat}(2^{k},\cc)$ holds. The classical are elements of $\cc^{2^{k-1}}$.
\end{itemize}

\begin{table}[H]
\begin{center}
\scalebox{1.2}{\begin{tabular}{c|c}
\rowcolor{Yellow!17}  & \\ \rowcolor{Yellow!17} 
$n$ even & $\cc^{2^{k-1}} \oplus \cc^{2^{k-1}}$                                                                               \\ 
\rowcolor{Yellow!17} & \\ \hline \\
$n$ odd             & $\cc^{2^{k}}$ \\                   \\          
\end{tabular}}  \caption{Classical Spinors Classification (Complex Case).}\label{table_complexspiasdg}
\end{center}
\end{table}

There is exactly one irreducible representation when $n$ is even, and two irreducible representations when $n$ is odd, which arises from the simplicity of the structure of complex Clifford algebras. Owing to this simplicity, the representations of the Spin group are frequently constructed by considering the corresponding complex Lie algebras, allowing for a more straightforward analysis of the group's representation theory.

\section{Spin Representation}\label{sec_spinrep}

\paragraph{ } From the representation theory point of view we have seen that a spinor is a element of the irreducible representation space of the double cover group of the rotation group $\SO(n)$ or more generally, $\SO_{+}(p,q)$. These double covers are \textit{Lie groups} and are called $\Spin(n)$ or $\Spin(p, q)$ groups. This double covering implies that for every rotation in $\SO(n)$, there are two corresponding elements in $\Spin(n)$, besides, $\SO(n)$ acts on vectors and $\Spin(n)$ through different actions. Despite these differences, the \textit{Lie algebras} of $\SO(n)$ and $\Spin(n)$ are isomorphic, that means that \textit{locally} they are the same. Let us present the definitions of Lie algebras and Lie groups \cite{liealg,liealglieg}.

\begin{definamarelo}
    A vector space $V$ over a field $\kk$ is said to be a \textbf{Lie algebra} if it is equipped with the bracket operation $[\,\cdot\,,\,\cdot\, ]: V \cross V \to V$, called the \textbf{Lie bracket}, with the
following properties:

\begin{itemize}
    \item [\textcolor{Goldenrod}{\textit{\textbf{1.}}}] $[\,\cdot\,,\,\cdot\, ]$ is  anti-symmetric, that is, for all $\mathbf{v},\mathbf{u} \in V$,
    \begin{equation}
        [\mathbf{v},\mathbf{u}] = -[\mathbf{u},\mathbf{v}].
    \end{equation}

    \item [\textcolor{Goldenrod}{\textit{\textbf{2.}}}] $[\,\cdot\,,\,\cdot\, ]$ is bilinear, that is,  , for all $\mathbf{v},\mathbf{u}, \mathbf{w} \in V$ and $a,b \in \kk$,

    \begin{align}
        \begin{aligned}
                [a\mathbf{v} + b\mathbf{u},\mathbf{w} ] &= a[\mathbf{v}, \mathbf{w} ] + b[\mathbf{u}, \mathbf{w} ],\\
                [\mathbf{v}, a\mathbf{u} + b\mathbf{w}] &= a[\mathbf{v}, \mathbf{u}] + b[\mathbf{v}, \mathbf{w}].
        \end{aligned}
    \end{align}
  \item [\textcolor{Goldenrod}{\textit{\textbf{3.}}}] $[\,\cdot\,,\,\cdot\, ]$  satisfies the Jacobi identity, that is for all $\mathbf{v},\mathbf{u}, \mathbf{w} \in V$,
    \begin{equation}
        [\mathbf{v}, [\mathbf{u}, \mathbf{w}]] + [\mathbf{u}, [\mathbf{w}, \mathbf{v}]] + [\mathbf{w}, [\mathbf{v}, \mathbf{u}]] = 0.
    \end{equation}
\end{itemize}
\end{definamarelo}

\begin{rem}
  The standard Lie bracket for matrix Lie algebras, called the commutator bracket, is defined for $A, B \in V$ as $[A, B] = AB - BA$ \cite{liealg}.
\end{rem}

\begin{definamarelo}
      A \textbf{Lie group} $G$ is a group which is also a manifold, such that the two structures are compatible in the sense that the multiplication and the inverse operations are smooth. 
\end{definamarelo}

\begin{ex}
{{\textcolor{yellow}{{$\blacktriangleright$\;}}}} The Lie group \normalfont$\text{GL}(n,\kk)$  \cite{Mastermath}. The group $\text{GL}(n,\kk)$ of $n \cross n$ invertible matrices with base field $\kk$ is a Lie group. Moreover, inside $\text{GL}(n,\kk)$ several interesting Lie groups that describes orthogonal transformations (Appendix \ref{app5}) can be found. Such as:
\begin{itemize}
    \item[\textcolor{Goldenrod}{\textbf{\textit{1.}}}] $\O(n) = \{A \in  \text{GL}(n,\rr) : A \cdot A^{T} = I \} $   (the orthogonal group);
     \item[\textcolor{Goldenrod}{\textbf{\textit{2.}}}] $\SO(n) = \{A \in  \O(n) : \det(A) = 1 \} $   (the special orthogonal group);
     \item[\textcolor{Goldenrod}{\textbf{\textit{3.}}}] $\text{SL}(n, \rr) = \{A \in  \text{GL}(n,\rr) : \det(A) = 1 \} $   (the special linear group);
     \item[\textcolor{Goldenrod}{\textbf{\textit{4.}}}] $\text{U}(n) = \{A \in  \text{GL}(n,\cc) : A \cdot A^{*} = I = 1 \} $   (the unitary linear group);
     \item[\textcolor{Goldenrod}{\textbf{\textit{5.}}}]$\text{SU}(n) = \{A \in  \text{U}(n) : \det(A) = 1 \} $   (the special unitary group). $\textcolor{yellow}{{\blacktriangleleft}}$
\end{itemize}\end{ex}

\begin{ex}
{{\textcolor{yellow}{{$\blacktriangleright$\;}}}} The Lie group \normalfont $\text{GL}(V)$. For any vector space $V$ over $\kk = \rr, \cc$, the general linear group associated to $V$ is given by,

\begin{equation}
    \text{GL}(V) = \{A : V \to V : A \text{ is a linear automorphism} \}.
\end{equation}

\noindent When $V$ is finite dimensional, say $\dim V = n$, then, as above, $\text{GL}(V)$ is a Lie group. As a matter of fact, $\text{GL}(V)$ is isomorphic to $\text{GL}(n,\kk)$, but since such isomorphism depends on the choice of basis of $V$, it is often useful to treat $\text{GL}(V)$ intrinsically rather than identifying it with $\text{GL}(n,\kk)$ \cite{Mastermath}. $\textcolor{yellow}{{\blacktriangleleft}}$\end{ex}

The relation of Lie Groups and Lie Algebras is the following \cite{liealglieg}:

 {\colorlet{shadecolor}{Yellow!17}\begin{shaded}  The tangent space of a matrix Lie group $G$ at a point $x$ in $G$ is the $n$-dimensional plane in $\rr^{n^{2}}$ tangent to $G$ at $x$. The tangent space at the identity equipped with
a Lie bracket operation is the Lie algebra $\mathbf{g}$ of $G$ \end{shaded}}

\begin{ex}
 \normalfont {{\textcolor{yellow}{{$\blacktriangleright$\;}}}}  $\text{SU}(2)/\zz_{2} \simeq \SO(3)$. In the Example \ref{ex_spintres} we presented the $\Spin(3)$ group together with the fact $\Spin(3)/\zz_{2} \simeq \SO(3)$ and we finish saying that $\Spin(3) \simeq \text{SU}(2)$. From Lie algebras, one can explicitly show the group homomorphism $\rho:  \text{SU}(2) \to \SO(3)$ with kernel $\ker \rho \simeq \mathbb{Z}_{2}$, the details are in the Appendix \ref{app6}. The reason for including this example and detailing this construction is also to demonstrate that everything that happens in the spinor space, with respect to symmetries with the spin group describing them, also occurs in the physical space with the special orthogonal group describing them. The common strategy to construct this explicit mapping comes from Lie algebras, the groups are said to be \textit{“locally isomorphic”}, since they have isomorphic Lie algebras. Considering one element $U \in \text{SU}(2) = \{U \in  \text{Mat}(2,\mathbb{C}): UU^{\dagger} = \mathbb{1}, \det(U) = 1 \}$ close to the identity, that is, $U = \mathbb{1} + i\varepsilon H$ for an small $\varepsilon$ and $H \in  \text{Mat}(2,\mathbb{C})$ from the fundamental condition $UU^{\dagger} = \mathbb{1}$ it follows that $H$ must be hermitian, indeed,

\begin{align}
    \begin{aligned}
        (\mathbb{1} + i\varepsilon H)(\mathbb{1} + i\varepsilon H)^{\dagger} = \mathbb{1} +i\varepsilon(H - H^{\dagger}) + \cancel{\mathcal{O}(\varepsilon^{2})} = \mathbb{1} \implies H = H^{\dagger}.
    \end{aligned}
\end{align}

\noindent From the second condition, $\det U = 1$ it follows that the trace of $H$ must vanish:

\begin{equation}
    \det(\mathbb{1} + i\varepsilon H) = 1 + \Tr(i\varepsilon H) +  \cancel{\mathcal{O}(\varepsilon^{2})} = 1 \implies \Tr H = 0.
\end{equation}

\noindent Therefore, the tangent vectors around the identity $(\mathbb{1} + i\varepsilon H)$ are traceless hermitian matrices in $ \text{Mat}(2,\mathbb{C})$ this set $\mathcal{H} = \{H \in  \text{Mat}(2,\mathbb{C}) : H = H^{\dagger}, \Tr H = 0\}$ has a Lie adjoint action $\text{SU}(2) \cross \mathcal{H} \to \mathcal{H}$, $(U,H) \mapsto \text{Ad}_{U} H = UHU^{-1} = UHU^{\dagger}$. Moreover, observe that $\mathcal{H}$ is generated by the Pauli matrices

\begin{align}
\begin{aligned}
\sigma_{1} = \begin{pmatrix} 
    0 & 1 \\
    1 & 0 
    \end{pmatrix}, \;\;\;\;\;\; \sigma_{2} = \begin{pmatrix} 
    0 & -i \\
    i & 0 
    \end{pmatrix}, \;\;\;\;\;\; \sigma_{3} = \begin{pmatrix} 
    1 & 0 \\
    0 & -1 
    \end{pmatrix},
\end{aligned}
\end{align}

\noindent satisfying $\sigma_j\sigma_k + \sigma_k\sigma_j = 2\delta_{ij}1$ $\dim_{\rr} \mathcal{H} = 3$, therefore,  $\rr^{3} \simeq \mathcal{H}$ as vector spaces. Choosing the standard basis $\{\'e_1, \'e_2,\'e_3\}$ on $\rr^{3}$ from Proposition \ref{propos_app6_inverse}  one has the linear isomorphism $\mu: \rr^{3} \to \mathcal{H}$, $x = (x_1,x_2,x_3) \mapsto  x \cdot \sigma = x_1\sigma_{1} + x_2\sigma_{2} + x_3\sigma_{3}$ which has inverse $\mu^{-1}: \mathcal{H} \to \rr^{3}$, $H \mapsto \frac{1}{2}\Tr(H\sigma^{i})e_{i}$.



In addition, from Proposition \ref{proposappp6_innerprodiso} it follows that $(\rr^{3}, \langle \,\cdot \,,\, \cdot\, \rangle_{\rr^{3}}) \simeq (\mathcal{H}, \langle \,\cdot \,,\, \cdot\, \rangle_{\mathcal{H}})$ as inner product spaces, where $\langle \,\cdot \,,\, \cdot\, \rangle_{\rr^{3}}$ is the standard inner product on $\rr^{3}$ and $\langle \,\cdot \,,\, \cdot\, \rangle_{\mathcal{H}}$ is an inner product defined by $\langle H_1, H_2 \rangle_{\mathcal{H}} := \frac{1}{2} \Tr(H_1H_2)$. This inner product defined on  $\mathcal{H}$ is $\text{SU}(2)$-invariant from Proposition \ref{propos_ik6gh5f}.

\begin{center}

\tikzset{every picture/.style={line width=0.75pt}} 

\begin{tikzpicture}[x=0.75pt,y=0.75pt,yscale=-1,xscale=1]

\draw [color={Goldenrod}  ,draw opacity=1 ]   (177.8,144.1) .. controls (187.8,130.6) and (171.8,131.6) .. (177.8,119.6) .. controls (183.5,108.2) and (181.98,110.79) .. (183.1,99.93) ;
\draw [shift={(183.3,98.1)}, rotate = 96.58] [color={Goldenrod}  ,draw opacity=1 ][line width=0.75]    (10.93,-3.29) .. controls (6.95,-1.4) and (3.31,-0.3) .. (0,0) .. controls (3.31,0.3) and (6.95,1.4) .. (10.93,3.29)   ;
\draw [color={Goldenrod}  ,draw opacity=1 ]   (320.8,142.6) .. controls (330.8,129.1) and (314.8,130.1) .. (320.8,118.1) .. controls (326.5,106.7) and (324.98,109.29) .. (326.1,98.43) ;
\draw [shift={(326.3,96.6)}, rotate = 96.58] [color={Goldenrod}  ,draw opacity=1 ][line width=0.75]    (10.93,-3.29) .. controls (6.95,-1.4) and (3.31,-0.3) .. (0,0) .. controls (3.31,0.3) and (6.95,1.4) .. (10.93,3.29)   ;

\draw (128.1,70.1) node [anchor=north west][inner sep=0.75pt]  [font=\normalsize]  {$\left(\mathbb{R}^{3} ,\ \langle \ \cdot \ ,\ \cdot \ \rangle _{\mathbb{R}^{3}}\right)$};
\draw (275.6,70.1) node [anchor=north west][inner sep=0.75pt]  [font=\normalsize]  {$\left(\mathcal{H} ,\ \langle \ \cdot \ ,\ \cdot \ \rangle _{\mathcal{H}}\right)$};
\draw (90.1,145.7) node [anchor=north west][inner sep=0.75pt]   [align=left] {{\footnotesize vector representation of SO(3)}};
\draw (277.6,145.7) node [anchor=north west][inner sep=0.75pt]   [align=left] {{\footnotesize adjoint representation of SU(2)}};
\draw (249.1,76.1) node [anchor=north west][inner sep=0.75pt]    {$\simeq $};

\end{tikzpicture}
\end{center}

\noindent Finally, Proposition \ref{propos_hjgfbdh} shows that there is a Lie group homomorphism $\rho: \text{SU}(2) \to \SO(3)$, $U \mapsto \rho(U): \rr^{3} \to \rr^{3} $ such that $\rho(U):= \frac{1}{2}\Tr(U\sigma_iU^{\dagger} \sigma^{j})e^{i} \tensor    e_{j}$ the kernel $\ker \rho \simeq \zz_2$ which again shows us that  $ \text{SU}(2)$ covers $\SO(3)$ twice, that is, $\Spin(3) \simeq \text{SU}(2)$. This example shows that $\text{SU}(2)$ and $\SO(3)$ are not global isomorphic but being isomorphic in any small neighborhood of $U \in  \text{SU}(2)$, manifesting as an isomorphism of their Lie algebras, in this sense, for a sufficiently small neighborhood around the identity, that neighborhood will include exactly one $U$ or $-U$ for all $U \in  \text{SU}(2)$$\; \textcolor{yellow}{{\blacktriangleleft}}$.
\end{ex}

The characteristics and applications of spinors originate in the Spin group, however, the idea is to study the well-known Lie algebra of the orthogonal group and its generators, which in turn can be defined in terms of Clifford algebras and exponentiated to obtain elements of the Spin group. Therefore, by exploring the classical definition of spinor, the space of spinors may also be defined from the \textit{spin representation} of the orthogonal Lie algebra. The Lie algebra of $\SO(n)$ and $\SO(p,q)$ is given by

\begin{align}
    \begin{aligned}
         \mathbf{so}(n) = \{A \in \text{Mat}(n,\rr) : A^{\intercal} = -A \},\\
         \mathbf{so}(p,q)  = \{A \in \text{Mat}(n,\rr) : A^{\intercal}J = -JA \}
     \end{aligned}\end{align}

\noindent with

\begin{equation}
    J = \begin{pmatrix}
        -I_{p} & 0 \\
        0 & I_{q}
    \end{pmatrix}.
\end{equation}

\noindent More generally, given a vector space $V$ (over a field with characteristic different than 2) with a nondegenerate symmetric bilinear form $g: V \cross V \to V$, the special orthogonal Lie algebra consists of traceless endomorphisms $A : V \to V$ which are skew-symmetric with respect to this bilinear form, that is:

\begin{equation}
    \textbf{so}(V) = \{ A :V \to V : g( A\mathbf{u}, \mathbf{v}) = -g(\mathbf{u}, A\mathbf{v}) \text{ for all } \mathbf{u}, \mathbf{v} \in V \}
\end{equation}

\noindent There is a natural isomorphism  $\bigwedge^{2} (V) \simeq \textbf{so}(V)$ such that \cite{spingeofig}

\begin{equation}
    \mathbf{v} \wedge \mathbf{u} \mapsto  \mathbf{v} \curlywedge  \mathbf{u}:=
    g(\mathbf{v}, \, \cdot \,)\mathbf{u} - g( \mathbf{u}, \, \cdot \,)\mathbf{v} 
\end{equation}

\noindent That is, $ \mathbf{v} \curlywedge  \mathbf{u}(\'x) = g(\mathbf{v}, \'x )\mathbf{u} - g( \mathbf{u}, \'x )\mathbf{v} \in \textbf{so}(V)$. Indeed,

\begin{align}
    \begin{aligned}
        g(  \mathbf{v} \curlywedge  \mathbf{u} (\mathbf{x}), \'y ) &= g(  g( \mathbf{v}, \'x )\mathbf{u} - g( \mathbf{u}, \'x )\mathbf{v} , \'y ) \\
        &=  g(\mathbf{v} , \'x ) g(\mathbf{u} , \'y )  - g(\mathbf{u} , \'x ) g(\mathbf{v} , \'y )\\
    &= - [ g(\mathbf{x} , \mathbf{u} ) g(\mathbf{v} , \'y )  - g(\mathbf{x} , \mathbf{v} ) g(\mathbf{u} , \'y )]\\
    &= - g(\mathbf{x} ,g( \mathbf{v}, \'y ) \mathbf{u} - g( \mathbf{u}, \'y ) \mathbf{v}\\
    &= - g( \'x, \mathbf{v} \curlywedge  \mathbf{u} (\mathbf{y}) ).
    \end{aligned}
\end{align}

\begin{tfpropos}
    The mapping \normalfont $\uplambda: \textbf{so}(V) \to \cl(V)$ \textit{given by}

    \begin{equation}
        \uplambda (\mathbf{v} \curlywedge  \mathbf{u}) = \frac{1}{4}(\mathbf{u}\mathbf{v} - \mathbf{v}\mathbf{u})
    \end{equation}

    \noindent \textit{is an injective Lie algebra homomorphism.}
\end{tfpropos}
\noindent \textcolor{Goldenrod}{\textit{Proof.}} Bivectors in the Clifford algebra $\bigwedge^{2}(V) \subset \cl(V,g)$  are closed under taking Lie commutators forming a Lie algebra and therefore contains
$\textbf{so}(V)$ as a Lie subalgebra via the embedding $\uplambda$ \cite{Gracia-Bondia:2001upu}. The commutator $[\mathbf{u} \curlywedge \mathbf{v},\'x \curlywedge \'y]$ leads to:

\begin{align}\label{eq_438ehufzwxerctvbyg}
    \begin{aligned}
        [\mathbf{u} \curlywedge \mathbf{v},\'x \curlywedge \'y] (\'z) &= \mathbf{u} \curlywedge \mathbf{v} (\'x \curlywedge \'y(\'z)) - \'x \curlywedge \'y( \mathbf{u} \curlywedge \mathbf{v} (\'z))\\
        &= \mathbf{u} \curlywedge \mathbf{v} (g(\'x,\'z)\'y - g(\'y,\'z)\'x) -  \'x \curlywedge \'y(g(\mathbf{v},\'z )\mathbf{u} - g(\mathbf{u},\'z)\mathbf{v})\\
        &= g(\'x,\'z)g(\mathbf{u},\'y)\mathbf{v} - g(\'x,\'z)g(\mathbf{v},\'y)\mathbf{u} 
        - g(\'y,\'z)g(\mathbf{u},\'x)\mathbf{v} + g(\'y,\'z)g(\mathbf{v},\'x)\mathbf{u}
\\&\quad+ g(\mathbf{v},\'z)g(\'x,\mathbf{u})\'y - g(\mathbf{v},\'z)g(\'y,\mathbf{u})\'x
- g(\mathbf{u},\'z)g(\'x,\mathbf{v})\'y + g(\mathbf{u},\'z)g(\'y,\mathbf{v})\'x\\
&= - g(\'y,\'z)g(\mathbf{u},\'x)\mathbf{v} + g(\mathbf{v},\'z)g(\mathbf{u},\'x)\'y
g(\'x,\'z)g(\mathbf{u},\'y)\mathbf{v} - g(\mathbf{v},\'z)g(\mathbf{u},\'y)\'x 
\\&\quad -  g(\'x,\'z)g(\mathbf{v},\'y)\mathbf{u}  + g(\mathbf{u},\'z)g(\mathbf{v},\'y)\'x
+ g(\'y,\'z)g(\mathbf{v},\'x)\mathbf{u} - g(\mathbf{u},\'z)g(\mathbf{v},\'x)\'y
\\&= g(\mathbf{u},\'x)\mathbf{v} \curlywedge \'y (\'z)
-g(\mathbf{u},\'y)\mathbf{v} \curlywedge \'x (\'z)
+g(\mathbf{v},\'y)\mathbf{u} \curlywedge \'x (\'z)
-g(\mathbf{v},\'x)\mathbf{u} \curlywedge \'y (\'z).
    \end{aligned}
\end{align}

\noindent In addition, the Clifford relation $\{\mathbf{v},\mathbf{u}\} = \mathbf{v}\mathbf{u} + \mathbf{u}\mathbf{v} = -2(\mathbf{v},\mathbf{u})1_V$ yields

\begin{align}\label{4t8hrufmx34d5fg}
    \begin{aligned}
        [\uplambda(\mathbf{u} \curlywedge \mathbf{v}), \mathbf{w}] &= 
        [\frac{1}{4}(\mathbf{u}\mathbf{v} - \mathbf{v} \mathbf{u}), \mathbf{w}] 
    =\frac{1}{4}[\mathbf{u}\mathbf{v},\mathbf{w}] - \frac{1}{4}[\mathbf{v}\mathbf{u},\mathbf{w}]\\
    &= \frac{1}{4}[\mathbf{u}\mathbf{v},\mathbf{w}] - \frac{1}{4}[- \mathbf{u}\mathbf{v} - 2(\mathbf{v},\mathbf{u})1_V, \mathbf{w}]
    = \frac{1}{2}[\mathbf{u}\mathbf{v},\mathbf{w}] + \frac{1}{2}g(\mathbf{v},\mathbf{u})[1_V, \mathbf{w}]\\
    &= \frac{1}{2}[\mathbf{u}\mathbf{v},\mathbf{w}]
    = \frac{1}{2} (\mathbf{u} {\mathbf{v},\mathbf{w}} -{\mathbf{u},\mathbf{w}}\mathbf{v})
    = g(\mathbf{v},\mathbf{w})\mathbf{u} - g(\mathbf{u},\mathbf{w})\mathbf{v}.
    \end{aligned}
\end{align}

\noindent Where we used the relation
$[AB,C] = A\{B,C\}- \{A,C\}B$ with respect to the anti-commutator \cite{Gracia-Bondia:2001upu}.

Moreover, by using Eq. \eqref{4t8hrufmx34d5fg} one has:
\begin{align}\label{eq_3456uytjg3588}
    \begin{aligned}
        [\uplambda(\mathbf{u} \curlywedge \mathbf{v}),\uplambda (\'x \curlywedge \'y)] &= [\uplambda(\mathbf{u} \curlywedge \mathbf{v}), \frac{1}{4}\'x\'y-\'y\'x]\\
        &= \frac{1}{4}[\uplambda(\mathbf{u} \curlywedge \mathbf{v}), [\'x,\'y]]\\
        &= \frac{1}{4}([\'x, [\'y, \uplambda(\mathbf{u} \curlywedge \mathbf{v})]]  + [\'y, [ \uplambda(\mathbf{u} \curlywedge \mathbf{v}),\'x]]) \\
        &= \frac{1}{4}(-[\'x, g(\mathbf{u},\'y)\mathbf{v} - g(\mathbf{v},\'y)u] +[\'y,  g(\mathbf{v},\'x)\mathbf{u} - g(\mathbf{u},\'x)\mathbf{v}])\\
        &=  \frac{1}{4}g(\mathbf{u},\'y)[\'x,\mathbf{v}] - \frac{1}{4}g(\mathbf{v},\'y)[\'x,\mathbf{u}] + \frac{1}{4}g(\mathbf{v},\'x)[\'y,\mathbf{u}] - \frac{1}{4}g(\mathbf{u},\'x)[\'y,\mathbf{v}]\\
&=  g(\mathbf{u},\'x)\uplambda(\mathbf{v}\curlywedge \'y) 
\!-\!g(\mathbf{u},\'y)\uplambda(\mathbf{v} \curlywedge \'x)
\!+\! g(\mathbf{v},\'y)\uplambda(\mathbf{u} \curlywedge \'x)
\!-\! g(\mathbf{v},\'x)\uplambda(\mathbf{u} \curlywedge \'y).\!
    \end{aligned}
\end{align}

\noindent We conclude by comparing Eqs. \eqref{eq_438ehufzwxerctvbyg} and \eqref{eq_3456uytjg3588} that $\uplambda$ defines an injective Lie algebra homomorphism. \textcolor{Goldenrod}{$\Box$}

\paragraph{ } The Lie algebra of the Spin group is the same Lie algebra of the special orthogonal group, therefore one has (for $n=p+q$) $\dim  \mathbf{spin}(n) = \dim \Spin(n) = \dim \SO(n) = \dim \mathbf{so}(n) = \frac{n(n - 1)}{2}$ = $\dim \bigwedge^{2}(\rr^{n})$\cite{spinazul}. The point is:

 {\colorlet{shadecolor}{Yellow!17}\begin{shaded}  Exponentiating $\textbf{so}(V)$ in $\text{End}(V)$ generates the identity component $\SO_{+}(V)$ of $\SO(V)$, whereas exponentiating $\uplambda(\textbf{so}(V))$ in $\cl(V,g)$ generates the Spin group $\Spin_{+}(V)$.
 \end{shaded}}

\begin{ex} \label{eq_example123456}{{\textcolor{yellow}{{$\blacktriangleright$\;}}}}  Exponentiating an element  \normalfont $\textbf{so}(V)$ \textit{in} $\text{End}(V)$ \textit{and in} $\uplambda(\textbf{so}(V))$ \textit{in} $\cl(V,g)$.
    Suppose that $\mathbf{e}_{1}, \mathbf{e}_{2} \in V$  are elements of an orthonormal basis of $V$. Therefore, $\mathbf{e}_{1} \wedge \mathbf{e}_{2} \in \bigwedge^{2} (V)$ and $\mathbf{e}_{1} \curlywedge \mathbf{e}_{2} \in \textbf{so}(V)$ is a antisymmetric rotation matrix of the form:

   \begin{equation}
       \begin{pmatrix}
        0& -1\\
        1&  0
    \end{pmatrix}
   \end{equation} 

\noindent whose exponential is

   \begin{equation}
     R_{\text{SO}}(\theta) = \exp(\theta (\mathbf{e}_{1} \curlywedge \mathbf{e}_{2} )) =   \begin{pmatrix}
        \cos \theta &  -\sin \theta\\
       \sin \theta&  \cos \theta
    \end{pmatrix}.
   \end{equation}

\noindent In particular, $  R_{\text{SO}}(2\pi)$ is the identity matrix $\mathbb{1}$. Conversely, by exponentiating the image of the
same Lie algebra element $\uplambda(\mathbf{e}_{1} \curlywedge \mathbf{e}_{2}) = \frac{1}{2}\mathbf{e}_{1}\mathbf{e}_{2} \in \cl(V,g)$ one gets

\begin{equation}\label{eq_nyhjtrgfjuytr}
    R_{\text{Spin}}(\theta) = \exp\left(\frac{1}{2}\right) = \cos\left(\frac{\theta}{2} \right)\mathbb{1} + \mathbf{e}_{1}\mathbf{e}_{2}\sin\left(\frac{\theta}{2} \right)
\end{equation}

Notably, $R_{\text{Spin}}(2\pi) = -\mathbb{1}$, so that the periodicity of $R_{\text{Spin}}(\theta)$ is $4\pi$. This is a fascinating and unique fact about spinors, $R_{\text{Spin}}$ acts on spinors by the left multiplication and unlike vectors and tensors, it takes a rotation of $720$ degrees for a spinor to go back to its original state. One can verify that acting on vectors via $\mathbf{v} \to R_{\text{Spin}}\mathbf{v} R_{\text{Spin}}^{-1}$ rotates the vector v by an angle $\theta$ in the plane defined by $\mathbf{e}_{1} \wedge \mathbf{e}_{2}$. Putting this all together, the group obtained by exponentiating $\textbf{so}(V) $ in terms of the Clifford algebra is precisely the double cover of $\SO(V)$, that is, the Spin group. This is an alternative way to obtain the Spin group, via Lie algebras, with no mention of the definitions of such group given in the previous Chapter \ref{chap_groups}.  $\; \textcolor{yellow}{{\blacktriangleleft}}$
\end{ex}

\begin{ex} \label{ex_re2654h56984ytrg}{{\textcolor{yellow}{{$\blacktriangleright$\;}}}}  Representations of  \normalfont SL$(2,\cc)$\textit{ and Weyl Spinors}. This example is quite long, but it has two objectives, which are to discuss $\text{SL}(2,\cc)$ representations and introduce spinors associated with the \textit{Spin representation} of $\text{SL}(2,\cc)$. When exponentiating $\textbf{sl}(2,\cc)$ in terms of $\cl_{3}$ the $\Spin_{+}(1,3)$ group and Weyl spinors arise.  We start from $\textbf{so}_{+}(1,3) \simeq \textbf{sl}(2,\cc)$ and write $\textbf{sl}(2,\cc)$ in terms of the Clifford algebra $\cl_{3}$ and we obtain a $2\times 2$ representation of the Spin group $\Spin_{+}(1,3) \simeq \text{SL}(2,\cc)$ that acts on $2 \times 2$ spinors called Weyl spinors. It is worth to emphasise first that the matrices in the Lie group $\text{SU}(2)$ are unitary, consequently, in the Lie algebra $\textbf{su}(2)$ the matrices are anti-hermitian.
\begin{align}
   \begin{aligned}
\color{Goldenrod} \begin{cases}
    \color{black} \text{Lie Group SU}(2) = \{U \in \cc^{2\times2}: U^{-1} = \color{black} U^{\dagger}, \det U = +1 \},\\\color{black} 
        \color{black}  \text{Lie Algebra }\textbf{su}(2) = \{H \in \cc^{2\times2}: H = -H^{\dagger} \tr H = 0\}. \color{black} 
\end{cases}\color{black} 
    \end{aligned}
\end{align}

\noindent The $\text{SL}(2,\cc)$ case is simply:

   \begin{align}
    \begin{aligned}
     \color{Goldenrod}  \begin{cases}
         \color{black}  \text{Lie Group SL}(2,\cc) = \{L \in \cc^{2\times2}: \det L = +1 \},\\\color{black} 
         \color{black} \text{Lie Algebra } \textbf{sl}(2,\cc) = \{M \in \cc^{2\times2}: \tr M = 0\}.\color{black} 
       \end{cases} \color{black} 
    \end{aligned}
\end{align}

\noindent In this case we have that $\textbf{sl}(2,\cc)$ is generated by three complex parameters or six real parameters. That is, $M \in \textbf{sl}(2,\cc)$ is such that for $a_{i} \in \rr$:

\begin{align}
    \begin{aligned}
      M \!=\!  a_{1}\!\begin{pmatrix} 
    0 & i \\
    i & 0 
    \end{pmatrix} \!+ a_{2}\!\begin{pmatrix} 
    0 & 1 \\
    -1 & 0 
    \end{pmatrix} \!+ a_{3}\!\begin{pmatrix} 
    i & 0 \\
    0 & -i 
    \end{pmatrix} \!+ a_{4}\!\begin{pmatrix} 
    0 & 1 \\
    1 & 0 
    \end{pmatrix} \!+ a_{5}\!\begin{pmatrix} 
    0 & -i \\
    i & 0 
    \end{pmatrix} \!+ a_{6}\!\begin{pmatrix} 
    1 & 0 \\
    0 & -1 
    \end{pmatrix}.
    \end{aligned}
\end{align}

\noindent Which corresponds with the three bivectors and the three vectors in the Clifford algebra $\cl_{3}$ represented by the Pauli matrices presented in the Example \ref{ex_reprec22}:

\begin{align}
\begin{aligned}
\mathbf{e}_{1} \leftrightarrow  \sigma_{1}, \;\;\;\;\;\;
\mathbf{e}_{2} \leftrightarrow  \sigma_{2}, \;\;\;\;\;\;
\mathbf{e}_{3} \leftrightarrow  \sigma_{3}, \;\;\;\;\;\;
\end{aligned}
\end{align}

\begin{center}
\begin{tabular}{cc|cc}
 \rowcolor{Yellow!17} & $\textbf{sl}(2,\mathbb{C})$  & $\clt$ &\; \; \tabularnewline 
 & $\sigma_{1}$,\;\;$\sigma_{2}$,\;\;$\sigma_{3}$ &  $\mathbf{e}_{1}$,\;\;$\mathbf{e}_{2}$,\;\;$\mathbf{e}_{3}$  \tabularnewline
 & $\sigma_{1}\sigma_{2}$,\;\;$\sigma_{1}\sigma_{3}$,\;\; $\sigma_{2}\sigma_{3}$ & $\mathbf{e}_{1} \mathbf{e}_{2}$,\;\;$\mathbf{e}_{1} \mathbf{e}_{3}$,\;\; $\mathbf{e}_{2} \mathbf{e}_{3}$  
\end{tabular}
\par \end{center}

\noindent That is,

\begin{align}
    \begin{aligned}
     \color{Goldenrod}\underbrace{\color{black} \begin{pmatrix} 
    0 & i \\
    i & 0 
    \end{pmatrix}}_{\color{black} \sigma_{2}\sigma_{3}}  \color{black}, \;\;
    \color{Goldenrod}\underbrace{\color{black} \begin{pmatrix} 
    0 & 1 \\
    -1 & 0 
    \end{pmatrix}}_{\color{black} \sigma_{3}\sigma_{1}}  \color{black}, \;\;
     \color{Goldenrod}\underbrace{\color{black} \begin{pmatrix} 
    i & 0 \\
    0 & -i 
    \end{pmatrix} }_{\color{black} \sigma_{1}\sigma_{2}}  \color{black},\;\;
    \color{Goldenrod} \underbrace{\color{black} \begin{pmatrix} 
    0 & 1 \\
    1 & 0 
    \end{pmatrix} }_{\color{black} \sigma_{1}}  \color{black},\;\;
      \color{Goldenrod} \underbrace{\color{black} \begin{pmatrix} 
    0 & -i \\
    i & 0 
    \end{pmatrix} }_{\color{black} \sigma_{2}}  \color{black},\;\;
    \color{Goldenrod}\underbrace{\color{black} \begin{pmatrix} 
    1 & 0 \\
    0 & -1 
    \end{pmatrix}}_{\color{black} \sigma_{2}}  \color{black}. 
    \end{aligned}
\end{align}

The anti-hermitians bivectors generate rotations and the hermitians vectors generate boosts \cite{LorentzG}. When multiplying the rotations $\sigma_{i}\sigma_{j}$ by the complex unity $i$ one gets the boosts generators. One can think about the six generators of $\textbf{sl}(2,\cc)$ as made out of two copies of $\textbf{su}(2)$: $\textbf{sl}(2,\cc) \simeq \textbf{su}(2) \oplus (\pm i) \textbf{su}(2)$. In addition, by defining the rotations generators $J_{ij}$ as \begin{align}
    \begin{aligned}
        J_{23} = -\frac{1}{2}\sigma_{2}\sigma_{3},\;\;\;
        J_{31} = -\frac{1}{2}\sigma_{3}\sigma_{1},\;\;\;
        J_{12} = -\frac{1}{2}\sigma_{1}\sigma_{2},
    \end{aligned}
\end{align}

\noindent and the boost generators as

\begin{align}
    \begin{aligned}
        K_{t1} = -\frac{1}{2}\sigma_{1},\;\;\;
        K_{t2} = -\frac{1}{2}\sigma_{2},\;\;\;
        K_{t3} = -\frac{1}{2}\sigma_{3},
    \end{aligned}
\end{align}

\noindent all the commutations relations are $(\textbf{so}_{+}({1,3}) \simeq \textbf{sl}(2,\cc))$

\begin{align}
\begin{aligned}
&[J_{13}, J_{31}] = J_{12}, &&[J_{12}, J_{23}] = J_{31}, &&[J_{31}, J_{12}] = J_{23},\\
&[J_{23}, K_{t2}] = K_{t3}, &&[J_{23}, K_{t3}] = -K_{t2}, &&[J_{31}, K_{t3}] = K_{t1},\\
&[J_{31}, K_{t1}] = K_{t2}, &&[J_{12}, K_{t1}] = K_{t3}, &&[J_{12}, K_{t2}] = -K_{t1},\\
&[K_{t1}, K_{t2}] = -J_{12}, &&[K_{t3}, K_{t1}] = -J_{31}, &&[K_{t1}, K_{t3}] = -J_{23}.
\end{aligned}
\end{align}

\noindent It is worth to emphasise that, one can also take the linear complex combinations of the rotations and boost generators and define 

\begin{equation}
    A_{i} = \frac{1}{2}(J_{i} + iK_{i}), \;\;\; B_{i} = \frac{1}{2}(J_{i} - iK_{i}).
\end{equation}

\noindent They satisfy the $\textbf{su}(2)_{\cc}$ algebra:

\begin{align}
    \begin{aligned}
    &[A_i,A_j] = i\epsilon_{ijk}A_k,\\
    &[B_i, B_j] = i\epsilon_{ijk}B_{k},\\
        &[A_i, B_j] = 0.
    \end{aligned}
\end{align}

\noindent By taking the complex $\textbf{sl}(2,\cc)_{\cc}$ algebra, we conclude that,

\begin{equation}
    \textbf{sl}(2,\cc)_{\cc} \simeq \textbf{su}(2)_{\cc} \oplus \textbf{su}(2)_{\cc}.
\end{equation}

Moreover, when taking the exponential of the generators the transformation matrices are:

\begin{align}
    \begin{aligned}\label{eq_gnfkjbdn6788}
R_{x}(\theta) &= \exp(\theta J_{23}) = \cos\left(\frac{\theta}{2}\right)\mathbb{1} - \sigma_{2}\sigma_{3}\sin\left(\frac{\theta}{2}\right) = \begin{pmatrix}
\cos\left(\frac{\theta}{2}\right) & -i \sin\left(\frac{\theta}{2}\right) \\
-i \sin\left(\frac{\theta}{2}\right) & \cos\left(\frac{\theta}{2}\right)
\end{pmatrix}, \\
R_{y}(\theta) &= \exp(\theta J_{31}) = \cos\left(\frac{\theta}{2}\right)\mathbb{1} - \sigma_{3}\sigma_{1}\sin\left(\frac{\theta}{2}\right) =  \begin{pmatrix}
\cos\left(\frac{\theta}{2}\right) & -\sin\left(\frac{\theta}{2}\right) \\
\sin\left(\frac{\theta}{2}\right) & \cos\left(\frac{\theta}{2}\right)
\end{pmatrix},\\
R_{z}(\theta) &= \exp(\theta J_{12}) = \cos\left(\frac{\theta}{2}\right)\mathbb{1} - \sigma_{1}\sigma_{2}\sin\left(\frac{\theta}{2}\right) =  \begin{pmatrix}
\text{exp}\left(\frac{-i\theta}{2}\right) & 0 \\
0 & \text{exp}\left(\frac{i\theta}{2}\right) 
\end{pmatrix},\\
    \end{aligned}
\end{align}

\noindent for the rotations and
\begin{align}
    \begin{aligned}
B_{x}(\eta) &= \exp(\theta K_{t1}) = \cosh\left(\frac{\eta}{2}\right)\mathbb{1} - \sigma_{1}\sinh\left(\frac{\eta}{2}\right) = \begin{pmatrix}
\cosh\left(\frac{\eta}{2}\right) & -\sinh\left(\frac{\eta}{2}\right) \\
-\sinh\left(\frac{\eta}{2}\right) & \cosh\left(\frac{\eta}{2}\right)
\end{pmatrix},\\
B_{y}(\eta) &= \exp(\theta K_{t2}) = \cosh\left(\frac{\eta}{2}\right)\mathbb{1} - \sigma_{2}\sinh\left(\frac{\eta}{2}\right) = \begin{pmatrix}
\cosh\left(\frac{\eta}{2}\right) & i \sinh\left(\frac{\eta}{2}\right) \\
-i \sinh\left(\frac{\eta}{2}\right) & \cosh\left(\frac{\eta}{2}\right)
\end{pmatrix},\\
B_{z}(\eta) &= \exp(\theta K_{t3})  = \cosh\left(\frac{\eta}{2}\right)\mathbb{1} - \sigma_{3}\sinh\left(\frac{\eta}{2}\right) = \begin{pmatrix}
\text{exp}\left(\frac{-\eta}{2}\right)  & 0 \\
0 & \text{exp}\left(\frac{\eta}{2}\right)
\end{pmatrix},
    \end{aligned}
\end{align}

\noindent stand for the boosts. Notice that a rotation $R \in \text{SL}(2,\cc) \simeq \Spin_{+}(1,3)$ acts on a spinor $\psi \to R\psi$ with
\begin{equation}\label{eq_4erdfjk}
    R(\theta)\psi = \cos\left(\frac{\theta}{2}\right)\mathbb{1} - \sigma \cdot n \sin\left(\frac{\theta}{2}\right)\psi
\end{equation}

\noindent Therefore, as seen before in the Example \ref{eq_example123456} a $2\pi$ rotation leads to $R = -\mathbb{1}$ and then $\psi \mapsto -\psi$ while a $4\pi$ rotation yield $\psi \mapsto \psi$. When considering two copies of $\textbf{su}(2)$ algebra, $\textbf{su}(2) \times \textbf{su}(2)$, the representations are labelled by pairs of half-integers $(j_1,j_2)$ (see Appendix \ref{app62} for more details). $\text{SL}(2,\cc)$ has two non equivalent $2\times2$ representations which are called the \textbf{left} and \textbf{right} representations, elements of the respective representation spaces are called left and right \textbf{Weyl spinors}. To understand why the space-time has \textbf{left} and \textbf{right} Spin$\frac{1}{2}$-representations, by considering the parity transformation $P$ in this case one has: $(\sigma_{1},\sigma_{2},\sigma_{3}) \to (-\sigma_{1},-\sigma_{2},-\sigma_{3})$. For instance, the result of the $xy$-rotation matrix $R_{z}(\theta)$ 

\begin{equation}
    R_{z}(\theta) = \cos\frac{\theta}{2} \mathbb{1} - \sin\frac{\theta}{2}\sigma_{x}\sigma_{y} =    
     \begin{pmatrix}
\text{exp}\left(\frac{-i\theta}{2}\right) & 0 \\
0 & \text{exp}\left(\frac{i\theta}{2}\right)  \end{pmatrix}
\end{equation}

\noindent under a parity transformation $P$ is still $R_{z}(\theta)$
\begin{equation}
    R_{z}(\theta)\longrightarrow_{P} \cos\frac{\theta}{2} \mathbb{1} - \sin\frac{\theta}{2}(-\sigma_{x})(-\sigma_{y}) =  \cos\frac{\theta}{2} \mathbb{1} - \sin\frac{\theta}{2}\sigma_{x}\sigma_{y} = R_{z}(\theta) \end{equation}

\noindent Which is indeed valid for any rotation generator $J_{ij} = -\frac{1}{2}\sigma_{i}\sigma{j}$. On the other hand, for the boosts the parity transformation does result in a sign change $K_{ti} \to_{P} -K_{ti}$ which results in a boost in a opposite direction.

\begin{equation}
    B_{i}(\eta) \longrightarrow_{P} -B_{i}(\eta)
\end{equation}

\noindent Therefore, the parity transformation give rise to two $\text{SL}(2,\cc)$ Spin$\frac{1}{2}$-representations with generators:

\begin{table}[H]\centering
\begin{tabular}{c|c}
\rowcolor{Yellow!17} 
\textbf{LEFT}-chiral Representation $\left(\frac{1}{2}, 0\right)$ & \textbf{RIGHT}-chiral Representation $\left(0, \frac{1}{2}\right)$ \\ \hline
$  J_{23} = -\frac{1}{2}\sigma_{2}\sigma_{3}$         & $  J_{23} = -\frac{1}{2}\sigma_{2}\sigma_{3}$           \\
$J_{31} = -\frac{1}{2}\sigma_{3}\sigma_{1}$         & $J_{31} = -\frac{1}{2}\sigma_{3}\sigma_{1}$           \\
$J_{12} = -\frac{1}{2}\sigma_{1}\sigma_{2}$          & $J_{12} = -\frac{1}{2}\sigma_{1}\sigma_{2}$            \\
$K_{t1} = -\frac{1}{2}\sigma_{1}$         & $K_{t1} = +\frac{1}{2}\sigma_{1}$            \\
$K_{t2} = -\frac{1}{2}\sigma_{2}$        & $K_{t2} = +\frac{1}{2}\sigma_{2}$            \\
$K_{t3} = -\frac{1}{2}\sigma_{3}$        & $K_{t3} = +\frac{1}{2}\sigma_{3}$           
\end{tabular}\caption{Left and Right $\text{SL}(2,\cc)$ Spin $\frac{1}{2}$-representations generators.}
\end{table}

\noindent and transformations:

\begin{table}[H]\centering
\begin{tabular}{
>{\columncolor{Yellow!17}}c |cc}
              & \multicolumn{1}{c|}{\cellcolor{Yellow!17}\textbf{LEFT}-chiral Representation $\left(\frac{1}{2}, 0\right)$ } & \cellcolor{Yellow!17}\textbf{RIGHT}-chiral Representation $\left(0, \frac{1}{2}\right)$  \\ \hline
$R_x(\theta)$ & $\begin{pmatrix}
\cos\left(\frac{\theta}{2}\right) & -i \sin\left(\frac{\theta}{2}\right) \\
-i \sin\left(\frac{\theta}{2}\right) & \cos\left(\frac{\theta}{2}\right)
\end{pmatrix}$\vspace{0.1cm}                                                       & $\begin{pmatrix}
\cos\left(\frac{\theta}{2}\right) & -i \sin\left(\frac{\theta}{2}\right) \\
-i \sin\left(\frac{\theta}{2}\right) & \cos\left(\frac{\theta}{2}\right)
\end{pmatrix}$                                  \\  
$R_y(\theta)$ & $ \begin{pmatrix}
\cos\left(\frac{\theta}{2}\right) & -\sin\left(\frac{\theta}{2}\right) \\
\sin\left(\frac{\theta}{2}\right) & \cos\left(\frac{\theta}{2}\right)
\end{pmatrix}$                                                       & $\begin{pmatrix}
\cos\left(\frac{\theta}{2}\right) & -\sin\left(\frac{\theta}{2}\right) \\
\sin\left(\frac{\theta}{2}\right) & \cos\left(\frac{\theta}{2}\right)
\end{pmatrix}$\vspace{0.1cm}                                     \\ 
$R_z(\theta)$ & $ \begin{pmatrix}
\text{exp}\left(\frac{-i\theta}{2}\right) & 0 \\
0 & \text{exp}\left(\frac{i\theta}{2}\right)  
\end{pmatrix}$                                                       & $\begin{pmatrix}
\text{exp}\left(\frac{-i\theta}{2}\right) & 0 \\
0 & \text{exp}\left(\frac{i\theta}{2}\right)  
\end{pmatrix}$ \vspace{0.1cm}                                    \\ 
$B_x(\eta)$   & $\begin{pmatrix}
\cosh\left(\frac{\eta}{2}\right) & -\sinh\left(\frac{\eta}{2}\right) \\
-\sinh\left(\frac{\eta}{2}\right) & \cosh\left(\frac{\eta}{2}\right)
\end{pmatrix}$                                                      & $\begin{pmatrix}
\cosh\left(\frac{\eta}{2}\right) & \sinh\left(\frac{\eta}{2}\right) \\
\sinh\left(\frac{\eta}{2}\right) & \cosh\left(\frac{\eta}{2}\right)
\end{pmatrix}$ \vspace{0.1cm}                                    \\ 
$B_y(\eta)$   & $\begin{pmatrix}
\cosh\left(\frac{\eta}{2}\right) & i \sinh\left(\frac{\eta}{2}\right) \\
-i \sinh\left(\frac{\eta}{2}\right) & \cosh\left(\frac{\eta}{2}\right)
\end{pmatrix}$                                                      & $\begin{pmatrix}
\cosh\left(\frac{\eta}{2}\right) & -i \sinh\left(\frac{\eta}{2}\right) \\
i \sinh\left(\frac{\eta}{2}\right) & \cosh\left(\frac{\eta}{2}\right)
\end{pmatrix}$  \vspace{0.1cm}                                   \\
$B_z(\eta)$   & $\begin{pmatrix}
\text{exp}\left(\frac{-\eta}{2}\right)  & 0 \\
0 & \text{exp}\left(\frac{\eta}{2}\right)
\end{pmatrix}$                                                       & $\begin{pmatrix}
\text{exp}\left(\frac{\eta}{2}\right)  & 0 \\
0 & \text{exp}\left(\frac{-\eta}{2}\right)
\end{pmatrix}$                                   
\end{tabular}\caption{Left and Right $\text{SL}(2,\cc)$ Spin $\frac{1}{2}$-representations.}\label{tab_11}
\end{table}

Recall that two non equivalent representations was expected from the previous discussion on complex spinor classification shown in the Table \ref{table_complexspiasdg}. The Table \ref{tab_11} shows that left-chiral ($\psi_{\text{Left}}$) and right-chiral ($\psi_{\text{Right}}$) representations of SL$(2,\mathbb{C})$ can be swapped by taking the Hermitian conjugate and the inverse of the transformation matrices. For a Lorentz transformation $L$ one has\cite{Zee:2003mt}:

\[
\color{Goldenrod} \begin{cases}
  \color{black}  \text{Left-chiral: } \psi_{\text{Left}} \rightarrow L \psi_{\text{Left}},\\\color{black}
\text{Right-chiral: } \psi_{\text{Right}} \rightarrow (L^\dagger)^{-1} \psi_{\text{Right}}.\color{black}
\end{cases}\color{black}
\]

\noindent For rotations, the SL$(2,\mathbb{C})$ matrices are unitary ($L^\dagger = L^{-1}$), meaning their Hermitian conjugate is their inverse. Therefore, left and right versions of the rotation matrices are equal.

\[\color{Goldenrod}\begin{cases}
   \color{black} \text{Left-chiral: } \psi_{\text{Left}} \rightarrow L \psi_{\text{Left}},\\ \color{black}
\text{Right-chiral: } \psi_{\text{Right}} \rightarrow L \psi_{\text{Right}}. \color{black}
\end{cases}\color{black}
\]

\noindent For boosts, the SL$(2,\mathbb{C})$ matrices are Hermitian, so the dagger operation does nothing. Therefore, the left and right versions of the boost matrices are inverses of each other.
\[\color{Goldenrod}\begin{cases}\color{black}
    \text{Left-chiral: } \psi_{\text{Left}} \rightarrow L \psi_{\text{Left}},\\
\color{black} \text{Right-chiral: } \psi_{\text{Right}} \rightarrow L^{-1} \psi_{\text{Right}}.\color{black}
\end{cases}\color{black}\]

This explains how the left-chiral and right-chiral representations of SL$(2,\mathbb{C})$ behave differently under rotations and boosts, and how their transformations can be related through Hermitian conjugation and inversion of the transformation matrices. These representations are not equivalent since they cannot be swapped by a change of basis $C$,  $(L^{\dagger})^{-1}_{\text{Right}} \neq C^{-1} L_{\text{Left}} C$. Since the left-chiral and right-chiral representations of SL$(2,\mathbb{C})$ follow different Lorentz transformation rules, they acts on Weyl spinors in column notation as:

\begin{equation}    \color{Goldenrod}\underbrace{\color{black} L\begin{bmatrix}
\psi^{1}\\ 
\psi^{2}
\end{bmatrix}}_{\color{black}\text{ Left-chiral}}  \color{black} \text{ and } \color{Goldenrod}\underbrace{\color{black} (L^{\dagger})^{-1}\begin{bmatrix}
\phi^{1}\\ 
\phi^{2}
\end{bmatrix}}_{\color{black}\text{ Right-chiral}}  \color{black}\end{equation}

\noindent By taking the direct sum of a left-chiral spinor and a right-chiral Weyl spinor one has the Dirac spinor (that was presented in the introduction in Eq. \eqref{eq_diracspinor}) and the $\left(\frac{1}{2},\frac{1}{2}\right) = \left(\frac{1}{2},0\right) \oplus \left(0, \frac{1}{2}\right) $ representation \cite{Zee:2003mt}

\begin{center}
{\colorlet{shadecolor}{Yellow!17}\begin{shaded} 
\begin{align}
    \begin{aligned}
         \color{Goldenrod}\underbrace{\color{black} \begin{bmatrix}
\psi^{1}\\ 
\psi^{2}
\end{bmatrix}}_{\color{black}\text{ Left-chiral Weyl spinor $\left(\frac{1}{2},0\right)$}}  \color{black} \oplus \color{Goldenrod}\underbrace{\color{black} \begin{bmatrix}
\phi^{1}\\ 
\phi^{2}
\end{bmatrix}}_{\color{black}\text{ Right-chiral Weyl spinor $\left(0, \frac{1}{2}\right)$}}  \color{black} =    \color{Goldenrod}\underbrace{\color{black} \begin{bmatrix}
\psi^{1}\\ 
\psi^{2} \\
\phi^{1}\\ 
\phi^{2} 
\end{bmatrix}.}_{\color{black}\text{Dirac spinor $\left(\frac{1}{2},\frac{1}{2}\right)$}}  \color{black}\\
    \end{aligned}
\end{align}
\end{shaded}}
\end{center}

\begin{center}
{\colorlet{shadecolor}{Yellow!17}\begin{shaded} 
\begin{align}
    \begin{aligned}
         \color{Goldenrod}\underbrace{\color{black} \begin{bmatrix}
L & \mathbb{O}\\ 
\mathbb{O} &  (L^{\dagger})^{-1}
\end{bmatrix}}_{\color{black}\text{ Representation $\left(\frac{1}{2},\frac{1}{2}\right)$}}  \color{black}   
\begin{bmatrix}
\psi^{1}\\ 
\psi^{2} \\
\phi^{1}\\ 
\phi^{2} 
\end{bmatrix}.
    \end{aligned}
\end{align}
\end{shaded}}
\end{center}

\noindent On the other hand the Dirac spinor $\psi$ can be splitted in two Weyl spinors $\psi_L$ and $\psi_R$ :

\begin{equation}
    \psi_L = P_L \psi = \frac{1}{2}(\mathbb{1} - \gamma_{5}) \psi, \;\;\;\psi_L = P_R \psi = \frac{1}{2}(\mathbb{1} + \gamma_{5}) \psi.
\end{equation}

\noindent where

\begin{equation}
    \gamma_{5} = \begin{bmatrix}
-\mathbb{1} & \mathbb{O}\\ 
\mathbb{O} & +\mathbb{1}
\end{bmatrix}.
\end{equation}

\noindent Finally, a vector $\left(\frac{1}{2}, \frac{1}{2} \right)$ can be realised as a \textit{spinor $\tensor$ spinor} \cite{Zee:2003mt}. The Pauli vector from the Pauli spinor (to be presented hereinafter in the Eq. \eqref{eq_gj5dfsv15}) is realised as:

\begin{align}
    \begin{aligned}
         \color{Goldenrod}\underbrace{\color{black} \begin{bmatrix}
\psi_{1}\\ 
\psi_{2}
\end{bmatrix} \otimes  \begin{bmatrix}
\psi_{1}^{*} & \psi_{2}^{*}
\end{bmatrix}}_{\color{black}\text{ Tensor product of Pauli Spinors}}  \color{black}
= \color{Goldenrod}\underbrace{\color{black} \begin{pmatrix}
    z & x - iy \\
    x + iy & -z
    \end{pmatrix}}_{\color{black}\text{ Pauli Vector }(\,\mathbf{v}\, \in\, \cl_{3})}  \color{black} 
    \end{aligned}
\end{align}

\noindent In this case,  $\mathbf{v} = x\sigma_{1} + y\sigma_{2} + z\sigma_{3}$ and $x = \frac{1}{2}(\psi_{1}\psi_{2}^{*} + \psi_{2}\psi_{1}^{*})$, $y = \frac{i}{2}(\psi_{1}\psi_{2}^{*} - \psi_{2}\psi_{1}^{*})$ and $z = \frac{1}{2}(||\psi_{1}||^2 - ||\psi_{2}||^2)$  as in the Eq. \eqref{eq_4ywvhuiodns}. The vector transforms with two $\textbf{su}(2)$ matrices under the $\left(\frac{1}{2},0\right) \tensor \left(0, \frac{1}{2}\right)$ representation and the spinor with one $\textbf{su}(2)$ matrix. For the space-time case, one has the realisation:

\begin{align}\label{eq_pauliy584thf}
    \begin{aligned}         \color{Goldenrod}\underbrace{\color{black} \begin{bmatrix}
\psi^{1}\\ 
\psi^{2}
\end{bmatrix} \otimes  \begin{bmatrix}
(\psi^{1})^{*} & (\psi^{2})^{*}
\end{bmatrix}}_{\color{black}\text{ Tensor product of Weyl Spinors}}  \color{black}
= \color{Goldenrod}\underbrace{\color{black} \begin{pmatrix}
    ct + z & x - iy \\
    x + iy & ct - z
    \end{pmatrix}}_{\color{black}\text{ Weyl Vector }(\,\mathbf{v}\, \in\, \cl_{1,3})}  \color{black} 
    \end{aligned}
\end{align}

\noindent Where $|\psi^{1}| =\sqrt{ct +z}, \;|\psi^{1}| =\sqrt{ct - z}$ and arg$(\psi^{1}/\psi^{2}) =$ arctan$(y/x)$. In summary, 

{\colorlet{shadecolor}{Yellow!17}\begin{shaded}
Representation of SL$(2, \cc)$

\begin{itemize}\centering    \item[] $(0, 0)$ : scalar representation: the trivial representation on $\cc$. 
\item[] $\left( \frac{1}{2} , 0\right)$ : left-handed Weyl spinor: the representation space is $\cc^{2}$.
\item[] $\left(0, \frac{1}{2} \right) $: right-handed Weyl spinor: the representation space is also $\cc^{2}$, but transforms independently with $\left( \frac{1}{2} , 0\right)$.
\item[] $\left(\frac{1}{2}  , \frac{1}{2} \right)$ : vector representation: transforms as
$\mathbf{v} \to L\mathbf{v}L^{\dagger}$ and induces a Lorentz transformation on $\rr^{1,3}$. $\; \textcolor{yellow}{{\blacktriangleleft}}$
\end{itemize}   \end{shaded}}
   
\end{ex}

\begin{ex}
{{\textcolor{yellow}{{$\blacktriangleright$\;}}}}    Majorana Representation. \normalfont In the Minkowski space-time $\rr^{1,3}$, Weyl and Dirac spinors come from complex representations and are elements of the complex representation space of two and four dimensions respectively. On the other hand, Majorana spinors come from real representations. Within the Minkowski space-time, consider the Clifford algebra $\cl_{1,3}$. From Theorem \ref{teo_casperiod1} it follows that

    \begin{equation}
        \cl_{1,3} \simeq \cl_{1,1} \tensor \cl_{0,2} \simeq \text{Mat}(2,\rr) \tensor \text{Mat}(2,\rr) \simeq \text{Mat}(4,\rr).
    \end{equation}

\noindent The real irreducible representation of $\cl_{1,3}$ on $\rr^{4}$ is called Majorana representation \cite{Konstantin, MajoranaJose, DeAndrade:1999xa}. Majorana spinors depends on the signature and does not exists in the $\cl_{3,1}$ space, since $\cl_{3,1} \simeq \text{Mat}(2,\hh)$ does not have a similar representation. $ \textcolor{yellow}{{\blacktriangleleft}}$
\end{ex}

\paragraph{ } A Lie algebra over $\rr$ leads to a complex Lie algebra by tensoring with $\cc$, this way, spin representations can be examined starting with a real representation of the Spin group and transforming it into a complex representation. As such, this representation extends to a complex representation of the special orthogonal group $\textbf{so}(n,\cc)$. Proceeding in reverse, complex representations $\textbf{spin}(n,\cc)$ can be constructed and then restricted to possible reductions to real representations.  Finite-dimensional irreducible representations of the complexified Lie algebra of the complexified double-covering  Lie group $\text{Spin}_+^{\mathbb{C}}(p, q)$ are characterised by their Lie algebra generators. They have the form:

\begin{equation}
    \varrho : \textbf{spin}_+^{\mathbb{C}}(p, q) \to \textbf{gl}_{\cc}(S)
\end{equation}

\noindent where $S = \cc^{r}$ for some $r>0$. The representations of the complexified Lie group $\text{Spin}_+^{\mathbb{C}}(p, q)$  is induced by the exponential map, namely

\begin{equation}
    \exp( \varrho) : \text{Spin}_+^{\mathbb{C}}(p, q) \to \text{GL}_{\cc}(S)
\end{equation}

\noindent are called \textbf{spin representations.} The associated complex representation spaces $S$ that the complexified Lie group acts on it are precisely the spinor spaces. In terms of complex algebra, we will have two types of spin representation depending on the parity of the dimension, as shown in the Table \ref{table_complexspiasdg}. When $n$ is odd, the spin representation is irreducible and when $n$ is even, there is a decomposition a direct sum of two inequivalent irreducible complex representations of the Spin group \cite{lund}, an example is the Left and Right chiral representation of the $\Spin_{+}(1,3)$ group drawn in the previous Example \ref{eq_example123456}.

Essentially, there are two types of representations of the Spin group, those that \textit{factor} in terms of $\SO(n)$ and those that do not factor.

\begin{definamarelo}    
A representation \normalfont \( \rho: G \to \text{GL}(V) \)\textit{ is said to \textbf{factor through} a representation of a subgroup} \( H \subseteq G \) \textit{if there exists a homomorphism} \(\phi: G \to H \) \textit{and a representation} \(\pi: H \to \text{GL}(V) \) \textit{such that:}  $\rho = \pi \circ \phi$. 

\begin{center}
\begin{tikzcd}
G \arrow[rd, "\phi"'] \arrow[rr, "\rho"] & & \text{GL}(V) \\
& H \arrow[ru, "\pi"'] & 
\end{tikzcd}
\end{center}
\end{definamarelo}

A representation $\lambda: \SO(n) \to \text{GL}(V)$ induces a representation of $\Spin(n)$ on $V$ by

\begin{equation}\label{eq_5ydrfhjknvm1}
    \lambda' := \lambda \circ \yhwidehat{\text{Ad}} : \Spin(n) \to \text{GL}(V)
\end{equation}

\noindent where $\yhwidehat{\text{Ad}}$ denotes the double covering Lie group homomorphism in Def. \ref{def_twofold2345}, the twisted adjoint representation. The representation $\lambda'$ of the form  $\lambda \circ \yhwidehat{\text{Ad}}$ must satisfy the following condition:

\begin{equation}
    \lambda'(-1) = \lambda( \yhwidehat{\text{Ad}}(-1)) = \lambda(1) = 1_{V}.
\end{equation}

 Therefore, if $\lambda' : \Spin(n) \to \text{GL}(V)$ is a representation of $\Spin(n)$ on $V$ such that  $ \lambda'(-1) = 1_{V}$, then $\ker \yhwidehat{\text{Ad}} = \{-1,+1\}\subset \ker \lambda'$. Since $\yhwidehat{\text{Ad}}$ is surjective, there is a map $\lambda : \SO(n) \to \text{GL}(V)$ such that $\lambda \circ \yhwidehat{\text{Ad}}$. In this case, with respect to the exact short sequence, the following diagram can be drawn:

\begin{center}

\tikzset{every picture/.style={line width=0.75pt}} 

\begin{tikzpicture}[x=0.75pt,y=0.75pt,yscale=-1,xscale=1]

\draw    (251.33,90) -- (251.33,144.13) ;
\draw [shift={(251.33,146.13)}, rotate = 270] [color={rgb, 255:red, 0; green, 0; blue, 0 }  ][line width=0.75]    (10.93,-3.29) .. controls (6.95,-1.4) and (3.31,-0.3) .. (0,0) .. controls (3.31,0.3) and (6.95,1.4) .. (10.93,3.29)   ;
\draw    (250.67,178.6) -- (250.67,227.8) ;
\draw [shift={(250.67,229.8)}, rotate = 270] [color={rgb, 255:red, 0; green, 0; blue, 0 }  ][line width=0.75]    (10.93,-3.29) .. controls (6.95,-1.4) and (3.31,-0.3) .. (0,0) .. controls (3.31,0.3) and (6.95,1.4) .. (10.93,3.29)   ;
\draw [color={Goldenrod}  ,draw opacity=1 ]   (286.4,165.33) -- (343.73,165.33) ;
\draw [shift={(345.73,165.33)}, rotate = 180] [color={Goldenrod}  ,draw opacity=1 ][line width=0.75]    (10.93,-3.29) .. controls (6.95,-1.4) and (3.31,-0.3) .. (0,0) .. controls (3.31,0.3) and (6.95,1.4) .. (10.93,3.29)   ;
\draw    (283.67,235.67) -- (347.46,182.88) ;
\draw [shift={(349,181.6)}, rotate = 140.39] [color={rgb, 255:red, 0; green, 0; blue, 0 }  ][line width=0.75]    (10.93,-3.29) .. controls (6.95,-1.4) and (3.31,-0.3) .. (0,0) .. controls (3.31,0.3) and (6.95,1.4) .. (10.93,3.29)   ;
\draw    (281,88.6) .. controls (254.14,79.64) and (298.55,107.32) .. (353.18,151.93) ;
\draw [shift={(354,152.6)}, rotate = 219.29] [color={rgb, 255:red, 0; green, 0; blue, 0 }  ][line width=0.75]    (10.93,-3.29) .. controls (6.95,-1.4) and (3.31,-0.3) .. (0,0) .. controls (3.31,0.3) and (6.95,1.4) .. (10.93,3.29)   ;

\draw (234.67,70.4) node [anchor=north west][inner sep=0.75pt]    {$\{\pm 1\}$};
\draw (231.33,234.4) node [anchor=north west][inner sep=0.75pt]    {$\SO(n)$};
\draw (224.33,155.73) node [anchor=north west][inner sep=0.75pt]    {$\Spin(n)$};
\draw (352.33,158.73) node [anchor=north west][inner sep=0.75pt]    {$\text{GL}(V)$};
\draw (214,189.4) node [anchor=north west][inner sep=0.75pt]    {$\yhwidehat{\text{Ad}}$};
\draw (314,218.4) node [anchor=north west][inner sep=0.75pt]    {$\lambda $};
\draw (300,142.4) node [anchor=north west][inner sep=0.75pt]    {$\lambda '$};

\end{tikzpicture}
\end{center}

However, some representations of $\Spin(n)$ are intrinsically of the spin group and they are not induced from $\SO(n)$. Within the context of spinors, this is the type of representation we will be interested in. For instance, given $\lambda' : \Spin(n) \to \text{GL}(\cl_{n,0})$ given by $\lambda'(a)(x) = ax$, the multiplication on $\cl_{n,0}$, is such that: 

\begin{equation}
    \lambda'(-1)(x) = -1x = -x
\end{equation}

\noindent which implies $  \lambda'(-1) = -1_{\cl_{n,0}}$. Thus, this representation cannot be induced by a representation of $\SO(n)$ on $\cl_{n,0}$, that is, this representation of the spin group does not factor through to a representation of the special orthogonal group.

{\colorlet{shadecolor}{Yellow!17}\begin{shaded} A \textbf{spin representation} is a representation of the Spin group that does not factor in terms of the special orthogonal group through the twisted adjoint representation. \end{shaded}}

This shows us the importance of Clifford algebras within the classical context of spinors, these algebras contains the groups Spin and Pin and so any representation of the algebra restricts to a representation of these groups which is non-trivial on the element $-1$ and therefore not induced from representations of $\O(n)$ and $\SO(n)$. It is fundamental to define spinors through a spin representation like that because spinors capture more information about the rotation structure than the vector representations of $\SO(n)$. For instance, for a vector $\mathbf{v}$ in the vector space, the transformation under $R_{\SO} \in \SO(n)$ is given by $\mathbf{v} \to R_{\SO}\mathbf{v}R_{\SO}^{-1}$. On the other hand, spinors transform directly by the action of $R_{\text{Spin}} \in \Spin(n)$  as $\psi \to R_{\text{Spin}}\psi$, such that the difference arises because spinors live in a representation space that is affected by left multiplication by the $\Spin(n)$ group. The other unique fact of the $\Spin$ group is that unlike vectors and tensors, it takes a rotation of $4\pi$ for a spinor to go back to its original state as seen in the Eq. \ref{eq_nyhjtrgfjuytr}. It is required to exclude then the possibility of the existence of any group homomorphisms $\lambda$ that makes the previous diagram commute since all these unique Spin group properties would not happen if considering representations that factor through the twisted adjoint representation $\yhwidehat{\text{Ad}} : \Spin_{+}(p,q) \to \SO_{+}(p,q)$.  One way to capture spin representations is to define them in terms of Clifford algebra, which, as seen, guarantees the condition $\lambda'(-1) \neq 1_{V} $. Therefore, spin representations are  representations of the $\Spin$ group that are found by restricting irreducible even Clifford algebra representations.

\section{Pauli and Dirac Spinors}

\paragraph{ } The objective of this section is to illustrate, through concrete examples, the concepts of classical and algebraic spinors that were defined in the previous sections and show how they appear through physical theories. We present the Pauli spinor, which is associated with the Clifford algebra of the Euclidean space $\clt$ , and the Dirac spinor, which is associated with the complex Clifford algebra of Minkowski spacetime $\cc \tensor \clm$. Many spinors are intrinsically linked to physical theories and several interesting developments have been reported in the literature with abundant applications in general relativity, gravitation and gauge field theory \cite{HoffdaSilva:2012uke, daRocha:2013qhu, daRocha:2011yr, Cavalcanti:2014uta, HoffdaSilva:2016gyy, Fabbri:2018dje, Fabbri:2017xlx, Ferrari:2016szq,Fabbri:2011mi, Bernardini:2007ex, Bernardini:2007ez, Rogerio:2023cwv}. In this section we approach the classical definition of the Pauli and Dirac spinors by introducing them from physical theories. We present the physical theory in which they are embedded and then we link it to the algebraic definition using minimal ideals of the Clifford algebra defined by specific idempotent elements.

By exploring these examples, we aim to provide a clearer understanding of the role and application of spinors within both physical and algebraic frameworks. This dual approach will not only illustrate their definitions but also demonstrate the practical significance of spinors in describing fundamental physical phenomena.

\subsection*{PAULI SPINOR}

\paragraph{ } The Pauli spinor is a spinor associated with the Clifford algebra $\clt$. It is thus an element of a irreducible representation space of the group $\Spin(3)$. We now introduce the classical definition of the Pauli spinor through the Pauli Theory following the approach in Ref. \cite{Lou01}. In classical mechanics, the total energy of a particle with mass $m$ is given by

\begin{equation}
    E = \frac{p^2}{2m} + W,
\end{equation}

\noindent such that $\'p = m\mathbf{v}$ is the momentum and $W= W(\'r)$ is potential energy. By inserting the differential operators for total energy and momentum, 
\begin{equation}
    E = i\hbar \frac{\partial}{\partial t}, \;\;\;\;\;\;\;\; \'p = -i\hbar\nabla.
\end{equation}

\noindent one has the Schrödinger equation which describes how a quantum state of a physical system (without magnetism and relativity) evolves with time for $\psi$ the wave function, with $\psi(\'r,t) \in \mathbb{C}$,

\begin{equation}
    i\hbar\frac{\partial\psi}{\partial t} = \frac{-\hbar^2}{2m} \nabla^2\psi + W\psi.
\end{equation}

Experiments conducted in the first half of the 20th century indicated that atoms responded to the presence of a magnetic field. Therefore, over the years, it was proposed that the electron possesses an intrinsic angular momentum, called \textit{spin} \cite{Lou01}. Considering that spin interacts with the magnetic field, in the presence of an electromagnetic field $E,B$ with potentials $V, A$, the Schrödinger equation becomes:

\begin{align}
\begin{aligned}\label{eq_espinordepauli1sch}
i\hbar\frac{\partial\psi}{\partial t} &=\frac{1}{2m}[(-i\hbar\nabla - e\'{A})^2]\psi - eV\psi \\
&= \frac{1}{2m} [-\hbar^2\nabla^2 + e^2\'{A}^2 + i\hbar e(\nabla \cdot \'{A} + \'{A} \cdot \nabla)]\psi - eV\psi.
\end{aligned} 
\end{align}

\noindent This equation still does not involve the electron's spin. Consider then the generalised momentum operator
\begin{align}
\begin{aligned}\label{eq_espinordepauli1sch2}
\boldsymbol{{\pi}} = \'p - e\'A, &&& \text{such that }  \'p = -i\hbar\nabla. 
\end{aligned} 
\end{align}

\noindent The components $\pi_{k} = p_{k} - eA_{k}$ of this operator satisfy

\begin{equation}
    \pi_1\pi_2 - \pi_2\pi_1 = \epsilon_{123}i\hbar e B_3,
\end{equation}

\noindent in which $(1,2,3)$ cyclically permutes, and this permutation is represented by the Levi-Civita symbol  $\epsilon_{123}$.
We recall the Pauli matrices $\sigma_{1}, \sigma_{2}, \sigma_{3}$, presented first in the Example \ref{ex1matripauli} and shown in the Example \ref{ex_reprec22} that $\mathcal{C}\ell_{3} \simeq  \text{Mat}(2,\mathbb{C})$. Those matrices satisfy the Clifford algebra fundamental relation: 

\begin{align}
\begin{aligned}\label{eq_espinordepauli1mapauli}
\sigma_{1}\sigma_{2} &= \epsilon_{123}i\sigma_{3},\\
\sigma_{i}\sigma_{j} &+  \sigma_{j}\sigma_{i} = 2\delta_{ij}I, \;\;\;\;\;\; (i,j \in \{1,2,3\}).
\end{aligned} 
\end{align}

\noindent Consider $\sigma \cdot \pi = \sigma_{1}\pi_{1} + \sigma_{2}\pi_{2} + \sigma_{3}\pi_{3}$. It follows that,

\begin{align}
\begin{aligned}\label{eq_espinordepaulisigmapi}
(\sigma \cdot \pi )^2 
&= (\sigma_{1}\pi_{1} + \sigma_{2}\pi_{2} + \sigma_{3}\pi_{3})\cdot(\sigma_{1}\pi_{1} + \sigma_{2}\pi_{2} + \sigma_{3}\pi_{3})\\
&= \sigma_1\pi_1\sigma_1\pi_1 + \sigma_1\pi_1\sigma_2\pi_2 + \sigma_1\pi_1\sigma_3\pi_3\\
&+ \sigma_2\pi_2\sigma_1\pi_1 + \sigma_2\pi_2\sigma_2\pi_2 + \sigma_2\pi_2\sigma_3\pi_3\\
&+\sigma_3\pi_3\sigma_1\pi_1 + \sigma_3\pi_3\sigma_2\pi_2 + \sigma_3\pi_3\sigma_3\pi_3\\
&= \sigma_1^2\pi_1^2 + \sigma_2^2\pi_2^2 + \sigma_3^2\pi_3^2 + \sigma_1\sigma_2\pi_1\pi_2 + \sigma_1\sigma_3\pi_1\pi_3 \\
&+ \sigma_2\sigma_1\pi_2\pi_1 + \sigma_2\sigma_3\pi_2\pi_3  + \sigma_3\sigma_1\pi_3\pi_1 + \sigma_3\sigma_2\pi_3\pi_2 \\
 &= {\pi}^2 + i\sigma_3(\pi_1\pi_2 -\pi_2\pi_1) + i\sigma_2(\pi_1\pi_3 -\pi_3\pi_1) + i\sigma_1(\pi_2\pi_3 -\pi_3\pi_2)\\
 &= {\pi}^2 - \hbar\sigma_{3}B_{3} - \hbar\sigma_{2}B_{2} - \hbar\sigma_{1}B_{1},
\end{aligned} 
\end{align}

\noindent That is, 
\begin{equation}\label{eq_espinpaulisigmab}
    (\sigma \cdot \pi )^2 = \pi^2 - \hbar e (\sigma \cdot \'B ),
\end{equation}

\noindent such that:

\begin{equation}
    \pi^2 = (\'p - e\'A)^2 = p^2 + e^2A^2 - e(\'p \cdot \'A + \'A \cdot \'p).
\end{equation}

\noindent Replacing $\pi^2 = (-i\hbar\nabla - e\'{A})^2$ for $(\sigma \cdot \pi )^2 = \pi^2 - \hbar e (\sigma \cdot \'B )$
in the Eq. \eqref{eq_espinordepauli1sch}, it follows that

\begin{align}
\begin{aligned}\label{eq_espinordepauli1schsubstituida}
i\hbar\frac{\partial\psi}{\partial t} &=\frac{1}{2m} [\pi^2 - \hbar e (\sigma \cdot \'B) ]\psi - eV\psi. \\
\end{aligned} 
\end{align}

\noindent The equation \eqref{eq_espinordepauli1schsubstituida} is called Schrödinger-Pauli Equation which describes the spin through the term:

\begin{equation}
\frac{\hbar e}{2m}(\sigma \cdot \'B).
\end{equation}

\noindent One also has that $(\sigma \cdot \'B) \in  \text{Mat}(2,\mathbb{C})$ with form:

\begin{equation}
   \sigma \cdot \'B = \sigma_1 B_1 + \sigma_2 B_2 + \sigma_3 B_3 =  \begin{pmatrix} 
    B_3 & B_1 - iB_2 \\
    B_1 - iB_2 & B_3
    \end{pmatrix}.
\end{equation}

\noindent Hence, the operator $(\sigma \cdot \'B)$ acts on a two-component column matrix, whose entries are in $\mathbb{C}$, which we denote as the \textit{Pauli spinor}. The wave function maps points in spacetime to Pauli spinors, meaning its values are given within the complex vector space $\mathbb{C}^2$, that is,
\begin{equation}
   \psi (\'r,t) = \begin{pmatrix} 
    \psi_{1}  \\
    \psi_{2}
    \end{pmatrix}, \;\;\;\;\;\;\;\; \psi_{1}, \psi_{2} \in \mathbb{C}. 
\end{equation}

A linear map $T$ from $\mathbb{C}^2$ to $\mathbb{C}^2$ is represented by $T \in  \text{Mat}(2,\mathbb{C})$, meaning such matrices represent endomorphisms of the vector space $\mathbb{C}^2$, which is the space of spinors in the case of the vector space $\mathbb{R}^3$. Thus, the \textbf{classical Pauli spinor} is defined as being an element of the form
\begin{equation}
    \psi =  \begin{pmatrix} 
    \alpha_{1} + i\beta_{1}  \\
    \alpha_{2} + i\beta_{2} 
    \end{pmatrix},
    \end{equation}
\noindent such that  $\alpha_{i},\beta_{i}  \in \rr$.  What differs the Pauli spinor from any arbitrary column vector is its transformation under the group $\text{SU}(2) \simeq \Spin(3)$. Consider a Pauli spinor $\psi$ given by
\begin{equation}
\psi = \begin{pmatrix}
\psi_{1} \\
\psi_{2}
\end{pmatrix},
\end{equation}

\noindent then the transformation $\psi \to \psi'$ is such that $\psi' = U\psi$ with $U \in $ $\text{SU}(2)$. For $\psi^{\dagger}$ given by
\begin{equation}
\psi^{\dagger} = \begin{pmatrix} \psi_{1}^{*} & \psi_{2}^{*}
\end{pmatrix},
\end{equation}

one has $\psi^{\dagger} \mapsto \psi^{\dagger'} = (U\psi)^{\dagger} = \psi^{\dagger}U^{\dagger}.$ So that the product $\psi^{\dagger}\psi$
\begin{equation}
\psi^{\dagger}\psi \to \psi^{\dagger'}\psi' = \psi^{\dagger}U^{\dagger}U\psi = \psi^{\dagger}I\psi = \psi^{\dagger}\psi,
\end{equation}

\noindent is invariant under $\text{SU}(2)$. On the other hand. The product $\psi \psi^{\dagger}$ transforms as a vector with respect to $\text{SU}(2)$:

\begin{equation}
    \psi \psi^{\dagger} \to U\psi \psi^{\dagger}U^{\dagger},
\end{equation}

This is expected according to the discussion in the Appendix \ref{app62} where a vector can be realised as \textit{'spinor $\tensor$ spinor'} and the vector transforms with two $su(2)$ matrices under the $(\frac{1}{2},0) \tensor (0, \frac{1}{2})$ representation and the spinor with one $su(2)$ matrix. In fact, comparing the $\psi \psi^{\dagger}$:

\begin{equation}
    \psi \psi^{\dagger} =   \begin{pmatrix} 
    \psi_{1}  \\
    \psi_{2}
    \end{pmatrix}\begin{pmatrix}     \psi_{1}^{*} &  \psi_{2}^{*}
    \end{pmatrix} = \begin{pmatrix}
    ||\psi_{1}||^2  &  \psi_{1}\psi_{2}^{*}\\
    \psi_{2}\psi_{1}^{*} & ||\psi_{2}||^2 
    \end{pmatrix} 
\end{equation}

\noindent with a Pauli vector $\mathbf{v} = (x,y,z) \in \rr^{3}$ (written in terms of $\matcomp$) 

\begin{align}
\begin{aligned}\label{eq_gj5dfsv15}
\psi \psi^{\dagger} =   \begin{pmatrix}
    ||\psi_{1}||^2  &  \psi_{1}\psi_{2}^{*}\\
    \psi_{2}\psi_{1}^{*} & ||\psi_{2}||^2 
    \end{pmatrix} \color{Goldenrod}\longleftrightarrow \color{black}\mathbf{v}  =  \begin{pmatrix}
    z  &  x - iy\\
    x + iy & -z
    \end{pmatrix},
\end{aligned} 
\end{align} 

\noindent the spinor components can be written in terms of the vector components 

\begin{align}
\begin{aligned} \label{eq_4ywvhuiodns}
\color{Goldenrod} \systeme*{ \color{black}
x = \color{black}\frac{1}{2}(\psi_{1}\psi_{2}^{*} + \color{black}\psi_{2}\psi_{1}^{*}), 
\color{black}y = \color{black}\frac{i}{2}(\psi_{1}\psi_{2}^{*} - \psi_{2}\psi_{1}^{*}),
\color{black}z = \color{black} \frac{1}{2}(||\psi_{1}||^2 - ||\psi_{2}||^2)}\color{black}.\\
\end{aligned} 
\end{align} 

\paragraph{ } Having introduced the classical definition of the Pauli spinor, we now proceed to present the algebraic Pauli spinor. In order to represent the Pauli spinor $\psi$ in terms of the Clifford algebra $\mathcal{C}\ell_{3}$, we must first represent the spinor in an isomorphic form, that is,

\begin{equation}
    \psi =  \begin{pmatrix} 
    \psi_{1} & 0 \\
    \psi_{2} & 0
    \end{pmatrix},
    \end{equation}

    \noindent where $\psi_{1},\psi_{2} \in \mathbb{C}$. This condition that only the first column is nonzero can be expressed by: 

    \begin{equation}\label{eq_espalg1mat}
    \psi \in \matcomp f_{0}, \;\;\;\; \text{such that} \;\;\;\;  f_{0} =  \begin{pmatrix} 
    1 & 0 \\
    0 & 0
    \end{pmatrix}.
\end{equation}

We recall that $\clt \simeq \matcomp$, one has the basis correspondence, $\mathbf{e}_{1} \leftrightarrow \sigma_{1}$, $\mathbf{e}_{2} \leftrightarrow \sigma_{2}$ and $\mathbf{e}_{3} \leftrightarrow \sigma_{3}$. Consider the idempotent $f_{0}$ given by
\begin{equation}
    f_{0} = \frac{1}{2}(1 + \mathbf{e}_{3}) \leftrightarrow \frac{1}{2}(I + \sigma_{3}) = \begin{pmatrix} 
    1 & 0 \\
    0 & 0
    \end{pmatrix} .
\end{equation}

\noindent The Eq. \eqref{eq_espalg1mat} represents that the product of any $M \in \matcomp$ with $f_{0}$ is:

\begin{equation}
    Mf_{0} = \begin{pmatrix} 
    \psi_{1} & \psi_{3} \\
    \psi_{2} & \psi_{4}
    \end{pmatrix} \begin{pmatrix} 
    1 & 0 \\
    0 & 0
    \end{pmatrix} = \begin{pmatrix} 
    \psi_{1} & 0 \\
    \psi_{2} & 0
    \end{pmatrix} \leftrightarrow \begin{pmatrix} 
    \psi_{1}  \\
    \psi_{2}
    \end{pmatrix} \in \cc^2.
\end{equation}

\noindent Hence, the set of elements $\psi \in \matcomp f_{0} $ forms a left ideal of the algebra $\matcomp$. Equivalently, in terms of the Clifford algebra $\mathcal{C}\ell_{3}$, it follows that $\mathcal{C}\ell_{3} f_{0}$ is a left ideal of $\mathcal{C}\ell_{3}$, meaning $A \psi \in S$ for all $A \in \mathcal{C}\ell_{3}$, for all $\psi \in S \subset \mathcal{C}\ell_{3}$. A \textbf{Pauli algebraic spinor} $\psi$ is defined as being an element of the left ideal $\mathcal{C}\ell_{3} f$. This left ideal contains no other left ideals besides itself and the zero ideal, that is, this type of left ideal is a \textit{minimal ideal} in $\mathcal{C}\ell_{3}$. As a vector space, $\mathcal{C}\ell_{3} f_{0}$ is generated by ${f_{0},f_{1},f_{2},f_{3}}$ such that \cite{Lou01}:

\begin{align}
\begin{aligned}\label{eq_espalgidealminibase}
f_{0} &= \frac{1}{2}(1 + \mathbf{e}_{3}) \leftrightarrow \begin{pmatrix} 
    1 & 0 \\
    0 & 0
    \end{pmatrix},
\\
f_{1} &= \frac{1}{2}(\mathbf{e}_{2}\mathbf{e}_{3} + \mathbf{e}_{2}) \leftrightarrow \begin{pmatrix} 
    0 & 0 \\
    i & 0
    \end{pmatrix},
\\
f_{2} &= \frac{1}{2}(\mathbf{e}_{3}\mathbf{e}_{1} - \mathbf{e}_{1}) \leftrightarrow \begin{pmatrix} 
    0 & 0 \\
    -1 & 0
    \end{pmatrix},
\\
f_{3} &= \frac{1}{2}(\mathbf{e}_{1}\mathbf{e}_{2} - \mathbf{e}_{1}\mathbf{e}_{2}\mathbf{e}_{3}) \leftrightarrow \begin{pmatrix} 
    i & 0 \\
    0 & 0
    \end{pmatrix}.
\end{aligned} 
\end{align}


\subsection*{DIRAC SPINOR}

\paragraph{ } The classical approach and definition of the Dirac spinor is already presented in the Introduction (Eq. \ref{eq_diracspinor}) as the solution of Dirac's equation. The Dirac equation is invariant under Lorentz transformations \cite{Dir28}, which are represented by the Lorentz group $\O(1,3)$, that describes orthogonal transformations of the Minkowski space-time $\rr^{1,3}$. The $\Spin_{+}(1,3)$ group is the double cover of the special orthogonal group $\SO_{+}(1,3)$ connected to the identity that describes rotations within the space-time. The classical Dirac spinor is then an element of a irreducible representation space of the group $\Spin_{+}(1,3)$. A spinorial representation of $\SO_{+}(1,3)$ can be constructed in terms $\text{SL}(2,\cc)$ and is called $\left(\frac{1}{2},0\right)$ representation.  The vector space upon which this representation acts is the set of two-component objects (complex $2 \times 1$ column vectors) called left-handed Weyl spinors. By taking
the complex conjugated matrices (or doing a parity transformation) this defines another inequivalent representation of $\SO_{+}(1,3)$
in terms of $\text{SL}(2,\cc)$ called $\left(0, \frac{1}{2}\right)$ and objects that are acted upon in this representation are now called right-handed Weyl spinors. These two representations are thus seen to be $2$-dimensional. By taking the direct sum
$\left(\frac{1}{2},0\right) \oplus \left(0, \frac{1}{2}\right)$ of the two representations,
a 4-dimensional (reducible) representation of the Lorentz group is obtained, which acts upon four-component objects called Dirac spinors; this discussion is detailed in the Appendix \ref{app62}. Therefore, Weyl spinors are classical spinors, that is, elements of a two-dimension irreducible representation space of the $\Spin_{+}(1,3)$ group and Dirac Spinor is an element of a four-dimension irreducible representation space of the $\Spin_{+}(1,3)$ group.

\paragraph{ } The Dirac algebraic spinor is an element of the minimal left ideal $ \text{Mat}(4,\cc)f$ such that

\begin{equation}
    f = \frac{1}{2}(1 + \gamma_{0})\frac{1}{2}(1 + i\gamma_1\gamma_2) = \begin{pmatrix} 
    1 & 0 & 0 & 0  \\
    0 & 0 & 0 & 0  \\
    0 & 0 & 0 & 0 \\
    0 & 0 & 0 & 0
    \end{pmatrix}.\end{equation}

\noindent One can express the algebraic Dirac spinor $\psi$ in terms of a basis $\{f_\alpha\} \subset$ $\matcompq f$:

\begin{equation}
    \psi = \psi_1f_1 + \psi_2f_2 + \psi_3f_3 + \psi_4f_4,
\end{equation}

\noindent such that\cite{Lou01}

\begin{align}
\begin{aligned}
f_1 &= \frac{1}{4}(1 + \gamma_0 + i \gamma_{12} + i\gamma_{012}), \\
f_2 &= \frac{1}{4}(-\gamma_{13} + i\gamma_{23} - \gamma_{013} + i\gamma_{023}), \\
f_3 &= \frac{1}{4}(\gamma_3 - \gamma_{03} + i\gamma_{123} - i\gamma_{0123}) ,\\
f_4 &= \frac{1}{4}(\gamma_1 - i\gamma_{2} - \gamma_{01} + i\gamma_{02}).
\end{aligned}
\end{align}

Therefore, the Dirac spinor $\psi$ might appear as a column spinor $\psi \in \cc^4$ or a square matrix spinor $\psi \in \matcompq f$.

\begin{equation}
  \psi =  \begin{pmatrix} 
    \psi_{1} \\
    \psi_{2} \\
    \psi_{3} \\
    \psi_{4}
    \end{pmatrix} \in \mathbb{C}^{4} \;\;\;\; \text{or} \;\;\;\;  \psi  =     \begin{pmatrix} 
    \psi_{1}& 0 & 0 & 0 \\
    \psi_{2}& 0 & 0 & 0 \\
    \psi_{3}& 0 & 0 & 0 \\
    \psi_{4}& 0 & 0 & 0
    \end{pmatrix}  \in \matcompq f.
\end{equation}

The advantage of expressing the Dirac spinor as the square matrix spinor is to be able to get everything - vectors, rotations, and spinors - represented within one mathematical system, namely, the Clifford algebra.

The Dirac spinor is crucial in describing various fermions with spin $1/2$, not just the electron. This includes other elementary particles such as quarks and massive neutrinos. In quantum field theory, the Dirac spinor formulates the Dirac field, essential for describing fermion particles and their interactions within the Standard Model of particle physics \cite{Zee:2003mt}.  Additionally, Dirac spinors can be decomposed into Weyl spinors as seen in the Appendix \ref{app62}, used for massless particles with defined helicity, and Majorana spinors, which solve the Dirac equation with a reality condition, having implications in neutrino theories and dark matter proposals \cite{HoffdaSilva:2012uke, daRocha:2011yr, Bernardini:2012sc, HoffdaSilva:2009is}. Dirac spinors exhibit intriguing physical and geometric properties, which are elucidated by the concept of \textcolor[rgb]{0.47,0.81,0.1}{bilinear covariants}, to be explored in the next Chapter \ref{chap_bil}. This concept not only encompasses Dirac spinors but also serves to unveil various types of spinors through the Lounesto's spinor classification based on the algebraic properties of spinors with physical meaning in the four-dimensional Minkowski space-time.

\colorlet{chapter}{green!50}
\chapter{Bilinear Covariants}\label{chap_bil}

\hypersetup{
  colorlinks = true,
  linkcolor  = ForestGreen,
  citecolor = LimeGreen,
}

\paragraph{} We have shown that the Dirac theory was developed within a new algebraic structure, transforming the concept of a \textit{wave function} into a specific column matrix called a \textit{spinor}. A natural question may arise: \textcolor{ForestGreen}{\textit{how physics could be conducted within this structure?}} The answer lies in the role of the \textit{bilinear covariants}. Bilinear covariants are quantities constructed from spinor fields and their adjoints that transform in a specific way under symmetry operations. These bilinear covariants can be understood within the mathematical framework of Clifford algebras and spin representations. In Minkowski space-time, they play crucial roles in describing physical phenomena, including probability currents, chirality, and electromagnetic interactions. The Clifford algebra provides tools to handle gamma matrices and their properties, forming the formalism used to describe and manipulate these covariants. In this four-dimensional space context, the bilinear covariants are 16 quantities with physical and geometrical nature that can be used not only to describe the physical features of fermionic particles but also to classify the spinors \cite{Lou01}. In this chapter, we introduce these quantities and Lounesto's spinor classification in this space. In Chapter \ref{chap_spinors}, we characterized spinors as classical or algebraic spinors, which were classified based on the periodicity of Clifford algebras. Here, as our main goal, we present another way to classify spinors which is based on their bilinear covariants. 

\paragraph{ }  Essentially, the bilinear covariants are nothing more than an \textit{inner product} defined in the space of spinors. Bilinear covariants can be seen as generalised inner products between spinor fields and their adjoints. They can be seen as generalised inner products between spinor fields and their adjoints. A classical spinor in a quadratic space, either $\rr^{p,q}$ or $\rr_{q,p}$, is an algebraic spinor (or an algebraic semispinor) in a quadratic space that is either $\rr^{q,p-1}$ or $\rr^{p,q-1}$. The \textbf{spinor inner product} is an inner product on the space $S$ of algebraic spinors with the property that the adjoint with respect to this inner product is the antiautomorphism in the corresponding Clifford algebra. Since there are two notions of antiautomorphisms, namely, reversion (Def. \ref{def16op}) and conjugation (Def. \ref{def17op}), two types of spinor inner products can be defined \cite{Roc16}:

 \begin{definverde}
 Let $S$ be the space of algebraic spinors. Given $x,y \in S$ and $a \in \clpq$ in such way that $\clpq \simeq$ \normalfont End$_\kk(S)$\textit{, the \textbf{spinor inner products} $\yhwidetilde{h} : S \cross S \to \kk$ and $\overline{h}: S \cross S \to \kk$ are defined as}
 
 \begin{align}
     \begin{aligned}
      \yhwidetilde{h}(ax,y) = \yhwidetilde{h}(x,\yhwidetilde{a}y),\\
      \overline{h}(ax,y) = \overline{h}(x,\overline{a}y).
     \end{aligned}
 \end{align}
 
\noindent \textit{Such that the adjoint corresponds to the reversion and the conjugation, respectively.}
 \end{definverde}
 
Let us denote by $\circ$ any one of those two antiautomorphisms.  That is, for $\psi \in \clpq$:

\begin{equation}
    \psi^{\circ} = \color{ForestGreen}\begin{cases}\color{black} \yhwidetilde{\psi}, \text{ if } \circ \text{ is the reversion,}\\\color{black}
      \overline{\psi}, \text{ if } \circ \text{ is the conjugation}.\color{black}
    \end{cases}\color{black}
\end{equation}

From the Appendix \ref{app7}, if $\clpq$ is a simple algebra, then $f\clpq f \simeq \kk$ whereas if $\clpq$ is a semisimple algebra,  $f\clpq f \simeq \kk \oplus \kk$, such that $\kk = \rr, \cc$ or $\hh$. In addition, regarding the fact that an algebraic spinor is an element of a minimal left ideal, we have that $S = \clpq f$. Therefore, for $\psi, \phi \in S$ we have $\psi f = \psi, \phi f = \phi$. Let us calculate the product $ \psi^{\circ} \phi $, since $\circ$ is an automorphism, it follows that:

\begin{equation}
    \psi^{\circ} \phi = (\psi f)^{\circ} \phi f = f^{\circ} \psi^{\circ} \phi f.
\end{equation}

\noindent Consequently, defining $h(\psi, \phi) =\psi^{\circ}\phi$ implies

\begin{equation}
    h(a\psi,\phi) = (a\psi)^{\circ} \phi = 
    \psi^{\circ}a^{\circ} \phi  = h(\psi,a^{\circ}\phi).
\end{equation}

\noindent Hence, $h$ as defined above corresponds to a spinor inner product if $f^{\circ} = f$. Although $(\clpq f)^{\circ} = f^{\circ} \clpq$ in general $f^{\circ} \neq f$. In this case, it is always possible to take an element $s \in \clpq^{*}$ such that $sf^{\circ}s^{-1} = f$ which allows us to define $s\psi^{\circ}\phi$ as a spinor scalar product \cite{Roc16}. This way, the mappings $\yhwidetilde{h}$ and $\overline{h}$, defined as

 \begin{align}
     \begin{aligned}
      \yhwidetilde{h}(\psi,\phi) =   s\yhwidetilde{\psi}\phi,\\
      \overline{h}(\psi,\phi) = r \overline{\psi} \phi,
     \end{aligned}
 \end{align}
 
\noindent with $s,r \in \clpq^{*}$ satisfying $s \yhwidetilde{f} f s^{-1} = f$ and  $r  \overline{f} f r^{-1} = f$ are spinor inner products. 

\paragraph{ } From the concept of the inner product in the space of spinors we now introduce the bilinear covariants that arise from them, in the Minkowski spacetime they can be categorised into several types and encode geometric and physical information about the spinor space.

\section{Bilinear Covariants in Minkowski Space-Time}

 \paragraph{ } We introduce bilinear covariants for the Minkowski space-time with some important consequences as the \textit{Fierz-Pauli-Kofink identities} and the \textit{Fierz aggregate} that brings a robust geometrical interpretation of such quantities. The \textit{Lounesto's spinors classification} is presented in the end as the main result of this Chapter. The main references of this section are Refs. \cite{Lou01, Roc16}.
 
 \paragraph{ }In terms of spin geometry, we consider a spin structure $(\Spin_{(1,3)}(\M),\pi)$ on a $4$-dimensional Lorentzian spacetime manifold $\M$ with respect to the $Spin(1,3)-$principal bundle $\pi_{Sp}: \Spin_{(1,3)}(\M) \to \M$ and an orthogonal frame bundle $\pi_{\SO}: \SO(\M) \to \M$ with a double cover $\pi:  \Spin_{(1,3)}(\M) \to \SO(\M)$ such that $\pi_{\SO} \circ \pi = \pi_{Sp}$. For the algebra $\cc \tensor \cl_{1,3}$ we consider the $\cc$-spinor bundle on $\M$ which in this case is the bundle $\Spin_{(1,3)}(\M) \cross_{\mu} \cc^{4}$ where $\mu = {\left(\frac12,0\right)}\oplus {\left(0,\frac12\right)}$ is the representation of the component of the Lorentz group connected to the identity that classical Dirac spinor fields carry. Therefore, spinor fields are elements of the sections of the $Spin_{(1,3)}(\M) \cross_{\mu} \cc^{4}$ bundle. We also recall the realisation of classical spinors in the Minkowski space-time as being an element of the representation space $\cc^4$, 

\begin{equation}
     \psi(x) = \begin{pmatrix} 
    \psi_{1}(x) \\
    \psi_{2}(x) \\
    \psi_{3}(x) \\
    \psi_{4}(x)
    \end{pmatrix} \in \mathbb{C}^{4},
\end{equation}

\noindent such that  $\psi_{i}: \rr^{1,3} \to \cc$ are scalar fields. In this case, the conjugate transpose $\psi^{\dagger}$ of $\psi$ is 

\begin{equation}
    \psi^{\dagger} = \begin{pmatrix} 
    \psi_{1}^{*} &
    \psi_{2}^{*} &
    \psi_{3}^{*} &
    \psi_{4}^{*}
    \end{pmatrix}.
\end{equation}

\noindent Now we consider the Dirac matrices $\{\gamma^{0}, \gamma^{1}, \gamma^{2}, \gamma^{3} \}$ from Eq. \eqref{eq_matdiracquatro}. The Dirac adjoint of a column spinor $\psi \in \cc^4$ is a row matrix

\begin{equation}
  \bar{\psi} = \psi^{\dagger}\gamma_{0} = \begin{pmatrix} 
    \psi_{1}^{*} &
    \psi_{2}^{*} &
    -\psi_{3}^{*} &
    -\psi_{4}^{*}
    \end{pmatrix} .
\end{equation}

{\colorlet{shadecolor}{green!15}\begin{shaded} The \textbf{bilinear covariants} are the following 16 quantities that determine the physical state of the electron \cite{Lou01}:

\begin{align}\label{eq_bilinearcovscoord}
\begin{aligned}
\sigma &= \bar{\psi}\psi, \\
J^{\mu} &=  \bar{\psi}\gamma^{\mu}\psi, \\
S^{\mu\nu} &= \frac{1}{2}i\bar{\psi}\gamma^{\mu\nu}\psi, \\
K^{\mu} &= i\bar{\psi}\gamma^{0123}\gamma^{\mu}\psi, \\
\omega &=  \bar{\psi}\gamma^{0123}\psi.
\end{aligned} 
\end{align}

\noindent such that $\mu,\nu = {0,1,2,3}$ and $\gamma^{0123} = \gamma^{0}\gamma^{1}\gamma^{2}\gamma^{3}$.\end{shaded}}

 \noindent The bilinear covariants can be realised as sections of the exterior bundle,

\begin{table}[H]\begin{center}
\scalebox{1.1}{\begin{tabular}{ccccc}
\rowcolor{green!15} 
$\Omega^{0}(\M)$ & $\Omega^{1}(\M)$ & $\Omega^{2}(\M)$ & $\Omega^{3}(\M)$ & $\Omega^{4}(\M)$ \\ \hline
$\sigma$ & $J_{\mu}\gamma^{\mu}$ & $S_{\mu\nu}\gamma^{\mu}\wedge\gamma^{\nu}$ & $K_{\mu}\gamma^{\nu}$ & $\omega$
\end{tabular}}\caption{Bilinear covariants as homogeneous sections of the exterior bundle.}
\end{center}
\end{table}

 \noindent That is, on its multivectorial structure the bilinear covariants  $\'J,\mathbf{S},\mathbf{K}$ are given by \cite{Lou01}:

\begin{align}
\begin{aligned}
\'J = J_{\mu}\'e^{\mu}, &&
\mathbf{S} = S_{\mu\nu}\'e_{\mu} \wedge \'e^{\nu}, &&
\mathbf{K} = K_{\mu}\'e^{\mu}.
\end{aligned} 
\end{align}

The expression of these quantities in terms of algebraic spinors can be found with details in the reference \cite{Lou01}. The bilinear covariants' physical interpretation in the Minkowski Space-Time is given by\cite{Roc16}:

 {\colorlet{shadecolor}{green!15}\begin{shaded}\begin{align}
\begin{aligned}
&eJ_{0}: &\textit{charge density}, \\
&ecJ_k: (k = 1,2,3)  &\textit{eletric current density}, \\
&\left ( \frac{e\hbar}{2mc} \right )S^{ij}: &\textit{magnetic moment density}, \\
&\left ( \frac{e\hbar}{2mc} \right )S^{0j}: &\textit{eletric moment density},\\
&\left ( \frac{\hbar}{2} \right )K_{\mu}: &\textit{spin density},  \\
&\rho^{2} = \sigma^{2} + \omega^{2}: ¨&\textit{probability density.}
\end{aligned} 
\end{align}\end{shaded}}

\paragraph{ } It is worth to mention the isomorphism $\mathbb{C} \otimes \clm \simeq  \text{Mat}(4,\mathbb{C})$ are not given just by the Dirac matrices. Another representation of  $\clm (\mathbb{C})$ in terms of $\matcompq$ is given also by the Weyl representation and it is also used to define the bilinear covariants. The Weyl representation is generated by the set

\begin{align}\label{eq_matweylquatro}
\begin{aligned}
&\gamma_{0} = \begin{pmatrix} 
    \;0 & \;0 & \;1 & \;0 \\
    \;0 & \;0 & \;0 & \;1\\
    \;1 & \;0 & \;0 & \;0\\
    \;0 & \;1 & \;0 & \;0
    \end{pmatrix}, &&&\gamma_{1} = \begin{pmatrix} 
    0 & 0 & 0 & -1 \\
    0 & 0 & -1 & 0\\
    0 & 1 & 0 & 0\\
    1 & 0 & 0 & 0
    \end{pmatrix}, \\
    &\gamma_{2} = \begin{pmatrix} 
    0 & 0 & 0 & i \\
    0 & 0 & -i & 0\\
    0 & -i & 0 & 0\\
    i & 0 & 0 & 0
    \end{pmatrix}, &&&\gamma_{3} = \begin{pmatrix} 
    0 & 0 & -1 & 0 \\
    0 & 0 & 0 & 1\\
    1 & 0 & 0 & 0\\
    0 & -1 & 0 & 0
    \end{pmatrix}.
\end{aligned}
\end{align}

\begin{tfpropos}\label{propos_weyltodirac}
The Dirac matrices are equivalent to the Weyl matrices by a similarity transformation
\end{tfpropos}
\noindent \textcolor{ForestGreen}{\textit{Proof.}} See the Proof \ref{propos_weyltodiracapp} in the Appendix \ref{app8}.\newline

\begin{tfpropos}\label{propos_asdfg65}
   The bilinear covariants are independent of the representation.
\end{tfpropos}  
\noindent \textcolor{ForestGreen}{\textit{Proof.}} See the Proof in \ref{propos_asdfg65app} in the Appendix \ref{app8}.

\begin{ex}\label{excalbi}
{\textcolor{ForestGreen}{{$\blacktriangleright$\;}}}
    Bilinear Covariants of the spinor:
    \begin{equation}
   \psi = \begin{pmatrix} 
    -i\beta^{*} \\
    i\alpha^{*} \\
    \alpha \\
    \beta 
    \end{pmatrix},
\end{equation}
\noindent with $\alpha,\beta : \rr^{1,3} \to \mathbb{C}$.
\end{ex}
\noindent For this spinor, we have that the bilinear covariants are  $\sigma = \omega =\mathbf{K} = 0$ and $\'J \neq 0,\;\mathbf{S} \neq 0.$ The complete computation of those bilinear covariants is presented in the Appendix \ref{app8}.
\; $\textcolor{ForestGreen}{{\blacktriangleleft}}$ 

\begin{ex} \label{ex_644t4e}
{\textcolor{ForestGreen}{{$\blacktriangleright$\;}}}
   Computing the bilinear Covariants for an arbitrary spinor. \normalfont This abstract and illustrative example aims to obtain the components of covariant bilinears in terms of any inputs of a spinor. Let us consider an abstract spinor $\psi$ and its conjugate $\psi^{\dagger}$ given by

\begin{align}
    \begin{aligned}
\psi = \begin{pmatrix} 
    a \\
    b \\
    c \\
    d 
    \end{pmatrix},  \;\;\;\; \psi^{\dagger} = \begin{pmatrix} 
    a^* & b^* & c^* & d^*
    \end{pmatrix}.
    \end{aligned}
\end{align}

\noindent For $\sigma = \bar\psi \psi$, one has

\begin{align}
    \begin{aligned}
\sigma = \bar\psi \psi = \psi^{\dagger} \gamma_0 \psi = c a^* + d b^* + a c^* + b d^*.
    \end{aligned}
\end{align}

\noindent For $J_{\mu} =  \bar{\psi}\gamma_{\mu}\psi$, one has

\begin{align}
    \begin{aligned}
    J_{0} &=  \bar{\psi}\gamma_{0}\psi = \psi^{\dagger}\gamma_{0}\gamma_{0}\psi = |a|^2 + |b|^2 + |c|^2 + |d|^2, \\
    J_{1} &=  \bar{\psi}\gamma_{0}\psi = \psi^{\dagger}\gamma_{0}\gamma_{1}\psi = b a^* + a b^* - d c^* - c d^*, \\
     J_{2} &=  \bar{\psi}\gamma_{0}\psi = \psi^{\dagger}\gamma_{0}\gamma_{2}\psi = i(- b a^* +  a b^* +  d c^* -  c d^*), \\
     J_{3} &=  \bar{\psi}\gamma_{0}\psi = \psi^{\dagger}\gamma_{0}\gamma_{3}\psi = |a|^2 - |b|^2 - |c|^2 + |d|^2.
    \end{aligned}
\end{align}

\noindent For $S_{\mu\nu} = \frac{1}{2}i\bar{\psi}\gamma_{\mu\nu}\psi$, one has

\begin{align}
    \begin{aligned}
S_{\mu\nu} &= \frac{1}{2}i\bar{\psi}\gamma_{\mu\nu}\psi = \frac{1}{2}i\psi^{\dagger}\gamma_0\gamma_{\mu\nu}\psi,\\
S_{00} &= \frac{1}{2}i\psi^{\dagger}\gamma_0\gamma_{00}\psi
=  \frac{1}{2}i (c a^* + d b^* + a c^* + b d^*), \\
S_{01} &= \frac{1}{2}i\psi^{\dagger}\gamma_0\gamma_{01}\psi = \frac{1}{2}i (-d a^* - c b^* + b c^* + a d^*),\\
S_{02} &= \frac{1}{2}i\psi^{\dagger}\gamma_0\gamma_{02}\psi = \frac{1}{2}i(i d a^* - i c b^* - i b c^* + i a d^*), \\
S_{03} &= \frac{1}{2}i\psi^{\dagger}\gamma_0\gamma_{03}\psi
= \frac{1}{2}i(-c a^* + d b^* + a c^* - b d^*),\\
    \end{aligned}
\end{align}
\begin{align}
    \begin{aligned}
S_{10} &= - S_{01}  \\
S_{11} &= \frac{1}{2}i\psi^{\dagger}\gamma_0\gamma_{11}\psi =  \frac{1}{2}i (-c a^* - d b^* - a c^* - b d^*), \\
S_{12} &= \frac{1}{2}i\psi^{\dagger}\gamma_0\gamma_{12}\psi =  \frac{1}{2} i(-i c a^* + i d b^* - i a c^* + i b d^*), \\
S_{13} &= \frac{1}{2}i\psi^{\dagger}\gamma_0\gamma_{13}\psi =  \frac{1}{2}i (d a^* - c b^* + b c^* - a d^*), \\
    \end{aligned}
\end{align}
\begin{align}
    \begin{aligned}
S_{20} &= - S_{02}, \\
S_{21} &= - S_{12}, \\
S_{22} &= \frac{1}{2}i\psi^{\dagger}\gamma_0\gamma_{22}\psi =   \frac{1}{2}i (-c a^* - d b^* - a c^* - b d^*), \\
S_{23} &= \frac{1}{2}i\psi^{\dagger}\gamma_0\gamma_{23}\psi =  \frac{1}{2} i(-i d a^* - i c b^* - i b c^* - i a d^*), \\
    \end{aligned}
\end{align}
\begin{align}
    \begin{aligned}
S_{30} &= - S_{03},\\
S_{31} &= - S_{13}, \\
S_{32} &= - S_{23}, \\
S_{33} &= \frac{1}{2}i\psi^{\dagger}\gamma_0\gamma_{33}\psi =  \frac{1}{2}i (-c a^* - d b^* - a c^* - b d^*).
    \end{aligned}
\end{align}

\noindent For $K_{\mu} = i\bar{\psi}\gamma_{0123}\gamma_{\mu}\psi$, one has
\begin{align}
    \begin{aligned}
    K_{0} &= i\bar{\psi}\gamma_{0123}\gamma_{0}\psi = i\psi^{\dagger}\gamma_0\gamma_{0123}\gamma_{0}\psi = -|a|^2 - |b|^2 + |c|^2 + |d|^2, \\
    K_{1} &= i\bar{\psi}\gamma_{0123}\gamma_{1}\psi = i\psi^{\dagger}\gamma_0\gamma_{0123}\gamma_{1}\psi = -b a^* - a b^* - d c^* - c d^*, \\
    K_{2} &= i\bar{\psi}\gamma_{0123}\gamma_{2}\psi =  i\psi^{\dagger}\gamma_0\gamma_{0123}\gamma_{2}\psi =   i( b a^* -  a b^* +  d c^* -  c d^*), \\
    K_{3} &= i\bar{\psi}\gamma_{0123}\gamma_{3}\psi = i\psi^{\dagger}\gamma_0\gamma_{0123}\gamma_{3}\psi =  -|a|^2 + |b|^2 - |c|^2 + |d|^2.
    \end{aligned}
\end{align}

\noindent Finally, for $\omega =  \bar{\psi}\gamma_{0123}\psi$, one has

\begin{equation}
    \omega =  \bar{\psi}\gamma_{0123}\psi = \psi^{\dagger}\gamma_0 \gamma_{0123} \psi = (i c a^* + i d b^* - i a c^* - i b d^*) = i( c a^* + d b^* -  a c^* -  b d^*).\;\textcolor{ForestGreen}{{\blacktriangleleft}} 
\end{equation}
\end{ex}

\section{Fierz-Pauli-Kofink Identities}
\paragraph{ } Based on the bilinear covariants $\sigma$ and $\omega$ one can define two types of spinors

\begin{definverde}
Let $\psi$ be a spinor. If  $\sigma = \omega = 0$, the spinor $\psi$ is called \textbf{singular spinor}. Otherwise, if $\sigma \neq 0$ or $\omega \neq 0$, $\psi$ is said to be a \textbf{regular spinor}.\end{definverde}

{\colorlet{shadecolor}{green!15}\begin{shaded} The \textbf{Fierz-Pauli-Kofink} (FPK) identities are quadratic algebraic relations between the bilinear covariants of a regular spinor \cite{Roc16}:

\begin{align}
    \begin{aligned}
   \mathbf{J}^2 &= \sigma^{2} + \omega^{2}, \\
   \mathbf{K}^{2} &= -\mathbf{J}^{2}, \\
   \mathbf{J} \cdot\mathbf{K} &= 0, \\
   \mathbf{J} \wedge\mathbf{K} &= -(\omega + \sigma\gamma_{0123} )\mathbf{S}.
    \end{aligned}
\end{align} \end{shaded}}

\noindent The computation of the first three identities can be seen in the Appendix \ref{fpkidapp8}. In coordinate form the FPK identities are as follows \cite{Lou01}:

\begin{align}
    \begin{aligned}
    J_{\mu}J^{\mu} &= \sigma^{2} + \omega^{2}, \\
    J_{\mu}J^{\mu} &= - K_{\mu}K^{\mu}, \\
    J_{\mu}K^{\mu} &= 0, \\
    J_{\mu}K_{\nu} - K_{\mu}J_{\nu},  
    &= -\omega S_{\mu\nu} + \sigma (\star S)_{\mu\nu}.
    \end{aligned}
\end{align}

\noindent such that $(\star S)_{\mu\nu} = - \frac{1}{2}\epsilon_{\mu\nu\alpha\beta}S^{\alpha\beta}$ (with $\epsilon_{0123} = 1$).\\

As a consequence of these identities, the bilinear covariants also satisfy \cite{Roc16}:

\begin{align}
    \begin{aligned}
       \mathbf{S} \lfloor\mathbf{J} &= \omega\mathbf{K}, \\
        \mathbf{S} \lfloor\mathbf{K} &= \omega\'J, \\
          (\gamma_{0123}\mathbf{S}) \lfloor\mathbf{K} &= \sigma\'J, \\
           \mathbf{S} \lfloor\mathbf{S} &= -\omega^2 + \sigma^2, \\
           \mathbf{S}\'J &= (\omega - \sigma \gamma_{0123})\mathbf{K},\\
           \mathbf{S}\mathbf{K} &= (\omega - \sigma \gamma_{0123)})\'J, \\
           \mathbf{S}^2 &= (\omega - \sigma \gamma_{0123)})^2 = \omega^2 - \sigma^2 - 2\omega \sigma \gamma_{0123},\\
           \mathbf{S}^{-1} &= -\mathbf{S} \frac{(\sigma - \omega \gamma_{0123)})}{(\omega^2 + \sigma^2)^2} = \frac{\mathbf{K}\mathbf{S}\mathbf{K}}{(\sigma^2 + \omega^2)^2}. 
    \end{aligned}
\end{align}

\paragraph{ }The FPK identities are a requirement for Lounesto's spinors classification which will be presented in the sequel. By the inversion theorem \cite{Takahashi:1982ib, Crawford:1985qg}, they are used to recover the original Dirac spinor from its bilinear covariants, up to a phase as shown in the Appendix \ref{recov_spinapp}. Considering a classical spinor $\zeta$ that satisfies $\zeta^{\dagger}\gamma_0\psi \neq 0$, the original spinor $\psi$ can be recovered from its aggregate $\mathbf{Z}$ provided by

\begin{equation}
    \mathbf{Z} = \sigma +\mathbf{J} + i\mathbf{S} + i\mathbf{K}\gamma_{0123} + \omega\gamma_{0123}.
\end{equation}

\noindent The spinor $\psi$ can be written as the multivector $\mathbf{Z}\zeta$ as

\begin{equation}
    \psi = \frac{1}{2\sqrt{\zeta^{\dagger}\gamma_{0}\mathbf{Z}\zeta}}e^{-i\theta}\mathbf{Z}\zeta,
\end{equation}

\noindent such that $e^{-i\theta} = 2(\zeta^{\dagger}\gamma_{0}\mathbf{Z}\zeta)^{1/2}\zeta^{\dagger}\gamma_0\psi \in U(1)$. The inversion theorem is extended in Ref. \cite{Rogerio:2023kcp}. Relevant ramifications have been reported in Refs. \cite{Fabbri:2020elt,HoffdaSilva:2022ixq,Rogerio:2020trs,Cavalcanti:2020obq}.

 {\colorlet{shadecolor}{green!15}\begin{shaded} When the bilinear covariants $\sigma,\,\mathbf{J} , \,\mathbf{S} , \,\mathbf{K} , \, \omega$ fulfil the FPK identities, the aggregate

\begin{equation}
\mathbf{Z} = \sigma +\mathbf{J} +\mathbf{S} + i\mathbf{K}\gamma_{0123} + \omega \gamma_{0123}
\end{equation}

\noindent is said to be a \textbf{Fierz aggregate}. Additionally, if $\gamma_{0}\mathbf{Z}^{\dagger}\gamma_0 = \mathbf{Z}$ then $\mathbf{Z}$ is said to be a \textbf{boomerang} \cite{Lou01}.  \end{shaded}}

The irreducible representation of the Lorentz group (Appendix \ref{app62}) can be understood in terms of these bilinear covariants.  The Fierz aggregate $\mathbf{Z}$ can be decomposed into homogeneous components that transform independently under the Lorentz group \cite{daRocha:2017vqh}.

\begin{equation*} {\displaystyle
\left[\left(0, \frac{1}{2}\right) \oplus \left(\frac{1}{2}, 0\right)\right] \otimes \left[\left(0, \frac{1}{2}\right) \oplus \left(\frac{1}{2}, 0\right)\right]  = 
 \color{ForestGreen}\underbrace{\color{black} (0,0)}_{\color{black} \sigma}  \color{black}
\oplus  \color{ForestGreen}\underbrace{\color{black} \left(\frac{1}{2},\frac{1}{2}\right)}_{\color{black} \mathbf{J}}  \color{black} \oplus 
    \color{ForestGreen}\underbrace{\color{black} (1, 0) \oplus (0, 1)}_{\color{black} \mathbf{S}}  \color{black} \oplus \color{ForestGreen}\underbrace{\color{black} \left(\frac{1}{2},\frac{1}{2}\right)}_{\color{black} \mathbf{K}}  \color{black} \oplus \color{ForestGreen}\underbrace{\color{black} (0,0)}_{\color{black} \omega}  \color{black}}
\end{equation*}

\noindent That is, the spinor representations are obtained from the fundamental
spinor representations $\left(\frac{1}{2},0 \right)  $ and $  \left(\frac{1}{2},0\right)$. The the set of boosts and rotations on the current density $\'J$ and the spin density $\'K$ are represented by $\left(\frac{1}{2},\frac{1}{2}\right)$.  The $(1, 0) \oplus (0, 1)$ representation is related to anti-symmetric irreducible
representations of the Lorentz group.

\paragraph{ } For singular spinors, the FPK identities are in general replaced by the following conditions\cite{Lou01,Roc16}:

\begin{align}
    \begin{aligned}
    \mathbf{Z}^2 &= 4 \sigma \mathbf{Z}, \\
    \mathbf{Z}\gamma_{\mu}\mathbf{Z} &= 4 J_{\mu} \mathbf{Z}, \\
    \mathbf{Z}i\gamma_{\mu \nu} \mathbf{Z} &= 4 S_{\mu\nu}\mathbf{Z},\\
    \mathbf{Z}i\gamma_{0123}\gamma_{\mu}\mathbf{Z} &= 4 K_{\mu} \mathbf{Z}, \\
    \mathbf{Z}\gamma_{0123}\mathbf{Z} &= - 4\omega \mathbf{Z}.
    \end{aligned}
\end{align}

\section{Lounesto's Spinor Classification}
\paragraph{ } In the \ref{chap_spinors}$^{\text{th}}$ chapter we presented the standard spinor classification that is  based on representation theory. In this chapter, we present a new classification for spinors \cite{Lou01}. Lounesto classified spinors in a new way by their bilinear covariants and since it is based on multivectors and physical observables, this new classification carries a huge geometrical and physical meaning. In addition, this classification reveals new spinors, called \textit{flag-dipole} spinors.\\

Dirac spinors are regular spinors that describe the electron. Weyl and Majorana spinors are singular spinors that describe the neutrino. For Weyl spinors, we have that the bilinear covariants satisfy $\mathbf{S} = 0, \,\mathbf{K} \neq 0$ and for Majorana spinors we have that $\mathbf{S} \neq 0, \,\mathbf{K} = 0$. Flag-dipole spinors are singular spinors such that $\mathbf{S} \neq 0, \,\mathbf{K} \neq 0$, therefore, they are not Dirac, Weyl, or Majorana spinors. Consequently, they cannot describe fermions, in particular, it has been conjectured that the flag-dipole spinors are related to the quark confinement \cite{Lou01}. \\

  {\colorlet{shadecolor}{green!15}\begin{shaded}Lounesto classified spinors fields into six disjoint classes:

\begin{align}
    \begin{aligned}
    &\textcolor{ForestGreen}{\textbf{1)}} \;\;\;\;\; \sigma \neq 0, \;\; &&\omega \neq 0, &&&\mathbf{S} \neq 0, &&&&\mathbf{K} \neq 0. \\
    &\textcolor{ForestGreen}{\textbf{2)}} \;\;\;\;\; \sigma \neq 0,\;\;  &&\omega = 0, &&&\mathbf{S} \neq 0, &&&&\mathbf{K} \neq 0. \\
    &\textcolor{ForestGreen}{\textbf{3)}} \;\;\;\;\; \sigma = 0, \;\; &&\omega \neq 0, &&&\mathbf{S} \neq 0, &&&&\mathbf{K} \neq 0. \\
    &\textcolor{ForestGreen}{\textbf{4)}} \;\;\;\;\; \sigma = 0, \;\; &&\omega = 0, &&&\mathbf{S} \neq 0, &&&&\mathbf{K} \neq 0. \\
    &\textcolor{ForestGreen}{\textbf{5)}} \;\;\;\;\; \sigma = 0,  \;\;&&\omega = 0, &&&\mathbf{S} \neq 0, &&&&\mathbf{K} = 0. \\
    &\textcolor{ForestGreen}{\textbf{6)}} \;\;\;\;\; \sigma = 0, \;\; &&\omega = 0, &&&\mathbf{S} = 0, &&&&\mathbf{K} \neq 0. \\
    \end{aligned}
\end{align}
 \end{shaded}}

\noindent The Lounesto's classifications reveals how different spinors can be \cite{HoffdaSilva:2017waf}.  The well-known Weyl, Dirac, and Majorana spinor fields occupy a place among the six disjoint classes of regular and singular spinors in Lounesto's classification \cite{Roc16}. In all the six cases, $\'J \neq 0$. More classes that are beyond Lounesto’s classification, that is, the case of non-trivial spinor fields corresponding $\'J = 0$,  playing the role of ghost spinor fields, was reported in Ref. \cite{CoronadoVillalobos:2015mns} and accounts for mass dimension one spinor fields \cite{HoffdaSilva:2019ykt, Rogerio:2022tsl}. According to the Lounesto spinor classification, spinors can be split into classes of charged and neutral ones under the under the $U(1)$ gauge symmetry, an extension to non-Abelian gauge symmetries is addressed in Ref. \cite{Fabbri:2017lvu}.  The Lounesto's classification considers the regular spinors \textcolor{ForestGreen}{\textbf{(1-3)}} and the singular spinors \textcolor{ForestGreen}{\textbf{(4-6)}}. The regular spinors of type \textcolor{ForestGreen}{\textbf{(1-3)}} are called Dirac spinors. The standard Dirac spinor, an eigenspinor of the parity operator as shown in the Appendix \ref{app62}, occupies a subset in the first class of regular spinors. Type-\textcolor{ForestGreen}{\textbf{(2)}}  spinors have been studied and even separated into subclasses in a  fermionic quantum field \cite{BuenoRogerio:2020bdm}. Spinors of type-\textcolor{ForestGreen}{\textbf{(4)}} are said to be flag-dipole, which in turn have been found to be solutions of the Dirac equation in an \( f(R) \)-background with torsion, in the context of Einstein-Sciama-Kibble gravity, marking the first explicit construction of flag-dipole spinors in physics \cite{daRocha:2013qhu, HoffdaSilva:2012uke}. 
 Spinors of type-\textcolor{ForestGreen}{\textbf{(5)}} are called flag-pole spinors. This fifth class, consisting of flagpole spinor fields, is particularly astonishing as it includes Majorana spinors, solutions of the Dirac equation in a Kerr black hole background \cite{daRocha:2016bil, deBrito:2016qzl}, and Elko (dark) spinor fields, that are eigenspinors of the charge conjugation operator with dual helicity and mass dimension one \cite{daRocha:2005ti, Ahluwalia:2023slc}. Hence, the fifth class contains both neutral and charged spinors, including fermions with mass dimension one. In this sense, mass dimension one fermions can occupy both classes four and five of spinor fields\cite{Cavalcanti:2015nna}. Moreover, in the example \ref{excalbi} we presented a type-\textcolor{ForestGreen}{\textbf{(5)}} spinor. Flagpoles, including mass dimension one Elko spinors, are candidates for solving the dark matter problem \cite{daRocha:2011yr, deGracia:2023yit, deGracia:2022enm}. Experimental signatures of type-\textcolor{ForestGreen}{\textbf{(5)}} flagpole dark spinors are proposed as byproducts of the Higgs bosonic field at LHC \cite{Dias:2010aa, Alves:2017joy}. Moreover, spinors of type-\textcolor{ForestGreen}{\textbf{(6)}} are called dipole,  the Weyl spinor is a particular dipole spinor in this sixth class.
There is a mapping that leads Dirac spinor fields to the ELKO spinor fields \cite{daRocha:2007pz} and type-\textcolor{ForestGreen}{\textbf{(4)}} spinor fields (corresponding to flag-dipoles) and type-\textcolor{ForestGreen}{\textbf{(5)}} ELKO spinor fields (corresponding to flagpoles) can be entirely described in terms of the Majorana and Weyl spinor fields corresponding to type-\textcolor{ForestGreen}{\textbf{(5)}} and -\textcolor{ForestGreen}{\textbf{(6)}} spinor fields \cite{daRocha:2008we}.

In addition, with respect to the geometrical meaning, since $\mathbf{S}$ is a bivector, which is geometrically an oriented plane fragment and $\'J,\mathbf{K}$ are both vectors, oriented lines segments, one may interpret $\mathbf{S}$ as a \textit{flag} and $\'J$ and $\mathbf{K}$ as the \textit{poles}, as we can see on the following Figure \ref{fig_flagpoles}:

\begin{figure}[H]
    \centering
    \includegraphics[scale=0.5]{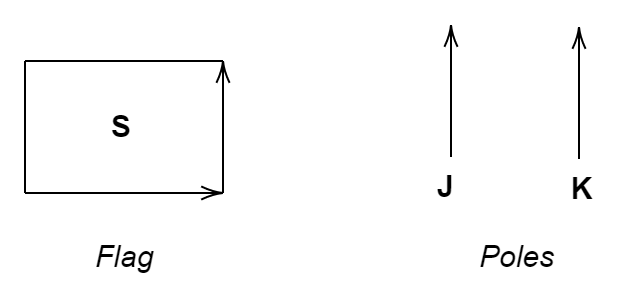}
    \caption{Interpretation of $\mathbf{S}, \, \mathbf{J}$ e $\mathbf{K}$ as flag and poles.}
    \label{fig_flagpoles}
\end{figure}

\noindent In this sense, the denomination of singular spinors of type-\textcolor{ForestGreen}{\textbf{(4)}}, type-\textcolor{ForestGreen}{\textbf{(5)}} and type-\textcolor{ForestGreen}{\textbf{(6)}} as flag-dipole, flag-pole e dipole, respectively, can be interpreted as in the Figure \ref{fig_esp456} above: 

\begin{figure}[H]
    \centering
    \includegraphics[scale=0.8]{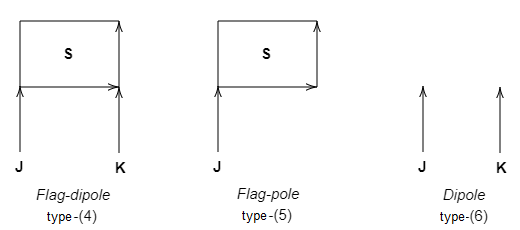}
    \caption{Representations of singular spinors as flag-dipole, flag-pole and dipole.}
    \label{fig_esp456}
\end{figure}

Other types of flagpole spinor fields and new types of pole and flag spinor fields are addressed in Ref. \cite{CoronadoVillalobos:2015mns}. Exotic dark spinor fields has been investigated in the context of non equivalent spin structures on arbitrary curved spacetimes in the literature \cite{DaRocha:2020oju, daRocha:2011xb} leading to many applications \cite{daRocha:2014dla, Cavalcanti:2015adn, Dantas:2015mfi}. In this context, the Lounesto's classification serves as a reference point for investigating extensions involving extended Clifford algebras, revealing an intriguing duality between traditional and extended classification schemes for exotic spinor fields \cite{daSilva:2023qnx, HoffDaSilva:2020uov, DaRocha:2020oju, HoffdaSilva:2019xvd} and also paving the way for other generalised spinor field classifications in other contexts \cite{Arcodia:2019flm, Cavalcanti:2014wia, Ablamowicz:2014rpa, HoffdaSilva:2022ixq}.  In general, the classification of spinor fields based on their bilinear covariants plays a crucial role for new fermions in quantum field theory, including mass dimension one quantum fields describing dark matter in a coherent manner \cite{deGracia:2024umr, Bernardini:2018uuy, Lee:2018ull,Rogerio:2022tsl,Ahluwalia:2022yvk,Ahluwalia:2022ttu,deGracia:2023yit,Fabbri:2010ws,Fabbri:2011mi,HoffdaSilva:2022mtq, fabbri, BuenoRogerio:2016seo, Cavalcanti:2014uta, Bernardini:2012sc, HoffdaSilva:2009is}. The approaches in Reference \cite{HoffdaSilva:2012uke} effectively derive the dynamics governing each subset of spinor fields within each category of the Lounesto classification. It is worth to emphasise that by investigating spinors fully defined by their bilinear covariants, one can examine the spinor representation space through its geometric, algebraic, and topological properties, thereby uniting the symmetries within the spinor space and the spinors as representatives of the classes in Lounesto's classification, this approach is shown in Reference \cite{HoffdaSilva:2017vic}.

\paragraph{ } Given that the bilinear covariants were defined in a very specific dimension, a natural question to ask is:
\textcolor{ForestGreen}{\textit{Is it possible to generalise and define the bilinear covariants for any dimension and signature?}} If so, \textcolor{ForestGreen}{\textit{could it also be possible to emulate a Lounesto-like classification of spinors for other dimensions and signatures?}}
As in the four-dimensional case, Lounesto's classification revealed new types of spinors, such as the type-\textcolor{ForestGreen}{\textbf{(5)}} flag-dipole. A similar classification in other dimensions and signatures could also reveal new classes of spinors in different spaces. The discovery of new spinors and their classification based on bilinear covariants have several important implications in theoretical physics. Not only allows for the study of intrinsic properties of spinors, such as charge, helicity, and symmetries, but also considers these newly found spinors as potential solutions in various applications that might have been previously overlooked. The discovery and classification of new spinors can provide solutions to open problems in theoretical physics. Thus, our objective becomes defining the bilinear covariants in any dimension and signature, as well as the Fierz identities that dictate the behaviour of these entities. To achieve this, we will adopt the same approach used in paper \cite{bab2}, constructing a geometric framework that bridges the algebraic and geometric concepts together in which such structures can be defined, namely, the \textcolor{blue}{\textit{Kähler-Atiyah bundle}}.

\part{\myred{G}\;\myorange{E}\;\myyellow{O}\;\mygreen{M}\;\mycyan{E}\;\myblue{T}\;\myviolet{R}\;\mymagenta{I}\;\myred{C}\;\;\;\myorange{A}\;\myyellow{P}\;\mygreen{P}\;\mycyan{R}\;\myblue{O}\;\myviolet{A}\;\mymagenta{C}\;\myred{H}\;\myorange{E}\;\myyellow{S}}
\colorlet{chapter}{Cerulean!50}
\chapter{Bundles}

\hypersetup{
  colorlinks = true,
  linkcolor  = Cerulean,
  citecolor = ProcessBlue,
}

\paragraph{ } We present now a foundational component for the dissertation. This chapter is expository and lays the foundations for the introduction of the \textcolor{blue}{\textit{Kähler-Atiyah bundle}} and the study of \textcolor[rgb]{0.74,0.06,0.88}{\textit{Spin Geometry}} to be presented in the sequel.

In this chapter, the tangent and the cotangent bundles are defined, which are essential for defining the Kähler-Atiyah bundle, as it is constructed from the exterior power of the cotangent bundle. Additionally, the concept of vector bundles are explored, given that the the tangent, the cotangent bundles and also the Kähler-Atiyah bundle are fundamentally vector bundles. This chapter further introduces differential forms, which can be defined within this framework and are significant for the thesis, since we aim to study the Clifford structures on the exterior algebra of differential forms with proper mathematical rigor, particularly focusing on the new geometric product acting on exterior differential forms that relies on the Kähler-Atiyah bundle.

Finally, in the last section the principal bundles are defined, a crucial concept when introducing spinors within a geometric approach. 
The definition of the principal bundles is necessary to understand the concept of a spin structure, which is essential for defining spinors in a geometric context in a coherent manner, particularly through the bundle of spinors. This comprehensive exploration of bundles in this Chapter lays the geometrical groundwork for the advanced topics that follow, ensuring a deep and rigorous understanding of the mathematical structures involved. The main references of this Chapter are Refs. \cite{LeeG, TuIntro}. Let us begin then with the concept of a manifold.

\begin{definazul}
    Suppose $\M$ is a topological space. $\M$ is said to be a \textbf{topological manifold of dimension $n$} if it has the following properties
\begin{itemize}
    \item[\textcolor{Cerulean}{\textbf{\textit{1.}}}] $\M$ is a \textbf{Hausdorff space}: for every pair of distinct points $x,y \in \M$ there are disjoint open subsets $U,V \subset \M$ such that $x \in U$ and $y \in V$.
    \item[\textcolor{Cerulean}{\textbf{\textit{2.}}}] $\M$ is \textbf{second-countable}: there exists a countable basis for the topology of $\M$.
    \item[\textcolor{Cerulean}{\textbf{\textit{3.}}}] $\M$ is \textbf{locally Euclidean of dimension $n$}: each point of $\M$ has a neighbourhood homeomorphic to an open subset of $\rr^{n}$. That is, each $x \in \M$ has a neighbourhood $U \subset \M$ such that there exists a homeomorphism $\varphi: U \to U' \subset \rr^{n}$.
\end{itemize}
\end{definazul}

\noindent The pair $(U, \varphi)$ is called a chart, the open set $U$ is called the coordinate neighbourhood and the mapping $\varphi$ is said to be coordinate map/system.

\begin{center}

\tikzset{every picture/.style={line width=0.75pt}} 

\begin{tikzpicture}[x=0.75pt,y=0.75pt,yscale=-1,xscale=1]

\draw   (145.3,48.04) .. controls (153.7,105.82) and (248.3,103.04) .. (199.3,136.04) .. controls (150.3,169.04) and (2.7,136.96) .. (12,100) .. controls (21.3,63.04) and (136.9,-9.74) .. (145.3,48.04) -- cycle ;
\draw  [line width=0.75] [line join = round][line cap = round] (92.3,60.04) .. controls (92.3,84.79) and (80.37,99.04) .. (55.3,99.04) ;
\draw  [line width=0.75] [line join = round][line cap = round] (90.9,65.96) .. controls (78.9,65.96) and (59.9,88.45) .. (59.9,98.96) ;
\draw  [fill={Cerulean!70}  ,fill opacity=1 ][dash pattern={on 4.5pt off 4.5pt}] (123.13,106.25) .. controls (116.08,91.34) and (119.66,79.26) .. (131.13,79.26) .. controls (142.6,79.26) and (157.62,91.34) .. (164.67,106.25) .. controls (171.72,121.16) and (168.14,133.25) .. (156.67,133.25) .. controls (145.2,133.25) and (130.18,121.16) .. (123.13,106.25) -- cycle ;
\draw  (263,127) -- (363,127)(273,37) -- (273,137) (356,122) -- (363,127) -- (356,132) (268,44) -- (273,37) -- (278,44)  ;
\draw  [fill={Cerulean!70}  ,fill opacity=1 ][dash pattern={on 4.5pt off 4.5pt}] (287.1,69.54) .. controls (300.95,60.88) and (313.04,67.51) .. (314.09,84.35) .. controls (315.14,101.19) and (304.77,121.86) .. (290.92,130.53) .. controls (277.06,139.19) and (264.98,132.56) .. (263.93,115.72) .. controls (262.87,98.88) and (273.25,78.2) .. (287.1,69.54) -- cycle ;
\draw  (263,127) -- (363,127)(273,37) -- (273,137) (356,122) -- (363,127) -- (356,132) (268,44) -- (273,37) -- (278,44)  ;
\draw    (168,92) .. controls (207.6,62.3) and (219.09,58.33) .. (261.03,94.16) ;
\draw [shift={(262.31,95.26)}, rotate = 220.71] [color={rgb, 255:red, 0; green, 0; blue, 0 }  ][line width=0.75]    (10.93,-3.29) .. controls (6.95,-1.4) and (3.31,-0.3) .. (0,0) .. controls (3.31,0.3) and (6.95,1.4) .. (10.93,3.29)   ;

\draw (209,45.4) node [anchor=north west][inner sep=0.75pt]    {$\varphi $};
\draw (123,59) node [anchor=north west][inner sep=0.75pt]   [align=left] {\textit{U}};
\draw (318,62) node [anchor=north west][inner sep=0.75pt]   [align=left] {\textit{U'}};

\end{tikzpicture}
\end{center}
A smooth structure on a topological manifold is fundamental in the study of topological manifolds since it allows us to define differentiable functions on the manifold, which in turn enables us to do calculus and differential geometry on it. Without a smooth structure, many important concepts and results in mathematics and physics would not be possible. A function $f: \M \to \rr$ is \textbf{smooth} if for each $\alpha \in \mathcal{I}, f \circ \varphi^{-1}_{\alpha}: \rr^{n} \to \rr$ is smooth as a function of $n$ real variables. Analogously, a function $f: \M \to \rr^{m}$ is smooth if each component $f_{i}: M \to \rr$, for $i = 1, \ldots, m,$ is smooth. The ring of smooth real valued functions over $\M$ shall be denoted by $\Cinf(\M)$ and functions of this set are said to be $\Cinf$.

\begin{definazul}
    Two charts $(U, \varphi), (V, \vartheta)$ of a topological manifold are said to be \textbf{compatible} if the two maps (called transition functions):

    \begin{align}
        \begin{aligned}
            \varphi \circ \vartheta^{-1} : \vartheta(U \cap V) \to \varphi (U\cap V),\\
            \vartheta \circ \varphi^{-1}: \vartheta(U \cap V) \to \varphi (U\cap V),
        \end{aligned}
    \end{align}

\noindent are $\Cinf$. 
\end{definazul}

\begin{center}

\tikzset{every picture/.style={line width=0.75pt}} 

\begin{tikzpicture}[x=0.75pt,y=0.75pt,yscale=-1,xscale=1]

\draw   (239.3,56.04) .. controls (247.7,113.82) and (342.3,111.04) .. (293.3,144.04) .. controls (244.3,177.04) and (96.7,144.96) .. (106,108) .. controls (115.3,71.04) and (230.9,-1.74) .. (239.3,56.04) -- cycle ;
\draw  [line width=0.75] [line join = round][line cap = round] (185.3,64.04) .. controls (185.3,88.79) and (173.37,103.04) .. (148.3,103.04) ;
\draw  [line width=0.75] [line join = round][line cap = round] (183.9,69.96) .. controls (171.9,69.96) and (152.9,92.45) .. (152.9,102.96) ;
\draw  [dash pattern={on 4.5pt off 4.5pt}] (216.13,110.25) .. controls (209.08,95.34) and (212.66,83.26) .. (224.13,83.26) .. controls (235.6,83.26) and (250.62,95.34) .. (257.67,110.25) .. controls (264.72,125.16) and (261.14,137.25) .. (249.67,137.25) .. controls (238.2,137.25) and (223.18,125.16) .. (216.13,110.25) -- cycle ;
\draw  (74,259) -- (174,259)(84,169) -- (84,269) (167,254) -- (174,259) -- (167,264) (79,176) -- (84,169) -- (89,176)  ;
\draw  [dash pattern={on 4.5pt off 4.5pt}] (98.1,201.54) .. controls (111.95,192.88) and (124.04,199.51) .. (125.09,216.35) .. controls (126.14,233.19) and (115.77,253.86) .. (101.92,262.53) .. controls (88.06,271.19) and (75.98,264.56) .. (74.93,247.72) .. controls (73.87,230.88) and (84.25,210.2) .. (98.1,201.54) -- cycle ;
\draw [draw opacity=0] (74,260) -- (174,260)(84,170) -- (84,270) (167,255) -- (174,260) -- (167,265) (79,177) -- (84,170) -- (89,177)  ;
\draw    (197.31,129.2) .. controls (130.67,133.12) and (113.01,155.29) .. (112.33,186.29) ;
\draw [shift={(112.31,188.2)}, rotate = 270] [color={rgb, 255:red, 0; green, 0; blue, 0 }  ][line width=0.75]    (10.93,-3.29) .. controls (6.95,-1.4) and (3.31,-0.3) .. (0,0) .. controls (3.31,0.3) and (6.95,1.4) .. (10.93,3.29)   ;
\draw  [dash pattern={on 4.5pt off 4.5pt}] (264.07,212.22) .. controls (281.03,206.55) and (292.64,214.13) .. (290,229.15) .. controls (287.36,244.16) and (271.46,260.92) .. (254.5,266.58) .. controls (237.53,272.25) and (225.92,264.67) .. (228.56,249.66) .. controls (231.21,234.64) and (247.1,217.88) .. (264.07,212.22) -- cycle ;
\draw  [fill={Cerulean!70}  ,fill opacity=1 ][dash pattern={on 4.5pt off 4.5pt}] (240.82,224.9) .. controls (248.94,217) and (257.14,216.35) .. (259.15,223.45) .. controls (261.16,230.55) and (256.2,242.7) .. (248.09,250.6) .. controls (239.98,258.5) and (231.77,259.15) .. (229.76,252.06) .. controls (227.76,244.96) and (232.71,232.8) .. (240.82,224.9) -- cycle ;
\draw  (227.45,261.75) -- (327.45,261.75)(237.45,171.75) -- (237.45,271.75) (320.45,256.75) -- (327.45,261.75) -- (320.45,266.75) (232.45,178.75) -- (237.45,171.75) -- (242.45,178.75)  ;
\draw  [fill={Cerulean!70}  ,fill opacity=1 ][dash pattern={on 4.5pt off 4.5pt}] (111.54,200.29) .. controls (118.21,196.12) and (124.27,203.31) .. (125.09,216.35) .. controls (125.91,229.39) and (121.16,243.34) .. (114.49,247.51) .. controls (107.82,251.68) and (101.75,244.49) .. (100.93,231.45) .. controls (100.12,218.42) and (104.87,204.46) .. (111.54,200.29) -- cycle ;
\draw  [dash pattern={on 4.5pt off 4.5pt}] (196.13,115.25) .. controls (189.08,100.34) and (192.66,88.26) .. (204.13,88.26) .. controls (215.6,88.26) and (230.62,100.34) .. (237.67,115.25) .. controls (244.72,130.16) and (241.14,142.25) .. (229.67,142.25) .. controls (218.2,142.25) and (203.18,130.16) .. (196.13,115.25) -- cycle ;
\draw  [fill={Cerulean!70}  ,fill opacity=1 ][dash pattern={on 4.5pt off 4.5pt}] (215.98,113.27) .. controls (209.71,100.02) and (209.64,89.28) .. (215.81,89.28) .. controls (221.98,89.28) and (232.06,100.02) .. (238.33,113.27) .. controls (244.59,126.51) and (244.67,137.25) .. (238.49,137.25) .. controls (232.32,137.25) and (222.24,126.51) .. (215.98,113.27) -- cycle ;
\draw    (266.31,136.2) .. controls (276.26,155.11) and (345.62,166.09) .. (293.11,209.54) ;
\draw [shift={(292.31,210.2)}, rotate = 320.83] [color={rgb, 255:red, 0; green, 0; blue, 0 }  ][line width=0.75]    (10.93,-3.29) .. controls (6.95,-1.4) and (3.31,-0.3) .. (0,0) .. controls (3.31,0.3) and (6.95,1.4) .. (10.93,3.29)   ;
\draw    (135.31,223.2) .. controls (164.86,206.46) and (189.56,214.94) .. (221.83,230.48) ;
\draw [shift={(223.31,231.2)}, rotate = 205.87] [color={rgb, 255:red, 0; green, 0; blue, 0 }  ][line width=0.75]    (10.93,-3.29) .. controls (6.95,-1.4) and (3.31,-0.3) .. (0,0) .. controls (3.31,0.3) and (6.95,1.4) .. (10.93,3.29)   ;

\draw (99,139.4) node [anchor=north west][inner sep=0.75pt]    {$\varphi $};
\draw (175,109) node [anchor=north west][inner sep=0.75pt]   [align=left] {\textit{U}};
\draw (321,165.34) node [anchor=north west][inner sep=0.75pt]    {$\vartheta $};
\draw (264,103.94) node [anchor=north west][inner sep=0.75pt]   [align=left] {\textit{V}};
\draw (292,232.55) node [anchor=north west][inner sep=0.75pt]    {$\vartheta $};
\draw (302,231.94) node [anchor=north west][inner sep=0.75pt]  [font=\small] [align=left] {\textit{(V)}};
\draw (33,203.4) node [anchor=north west][inner sep=0.75pt]    {$\varphi $};
\draw (45,205.94) node [anchor=north west][inner sep=0.75pt]   [align=left] {\textit{(U)}};
\draw (150.4,190.72) node [anchor=north west][inner sep=0.75pt]  [font=\small]  {$\vartheta \ \circ \ \varphi ^{-1}$};

\end{tikzpicture}
    
\end{center}

\begin{definazul}
    An \textbf{atlas} on a locally Euclidean space $\M$ is a collection $\mathcal{U} = \{(U_{\alpha},\varphi_{\alpha})\}$ of pairwise compatible charts that cover $\M$, that is $\M = \bigcup_{\alpha} U_{\alpha}$. 
\end{definazul}

\noindent  An atlas $\mathcal{U}$ on $\M$ is called a \textbf{smooth atlas} if any two charts of it are smoothly compatible with each other. A smooth atlas is said to be \textbf{maximal} if it is not properly contained in any larger smooth atlas. A \textbf{smooth structure on $\M$} is a maximal smooth atlas. Finally, a \textbf{smooth manifold} is a pair $(\M, \mathcal{U})$, where $\M$ is a topological manifold and $\mathcal{U}$ is a smooth structure on $\M$.

 The terms smooth and $\Cinf$ are equivalent and are used interchangeably, however, from now on, a manifold, will mean a smooth manifold. In this context, the standard coordinates of $\rr^{n}$ are denoted by $r^{1}, \ldots, r^{n}$. Considering a chart $(U,\varphi : U \to \rr^{n})$ of a manifold $\M$, one sets $t^{i} = r^{i} \circ \varphi$ to be the $i$th component of $\varphi$ which in turn can be written as $\varphi = (t^{1}, \ldots, t^{n})$, hence, $(U,\varphi) = (U,t^{1}, \ldots, t^{n})$. When considering $x \in U$, one has $(t^1(x), \ldots, t^{n}(x)) \in \rr^{n}$. The functions $t^{1}, \ldots, t^{n}$ are said to be coordinates or local coordinates on $U$ \cite{TuIntro}.
Having established the main concepts about smooth manifolds, we are now in a position to advance to the study of fibre bundles starting with two important bundles: the tangent and the cotangent bundle.

\section{Tangent and Cotangent Bundle}

\paragraph{ } To begin discussing the tangent and cotangent bundles, it is essential to have a well-established understanding of the tangent space. The tangent space at a point is the vector space of all tangent vectors at that point. Geometrically speaking it represents all the possible directions that a curve can pass through that point on a manifold. The tangent space is essential to understand the local behaviour of a manifold, it allow us to define derivations, study curves and surfaces, make calculations and define the Clifford bundle.

The tangent vectors and the tangent space can be defined through many approaches \cite{LeeG}. The one we present here is based on the notion of germs of smooth functions. The \textit{germ} of a $\Cinf$ function at a point $x \in \M$ is defined as being an equivalence class of $\Cinf$ functions defined in a neighbourhood of $x$. That is, given a point $x \in \M$ and a smooth function element $(f,U)$ where $U \subset \M$ is an open subset and $f: U \to \rr$ is a smooth function, an equivalence relation on the set of all $\Cinf$ function elements whose domains contain $x$ can be defined by setting $(f,U) \sim (g,V)$ if $f \equiv g$ on some neighbourhood of $x$. The set of germs of $\Cinf$ real-valued functions at $x \in \M$ is denoted by
 $\Cinf_{x}(\M)$, which is an algebra over $\rr$ when endowed with the addition and multiplication of functions and the scalar multiplication by real numbers. 
 
 Since germs refer to the infinitesimal neighbourhood around a point on a manifold the main advantage of the germ approach is that the local nature of the tangent space is naturally embedded on the definition.

\begin{definazul}
    The \textbf{derivation at a point} in a manifold $\M$ is a linear map $D: \Cinf_{x}(\M) \to \rr$ such that

    \begin{equation}
        D(fg) = (Df)g(x) + f(x)Dg.
    \end{equation}
\end{definazul}

\begin{definazul}
    A \textbf{tangent vector at a point $x$} in a manifold $\M$ is a derivation at $x$.
\end{definazul}
  {\colorlet{shadecolor}{Cyan!15}\begin{shaded}The tangent vectors at $x$ form a vector space $T_{x}M$, called the \textbf{tangent space} of $\M$ at $x$.\end{shaded}}

\begin{rem}
    (Tangent space to an open subset) \cite{LeeG} \normalfont. The algebra $\Cinf_{x}(U)$, where $U \subset \M$ is an open neighbourhood of $x$, is the same as $\Cinf_{x}(\M)$. Therefore, $T_{x}U = T_{x}\M$.
\end{rem}

Given a chart $(U,\varphi) = (U, t^{1}, \ldots, t^{n})$ for a point $x \in \M$ and $r^{1},\ldots,r^{n}$ the standard coordinates on $\rr^{n}$ one has $t^{i} = r^{i} \circ \varphi : U \to \rr$. In this sense, for a smooth function $f$ in a neighbourhood of $x$, a vector (derivation) is defined by

\begin{align}
    \begin{aligned}
        \frac{\partial}{\partial t^{i}}\Bigg|_{x} f  =  \frac{\partial}{\partial r^{i}}\Bigg|_{\varphi(x)} (f \circ \varphi^{-1})  \in \rr
    \end{aligned}
\end{align}

\noindent which satisfies the derivation property and so is a tangent vector at $x$. It is customary to write $\frac{\partial}{\partial t^{i}}$ as $\partial^{i}$.

\begin{tfpropos}
    If $\dim \M = n$, then for each $x \in \M$, the tangent space $T_x\M$ is an $n$-dimensional vector space and for any smooth chart $(U,t^{1}, \ldots, t^{n})$ containing $x$, the coordinate vectors $\{\partial^{1}, \ldots, \partial^{n}\}$ at $x$ form a basis for $T_x\M$, called a coordinate basis. \cite{LeeG}
\end{tfpropos}



\begin{definazul}
    Given a smooth manifold $\M$, the \textbf{tangent bundle}, denoted by $T\M$, is defined as being the disjoint union of the tangent spaces of all points of $\M$.
    \begin{equation}
        T\M = \bigsqcup_{x \in \M} T_x\M =\bigsqcup_{x \in \M} (\{x\} \cross T_x\M) .
    \end{equation}
\end{definazul}
 
Usually an element of this disjoint union is written as a pair $(x,v)$  with $x \in \M$ and $v \in T_x\M$. There is a natural projection $\pi : T\M \to \M$ sending each vector in $T_x\M$ to the point $x$ at which it is tangent: $\pi(x,v) = x$.

\begin{center}
\tikzset{every picture/.style={line width=0.75pt}} 

\begin{tikzpicture}[x=0.75pt,y=0.75pt,yscale=-1,xscale=1]

\draw    (62.5,99.04) .. controls (72.13,111.8) and (121.83,103.27) .. (120.98,99.04) .. controls (120.13,94.8) and (164.13,85.8) .. (179.47,99.04) ;
\draw    (170.4,178.8) .. controls (179.07,180.13) and (179.73,198.8) .. (169.73,198.8) ;
\draw    (67.4,179.47) .. controls (57.4,180.13) and (58.07,199.47) .. (66.73,199.47) ;
\draw    (170.52,67.82) -- (170.52,121) ;
\draw  [fill={Cerulean!70}  ,fill opacity=0.5 ][dash pattern={on 4.5pt off 4.5pt}] (62.5,72.3) .. controls (62.5,75.99) and (75.59,78.98) .. (91.74,78.98) .. controls (107.89,78.98) and (120.98,75.99) .. (120.98,72.3) .. controls (120.98,68.61) and (134.08,65.62) .. (150.22,65.62) .. controls (166.37,65.62) and (179.47,68.61) .. (179.47,72.3) -- (179.47,125.77) .. controls (179.47,122.08) and (166.37,119.09) .. (150.22,119.09) .. controls (134.08,119.09) and (120.98,122.08) .. (120.98,125.77) .. controls (120.98,129.46) and (107.89,132.45) .. (91.74,132.45) .. controls (75.59,132.45) and (62.5,129.46) .. (62.5,125.77) -- cycle ;
\draw [line width=1.5]    (60.17,190.97) .. controls (69.8,203.73) and (119.5,195.2) .. (118.65,190.97) .. controls (117.8,186.73) and (161.8,177.73) .. (177.13,190.97) ;
\draw [line width=0.75]    (28.47,190.07) .. controls (40.13,196.73) and (52.47,183.73) .. (60.17,190.97) ;
\draw    (177.13,190.97) .. controls (188.8,197.63) and (201.13,184.63) .. (208.83,191.87) ;
\draw    (179.47,99.04) .. controls (191.13,105.7) and (203.47,92.7) .. (211.17,99.94) ;
\draw    (30.8,98.14) .. controls (42.47,104.8) and (54.8,91.8) .. (62.5,99.04) ;
\draw    (160.27,65.82) -- (160.27,119) ;
\draw    (150.22,65.9) -- (150.22,119.09) ;
\draw    (140.52,65.82) -- (140.52,119) ;
\draw    (130.02,67.57) -- (130.02,120.75) ;
\draw    (120.27,73.82) -- (120.27,127) ;
\draw    (110.52,77.32) -- (110.52,130.5) ;
\draw    (100.77,79.07) -- (100.77,132.25) ;
\draw    (89.77,79.32) -- (89.77,132.5) ;
\draw    (80.52,79.07) -- (80.52,132.25) ;
\draw    (70.02,77.32) -- (70.02,130.5) ;
\draw    (170.52,67.82) -- (170.52,121) ;
\draw    (241.1,190.9) -- (418.6,190.9) ;
\draw [shift={(420.6,190.9)}, rotate = 180] [color={rgb, 255:red, 0; green, 0; blue, 0 }  ][line width=0.75]    (10.93,-4.9) .. controls (6.95,-2.3) and (3.31,-0.67) .. (0,0) .. controls (3.31,0.67) and (6.95,2.3) .. (10.93,4.9)   ;
\draw [shift={(239.1,190.9)}, rotate = 0] [color={rgb, 255:red, 0; green, 0; blue, 0 }  ][line width=0.75]    (10.93,-3.29) .. controls (6.95,-1.4) and (3.31,-0.3) .. (0,0) .. controls (3.31,0.3) and (6.95,1.4) .. (10.93,3.29)   ;
\draw    (384.9,181.3) .. controls (393.57,182.63) and (394.23,201.3) .. (384.23,201.3) ;
\draw    (276.9,180.97) .. controls (266.9,181.63) and (267.57,200.97) .. (276.23,200.97) ;
\draw    (241.6,99.9) -- (419.1,99.9) ;
\draw [shift={(421.1,99.9)}, rotate = 180] [color={rgb, 255:red, 0; green, 0; blue, 0 }  ][line width=0.75]    (10.93,-4.9) .. controls (6.95,-2.3) and (3.31,-0.67) .. (0,0) .. controls (3.31,0.67) and (6.95,2.3) .. (10.93,4.9)   ;
\draw [shift={(239.6,99.9)}, rotate = 0] [color={rgb, 255:red, 0; green, 0; blue, 0 }  ][line width=0.75]    (10.93,-3.29) .. controls (6.95,-1.4) and (3.31,-0.3) .. (0,0) .. controls (3.31,0.3) and (6.95,1.4) .. (10.93,3.29)   ;
\draw  [color={rgb, 255:red, 0; green, 0; blue, 0 }  ,draw opacity=1 ][fill={Cerulean!70}  ,fill opacity=0.5 ] (270,72.7) -- (390.7,72.7) -- (390.7,127.1) -- (270,127.1) -- cycle ;
\draw    (280.02,72.82) -- (280.02,126.48) ;
\draw    (290.17,73.17) -- (290.17,126.83) ;
\draw    (300.02,73.07) -- (300.02,126.73) ;
\draw    (310.17,73.25) -- (310.17,127.08) ;
\draw    (319.77,72.82) -- (319.77,126.48) ;
\draw    (329.92,73.17) -- (329.92,126.83) ;
\draw    (339.77,73.07) -- (339.77,126.73) ;
\draw    (349.92,73.25) -- (349.92,127.08) ;
\draw    (359.52,72.82) -- (359.52,126.48) ;
\draw    (369.67,73.17) -- (369.67,126.83) ;
\draw    (379.52,73.07) -- (379.52,126.73) ;
\draw [line width=0.75]    (148.7,180.3) .. controls (186.81,139.21) and (267.07,148.12) .. (298.27,182.26) ;
\draw [shift={(299.2,183.3)}, rotate = 228.93] [color={rgb, 255:red, 0; green, 0; blue, 0 }  ][line width=0.75]    (8.74,-2.63) .. controls (5.56,-1.12) and (2.65,-0.24) .. (0,0) .. controls (2.65,0.24) and (5.56,1.12) .. (8.74,2.63)   ;
\draw    (98.5,177.9) -- (98.5,141.8) ;
\draw [shift={(98.5,180.9)}, rotate = 270] [fill={rgb, 255:red, 0; green, 0; blue, 0 }  ][line width=0.08]  [draw opacity=0] (6.25,-3) -- (0,0) -- (6.25,3) -- cycle    ;
\draw    (356,177.4) -- (356,141.3) ;
\draw [shift={(356,180.4)}, rotate = 270] [fill={rgb, 255:red, 0; green, 0; blue, 0 }  ][line width=0.08]  [draw opacity=0] (6.25,-3) -- (0,0) -- (6.25,3) -- cycle    ;
\draw [line width=1.5]    (269.5,191.5) -- (391.7,191.5) ;

\draw (79.5,152.3) node [anchor=north west][inner sep=0.75pt]  [font=\small]  {${\displaystyle \pi }$};
\draw (128.98,128.67) node [anchor=north west][inner sep=0.75pt]  [font=\scriptsize]  {$\pi ^{-1}(U)$};
\draw (215,128.4) node [anchor=north west][inner sep=0.75pt]    {$\varphi $};
\draw (26,201.9) node [anchor=north west][inner sep=0.75pt]    {$\mathcal{M}$};
\draw (99,35.9) node [anchor=north west][inner sep=0.75pt]    {$T\mathcal{M}$};
\draw (320,33.9) node [anchor=north west][inner sep=0.75pt]    {$\mathbb{R}^{2n}$};
\draw (285.5,135.4) node [anchor=north west][inner sep=0.75pt]  [font=\scriptsize]  {$\varphi (U) \ \times \ \mathbb{R}^{n}$};
\draw (350.5,201.4) node [anchor=north west][inner sep=0.75pt]  [font=\scriptsize]  {$\varphi (U)$};
\draw (150.5,196.5) node [anchor=north west][inner sep=0.75pt]  [font=\footnotesize] [align=left] {$U$};

\end{tikzpicture}\end{center}

\begin{ex}
   {{\textcolor{Cerulean}{{$\blacktriangleright$\;}}}}  $(\M = \rr^{n})$. \normalfont The tangent bundle of $\rr^{n}$ is just the Cartesian product of $\rr^{n}$ with itself.
    \begin{equation}
        T\rr^{n} = \bigsqcup_{x \in \rr^{n}} (\{x\} \cross T_x\rr^{n}) = \rr^{n} \cross \rr^{n}.\;\; \textcolor{Cerulean}{{\blacktriangleleft}}    \end{equation}
\end{ex}

\paragraph{ } When $\M$ is a smooth manifold, the tangent bundle is not just a disjoint union of vector spaces, $T\M$ inherits the topology and the smooth structure of $\M$ and it is a smooth manifold itself.

\begin{tfpropos}
    For any smooth manifold $\M$ with dimension $n$, the tangent bundle $T\M$ has a natural topology and smooth structure that make it into $2n$-dimensional smooth manifold. With respect to this structure, the projection $\pi: T\M \to \M$ is smooth \cite{LeeG}.
\end{tfpropos}

\begin{definazul}\label{defvecfi}
    An application $\mathcal{X} : \M \to T\M$ is said to be a \textbf{vector field} if $\mathcal{X}$ is smooth and \normalfont$\pi \circ \mathcal{X} = \text{Id}$\textit{, that is, $\mathcal{X}(x) \in T_x\M$ for each $x \in \M$}.
\end{definazul}

\noindent If $(U, \varphi) = (U, t^{1}, \ldots, t^{n})$ is a chart for $\M$, othe value of $\mathcal{X}$ at any point $x \in U$ can be written in terms of the coordinate basis vectors:

\begin{equation}
    \mathcal{X}_{x} = \sum^{n}_{i=1} \mathcal{X}^{i}(x)     \frac{\partial}{\partial t^{i}}\Bigg|_{x} . 
\end{equation}

\noindent This defines $n$ functions $\mathcal{X}^{i}: U \to \rr$, namely, the component functions of $\mathcal{X}$ with respect to the given chart. If those components are smooth, then $\mathcal{X}: U \to \M$ is smooth\cite{LeeG}. 

{\colorlet{shadecolor}{Cyan!15}\begin{shaded}With the definition of the tangent space on hands, now, for each $x \in \M$, the \textbf{cotangent space at $x$}, denoted by $ T^{*}_{x}\M$, is defined as being the dual space to $T_{x}M$.

\begin{equation}
    T_x^{*}\M = (T_x\M)^{*}. 
\end{equation}\end{shaded}}

\noexpand The elements of $T_x^{*}\M$ are called \textbf{tangent covectors at $x$} or just \textbf{covectors at $x$}.

\begin{definazul}
    For any smooth manifold $\M$, the \textbf{cotangent bundle} is defined as being the disjoint union

    \begin{equation}
        T^{*}\M = \bigsqcup_{x \in \M} T_x^{*}\M =\bigsqcup_{x \in \M} (\{x\} \cross T_x^{*}\M) .
    \end{equation}
\end{definazul}

\noindent It also has a natural projection map $\pi: T^{*}\M \to \M$ sending $(x,\upomega) \in \{x\} \times T_x^{*}\M$ to $x \in \M$.

\paragraph{ } The tangent and cotangent bundles have an important property related to the projection maps $\pi : T\M \to \M$ and $\pi^{'}: T^{*}\M \to \M$. Specifically, the sets $\pi^{-1}(x) = T_{x}\M$ and $\pi^{'-1}(x) = T^{*}_{x}\M$ are vector spaces for every $x \in \M$. While this may seem obvious, it is a crucial property that forms the basis for understanding a more general class of bundles, namely, the vector bundles.


\section{Vector Bundles}

\paragraph{ } Given an open subset $U$ of a manifold $\M$, one can think of $U \cross \rr^{n}$ as a family of vector spaces $\rr^{n}$ parameterised by the points in $U$. A \textbf{vector bundle}, roughly speaking, is a family of vector spaces that locally \textit{looks like} $U \cross \rr^{n}$. 

 \begin{definazul}
     A (real) \textbf{vector bundle} of rank $n$ over a manifold $\M$, consists of a triple $(E, \pi, \M)$ such that:

     \begin{itemize}
        \item[\textcolor{Cerulean}{\textbf{\textit{1.}}}]The set $E$ is a manifold, called the \textbf{total space}.
        \item[\textcolor{Cerulean}{\textbf{\textit{2.}}}] The application $\pi : E \to \M$ is a smooth surjective map; for every $x \in \M$ the set $E_x := \pi^{-1}(x)$ is a (real) vector space of dimension $n$ called \textbf{fibre}.
         \item[\textcolor{Cerulean}{\textbf{\textit{3.}}}] The \textbf{local triviality condition} is satisfied: for each $x_0 \in \M$, there exists an open neighbourhood $U$ of $x_0$ and a diffeomorphism 

         \begin{equation}
             \varphi_U : U \cross \rr^{n}  \to \pi^{-1}(U)
         \end{equation}

         \noindent that maps each fibre $E_x$ to the fibre $\{x\} \cross \rr^{n}$ linearly. Moreover, $(\pi \circ \varphi_U)(x,v) = x$ for all $v \in \rr^{n}$ and the map $v \mapsto \varphi_U(x, v)$ is a linear isomorphism between the vector spaces $\rr^{n}$ and the fibre $\pi^{-1}(x) = E_x$.
          \[
\begin{tikzcd}
    U \cross \rr^{n} \arrow{rr}{\varphi_U} \arrow[swap]{dr}{\text{pr}_1} && \pi^{-1}(U) \arrow{dl}{\pi} \\[10pt]
    & U
\end{tikzcd}
\]
     \end{itemize}
 \end{definazul}

Complex vector bundles are similarly defined with $\cc^{n}$ taking the role of $\rr^{n}$ above. The real and complex cases are quite similar, so in various definitions the only real case is mentioned, with the implicit understanding that the complex case goes the same way. Instead of using the triple notation $(E,\pi,\M)$ we adopt a simpler terminology to refer to vector bundles, we mention the mapping $\pi: E \to \M$ or even just total space $E$, by which it is self understood that $E$ comes with a map onto $\M$, generically denoted $\pi_E$, called the projection of the vector bundle $E$. If $\M$ and $E$ are smooth manifolds then $E$ is called a smooth vector bundle. As with manifolds, all vector bundles here are considered to be smooth.

\begin{definazul}
    A vector bundle $\pi: E \to \M$ is said to be \textbf{trivial} if it is isomorphic to a product bundle $\M \cross \rr^{n} \to \M$ over $\M$.
\end{definazul}

One should think about $E$ as the union of all the fibres 

\begin{equation}
    E = \bigsqcup_{x \in \M} E_x.
\end{equation}

Note that $E|_U := \pi^{-1} (U)$ gives rise to a vector bundle $\pi_U : E|_U \to U$ over $U$ which is trivial by definition.

\begin{ex}
\textcolor{purple}{{\textcolor{Cerulean}{{$\blacktriangleright$\;}}}} Vector spaces. \normalfont A vector space is a vector bundle over a point.\; $\textcolor{Cerulean}{{\blacktriangleleft}} $
\end{ex}

\begin{ex}
\textcolor{purple}{{\textcolor{Cerulean}{{$\blacktriangleright$\;}}}}     The product bundle.\normalfont\; Let $V$  be  a vector space of dimension $n$, the projection $\pi : \M \cross V \to \M$ is a vector bundle of rank $n$, called the product bundle. For instance, via $\pi : S^{1} \cross \rr \to S^{1}$, the cylinder $S^{1} \cross \rr$ is a product bundle over the circle $S^{1}$.\; $\textcolor{Cerulean}{{\blacktriangleleft}}$\end{ex}

\paragraph{ } The tangent bundle $T\M$ is indeed a vector bundle. Both the topological smooth structure and the local triviality condition is shown respectively in the \textit{Proposition 3.18} and \textit{10.4} of Reference \cite{LeeG}. 

\begin{definazul}
    Given two bundles $E$ and $F$ over $\M$ a \textbf{ vector bundle morphism} from $E$ to $F$ is a smooth map $h : E \to F$ such that the diagram

 \[
\begin{tikzcd}
    E \arrow{rr}{f} \arrow[swap]{dr}{\pi_E} && F \arrow{dl}{\pi_F} \\[10pt]
    & \M
\end{tikzcd}
\]
    
\noindent commutes together with the property that for each $x \in \M$, $f$ sends $E_x$ to $F_x$ whose restriction

    \begin{equation}
        f_x := f_{E_x} : E_x \to F_x
    \end{equation}

    \noindent is linear.
\end{definazul}

\noindent If each $f_x$ is an isomorphism then $f$ is an isomorphism of vector bundles over $\M$ (or, equivalently, if $f$ is also a diffeomorphism).
\subsection*{GLUINGS AND CONSTRUCTION FROM LOCAL DATA}

\paragraph{ } Here we present a way to construct vector bundles via '\textit{gluing functions}' that satisfy '\textit{cocycle condition}'. Let us consider $\mathcal{U} = \{U_{\a}\}_{\a \in A}$ a cover for $\M$. With respect to $(E, \pi,\M)$, for each $a \in A$,  the pair $(U_{\a},\phi_{\a})$, with $\phi_{\a} : U_{\a} \cross \rr^{n} \to \pi^{-1}(U_{\a})$, is called a local trivialisation and $\mathcal{U}$ is the trivialising cover. For $\a,\b \in A$, let $U_{\a\b} = U_{\a} \inter U_{\b}$.

   {\colorlet{shadecolor}{Cyan!15}\begin{shaded} The maps,

\begin{equation}
     \phi_{\b} \circ \phi_{\a}^{-1} : U_{\a\b} \cross \rr^{n} \to U_{\a\b} \cross \rr^{n}
\end{equation}

\noindent satisfy

\begin{equation}\label{eq_retgfjndfbsa3z12}
    \phi_{\b} \circ \phi_{\a}^{-1}(x,v) =  (x,g_{\a\b}(x)v)
\end{equation}

\noindent such that $g_{\a\b}: U_{\a\b} \to \text{GL}(n,\rr)$ are the \textbf{gluing maps} or \textbf{transition functions}.\end{shaded}}

 \noindent From construction, they satisfy for all $\a,\b $ and $\gamma \in A$:

\begin{flalign}
    \begin{aligned}
        &\textcolor{Cerulean}{\textbf{(i)}}\;\;\;  g_{\a \a} (x) = \text{id}, \;\; \text{ for all } x \in U_\a;\\
     &\textcolor{Cerulean}{\textbf{(ii)}}\;\; g_{\a \b}(x) = g^{-1}_{\b \a}, \;\; \text{ for all } x \in U_{\a} \inter U_{\b};\\
      &\textcolor{Cerulean}{\textbf{(iii)}}\;g_{\b \delta}(x)g_{\a \b}(x) =   g_{\a \delta}(x)  \;\; \text{ for all } x \in U_{\a} \inter U_{\b} \inter U_{\delta} \textit{\;\;(cocycle condition)}.
    \end{aligned}&&
\end{flalign}

\noindent The local trivialisations $(U_{\a}, \phi_{\a})$ glue to define a diffeomorphism

\begin{equation}
    E \simeq \left(\bigsqcup_{\a \in A} U_{\a} \cross \rr^{n} \right) \Big/ \sim
\end{equation}

\noindent where $(x,v) \sim (a,b)$ if $a = x$ and $b = g_{\a\b}(x)v$. Hence, any open covering $\mathcal{U}$ for $\M$ with a family of gluing maps $\{g_{\a\b}\}$, defines a unique vector bundle $E = (\mathcal{U}, g_{\a\b})$ up to isomorphism \cite{Morita}.


With a vector bundle in hands it is possible to construct new vector bundles from this one. All natural constructions of multilinear algebra, such as operations with vector spaces: dualising, direct sums, tensor products, etc, give rise to a similar construction on vector bundles. Let $E = (\mathcal{U},g_{\a\b})$ and $F = (\mathcal{U},h_{\a\b})$ be two vector bundles over the same trivialising cover. The following bundles can be defined \cite{Konstantin}:

\begin{itemize}
    \item [\textcolor{Cerulean}{\textbf{\textit{1.}}}] The dual bundle $E^{*}$ by
\begin{equation}
    E^{*} = (\mathcal{U}, (g_{\a\b}^{*})^{-1}).
\end{equation}
\item [\textcolor{Cerulean}{\textbf{\textit{2.}}}] The direct sum $E \oplus F$ by

\begin{equation}
    E \oplus F = (\mathcal{U},g_{\a\b}) \oplus h_{\a\b}.
\end{equation}
\item [\textcolor{Cerulean}{\textbf{\textit{3.}}}]The tensor product $E \tensor F$ by

\begin{equation}
    E \tensor F = (\mathcal{U},g_{\a\b}) \tensor h_{\a\b}.
\end{equation}

\item [\textcolor{Cerulean}{\textbf{\textit{4.}}}]The symmetric and exterior powers $\text{Sym}^{k}E$ and $\bigwedge^{k} E$ by

\begin{align}
    \begin{aligned}
        \text{Sym}^{k}E = (\mathcal{U}, \text{Sym}^{k}g_{\a\b}),\\
        \bigwedge^{k} E = \big(\mathcal{U}, \bigwedge^{k}g_{\a\b}\big).
    \end{aligned}
\end{align}

\item [\textcolor{Cerulean}{\textbf{\textit{5.}}}] The Hom-bundle $\text{Hom}(E,F)$ 
\begin{equation}
    \text{Hom}(E,F) \simeq E^{*} \tensor F = (\mathcal{U}, (g_{\a\b}^{*})^{-1}) \tensor h_{\a\b}.  
\end{equation}
    
\end{itemize}

\subsection*{SECTIONS AND FRAMES}

\paragraph{ } One of the main objects associated to vector bundles are their (local) \textit{sections}. Sections of vector bundles are crucial objects in differential geometry, representing smooth choices of vectors from each fibre of a vector bundle over a manifold. They allow the analysis and the description of geometric structures and physical fields in a in a coordinate-free manner. For instance: differential forms can be characterise as being sections of the exterior bundle of the cotangent space; spinors are also characterised as being sections of the bundle of spinors. In this sense, understanding sections is fundamental for exploring more complex geometric and topological properties of manifolds and their associated bundles.

\begin{definazul}
Given a vector bundle $\pi: E \to \M$, a \textbf{section} of $E$ is a smooth map $s: \M \to E$ satisfying \normalfont $\pi \circ s = \id$ \textit{, that is, for any} $x \in \M$, $s(x) \in E_x$.    
\end{definazul}

\begin{center}
\tikzset{every picture/.style={line width=0.75pt}} 

\begin{tikzpicture}[x=0.75pt,y=0.75pt,yscale=-1,xscale=1]

\draw    (66.1,177.41) -- (198,177.89) ;
\draw [shift={(200,177.9)}, rotate = 180.21] [color={rgb, 255:red, 0; green, 0; blue, 0 }  ][line width=0.75]    (10.93,-4.9) .. controls (6.95,-2.3) and (3.31,-0.67) .. (0,0) .. controls (3.31,0.67) and (6.95,2.3) .. (10.93,4.9)   ;
\draw [shift={(64.1,177.4)}, rotate = 0.21] [color={rgb, 255:red, 0; green, 0; blue, 0 }  ][line width=0.75]    (10.93,-3.29) .. controls (6.95,-1.4) and (3.31,-0.3) .. (0,0) .. controls (3.31,0.3) and (6.95,1.4) .. (10.93,3.29)   ;
\draw  [color={rgb, 255:red, 0; green, 0; blue, 0 }  ,draw opacity=1 ][fill={Cerulean}  ,fill opacity=0.5 ] (78,70.75) -- (194.97,70.75) -- (194.97,125.32) -- (78,125.32) -- cycle ;
\draw [line width=1.5]  [dash pattern={on 1.69pt off 2.76pt}]  (78,98.04) .. controls (103,139.9) and (113.97,74.67) .. (136.48,98.04) .. controls (159,121.4) and (179.63,84.8) .. (194.97,98.04) ;
\draw    (203,116.9) .. controls (228.38,123.95) and (213.11,151.76) .. (202.05,163.8) ;
\draw [shift={(200,165.9)}, rotate = 316.33] [fill={rgb, 255:red, 0; green, 0; blue, 0 }  ][line width=0.08]  [draw opacity=0] (8.93,-4.29) -- (0,0) -- (8.93,4.29) -- cycle    ;
\draw    (67.82,115.02) .. controls (40.18,121.62) and (56.42,151.82) .. (68,163.4) ;
\draw [shift={(71,114.4)}, rotate = 171.12] [fill={rgb, 255:red, 0; green, 0; blue, 0 }  ][line width=0.08]  [draw opacity=0] (8.93,-4.29) -- (0,0) -- (8.93,4.29) -- cycle    ;

\draw (223.5,130.8) node [anchor=north west][inner sep=0.75pt]  [font=\small]  {${\displaystyle \pi }$};
\draw (35.5,128.9) node [anchor=north west][inner sep=0.75pt]    {$s$};
\draw (52,189.4) node [anchor=north west][inner sep=0.75pt]    {$\mathcal{M}$};
\draw (80,74.15) node [anchor=north west][inner sep=0.75pt]    {$E$};
\draw (151.5,109.4) node [anchor=north west][inner sep=0.75pt]  [font=\footnotesize]  {$s( x)$};
\draw (157,184.4) node [anchor=north west][inner sep=0.75pt]  [font=\footnotesize]  {$x$};
\draw (150,166.7) node [anchor=north west][inner sep=0.75pt]   [font=\fontsize{3em}{4em}\selectfont]  {$\cdot $};
\draw (150,94.2) node [anchor=north west][inner sep=0.75pt] [font=\fontsize{3em}{4em}\selectfont]   {$\cdot $};
\end{tikzpicture}\end{center}

\begin{ex}\label{extrivbsec}
{{\textcolor{Cerulean}{{$\blacktriangleright$\;}}}} Section of the trivial bundle. \normalfont If $E = \M \cross \rr^{n}$ is the trivial bundle, a section of $E$ is simply a function from $\M$ to $\rr^{n}$. In other words, given a section $s: \M \to E$, there is a continuous function $f: \M \to \rr^{n}$ such that $s(x) = (x,f(x)) \in E_{x}.$ $\,\textcolor{Cerulean}{{\blacktriangleleft}}$
\end{ex}

\begin{ex}
{{\textcolor{Cerulean}{{$\blacktriangleright$\;}}}} Section of tangent bundle. \normalfont Recall from Definition \ref{defvecfi} that a vector field is a function $\mathcal{X}$ that assigns a tangent vector $\mathcal{X}_{x} \in T_{x}\M$ for each point $x \in \M$. If $E = T\M$ is the tangent bundle, a section $s: \M \to T\M$ is a vector field. $\;\textcolor{Cerulean}{{\blacktriangleleft}}$
\end{ex}

\begin{tfpropos}\label{propossec}
    Let $s$ and $s'$ be sections of the vector bundle $\pi : E \to \M$, $f$ be a real-valued smooth function on $\M$ and $\lambda \in \rr$ a scalar. It follows that \cite{TuIntro}:

\begin{itemize}
    \item[\textcolor{Cerulean}{\textbf{\textit{1.}}}] The sum $s + s': \M \to E$ is a section of $E$ and is defined by adding the sections pointwise

\begin{equation}
    (s + s')(x) := s(x) + s'(x) \in E_{x}, \; x \in \M.
\end{equation}

 \item[\textcolor{Cerulean}{\textbf{\textit{2.}}}] The scalar multiplication $\lambda s: \M \to E$ is a section of $E$ given by,

\begin{equation}
    (\lambda s)(x) := \lambda s(x) \in E_{x}, \; x \in \M.
\end{equation}

 \item[\textcolor{Cerulean}{\textbf{\textit{3.}}}] The product  $fs: \M \to E$ is a section of $E$ and is defined by the pointwise multiplication

 \begin{equation}
     (fs)(x) := f(x)s(x) \in E_{x},\; x \in \M.
 \end{equation}
\end{itemize}\end{tfpropos}

 The set of all \textit{global sections} of $E$ is denoted by
\begin{equation}
    \Gamma(\M,E) = \Gamma(E).
\end{equation}

\noindent It follows from Proposition \ref{propossec} that $ \Gamma(E)$ is not only a real vector space, but also a module over the ring $\Cinf(\M)$. For $U \subset \M$ open, the set of all \textit{local sections} of $E$ defined over $U$ is denoted by

\begin{equation}
    \Gamma(U,E) = \Gamma(E|_{U}).
\end{equation}

\begin{ex}
{{\textcolor{Cerulean}{{$\blacktriangleright$\;}}}} The space of sections of the trivial bundle. \normalfont From the Example \ref{extrivbsec} one has that $\Gamma(\M,\rr^{n}) \simeq \Cinf(\M)$.  $\,\textcolor{Cerulean}{{\blacktriangleleft}}$
\end{ex}

\paragraph{ } From the concept of \textit{sections}, the \textit{frame} of a vector bundle can be defined. Frames provide a structured way to describe bases for the fibres at each point of the manifold. They offer a local trivialisation of the bundle, simplifying the study of its geometric and topological properties. Understanding frames is essential for constructing and working with sections, as they facilitate computations and provide insight into the local and global structure of vector bundles.

\begin{definazul}
    Let $\pi: E \to \M$ be a vector bundle over $\M$. A \textbf{frame} of $E$ is a collection 
$s = (s_1, \ldots, s_n)$ of sections $s_i$ of $E$ such that for each $x \in \M$, $(s_1(x), \ldots, s_n(x))$ is a basis of the vector space $E_x$.
\end{definazul}

\noindent A \textit{local frame} of $E$ over $U$ is a frame $s$ of $E|_{U}$ for some open set $U \subset \M$. In terms of terminology, the global or local frames for $E$ (over $\M$) could be also said global or local frames for $\M$, meaning the same thing. 

\begin{ex}
\textcolor{purple}{{\textcolor{Cerulean}{{$\blacktriangleright$\;}}}} Frame of $T\rr^{3}$. \normalfont The collection of vector fields $\{\frac{\partial}{\partial x}, \frac{\partial}{\partial y},\frac{\partial}{\partial z}\}$ is a frame of the tangent bundle $T\rr^{3}$.\;  $\textcolor{Cerulean}{{\blacktriangleleft}}$
\end{ex}

\begin{ex}
\textcolor{purple}{{\textcolor{Cerulean}{{$\blacktriangleright$\;}}}} Frame of the product bundle. \normalfont Let $\M \cross \rr^{n} \to \M$ be the product bundle of a manifold $\M$. Consider the standard basis $e_{1},\ldots,e_{n}$ for $\rr^{n}$ . Let us consider the sections $e'_{i}: \M \to  \M \cross \rr^{n}$ by $e'_{i}(x) = (x,e_{i}).$ Thus, $(e'_{1}, \ldots, e'_{n})$ is a frame for the trivial product bundle. $\textcolor{Cerulean}{{\blacktriangleleft}}$
\end{ex}

\begin{ex}
\textcolor{purple}{{\textcolor{Cerulean}{{$\blacktriangleright$\;}}}} Frame of a trivialisation. \normalfont For a vector bundle $\pi: E \to \mathcal{M}$ of rank $n$, let us consider $\phi: U \times \mathbb{R}^{n} \to \pi^{-1}(U)$, a trivialisation of $E$ over an open set $U \subset \mathcal{M}$. The mapping $\phi$ being a diffeomorphism, takes basis to basis. Then it carries the frame $(e'_1, \ldots, e'_n)$ of the product bundle to a frame $(s_1, \ldots, s_n)$ for $E$ over $U$: \begin{equation}
    s_{i}(x) = \phi(e'_{i}(x)) = \phi(x,e_{i}), \; x \in U. 
\end{equation}

\[
\begin{tikzcd}[row sep=huge, column sep=huge]
    & U \arrow[Cerulean, rd, swap, "e'_{i}"] \arrow[Cerulean, ld, "s_{i}"] & \\
   \pi^{-1}(U) \arrow[bend left, "\pi"]{ru} & & U \cross \rr^{n} \arrow[bend right, swap, "\text{pr}_1"]{lu} \arrow[ll, "\phi"]  \\
\end{tikzcd}
\]

\noindent $(s_{1}, \ldots, s_{n})$ is called the frame over $U$ of the trivialisation $\phi$. Every smooth local frame for a smooth vector bundle is associated with a smooth local trivialisation \cite{LeeG}.  $\;\textcolor{Cerulean}{{\blacktriangleleft}}$
\end{ex}

\paragraph{ } In particular, the concept of frames provides the framework for introducing the notion of spin structures later on. This is because specific bundles will be constructed as frames being their elements, such that symmetry groups such as the orthogonal group can act upon them, laying the foundation for defining and working with spin structures.

\section{Differential Forms}

\paragraph{ } Differential forms play a central role in modern differential geometry and mathematical physics, serving as a versatile tool for integrating functions over manifolds and expressing various physical laws in a coordinate-free manner. Differential forms are an essential generalisation of real-valued functions on a manifold, assigning $k$-covectors to each point instead of numbers. For $k = 0$ and $1$, they correspond to functions and covector fields, respectively. Crucially, differential forms are intrinsic to manifolds and possess a rich algebraic structure. For instance, unlike vector fields, differential forms benefit from the wedge product, grading, and exterior derivative, forming a differential graded algebra. They are not just important in the algebraic context but also in the field of topology and calculus on manifolds. This structure facilitates the de Rham complex, a contravariant functor, and leads to the de Rham cohomology, pivotal for manifold theory \cite{TuIntro}. The integration theory on manifolds relies on differential forms of top degree, ensuring coordinate invariance. This way we see that differential forms are fundamental in many context in mathematics and physics. However, we are just interested here on the algebra of differential forms.

In this section, we introduce differential forms from the vector bundle perspective.  Originating from the works of Henri Poincaré and Élie Cartan, differential forms have evolved significantly, now viewed as sections of the exterior power of the cotangent bundle. Understanding differential forms is essential for our subsequent discussions on the Kähler-Atiyah bundle as it is nothing more than a Clifford structure on the exterior algebra of
differential forms. We present just the definition in the context of vector bundles here whereas the theory of exterior algebra is presented in the Appendix \ref{app3}.

\subsection*{DIFFERENTIAL $1$-FORMS}

\paragraph{ } Let us consider a manifold $\M$ and a point $x \in \M$. We recall the cotangent space of $\M$ at $p$, the dual space of the tangent space $T_{x}\M$:

\begin{equation}
    T_{x}^{*}\M = \text{Hom}(T_{x}\M, \rr)
\end{equation}

\noindent An element of the cotangent space $\uptheta_{x} \in T_{x}^{*}\M$ is then a linear function

\begin{equation}
    \uptheta_{x}: T_{x}\M \to \rr
\end{equation}

\noindent called a covector field at $x$. 

\begin{definazul}
    A differential $1$-form in a open set $U \subset \M$ is a function $\uptheta$ that assigns for each point $x \in \M$, a covector $\uptheta_{x} \in T_{x}^{*}\M$:

    \begin{align}
    \begin{aligned}
       \uptheta : U &\to \bigsqcup_{x \in \M} T_x^{*}\M =  T^{*}\M ,\\
x &\mapsto \uptheta_x \in T_{x}^{*}\M.       
    \end{aligned}
\end{align}
\end{definazul}

It is possible to construct a $1$-form from functions. If $f$ is a $\Cinf$ real valued function on $\M$, its \textit{differential} is defined to be the $1$-form $df$ on $\M$ such that for any $x \in \M$ and $\mathcal{X}_{x} \in T_{x}\M$,

\begin{equation}
    (df)_x (\mathcal{X}_{x}) = \mathcal{X}_{x} f.
\end{equation}

\noindent With respect to a coordinate chart $(U,\phi) = (U, t^{1}, \ldots, t^{n})$ on $\M$, the differentials $dt^{1}, \ldots, dt^{n}$ are $1$-forms on $U$. At each point $x \in \M$ the coordinate bases $(\partial^{i}, \ldots, \partial^{n})_{x}$ gives rise to a dual basis $(dt^{1}, \ldots, dt^{n})_{x}$ for $T_x^{*}\M$,

\begin{equation}
(dt^{i})_x\left(\frac{\partial}{\partial t_j}\right)\bigg\rvert_x  = \frac{\partial t_i}{\partial t_j} (x) = \delta^{i}_{j}.
\end{equation}

Thus, every $1$-form $\uptheta$ on $U$ can be written as a linear combination

\begin{equation}
    \uptheta = \sum a_{i}dt^{i}
\end{equation}

\noindent in which the coefficients $a_{i}$ are functions on $\M$. This way, $\uptheta$ is smooth if  $a_{i}$ are smooth. The set of all smooth $1$-forms on $\M$ has the structure of a vector space, denoted by $\Omega^{1}(\M).$ The value of the $1$-form at $x \in U$ is a linear combination \cite{TuIntro}

\begin{equation}
    \uptheta_{x} = \sum a_{i}(x)dt^{i}|_{x}.
\end{equation}

Moreover,  if $f$ is a $\Cinf$ function on $\M$, then the restriction of $df$ to $U$ is a linear combination

\begin{equation}
    df = \sum a_{i}dt^{i}.
\end{equation}

\noindent To have a local expression for  $df$ we apply both sides to $\partial^{j}$

\begin{equation}
    df \left ( \frac{ \partial }{\partial t^j}\right ) = \sum_i a_i dt^{i} \left ( \frac{ \partial }{\partial t^j}\right ) = \sum_{i} a_i \delta^{i}_{j} = a_j.
\end{equation}

\noindent Consequently,

\begin{equation}
    df = \sum \frac{ \partial f }{\partial t^j}dt^{i}.
\end{equation}

Note that with respect to the cotangent bundle $T^{*}\M$, a $1$-form $\uptheta: \M \to T^{*}\M$ is simply a section of $T^{*}\M$. That is, $\Omega^{1}(\M) = \Gamma(T^{*}\M)$.

\subsection*{DIFFERENTIAL $k$-FORMS}

\paragraph{ } Consider a finite-dimensional (real) vector space $V$. From Definition \ref{def_aaalt} in the Exterior Algebra Appendix \ref{app3}, an $k$-covariant tensor on $V$ is termed alternating if its value undergoes a sign change upon interchanging any two of its arguments. Equivalently, if any permutation of the arguments causes its value to be multiplied by the sign of the permutation. The space of the multicovectors or $k$-covectors in $V$ is denoted by $\bigwedge^{k}(V^{*})$. Considering $V = T_{x}\M$ for $x \in U \subset \M$.

The space $\bigwedge^{k}(T_{x}^{*}\M)$ is the space of all alternating $k$-tensors of the tangent space $T_{x}\M$.

\begin{definazul}
    A \textbf{differential $k$-form} is a $k$-covector field, that is, a function $\upomega$ that assigns to each point $x \in \M$ a $k$-covector $\upomega_{x} \in \bigwedge^{k}(T_{x}^{*}\M)$.
\end{definazul}

The space of the smooth differential $k$-forms is given by $\Omega^{k}(\M)$. In terms of the exterior bundle $\bigwedge^{k}(T_{x}^{*}\M)$ we have that

\begin{equation}
    \Omega^{k}(\M) = \Gamma\Big(\bigwedge^{k}(T_{x}^{*}\M) \Big)
\end{equation}

\noindent Moreover, if $\dim \M = n$ the space of all smooth differential forms over $\M$ is given by:

\begin{equation}
    \Omega(\M) = \bigoplus_{k=0}^{n} \Omega^{k}(\M) = \Gamma\Big(\bigwedge (T_{x}^{*}\M) \Big) 
\end{equation}

The exterior product is generally defined in the Appendix \ref{sec_app_extrprod}. In particular, given a $k$-form $\upomega$ and a $r$-form $\upbeta$ in the open set $U \subset \M$, the exterior product  $\wedge : \Omega^{k} (U) \cross \Omega^{r} (U) \to \Omega^{k + r} (U)$ between those forms is given by 

\begin{equation}
     \upomega \wedge \upbeta = \frac{(k+l)!}{k!l!} \text{Alt} (\upomega \tensor \upbeta)
\end{equation}

\noindent being defined pointwise

\begin{equation}
    ( \upomega \wedge \upbeta)_x = \upomega_x \wedge \upbeta_x, \;\;\; x \in U.
\end{equation}



\noindent Where Alt is the alternator operator presented in the Eq. \eqref{eq_assd4123a}. 

  {\colorlet{shadecolor}{Cyan!15}\begin{shaded}The space $(\Omega(\M), \wedge)$ is the exterior algebra of differential forms on $\M$. \end{shaded}}

All the properties of $(\Omega(\M), \wedge)$ follow the same construction as the ones made for exterior algebra of multivectors that is presented with details in the Appendix \ref{app3}. 

\section{Principal Bundles}\label{sec_principalbundles}

\paragraph{ } Principal bundles are structures that allow a manifold to carry an action of a Lie group in a smooth and continuous way. Principal bundles are particularly significant because they provide a natural setting for defining spin structures which are essential for integrating the algebraic concept of spinors into the geometric framework of manifolds. We begin this section by discussing Lie groups and their actions, which are essential for understanding the framework of principal bundles. After laying this groundwork, we define what a principal $G$-bundle is.  We conclude this section by examining the connection between vector bundles and principal bundles. Although they are distinct objects, they are deeply interconnected and can be viewed as two sides of the same coin. 

\subsection*{LIE GROUPS AND ACTIONS}

\begin{definazul}
    Given a Lie group $G$ and a manifold $\M$, a \textbf{right} action of the Lie group $G$ on $\M$ is a smooth map 

\begin{equation}
    m: \M \cross G \to \M,\;\;\;  m(x,g) = x \cdot g
\end{equation}

\noindent which satisfy the axioms for an action:

\begin{equation}
    (x \cdot h) \cdot g = x \cdot (hg),\;\;\; x \cdot 1_{G} = x
\end{equation}

 \noindent for all $g, h \in G, x \in \M$, where $1_{G} \in G$ is the unit of the group. The notion of the left action is defined similarly.  
\end{definazul}

\begin{ex} {{\textcolor{Cerulean}{{$\blacktriangleright$\;}}}}  
    Left action defined by a representation. \normalfont Representations of a Lie group on some vector space area a very special case of left $G$-actions. Let $G$ be a Lie group, $\M = V$ the representation space of $G$ with $\rho : G \to \text{GL}(V)$ a linear representation. A left action $m: G \cross V \to V$ can be defined by setting $m(g,v) := \rho(g)(v). \; \textcolor{Cerulean}{{\blacktriangleleft}}$ 
\end{ex}

\begin{ex}{{\textcolor{Cerulean}{{$\blacktriangleright$\;}}}}  
    Right action defined by a left action. \normalfont  Given a left action $m_l : G \cross \M \to \M$, a right action $m_r :\M \cross G \to \M$ can be defined by setting $m_r(x,g) := m_l(g^{-1},x).$ In fact, one has that $m_r$ is smooth since in a Lie group the inverse operation is smooth. In addition, $m_r$ satisfy the axioms of an action: 
    \begin{align}
        \begin{aligned}
       m_r(m_r(x,h), g) = m_l(g^{-1}, m_l(h^{-1},x)) &= m_l(g^{-1}h^{-1},x) =m_l((hg)^{-1},x) = m_r(x,hg), 
       \\
            m_r(x,1_G) &= m_l(1_G, x) = x.\; \textcolor{Cerulean}{{\blacktriangleleft}}
        \end{aligned}
    \end{align}   
\end{ex}

\begin{ex}{{\textcolor{Cerulean}{{$\blacktriangleright$\;}}}} 
    Multiplication on a Lie group. \normalfont The multiplication in a Lie group $G$ can be seen as both as left, as well as right, action of $G$ on itself. Giving rise to:

\begin{itemize}

    \item[\textcolor{Cerulean}{\textbf{\textit{1.}}}] \textit{right translations}: any $a \in G$ induces the right translation $R_{a}: G \to G,\;\;\; R_{a}(g) = g \cdot a$.
    \item[\textcolor{Cerulean}{\textbf{\textit{2.}}}] \textit{left translations}:  $L_{a}: G \to G,\;\;\; L_{a}(g) = a \cdot g$.
    \item[\textcolor{Cerulean}{\textbf{\textit{3.}}}] \textit{(left) adjoint action}: $Ad: G \cross G \to G$, \;\;\; $Ad(a,g) = a \cdot g \cdot a^{-1}$. By fixing $a \in G$, one has $Ad_{a} = L_aR_{a^{-1}}: G \to G, \;\;\; Ad_a = a \cdot g \cdot a^{-1}$ which is the adjoint action of $a$ on $G$. $\textcolor{Cerulean}{{\blacktriangleleft}}$
\end{itemize}
\end{ex}

\begin{definazul}\label{defiequivi}
    Let $G$ and $H$ be two Lie groups and $f: G \to H$ a Lie group homomorphism and $\M$ and $\mathcal{N}$ be two smooth manifolds. Consider $m_{G}: G \cross \M \to \M$, $m_{H}: H \cross \mathcal{N} \to \mathcal{N}$ two left actions and $F: \M \to \mathcal{N}$ a smooth map. Then $F$ is called $\mathbf{f}-$\textbf{equivariant} if the following diagram commutes:

    \[ \begin{tikzcd}
G \cross \M \arrow[r, "f \cross F"] \arrow[d, "m_{G}"] & H \cross \mathcal{N} \arrow{d}{m_{H}}   \\
\M \arrow[r, "F"]          & \mathcal{N}                                      
\end{tikzcd} 
    \]
\noindent That is, the map $F$ between manifolds preserves the left actions   $F(m_G(g,x)) = m_H(f(g),F(x))$.
\end{definazul}

\begin{definazul}
    Given a (left, say) Lie group action $m: G \cross \M \to \M $, one has:

    \begin{itemize}
       \item[\textcolor{Cerulean}{\textbf{\textit{1.}}}] For any $x \in \M$ the \textbf{orbit} under the action is defined as being the set

        \begin{equation}
            \mathcal{O}_x = \{y \in \M : \exists g \in G, m(g,x) = y\}.
        \end{equation}

        \item[\textcolor{Cerulean}{\textbf{\textit{2.}}}] For any $x \in \M$ the \textbf{stabiliser} is defined as being the set

        \begin{equation}
            S_{x} = \{g \in G : m(g,x) = x \}.
        \end{equation}
    \end{itemize}
\end{definazul}

The orbit is the set of all images of a point under the action by elements of $G$ and the stabiliser  is the set of elements of $G$ that fix the point. Note that the stabiliser is a subset of the group whereas the orbit is a subset of the manifold. An equivalence relation of points that relies in the same orbit can be defined: $x \sim y$ if there exists $g \in G : y = m(g,x)$. The set of orbits is the quotient set $\M/\sim\; \equiv \M/G$; with the quotient topology it is called the \textit{orbit space} of the action \cite{LeeG}.

\begin{ex}{{\textcolor{Cerulean}{{$\blacktriangleright$\;}}}}
    Stabiliser and orbit of points on the plane with respect to rotations. \normalfont Let us consider $\M = \rr^{2}$ and $G = \text{SO}(2)$. The group $\text{SO}(2)$ acts on the plane $\rr^{2}$ by rotations,

    \begin{equation}
        \text{SO}(2) = \left\{ \begin{bmatrix}
\cos \theta & -\sin \theta \\ 
\sin \theta & \cos \theta
\end{bmatrix} : \theta \in [0,\pi) \right\}. 
    \end{equation}
\end{ex}

Every point of the plane relies in some orbit. In fact, given any non zero point $x = (x_1,x_2)$ the orbit are circles around the origin at the point $(0,0)$:

\tikzset{every picture/.style={line width=0.75pt}} 
\begin{center}

\begin{tikzpicture}[x=0.75pt,y=0.75pt,yscale=-1,xscale=1]

\draw    (280,60.5) -- (280,170.15) ;
\draw    (339.4,120.65) -- (221,120.65) ;
\draw  [color={Cerulean}  ,draw opacity=1 ] (255.2,120.65) .. controls (255.2,106.84) and (266.39,95.65) .. (280.2,95.65) .. controls (294.01,95.65) and (305.2,106.84) .. (305.2,120.65) .. controls (305.2,134.46) and (294.01,145.65) .. (280.2,145.65) .. controls (266.39,145.65) and (255.2,134.46) .. (255.2,120.65) -- cycle ;
\draw  [fill={rgb, 255:red, 0; green, 0; blue, 0 }  ,fill opacity=1 ] (300.4,104.07) .. controls (300.4,103.21) and (299.69,102.5) .. (298.82,102.5) .. controls (297.96,102.5) and (297.25,103.21) .. (297.25,104.07) .. controls (297.25,104.94) and (297.96,105.65) .. (298.82,105.65) .. controls (299.69,105.65) and (300.4,104.94) .. (300.4,104.07) -- cycle ;

\draw (305.4,93.9) node [anchor=north west][inner sep=0.75pt]  [font=\scriptsize]  {$x$};
\draw (295.9,148.4) node [anchor=north west][inner sep=0.75pt]  [font=\scriptsize,color={Cerulean}  ,opacity=1 ]  {$\mathcal{O}_{x}$};

\end{tikzpicture}
\end{center}

\noindent Therefore, the orbits of two different points are the same if and only if the points are the same distance from the origin and different orbits of this group action do not overlap. The orbits of two different points are the same if and only if the points are the same distance from the origin. One can also notice that the stabiliser of $x$ is given by $S_{x} = \{1_G\}$, in this case $1_G$ is the $2\times 2$ identity matrix. In the other hand, the stabiliser of the origin $(0,0)$ is the entire group $S_{(0,0)} = \text{SO}(2)$.$\,\textcolor{Cerulean}{{\blacktriangleleft}}$ 
\begin{definazul}
    A (right, say) Lie group action  $ \M \cross G \to \M$ of $G$ on a manifold $\M$, is called \textbf{free} if $x \cdot g = x \cdot h$ for some $x \in \M$ implies $g = h$. If for every pair of elements $x, y$ there is a group element $g$ such that $x \cdot g = y$ the group action is called \textbf{transitive}.
\end{definazul}

\noindent Transitivity means that from any starting point on the set one gets anywhere, since orbits of the group are precisely sets of points that is the image of the group action. Transitivity is equivalent as stating that there is just one orbit which is all of $\M$. With respect to the stabiliser, the action is free if for all $x \in \M$,  $S_{x} = \{1_G\}$ since $x \cdot g = x \cdot 1_G$ would imply $ g = 1_G$.

\subsection*{PRINCIPAL G-BUNDLES}

We are in a position now to define principal bundles.

\begin{definazul}
    Let $\M$ be a manifold and $G$ a Lie group. A \textbf{principal $G$-bundle} over $\M$ consists of a manifold $P$ together with:

\begin{itemize}
    \item[\textcolor{Cerulean}{\textbf{\textit{1.}}}] A right free and transitive action of $G$ on $P$, $P \cross G \to P$, $(x,g) \mapsto xg$.
    \item[\textcolor{Cerulean}{\textbf{\textit{2.}}}] A surjective map $\pi_P: P \to \M$ which is $G$-invariant, that is, $\pi_P(xg) = \pi_P(x)$ for all $x \in \M$ and $g \in G$.
    \item[\textcolor{Cerulean}{\textbf{\textit{3.}}}] The local triviality condition: for each $x \in \M$, there exists an open neighbourhood $U$ of $x$ and a diffeomorphism

    \begin{equation}
        \Phi: \Phi_U := \pi_P^{-1}(U) \to U \cross G
    \end{equation}
    \noindent which maps each fibre $\pi_P^{-1}(x)$ to the fibre $\{x\} \cross G$ and which is $G$-equivariant; here the right action of $G$ on $U \cross G$ is the one on the last factor: $(x,a)g = (x,ag)$. That is, the following diagram

              \[
\begin{tikzcd}
    \pi_P^{-1}(U) \arrow{rr}{\Phi_U} \arrow[swap]{dr}{\pi_P} && U \cross G \arrow{dl}{\text{pr}_1} \\[10pt]
    & U
\end{tikzcd}
\]
\noindent commutes and $\Phi_U (xg) = \Phi (x)g$.
\end{itemize}

\end{definazul}

We call $\M$ the base space and $P$ the total space of the bundle. The Lie group $G$ is called the structure group of the bundle. As for vector bundles,  the local triviality condition says that any point in $\M$ has a neighbourhood $U$ such that $P|_U$ is isomorphic to the trivial principal $G$-bundle.

\begin{rem}
    The trivial principal $G$-bundle over a manifold $\M$ is $\M \cross G$, endowed with the second projection and the action of $G$ given by $(x,a)g = (x,ag)$.
\end{rem}
\begin{rem}
    Given two principal $G$-bundles $\pi_i : P_i \to \M$, $i \in \{1,2 \}$, a morphism between them is a smooth map $F: P_1 \to P_2$ which commutes with the projections, $\pi_2 \circ F = \pi_1 $
    and which is $G$-equivariant, $F(pg) = F(p)g$ for all $p \in P_1$ and $g \in G$. Since the group action is free and transitive and $F$ is $G$-equivariant, any morphism of principal $G$-bundles over the same base space is an isomorphism.
\end{rem}

\begin{rem}
    The principal $G$-bundle has fibre $P_x := \pi^{-1}(x)$ above any point $x \in \M$. Now, the fibre is no longer a vector space but a $G$-manifold with respect the (right) action of $G$.
\end{rem}

Similarly, the gluing maps are also defined in the context of the principal bundles:

\begin{equation}
    \Phi_{\beta} \circ \Phi^{-1}_{\alpha} : U_{\a \b} \cross G \to U_{\a \b} \cross G
\end{equation}

\noindent which are given by

\begin{equation}
    \Phi_{\beta} \circ \Phi^{-1}_{\alpha}(x,h) = (x, g_{\a \b}(x)h).
\end{equation}

\noindent The gluing maps  $ g_{\a \b}: U_{\a \b} \to G$ satisfy

\begin{flalign}\label{eq_e54948543798t4rg651}
    \begin{aligned}
        &\textcolor{Cerulean}{\textbf{(i)}}\;\;\;  g_{\a \a} (x) = 1_G, \;\; \text{ for all } x \in U_\a;\\
     &\textcolor{Cerulean}{\textbf{(ii)}}\;\; g_{\a \b}(x) = g^{-1}_{\b \a}, \;\; \text{ for all } x \in U_{\a} \inter U_{\b};\\
      &\textcolor{Cerulean}{\textbf{(iii)}}\;g_{\b \delta}(x)g_{\a \b}(x) =   g_{\a \delta}(x)  \;\; \text{ for all } x \in U_{\a} \inter U_{\b} \inter U_{\delta} \textit{\;\;(cocycle condition)}.
    \end{aligned}&&
\end{flalign}

\noindent The gluing maps together with an open cover $\mathcal{U}$ for $\M$ give rises to a principal $G$-bundle $P$ over $\M$ with trivialising cover $\mathcal{U}$ and gluing maps $g_{\a \b}: U_{\a \b} \to G$ which is unique up to isomorphism.

The next definition is analogous of the equivariance Definition \ref{defiequivi} but in the context of principal bundles. It gives a generelisation  of the notion of G-principal fibre bundle morphism.

\begin{definazul}\label{def_fred}
    Let $(P, \pi_P, G)$ be a principal $G$-bundle over $\M$ and $f: H \to G$ a Lie group homomorphism. An \textbf{$f$-reduction of $P$} is a pair $(Q, F)$ consisting of an  principal $H$-bundle $(Q, \pi_Q, H)$ over $\M$ and a smooth map $F: Q \to P$ such that

\begin{itemize}
   \item[\textcolor{Cerulean}{\textbf{\textit{1.}}}] $\pi_P \circ F = \pi_Q$ and
    \item[\textcolor{Cerulean}{\textbf{\textit{2.}}}] $F(q\, \textcolor{cyan}{\bullet}\, h) = F(q)\, \textcolor{ProcessBlue}{\circ}\, f(h)$ for all $q \in Q, \, h \in H$. That is, the following diagram
   
\end{itemize}
    \[ \begin{tikzcd}
Q \cross H \arrow[r, "\textcolor{cyan}{\bullet}"] \arrow[d, swap, "F \cross f"] & Q \arrow{d}{F}  \arrow[dr, "\pi_Q"]  \\
P \cross G \arrow[r, "\textcolor{ProcessBlue}{\circ}"]          & P \arrow[r, "\pi_P"]                                         & \M
\end{tikzcd} 
    \]
\noindent is commutative.\end{definazul}

\noindent The bullets, $\textcolor{cyan}{\bullet }: Q \cross H \to Q$ and $\textcolor{ProcessBlue}{\circ }: P \cross G \to P$, represents the right action of the Lie groups. In particular, if $H \subset G$ is a Lie subgroup and $f = \iota : H \xhookrightarrow{} G$ is the inclusion, then the $f$-reduction is also called an $H$-reduction of $P$ or a reduction of $P$ to $H$.    The Definition \ref{def_fred} plays a crucial role in the context of spin structure which is basically an reduction of the bundle of positively oriented orthonormal frames from $\SO(n)$ to the spin group $\Spin(n)$ which maps to $\SO(n)$ as a double cover.

\subsection*{VECTOR BUNDLES AND PRINCIPAL BUNDLES}

\paragraph{ } We finish the chapter by showing that vector bundles and principal bundles can be seen as a two side of the same coin even though they are different objects. For any real $n$-dimensional vector bundle, there is a natural action of $\text{GL}(n,\rr)$ on the fibres since they are vector spaces. However, this action is not necessarily transitive, free or even fibre preserving. Still, we can go back and forth between vector and principal bundles via two natural constructions:

\[
\begin{tikzcd}[column sep=4em]
    \text{Vector Bundles \;\;\;} \arrow[Cerulean, r, shift left=1ex, "bundle\ of\ frames", -stealth', shorten >=-5pt, shorten <=-5pt] & \text{\;\;\;Principal Bundles} \arrow[Cerulean, l, shift left=1ex, "associated\ bundle", -stealth, shorten >=-5pt, shorten <=-5pt] \\
\end{tikzcd}\]

The \textbf{frame bundle} $\pi: \text{Fr}(E) \to \M$ is an object related to a vector bundle $E$ of rank $n$. The total space $\text{Fr}(E)$ of the frame bundle is the set of all ordered bases (frames) in the fibres of the vector bundle $E \to \M$, with a suitable topology and manifold structure \cite{Tu2}.

\begin{equation}
    \text{Fr}(E) = \{(x,u) : x \in \M, u - \text{a frame of } E_x \} = \bigsqcup_{x \in \M} \text{Fr}(E_x) 
\end{equation}

\noindent The general linear group acts naturally on the frame bundle via a change of basis, giving the frame bundle the structure of a principal bundle. There is a natural projection map $\pi: \text{Fr}(E) \to \M$ that maps $\text{Fr}(E_x)$ to $\{x\}$. A local trivialisation $\phi_U : E|_U := \pi^{-1}(U) \to U \cross \rr^n$ induces a bijection

\begin{equation}
    \phi ' : \text{Fr}(E)|_U \to U \cross \text{Fr}(\rr^{n})
\end{equation}

\noindent such that for each $x \in U$ one considers the linear isomorphisms $\phi_{U,x} : E_x \to \{x\} \cross \rr^{n}$:

\begin{equation}
    [v_1 \cdots v_n] \in \text{Fr}(E_x) \mapsto (x, [\phi_{U,x}(v_1) \cdots \phi_{U,x}(v_n) ]).
\end{equation}

\noindent The topology and manifold structure of $\text{Fr}(E)|_U$ that make $\pi: \text{Fr}(E) \to \M$ locally trivial with fibre $\text{Fr}(\rr^{n})$ is given from $U \cross \text{Fr}(\rr^{n})$ by $ \phi ' .$ The frame bundle  $\text{Fr}(\rr^{n})$ has a natural right action by $\text{GL}(n, \rr)$. Indeed, in each point a frame $v = (v_1, \cdots, v_n)$ can be viewed as a linear isomorphism $v : \rr^{n} \to E_x$ that sends the component $e_{i}$ from the standard basis of $\rr^{n}$, to $v_{i}$ for each $i$. Hence a group element $g \in \text{GL}(n, \rr) $ acts on the frame $v$ via composition to give a new frame 

\begin{equation}\label{eq_347dvnjacvb}
    \text{Fr}(\rr^{n}) \cross \text{GL}(n, \rr) \to  \text{Fr}(\rr^{n}), \;\; (v,g) \mapsto  v \circ g : \rr^{n} \to E_x.
\end{equation}

\noindent The frame bundle is then a principal $\text{GL}(n, \rr)$-bundle. It is worth to emphasise that the group structure can be changed by restricting the group action of $\text{GL}(r, \rr)$ to $\O(p,q)$ or $\SO(p,q)$ (or $\text{U}(n);\; \text{SU}(n)$ in the complex case). Conversely, now let $\pi: P \to \M$ be a principal $G$-bundle and $\rho : G \to \text{GL}(V)$ a representation of $G$ on a finite-dimensional vector space $V$. We write $\rho(g)v$ as $g \cdot v$ or even $gv$. The \textbf{associated bundle} $ P \cross_\rho V$ is defined as the set of equivalent classes of pairs $[x, v] \in  P \cross V$ under the equivalence relation \cite{Mastermath, Tu2}

\begin{equation}
    [x, v] \sim [xg,g^{-1}v]
\end{equation}

\noindent for $g \in G$. The bundle $P \cross_\rho V$ is endowed with the natural projection $\pi': P \cross_\rho V \to \M$ such that

\begin{equation}\label{eq_assproj}
   \pi'([x,v]) =   \pi'([xg,g^{-1}v]) =   \pi(xg) = \pi(x)
\end{equation}

\noindent If one considers any manifold $U$, there exists a fibre-preserving diffeomorphism $\phi: (U \cross G) \cross_{\rho} V \to U \cross V$ such that $\phi([(x,g),v]) = (x, gv)$ \cite{Tu2}. Therefore, since the principal bundle $\pi: P \to \M$ is locally $U \cross G$, the associated bundle $\pi': P \cross_{\rho} V \to \M$ is locally trivial with fibre $V$. Then, the vector space structure on $V$,

\begin{align}
    \begin{aligned}
        [x,v_1] + [x, v_2] &= [x, v_1 + v_2]\\
        \lambda[x,v] &= [x, \lambda v], \;\; \lambda \in \rr,
    \end{aligned}
\end{align}

\noindent makes $P \cross_{\rho} V$ into a vector bundle over $\M$.

\colorlet{chapter}{blue!50}
\chapter{The Kähler-Atiyah Bundle}\label{chap3}

\hypersetup{
  colorlinks = true,
  linkcolor  = RoyalBlue,
  citecolor = Blue,
}

\paragraph{ } We have now arrived at the central chapter where we delve into the \textcolor[rgb]{0.82,0.01,0.11}{Clifford algebras} on \textcolor[rgb]{0.41,0.67,0.97}{bundles} from the formalism of the \textcolor[rgb]{0.08,0.18,0.88}{Kähler-Atiyah bundle}. This bundle encapsulates the essence of geometric algebra, offering a profound insight into the algebraic structures underlying a geometric framework. The main reference of this discussion are the Refs. \cite{bab2, Lopes:2017exe}. By exploring the Kähler-Atiyah bundle, we aim to uncover its intricate properties. The non-commutative multiplication on this bundle induces an non-commutative associative  multiplication on inhomogeneous differential forms, called geometric product denoted by $\diamond$ in which we deeply explore. We also define the volume form and delve into its properties and computations. We explore three important subalgebras and we finish by showing that they are isomorphic. It is worth to emphasise the significance of the Kähler-Atiyah bundle in bridging the gap between geometric and algebraic concepts, paving the way for a more comprehensive understanding of \textcolor[rgb]{0.74,0.06,0.88}{\text{Spin Geometry}} and \textcolor[rgb]{0.88,0.82,0.11}{spinors} within a geometric framework. The Kähler-Atiyah Bundle plays a fundamental role in the essence of the bundle of spinors. The typical geometric algebraic approach to spinors is viewing a bundle $\mathcal{S}$ of spinors over a pseudo-Riemannian manifold $(\M, g)$ as a bundle of modules over the real Kähler-Atiyah bundle with module structure specified by a particular morphism of bundles of algebras. The Kähler-Atiyah Bundle is a Clifford bundle and we start by defining Clifford bundles in general.

\begin{definanil}
    Let $E \to \mathcal{M}$ be a smooth vector bundle and $g$ a smooth symmetric bilinear form on $E$. The \textbf{Clifford bundle} of $E$ is the algebra bundle $\cl(E) \to \mathcal{M}$ whose fibres are Clifford algebra over the fibres of $E$,

    \begin{equation}
        \cl(E) = \bigsqcup_{x \in \mathcal{M}} \cl(E_x, g_x),
    \end{equation}

\noindent such that $E_x$ is a fibre and $g_x = g|_{E_x}$.
\end{definanil}

$E$ is the bundle of spaces with $g$ as inner product and $\cl(E)$ is just the associated bundle of Clifford algebras.
The Clifford bundle over $\M$ equals to the exterior bundle over $\M$ as a vector bundle. However, it is endowed with the Clifford product on its fibres, thereby transforming it into a bundle of associative algebras over $\M$. In fact, $\cl(E)$ could be defined as the quotient bundle \cite{clifbundle}

\begin{equation}
    \cl(E) = \mathscr{T}(E)/\mathscr{I}(E)
\end{equation}

\noindent where $\mathscr{T}(E)$ is the bundle of tensor algebras of $E$ and $\mathscr{I}(E)$ is the bundle of ideals generated by the elements $\boldsymbol{\uptau} \tensor \boldsymbol{\uptau} - g(\boldsymbol{\uptau},\boldsymbol{\uptau})1_{T} $ for $\boldsymbol{\uptau} \in E$. 

Defining the Clifford bundle from the cotangent bundle allows us to establish a connection between geometric algebra and differential forms. This connection is significant because many modern theories can be described using differential forms and geometric algebra provides a powerful language and efficient techniques to approaches. 

Let $(\mathcal{M},g)$ be a pseudo-Riemannian manifold of dimension $n$. We endow the cotangent bundle $T^{*}M$ with the metric $g^{*}$ induced by $g$. The Clifford bundle of differential forms of the pair $(\mathcal{M},g)$ is

  \begin{equation}
        \cl(T^{*}\M) = \bigsqcup_{x \in \mathcal{\M}} \cl(T^{*}_{x}\M, g^{*}_x).
    \end{equation}

\noindent At $x\in\M$, the Clifford algebras on the cotangent bundle  $\mathcal{C}\ell(T^{*}_{x}\M, g^{*}_x)$ are also denoted by $\mathcal{C}\ell_{p,q}$, with $(p,q)$ being the signature of $g$. The Clifford bundle $\mathcal{C}\ell(T^{*}\M)$ is identified with the exterior bundle $\bigwedge T^{*}\M$ endowed with the geometric product $\diamond: \bigwedge T^{*}\M \times \bigwedge T^{*}\M \to \bigwedge T^{*}\M$ whose induced action on sections $\Gamma(\M,\bigwedge T^{*}\M)$, which is again
denote by $\diamond$ for simplicity, satisfies the following relations for every $1$-form $\uptheta \in \Omega^{1}(\M)$ and $k$-form $\upomega \in \Omega^{k}(\M)$
\begin{equation}\label{Eq_cliffproduct}
      \uptheta \diamond \upomega = \uptheta \wedge \upomega + \sharp(\uptheta) \rfloor  \upomega,\hspace{0.5cm}         \upomega \diamond \uptheta = (-1)^{k} ( \uptheta \wedge \upomega - \sharp(\uptheta) \rfloor  \upomega),
\end{equation}
\noindent where $\rfloor$ is the left contraction (Def. \ref{def_fd65g415sad}) such that $g(\upomega_1\rfloor\upomega_2,\upomega_3)=g(\upomega_1,\upomega_2\wedge\upomega_3)$, for all $\upomega_1, \upomega_2, \upomega_3\in \Omega(\M)$ and $\sharp(\uptheta)$ is the musical isomorphism (Def. \ref{defin_w23485rte}) $\sharp : \Gamma (\M, T^{*}\M) \to \Gamma (\M, T\M),$ with inverse $ \sharp^{-1}\equiv \flat: \Gamma (\M, T\M) \to \Gamma (\M, T^{*}\M)$, induced by the metric $g$,  raising and lowering  indexes, as $\sharp(\uptheta) = \sharp(\uptheta_{i}e^{i}) = g^{ij}\uptheta^{j}\mathbf{e}_{j}.$ One can notice that the Eq.\eqref{Eq_cliffproduct} is precisely the Clifford product settled in the Eq. \eqref{eq_345esaghnb} but in the context of differential forms.

A section of the cotangent bundle represented by $e^j$ is compressed proportionally to 1 by a section $e_i$ of the tangent bundle in the following manner:

\begin{equation}
    \mathbf{e}_{i} \rfloor e^{j} = e^{j} (\mathbf{e}_{i}) = \delta^{i}_{j} 1 = \color{RoyalBlue}\begin{cases} \color{black} 1, \;\;\; \text{if} \; i = j \color{black}\\\color{black}
   \color{black} 0, \;\;\; \text{if} \; i\color{black} \neq j \end{cases}\color{black}.
\end{equation}

{\colorlet{shadecolor}{RoyalBlue!20}\begin{shaded}The bundle of algebras $(\bigwedge T^{*}\M, \diamond)$ is called the \textbf{Kähler--Atiyah bundle} of $(\M,g) $.\end{shaded}}

{\colorlet{shadecolor}{RoyalBlue!20}\begin{shaded} The space $\Omega(\M)$ of all inhomogeneous smooth forms on $\M$, endowed with the geometric product $\diamond$, is an associative algebra with unity over the ring $\Cinf(\M,\rr)$, known as \textbf{Kähler--Atiyah algebra} of $(\M,g)$.  It  satisfies the isomorphisms  

\begin{equation}
    (\Omega(\M), \diamond) \simeq \Gamma(\M, \mathcal{C}\ell(T^{*}\M)) \simeq \Gamma(\M,\bigwedge T^{*}\M).
\end{equation}
 \end{shaded}}

The Kähler--Atiyah algebra is also called Graf--Clifford algebra \cite{Lopes:2017exe}. Each of the  intrinsic notions to Clifford algebras defined in the Chapter \ref{chap1} carries over to Clifford bundles. For instance, following from the Def. \ref{subsp_defq5asw}, the even-odd decomposition of the $\mathbb{Z}_{2}$-graded algebra,
\begin{equation}
    \bigwedge T^{*}\M = \bigwedge T^{*}\M^{\textsc{even}} \oplus \bigwedge T^{*}\M^{\textsc{odd}},
\end{equation}
\noindent as well as the grade involution, reversion, and conjugation operators presented in the Defs. \ref{def15op}, \ref{def16op} and \ref{def17op}, respectively given for $\upalpha \in \Omega^{k}({\M})$ by
\begin{equation}\label{involution}
    \yhwidehat{\upalpha} = (-1)^{k} \upalpha, \hspace{0.5cm}\qquad \yhwidetilde{\upalpha} = (-1)^{\frac{k(k-1)}{2}} \upalpha, \hspace{0.5cm} \qquad \overline{\upalpha} = \yhwidehat{\yhwidetilde{\upalpha}} =  (-1)^{\frac{k(k+1)}{2}}\upalpha.
\end{equation}

We recall that the Clifford structures is constructed under the sections of exterior bundle regarding a local coframe, that is, all considered sections are local. The geometric product between forms of arbitrary degree is constructed by repeated application of Eq. \eqref{Eq_cliffproduct}.  To express this product concisely, Chevalley and Riesz \cite{Chevalley, Riesz} introduced the contracted wedge product.

\begin{definanil}\label{defincontractedwe}
    The \textbf{contracted wedge product of order $d$} between two arbitrary forms $\upalpha, \upbeta \in \Omega(\M)$ is defined inductively by

\begin{align}
    \begin{aligned}
          \upalpha \wedge_{0} \upbeta &= \upalpha \wedge \upbeta,\\
     \upalpha \wedge_{1} \upbeta &= \sum_{i_{1}, j_{1} = 1}^{n} g^{i_{1}j_{1}} (\mathbf{e}_{i_{1}}\rfloor \upalpha) \wedge_{0} (\mathbf{e}_{j_{1}} \rfloor \upbeta),\\
     &\vdots\\
\upalpha \wedge_{d} \upbeta &= \sum_{i_{d}, j_{d} = 1}^{n} g^{i_{d}j_{d}} (\mathbf{e}_{i_{d}}\rfloor \upalpha) \wedge_{d-1} (\mathbf{e}_{j_{d}} \rfloor \upbeta).
    \end{aligned}
\end{align}
\end{definanil}

Given a $k$-form $\upalpha$ and a $l$-form $\upbeta$, the contracted product of order $1$ is lies in the space  $\Gamma(U,\bigwedge^{k -1 + l -1} T^{*}\M)$ and is given by

\begin{equation}
     \upalpha \wedge_{1} \upbeta = \sum_{i_{1}, j_{1} = 1}^{n} g^{i_{1}j_{1}} (\mathbf{e}_{i_{1}}\rfloor \upalpha) \wedge (\mathbf{e}_{j_{1}} \rfloor \upbeta).
\end{equation}

\noindent The contracted product of order $2$ is given as follows:

\begin{align}
    \begin{aligned}
        \upalpha \wedge_{2} \upbeta &= \sum_{r, s = 1}^{n} g^{kl} (\mathbf{e}_{k}\rfloor \upalpha) \wedge_{1} (\mathbf{e}_{l} \rfloor \upbeta) \\
        &= \sum_{r, s = 1}^{n} g^{kl} \left (\sum_{i, j = 1}^{n} (\mathbf{e}_{i} \rfloor  \mathbf{e}_{r}\rfloor \upalpha) \wedge_{1} (\mathbf{e}_{j} \rfloor \mathbf{e}_{s} \rfloor \upbeta) \right),
    \end{aligned}
\end{align}

\noindent which lies in the space $\Gamma(U,\bigwedge^{k-2+l-2} \Tmd)$. With respect to an orthonormal coframe, the contracted wedge product of order $l$ is

\begin{equation}
    \upalpha \wedge_l \upbeta = \sum^{n}_{i=1}  g^{ii} (\mathbf{e}_{i} \rfloor \upalpha) \wedge_{l-1} (\mathbf{e}_{i}\rfloor \upbeta).
\end{equation}

\noindent In all future considerations, every coframe will be orthonormal.

\section{The Geometric Product}
\paragraph{ } In this section we deeply explore the geometric product of the Kähler--Atiyah algebra. We are now in a position to present the geometric product between two arbitrary forms \cite{Lopes:2017exe}

\begin{definanil}\label{defin_graf1}
    \noindent For a $k$-form $\upalpha \in \Omega^{k}(\M)$ and a $l$-form $\upbeta \in \Omega^{l}(\M)$ with $k\leq l$, the geometric product $\diamond$ between $\upalpha$ and $\upbeta$ is defined as follows 
\begin{align}
    \begin{aligned}
          \upalpha \diamond \upbeta &= \sum^{k}_{d=0} \frac{(-1)^{d(k-d) + \left [  \frac{d}{2} \right]}}{d!}  \upalpha \wedge_d \upbeta,\\
           \upbeta \diamond \upalpha &= (-1)^{kl} \sum^{k}_{d=0} \frac{(-1)^{d(k-d+1) + \left [  \frac{d}{2} \right]}}{d!}  \upalpha \wedge_d \upbeta,
    \end{aligned}
\end{align}
\noindent where  $\left [  \frac{d}{2} \right]$ represents the integer part of $\frac{d}{2}$ .
\end{definanil}

\begin{tftheorem}
    $(\Omega(\M), \diamond)$ is a Clifford algebra.
\end{tftheorem}
\noindent \textcolor{RoyalBlue}{\textit{Proof.}} The square of any basis element $e^{a}$ in an orthonormal coframe $\{e^{1},\ldots, e^{n}\}$ with respect to the geometric product is such that

\begin{align}
    \begin{aligned}
        (e^{a})^{2} = e^{a} \diamond e^{a} 
        &= \sum^{1}_{d=0} \frac{(-1)^{d(1-d) + \left [  \frac{d}{2} \right]}}{d!}  e^{a}\wedge_d e^{a}\\
        &= e^{a} \wedge_0 e^{a} + e^{a} \wedge_1 e^{a} \\
        &= e^{a} \wedge e^{a} + \sum_{i = 1}^{n} g^{ii} (\mathbf{e}_{i}\rfloor e^a) \wedge (\mathbf{e}_{i} \rfloor e^a)\\
        &= \sum_{i = 1}^{n} g^{ii}\delta_{i}^{a}\delta_{i}^{a}1\\
        &= g^{aa}1 = g^{*}(e^{a},e^{a})1. 
    \end{aligned}
\end{align}

\noindent By Eq. \eqref{eq_clifrelideal} we conclude that $(\Omega(\M), \diamond)$ is a Clifford algebra. \textcolor{RoyalBlue}{$\Box$}

\begin{ex}\label{extruncated}
{{\textcolor{Blue}{{$\blacktriangleright$\;}}}}  The geometric product on a 6-dimensional manifold.\normalfont{ Let us consider $\upalpha = e^{1} + e^{36},\, \upbeta = e^{1} + e^{2} + e^{14} + e^{25} \in \Omega(\M)$ on a $6$-dimensional manifold with signature $(4,2)$ and an orthonormal coframe $\{e^{1},e^{2}, e^{3}, e^{4}, e^{5}, e^{6} \}$.} Let us compute the product $\upalpha \diamond \upbeta$\end{ex}
\begin{align}
    \begin{aligned}
        \upalpha \diamond \upbeta &= (e^{1} + e^{36})(e^{1} + e^{2} + e^{14} + e^{25})\\
        &= e^{1} \diamond e^{1} + e^{1} \diamond e^{2} +e^{1} \diamond e^{14} + e^{1} \diamond e^{25} + e^{36} \diamond e^{1} +e^{36} \diamond e^{2} + e^{36} \diamond e^{14} + e^{36} \diamond e^{25}.
    \end{aligned}
\end{align}
\noindent We now compute each one of the eight diamond products above
\begin{flalign}
\begin{aligned}
\textit{\textcolor{RoyalBlue}{\textbf{\textit{1)}}}\;\;}e^1 \diamond e^1 &= \sum^{1}_{d=0} \frac{(-1)^{d(1-d) + \left [ \frac{d}{2} \right]}}{d!} e^1 \wedge_d e^1 \\
&= \left( \frac{(-1)^{0(1-0) + \left [ \frac{0}{2} \right]}}{0!} e^1 \wedge_0 e^1 \right) + \left( \frac{(-1)^{1(1-1) + \left [ \frac{1}{2} \right]}}{1!} e^1 \wedge_1 e^1 \right)\\
&= e^{1}\wedge_{0} e^{1} + e^{1} \wedge_{1} e^{1}\\
&= 0 + \sum_{i= 1}^{6} g^{ii} (\mathbf{e}_{i}\rfloor e^1) \wedge_{0} (\mathbf{e}_i \rfloor e^1) \\
&= g^{11} (\mathbf{e}_{1}\rfloor e^1) \wedge_{0} (\mathbf{e}_{1} \rfloor e^1) \\
&= g^{11}1\\
&= 1.
\end{aligned} &&
\end{flalign}

\begin{flalign}
\begin{aligned}
\textit{\textcolor{RoyalBlue}{\textbf{\textit{2)}}}\;\;}e^{1} \diamond e^{2} &= e^{1} \wedge_0 e^{2} + e^{1} \wedge_{1} e^{2}\\
&= e^{1}\wedge e^{2} + \sum_{i=1}^{6} g^{ii} (\mathbf{e}_{i}\rfloor e^1) \wedge_{0} (\mathbf{e}_{i} \rfloor e^2) \\
&= e^{1}\wedge e^{2} + \sum_{i=1}^{6} g^{ii} \delta_{i}^{1}\delta_{i}^{2} \\
&= e^{1}\wedge e^{2}.
\end{aligned} &&
\end{flalign}

\begin{flalign}
\begin{aligned}
\textit{\textcolor{RoyalBlue}{\textbf{\textit{3)}}}\;\;}e^{1} \diamond (e^{1} \wedge e^{4}) &= e^{1} \wedge (e^{1} \wedge e^{4}) + e^{1}\wedge_{1} (e^{1} \wedge e^{4})  \\
&= \sum_{i=1}^{6} g^{ii} (\mathbf{e}_{i}\rfloor e^1) \wedge_{0} (\mathbf{e}_{i} \rfloor (e^{1} \wedge e^{4})) \\
&=  \sum_{i=1}^{6} g^{ii} (\mathbf{e}_{i}\rfloor e^1) \wedge ((\mathbf{e}_{i} \rfloor e^{1})e^{4} - e^{1}(\mathbf{e}_{i}\rfloor e^{4}))\\
&= \sum_{i=1}^{6} g^{ii} \delta_{i}^{1} \wedge (\delta_{i}^{1} e^{4} - e^{1}\delta_{i}^{4})\\
&= \sum_{i=1}^{6} g^{ii} \delta_{i}^{1}\delta_{i}^{1} e^{4} - e^{1}\delta_{i}^{1}\delta_{i}^{4}\\
&= g^{11}e^{4} = e^{4}.
\end{aligned}  &&
\end{flalign}

\begin{flalign}
\begin{aligned}
\textit{\textcolor{RoyalBlue}{\textbf{\textit{4)}}}\;\;}e^{1} \diamond (e^{2} \wedge e^{5}) &= e^{1} \wedge_0 (e^{2} \wedge e^{5}) + e^{1}\wedge_{1} (e^{2} \wedge e^{5})  \\
&= e^{1} \wedge e^{2} \wedge e^{5} + \sum_{i=1}^{6} g^{ii} (\mathbf{e}_{i}\rfloor e^1) \wedge_{0} (\mathbf{e}_{i} \rfloor (e^{2} \wedge e^{5})) \\
&=  e^{1} \wedge e^{2} \wedge e^{5} +\sum_{i=1}^{6} g^{ii} (\mathbf{e}_{i}\rfloor e^1) \wedge ((\mathbf{e}_{i} \rfloor e^{2})e^{5} - e^{2}(\mathbf{e}_{i}\rfloor e^{5}))\\
&= e^{1} \wedge e^{2} \wedge e^{5} +\sum_{i=1}^{6} g^{ii} \delta_{i}^{1} \wedge (\delta_{i}^{2} e^{5} - e^{2}\delta_{i}^{5})\\
&=e^{1} \wedge e^{2} \wedge e^{5} + \sum_{i=1}^{6} g^{ii} \delta_{i}^{1}\delta_{i}^{2} e^{5} - e^{2}\delta_{i}^{1}\delta_{i}^{5}\\
&= e^{1} \wedge e^{2} \wedge e^{5} + 0.
\end{aligned} &&
\end{flalign}

\begin{flalign}
    \begin{aligned}
        \textit{\textcolor{RoyalBlue}{\textbf{\textit{5)}}}\;\;}e^{36} \diamond e^{1} &= (-1)^{1\cdot 2} \sum^{1}_{d=0} \frac{(-1)^{d(1-d+1) + \left [ \frac{d}{2} \right]}}{d!} e^{1} \wedge_d e^{36}\\
        &=  e^{1} \wedge_0 e^{36} - e^{1} \wedge_1 e^{36}  \\
        &= e^{1} \wedge e^{3}\wedge e^{6} - \sum_{i=1}^{6} g^{ii} \delta_{i}^{1}\delta_{i}^{3} e^{6} - e^{3}\delta_{i}^{1}\delta_{i}^{6}\\
        &=  e^{1} \wedge e^{3}\wedge e^{6}.
    \end{aligned} &&
\end{flalign}

\begin{flalign}
    \begin{aligned}
       \textit{\textcolor{RoyalBlue}{\textbf{\textit{6)}}}\;\;} e^{36} \diamond e^{2} =  e^{2} \wedge e^{3}\wedge e^{6} .
    \end{aligned}  &&
\end{flalign}

\begin{flalign}
    \begin{aligned}
       \textit{\textcolor{RoyalBlue}{\textbf{\textit{7)}}}\;\;} e^{36} \diamond e^{14} &= \sum^{2}_{d=0} \frac{(-1)^{d(2-d) + \left [  \frac{d}{2} \right]}}{d!}   e^{36} \wedge_d e^{14}\\
        &= e^{36} \wedge_0 e^{14} - e^{36} \wedge_{1} e^{14} - \frac{1}{2} e^{36} \wedge_2 e^{14}  \\
        &=  e^{3} \wedge e^{6} \wedge e^{1} \wedge e^{4} = - e^{1} \wedge e^{3} \wedge e^{4} \wedge e^{6}
        \end{aligned} &&
\end{flalign}  
\noindent Such that
\begin{flalign}
    \begin{aligned}
         e^{36} \wedge_{1} e^{14} &= \sum^{6}_{i=1}g^{ii} (\mathbf{e}_{i} \rfloor e^{3} \wedge e^{6}) \wedge_0 (\mathbf{e}_{i} \rfloor e^{1} \wedge e^{4})\\
         &= \sum^{6}_{i=1}g^{ii} (\delta_{i}^{3} e^{6} - e^{3}\delta^{6}_{i}) \wedge (\delta_{i}^{1}e^{4} - e^{1}\delta_{i}^{4})\\
         &= \sum^{6}_{i=1}g^{ii} (\delta_{i}^{3}\delta_{i}^{1} e^{6} \wedge e^{4} - \delta_{i}^{3}\delta_{i}^{4} e^{6} \wedge e^{1} - \delta_{i}^{6}\delta_{i}^{1}e^{3} \wedge e^{4} + \delta_{i}^{6}\delta_{i}^{4}e^{3} \wedge e^{1} )
         = 0.
        \end{aligned} &&
\end{flalign}

\begin{flalign}
    \begin{aligned}
        e^{36} \wedge_{2} e^{14} &= \sum^{6}_{i=1}g^{ii} (\mathbf{e}_{i} \rfloor e^{3} \wedge e^{6}) \wedge_1 (\mathbf{e}_{i} \rfloor e^{1} \wedge e^{4})\\
         &= \sum^{6}_{i=1}g^{ii} (\delta_{i}^{3} e^{6} - e^{3}\delta^{6}_{i}) \wedge_1 (\delta_{i}^{1}e^{4} - e^{1}\delta_{i}^{4})\\
         &= \sum^{6}_{i=1}g^{ii} 
         \left( \sum^{6}_{j=1}g^{jj}  (\mathbf{e}_j \rfloor (\delta_{i}^{3} e^{6} - e^{3}\delta^{6}_{i})) \wedge_0 (\mathbf{e}_j \rfloor (\delta_{i}^{1}e^{4} - e^{1}\delta_{i}^{4}))  \right)\\
         &= \sum^{6}_{i=1}g^{ii} 
         \left( \sum^{6}_{j=1}g^{jj}  (\delta_{i}^{3} \delta_{j}^{6} - \delta_{j}^{3}\delta^{6}_{i}) \wedge (\delta_{i}^{1}\delta_{j}^{4} - \delta_{j}^{1}\delta_{i}^{4})\right) \\
         &= \sum^{6}_{i=1}g^{ii} 
         \left( \sum^{6}_{j=1}g^{jj}  (\delta_{i}^{3} \delta_{j}^{6} - \delta_{j}^{3}\delta^{6}_{i}) \wedge (\delta_{i}^{1}\delta_{j}^{4} - \delta_{j}^{1}\delta_{i}^{4})\right) \\
         &= \sum^{6}_{i=1}g^{ii} 
         \left( \sum^{6}_{j=1}g^{jj}  
         (\delta_{i}^{3} \delta_{i}^{1} \delta_{j}^{6}\delta_{j}^{4} - \delta_{i}^{3}\delta_{i}^{4} \delta_{j}^{6}\delta_{j}^{1} - \delta_{j}^{3}\delta_{j}^{4}\delta^{6}_{i}\delta_{i}^{1} + \delta_{j}^{3}\delta_{j}^{1}\delta^{6}_{i}\delta_{i}^{4})\right)\\
         &= \sum^{6}_{i=1}g^{ii} (0) = 0.
        \end{aligned}  &&
\end{flalign}

\begin{flalign}
\begin{aligned}
   \textit{\textcolor{RoyalBlue}{\textbf{\textit{8)}}}\;\;}  e^{36} \diamond e^{25} = e^{3} \wedge e^{6} \wedge e^{2} \wedge e^{5} = - e^{2} \wedge e^{3} \wedge e^{5} \wedge e^{6}.
    \end{aligned}  &&
\end{flalign}

Therefore, the final result is given by: 
\begin{align}
    \begin{aligned}
       \upalpha \diamond \upbeta &= (e^{1} + e^{36})(e^{1} + e^{2} + e^{14} + e^{25})\\&= 1 + e^{4} + e^{12} + e^{125}+ e^{136} + e^{236} - e^{1346} - e^{2356}. \;\textcolor{Blue}{{\blacktriangleleft}}
    \end{aligned} &&
\end{align}

One can notice from Example \ref{extruncated} that the contracted wedge product with order $d>0$ for elements $e^{a},e^{b}$ with $a \neq b$ is such that $e^{a} \wedge_{d} e^{b} = 0$. The next proposition generalises this fact and helps with the computation of the diamond product.

\begin{tfpropos}\label{propos3201r}
    Given a $k$-form $\upalpha^{I}$ and a $l$-form $\upbeta^{J}$ such that $I \cap J = \emptyset$, then $\upalpha \wedge_{d} \upbeta = 0$ for all $d>0$.
\end{tfpropos}
\noindent \textcolor{RoyalBlue}{\textit{Proof.}} Without loss of generality, let us consider an orthonormal coframe  $\mathcal{C} = \{e^{1},\ldots,e^{n}\}$ and $I = \{e^{i_{1}}, \ldots, e^{i_{k}}\},J= \{e^{j_{1}}, \ldots, e^{j_{l}}\} \subset \mathcal{C}$ with $I \cap J = \emptyset$ and  $\upalpha = e^{{i_{1}} \cdots{i_{k}}}$, $\upbeta = e^{{j_{1}} \cdots{j_{l}}}$. For $d=1$ it follows that:

\begin{align}
    \begin{aligned}
        \upalpha \wedge_{1} \upbeta &= \sum^{n}_{a=0} g^{aa} (\mathbf{e}_{a}\rfloor \upalpha) \wedge (\mathbf{e}_{a} \rfloor \upbeta)\\
        &= \sum^{n}_{a=0} g^{aa} (\mathbf{e}_{a}\rfloor e^{{i_{1}} \cdots{i_{k}}}) \wedge (\mathbf{e}_{a} \rfloor e^{{j_{1}} \cdots{j_{l}}})\\
        &=\sum^{n}_{a=0} g^{aa} (
        (\mathbf{e}_{a}\rfloor e^{i_{1}})e^{{i_{2}} \cdots{i_{k}}} - e^{i_{1}} \wedge \!(\mathbf{e}_{a}\rfloor e^{{i_{2}} \cdots{i_{k}}})) \wedge ((\mathbf{e}_{a}\rfloor e^{j_{1}})e^{{j_{2}} \cdots{j_{k}}} - e^{j_{1}} \wedge\! (\mathbf{e}_{a}\rfloor e^{{j_{2}} \cdots{j_{k}}}))\\
        &=\sum^{n}_{a=0} g^{aa} (
        \delta_{a}^{i_{1}}e^{{i_{2}} \cdots{i_{k}}} - e^{i_{1}} \wedge (\mathbf{e}_{a}\rfloor e^{{i_{2}} \cdots{i_{k}}})) \wedge (\delta_{a}^{j_{1}}e^{{j_{2}} \cdots{j_{k}}} - e^{j_{1}} \wedge (\mathbf{e}_{a}\rfloor e^{{j_{2}} \cdots{j_{k}}}))\\
        &=\sum^{n}_{a=0}\! g^{aa}\! (\delta_{a}^{i_{1}}e^{i_{2}\cdots i_{k}} \!+\! \cdots \!+\! (-1)^{i_{k}+1}\delta_{a}^{i_{k}}e^{i_{1}i_{2}\cdots i_{k-1}})\! \wedge\! (\delta_{a}^{j_{1}}e^{j_{2}\cdots j_{k}} \!+\! \cdots \!+\! (-1)^{j_{k}+1}\delta_{a}^{j_{k}}e^{j_{1}j_{2}\cdots j_{l-1}}).\!
    \end{aligned}
\end{align}

\noindent The Eq. \ref{eq_contracttermsap} was used in the above computations. Since $I \cap J = \emptyset$ each one of the products $\delta_{a}^{i_{s}}\delta_{a}^{j_{r}}$ equals zero for any $i_{s} \in I, j_{r} \in J$. Hence, $\upalpha \wedge_{1} \upbeta = 0$. Assuming $\upalpha \wedge_{m} \upbeta = 0$ for $m>1$ we claim that $\upalpha \wedge_{m+1} \upalpha = 0$. In fact,

\begin{align}
    \begin{aligned}
        \upalpha \wedge_{m+1} \upbeta = \sum^{n}_{a=0} g^{aa} (\mathbf{e}_{a}\rfloor \upalpha) \wedge_{m} (\mathbf{e}_{a} \rfloor \upbeta).
    \end{aligned}
\end{align}

\noindent The element $(\mathbf{e}_{a}\rfloor \upalpha) \wedge_{m} (\mathbf{e}_{a} \rfloor \upbeta)$ is a $\wedge_{m}$-product of a $(k-1)$-form and a $(l-1)$-form with disjoint indices, hence by the inductive hypothesis $(\mathbf{e}_{a}\rfloor \upalpha) \wedge_{m} (\mathbf{e}_{a} \rfloor \upbeta) = 0$ and $\upalpha \wedge_{m+1} \upbeta = 0$ as we wished. Therefore, our proof is conclude by induction. \textcolor{RoyalBlue}{$\Box$}\newline

\begin{tfcor}
    On the conditions of the Proposition \ref{propos3201r} the geometric product coincides with the wedge product.
\end{tfcor}
\noindent \textcolor{RoyalBlue}{\textit{Proof.}} For any $k$-form $\upalpha \in \Omega^{k}(\M)$ and a $l$-form $\upbeta \in \Omega^{l}(\M)$ the geometric product yields

\begin{align}
    \begin{aligned}
        \upalpha \diamond \upbeta &= \sum^{k}_{d=0} \frac{(-1)^{d(k-d) + \left [  \frac{d}{2} \right]}}{d!}  \upalpha \wedge_d \upbeta\\
           &= \upalpha \wedge \upbeta + \sum^{k}_{d=1} \frac{(-1)^{d(k-d) + \left [  \frac{d}{2} \right]}}{d!}  \upalpha \wedge_{d} \upbeta\\
           &= \upalpha \wedge \upbeta + 0\\
           &= \upalpha \wedge \upbeta. \; \textcolor{RoyalBlue}{\Box}
    \end{aligned}
\end{align}

\section{The Volume Form}

\paragraph{ } In this section, the \textit{volume form} is defined. It is an important element of the Kähler--Atiyah algebra. We explore the geometric product involving this element in this section and use it to define important operators and subspaces in the next section. 

\begin{definanil}
    The \textbf{volume form} denoted by $\boldsymbol{\uptau}$ is defined as being the following $n$-form:

\begin{equation}
    \boldsymbol{\uptau} = \normalfont \text{vol}(\Omega(\M)) = e^{12\ldots n} = e^{1} \wedge e^{2} \wedge \cdots \wedge e^{n}
\end{equation}

\end{definanil}

\begin{ex}\label{exvolumeprod}
    {{\textcolor{Blue}{{$\blacktriangleright$\;}}}} The geometric product of the volume form. \normalfont{ Let us consider a manifold $(\M, g)$ with signature (1,2) with local orthonormal coframe $\{e^{1}, e^{2}, e^{3}\}$. Let us compute the product $\boldsymbol{\uptau} \diamond \boldsymbol{\uptau}$.}
\end{ex}
\noindent The product yields:

\begin{align}
    \begin{aligned}
      \boldsymbol{\uptau} \diamond \boldsymbol{\uptau} &= \sum^{3}_{d=0} \frac{(-1)^{d(3-d) + \left [ \frac{d}{2} \right]}}{d!} \boldsymbol{\uptau} \wedge_d \boldsymbol{\uptau} = \boldsymbol{\uptau} \wedge \boldsymbol{\uptau} + \boldsymbol{\uptau} \wedge_{1} \boldsymbol{\uptau} - \frac{1}{2!} \boldsymbol{\uptau} \wedge_{2} \boldsymbol{\uptau} - \frac{1}{3!}\boldsymbol{\uptau} \wedge_{3} \boldsymbol{\uptau}.
    \end{aligned}
\end{align}

\noindent Let us now compute each one of the four wedge products.

\begin{flalign}
    \begin{aligned}
    {\textcolor{RoyalBlue}{\textbf{\textit{1)}}}}\;\; \boldsymbol{\uptau} \wedge \boldsymbol{\uptau} = 0
    \end{aligned} &&
\end{flalign}


\begin{flalign}
    \begin{aligned}
        \textit{\textcolor{RoyalBlue}{\textbf{\textit{2)}}}}\;\; \boldsymbol{\uptau} \wedge_{1} \boldsymbol{\uptau} &= \sum^{3}_{i=0} g^{ii} (\mathbf{e}_{i} \rfloor \boldsymbol{\uptau}) \wedge (\mathbf{e}_{i} \rfloor \boldsymbol{\uptau}) \\
        &= \sum^{3}_{i=0} g^{ii} (\mathbf{e}_{i} \rfloor e^{1}\wedge e^{3}\wedge e^{3}) \wedge (\mathbf{e}_{i} \rfloor e^{1}\wedge e^{3}\wedge e^{3}) \\
        &=  \sum^{3}_{i=0} g^{ii} ( \delta_{i}^{1} e^{2}\!\wedge e^{3} \!- \delta_{i}^{2} e^{1}\!\wedge e^{3} \!+ \delta_{i}^{3}e^{1}\! \wedge e^{2}) \wedge ( \delta_{i}^{1} e^{2}\!\wedge e^{3} \!- \delta_{i}^{2} e^{1}\!\wedge e^{3} \!+ \delta_{i}^{3}e^{1}\! \wedge e^{2})\\
        &= \sum^{3}_{i=0} g^{ii} (\delta_{i}^{1}\delta_{i}^{1}(e^{2}\!\wedge e^{3})^{2} - \delta_{i}^{1}\delta_{i}^{2}e^{2}\!\wedge e^{3}\wedge e^{1}\!\wedge e^{3} + \delta_{i}^{1}\delta_{i}^{3}e^{2}\!\wedge e^{3}\wedge e^{1}\!\wedge e^{2}
        \\&\qquad\qquad + (\delta_{i}^{2}\delta_{i}^{2}(e^{1}\!\wedge e^{3})^{2} - \delta_{i}^{2}\delta_{i}^{1}e^{1}\!\wedge e^{3}\wedge e^{2}\!\wedge e^{3} - \delta_{i}^{2}\delta_{i}^{3}e^{1}\!\wedge e^{3}\wedge e^{1}\!\wedge e^{2}\\
        &\qquad\qquad\quad +  (\delta_{i}^{3}\delta_{i}^{3}(e^{1}\!\wedge e^{2})^{2} + \delta_{i}^{3}\delta_{i}^{1}e^{1}\!\wedge e^{2}\wedge e^{2}\!\wedge e^{3} - \delta_{i}^{3}\delta_{i}^{2}e^{1}\!\wedge e^{2}\wedge e^{1}\!\wedge e^{3})\\
        &=\sum^{3}_{i=0} g^{ii} 0 = 0.
    \end{aligned} &&
\end{flalign}


\begin{flalign}
    \begin{aligned}
          \textit{\textcolor{RoyalBlue}{\textbf{\textit{3)}}}}\;\; \boldsymbol{\uptau} \wedge_{2} \boldsymbol{\uptau} &= \sum^{3}_{i=0} g^{ii} (\mathbf{e}_{i} \rfloor \boldsymbol{\uptau}) \wedge_{1} (\mathbf{e}_{i} \rfloor \boldsymbol{\uptau}) \\
          &= \sum^{3}_{j=0} g^{jj} \left(\sum^{3}_{i=0} g^{ii} (\mathbf{e}_{j} \rfloor \mathbf{e}_{i} \rfloor \boldsymbol{\uptau}) \wedge (\mathbf{e}_{j} \rfloor \mathbf{e}_{i} \rfloor \boldsymbol{\uptau}) \right)
    \end{aligned} &&
\end{flalign}

\noindent such that

\begin{flalign}
    \begin{aligned}
        (\mathbf{e}_{j} \rfloor \mathbf{e}_{i} \rfloor \boldsymbol{\uptau}) &=   (\mathbf{e}_{j} \rfloor \mathbf{e}_{i} \rfloor e^{1} \wedge e^{2} \wedge e^{3}) \\
        &=  \mathbf{e}_{j} \rfloor ( \delta_{i}^{1} e^{2}\!\wedge e^{3} - \delta_{i}^{2} e^{1}\!\wedge e^{3} + \delta_{i}^{3}e^{1}\wedge e^{2})\\
        &= \delta_{i}^{1} \mathbf{e}_{j} \rfloor(e^{2}\!\wedge e^{3}) - \delta_{i}^{2} \mathbf{e}_{j} \rfloor(e^{1}\wedge e^{3}) + \delta_{i}^{3}\mathbf{e}_{j} \rfloor(e^{1} \wedge e^{2})\\
        &= \delta_{i}^{1}(\delta_{j}^{2}e^{3} - \delta_{j}^{3}e^{2})
        - \delta_{i}^{2}(\delta_{j}^{1}e^{3} - \delta_{j}^{3}e^{1}) + \delta_{i}^{3}(\delta_{j}^{1}e^{2} - \delta_{j}^{2}e^{1})\\
        &= \delta_{i}^{1}\delta_{j}^{2}e^{3} - \delta_{i}^{1}\delta_{j}^{3}e^{2}
        - \delta_{i}^{2}\delta_{j}^{1}e^{3} + \delta_{i}^{2}\delta_{j}^{3}e^{1}
        +\delta_{i}^{3}\delta_{j}^{1}e^{2} - \delta_{i}^{3}\delta_{j}^{2}e^{1}
    \end{aligned} &&
\end{flalign}

Therefore, 

\begin{flalign}
    \begin{aligned}
         \boldsymbol{\uptau} \wedge_{2} \boldsymbol{\uptau} &= \sum^{3}_{j=0} g^{jj} \Bigg(\sum^{3}_{i=0} g^{ii} (\delta_{i}^{1}\delta_{j}^{2}e^{3} - \delta_{i}^{1}\delta_{j}^{3}e^{2}
        - \delta_{i}^{2}\delta_{j}^{1}e^{3} + \delta_{i}^{2}\delta_{j}^{3}e^{1}
        +\delta_{i}^{3}\delta_{j}^{1}e^{2}\\&\qquad\qquad - \delta_{i}^{3}\delta_{j}^{2}e^{1}) \!\wedge \!(\delta_{i}^{1}\delta_{j}^{2}e^{3} - \delta_{i}^{1}\delta_{j}^{3}e^{2}
        - \delta_{i}^{2}\delta_{j}^{1}e^{3} + \delta_{i}^{2}\delta_{j}^{3}e^{1}
        +\delta_{i}^{3}\delta_{j}^{1}e^{2} - \delta_{i}^{3}\delta_{j}^{2}e^{1})
        \Bigg)\\
        &= \sum^{3}_{j=0} g^{jj} \Bigg(\sum^{3}_{i=0} g^{ii} \delta_{i}^{1}\delta_{j}^{2}\delta_{i}^{1}\delta_{j}^{2}e^{3} \wedge e^{3} - \delta_{i}^{1}\delta_{j}^{2}\delta_{i}^{1}\delta_{j}^{3}e^{3} \wedge e^{2}  - \delta_{i}^{1}\delta_{j}^{2}\delta_{i}^{2}\delta_{j}^{1}e^{3} \wedge e^{3} \\&\qquad+ \delta_{i}^{1}\delta_{j}^{2}\delta_{i}^{2}\delta_{j}^{3}e^{3} \wedge e^{1}  + \delta_{i}^{1}\delta_{j}^{2}\delta_{i}^{3}\delta_{j}^{1}e^{3} \wedge e^{2} - \delta_{i}^{1}\delta_{j}^{2}\delta_{i}^{3}\delta_{j}^{2}e^{3} \wedge e^{1} - \delta_{i}^{1}\delta_{j}^{3}\delta_{i}^{1}\delta_{j}^{2}e^{2} \wedge e^{3} \\&\qquad+ \delta_{i}^{1}\delta_{j}^{3}\delta_{i}^{1}\delta_{j}^{3}e^{2} \wedge e^{2}  - \delta_{i}^{1}\delta_{j}^{3}\delta_{i}^{2}\delta_{j}^{1}e^{2} \wedge e^{3} + \delta_{i}^{1}\delta_{j}^{3}\delta_{i}^{2}\delta_{j}^{3}e^{2} \wedge e^{1}  + \delta_{i}^{1}\delta_{j}^{3}\delta_{i}^{3}\delta_{j}^{1}e^{2} \wedge e^{2} \\&\qquad- \delta_{i}^{1}\delta_{j}^{3}\delta_{i}^{3}\delta_{j}^{2}e^{2} \wedge e^{1}  - \delta_{i}^{2}\delta_{j}^{1}\delta_{i}^{1}\delta_{j}^{2}e^{3} \wedge e^{3} + \delta_{i}^{2}\delta_{j}^{1}\delta_{i}^{1}\delta_{j}^{3}e^{3} \wedge e^{2}  - \delta_{i}^{2}\delta_{j}^{1}\delta_{i}^{2}\delta_{j}^{1}e^{3} \wedge e^{3} \\&\qquad+ \delta_{i}^{2}\delta_{j}^{1}\delta_{i}^{2}\delta_{j}^{3}e^{3} \wedge e^{1} + \delta_{i}^{2}\delta_{j}^{1}\delta_{i}^{3}\delta_{j}^{1}e^{3} \wedge e^{2} - \delta_{i}^{2}\delta_{j}^{1}\delta_{i}^{3}\delta_{j}^{2}e^{3} \wedge e^{1}  + \delta_{i}^{2}\delta_{j}^{3}\delta_{i}^{1}\delta_{j}^{2}e^{1} \wedge e^{3} \\&\qquad- \delta_{i}^{2}\delta_{j}^{3}\delta_{i}^{1}\delta_{j}^{3}e^{1} \wedge e^{2}  + \delta_{i}^{2}\delta_{j}^{3}\delta_{i}^{2}\delta_{j}^{1}e^{1} \wedge e^{3} - \delta_{i}^{2}\delta_{j}^{3}\delta_{i}^{2}\delta_{j}^{3}e^{1} \wedge e^{1}  + \delta_{i}^{2}\delta_{j}^{3}\delta_{i}^{3}\delta_{j}^{1}e^{1} \wedge e^{2} \\&\qquad- \delta_{i}^{2}\delta_{j}^{3}\delta_{i}^{3}\delta_{j}^{2}e^{1} \wedge e^{1}  + \delta_{i}^{3}\delta_{j}^{1}\delta_{i}^{1}\delta_{j}^{2}e^{2} \wedge e^{3} - \delta_{i}^{3}\delta_{j}^{1}\delta_{i}^{1}\delta_{j}^{3}e^{2} \wedge e^{2}  + \delta_{i}^{3}\delta_{j}^{1}\delta_{i}^{2}\delta_{j}^{1}e^{2} \wedge e^{3} \\&\qquad- \delta_{i}^{3}\delta_{j}^{1}\delta_{i}^{2}\delta_{j}^{3}e^{2} \wedge e^{1}  + \delta_{i}^{3}\delta_{j}^{1}\delta_{i}^{3}\delta_{j}^{1}e^{2} \wedge e^{2} - \delta_{i}^{3}\delta_{j}^{1}\delta_{i}^{3}\delta_{j}^{2}e^{2} \wedge e^{1}  - \delta_{i}^{3}\delta_{j}^{2}\delta_{i}^{1}\delta_{j}^{2}e^{1} \wedge e^{3} \\&\qquad+ \delta_{i}^{3}\delta_{j}^{2}\delta_{i}^{1}\delta_{j}^{3}e^{1} \wedge e^{2}  - \delta_{i}^{3}\delta_{j}^{2}\delta_{i}^{2}\delta_{j}^{1}e^{1} \wedge e^{3} + \delta_{i}^{3}\delta_{j}^{2}\delta_{i}^{2}\delta_{j}^{3}e^{1} \wedge e^{1}  + \delta_{i}^{3}\delta_{j}^{2}\delta_{i}^{3}\delta_{j}^{1}e^{1} \wedge e^{2} \\&\qquad- \delta_{i}^{3}\delta_{j}^{2}\delta_{i}^{3}\delta_{j}^{2}e^{1} \wedge e^{1}\Bigg)\\
        &= \sum^{3}_{j=0} g^{jj} \Bigg(\sum^{3}_{i=0} g^{ii}  (- \delta_{i}^{1}\delta_{i}^{1}\delta_{j}^{2}\delta_{j}^{3}e^{3} \wedge e^{2}   + \delta_{i}^{1}\delta_{i}^{2}\delta_{j}^{2}e^{3}\delta_{j}^{3} \wedge e^{1}  + \delta_{i}^{1}\delta_{i}^{3}\delta_{j}^{2}\delta_{j}^{1}e^{3} \wedge e^{2} \\&\qquad- \delta_{i}^{1}\delta_{i}^{3}\delta_{j}^{2}\delta_{j}^{2}e^{3} \wedge e^{1} - \delta_{i}^{1}\delta_{i}^{1}\delta_{j}^{3}\delta_{j}^{2}e^{2} \wedge e^{3} - \delta_{i}^{1}\delta_{i}^{2}\delta_{j}^{3}\delta_{j}^{1}e^{2} \wedge e^{3} + \delta_{i}^{1}\delta_{i}^{2}\delta_{j}^{3}\delta_{j}^{3}e^{2} \wedge e^{1}   \\&\qquad- \delta_{i}^{1}\delta_{i}^{3}\delta_{j}^{3}\delta_{j}^{2}e^{2} \wedge e^{1}   + \delta_{i}^{2}\delta_{i}^{1}\delta_{j}^{1}\delta_{j}^{3}e^{3} \wedge e^{2}  + \delta_{i}^{2}\delta_{i}^{2}\delta_{j}^{1}\delta_{j}^{3}e^{3} \wedge e^{1} + \delta_{i}^{2}\delta_{i}^{3}\delta_{j}^{1}\delta_{j}^{1}e^{3} \wedge e^{2} \\&\qquad- \delta_{i}^{2}\delta_{i}^{3}\delta_{j}^{1}\delta_{j}^{2}e^{3} \wedge e^{1}  + \delta_{i}^{2}\delta_{i}^{1}\delta_{j}^{3}\delta_{j}^{2}e^{1} \wedge e^{3} - \delta_{i}^{2}\delta_{i}^{1}\delta_{j}^{3}\delta_{j}^{3}e^{1} \wedge e^{2}  + \delta_{i}^{2}\delta_{i}^{2}\delta_{j}^{3}\delta_{j}^{1}e^{1} \wedge e^{3}   \\&\qquad+ \delta_{i}^{2}\delta_{i}^{3}\delta_{j}^{3}\delta_{j}^{1}e^{1} \wedge e^{2}   + \delta_{i}^{3}\delta_{i}^{1}\delta_{j}^{1}\delta_{j}^{2}e^{2} \wedge e^{3}   + \delta_{i}^{3}\delta_{i}^{2}\delta_{j}^{1}\delta_{j}^{1}e^{2} \wedge e^{3} - \delta_{i}^{3}\delta_{i}^{2}\delta_{j}^{1}\delta_{j}^{3}e^{2} \wedge e^{1}   \\&\qquad- \delta_{i}^{3}\delta_{i}^{3}\delta_{j}^{1}\delta_{j}^{2}e^{2} \wedge e^{1}  - \delta_{i}^{3}\delta_{i}^{1}\delta_{j}^{2}\delta_{j}^{2}e^{1} \wedge e^{3} + \delta_{i}^{3}\delta_{i}^{1}\delta_{j}^{2}\delta_{j}^{3}e^{1} \wedge e^{2}  - \delta_{i}^{3}\delta_{i}^{2}\delta_{j}^{2}\delta_{j}^{1}e^{1} \wedge e^{3}   \\&\qquad+ \delta_{i}^{3}\delta_{i}^{3}\delta_{j}^{2}\delta_{j}^{1}e^{1} \wedge e^{2}) \Bigg)
    \end{aligned} &&
\end{flalign}

\noindent The terms with $\delta_{i}^{r}\delta_{i}^{s}$ where $r \neq s$ will vanish when the summation runs over the index $i$. However, the remaining terms with $\delta_{i}^{k}\delta_{i}^{k}$ are multiplied by terms of the form $\delta_{j}^{r}\delta_{i}^{s}$ with $r \neq s$ that will vanish when the sum is carried out over the index $j$. Consequently, $\boldsymbol{\uptau} \wedge_{2} \boldsymbol{\uptau} = 0$.


\begin{flalign}
    \begin{aligned}
        \textit{\textcolor{RoyalBlue}{\textbf{\textit{4)}}}}\;\; \boldsymbol{\uptau} \wedge_{3} \boldsymbol{\uptau} &= \sum^{3}_{i=0} g^{ii} (\mathbf{e}_{i} \rfloor \boldsymbol{\uptau}) \wedge_{2} (\mathbf{e}_{i} \rfloor \boldsymbol{\uptau}) \\
          &= \sum^{3}_{k=0} g^{kk}\left ( \sum^{3}_{j=0} g^{jj} \left(\sum^{3}_{i=0} g^{ii} (\mathbf{e}_{k} \rfloor \mathbf{e}_{j} \rfloor \mathbf{e}_{i} \rfloor \boldsymbol{\uptau}) \wedge (\mathbf{e}_{k} \rfloor \mathbf{e}_{j} \rfloor \mathbf{e}_{i} \rfloor \boldsymbol{\uptau}) \right)\right)
    \end{aligned} &&
\end{flalign}

\noindent where

\begin{flalign} 
    \begin{aligned}
         (\mathbf{e}_{k} \rfloor \mathbf{e}_{j} \rfloor \mathbf{e}_{i} \rfloor \boldsymbol{\uptau}) &= (\mathbf{e}_{k} \rfloor (\delta_{i}^{1}\delta_{j}^{2}e^{3} - \delta_{i}^{1}\delta_{j}^{3}e^{2}
        - \delta_{i}^{2}\delta_{j}^{1}e^{3} + \delta_{i}^{2}\delta_{j}^{3}e^{1}
        +\delta_{i}^{3}\delta_{j}^{1}e^{2} - \delta_{i}^{3}\delta_{j}^{2}e^{1})\\
        &= \delta_{i}^{1}\delta_{j}^{2}\delta_{k}^{3} - \delta_{i}^{1}\delta_{j}^{3}\delta_{k}^{2}
        - \delta_{i}^{2}\delta_{j}^{1}\delta_{k}^{3} + \delta_{i}^{2}\delta_{j}^{3}\delta_{k}^{1}
        +\delta_{i}^{3}\delta_{j}^{1}\delta_{k}^{2} - \delta_{i}^{3}\delta_{j}^{2}\delta_{k}^{1}.
    \end{aligned} &&
\end{flalign}

\noindent Hence

\begin{flalign}
    \begin{aligned}
       \boldsymbol{\uptau} \wedge_{3} \boldsymbol{\uptau}
          &= \sum^{3}_{k=0} g^{kk}\left ( \sum^{3}_{j=0} g^{jj} \left(\sum^{3}_{i=0} g^{ii} (\mathbf{e}_{k} \rfloor \mathbf{e}_{j} \rfloor \mathbf{e}_{i} \rfloor \boldsymbol{\uptau}) \wedge (\mathbf{e}_{k} \rfloor \mathbf{e}_{j} \rfloor \mathbf{e}_{i} \rfloor \boldsymbol{\uptau}) \right)\right)\\
          &= \sum^{3}_{k=0} g^{kk}\Bigg( \sum^{3}_{j=0} g^{jj} \Bigg(\sum^{3}_{i=0} g^{ii} (\delta_{i}^{1}\delta_{j}^{2}\delta_{k}^{3} - \delta_{i}^{1}\delta_{j}^{3}\delta_{k}^{2}
        - \delta_{i}^{2}\delta_{j}^{1}\delta_{k}^{3} + \delta_{i}^{2}\delta_{j}^{3}\delta_{k}^{1}
        +\delta_{i}^{3}\delta_{j}^{1}\delta_{k}^{2}\\&\qquad - \delta_{i}^{3}\delta_{j}^{2}\delta_{k}^{1}) \wedge (\delta_{i}^{1}\delta_{j}^{2}\delta_{k}^{3} - \delta_{i}^{1}\delta_{j}^{3}\delta_{k}^{2}
        - \delta_{i}^{2}\delta_{j}^{1}\delta_{k}^{3} + \delta_{i}^{2}\delta_{j}^{3}\delta_{k}^{1}
        +\delta_{i}^{3}\delta_{j}^{1}\delta_{k}^{2} - \delta_{i}^{3}\delta_{j}^{2}\delta_{k}^{1}) \Bigg)\Bigg).
    \end{aligned} &&
\end{flalign}

\noindent Note that when computing the product the only terms that will not vanish when the sum runs over all the indexes will be the square terms: $(\delta_{i}^{1}\delta_{j}^{2}\delta_{k}^{3})(\delta_{i}^{1}\delta_{j}^{2}\delta_{k}^{3}),$ $ (\delta_{i}^{1}\delta_{j}^{3}\delta_{k}^{2})(\delta_{i}^{1}\delta_{j}^{3}\delta_{k}^{2}),$ $(\delta_{i}^{2}\delta_{j}^{1}\delta_{k}^{3})(\delta_{i}^{2}\delta_{j}^{1}\delta_{k}^{3}),$ $ (\delta_{i}^{2}\delta_{j}^{3}\delta_{k}^{1})(\delta_{i}^{2}\delta_{j}^{3}\delta_{k}^{1}),$ $ (\delta_{i}^{3}\delta_{j}^{1}\delta_{k}^{2})(\delta_{i}^{3}\delta_{j}^{1}\delta_{k}^{2}),$ $  (\delta_{i}^{3}\delta_{j}^{2}\delta_{k}^{1})(\delta_{i}^{3}\delta_{j}^{2}\delta_{k}^{1})$. Consequently,

\begin{flalign}
    \begin{aligned}
       \boldsymbol{\uptau} \wedge_{3} \boldsymbol{\uptau}
          &= g^{11}g^{22}g^{33}
        +  g^{11}g^{33}g^{22}
        + g^{22}g^{11}g^{33}
        +  g^{22}g^{33}g^{11}
        + g^{33}g^{11}g^{22}
        + g^{33}g^{22}g^{11}= 6.
    \end{aligned} &&
\end{flalign}

Finally,

\begin{align}
    \begin{aligned}
      \boldsymbol{\uptau} \diamond \boldsymbol{\uptau} &= \boldsymbol{\uptau} \wedge \boldsymbol{\uptau} + \boldsymbol{\uptau} \wedge_{1} \boldsymbol{\uptau} - \frac{1}{2!} \boldsymbol{\uptau} \wedge_{2} \boldsymbol{\uptau} - \frac{1}{3!}\boldsymbol{\uptau} \wedge_{3} \boldsymbol{\uptau} = 0 + 0 - 0 - \frac{1}{6}6 = -1.\;\; \textcolor{Blue}{{\blacktriangleleft}}
    \end{aligned}
\end{align}

Our next goal is to generalise the Example \ref{exvolumeprod} studying the behaviour of $\boldsymbol{\uptau} \diamond \boldsymbol{\uptau}$ for arbitrary dimensions and signatures. The next Lemma will be of very importance to do so.

\begin{tflemma}\label{lema21} With respect to the volume form $\boldsymbol{\uptau}$, the contracted wedge product satisfies:

\begin{align}
    \begin{aligned}
        \boldsymbol{\uptau} \wedge_{d} \boldsymbol{\uptau} &= 0 \text{ for all } 0 \leq d < n,\\
        \boldsymbol{\uptau} \wedge_{n} \boldsymbol{\uptau} &= n! \,g^{11}\!\!\ldots g^{nn}1
    \end{aligned}
\end{align}
\end{tflemma}
\noindent \textcolor{RoyalBlue}{\textit{Proof.}} Firstly, $\boldsymbol{\uptau} \wedge \boldsymbol{\uptau} = 0$. Now, for any $d \in \mathbb{N}$ given such that $d<n$, with respect to $\boldsymbol{\uptau} \wedge_{d} \boldsymbol{\uptau}$ we inspect three cases: $n> d+1, n = d+1$ and $n = d$. Note that by Definition \ref{defincontractedwe}, $\boldsymbol{\uptau} \wedge_{d} \boldsymbol{\uptau} =0$ if $d > n$. Therefore the case $n < d+1$  is not considered. {\textcolor{RoyalBlue}{\textbf{\textit{1) Case:}}}} $n> d+1$. The contraction of $\boldsymbol{\uptau}$ by $ \mathbf{e}_{i_{1}}$ yields
\begin{align} \label{eq_126586765}
    \begin{aligned}
        \mathbf{e}_{i_{1}}\rfloor\boldsymbol{\uptau} &= (\mathbf{e}_{i_{1}}\rfloor e^{1} \wedge e^{2} \wedge \cdots \wedge e^{n}) \\
        &= \delta_{i_{1}}^{1} e^{2} \wedge  \cdots\wedge  e^{n} - \delta_{i_{1}}^{2} e^{1} \wedge e^{3}\! \wedge \cdots \wedge e^{n}+ \delta_{i_{1}}^{3} e^{2} \wedge e^{4} \wedge \cdots \wedge e^{n}
        \\&\qquad + \cdots + (-1)^{n+1}\delta_{i_{1}}^{n} e^{1} 
        \wedge \cdots \wedge e^{n-1}.\\
    \end{aligned}
\end{align} 

\noindent The result of Eq. \eqref{eq_126586765} is a sum of $(n-1)$-forms. Therefore, when computing $\boldsymbol{\uptau} \wedge_{1} \boldsymbol{\uptau}$:

\begin{align}
    \begin{aligned}
        \boldsymbol{\uptau} \wedge_{1} \boldsymbol{\uptau} &= \sum^{n}_{i_{1}=0} g^{i_{1}i_{1}} (\mathbf{e}_{i_{1}} \rfloor \boldsymbol{\uptau}) \wedge (\mathbf{e}_{i_{1}} \rfloor \boldsymbol{\uptau}) \\
       &= \sum^{n}_{i_{1}=0} g^{i_{1}i_{1}} (\delta_{i_{1}}^{1} e^{2} \wedge \! \cdots\! \wedge e^{n} + \!\cdots\! + (-1)^{n+1}\delta_{i_{1}}^{n} e^{1} 
        \wedge\! \cdots\! \wedge e^{n-1}) \wedge (\delta_{i_{1}}^{1} e^{2} \wedge \! \cdots\! \wedge e^{n}\\
        &\qquad\qquad\qquad\qquad\qquad\qquad\qquad\qquad\qquad\qquad+ \!\cdots\! + (-1)^{n+1}\delta_{i_{1}}^{n} e^{1} 
        \wedge\! \cdots\! \wedge e^{n-1})\\
    \end{aligned}
\end{align}

\noindent the result equals zero if $n > 2$ since in each term of the wedge product, one has $e^{j} \wedge e^{j}$ for some $j$. The contraction of $(\mathbf{e}_{i_{1}}\rfloor\boldsymbol{\uptau})$ by $ \mathbf{e}_{i_{2}}$ yields

\begin{align}
    \begin{aligned}\label{eqe2e1v}
        \mathbf{e}_{i_{2}}\rfloor \mathbf{e}_{i_{1}}\rfloor\boldsymbol{\uptau} 
        &= \delta_{i_{1}}^{1} (\delta_{i_{2}}^{2} e^{3} \wedge \cdots \wedge e^{n} - \delta_{i_{2}}^{3} e^{2} \wedge e^{4} \cdots \wedge e^{n} + \cdots + (-1)^{n+1}\delta_{i_{2}}^{n} e^{2} \wedge \cdots \wedge e^{n-1}) \\
        &\quad - \delta_{i_{1}}^{2} (\delta_{i_{2}}^{1} e^{3} \wedge \! \cdots\! \wedge e^{n} - \delta_{i_{2}}^{3} e^{1} \wedge e^{4}\! \cdots\! \wedge e^{n} + \!\cdots\! + (-1)^{n+1}\delta_{i_{2}}^{n} e^{1} \!\wedge e^{3} \!\wedge\! \cdots\! \wedge e^{n-1})   \\
        &\quad + \cdots + (-1)^{n+1}\delta_{i_{1}}^{n} (\delta_{i_{2}}^ {1} e^{2} \wedge \cdots \wedge e^{n-1} + \cdots + (-1)^{n}\delta_{i_{2}}^{n-1} e^{1} \wedge \cdots \wedge e^{n-2}).  
        \end{aligned}
\end{align}

\noindent The first line of Eq. \eqref{eqe2e1v} is the term $\delta_{i_{1}}^{1} \mathbf{e}_{i_{2}}\rfloor e^{2} \wedge \cdots \wedge e^{n}$, the second is the term $\delta_{i_{1}}^{2} \mathbf{e}_{i_{2}}\rfloor e^{1} \wedge e^{3} \wedge \cdots \wedge e^{n}$ and the third is the term $\delta_{i_{1}}^{n} \mathbf{e}_{i_{2}}\rfloor e^{1} \wedge \cdots \wedge e^{n-1}$. Each one is a sum of $(n-2)$-forms. The contracted wedge product of order $2$ of $(\boldsymbol{\uptau},\boldsymbol{\uptau})$ is

\begin{align}
    \begin{aligned}
         \boldsymbol{\uptau} \wedge_{2} \boldsymbol{\uptau} &=  \sum^{n}_{i_{1}=0} g^{i_{1}i_{1}} (\mathbf{e}_{i_{1}} \rfloor \boldsymbol{\uptau}) \wedge_{1} (\mathbf{e}_{i_{1}} \rfloor \boldsymbol{\uptau}) = \sum^{n}_{i_{1}=0} g^{i_{1}i_{1}} \left(  \sum^{n}_{i_{2}=0}  g^{i_{2}i_{2}} (\mathbf{e}_{i_{2}} \rfloor \mathbf{e}_{i_{1}} \rfloor \boldsymbol{\uptau}) \wedge (\mathbf{e}_{i_{2}} \rfloor \mathbf{e}_{i_{1}} \rfloor \boldsymbol{\uptau}) \right).
    \end{aligned}
\end{align}

\noindent By looking the Eq. \eqref{eqe2e1v} if $n>3$, the product equals zero since only terms of type $e^{a} \cdots \wedge e^{j} \wedge e^{j} \wedge \cdots e^{b}$ for some $j$ will appear. Therefore, doing the same steps for $e_{i_{d}} \rfloor \cdots \mathbf{e}_{i_{2}}\rfloor \mathbf{e}_{i_{1}}\rfloor\boldsymbol{\uptau}$, which will be a sum of $(n-d)$-forms, we conclude that $\boldsymbol{\uptau} \wedge_{d} \boldsymbol{\uptau} = 0$ for the case $n > d+1$. {\textcolor{RoyalBlue}{\textbf{\textit{2) Case:}}} $n = d+1$}. In this case, $e_{i_{d}} \rfloor \cdots \mathbf{e}_{i_{2}}\rfloor \mathbf{e}_{i_{1}}\rfloor\boldsymbol{\uptau} = \mathbf{e}_{i_{n-1}} \rfloor \cdots \mathbf{e}_{i_{2}}\rfloor \mathbf{e}_{i_{1}}\rfloor\boldsymbol{\uptau}$ will be a sum of $1$-forms. Therefore, $e_{i_{n-1}} \rfloor \cdots \mathbf{e}_{i_{2}}\rfloor \mathbf{e}_{i_{1}}\rfloor\boldsymbol{\uptau} \wedge \mathbf{e}_{i_{n-1}} \rfloor \cdots \mathbf{e}_{i_{2}}\rfloor \mathbf{e}_{i_{1}}\rfloor\boldsymbol{\uptau} \neq 0$ and we need to inspect the summations $\sum_{j=0}^{n} g^{i_j i_j}$, $j = \{i_{1}, \ldots, i_{n-1}\}$

\begin{align}
    \begin{aligned}
        \boldsymbol{\uptau} \!\wedge_{d} \boldsymbol{\uptau} \!=\! \sum^{n}_{i_{1}=0}  g^{i_{1}i_{1}} \!\left(\sum^{n}_{i_{2}=0}  g^{i_{2}i_{2}} \left( \!\cdots\! \left( \sum^{n}_{i_{d}=0}  g^{i_{d}i_{d}} (\mathbf{e}_{i_{d}} \rfloor \!\cdots \! \mathbf{e}_{i_{2}}\rfloor \mathbf{e}_{i_{1}}\rfloor\boldsymbol{\uptau}) \!\wedge\! (\mathbf{e}_{i_{d}} \rfloor \!\cdots\! \mathbf{e}_{i_{2}}\rfloor \mathbf{e}_{i_{1}}\rfloor\boldsymbol{\uptau}) \! \right) \!\cdots \!\right)\! \right)\!. 
    \end{aligned}
\end{align}

\noindent Firstly, note that, $ \mathbf{e}_{i_{n-1}} \rfloor \cdots \mathbf{e}_{i_{2}}\rfloor \mathbf{e}_{i_{1}}\rfloor\boldsymbol{\uptau}$ is a sum of $n!$ terms, the first one is $\delta_{i_{1}}^{1}\delta_{i_{2}}^{2}\cdots\delta_{i_{n-1}}^{n-1}e^{n}$. Generically, the terms are a combination of indexes of the form: $\delta_{i_{1}}^{k_{1}}\delta_{i_{2}}^{k_{2}}\cdots\delta_{i_{n-1}}^{k_{n-1}}e^{k_{n}}$ such that $i_{a},k_{s} \in \{1,\ldots,n \}$ and $k_{s} \neq k_{r}$ if $r \neq s$ and $i_{a} \neq i_{b}$ if $a \neq b$. Therefore, after computing the wedge product $ \mathbf{e}_{i_{n-1}} \rfloor \cdots \mathbf{e}_{i_{2}}\rfloor \mathbf{e}_{i_{1}}\rfloor\boldsymbol{\uptau}$ $\wedge  \mathbf{e}_{i_{n-1}} \rfloor \cdots \mathbf{e}_{i_{2}}\rfloor \mathbf{e}_{i_{1}}\rfloor\boldsymbol{\uptau}$, all the
mixed terms will vanish when the sum runs over all the indexes. The squares $(\delta_{i_{1}}^{k_{1}}\delta_{i_{2}}^{k_{2}}\cdots\delta_{i_{n-1}}^{k_{n-1}}e^{k_{n}})\wedge(\delta_{i_{1}}^{k_{1}}\delta_{i_{2}}^{k_{2}}\cdots\delta_{i_{n-1}}^{k_{n-1}}e^{k_{n}})$ will also vanish since $e^{k_{n}} \wedge e^{k_{n}} = 0$. This way we conclude that $ \boldsymbol{\uptau} \!\wedge_{d} \boldsymbol{\uptau} =  \boldsymbol{\uptau} \!\wedge_{n-1} \boldsymbol{\uptau} = 0$. \textit{{\textcolor{RoyalBlue}{\textbf{\textit{3) Case:}}}} $n = d$.} One has:

\begin{align}
    \begin{aligned}
        \boldsymbol{\uptau} \!\wedge_{n} \boldsymbol{\uptau} \!=\! \sum^{n}_{i_{1}=0}  g^{i_{1}i_{1}} \!\left(\sum^{n}_{i_{2}=0}  g^{i_{2}i_{2}} \left( \!\cdots\!\! \left( \sum^{n}_{i_{n}=0}  g^{i_{n}i_{n}} (\mathbf{e}_{i_{n}} \rfloor \!\cdots \! \mathbf{e}_{i_{2}}\rfloor  \mathbf{e}_{i_{1}}\rfloor\boldsymbol{\uptau}) \!\wedge\! (\mathbf{e}_{i_{n}} \rfloor \!\cdots\!  \mathbf{e}_{i_{2}}\rfloor  \mathbf{e}_{i_{1}}\rfloor\boldsymbol{\uptau}) \! \right) \!\!\cdots \!\right)\! \right)\!. 
    \end{aligned}
\end{align}

\noindent In this case, $(\mathbf{e}_{i_{n}} \rfloor \cdots  \mathbf{e}_{i_{2}}\rfloor  \mathbf{e}_{i_{1}}\rfloor\boldsymbol{\uptau})$ is a sum of $n!$ $0$-forms of type $\delta_{i_{1}}^{k_{1}}\delta_{i_{2}}^{k_{2}}\cdots\delta_{i_{n}}^{k_{n}}$ such that $i_{a},k_{s} \in \{1,\ldots,n \}$ and $k_{s} \neq k_{r}$ if $r \neq s$ and $i_{a} \neq i_{b}$ if $a \neq b$. Therefore, the only terms that will not vanish after the sum runs over all the indexes will be the squares $(\delta_{i_{1}}^{k_{1}}\delta_{i_{2}}^{k_{2}}\cdots\delta_{i_{n}}^{k_{n}})(\delta_{i_{1}}^{k_{1}}\delta_{i_{2}}^{k_{2}}\cdots\delta_{i_{n}}^{k_{n}})$, leading to a sum of combinations of type $g^{k_{1}k_{1}}g^{k_{2}k_{2}}\cdots g^{k_{n}k_{n}}$ with $k_{s} \in \{1,\ldots,n \}$ and $k_{s} \neq k_{r}$ if $r \neq s$ which results on $g^{11}g^{22}\cdots g^{nn}$ $n!$ times. Therefore, we conclude that $ \boldsymbol{\uptau} \!\wedge_{n} \boldsymbol{\uptau} = n! g^{11}g^{22}\cdots g^{nn}$. \textcolor{RoyalBlue}{$\Box$}\\

 We are now in a position to present the main result concerning the volume form. Before that, it is important to emphasise the following remark.

\begin{rem}\label{rem21} If $p - q = m \mod 8$ is odd (even), then $n = p + q$ must be odd (even). \normalfont{Indeed, observe that}

\begin{align}
    \begin{aligned}
         p - q =\! 2k + 1 \!\!\!\mod\! 8\! &\implies\! p - q = 8l + 2k + 1 \!\implies\! n = p - q +2q = 2(k + 4l + q) + 1\\
         p - q =\! 2k \!\!\!\mod\! 8\! &\implies\! p - q = 8l + 2k  \!\implies\! n = p - q +2q = 2(k + 4l + q)
    \end{aligned}
\end{align}

\noindent for some $k,l \in \mathbb{N}.$
\end{rem}

\begin{tfpropos} The volume element $\boldsymbol{\uptau}$ satisfies:

\begin{equation}
     \boldsymbol{\uptau} \diamond \boldsymbol{\uptau} = \color{RoyalBlue}\begin{cases}\color{black}
        +1, \text{ if } p - q \equiv_8 \color{black}0,1,4,5\\\color{black}
        \color{black}-1, \text{ if } p - q \equiv_8 2,3,6,7\color{black}
    \end{cases}\color{black}.
\end{equation}
\end{tfpropos}
\noindent \textcolor{RoyalBlue}{\textit{Proof.}} It follows from Lemma \ref{lema21} that   

\begin{align}
    \begin{aligned}
      \boldsymbol{\uptau} \diamond \boldsymbol{\uptau} &= \sum^{n}_{d=0} \frac{(-1)^{d(n-d) + \left [ \frac{d}{2} \right]}}{d!} \boldsymbol{\uptau} \wedge_d \boldsymbol{\uptau} = \frac{(-1)^{\left [ \frac{d}{2} \right]}}{n!}   
      \boldsymbol{\uptau} \wedge_{n} \boldsymbol{\uptau} = (-1)^{\left [ \frac{n}{2} \right]} g^{11}g^{22}\cdots g^{nn}.
    \end{aligned}
\end{align}

\noindent Since $g^{11}g^{22}\cdots g^{nn} = (-1)^{q}$, it follows that

\begin{align}
    \begin{aligned}
      \boldsymbol{\uptau} \diamond \boldsymbol{\uptau} &= (-1)^{q + \left [ \frac{n}{2} \right]}  = \color{RoyalBlue}\begin{cases}\color{black}
        (-1)^{q +  \frac{n}{2}} , \text{ if } n \text{ is even}\color{black}\\
        \color{black}(-1)^{q +  \frac{n-1}{2}} , \text{ if } n \text{ is odd}\color{black}
    \end{cases}\color{black}.
    \end{aligned}
\end{align}

\noindent The next step is to analyse for which values of $p-q \mod 8$ the index $q + \left [ \frac{n}{2} \right]$ is even or odd. Since when  $q + \left [ \frac{n}{2} \right]$ is odd (even), $\boldsymbol{\uptau} \diamond \boldsymbol{\uptau} = -1\; (\boldsymbol{\uptau} \diamond \boldsymbol{\uptau} = 1)$.  One has

\begin{align}
    \begin{aligned}
        q + \frac{n}{2} &=  \frac{p + 3q}{2} = \frac{p - q}{2} + 2q,\\
        q +  \frac{n - 1}{2} &=  \frac{p + 3q -1}{2} = \frac{p - q - 1}{2} + 2q.\\
    \end{aligned}
\end{align}

\noindent Two cases are identified: $p - q \equiv_{8} m$ with $m$ odd or $m$ even. {\textcolor{RoyalBlue}{\textbf{\textit{1) Case:}}}} $m$ odd. This case leads us to $n$ odd by Remark \ref{rem21}, therefore $q + \left [ \frac{n}{2} \right] =  \frac{p - q - 1}{2} + 2q.$ For this case, it follows that:
\begin{align}
    \begin{aligned}
        p - q &\equiv_{8} 1 \!\implies\!\! p \!-\! q = 8k_1 + 1 \!\implies\!\! \frac{p - q - 1}{2} + 2q = \frac{8k_1 + 1 - 1}{2} +2q = 4k_1 + 2q \\
        p - q &\equiv_{8} 3 \!\implies\!\! p \!-\! q = 8k_3 + 3 \!\implies\!\! \frac{p - q - 1}{2} + 2q = \frac{8k_3 + 3 - 1}{2} +2q = 4k_3 + 2q + 1\\
        p - q &\equiv_{8} 5 \!\implies\!\! p \!-\! q = 8k_5 + 5 \!\implies\!\! \frac{p - q - 1}{2} + 2q = \frac{8k_5 + 5 - 1}{2} +2q = 4k_5 + 2q + 2\\
        p - q &\equiv_{8} 7 \!\implies\!\! p \!-\! q = 8k_7 + 7 \!\implies\!\! \frac{p - q - 1}{2} + 2q = \frac{8k_7 + 7 - 1}{2} +2q = 4k_7 + 2q + 3 
    \end{aligned}
\end{align}

\noindent In this case, $q + \left [ \frac{n}{2} \right]$ is odd for $p - q \equiv_{8} 3,7$ and even for $p - q \equiv_{8} 1,5$. {\textcolor{RoyalBlue}{\textbf{\textit{2) Case:}}}} $m$ even.  This case leads us to $n$ even by Remark \ref{rem21}, therefore $q + \left [ \frac{n}{2} \right] =  \frac{p - q}{2} + 2q.$ As a result:

\begin{align}
    \begin{aligned}
         p - q &\equiv_{8} 0 \implies p - q = 8k_0 \implies \frac{p - q}{2} + 2q = \frac{8k_0}{2} +2q = 4k_0 +2q\\
         p - q &\equiv_{8} 2\implies p - q = 8k_2 + 2 \implies \frac{p - q}{2} + 2q = \frac{8k_2 + 2}{2} +2q = 4k_2 +2q + 1\\
         p - q &\equiv_{8} 4 \implies p - q = 8k_4 + 4 \implies \frac{p - q}{2} + 2q = \frac{8k_4 + 4}{2} +2q = 4k_4 +2q + 2\\
         p - q &\equiv_{8} 6 \implies p - q = 8k_6 \implies \frac{p - q}{2} + 2q = \frac{8k_6 + 6}{2} +2q = 4k_6 +2q + 3\\
    \end{aligned}
\end{align}

\noindent In this case, $q + \left [ \frac{n}{2} \right]$ is odd for $p - q \equiv_{8} 2,6$ and even for $p - q \equiv_{8} 0,4$. We conclude that

\begin{equation}
    \boldsymbol{\uptau} \diamond \boldsymbol{\uptau} = \color{RoyalBlue}\begin{cases}\color{black}
        +1, \text{ if } p - q \equiv_8 \color{black}0,1,4,5\\\color{black}
        \color{black}-1, \text{ if } p - q \equiv_8 2,3,6,7\color{black}
    \end{cases}\color{black}. \;\;\; \textcolor{RoyalBlue}{\Box}
\end{equation}

\begin{ex}\label{exkformvolumeprod}
    {{\textcolor{Blue}{{$\blacktriangleright$\;}}}} The geometric product between an homogeneous $k$-form and the volume form. \normalfont{ Let us consider a manifold $(\M, g)$ with signature (1,2) with local orthonormal coframe $\{e^{1}, e^{2}, e^{3}\}$. Given $\alpha \in \Omega^{2}(\M)$ such that $\upalpha = 2e^{12} +  e^{23}$, let us compute the product $\upalpha \diamond \boldsymbol{\uptau}$.}
\end{ex}
\noindent By Definition \ref{defin_graf1}, one has:

\begin{equation}
    \upalpha \diamond \boldsymbol{\uptau} = \sum^{2}_{d=0} \frac{(-1)^{d(2-d) + \left [ \frac{d}{2} \right]}}{d!} \upalpha \wedge_d \boldsymbol{\uptau} = \upalpha \wedge \boldsymbol{\uptau} - \upalpha \wedge_{1} \boldsymbol{\uptau} - \frac{1}{2!} \upalpha \wedge_{2} \boldsymbol{\uptau}
\end{equation}

\noindent Let us compute the three wedge products above.

\begin{flalign}
    \begin{aligned}
        \textcolor{RoyalBlue}{\textbf{\textit{1)}}}\;\; \upalpha \wedge \boldsymbol{\uptau} &=  (2e^{1} \wedge e^{2} +  e^{2} \wedge e^{3})\wedge (e^{1} \wedge e^{2} \wedge e^{3}) = 0. 
    \end{aligned} &&
\end{flalign}

\begin{flalign}
    \begin{aligned}
          \textcolor{RoyalBlue}{\textbf{\textit{2)}}}\;\; \upalpha \wedge_{1} \boldsymbol{\uptau} &= \sum^{3}_{i=0} g^{ii} (\mathbf{e}_{i} \rfloor \upalpha) \wedge (\mathbf{e}_{i} \rfloor \boldsymbol{\uptau})\\
        &= \sum^{3}_{i=0} g^{ii} (\mathbf{e}_{i} \rfloor (2e^{1} \wedge e^{2} +  e^{2} \wedge e^{3})) \wedge (\mathbf{e}_{i} \rfloor e^{1} \wedge e^{2} \wedge e^{3})\\
        &= \sum^{3}_{i=0} g^{ii} ( 2\delta_{i}^{1}e^{2} 
        - 2\delta_{i}^{2}e^{1} 
        + \delta_{i}^{2}e^{3} 
        - \delta_{i}^{3}e^{2}) \wedge (\delta_{i}^{1} e^{2}\wedge e^{3} 
        - \delta_{i}^{2} e^{1}\wedge e^{3} 
        + \delta_{i}^{3}e^{1} \wedge e^{2})\\
        &= \sum^{3}_{i=0} g^{ii} 
        (-2\delta_{i}^{1}\delta_{i}^{2}e^{2}\wedge e^{1}\wedge e^{3}  
        - 2\delta_{i}^{2}\delta_{i}^{1}e^{1}\wedge e^{2}\wedge e^{3}  
        + \delta_{i}^{2}\delta_{i}^{3} e^{3}\wedge e^{1} \wedge e^{2} 
        \\&\qquad\qquad+ \delta_{i}^{3}\delta_{i}^{2}e^{2}\wedge e^{1} \wedge e^{3}) = 0.
    \end{aligned} &&
\end{flalign}

\begin{flalign}
    \begin{aligned}
  \textcolor{RoyalBlue}{\textbf{\textit{3)}}}\;\; \upalpha \wedge_{2} \boldsymbol{\uptau} &= \sum^{3}_{i=0} g^{ii} (\mathbf{e}_{i} \rfloor \upalpha) \wedge_{1} (\mathbf{e}_{i} \rfloor \boldsymbol{\uptau})\\
        &= \sum^{3}_{j=0} g^{jj} \Bigg( \sum^{3}_{i=0} g^{ii} (\mathbf{e}_{j} \rfloor \mathbf{e}_{i} \rfloor (2e^{1} \wedge e^{2} +  e^{2} \wedge e^{3})) \wedge (\mathbf{e}_{j} \rfloor \mathbf{e}_{i} \rfloor e^{1} \wedge e^{2} \wedge e^{3})\Bigg)\\
        &= \sum^{3}_{j=0} g^{jj} \Bigg( \sum^{3}_{i=0} g^{ii} 
        (2\delta_{i}^{1}\delta_{j}^{2} 
        - 2\delta_{i}^{2}\delta_{j}^{1} 
        + \delta_{i}^{2}\delta_{j}^{3} 
        - \delta_{i}^{3}\delta_{j}^{2}) \wedge (\delta_{i}^{1}\delta_{j}^{2}e^{3} 
- \delta_{i}^{1}\delta_{j}^{3}e^{2}
\\&\qquad\qquad- \delta_{i}^{2}\delta_{j}^{1}e^{3} 
+ \delta_{i}^{2}\delta_{j}^{3}e^{1}
+\delta_{i}^{3}\delta_{j}^{1}e^{2} 
- \delta_{i}^{3}\delta_{j}^{2}e^{1})\Bigg)\\
&= \sum^{3}_{j=0} g^{jj} \Bigg( \sum^{3}_{i=0} g^{ii}
2\delta_{i}^{1}\delta_{i}^{1}\delta_{j}^{2}\delta_{j}^{2}e^{3} 
\!- 2\delta_{i}^{1}\delta_{i}^{1}\delta_{j}^{2}\delta_{j}^{3}e^{2}
\!- 2\delta_{i}^{1}\delta_{i}^{2}\delta_{j}^{2}\delta_{j}^{1}e^{3} 
\!+ 2\delta_{i}^{1}\delta_{i}^{2}\delta_{j}^{2}\delta_{j}^{3}e^{1}
\\&\quad+2\delta_{i}^{1}\delta_{i}^{3}\delta_{j}^{2}\delta_{j}^{1}e^{2} 
\!- 2\delta_{i}^{1}\delta_{i}^{3}\delta_{j}^{2}\delta_{j}^{2}e^{1}
\!+2\delta_{i}^{2}\delta_{i}^{1}\delta_{j}^{1}\delta_{j}^{2}e^{3} 
\!+2\delta_{i}^{2}\delta_{i}^{1}\delta_{j}^{1}\delta_{j}^{3}e^{2}
\!+ 2\delta_{i}^{2} \delta_{i}^{2}\delta_{j}^{1}\delta_{j}^{1}e^{3} 
\\&\quad- 2\delta_{i}^{2}\delta_{i}^{2}\delta_{j}^{1} \delta_{j}^{3}e^{1}
- 2\delta_{i}^{2}\delta_{i}^{3}\delta_{j}^{1}\delta_{j}^{1}e^{2} 
+ 2\delta_{i}^{2}\delta_{i}^{3}\delta_{j}^{1}\delta_{j}^{2}e^{1}
+\delta_{i}^{2}\delta_{i}^{1}\delta_{j}^{3}\delta_{j}^{2}e^{3} 
- \delta_{i}^{2}\delta_{i}^{1}\delta_{j}^{3}\delta_{j}^{3}e^{2}
\\&\quad- \delta_{i}^{2}\delta_{i}^{2}\delta_{j}^{3}\delta_{j}^{1}e^{3} 
+ \delta_{i}^{2}\delta_{i}^{2}\delta_{j}^{3}\delta_{j}^{3}e^{1}
+\delta_{i}^{2}\delta_{i}^{3}\delta_{j}^{3}\delta_{j}^{1}e^{2} 
- \delta_{i}^{2}\delta_{i}^{3}\delta_{j}^{3}\delta_{j}^{2}e^{1}
- \delta_{i}^{3}\delta_{i}^{1}\delta_{j}^{2}\delta_{j}^{2}e^{3} 
\\&\quad+ \delta_{i}^{3}\delta_{i}^{1}\delta_{j}^{2}\delta_{j}^{3}e^{2}
+ \delta_{i}^{3}\delta_{i}^{2}\delta_{j}^{2}\delta_{j}^{1}e^{3} 
- \delta_{i}^{3}\delta_{i}^{2}\delta_{j}^{2}\delta_{j}^{3}e^{1}
-\delta_{i}^{3}\delta_{i}^{3}\delta_{j}^{2}\delta_{j}^{1}e^{2} 
+\delta_{i}^{3}\delta_{i}^{3}\delta_{j}^{2}\delta_{j}^{2}e^{1}\Bigg)\\
    \end{aligned} &&
\end{flalign}

\noindent Observe that the only non vanish terms after the sum runs over both indexes are $2\delta_{i}^{1}\delta_{i}^{1}\delta_{j}^{2}\delta_{j}^{2}e^{3}, $
$2\delta_{i}^{2} \delta_{i}^{2}\delta_{j}^{1}\delta_{j}^{1}e^{3}, $ 
$\delta_{i}^{2}\delta_{i}^{2}\delta_{j}^{3}\delta_{j}^{3}e^{1}, $
$\delta_{i}^{3}\delta_{i}^{3}\delta_{j}^{2}\delta_{j}^{2}e^{1}.$

\begin{flalign}
    \begin{aligned}
       \upalpha \wedge_{2} \boldsymbol{\uptau} =  
g^{11}
g^{22}2e^{3}
\!+ g^{22}(g^{11}2e^{3}
\!+g^{33}e^{1})
\!+g^{33}g^{22}e^{1}
\!= -2e^{3} \!- 2e^{3} \!+ e^{1}\! + e^{1}
\!= 2e^{1} - 4e^{3}
    \end{aligned} &&
\end{flalign}

\noindent Finally,

\begin{align}
    \begin{aligned}
    \upalpha \diamond \boldsymbol{\uptau} &= \upalpha \wedge \boldsymbol{\uptau} - \upalpha \wedge_{1} \boldsymbol{\uptau} - \frac{1}{2!} \upalpha \wedge_{2} \boldsymbol{\uptau} = - \frac{1}{2!} \upalpha \wedge_{2} \boldsymbol{\uptau} =  2e^{1} - 4e^{3}.\; \textcolor{Blue}{{\blacktriangleleft}}  
    \end{aligned}
\end{align}

\begin{tflemma}\label{lemma3202}
    A homogeneous $k$-form $\upalpha$ satisfies

    \begin{equation}
        \upalpha \diamond \boldsymbol{\uptau} = \frac{1}{k!}(-1)^{\left [ \frac{k}{2} \right]} \upalpha \wedge_{k} \boldsymbol{\uptau}.
    \end{equation}
\end{tflemma}
\noindent \textcolor{RoyalBlue}{\textit{Proof.}}  By Definition, given $\upalpha \in \Omega^{k}(\M)$, the product $\upalpha \diamond \boldsymbol{\uptau}$ is given by

\begin{equation}
    \upalpha \diamond \boldsymbol{\uptau} = \sum^{k}_{d=0} \frac{(-1)^{d(k-d) + \left [ \frac{d}{2} \right]}}{d!} \upalpha \wedge_d \boldsymbol{\uptau} .
\end{equation}

\noindent Without loss of generality, let us fix $\upalpha = e^{m_{1}} \wedge \cdots \wedge e^{m_{k}}$, $m_{t}\in \{1,\ldots, n\}$, $t \in \{1,\ldots, k\}$.
First it is straightforward to notice that for $d= 0$ , $\upalpha \wedge \boldsymbol{\uptau} = 0$ since $\boldsymbol{\uptau} = e^{1} \wedge \cdots \wedge e^{n}$ is the volume form which is composed by all the indexes that $\upalpha$ has. Let us consider $1 \leq d<k$. It follows that,

\begin{align}
    \begin{aligned}\label{eqalphavd}
        \upalpha \!\wedge_{d} \boldsymbol{\uptau} \!=\! \sum^{n}_{i_{1}=0}  g^{i_{1}i_{1}} \!\left(\sum^{n}_{i_{2}=0}  g^{i_{2}i_{2}}\! \left( \!\cdots\! \left( \sum^{n}_{i_{d}=0}  g^{i_{d}i_{d}} (\mathbf{e}_{i_{d}} \rfloor \!\cdots \! \mathbf{e}_{i_{2}}\rfloor  \mathbf{e}_{i_{1}}\rfloor \upalpha) \!\wedge\! (\mathbf{e}_{i_{d}} \rfloor \!\cdots\!  \mathbf{e}_{i_{2}}\rfloor  \mathbf{e}_{i_{1}}\rfloor\boldsymbol{\uptau}) \! \right) \!\cdots \!\right)\! \!\right)\!. 
    \end{aligned}
\end{align}

As a result $e_{i_{d}} \rfloor\cdots \rfloor  \mathbf{e}_{i_{1}}\rfloor\boldsymbol{\uptau}$ is a sum of $\frac{n!}{(n-d)!}$ $(n-d)$-forms of type $(\delta_{i_{1}}^{r_{1}} \cdots \delta_{i_{d}}^{r_{d}})u_\lambda$, $u_\lambda \in \Omega^{n-d}(\M)$; such that  $i_{a},r_{l} \in \{1,\ldots,n \}$ and $r_{l} \neq r_{h}$ if $h \neq l$ and $i_{a} \neq i_{b}$ if $a \neq b$,  $\lambda \in \{1, \ldots, \frac{n!}{(n-d)!} \}$ and $\mu \in \{1, \ldots, \frac{k!}{(k-d)!} \}$. On the other hand, $e_{i_{d}} \rfloor \cdots  \mathbf{e}_{i_{1}}\rfloor \upalpha$ is a sum of $\frac{k!}{(k-d)!}$  $(k-d)$-forms of type  $(\delta_{i_{1}}^{s_{1}} \cdots \delta_{i_{d}}^{s_{d}})w_\mu$, $w_\mu \in \Omega^{k-d}(\M)$, $\{ s_{1},\ldots, s_{d}\} \subset \{m_{1}, \ldots, m_{k}\}$, $s_{l} \neq s_{h}$ if $h \neq l$  and $i_{a} \neq i_{b}$ if $a \neq b$, $\mu \in \{1, \ldots, \frac{k!}{(k-d)!} \}$.

Suppose that $(\mathbf{e}_{i_{d}} \rfloor\cdots \rfloor  \mathbf{e}_{i_{1}}\rfloor\boldsymbol{\uptau}) \wedge (\mathbf{e}_{i_{d}} \rfloor \cdots  \mathbf{e}_{i_{1}}\rfloor \upalpha) = 0$, since the result is given by different combinations of $u_\lambda \wedge w_\mu$, they do not cancel with each other and consequently each combination $u_\lambda \wedge w_\mu = 0$. Each $\lambda$ generates a combination of $(n-d)$ indexes from $\{1,\ldots, n\}$ of $\boldsymbol{\uptau}$ that compose the basis elements of the remaining $(n-d)$-form $u_\lambda$, which compose the sum of the contracted $\boldsymbol{\uptau}$ $d$ times, fix $\mathcal{R}_{\lambda}$ the set of those indexes. For instance, for the case $d=1$

\begin{align} 
    \begin{aligned}
         \mathbf{e}_{i_{1}}\rfloor\boldsymbol{\uptau} &= (\mathbf{e}_{i_{1}}\rfloor e^{1} \wedge e^{2}\! \wedge \cdots \wedge e^{n}) \\
        &= \delta_{i_{1}}^{1} e^{2} \wedge \! \cdots\! \wedge e^{n} - \delta_{i_{1}}^{2} e^{1} \wedge e^{3}\! \wedge \!\cdots\! \wedge e^{n} + \cdots + (-1)^{n+1}\delta_{i_{1}}^{n} e^{1} 
        \wedge\! \cdots\! \wedge e^{n-1}\\
       &= \delta_{i_{1}}^{1}u_{1} - \delta_{i_{1}}^{2}u_{2} + \cdots + (-1)^{n+1}\delta_{i_{1}}^{n} u_{n}
    \end{aligned}
\end{align}

\noindent where $\mathcal{R}_{1}=\{2,\ldots, n\}$, $\mathcal{R}_{2}=\{1,3, \ldots, n\}$, $\mathcal{R}_{n}=\{1,\ldots, n-1\}$. Analogously, each $\mu$  generates a combination of $(k-d)$ indexes for $w_\mu$, that come from the indexes from $\upalpha$, that is, $\{m^{1}, \ldots, m^{k}\}$, fix $\mathcal{R}_{\mu}$ the set of those indexes. It follows that $u_\lambda \wedge w_\mu = 0$ for all $\lambda \in \{1, \ldots, \frac{n!}{(n-d)!} \}$ and $\mu \in \{1, \ldots, \frac{k!}{(k-d)!} \}$ if and only if $\mathcal{R}_{\lambda} \cap \mathcal{R}_{\mu} \neq \emptyset$ for all $\lambda,\mu$. Summarising,
\begin{align}
    \begin{aligned}
        (\mathbf{e}_{i_{d}} \rfloor\cdots \rfloor  \mathbf{e}_{i_{1}}\rfloor\boldsymbol{\uptau}) \wedge (\mathbf{e}_{i_{d}} \rfloor \cdots  \mathbf{e}_{i_{1}}\rfloor \upalpha) = 0 &\iff  u_\lambda \wedge w_\mu = 0 \text{ for all } \lambda, \mu\\
        &\iff \mathcal{R}_{\lambda} \cap \mathcal{R}_{\mu} \neq \emptyset \text{ for all }\lambda,\mu.
    \end{aligned}
\end{align}

Suppose now that there exists $\lambda^{*} \in \{1, \ldots, \frac{n!}{(n-d)!} \}$ and $\mu^{*} \in \{1, \ldots, \frac{k!}{(k-d)!} \}$ such that $\mathcal{R}_{\lambda^{*}} \cap \mathcal{R}_{\mu^{*}} = \emptyset$, consequently, $u_{\lambda^{*}} \wedge w_{\mu^{*}} \neq 0$ and $ (\mathbf{e}_{i_{d}} \rfloor\cdots \rfloor  \mathbf{e}_{i_{1}}\rfloor\boldsymbol{\uptau}) \wedge (\mathbf{e}_{i_{d}} \rfloor \cdots  \mathbf{e}_{i_{1}}\rfloor \upalpha) \neq 0$. \noindent We claim that if $u_{\lambda^{*}} \wedge w_{\lambda^{*}} \neq 0$ at least one of the $d$ contractions will contract different indexes of the element basis that compose $\boldsymbol{\uptau}$ and $\upalpha$. In fact, suppose the contrary, that is, all of the $d$ contractions $ (\mathbf{e}_{i_{d}} \rfloor\cdots \rfloor  \mathbf{e}_{i_{1}}\rfloor\boldsymbol{\uptau}), (\mathbf{e}_{i_{d}} \rfloor \cdots  \mathbf{e}_{i_{1}}\rfloor \upalpha)$  that rise respectively the terms $u_{\lambda^{*}}$ and $ w_{\lambda^{*}}$ acted on the same index simultaneously. To facilitate our calculations, observe that  the indexes $\{1,\ldots, n\}$ of $\boldsymbol{\uptau}$ can also be written in terms of the ones of $\upalpha$ as $\{m^{1}, \ldots, m^{k}, \ldots, m^{n-k}\}$. Therefore, $\boldsymbol{\uptau}$ is $e^{m^{1}} \wedge \cdots \wedge m^{k} \wedge \cdots \wedge e^{n-k}$ up to a signal, due the permutation, that will not interfere in the result and it can be ignored. Therefore, representing the $d$ contractions on the indexes of $\upalpha$ and $\boldsymbol{\uptau}$ as follows

\begin{align}
    \begin{aligned}
        \{m^{1}, m^{2}, \ldots, m^{d}, \ldots, m^{k}\} &\overset{\rfloor}{\rightarrow}   \{\cancel{m^{1}}, \cancel{m^{2}}, \ldots,\cancel{m^{d}}, \ldots, m^{k}\} \\ \{m^{1}, m^{2}, \ldots, m^{d}, \ldots, m^{k}, \ldots, m^{n-k}\}
      &\overset{\rfloor}{\rightarrow}    \{\cancel{m^{1}}, \cancel{m^{2}}, \ldots,\cancel{m^{d}}, \ldots, m^{k}, \ldots, m^{n-k}\}, 
    \end{aligned}
\end{align}

\noindent where $\cancel{m^{l}}$ represents $e_{i_{j}}\rfloor e^{m^{l}}$, one has $\mathcal{R}_{\lambda^{*}} = \{m^{d+1}, \ldots, m^{n-k}\} $ and $\mathcal{R}_{\mu^{*}} = \{m^{d+1}, \ldots, m^{k}\}$, which is a contradiction with the fact $\mathcal{R}_{\lambda^{*}} \cap \mathcal{R}_{\mu^{*}} = \emptyset$. We conclude that at least one of the $d$ contractions will contract different basis elements from $\boldsymbol{\uptau}$ and $\upalpha$. Let us consider the sets

\begin{align}
    \begin{aligned}
         \mathcal{C}_{\mu^{*}} &= \{m^{1}, m^{2}, \ldots, m^{d}, \ldots, m^{k}\} \setminus \mathcal{R}_{\mu^{*}}\\        
        \mathcal{C}_{\lambda^{*}} &= \{m^{1}, m^{2}, \ldots, m^{d}, \ldots, m^{k}, \ldots, m^{n-k}\} \setminus \mathcal{R}_{\lambda^{*}}
    \end{aligned}
\end{align}

\noindent $\mathcal{C}_{\mu^{*}}$ is the sets of the indexes that have been contracted from $\boldsymbol{\uptau}$ with respect to the $u_{\lambda^{*}}$ and $ \mathcal{C}_{\lambda^{*}}$ is the of the indexes that have been contracted from $\upalpha$ with respect to the $w_{\mu^{*}}$.  Therefore, for a contraction $e_{i_{x}}\rfloor$ that contracted different indexes, there exists $y \in \mathcal{C}_{\mu^{*}}$ and $y' \in \mathcal{C}_{\lambda^{*}}$ with $y \neq y'$. This means that $w_{\mu^{*}}$ is multiplied by $\delta_{i_{x}}^{y}$ and $u_{\lambda^{*}}$ is multiplied by $\delta_{i_{x}}^{y'}$. Therefore, $w_{\mu^{*}} \wedge u_{\lambda^{*}}$ is multiplied by  $\delta_{i_{x}}^{y}\delta_{i_{x}}^{y'}$, consequently, when the sum runs over the index $i_{x}$ it follows that $\delta_{i_{x}}^{y}\delta_{i_{x}}^{y'}$ will vanish. Hence, the non-null terms that constitute $(\mathbf{e}_{i_{d}} \rfloor\cdots \rfloor  \mathbf{e}_{i_{1}}\rfloor\boldsymbol{\uptau}) \wedge (\mathbf{e}_{i_{d}} \rfloor \cdots  \mathbf{e}_{i_{1}}\rfloor \upalpha)$ are composed by elements in which at least one contraction acts simultaneously on the same basis element leading to products of $\delta$ that will vanish the term when the sum runs over the respective index. We conclude that for $d<k$ if $(\mathbf{e}_{i_{d}} \rfloor\cdots \rfloor  \mathbf{e}_{i_{1}}\rfloor\boldsymbol{\uptau}) \wedge (\mathbf{e}_{i_{d}} \rfloor \cdots  \mathbf{e}_{i_{1}}\rfloor \upalpha) = 0 $ then $\upalpha \wedge_{d} \boldsymbol{\uptau} = 0$ by Eq. \eqref{eqalphavd}. If $(\mathbf{e}_{i_{d}} \rfloor\cdots \rfloor  \mathbf{e}_{i_{1}}\rfloor\boldsymbol{\uptau}) \wedge (\mathbf{e}_{i_{d}} \rfloor \cdots  \mathbf{e}_{i_{1}}\rfloor \upalpha) \neq 0$ then, by the last discussion, the non-null terms will vanish due the sums, resulting $\upalpha \wedge_{d} \boldsymbol{\uptau} = 0$. Therefore,

\begin{align}
    \begin{aligned}
        \upalpha \diamond \boldsymbol{\uptau} = \sum^{k}_{d=0} \frac{(-1)^{d(k-d) + \left [ \frac{d}{2} \right]}}{d!} \upalpha \wedge_d \boldsymbol{\uptau}  = \frac{(-1)^{k(k-k) + \left [ \frac{k}{2} \right]}}{k!} \upalpha \wedge_k \boldsymbol{\uptau} = \frac{1}{k!}(-1)^{\left [ \frac{k}{2} \right]} \upalpha \wedge_k \boldsymbol{\uptau}. \;\textcolor{RoyalBlue}{\Box}
    \end{aligned}
\end{align}

\begin{tfpropos}
    If $n = \dim M$ is odd, then $\boldsymbol{\uptau}$ is central on the Kähler-Atiyah bundle.
\end{tfpropos}
\noindent \textcolor{RoyalBlue}{\textit{Proof.}} Given $\upalpha \in \Omega^{k}(\M)$, from Lemma \ref{lemma3202} it follows that 

\begin{align}
    \begin{aligned}\label{eq270g}
        \upalpha \diamond \boldsymbol{\uptau}  = \frac{1}{k!}(-1)^{\left [ \frac{k}{2} \right]} \upalpha \wedge_k \boldsymbol{\uptau}. 
    \end{aligned}
\end{align}

\noindent On the other hand, from Definition \ref{defin_graf1} and from Lemma \ref{lemma3202} one has

\begin{align}
    \begin{aligned}\label{eq271g}
       \boldsymbol{\uptau} \diamond \upalpha = (-1)^{kn} \sum^{k}_{d=0} \frac{(-1)^{d(k-d+1) + \left [  \frac{d}{2} \right]}}{d!}  \upalpha \wedge_d \upbeta =\frac{1}{k!}   (-1)^{k(n+1) + \left [  \frac{k}{2} \right]}  
 \upalpha \wedge_d \upbeta .
    \end{aligned}
\end{align}

\noindent Therefore, Eqs. \eqref{eq270g} and \eqref{eq271g} yield

\begin{align}
    \begin{aligned}
        \upalpha \diamond \boldsymbol{\uptau} = (-1)^{k(n+1)} \boldsymbol{\uptau} \diamond \upalpha.
    \end{aligned}
\end{align}

\noindent If $n$ is odd then $k(n+1)$ is even for any $k \in \mathbb{N}$, consequently

\begin{align}
    \begin{aligned}
        \upalpha \diamond \boldsymbol{\uptau} =  \boldsymbol{\uptau} \diamond \upalpha.
    \end{aligned}
\end{align}

\noindent That is, $\boldsymbol{\uptau}$ commutes with any $\upalpha \in \Omega^{k}(\M)$. Then, by linearity of the $\diamond$ product, we conclude that $\boldsymbol{\uptau}$ commutes with any $\upalpha \in \Omega(\M)$ whenever $\dim \M = n$ is odd. \textcolor{RoyalBlue}{$\Box$}

\section{The Hodge Duality Operator}
\paragraph{ } From the volume form, in this section we define the Hodge operator and explore important consequences that this operator leads in the Kähler-Atiyah algebra.

\begin{definanil}\label{def3204}
    The \textbf{Hodge operator} is defined as being the following mapping

    \begin{align}
        \begin{aligned}
            \star : \;\; \Omega^{k}(\M)\;\;\; &\to \;\;\;\Omega^{n-k}(\M)\\
            \upalpha \;\;\;\;\;&\mapsto\;\;\;\;\; \upalpha \diamond \boldsymbol{\uptau}.
        \end{aligned}
    \end{align}
\end{definanil}

This operator is well-defined since from Lemma \ref{lemma3202} one has

\begin{align}
    \begin{aligned}
        \star\upalpha = \upalpha \diamond \boldsymbol{\uptau}  = \frac{1}{k!}(-1)^{\left [ \frac{k}{2} \right]} \upalpha \wedge_k \boldsymbol{\uptau}. 
    \end{aligned}
\end{align}

From Definition \ref{def3204}, the next identities hold:

\begin{align}
    \begin{aligned}
        \star 1 &= 1 \diamond \boldsymbol{\uptau} = \frac{(-1)^{0(0-0)+\left [  \frac{0}{2} \right]}}{0!}1 \wedge_{0} \boldsymbol{\uptau} = 1 \diamond \boldsymbol{\uptau} = \boldsymbol{\uptau},\\
        \star \boldsymbol{\uptau} &=     \boldsymbol{\uptau} \diamond \boldsymbol{\uptau} = \color{RoyalBlue}\begin{cases}
       \color{black}  +1, \text{ if } p - q \equiv_8 0,1,4,5\\\color{black}
        -1, \text{ if } p - q \equiv_8 2,3,6,7\color{black}
    \end{cases}\color{black}.
    \end{aligned}
\end{align}

Let $\upalpha$ be an $k$-form in $\Omega^{k}(\M)$. Then, applying the Hodge operator twice results in the following

\begin{align}\label{eq_doublestar}
    \begin{aligned}
        \star^{2} (\upalpha) &= \star (\star \upalpha)\\
        &= \star (\upalpha \diamond \boldsymbol{\uptau})\\
        &= (\upalpha \diamond \boldsymbol{\uptau}) \diamond \boldsymbol{\uptau}\\
        &= \upalpha \diamond (\boldsymbol{\uptau} \diamond \boldsymbol{\uptau})\\
        &= (-1)^{k \cdot 0}(\boldsymbol{\uptau} \diamond \boldsymbol{\uptau}) \wedge \upalpha\\
        &=\color{RoyalBlue}\begin{cases}\color{black}
        1 \wedge \upalpha = \upalpha, \text{ if } p - q \equiv_{8} 0,1,4,5 \\\color{black}
        -1 \wedge \upalpha = -\upalpha, \text{ if } p - q \equiv_8 2,3,6,7
    \end{cases}\color{black}.
    \end{aligned}
\end{align}

\noindent This result can be applied to any arbitrary form $\upalpha \in \Omega(\M)$ by linearity. From the Hodge operator, the following two sets can be defined 

\begin{equation}
    \Omega^{\pm}_{\star^{2}} = \{\upalpha \in \Omega(\M): \star^{2}\upalpha = \pm \upalpha\}.
\end{equation}

\noindent  By doing so, we obtain a $\mathbb{Z}_2$-grading for the geometric algebra of forms, as shown below

\begin{equation}
    \Omega(\M) = \Omega^{+}_{\star^{2}} \oplus \Omega^{-}_{\star^{2}}.
\end{equation}

Consider the elements

\begin{equation}
    \uprho_{\pm} := \frac{1}{2} (1 \pm \boldsymbol{\uptau}) \in  \Omega^{0}(\M) \oplus \Omega^{n}(\M).
\end{equation}

\noindent These algebraic objects  satisfy the following properties:
\begin{itemize}
    \item [\textcolor{RoyalBlue}{\textbf{\textit{1.}}}] $\uprho_{+} + \uprho_{-} = 1,$
     \item [\textcolor{RoyalBlue}{\textbf{\textit{2.}}}] $ \uprho_{\pm} \diamond  \uprho_{\pm} = \frac{1}{4} (1 \pm \boldsymbol{\uptau})^{2} =  \frac{1}{4} (1 \pm 2\boldsymbol{\uptau} + \boldsymbol{\uptau} \diamond \boldsymbol{\uptau}) 
     = \color{RoyalBlue}\begin{cases}
        \color{black} \frac{1}{2} (1 \pm \boldsymbol{\uptau}), \text{ if } p - q \equiv_8 0,1,4,5 \\\color{black}
       \pm \frac{1}{2} \boldsymbol{\uptau}, \text{ if } p - q \equiv_8 2,3,6,7 
    \end{cases}\color{black},$ 
    \item [\textcolor{RoyalBlue}{\textbf{\textit{3.}}}] $\uprho_{\pm} \diamond \uprho_{\mp} = \frac{1}{4} (1 \pm \boldsymbol{\uptau}) (1 \mp \boldsymbol{\uptau})= \frac{1}{4} (1 - \boldsymbol{\uptau} +\boldsymbol{\uptau} - \boldsymbol{\uptau} \diamond \boldsymbol{\uptau}) 
    = \color{RoyalBlue}\begin{cases}\color{black}
         0, \text{ if } p - q \equiv_8 0,1,4,5\\\color{black}
        \frac{1}{2} \boldsymbol{\uptau}, \text{ if } p - q \equiv_8 2,3,6,7
    \end{cases}\color{black}.$
\end{itemize}

Two operators can be defined through right $\diamond$-multiplication with these elements by setting for $\upalpha \in \Omega(\M):$

\begin{align}
    \begin{aligned}
        P_{\pm} (\upalpha) &:= \upalpha \diamond \rho_{\pm} = \frac{1}{2}(\upalpha \diamond 1 \pm \upalpha \diamond \boldsymbol{\uptau})= \frac{1}{2}(\upalpha \pm \upalpha \diamond \boldsymbol{\uptau})= \frac{1}{2}( \upalpha \pm \star \upalpha).
    \end{aligned}
\end{align}

\noindent That is,

   {\colorlet{shadecolor}{RoyalBlue!20}\begin{shaded}\begin{equation}\label{eq_asdfgh62525234}
    P_{\pm} = \frac{1}{2} (1 \pm \star).
\end{equation}\end{shaded}}

\noindent Note that if $p - q \equiv_8 0,1,4,5$, then the $\uprho_{\pm}$ are two mutually orthogonal idempotents in $\Omega(\M)$. Furthermore, we conclude that whenever $p - q \equiv_8 0,1,4,5$, the elements $P_{\pm}$ are complementary mutually orthogonal idempotents:

\begin{itemize}
    \item [\textcolor{RoyalBlue}{\textbf{\textit{1.}}}] $P_{\pm}^{2} = P_{\pm}$,
     \item [\textcolor{RoyalBlue}{\textbf{\textit{2.}}}] $P_{\pm} \circ P_{\mp} = 0$, 
     \item [\textcolor{RoyalBlue}{\textbf{\textit{3.}}}] $P_{+} + P_{-} = 1$.
\end{itemize}

The images

   {\colorlet{shadecolor}{RoyalBlue!20}\begin{shaded}\begin{equation}
    P_{\pm}\Omega := P_{\pm} \left( \Omega(\M)\right) = \Omega(\M) \diamond \rho_{\pm}
\end{equation}\end{shaded}}

\noindent give rise to the next proposition

\begin{tfpropos}\label{propos3301}
    If $p - q \equiv_8 0,1,4,5;$  the direct sum decomposition $\Omega(\M)=   P_{+}\Omega\oplus    P_{-}\Omega$ holds.
\end{tfpropos}
\noindent \textcolor{RoyalBlue}{\textit{Proof.}} Take $\upalpha \in \Omega(\M)$ and define $\upalpha_{+}$ and $\upalpha_{-}$ as

\begin{align}
\begin{aligned}
    \upalpha_{\pm} = \upalpha \diamond \rho_{\pm} = \frac{1}{2}(\upalpha \pm \star \upalpha) \in  P_{\pm}\Omega.
\end{aligned}
\end{align}

\noindent Both $\upalpha$ and $\star \upalpha$ can be written uniquely in terms of $\upalpha_{\pm}$ as

\begin{align}
\begin{aligned}\label{eq_fpfmfs}
     \upalpha &= \upalpha_{+} + \upalpha_{-} \in P_{+}\Omega +     P_{-}\Omega \\
     \star \upalpha &= \upalpha_{+} - \upalpha_{-}.
\end{aligned}
\end{align}

\noindent We claim that $P_{+}\Omega\cap    P_{-}\Omega = \{0\}$ holds only for $p - q \equiv_8 0,1,4,5$. In fact, consider $\upbeta \in P_{+}\Omega\cap    P_{-}\Omega$. Hence there are $x,y \in \Omega(\M)$ such that

\begin{align}
\begin{aligned}
    \upbeta  = x \diamond p_{+} = \frac{1}{2}(x + \star x)\\
    \upbeta  = y \diamond p_{-} = \frac{1}{2}(y - \star y)
\end{aligned}
\end{align}

\noindent That is,  $x + \star x = y - \star y$. By applying the Hodge operator considering $p - q \equiv_8 0,1,4,5$ and Eq. \eqref{eq_doublestar} it follows that

\begin{align}
\begin{aligned}
     \star x + \star^{2}x &= \star y - \star^{2}y \\  \star x + x &= \star y - y \\
    \frac{1}{2} (\star x + x)&= \frac{1}{2} (- y + \star y ) \\
      \upbeta &= -\upbeta \implies \upbeta = 0.
\end{aligned}
\end{align}

\noindent For the case $p - q \equiv_8 2,3,6,7 $ it is not possible to conclude that $\upbeta = 0$. Therefore, $\Omega(\M)= P_{+}\Omega +    P_{-}\Omega$ and for $p - q \equiv_8 0,1,4,5$ one has $P_{+}\Omega \cap    P_{-}\Omega = \{0\}$. We conclude then $\Omega(\M)=   P_{+}\Omega\oplus    P_{-}\Omega$ holds for $p - q \equiv_8 0,1,4,5.\;\; \textcolor{RoyalBlue}{\Box}$

\paragraph{ } Let us make a consideration about the behaviour of $\star \upa$ of a $\upa \in P_{\pm}\Omega$. Consider $\upa \in  P_{+}\Omega$ and $\upb \in P_{-}\Omega$ such that:

    \begin{align}
\begin{aligned}
    \upa = \upomega \diamond \rho_{+} = \frac{1}{2}(\upomega + \star \upomega)\\
    \upb = \upomega \diamond \rho_{-} = \frac{1}{2}(\upomega - \star \upomega)
\end{aligned}
\end{align}

For $\upomega \in \Omega(\M)$. From Eq. \eqref{eq_doublestar} one has two cases:

  \begin{align}
\begin{aligned}\label{eq_starzero0145}
\textcolor{RoyalBlue}{\textbf{\textit{1)}}}\;\; p - q \equiv_8 0,1,&4,5:\\
   \star \upa  &= \frac{1}{2}(\star \upomega + \star^2 \upomega) = \frac{1}{2}(\star \upomega +  \upomega) = \upa, \\
    \star \upb  &= \frac{1}{2}(\star \upomega - \star^{2} \upomega) = \frac{1}{2}(\star \upomega -  \upomega)  = -\upb.
\end{aligned}
\end{align}

   \begin{align}
\begin{aligned}
\textcolor{RoyalBlue}{\textbf{\textit{2)}}}\;\; p - q \equiv_8 2,3,&6,7:\\
   \star \upa  &= \frac{1}{2}(\star \upomega + \star^2 \upomega) = \frac{1}{2}(\star \upomega - \upomega) = -\upb, \\
    \star \upb  &= \frac{1}{2}(\star \upomega - \star^{2} \upomega) = \frac{1}{2}(\star \upomega + \upomega)  = \upa.
\end{aligned}
\end{align}

\begin{tfpropos}\label{propos3302}
    If $p - q \equiv_8 0,1,4,5$  and $n = p+q$ is odd, then $P_{\pm}\Omega$ are subalgebras of  $(\Omega(\M), \diamond)$  with respective units $\rho_{\pm}$.
\end{tfpropos}

\noindent \textcolor{RoyalBlue}{\textit{Proof.}} $\upalpha, \upbeta \in P_{\pm}\Omega$. Therefore, there are $x_{\pm}, y_{\pm} \in \Omega(\M)$ such that

\begin{align}
    \begin{aligned}
        \upalpha \diamond \upbeta = (x_{\pm} \diamond \rho_{\pm}) \diamond (y_{\pm} \diamond \rho_{\pm}).
    \end{aligned}
\end{align}

\noindent Due the associativity of the $\diamond$ product, the fact that $\rho_{\pm}$ is central and idempotent since $n$ is odd and $p - q \equiv_8 0,1,4,5$, it follows that

\begin{align}
    \begin{aligned}
        \upalpha \diamond \upbeta &= (x_{\pm} \diamond \rho_{\pm}) \diamond (y_{\pm} \diamond \rho_{\pm})\\
        &= (x_{\pm}  \diamond y_{\pm})\diamond (\rho_{\pm} \diamond \rho_{\pm})\\
        &= (x_{\pm}  \diamond y_{\pm})\diamond \rho_{\pm}.
    \end{aligned}
\end{align}

\noindent Which implies that $ \upalpha \diamond \upbeta \in P_{\pm}\Omega$, that is, those subsets are closed with respect to the $\diamond$ product. In addition, since $\rho_{\pm}$ is idempotent,

\begin{align}
    \begin{aligned}\label{eqsalvastar}
        P_{\pm} (\upalpha) = \upalpha \diamond \rho_{\pm} = (x_{\pm} \diamond \rho_{\pm}) \diamond \rho_{\pm} =x_{\pm} \diamond \rho_{\pm}= \upalpha. 
    \end{aligned}
\end{align}

\noindent One also has that $\upalpha \diamond \rho_{\pm} = \rho_{\pm} \diamond  \upalpha = \upalpha$ due the fact that $\rho_{\pm}$ is  central. That means that $\rho_{\pm}$ is the identity of the subalgebras $(P_{\pm}\Omega, \diamond)$. \textcolor{RoyalBlue}{$\Box$}

\begin{tfpropos}\label{cor_projendo} If $p - q \equiv_8 0,1,4,5$  and $n = p+q$ is odd, the operators $P_{\pm}: \Omega(\M) \to P_{\pm}\Omega \subset \Omega(\M) $ are homomorphisms.
\end{tfpropos}
\noindent \textcolor{RoyalBlue}{\textit{Proof.}} Since $p - q \equiv_8 0,1,4,5$  and $n$ is odd, the element $\rho_{\pm}$ is central.  For $\upa , \upb \in \Omega(\M)$, it follows that

\begin{equation}
   P_{\pm}(\upa) \diamond P_{\pm} (\upb) = (\upa \diamond \rho_{\pm}) \diamond (\upb \diamond \rho_{\pm}) = (\upa \diamond \upb) \diamond \rho_{\pm} = P_{\pm} (\upa \diamond \upb). \;\;\textcolor{RoyalBlue}{\Box}
\end{equation}

\section{The Truncated Algebra}

\paragraph{ } In this section we present a first model for the algebra $P_{\pm}\Omega$, \textit{the truncated algebra}. We start by considering the sets

\begin{align}
    \begin{aligned}
        \Omega_{L} &:= \bigoplus_{k=0}^{\left [  \frac{n}{2} \right]} \Omega^{k}(\M),\\ 
         \Omega_{U} &:= \bigoplus_{k=\left [  \frac{n}{2} \right] +1}^{n } \Omega^{k}(\M),
    \end{aligned}
\end{align}

\noindent that gives the splitting
    $\Omega(\M) = \Omega_{L} + \Omega_{U}.$

    \begin{definanil}\label{def3301}
    The \textbf{lower truncation} is defined by the mapping 
    
    \begin{align}
        \begin{aligned}
            P_{L} : \;\; \Omega(\M) \;\;\; &\to \;\;\;\bigoplus_{k=0}^{\left [  \frac{n}{2} \right]} \Omega^{k}(\M)\\
            \upa = \sum_{k=0}^{n} \upa^{j}_{I_{k}} e^{I_{k}} \;\;\;\;\;&\mapsto\;\;\;\;\;  \upa_{L} = \sum_{k=0}^{\left [  \frac{n}{2} \right]} \upa^{j}_{I_{k}} e^{I_{k}}
        \end{aligned}
    \end{align}
      whereas \textbf{upper truncation} is defined by 
  \begin{align}
        \begin{aligned}
             P_{U} : \;\; \Omega(\M) \;\;\; &\to \;\;\;\bigoplus_{k=\left [  \frac{n}{2} \right] +1}^{n } \Omega^{k}(\M)\\
            \upa = \sum_{k=0}^{n} \upa^{j}_{I_{k}} e^{I_{k}} \;\;\;\;\;&\mapsto\;\;\;\;\;  \upa_{U} = \sum_{k=\left [  \frac{n}{2} \right]+1}^{n} \upa^{j}_{I_{k}} e^{I_{k}}.
        \end{aligned}
    \end{align}
\end{definanil}

\begin{tfpropos}\label{propos3303}
    Given $\upalpha \in P_{\pm}(\M)$, if $p - q \equiv_8 0,1,4,5$ it follows that
    \begin{equation}
        2P_{\pm} (P_{L}(\upa)) = \upa.
    \end{equation}
\end{tfpropos}
\noindent \textcolor{RoyalBlue}{\textit{Proof.}} Write the element $\upa \in P_{\pm}\Omega$ as $\upa = \upalpha_L + \upalpha_U$. By Eq. \eqref{eq_starzero0145} for $\upa \in P_{\pm}\Omega$, if $p - q \equiv_8 0,1,4,5$ then $\star\upa = \pm \upa$. That is,

\begin{align}
    \begin{aligned}
      \pm (\upalpha_L + \upalpha_U) =  \pm \upa = \star \upa = \star (\upalpha_L + \upalpha_U)   = \star \upalpha_L + \star  \upalpha_U 
            \end{aligned}
\end{align}

 On the other hand, one can notice that for $\upalpha_L \in \Omega_{L}$ and $\upalpha_U \in \Omega_{U}$, due the nature of the Hodge operator: $\star \upalpha_L \in \Omega_{U}$, $\star \upalpha_U \in \Omega_{L}$. Consequently,

 \begin{align}
     \begin{aligned}
         \upalpha_U = \pm \star \upalpha_L,\\
         \upalpha_L = \pm \star \upalpha_U.
     \end{aligned}
 \end{align}

 Finally, 

 \begin{align}
     \begin{aligned}
         \upa &= \upalpha_L + \upalpha_U \\
         &= \upalpha_L \pm \star \upalpha_L\\
         &= 2 P_{\pm} (\upalpha_L)\\
         &= 2 P_{\pm}(P_L(\upa))
     \end{aligned}
 \end{align}

\noindent concluding the proof. \textcolor{RoyalBlue}{$\Box$}

\begin{definanil}\label{def3302}
    The \textbf{truncated product} is defined as follows

  \begin{align}
        \begin{aligned}
            \bomdia_{\pm} : \;\; \Omega(\M) \cross \Omega(\M) \; &\to \;\;\;\;\; \Omega_{L} \subset \Omega(\M) \\
            (\upa,\upb)\;\;\;\;\;\;\;\;\;&\mapsto\;\;  2 P_L(P_{\pm}(\upa) \diamond P_{\pm}(\upb)).
        \end{aligned}
    \end{align}
\end{definanil}

 Observe that from Example \ref{extruncated}, for the 6-dimensional manifold, $\upalpha = e^{1} + e^{36} \in \Omega_{L}$ and $\upbeta = e^{1} + e^{2} + e^{14} + e^{25} \in \Omega_{L}$, however, $\upalpha \diamond \upbeta = e^{4} + e^{12} + e^{125} + e^{136} - e^{1346} - e^{2356} \notin  \Omega_{L}$.  Which means that $P_L$ does not preserve the geometric product on its entire domain of definition $\Omega(\M)$.  Therefore $(\Omega_{L},\diamond)$ is not a subalgebra of $(\Omega(\M), \diamond)$ while $(\Omega_{L}, \bomdia_{\pm})$ is a subalgebra of the algebra $(\Omega(\M), \bomdia_{\pm})$ called \textit{truncated algebra}.

 We assume hereinafter that we are in the case with $n$ odd and $p - q \equiv_8 0,1,4,5$. In these conditions, from Proposition \ref{cor_projendo} the truncated product becomes $2 P_L(P_{\pm}(\upa) \diamond P_{\pm}(\upb)) = 2 P_L(P_{\pm}(\upa \diamond \upb))$. \newline

 \begin{tfpropos}\label{propos3304}
    $P_\pm$ is an algebra homomorphism between $(\Omega(\M), \diamond)$ and $(\Omega(\M), \bomdia_{\pm})$.
\end{tfpropos}
\noindent \textcolor{RoyalBlue}{\textit{Proof.}} In fact, consider $\upa, \upb \in \Omega(\M)$, it follows that

\begin{align}
    \begin{aligned}
        P_{\pm} (\upa \bomdia_{\pm} \upb) &= P_{\pm}(2 P_L(P_{\pm}(\upa \diamond \upb)))
    \end{aligned}
\end{align}

\noindent By Proposition \ref{propos3302}, the element $P_{\pm}(\upa \diamond \upb)$ lies on the subalgebra $ P_{\pm}\Omega$, therefore, Proposition \ref{propos3303} yields

\begin{align}
    \begin{aligned}
        P_{\pm} (\upa \bomdia_{\pm} \upb) &= P_{\pm}(2 P_L(P_{\pm}(\upa \diamond \upb))) = P_{\pm}(\upa \diamond \upb) = P_{\pm}(\upa) \diamond P_{\pm}(\upb). \;\;\textcolor{RoyalBlue}{\Box}
    \end{aligned}
\end{align}

\begin{tfcor}
For $\upa, \upb \in \Omega(\M)$,    $\upa \bomdia_{\pm} \upb = P_{\pm} (\upa)  \bomdia_{\pm} P_{\pm}(\upb)$.
\end{tfcor}
\noindent \textcolor{RoyalBlue}{\textit{Proof.}} Let us consider $\upa, \upb \in \Omega(\M)$, since $P_{\pm}$ are idempotents, 
\begin{align}
    \begin{aligned}
         P_{\pm} (\upa)  \bomdia_{\pm} P_{\pm}(\upb) &=  2 P_L(P_{\pm}(P_{\pm}(\upa) \diamond P_{\pm}(\upb))) \\
         &= 2 P_L(P_{\pm}(P_{\pm}(\upa \diamond \upb)))\\
         &= 2 P_L(P_{\pm}(\upa \diamond \upb))\\
         &= 2 P_L( P_{\pm} (\upa \bomdia_{\pm} \upb))\\    
    \end{aligned}
\end{align}

\noindent Applying $(P_{\pm}\circ P_{\pm})^{2} = 1$ in both sides and using Proposition \ref{propos3303}, 

\begin{align}
    \begin{aligned}
        P_{\pm} (\upa)  \bomdia_{\pm} P_{\pm}(\upb) &= (P_{\pm}\circ P_{\pm}) 2 P_L( P_{\pm} (\upa \bomdia_{\pm} \upb))\\
        &= P_{\pm} (P_{\pm} ( 2 P_L( P_{\pm} (\upa \bomdia_{\pm} \upb))))\\
        &= P_{\pm}(P_{\pm} (\upa \bomdia_{\pm} \upb))\\
        &=\upa \bomdia_{\pm} \upb.\;\; \textcolor{RoyalBlue}{\Box}
    \end{aligned}
\end{align}

\begin{tfpropos} $(\Omega(\M), \diamond)$ and $(\Omega(\M), \bomdia_{\pm})$ have the same unit $1$.
\end{tfpropos}

\noindent \textcolor{RoyalBlue}{\textit{Proof.}} Let us consider $\upa \in \Omega_{L}$. Therefore,
\begin{align}
    \begin{aligned}
        \upa \bomdia_{\pm} 1&= 2 P_{L} (P_{\pm} (\upa \diamond 1))\\
        &= 2 P_{L} (P_{\pm} (\upa))\\
               &=  P_{L} (\upa \pm \upa \diamond \boldsymbol{\uptau})\\
        &=   P_{L} (\upa) \pm P_L (\upa \diamond \boldsymbol{\uptau})\\
       &= \upa.
    \end{aligned}
\end{align}

\noindent We conclude that, $1 \bomdia_{\pm} \upa = 2 P_{L} (P_{\pm} (1 \diamond \upa))  = 2 P_{L} (P_{\pm} (\upa \diamond 1))  = \upa \bomdia_{\pm}  1 = \upalpha$. \textcolor{RoyalBlue}{$\Box$}

\begin{tfpropos}\label{propos3306}
  $(\Omega_{L}, \bomdia_{\pm}) \simeq (P_{\pm}\Omega, \diamond).$
\end{tfpropos}
\noindent \textcolor{RoyalBlue}{\textit{Proof.}} Consider the following mappings \begin{align}
    \begin{aligned}
P_{\pm}\!\restriction_{\Omega_{L}} : (\Omega_{L}, \bomdia_{\pm}) \to (P_{\pm}\Omega, \diamond),\\
2P_{L}\!\restriction_{P_{\pm}\Omega} : (P_{\pm}\Omega, \diamond) \to (\Omega_{L}, \bomdia_{\pm}).
    \end{aligned}
\end{align}

\noindent We claim that they are homomorphisms of subalgebras and inverses of each other. In fact, for $\upa,\upb \in \Omega_{L}$, by Propositions \ref{cor_projendo} and \ref{propos3303}, it follows that

\begin{align}
    \begin{aligned}\label{eq2109}
P_{\pm}\!\restriction_{\Omega_{L}} (\upa   \bomdia_{\pm} \upb) &=   P_{\pm}\!\restriction_{\Omega_{L}} (2 P_L  (P_\pm (\upa \diamond \upb)) \\
&= P_{\pm} (\upa \diamond \upb)\\
&= P_{\pm}(\upa) \diamond P_{\pm} ( \upb)\\
&=  P_{\pm}\!\restriction_{\Omega_{L}}  (\upa) \diamond P_{\pm}\!\restriction_{\Omega_{L}}  (\upb).
\end{aligned}
\end{align}

\noindent Moreover, for $\upomega,\upzeta \in P_{\pm}\Omega$,

\begin{align}
    \begin{aligned}\label{eq2110}
        2P_{L}\!\restriction_{P_{\pm}\Omega} (\upomega \diamond \upzeta) &= 2 P_{L}\!\restriction_{P_{\pm}\Omega} (2 P_{\pm}(P_{L}\!\restriction_{P_{\pm}\Omega}(\upomega)) \diamond 2P_{\pm}( P_{L}\!\restriction_{P_{\pm}\Omega}(\upzeta)))\\
        &= 4 (2P_{L}\!\restriction_{P_{\pm}\Omega}( P_{\pm} (P_{L}\!\restriction_{P_{\pm}\Omega}(\upomega) \diamond P_{L}\!\restriction_{P_{\pm}\Omega}(\upzeta))))\\
        &= 4 ( P_{L}\!\restriction_{P_{\pm}\Omega}(\upomega) \bomdia_{\pm} P_{L}\!\restriction_{P_{\pm}\Omega}(\upzeta))\\
        &=   2P_{L}\!\restriction_{P_{\pm}\Omega}(\upomega) \bomdia_{\pm} 2P_{L}\!\restriction_{P_{\pm}\Omega}(\upzeta).
    \end{aligned}
\end{align}

\noindent Consequently, by Eq.\eqref{eq2109}:

\begin{align}
    \begin{aligned}         2P_{L}\!\restriction_{P_{\pm}\Omega} (P_{\pm}\!\restriction_{\Omega_{L}} (\upa   \bomdia_{\pm} \upb)) &= 2P_{L}\!\restriction_{P_{\pm}\Omega} ( P_{\pm}\!\restriction_{\Omega_{L}}  (\upa) \diamond P_{\pm}\!\restriction_{\Omega_{L}}  (\upb)) = \upa   \bomdia_{\pm} \upb.
    \end{aligned}
\end{align}

\noindent Thus, the mapping $(2P_{L}\!\restriction_{P_{\pm}\Omega}) \circ (P_{\pm}\!\restriction_{\Omega_{L}})$ is the identity map in $(\Omega_{L}, \bomdia_{\pm})$. In addition, from Eq. \eqref{eq2110} and Propositions \ref{cor_projendo}, \ref{propos3304} and \ref{propos3303} it follows that

\begin{align}
    \begin{aligned}
        P_{\pm}\!\restriction_{\Omega_{L}} (2P_{L}\!\restriction_{P_{\pm}\Omega} (\upomega \diamond \upzeta)) &=
 P_{\pm}\!\restriction_{\Omega_{L}} (2P_{L}\!\restriction_{P_{\pm}\Omega}(\upomega) \bomdia_{\pm} 2P_{L}\!\restriction_{P_{\pm}\Omega}(\upzeta)) \\
 &=P_{\pm}\!\restriction_{\Omega_{L}}((2P_{L}\!\restriction_{P_{\pm}\Omega}(\upomega)) \diamond  2P_{L}\!\restriction_{P_{\pm}\Omega}(\upzeta) )\\
 &= P_{\pm}\!\restriction_{\Omega_{L}}(2P_{L}\!\restriction_{P_{\pm}\Omega}(\upomega)) \diamond  P_{\pm}\!\restriction_{\Omega_{L}}(2P_{L}\!\restriction_{P_{\pm}\Omega}(\upzeta))\\
 &= \upomega \diamond \upbeta,
    \end{aligned}
\end{align}

\noindent showing that the mapping $(P_{\pm}\!\restriction_{\Omega_{L}}) \circ (2P_{L}\!\restriction_{P_{\pm}\Omega})$ is the identity map in $(P_{\pm}\Omega, \diamond)$. Therefore, we conclude that those mapping are inverses of each other establishing the isomorphism $(\Omega_{L}, \bomdia_{\pm}) \simeq (P_{\pm}\Omega, \diamond). \;\;\textcolor{RoyalBlue}{\Box}$

{\colorlet{shadecolor}{RoyalBlue!20}\begin{shaded} 
 Henceforth, the ensuing identification has been ascertained in the following diagram.
\[ 
\begin{tikzcd}[row sep=4.5em, xscale=2, yscale=2, scale=0.5, transform shape]
(\Omega_{L}, \bomdia_{\pm}) \arrow[bend left=30, RoyalBlue]{r}{\color{black} P_{\pm}\!\restriction_{\Omega_{L}}} & (P_{\pm}\Omega, \diamond) \arrow[bend left=30, RoyalBlue]{l}{\color{black}2P_{L}\!\restriction_{P_{\pm}\Omega}}
\end{tikzcd}\]\end{shaded}}

The problem about $(\Omega_L, \diamond)$ not being a subalgebra is then solved by transferring the geometric multiplication $\diamond$ from $P_{\pm}\Omega$ to an associative an unital multiplication $\bomdia_{\pm}$ defined on $\Omega_{L}$. The truncated product is very important to find new spinor classes as shown in the Reference \cite{Lopes:2018cvu}.

\section{Orthogonality and Parallelism}

\paragraph{ } In this section we explore some geometrical aspects of the Kähler--Atiyah algebra, the orthogonality and parallelism concepts. In addition, from this concepts we finish the chapter by presenting another model for the algebra $P_{\pm}\Omega$. We begin by fixing $\uptheta \in \Omega^{1}(\M)$ an $1$- form that satisfies the normalisation condition, $g^{*}(\uptheta, \uptheta) = 1$, that is, $\sharp(\uptheta) \rfloor  \uptheta = 1$. Therefore, since $\uptheta \wedge \uptheta = 0$, Eq. \eqref{Eq_cliffproduct} yields $\uptheta \diamond \uptheta = 1$. Consider the linear operators $\uptheta^{\sharp} _{\rfloor} : \Omega^{k}(\M) \to \Omega^{k-1}(\M)$ to and $\wedge_{\uptheta}: \Omega(\M) \to \Omega(\M)$ defined by setting for $\upomega \in \Omega(\M)$

\begin{align}
   \uptheta^{\sharp}_\rfloor  (\upomega) &= \sharp(\uptheta) \rfloor  \upomega \\
   \wedge_{\uptheta} (\upomega) &= \uptheta \wedge \upomega
\end{align}

\begin{definanil}
   An inhomogeneous form $\upomega \in \Omega(\M)$ is said to be \textbf{parallel} to $\uptheta$, denoted by $\theta \parallel \upomega$, if $\uptheta \wedge \upomega =0$ and \textbf{orthogonal} to $\theta$, denoted by $\uptheta \perp \upomega$, if $\sharp(\uptheta) \rfloor  \upomega = 0$, that is,

\begin{align}
    \begin{aligned}
        \uptheta \parallel \upomega &\iff \upomega \in \ker(\wedge_{\uptheta}),\\
        \uptheta \perp \upomega &\iff \upomega \in \ker(\uptheta^{\sharp}_\rfloor).
    \end{aligned}
\end{align}
 
\end{definanil}

\begin{tfpropos}
    Every inhomogeneous differential form $\upomega \in \Omega(\M)$ decomposes uniquely as 

    \begin{equation}
        \upomega = \upomega_{\parallel} + \upomega_{\perp}
    \end{equation}
  \noindent such that $\uptheta \parallel  \upomega_{\parallel}$ and  $\uptheta \perp \upomega_{\perp}$. In addition, the parallel and orthogonal parts of $\upomega$ are given by:

  \begin{align}
      \begin{aligned}
          \upomega_{\parallel} &= 
          (\wedge_{\uptheta} \circ \uptheta^{\sharp}_\rfloor)(\upomega),\\          
          \upomega_{\perp} &= (\uptheta^{\sharp}_\rfloor \circ \wedge_{\uptheta})(\upomega).
      \end{aligned}
  \end{align}
\end{tfpropos}
\noindent \textcolor{RoyalBlue}{\textit{Proof.}} From the properties of the contraction and the normalisation condition for $\uptheta$, 

\begin{align}
    \begin{aligned}
        \uptheta^{\sharp}_\rfloor (\uptheta \wedge \upomega) = \uptheta^{\sharp}_\rfloor (\uptheta)\wedge\upomega - \uptheta\wedge\uptheta^{\sharp}_\rfloor (\upomega) =  \upomega - \uptheta\wedge\uptheta^{\sharp}_\rfloor (\upomega).
    \end{aligned}
\end{align}

\noindent Therefore, $ \upomega$ can be written uniquely as

\begin{align}
    \begin{aligned}
        \upomega = \uptheta\wedge\uptheta^{\sharp}_\rfloor (\upomega)+ \uptheta^{\sharp}_\rfloor (\uptheta \wedge \upomega).
    \end{aligned}
\end{align}

\noindent We claim that  $\upomega_{\parallel} = \uptheta\wedge\uptheta^{\sharp}_\rfloor (\upomega)$, $\upomega_{\perp} = \uptheta^{\sharp}_\rfloor (\uptheta \wedge \upomega)$. That is,
 $\uptheta \parallel  \upomega_{\parallel} = \uptheta\wedge\uptheta^{\sharp}_\rfloor (\upomega)$ and  $\uptheta \perp \upomega_{\perp} = \uptheta^{\sharp}_\rfloor (\uptheta \wedge \upomega)$. Indeed, it follows that

 \begin{align}
     \begin{aligned}
         \uptheta \wedge \upomega_{\parallel} = \uptheta \wedge (\uptheta\wedge\uptheta^{\sharp}_\rfloor (\upomega))= (\uptheta \wedge \uptheta)\wedge\uptheta^{\sharp}_\rfloor (\upomega)) = 0.
     \end{aligned}
 \end{align}

 \noindent That is, $\uptheta \parallel  \upomega_{\parallel}$. Moreover,

 \begin{align}
     \begin{aligned}
        \uptheta^{\sharp}_\rfloor (\upomega_{\perp}) =   \uptheta^{\sharp}_\rfloor(\uptheta^{\sharp}_\rfloor (\uptheta \wedge \upomega)) = (\uptheta^{\sharp}_\rfloor \circ \uptheta^{\sharp}_\rfloor) (\uptheta \wedge \upomega) = 0.
     \end{aligned}
 \end{align}

\noindent Thus, we conclude that

\begin{equation}
    \upomega =  \uptheta\wedge\uptheta^{\sharp}_\rfloor (\upomega)+ \uptheta^{\sharp}_\rfloor (\uptheta \wedge \upomega) = (\wedge_{\uptheta} \circ \uptheta^{\sharp}_\rfloor)(\upomega) +  (\uptheta^{\sharp}_\rfloor \circ \wedge_{\uptheta})(\upomega) = \upomega_{\parallel} +   \upomega_{\perp}. \;\; \textcolor{RoyalBlue}{\Box}
\end{equation}

\begin{tfpropos}\label{proposmutidort}
    The linear operators

\begin{align}
    \begin{aligned}
        P_{\parallel} &:= \wedge_{\uptheta} \circ \uptheta^{\sharp}_\rfloor\\
        P_{\perp} &:= \uptheta^{\sharp}_\rfloor \circ \wedge_{\uptheta}
    \end{aligned}
\end{align}

\noindent are complementary mutual idempotents. \end{tfpropos}

\noindent \textcolor{RoyalBlue}{\textit{Proof.}} In fact, given $\upomega \in \Omega(\M)$

\begin{align}
    \begin{aligned}
        P_{\parallel} + P_{\perp}(\upomega) 
        = \upomega_{\parallel} + \upomega_{\perp} = \upomega.
    \end{aligned}
\end{align}

\begin{align}
    \begin{aligned}
    P_{\parallel} \circ P_{\perp}  &= \wedge_{\uptheta} \circ \uptheta^{\sharp}_\rfloor \circ \uptheta^{\sharp}_\rfloor \circ \wedge_{\uptheta} = 0\\
    P_{\perp} \circ P_{\parallel}  &= \uptheta^{\sharp}_\rfloor \circ \wedge_{\uptheta} \circ \wedge_{\uptheta} \circ \uptheta^{\sharp}_\rfloor = 0.
    \end{aligned}
\end{align}

\begin{align}
    \begin{aligned}
        P_{\parallel} \circ P_{\parallel}(\upomega) &= \wedge_{\uptheta} \circ \uptheta^{\sharp}_\rfloor \circ \wedge_{\uptheta} \circ \uptheta^{\sharp}_\rfloor (\upomega)\\
        &= \wedge_{\uptheta} \circ \uptheta^{\sharp}_\rfloor(\uptheta\wedge\uptheta^{\sharp}_\rfloor (\upomega))\\
        &=  \wedge_{\uptheta} (       \uptheta^{\sharp}_\rfloor(\uptheta) \wedge \uptheta^{\sharp}_\rfloor(\upomega) - \uptheta \wedge \uptheta^{\sharp}_\rfloor(\uptheta^{\sharp}_\rfloor(\upomega)))\\
        &= \wedge_{\uptheta} (\uptheta^{\sharp}_\rfloor(\upomega))\\
        &=  (\wedge_{\uptheta} \circ \uptheta^{\sharp}_\rfloor)  (\upomega)\\
        &= P_{\parallel}(\upomega).
    \end{aligned}
\end{align}

\begin{align}
    \begin{aligned}
        (P_{\perp} \circ P_{\perp})(\upomega) &= (\uptheta^{\sharp}_\rfloor \circ \wedge_{\uptheta} \circ \uptheta^{\sharp}_\rfloor \circ \wedge_{\uptheta}) (\upomega)\\
        &= (\uptheta^{\sharp}_\rfloor \circ \wedge_{\uptheta})(\uptheta^{\sharp}_\rfloor(\uptheta \wedge \upomega))\\
        &= (\uptheta^{\sharp}_\rfloor \circ \wedge_{\uptheta})(\uptheta^{\sharp}_\rfloor(\uptheta)\wedge\upomega - \uptheta \wedge \uptheta^{\sharp}_\rfloor(\upomega))\\
        &=\uptheta^{\sharp}_\rfloor (\uptheta \wedge \upomega - \uptheta \wedge \uptheta \wedge \uptheta^{\sharp}_\rfloor(\upomega))\\
        &= \uptheta^{\sharp}_\rfloor (\uptheta \wedge \upomega)\\
        &= (\uptheta^{\sharp}_\rfloor \circ \wedge_{\uptheta})(\upomega)\\
        &= P_{\perp} (\upomega). \;\; 
    \end{aligned}
\end{align}

\noindent That is,

\begin{itemize}
    \item [\textcolor{RoyalBlue}{\textbf{\textit{1.}}}]$ P_{\parallel} + P_{\perp} = 1.$
    \item [\textcolor{RoyalBlue}{\textbf{\textit{2.}}}] $ P_{\parallel} \circ P_{\perp} = P_{\perp} \circ P_{\parallel} = 0.$
\item [\textcolor{RoyalBlue}{\textbf{\textit{3.}}}]  $P_{\parallel} \circ P_{\parallel} = P_{\parallel},\;P_{\perp} \circ P_{\perp} = P_{\perp}. \;\textcolor{RoyalBlue}{\Box}$
\end{itemize}

\noindent From those operators, one has $\upomega = P_{\parallel}\upomega + P_{\perp}\upomega$. Note that from the previous computations one has $\uptheta^{\sharp}_{\rfloor}$ and $\wedge_{\uptheta}$ preserved respectively by $P_{\parallel}$ and $P_{\perp}$. Indeed,

\begin{align}
    \begin{aligned}\label{eqortpreserv}
        \uptheta^{\sharp}_{\rfloor} P_{\parallel}\upomega &= \uptheta^{\sharp}_{\rfloor}(\wedge_{\uptheta} \circ \uptheta^{\sharp}_\rfloor )(\upomega) = \uptheta^{\sharp}_\rfloor(\upomega),\\
        \uptheta^{\sharp}_{\rfloor}P_{\perp}\upomega &=  \uptheta^{\sharp}_{\rfloor}(\uptheta^{\sharp}_\rfloor \circ \wedge_{\uptheta})(\upomega) =0,\\
        \wedge_{\uptheta}P_{\parallel}\upomega &= \wedge_{\uptheta}(\wedge_{\uptheta} \circ \uptheta^{\sharp}_\rfloor )(\upomega) = 0,\\
          \wedge_{\uptheta}P_{\perp}\upomega &= \wedge_{\uptheta}(\uptheta^{\sharp}_\rfloor \circ \wedge_{\uptheta})(\upomega) = \wedge_{\uptheta}
          (\upomega).
    \end{aligned}
\end{align}

\begin{tfcor}
    $\ker(\uptheta^{\sharp}_\rfloor) = \normalfont{\text{im}}(\uptheta^{\sharp}_\rfloor)$ and $\ker(\wedge_{\uptheta}) = \normalfont{\text{im}}(\wedge_{\uptheta})$.
\end{tfcor}

\noindent \textcolor{RoyalBlue}{\textit{Proof.}} \textcolor{Blue}{$(\supset )$} 
Take $\upomega \in \text{im}(\uptheta^{\sharp}_\rfloor)$, therefore, $ \upomega = (\uptheta^{\sharp}_\rfloor)(\upa)$. Consequently, $(\uptheta^{\sharp}_\rfloor)(\upomega) = (\uptheta^{\sharp}_\rfloor \circ \uptheta^{\sharp}_\rfloor)(\upa) = 0$, which implies  $\upomega \in \ker(\uptheta^{\sharp}_\rfloor)$ and $\ker(\uptheta^{\sharp}_\rfloor) \supset \normalfont{\text{im}}(\uptheta^{\sharp}_\rfloor)$. Analogously, $\ker(\wedge_{\uptheta}) \supset \normalfont{\text{im}}(\wedge_{\uptheta})$ since $\wedge_{\uptheta} \circ \wedge_{\uptheta} = \uptheta^{\sharp}_\rfloor \circ \uptheta^{\sharp}_\rfloor = 0$.\\
\textcolor{Blue}{$(\subset )$}  Take $\upomega \in \ker(\uptheta^{\sharp}_\rfloor) $. Therefore, $\uptheta^{\sharp}_\rfloor(\upomega) = 0$ implying $\uptheta \perp \upomega$, that is, $\upomega = \upomega_{\perp} = \uptheta^{\sharp}_\rfloor (\uptheta \wedge \upomega)$, hence, $\upomega \in\text{im}(\uptheta^{\sharp}_\rfloor)$. For $\upomega \in \ker(\wedge_{\uptheta})$, it follows that: $\uptheta \wedge \upomega = 0$, then $\upomega = \upalpha_{\parallel} = \uptheta \wedge \uptheta^{\sharp}_\rfloor(\upomega)$, that is, $\upomega \in \text{im}(\wedge_{\uptheta})$. We conclude that $\ker(\uptheta^{\sharp}_\rfloor) = \normalfont{\text{im}}(\uptheta^{\sharp}_\rfloor)$ and $\ker(\wedge_{\uptheta}) = \normalfont{\text{im}}(\wedge_{\uptheta})$. \;\textcolor{RoyalBlue}{$\Box$}

\paragraph{ } The orthogonality and parallelism conditions can also be written in terms of the geometric product. 

 {\colorlet{shadecolor}{RoyalBlue!20}\begin{shaded}Consider the right and left operators with respect to the $\diamond$-product given respectively by:

\begin{align}
    \begin{aligned}
        R_{\upa}(\upomega) &:=  \upomega \diamond \upa,\\
        L_{\upa}(\upomega) &:= \upa \diamond \upomega,
    \end{aligned}
\end{align}

\noindent for all $\upa, \upomega \in \Omega(\M)$.\end{shaded}}

\begin{tfpropos}
    The right and left operators satisfy
\begin{itemize} 
\item [\textcolor{RoyalBlue}{\textbf{\textit{1.}}}]  $ L_{\upa_1} \circ R_{\upa_2} = R_{\upa_2} \circ L_{\upa_1} $
  \item [\textcolor{RoyalBlue}{\textbf{\textit{2.}}}]  $L_{\upa_1} \circ L_{\upa_2} = L_{\upa_1 \diamond \upa_2}, $ 
    \item [\textcolor{RoyalBlue}{\textbf{\textit{3.}}}]   $R_{\upa_1} \circ R_{\upa_2} = R_{\upa_2 \diamond \upa_1}. $
    \item [\textcolor{RoyalBlue}{\textbf{\textit{4.}}}] $L_{\upa} \circ\, \yhwidehat{\cdot} =\, \yhwidehat{\cdot}\, \circ L_{\yhwidehat{\upa}}.$
    \item [\textcolor{RoyalBlue}{\textbf{\textit{5.}}}]  $R_{\upa} \circ\, \yhwidehat{\cdot} =\, \yhwidehat{\cdot}\, \circ R_{\yhwidehat{\upa}}.$
     \item [\textcolor{RoyalBlue}{\textbf{\textit{6.}}}]  $L_{\upa} \circ \, \yhwidetilde{\cdot}\, = \, \yhwidetilde{\cdot}\, \circ R_{\yhwidetilde{\upa}}$.
    \item [\textcolor{RoyalBlue}{\textbf{\textit{7.}}}]  $\, \yhwidetilde{\cdot}\, 
 \circ L_{\upa} = R_{\yhwidetilde{\upa}} \circ \, \yhwidetilde{\cdot}\,.$
\end{itemize}

\noindent for all $\upa, \upa_1, \upa_2 \in \Omega(\M)$ where $\;\yhwidehat{\cdot}\;$ and $\;\yhwidetilde{\cdot}\;$ are the grade involution and reversion operators as shown in Eq. \eqref{involution}.
\end{tfpropos}
\noindent \textcolor{RoyalBlue}{\textit{Proof.}} First, note that $\yhwidetilde{\cdot}^{2}\,=\, \yhwidehat{\cdot}^{2}\,=\text{id}$. Given $\upomega \in \Omega(\M)$, due the associativity of the geometric product, it follows that

\begin{flalign}
    \begin{aligned}
        \textcolor{RoyalBlue}{\textbf{\textit{1.}}}\;\;\; L_{\upa_1} \circ R_{\upa_2}(\upomega) &=  L_{\upa_1}(\upomega \diamond \upa_2) = \upa_1 \diamond \upomega \diamond \upa_2 = R_{\upa_2} (\upa_1 \diamond \upomega) =  R_{\upa_2} \circ L_{\upa_1}(\upomega).  \end{aligned}&&
\end{flalign}

\begin{flalign}
    \begin{aligned}
          \textcolor{RoyalBlue}{\textbf{\textit{2.}}}\;\;\; L_{\upa_1} \circ L_{\upa_2} (\upomega) = L_{\upa_1}(\upa_2 \diamond \upomega) = \upa_1 \diamond \upa_2 \diamond \upomega = L_{\upa_1 \diamond \upa_2}(\upomega). 
    \end{aligned}&&
\end{flalign}

\begin{flalign}
    \begin{aligned}
          \textcolor{RoyalBlue}{\textbf{\textit{3.}}}\;\;\; R_{\upa_1} \circ R_{\upa_2} (\upomega) =R_{\upa_1}(\upomega \diamond \upa_2) = \upomega \diamond \upa_2 \diamond \upa_1 =  L_{\upa_2 \diamond \upa_1}(\upomega). 
    \end{aligned}&&
\end{flalign}

\begin{flalign}
    \begin{aligned}
    \textcolor{RoyalBlue}{\textbf{\textit{4.}}}\;\;\; (L_{\upa} \circ\, \yhwidehat{\cdot}\;)(\upomega) = \upa \diamond \yhwidehat{\upomega} =  \yhwidehat{\yhwidehat{\upa}} \diamond \yhwidehat{\upomega} = \yhwidehat{\yhwidehat{\upa} \diamond \upomega} = \, \yhwidehat{\cdot}\;  (\yhwidehat{\upa} \diamond \upomega) = \, \yhwidehat{\cdot}\; \circ L_{\yhwidehat{\upa} }\,(\upomega).
    \end{aligned}&&
\end{flalign}

\begin{flalign}
    \begin{aligned}
       \textcolor{RoyalBlue}{\textbf{\textit{5.}}}\;\;\; (R_{\upa} \circ\, \yhwidehat{\cdot}\;)(\upomega) =  \yhwidehat{\upomega} \diamond \upa =  \yhwidehat{\upomega} \diamond \yhwidehat{\yhwidehat{\upa}} = \yhwidehat{\upomega \diamond \yhwidehat{\upa} } = \, \yhwidehat{\cdot}\;  ( \upomega  \diamond \yhwidehat{\upa}) = \, \yhwidehat{\cdot}\; \circ R_{\yhwidehat{\upa} }\,(\upomega).
    \end{aligned}&&
\end{flalign}

\begin{flalign}
    \begin{aligned}
         \textcolor{RoyalBlue}{\textbf{\textit{6.}}}\;\;\; L_{\upa} \circ \, \yhwidetilde{\cdot}\,(\upomega) = \upa \diamond  \yhwidetilde{\upomega} = \yhwidetilde{\yhwidetilde{\upa}}  \yhwidetilde{\upomega} = \yhwidetilde{\upomega \diamond \yhwidetilde{\upa}} =   \, \yhwidetilde{\cdot}\;(\upomega \diamond \yhwidetilde{\upa})= 
         \, \yhwidetilde{\cdot}\, \circ R_{\yhwidetilde{\upa}}\,(\upomega).
    \end{aligned}&&
\end{flalign}

\begin{flalign}
    \begin{aligned}
        \textcolor{RoyalBlue}{\textbf{\textit{7.}}}\;\;\;\, \yhwidetilde{\cdot}\, 
 \circ L_{\upa} (\upomega)= \yhwidetilde{\cdot}\;(\upa \diamond \upomega) = \yhwidetilde{\upa \diamond \upomega} = \yhwidetilde{\upomega} \diamond \yhwidetilde{\upa} = R_{\yhwidetilde{\upa}}(\yhwidetilde{\upomega}) = 
 R_{\yhwidetilde{\upa}} \circ \, \yhwidetilde{\cdot}\;(\upomega).\;\; \textcolor{RoyalBlue}{\Box}
    \end{aligned}&&
\end{flalign}

The geometric product,
\begin{equation}\label{Eq_cliffproduct22}
      \uptheta \diamond \upomega = \uptheta \wedge \upomega + \sharp(\uptheta) \rfloor  \upomega,\hspace{0.5cm}         \upomega \diamond \uptheta = (-1)^{k} ( \uptheta \wedge \upomega - \sharp(\uptheta) \rfloor  \upomega),
\end{equation}

\noindent  can be then written as

\begin{align}
    \begin{aligned}
        L_{\uptheta} = \wedge_{\theta} + \uptheta^{\sharp}_{\rfloor}, \hspace{0.5cm} R_{\uptheta} \circ\, \yhwidehat{\cdot}\;  = \wedge_{\theta} + \uptheta^{\sharp}_{\rfloor},
    \end{aligned}
\end{align}

\noindent for all $\uptheta \in \Omega^{1}(\M)$. Moreover,

\begin{align}
    \begin{aligned}
        \wedge_{\theta} = \frac{1}{2}( L_{\uptheta} + R_{\uptheta} \circ\, \yhwidehat{\cdot}\;), \hspace{0.5cm}
        \uptheta^{\sharp}_{\rfloor} = \frac{1}{2}( L_{\uptheta} - R_{\uptheta} \circ\, \yhwidehat{\cdot}\;).
    \end{aligned}
\end{align}

\noindent We are now in position to present the following characterisations for the orthogonality and parallelism properties:

{\colorlet{shadecolor}{RoyalBlue!20}\begin{shaded}\begin{align}
    \begin{aligned}
        \uptheta\parallel \upomega &\iff   \uptheta \wedge \upomega = 0\\
    &\iff \upomega \in \ker(\wedge_{\uptheta})\\
    &\iff \upomega \in \ker(L_{\uptheta} + R_{\uptheta} \circ\, \yhwidehat{\cdot}\;)\\
    &\iff  \upomega \in \text{im}(\wedge_{\uptheta})\\
    &\iff \upomega = \uptheta \wedge \upalpha,\; \upa \in \Omega(\M).\\
    &\iff \uptheta \diamond \upomega = -\yhwidehat{\upomega} \diamond \uptheta.
    \end{aligned}
\end{align}\end{shaded}}

{\colorlet{shadecolor}{RoyalBlue!20}\begin{shaded}\begin{align}
    \begin{aligned}
        \uptheta\perp \upomega &\iff    \sharp(\uptheta) \rfloor \upomega = 0\\
    &\iff \upomega \in \ker( \uptheta^{\sharp}_{\rfloor})\\
     &\iff \upomega \in \ker(L_{\uptheta} - R_{\uptheta} \circ\, \yhwidehat{\cdot}\;)\\
    &\iff  \upomega \in \text{im}( \uptheta^{\sharp}_{\rfloor})\\
    &\iff \upomega = \sharp(\uptheta) \rfloor  \upb,\; \upb \in \Omega(\M).\\
    &\iff \uptheta \diamond \upomega = \yhwidehat{\upomega} \diamond \uptheta.
    \end{aligned}
\end{align}\end{shaded}}

\noindent We conclude that $ \uptheta\parallel \upomega$ iff. $\upomega$ graded anticommutes with $\uptheta$ and $\uptheta\perp \upomega$ iff. $\upomega$ graded commutes with $\uptheta$ in the Kähler-Atiyah algebra. In addition, it follows from those characterisations that the grade involution and the reversion preserve parallelism and orthogonality with respect to $\uptheta$. That is,

\begin{align}
    \begin{aligned}
        \uptheta \parallel \upomega &\implies \uptheta \parallel \yhwidehat{\upomega }\text{ and } \uptheta \parallel \yhwidetilde{\upomega},\\
            \uptheta \perp \upomega &\implies \uptheta \perp \yhwidehat{\upomega }\text{ and } \uptheta \perp \yhwidetilde{\upomega}.
    \end{aligned}
\end{align}

  {\colorlet{shadecolor}{RoyalBlue!20}\begin{shaded}Consider the following subsets,

\begin{align}
    \begin{aligned}
        \Omega^{\parallel} &:= \{\upomega \in \Omega(\M) : \uptheta \parallel \upomega\},\\
        \Omega^{\perp} &:= \{\upomega \in \Omega(\M) : \uptheta \perp \upomega\}.
    \end{aligned}
\end{align}\end{shaded}}

\begin{tfpropos}\label{proposinclort}
    The following inclusions hold
\begin{itemize}
    \item [\textit{1.}] $\Omega^{\parallel} \diamond\, \Omega^{\parallel} \subset \Omega^{\perp},$
     \item [\textit{2.}] $\Omega^{\perp} \diamond\, \Omega^{\perp} \subset \Omega^{\perp},$
     \item [\textit{3.}] $\Omega^{\perp} \diamond\, \Omega^{\parallel} \subset \Omega^{\parallel},$
      \item [\textit{4.}] $\Omega^{\parallel}\diamond\, \Omega^{\perp} \subset \Omega^{\parallel}.$ 
\end{itemize}
\end{tfpropos}

\noindent \textcolor{RoyalBlue}{\textit{Proof.}} Take $\uptheta \in \Omega^{1}(\M)$ and $ \upzeta, \upomega \in \Omega(\M).$ The orthogonality and parallelism characterisation will be used to prove all the inclusions. To prove the first inclusion, suppose that $\uptheta \parallel \upzeta$ and $\uptheta \parallel \upomega$. It follows that,

\begin{align}
    \begin{aligned}
   \uptheta \diamond (\upzeta \diamond \upomega) =  (\uptheta \diamond \upzeta) \diamond \upomega=  -(\yhwidehat{\upzeta} \diamond \uptheta  ) \diamond \upomega= -\yhwidehat{\upzeta} \diamond (\uptheta   \diamond \upomega)= \yhwidehat{\upzeta} \diamond (\yhwidehat{\upomega}  \diamond \uptheta   )= (\yhwidehat{\upzeta \diamond \upomega}) \diamond \uptheta.
    \end{aligned}
\end{align}

\noindent therefore, $\uptheta \perp (\upzeta \diamond \upomega)$ and the first inclusion holds. For the second one, suppose now that $\uptheta \perp \upzeta$ and $\uptheta \perp \upomega$, hence

\begin{align}
    \begin{aligned}
        \uptheta \diamond (\upzeta \diamond \upomega) =  (\uptheta \diamond \upzeta) \diamond \upomega=  (\yhwidehat{\upzeta} \diamond \uptheta  ) \diamond \upomega= \yhwidehat{\upzeta} \diamond (\uptheta   \diamond \upomega)= \yhwidehat{\upzeta} \diamond (\yhwidehat{\upomega}  \diamond \uptheta   )= (\yhwidehat{\upzeta \diamond \upomega}) \diamond \uptheta.
    \end{aligned}
\end{align}

\noindent Thus $\upomega \perp (\upzeta \diamond \upomega)$ and $\Omega^{\perp} \diamond\, \Omega^{\perp} \subset \Omega^{\perp}$. For the third and the fourth inclusion let us consider now $\uptheta \parallel \upzeta$ and $\uptheta \perp \upomega$. It follows that,

\begin{align}
    \begin{aligned}
        &\uptheta \diamond (\upomega \diamond \upzeta) = (\yhwidehat{\upomega} \diamond \uptheta) \diamond \upzeta = - \yhwidehat{\upomega} \diamond (\yhwidehat{\upzeta} \diamond \upomega) = -(\yhwidehat{\upomega \diamond \upzeta}) \diamond \upomega,\\
    &\uptheta \diamond (\upzeta \diamond \upomega) = - (\yhwidehat{\upzeta}  \diamond \uptheta) \diamond \upomega =  - \yhwidehat{\upzeta} \diamond (\yhwidehat{\upomega} \diamond \uptheta) =-(\yhwidehat{\upzeta \diamond  \upomega }) \diamond \upomega.
    \end{aligned}
\end{align}

\noindent We conclude that $\uptheta \parallel (\upomega \diamond \upzeta)$ and $\uptheta \parallel (\upzeta \diamond \upomega)$, which proves the last inclusions. \textcolor{RoyalBlue}{$\Box$}

\paragraph{ } In other words, one has

\begin{align}\label{eqregrasprodpar}
    \begin{aligned}\uptheta \parallel \upomega \text{ and }  \uptheta \parallel \upzeta &\implies \uptheta \perp(\upomega \diamond \upzeta), \\
         \uptheta \perp \upomega \text{ and }  \uptheta \perp \upzeta &\implies \uptheta \perp(\upomega \diamond \upzeta), \\
        \uptheta \parallel \upomega \text{ and }  \uptheta \perp \upzeta &\implies  \uptheta \parallel(\upomega \diamond \upzeta) \text{ and }  \uptheta \parallel(\upzeta \diamond \upomega).         
    \end{aligned}
\end{align}

\noindent Moreover, for $\upomega, \upzeta \in (\Omega(\M),\diamond)$ one has

\begin{align}
    \begin{aligned}
        (\upomega \diamond \upzeta) &= ((\upomega_\perp + \upomega_\parallel)\diamond(\upzeta_\perp + \upzeta_\parallel))= \upomega_\perp\diamond\upzeta_\perp +  \upomega_\perp\diamond\upzeta_\parallel + \upomega_\parallel\diamond\upzeta_\perp + \upomega_\parallel\diamond\upzeta_\parallel
    \end{aligned}
\end{align}

\noindent From Eq. \eqref{eqregrasprodpar}, we conclude:

\begin{align}
    \begin{aligned}\label{eq2149a}
         (\upomega \diamond \upzeta)_{\perp} &= \upomega_\perp\diamond\upzeta_\perp + \upomega_\parallel\diamond\upzeta_\parallel,\\
        (\upomega \diamond \upzeta)_\parallel &= \upomega_\perp\diamond\upzeta_\parallel + \upomega_\parallel\diamond\upzeta_\perp .
    \end{aligned}
\end{align}

One may notice that $\uptheta \perp 1$ since $\sharp(\uptheta)\rfloor1=0.$ Therefore, the second inclusion $\Omega^{\perp} \diamond\, \Omega^{\perp} \subset \Omega^{\perp}$ from Proposition \ref{proposinclort}, together with the identity $1 \in \Omega^{\perp}$, shows that $\Omega^{\perp}$ is a unital subalgebra of $(\Omega(\M),\diamond)$.

\paragraph{ } The top component of an inhomogeneous form  $\upomega \in \Omega(\M)$ can be introduced as

\begin{equation}
    \upomega_{\top} = \uptheta^{\sharp}_{\rfloor}(\upomega) \in \Omega^{\perp}.
\end{equation}

\noindent This way, the parallel part of $\upomega$ can be written as $\upomega_{\parallel} = \uptheta \wedge \upomega_{\top}$ and $\upomega$ is written uniquely as

\begin{equation}
    \upomega = \uptheta \wedge \upomega_{\top} + \upomega_{\perp},
\end{equation}

\noindent that is, $\upomega$ determines and is determined by two inhomogeneous form $\upomega_{\top}, \upomega_{\perp} \in \Omega^{\perp}$. 
\begin{tfpropos}
   $\Omega(\M) \simeq \Omega^{\perp} \oplus\, \Omega^{\perp} $.
\end{tfpropos}
\noindent \textcolor{RoyalBlue}{\textit{Proof.}} Consider the linear mapping 

\begin{equation}
    \uptheta^{\sharp}_{\rfloor} + P_{\perp} : \Omega(\M) \to \Omega^{\perp} \oplus \Omega^{\perp}
\end{equation}

\noindent Therefore, from Proposition \ref{proposmutidort} and Eq. \eqref{eqortpreserv} it follows that for any $\upomega \in \Omega(\M)$

\begin{align}
    \begin{aligned}        \uptheta^{\sharp}_{\rfloor} + P_{\perp}(\upomega) &=  \uptheta^{\sharp}_{\rfloor} + P_{\perp} (P_{\parallel}\upomega + P_{\perp}\upomega)\\
    &= (\uptheta^{\sharp}_{\rfloor} + P_{\perp})(P_{\parallel}\upomega) + (\uptheta^{\sharp}_{\rfloor} + P_{\perp})(P_{\perp}\upomega) \\
    &= \uptheta^{\sharp}_{\rfloor}P_{\parallel}\upomega + \uptheta^{\sharp}_{\rfloor}P_{\perp}\upomega + P_{\perp}\upomega\\
    &= \uptheta^{\sharp}_{\rfloor}\upomega + P_{\perp}\upomega\\
    &= \upomega_{\top} + \upomega_{\perp}.
    \end{aligned}
\end{align}

\noindent That is, $\upomega \mapsto (\upomega_{\top}, \upomega_{\perp}) \in \Omega^{\perp} \oplus\, \Omega^{\perp}$. The inverse map, $ (\uptheta^{\sharp}_{\rfloor} + P_{\perp}(\upomega))^{-1}$ is given straightforward from the orthogonality characterisations since any pair $(\upa, \beta) \in \Omega^{\perp} \oplus\, \Omega^{\perp}$ uniquely determines a  $\upzeta \in \Omega(\M)$ by sending $(\upa, \beta) \mapsto \uptheta \wedge \upa + \beta = \upzeta$. \textcolor{RoyalBlue}{$\Box$}

\paragraph{ } The inhomogeneous form $\upomega$ can be decomposed in terms of the geometric product. Since $\upomega_{\top} \in \Omega^{\perp}$, $\uptheta \diamond \upomega_{\top} = \uptheta \wedge \upomega_{\top}$, therefore,

\begin{equation}
    \upomega = \uptheta \diamond \upomega_{\top} + \upomega_{\perp}.
\end{equation}

\noindent Moreover, given $\upomega, \upzeta \in \Omega(\M)$ it follows that

\begin{align}
    \begin{aligned}
        (\upomega \diamond \upzeta) &= (\uptheta \diamond \upomega_{\top} + \upomega_{\perp}) \diamond (\uptheta \diamond \upzeta_{\top} + \upzeta_{\perp})\\
        &= \uptheta \diamond \upomega_{\top}  \diamond \uptheta \diamond \upzeta_{\top} + \uptheta \diamond \upomega_{\top}  \diamond \upzeta_{\perp} + \upomega_{\perp} \diamond \uptheta \diamond \upzeta_{\top} +\upomega_{\perp} \diamond \upzeta_{\perp}\\
&= \uptheta \diamond   \uptheta \diamond \yhwidehat{\upomega_{\top}} \diamond \upzeta_{\top}         
+ \uptheta \diamond \upomega_{\top}  \diamond \upzeta_{\perp} +  \uptheta \diamond \yhwidehat{\upomega_{\perp}}\diamond \upzeta_{\top} 
+\upomega_{\perp} \diamond \upzeta_{\perp}\\
&= \uptheta \diamond ( \upomega_{\top}  \diamond \upzeta_{\perp} + \yhwidehat{\upomega_{\perp}}\diamond \upzeta_{\top} )
+\upomega_{\perp} \diamond \upzeta_{\perp} + \yhwidehat{\upomega_{\top}} \diamond \upzeta_{\top}.
    \end{aligned}
\end{align}

\noindent That is,

\begin{align}
    \begin{aligned}
  (\upomega \diamond \upzeta)_{\top} &=  \upomega_{\top}  \diamond \upzeta_{\perp} + \yhwidehat{\upomega_{\perp}}\diamond \upzeta_{\top}, \\
 (\upomega \diamond \upzeta)_{\perp} &= \upomega_{\perp} \diamond \upzeta_{\perp} + \yhwidehat{\upomega_{\top}} \diamond \upzeta_{\top}.
    \end{aligned}
\end{align}

\paragraph{ } Let us now assume that $n$ is odd and $p - q \equiv_8 0,1,4,5$ and a distinguish $1$-form $\uptheta \in \Omega^{1}(\M)$ satisfying the normalisation condition $\uptheta^{\sharp}_{\rfloor}(\uptheta)=1$. Since $\uptheta \wedge \boldsymbol{\uptau} = 0$, one has $\uptheta \parallel \boldsymbol{\uptau}$. Take $\upomega \in \Omega^{\parallel}$ and $\upzeta \in \Omega^{\perp}$ it follows from Eq. \eqref{eqregrasprodpar}:

\begin{align}
    \begin{aligned}
     \uptheta  \parallel \upomega &\implies  \uptheta \perp (\upomega \diamond \boldsymbol{\uptau}) \implies \uptheta \perp \star\upomega,  \\
       \uptheta  \perp \upzeta  &\implies \uptheta \parallel (\upzeta\diamond \boldsymbol{\uptau}) \implies \uptheta \parallel \star\upzeta .
    \end{aligned}
\end{align}

\noindent That is, for any $\upa = \upa_{\parallel} + \upa_{\perp} \in \Omega(\M)$,

\begin{align}
    \begin{aligned}
       \star \upa = \star\upa_{\parallel} + \star\upa_{\perp} \implies \uptheta \perp \star\upa_{\parallel} \text{ and } \uptheta \parallel \star\upa_{\perp}.
    \end{aligned}
\end{align}

\noindent Moreover, since $p-q \equiv_{8} 0,1,4,5$ one has $\star\upomega = \pm \upomega$ for any inhomogeneous form $\upomega \in P_{\pm}\Omega$, therefore, 

\begin{align}
    \begin{aligned}\label{eq2159star}
        \star \upa_{\parallel} = \pm \upa_{\perp},\\
        \star\upa_{\perp} = \pm \upa_{\parallel}.
    \end{aligned}
\end{align}
\begin{tfpropos}\label{proposisoperp}
    $(P_{\pm}\Omega, \diamond) \simeq (\Omega^{\perp}, \diamond)$
\end{tfpropos}
\noindent \textcolor{RoyalBlue}{\textit{Proof.}} Consider $\upomega \in P_{\pm}\Omega$, from Eq \eqref{eqsalvastar} one has $\upomega = P_{\pm}(\upomega)$. By using Eq. \eqref{eq2159star} and considering the linear mapping $2P_{\perp}$, it follows that 

\begin{align}
    \begin{aligned}
        2P_{\perp}(\upomega) &=   2P_{\perp}(\frac{1}{2}\upomega \pm \star\upomega) \\
        &= P_{\perp}(\upomega_{\parallel} + \upomega_{\perp} \pm \star\upomega_{\parallel} + \star\upomega_{\perp})\\
        &= P_\perp(\upomega_{\parallel}) +  P_\perp(\upomega_{\perp}) \pm (P_\perp(\star\upomega_{\parallel}) + P_\perp(\star\upomega_{\perp}))\\
        &= \upomega_{\perp} \pm \star\upomega_{\parallel}\\ 
       &=  \upomega_{\perp} \pm (\pm  \upomega_{\perp})\\
       &= 2\upomega_{\perp} \in \Omega^{\perp}.
    \end{aligned}
\end{align}

\noindent Observe that

\begin{align}
    \begin{aligned}
       (P_{\pm} \circ 2P_{\perp})(\upomega) = P_{\pm}(2\upomega_{\perp}) 
       = (\upomega_{\perp} \pm \star \upomega_{\perp})= \upomega_{\perp} \pm (\pm \upomega_{\parallel})= \upomega.
    \end{aligned}
\end{align}

\noindent For $\upomega, \upzeta \in P_{\pm}\Omega$, it follows that from Eq. \eqref{eq2149a}

\begin{align}
    \begin{aligned}
        2P_{\perp} (\upomega \diamond \upzeta) &=  2 (\upomega_\perp\diamond\upzeta_\perp + \upomega_\parallel\diamond\upzeta_\parallel)\\
        &= 2(\upomega_\perp\diamond\upzeta_\perp + ((\pm \star\upomega_{\perp})\diamond(\pm \star\upzeta_{\perp})))\\
        &= 2(\upomega_\perp\diamond\upzeta_\perp + ((\upomega_\perp \diamond \boldsymbol{\uptau})\diamond(\upzeta_\perp \diamond \boldsymbol{\uptau})))\\ 
        &= 2(\upomega_\perp\diamond\upzeta_\perp + \upomega_{\perp} \diamond \upzeta_{\perp})\\
        &= 4\upomega_\perp\diamond\upzeta_\perp\\              &=2\upomega_\perp\diamond2\upzeta_\perp\\
        &= 2P_{\perp}(\upomega) \diamond 2P_{\perp}(\upzeta).
    \end{aligned}
\end{align}

\noindent Moreover, since $\uptheta \perp 1$ and $\uptheta \parallel \boldsymbol{\uptau}$:

\begin{align}
    \begin{aligned}
        2P_{\perp}(\rho_{\pm}) = P_{\perp}(1 \pm \boldsymbol{\uptau}) =P_{\perp}(1) \pm P_{\perp}(\boldsymbol{\uptau}) = 1.
    \end{aligned}
\end{align}

\noindent Therefore, we conclude that $2P_{\perp}: (P_{\pm}\Omega, \diamond) \overset{\sim}{\rightarrow} (\Omega^{\perp}, \diamond)$ is a unital isomorphism of algebras with inverse $P_{\pm}$ (restricted to $\Omega^{\perp}$). \textcolor{RoyalBlue}{$\Box$}

 {\colorlet{shadecolor}{RoyalBlue!20}\begin{shaded}Combining Propositions \ref{propos3306} and \ref{proposisoperp}, we finish the Chapter with two
isomorphic models for the unital subalgebra $(P_{\pm}\Omega, \diamond).$\\

 \[ 
\begin{tikzcd}[row sep=4.5em, xscale=2, yscale=2, scale=0.5, transform shape]
(\Omega_{L}, \bomdia_{\pm}) \arrow[bend left=20, RoyalBlue]{r}{\color{black} P_{\pm}\!\restriction_{\Omega_{L}}} & (P_{\pm}\Omega, \diamond) \arrow[bend left=20, RoyalBlue]{l}{\color{black}2P_{L}\!\restriction_{P_{\pm}\Omega}} \arrow[bend left=20, RoyalBlue]{r}{\color{black} 2P_{\perp}\!\restriction_{P_{\pm}\Omega}} & (\Omega^{\perp}, \diamond) \arrow[bend left=20, RoyalBlue]{l}{\color{black}P_{\pm}\!\restriction_{\Omega^{\perp}}}
\end{tikzcd}\]
\end{shaded}}

\colorlet{chapter}{DarkOrchid!70}
\chapter{Spin Geometry}\label{chap_spingeo}

\hypersetup{
  colorlinks = true,
  linkcolor  = DarkOrchid,
  citecolor = [rgb]{0.74,0.06,0.88},
}

 \paragraph{ } In the geometric context of bundles, \textbf{\textcolor{DarkOrchid}{what is a spinor?}} In this context, a spinor is a section of the \textit{bundle of spinors} over a manifold  $(\M,g)$. However, it is only possible to define the bundle of spinors if 
$(\M,g)$ has a \textcolor[rgb]{0.74,0.06,0.88}{\textit{spin structure}}, since the bundle of spinors is the vector bundle associated with the principal bundle of the spin structure. Our aim is to present these two concepts, this study is part of the broader field of Spin geometry which is a special topic in Riemannian geometry that bridges the algebraic world of \textcolor[rgb]{0.88,0.82,0.11}{spinors} and \textcolor[rgb]{0.82,0.01,0.11}{Clifford algebras} with the geometric structure of \textcolor[rgb]{0.41,0.67,0.97}{bundles}. In this chapter, we introduce the concept of a spin structure on a manifold, which is essential to consider spinor fields in a geometrically coherent manner. 

To define the spin structure on a manifold we first introduce the fibre bundle of positively oriented orthonormal frames of a smooth $n$-dimensional manifold $\M$. Recall that a vector bundle $\pi: E \to \M$ of rank $n$ is determined by its gluing maps  $g_{\alpha \beta}$ (Eq. \eqref{eq_retgfjndfbsa3z12}) mapping into
the general linear group $\text{GL}(n,\rr)$. Further knowing that the manifold is equipped with a continuous choice of inner product on its fibres, the transition gluing maps can be mapped into the orthogonal group $\O(n)$. Orientability corresponds to reducing the structure group to $\SO(n)$ and a spin structure corresponds to then reducing the structure group to the universal covering group $\Spin(n) \to \SO(n)$ through the twisted adjoint representation $\yhwidehat{\text{Ad}}$. Therefore, first we define a fibre of \textit{the frame bundle of $\M$} which is defined for $x \in \M$ as

\begin{equation}
    \text{GL}(\M)_{x} := \{v_{x} = (v_1, \ldots, v_{n}) : v_{x} \text{ is a basis of } T_{x}\M\} 
\end{equation}

\noindent Hence, 

\begin{equation}
     \text{GL}(\M)  = \bigsqcup_{x \in \M} \text{GL}(\M)_{x}.
\end{equation}

\noindent The projection is defined via $\pi_{\text{GL}(\M)}: \text{GL}(\M) \to \M$, $v_{x} \mapsto x$. Note that if $(U, \varphi = (x^{1}, \ldots, x^{n}))$ is a coordinate chart for $\M$, then for every $x \in U$ the associated frame $(\partial_{1}(x),\ldots, \partial_{n}(x)) \in \text{GL}(\M)_{x}$ and if every such coordinate frames are smooth, so is $\text{GL}(\M)$. As we know from Section \ref{sec_principalbundles} in the Eq. \eqref{eq_347dvnjacvb}, the natural right action $\text{GL}(n,\rr)$ on $\text{GL}(\M)_{x}$ turns $\text{GL}(\M)$ into a $\text{GL}(n,\rr)$-principal bundle called \textit{the frame bundle of $\M$}. 

 {\colorlet{shadecolor}{DarkOrchid!15}\begin{shaded}
By defining the set

    \begin{equation*}
        \SO(\M) \!=\! \SO(\M,g) \!:= \{v_{x} \in \text{GL}(\M)_{x} : v_{x} \text{ is a positively oriented orthonormal basis of } T_{x}\M\}
    \end{equation*}

\noindent and defining a $\SO(n)$-right action on $\SO(\M)$ by restricting the $\text{GL}(n,\rr)$ action on $\text{GL}(\M)_{x}$. Then it turns $\SO(\M) $ into a $\SO(n)$-principal fibre bundle. \end{shaded}}

 From a representation $\lambda: \SO(n) \to \text{GL}(V)$ of $\SO(n)$ on a $\kk$-vector space $V$, it is possible to construct a vector bundle by gluing the vector space $V$ onto the fibres of the frame bundle. The \textit{associated bundle} to $\SO(\M)$ over $\M$ is defined as $\SO(\M) \cross_{\lambda} V := \SO(\M) \cross_{\lambda} V /\sim $ such that the equivalence class $\sim$ is defined as

\begin{equation}
    (h_1,v_1) \sim (h_2,v_2) \Leftrightarrow \exists A \in \SO(n) : h_2 = h_1 \circ A \text{ and } v_2 = \lambda(A^{-1})v_1.
\end{equation}

The associated projection is defined as Eq. \eqref{eq_assproj} and is well defined. For the standard representation $\lambda: \SO(n) \to \text{GL}(n,\rr)$, the map 

    \begin{align}
        \begin{aligned}
             \SO(\M)\cross_{\lambda} \rr^{n} \;\;\; &\to \;\;\; T\M\\
            [h,v] \;\;\;\;\;&\mapsto\;\;\;\;\; h(v)
        \end{aligned}
    \end{align}

\noindent is an isomorphism of vector bundles. There are the following isomorphisms of vector bundles, associated to representations of $\SO(n)$ \cite{spinazul}.

\begin{table}[H]
\begin{tabular}{c|c|c}
\rowcolor{BlueViolet!15} 
\multicolumn{1}{l|}{\cellcolor{BlueViolet!15}vector space $V$} & \multicolumn{1}{l|}{\cellcolor{BlueViolet!15}representation $\lambda$ of $\SO(n)$ on $V$} & \multicolumn{1}{l}{\cellcolor{BlueViolet!15}$\SO(\M) \cross_{\lambda} V$} \\ \hline
$\rr$                                                         & id$_{V}$                                                                                 & $\M \cross \rr$                                                          \\
$\rr^{n}$                                                     & standard representation                                                                  & $T\M$                                                                    \\
$(\rr^{n})^{*}$                                               & dual of standard representation                                                          & $T^{*}\M$                                                                \\
$\bigwedge^{k}(\rr^{n})^{*}$                                  & $\bigwedge^{k}$ (dual of standard representation)                                        & $\bigwedge^{k} T^{*} \M$                                                 \\
$(\tensor ^{k}\rr^{n}) \tensor (\tensor^{l} (\rr^{n})^{*})$   & $(\tensor ^{k}$(std rep.)) $\tensor (\tensor^{l}$ (dual of std. rep.))                     & $(\tensor ^{k}T\M) \tensor (\tensor^{l} T^{*}\M)$                        
\end{tabular}
\end{table}

The Clifford bundle $\cl(T^{*}\M)$  is a vector bundle associated with the bundle of oriented
orthonormal frames \cite{Konstantin}. In fact, consider $\rho: \SO(n) \to \text{Aut}(\cl_{n})$ with $\rho(A): \cl_n \to \cl_n$ for $A \in \SO(n)$ defined on generators $\'v \in \rr^{n}$ of $\cl_{n}$ by $\'v \mapsto \rho(A)\'v = A\'v$. This map defines an algebra
homomorphism since

\begin{equation}
    \rho(A)(\'v\'u + \'u\'v) = A\'vA\'u + A\'u A\'v = -2g( A\'v,A\'u) = -2g(\'v,\'u).
\end{equation}

\noindent Therefore, the Clifford bundle can be realised as $\cl(T^{*}\M) = \SO(\M) \cross_{\rho} \cl_n$. The goal now is to construct a Spin$(n)$-principal bundle $\Spin(\M)$ such that for any representation $\lambda' : \Spin(n) \to \text{GL}(V)$  (recall the Eq. \eqref{eq_5ydrfhjknvm1}) of the form $\lambda': \lambda  \circ \yhwidehat{\text{Ad}}$, with $\lambda : \SO(n) \to \text{GL}(V)$, the associated vector bundles $\Spin(\M) \cross_\lambda' V$ and $\SO(\M) \cross_\lambda V$ coincide. This happens through the concept of spin structure.

\section{Spin Structures}

\paragraph{ } A spin structure on an orientable Riemannian manifold $(\M, g)$ allows one to define the bundle of spinors, giving rise to the notion of a spinor in differential geometry. It is worth to emphasise that besides the case of Riemannian manifold is developed here the same definition of a spin structure applies to the case of pseudo-Riemannian manifolds. 

\begin{definvioleta}
    Let $(\M,g)$ be an oriented Riemannian manifold with dimension $n$. A \textbf{Spin structure on $\M$} \textit{is a pair} \normalfont $(\Spin(\M),\pi)$ \textit{consisting of a} $\Spin(n)$\textit{-principal fibre bundle} $(\Spin(\M), \pi_{\text{Sp}}, \Spin(n))$ \textit{over} $\M$ \textit{and a smooth map} $\pi: \Spin(\M) \to \SO(\M)$ \textit{such that}

    \begin{itemize}
        \item [\textbf{\textcolor{DarkOrchid}{\textit{1.}}}] $\pi_{\SO} \circ \pi = \pi_{\text{Sp}}$
        \item [\textbf{\textcolor{DarkOrchid}{\textit{2.}}}]  $\pi(x \, \textcolor{DarkOrchid}{\bullet} \, h) = \pi(x)\, \textcolor{violet}{\circ}\, \yhwidehat{\text{Ad}}(h)$ \textit{for all} $x \in \Spin(\M)$ \textit{with} $\yhwidehat{\text{Ad}}: \Spin(n) \to \SO(n)$ \textit{the double covering Lie group homomorphism in Def. \ref{def_twofold2345} (twisted adjoint representation).}
    \end{itemize}

    \noindent \textit{That is, the following diagram commutes:}

    \[ \begin{tikzcd}
\Spin(\M) \cross \Spin(n) \arrow[r, "\textcolor{DarkOrchid}{\bullet}"] \arrow[d, "\pi \cross \yhwidehat{\text{Ad}}"] & \Spin(\M) \arrow{d}{\pi}  \arrow[dr, "\pi_{Sp}"]  \\
\SO(\M) \cross \SO(n) \arrow[r,"\textcolor{violet}{\circ}"]          & \SO(\M) \arrow[r, "\pi_{\SO}"]                                         & \M
\end{tikzcd} 
    \]

  \end{definvioleta}

\noindent Such that the bullets $\textcolor{DarkOrchid}{\bullet\,}: \Spin(\M) \cross \Spin(n) \to \Spin(\M)$ and $\textcolor{violet}{\circ\,}: \SO(\M) \cross \SO(n) \to \SO(\M)$ are the right Lie actions. In other words, a $\Spin$-structure on $\M$ is a $\yhwidehat{\text{Ad}}$-reduction (Def. \ref{def_fred}) of the bundle $\SO(\M)$ of oriented orthonormal frames of $\M$. It is also often said to be a lift of $\SO(n)$ to $\Spin(n)$. We recall that the horizontal maps the group actions on the principal
bundles. 

If $\lambda' = \lambda \circ \yhwidehat{\text{Ad}} : \Spin(n) \to \text{GL}(V)$ is a representation induced by $\lambda: \SO(n) \to \text{GL}(V)$ it follows that

    \begin{align}
        \begin{aligned}
             \Spin(\M)\cross_{\lambda'} V \;\;\; &\to \;\;\; \SO(\M) \cross_{\lambda} V\\
            [x,v] \;\;\;\;\;&\mapsto\;\;\;\;\; [\pi(x), v].
        \end{aligned}
    \end{align}

\noindent is well defined. Indeed, for $a \in \Spin(n)$ one has

\begin{equation}
    [xa, \lambda(a^{-1})v] \mapsto [\pi(xa),\lambda(a^{-1})v] = [\pi(x)\yhwidehat{\text{Ad}}(a),\lambda'(\yhwidehat{\text{Ad}}(a^{-1})v]=[\pi(x),v)].
\end{equation}

\noindent This map is an isomorphism of vector bundles \cite{spinazul}. By considering the representation $\text{Ad}': \Spin(n) \to \text{Aut}\cl_{n}$ given by $\text{Ad}_{a}(\'v) = a\'va^{-1}$ for $a \in \Spin(n) \subset \cl_{n}$. One has$\text{Ad}_{-1} = $ identity, thus this representation descends to a representation $\text{Ad}$ of $\SO(n)$.  The Clifford bundle $\cl(T^{*}\M)$ can be realised as $\cl(T^{*}\M) = \Spin(\M) \cross_{\text{Ad}'} \cl_{n}$ \cite{SpinGeo}.

The existence of a spin structure $\Spin(\M)$ implies the existence of an $\SO(n)$-structure, one can just consider the bundle of orthonormal frames if $\M$ is equipped with an Riemannian metric $g$. Therefore, unlike the tangent and cotangent bundles, the very definition and structure of the spinor bundle are intrinsically linked on the choice Riemannian metric of the manifold. Conversely, we recall that given an oriented Riemannian manifold $\M$ with tangent bundle $T\M = (U, g_{\alpha \beta})$ in which each trivialising covering of $\M$ induces the gluing maps $g_{\alpha \beta}: U_{\alpha \beta} \to \SO(n)$ satisfying the the relations \eqref{eq_e54948543798t4rg651}, that is,

\begin{flalign}\label{eq_e54944356ytgrewrt}
    \begin{aligned}
        &\textcolor{DarkOrchid}{\textbf{(i)}}\;\;\;  g_{\a \a} = 1_{\SO(n)},\\
     &\textcolor{DarkOrchid}{\textbf{(ii)}}\;\; g_{\a \b}(x)g_{\b \a} = 1_{\SO(n)}, \;\; \text{ for all } x \in U_{\a} \inter U_{\b};\\
      &\textcolor{DarkOrchid}{\textbf{(iii)}}\;g_{\a \delta}(x)g_{\delta \b}(x)g_{\b \a}(x) =     1_{\SO(n)} \;\; \text{ for all } x \in U_{\a} \inter U_{\b} \inter U_{\delta} \textit{\;\;(cocycle condition)}.
    \end{aligned}&&
\end{flalign}

\noindent A spin structure is a collection of lifts $f_{\alpha \beta}$ such that the following diagram commutes

\[
  \begin{tikzcd}
    U_{\alpha \beta} \arrow{r}{f_{\alpha \beta}} \arrow[DarkOrchid, swap, dashed]{dr}[black]{g_{\alpha \beta}} & [20pt] \Spin(n) \arrow{d}{\yhwidehat{\text{Ad}}} \\[20pt]
     & \SO(n)
  \end{tikzcd}
\]

\noindent Lifting the frame bundle $\SO(\M)$ to a $\Spin(n)$-bundle then requires that the gluing functions $f_{\alpha \beta}: U_{\alpha \beta} \to  \Spin(n)$ with $\yhwidehat{\text{Ad}}(f_{\alpha \beta}) = g_{\alpha \beta}$ must also satisfy the relations \eqref{eq_e54944356ytgrewrt}  as the gluings $g_{\alpha \beta}$. When do such lifts exist? From covering theory, one can conclude that if $U_{\a \b}$ is simply connected, then lifts satisfying \textcolor{DarkOrchid}{\textbf{(i)}} and \textcolor{DarkOrchid}{\textbf{(ii)}} do always exist \cite{spingluing}. The non-trivial question is whether one can find a lift satisfying \textcolor{DarkOrchid}{\textbf{(iii)}} called the cocycle condition. Since $\yhwidehat{\text{Ad}}(f_{\a \delta}f_{\delta \b}f_{\b \a}) = g_{\a \delta}g_{\delta \b}g_{\b \a} = 1$, one has $f_{\a \delta}f_{\delta \b}f_{\b \a} = \pm1$. Namely, it may happen that $f_{\a \delta}(x)f_{\delta \b}(x)$ is distinct from $f_{\b \a}(x)$ due to the nature of the non-trivial double cover $\yhwidehat{\text{Ad}} : \Spin(n) \to \SO(n)$. Thus the cocycle condition can fail for the lifts $\{f_{\a\b}\}$. The possibility of such a lift, that is,  the existence of a spin structure on the manifold relies on the field of algebraic topology and the concept of Cech cohomology. Therefore, spin structures need not exist and even if when they do they may not be unique. The existence depends on a topological condition, the vanishing of the so-called second Stiefel-Whitney class $w_{2}(\M) \in H^{2}(\M,\zz_{2})$. Here, however, we will not define these topological concepts. A complete discussion of existence and uniqueness of spin structures on oriented Riemannian manifolds can be found in Chapter II of reference \cite{SpinGeo}. The definition of a spin structure for orientable pseudo-Riemannian manifolds is the same, what differs is that the conditions for existence of spinor bundles in indefinite signature are different from those in Riemannian signature, since $\Spin(p,q)$ groups with $pq\neq0$ are disconnected, unlike the $\Spin(n)$ groups. The topological conditions for existence of $\Spin_{+}(p, q)$ structures involve the modified second Stiefel-Whitney class \cite{Karoubi, Lazaroiu:2016vov}.

\begin{ex}
 {{\textcolor{DarkOrchid}{{$\blacktriangleright$\;}}}}    Spin structures on $\M = S^{1}$ \cite{spinazul}. \normalfont \textcolor{DarkOrchid}{\textbf{\textit{1) Case}}} Trivial spin structure: in this case one has $\SO(1) = \{1\}$, $\Spin(1) = \zz_2$ and $\SO(\M) = S^{1}$. Hence a spin structure on $S^{1}$ is a two-fold covering of $S^{1}$. The \textit{trivial} spin structure is the trivial covering $\Spin(S^{1}) = S^{1} \cross \zz_{2}$. Let $\pi: \Spin(S^{1}) = S^{1} \cross \zz_{2} \to \SO(S^{1}) = S^{1}$ be the projection on the first factor denoted by $\text{pr}_{1}$. It is straightforward that the following diagram commutes hence it defines a spin structure:

    \[ \begin{tikzcd}
S^{1} \cross \zz^{2} \cross \zz^{2} \arrow[r] \arrow[d, "\text{pr}_{1} \cross \rho"] & S^{1} \cross \zz^{2} \arrow{d}{\text{pr}_{1}}  \arrow[dr, "\text{pr}_{1}"]  \\
S^{1} \cross \{1\} \arrow[r]          & S^{1} \arrow[r, "\text{id}"]                                         & S^{1}
\end{tikzcd} 
    \]

    \noindent \textcolor{DarkOrchid}{\textbf{\textit{2) Case}}}  Non-trivial spin structure on $S^{1}$: the non-trivial spin structure is the non-trivial covering $\Spin(S^{1}) = S^{1}$ with $\pi: S^{1} \to S^{1}$, $z \mapsto z^{2}$. The action of $\Spin({1}) = \zz^{2}$ on $\Spin(S^{1}) = S^{1}$ is given by $z \mapsto -z$. This way one has $\pi = \pi_{Sp}$ and that the conditions $\pi_{\SO} \circ \pi = \text{id} \circ \pi = \pi_{Sp}$ and $\pi(p\cdot h) = \pi(p) \cdot \rho(h) = p^{2}$ are satisfied, that is:

        \[ \begin{tikzcd}
S^{1}  \cross \zz^{2} \arrow{r}{(z,a) \mapsto az} \arrow{d}{(z,a) \mapsto (z^{2},1)} & S^{1} \arrow{d}{z \mapsto z^{2}}  \arrow[dr, "z \mapsto z^{2}"]  \\
S^{1} \cross \{1\} \arrow{r}{\text{pr}_1}          & S^{1} \arrow[r, "\text{id}"]                                         & S^{1}
\end{tikzcd} 
    \]

     \noindent since the diagram commutes it defines a spin structure on $S^{1}$. $\;\textcolor{DarkOrchid}{{\blacktriangleleft}}$
\end{ex}

\section{The Bundle of Spinors}
 \paragraph{ } The spin structure give rise to the bundle of spinors.

\begin{definvioleta}
    The \textbf{bundle of spinors} of $\M$ for the spin structure $(\Spin(\M),\pi)$ is the associated vector bundle $\pi' : \Spin(\M)\cross_{\varrho} S \to \M$ with respect to a spin representation $\varrho$ (Table \ref{table7_spin}). A \textbf{spinor} $\psi$ is a section of the bundle of spinors, $\psi \in \Gamma(\M, Spin(\M) \cross_{\varrho} S).$ 
\end{definvioleta}

\begin{center}

\tikzset{every picture/.style={line width=0.75pt}} 

\begin{tikzpicture}[x=0.75pt,y=0.75pt,yscale=-1,xscale=1]

\draw  [draw opacity=0][fill={rgb, 255:red, 248; green, 231; blue, 28 }  ,fill opacity=0.39 ] (422.33,34.7) -- (559.33,34.7) -- (559.33,126.32) -- (422.33,126.32) -- cycle ;
\draw [color={rgb, 255:red, 195; green, 181; blue, 11 }  ,draw opacity=1 ][line width=1.5]  [dash pattern={on 1.69pt off 2.76pt}]  (422.17,48.11) .. controls (436.18,72.5) and (446.43,67.03) .. (456.2,57.81) .. controls (466.84,47.77) and (476.92,33.28) .. (490.67,48.11) .. controls (517.04,76.55) and (541.21,32) .. (559.17,48.11) ;
\draw  [draw opacity=0][fill={rgb, 255:red, 240; green, 224; blue, 247 }  ,fill opacity=1 ] (73.9,44.7) .. controls (73.9,28.85) and (86.75,16) .. (102.6,16) -- (313.1,16) .. controls (328.95,16) and (341.8,28.85) .. (341.8,44.7) -- (341.8,130.8) .. controls (341.8,146.65) and (328.95,159.5) .. (313.1,159.5) -- (102.6,159.5) .. controls (86.75,159.5) and (73.9,146.65) .. (73.9,130.8) -- cycle ;
\draw    (287.12,104.73) .. controls (304.93,92.13) and (298,78.86) .. (286,64.55) ;
\draw [shift={(284.07,62.28)}, rotate = 49.15] [fill={rgb, 255:red, 0; green, 0; blue, 0 }  ][line width=0.08]  [draw opacity=0] (8.93,-4.29) -- (0,0) -- (8.93,4.29) -- cycle    ;
\draw    (185.87,123.07) -- (221.4,123.07) ;
\draw [shift={(224.4,123.07)}, rotate = 180] [fill={rgb, 255:red, 0; green, 0; blue, 0 }  ][line width=0.08]  [draw opacity=0] (8.93,-4.29) -- (0,0) -- (8.93,4.29) -- cycle    ;
\draw    (120.87,101.23) .. controls (103.2,88.68) and (113.2,74.31) .. (119.52,62.78) ;
\draw [shift={(120.87,60.23)}, rotate = 116.76] [fill={rgb, 255:red, 0; green, 0; blue, 0 }  ][line width=0.08]  [draw opacity=0] (8.93,-4.29) -- (0,0) -- (8.93,4.29) -- cycle    ;
\draw [color={rgb, 255:red, 245; green, 166; blue, 35 }  ,draw opacity=1 ]   (184.53,138) -- (208.57,138) -- (223.2,138) ;
\draw [shift={(225.2,138)}, rotate = 180] [color={rgb, 255:red, 245; green, 166; blue, 35 }  ,draw opacity=1 ][line width=0.75]    (10.93,-3.29) .. controls (6.95,-1.4) and (3.31,-0.3) .. (0,0) .. controls (3.31,0.3) and (6.95,1.4) .. (10.93,3.29)   ;
\draw  [draw opacity=0][fill={rgb, 255:red, 75; green, 181; blue, 244 }  ,fill opacity=0.18 ] (133.66,46.46) .. controls (133.66,43.22) and (136.28,40.6) .. (139.52,40.6) -- (271.14,40.6) .. controls (274.38,40.6) and (277,43.22) .. (277,46.46) -- (277,64.03) .. controls (277,67.26) and (274.38,69.89) .. (271.14,69.89) -- (139.52,69.89) .. controls (136.28,69.89) and (133.66,67.26) .. (133.66,64.03) -- cycle ;
\draw  [draw opacity=0][fill={rgb, 255:red, 240; green, 224; blue, 247 }  ,fill opacity=1 ] (341.8,85.37) -- (383.8,85.37) -- (383.8,75.37) -- (411.8,95.37) -- (383.8,115.37) -- (383.8,105.37) -- (341.8,105.37) -- cycle ;
\draw  [draw opacity=0][fill={rgb, 255:red, 181; green, 134; blue, 230 }  ,fill opacity=0.85 ] (231.07,119.07) .. controls (231.07,114.65) and (234.65,111.07) .. (239.07,111.07) -- (293.07,111.07) .. controls (297.48,111.07) and (301.07,114.65) .. (301.07,119.07) -- (301.07,143.07) .. controls (301.07,147.48) and (297.48,151.07) .. (293.07,151.07) -- (239.07,151.07) .. controls (234.65,151.07) and (231.07,147.48) .. (231.07,143.07) -- cycle ;
\draw  [draw opacity=0][fill={rgb, 255:red, 181; green, 134; blue, 230 }  ,fill opacity=1 ] (111.37,119.07) .. controls (111.37,114.65) and (114.95,111.07) .. (119.37,111.07) -- (173.37,111.07) .. controls (177.78,111.07) and (181.37,114.65) .. (181.37,119.07) -- (181.37,143.07) .. controls (181.37,147.48) and (177.78,151.07) .. (173.37,151.07) -- (119.37,151.07) .. controls (114.95,151.07) and (111.37,147.48) .. (111.37,143.07) -- cycle ;
\draw [color={rgb, 255:red, 222; green, 207; blue, 24 }  ,draw opacity=1 ]   (278.55,37.33) .. controls (293.35,26.23) and (316.35,21.29) .. (340.99,20.93) .. controls (382.1,20.35) and (427.76,32.57) .. (447.37,50.37) ;
\draw [shift={(448.53,51.47)}, rotate = 224.21] [color={rgb, 255:red, 222; green, 207; blue, 24 }  ,draw opacity=1 ][line width=0.75]    (10.93,-3.29) .. controls (6.95,-1.4) and (3.31,-0.3) .. (0,0) .. controls (3.31,0.3) and (6.95,1.4) .. (10.93,3.29)   ;
\draw    (292.62,54.13) .. controls (325.17,59.33) and (388.75,39.92) .. (430.53,70.13) ;
\draw [shift={(289.2,53.47)}, rotate = 13.17] [fill={rgb, 255:red, 0; green, 0; blue, 0 }  ][line width=0.08]  [draw opacity=0] (8.93,-4.29) -- (0,0) -- (8.93,4.29) -- cycle    ;

\draw (115.5,79.4) node [anchor=north west][inner sep=0.75pt]    {$\pi _{sp}$};
\draw (302,88.67) node [anchor=north west][inner sep=0.75pt]  [font=\small,color={rgb, 255:red, 237; green, 237; blue, 237 }  ,opacity=1 ] [align=left] {\textcolor[rgb]{0.73,0.07,1}{\textbf{Spin Structure}}};
\draw (116.37,113.47) node [anchor=north west][inner sep=0.75pt]  [font=\footnotesize]  {$ \begin{array}{l}
\textcolor[rgb]{1,1,1}{\Spin}\textcolor[rgb]{1,1,1}{(}\mathcal{\textcolor[rgb]{1,1,1}{M}}\textcolor[rgb]{1,1,1}{)}\\
\textcolor[rgb]{0.98,0.85,0.65}{\Spin(n)}
\end{array}$};
\draw (241.07,113.47) node [anchor=north west][inner sep=0.75pt]  [font=\footnotesize]  {$ \begin{array}{l}
\textcolor[rgb]{1,1,1}{\SO}\textcolor[rgb]{1,1,1}{(}\mathcal{\textcolor[rgb]{1,1,1}{M}}\textcolor[rgb]{1,1,1}{)}\\
\textcolor[rgb]{0.98,0.85,0.65}{\SO(n)}
\end{array}$};
\draw (268.12,80.63) node [anchor=north west][inner sep=0.75pt]    {$\pi _{so}$};
\draw (197.17,104.07) node [anchor=north west][inner sep=0.75pt]    {$\pi $};
\draw (190.67,144.4) node [anchor=north west][inner sep=0.75pt]  [font=\footnotesize,color={rgb, 255:red, 184; green, 233; blue, 134 }  ,opacity=1 ]  {${\textstyle \textcolor[rgb]{0.96,0.65,0.14}{\yhwidehat{\text{Ad}}}}$};
\draw (348.83,31.23) node [anchor=north west][inner sep=0.75pt]    {$\pi '$};
\draw (447.67,88.97) node [anchor=north west][inner sep=0.75pt]  [font=\footnotesize]  {$\Spin(\mathcal{M}) \times _{\varrho } S$};
\draw (488.47,64.5) node [anchor=north west][inner sep=0.75pt]  [font=\footnotesize] [align=left] {\textcolor[rgb]{0.76,0.71,0.04}{spinor field}};
\draw (354.03,1.07) node [anchor=north west][inner sep=0.75pt]  [color={rgb, 255:red, 222; green, 207; blue, 24 }  ,opacity=1 ]  {$\psi $};
\draw (442.17,106.2) node [anchor=north west][inner sep=0.75pt]  [font=\footnotesize,color={rgb, 255:red, 0; green, 0; blue, 0 }  ,opacity=1 ] [align=left] {\textit{Bundle of Spinors}};
\draw (179.17,44.61) node [anchor=north west][inner sep=0.75pt]  [color={rgb, 255:red, 0; green, 0; blue, 0 }  ,opacity=1 ]  {$(\mathcal{M} ,g)$};

\end{tikzpicture}
\end{center}

\noindent The representations $\lambda: \SO(n) \to \text{GL}(V)$ and $\lambda': \Spin(n) \to \text{GL}(V)$ were used within the construction of the spin structure since it is a principal bundle with fibre $\Spin(n)$ for which the quotient of each fibre by the centre is $\pm1$ is isomorphic to the frame bundle $\SO(\M)$ of $\M$. However, for the construction of the spinor bundle the $\lambda'$ shall not be considered, since it is induced from $\SO(n)$ and it is not a spin representation as discussed in the Section \ref{sec_spinrep}. Thus, the associated bundle $\Spin(\M) \cross_{\lambda'} V$ does not define a bundle of spinors. For the construction of the bundle of spinors a spin representation is required to be considered. Therefore, with respect to a spin representation $\varrho$ a real spinor bundle $\mathcal{S}$ is then a bundle of the form 

\begin{equation}
    \mathcal{S} = \Spin(\M) \cross_{\varrho} S
\end{equation}

\noindent where $S$ is a left module for the real Clifford algebra $\cl_{p,q}$, whereas a a complex spinor bundle is a bundle of the form  

\begin{equation}
    \mathcal{S}_{\cc} = \Spin(\M) \cross_{\varrho} S_{\cc}
\end{equation}

\noindent where $S$ is a left module for the complex Clifford algebra $\cc \tensor \cl_{p,q}$ \cite{SpinGeo}. At this point, we have seen three definitions of spinors. Namely:

\begin{itemize}
    \item[\textcolor{Goldenrod}{$\triangleright$}] \textbf{\textcolor{Goldenrod}{classical}}: spinor is an element of the irreducible representation space  $S$ of the Spin group with respect to the mapping $\yhwidehat{\text{Ad}}|_{\Spin_{+}(p,q)}:  \Spin_{+}(p,q) \to \text{GL}(S)$.
    \item[\textcolor{Goldenrod}{$\triangleright$}]\textbf{\textcolor{Goldenrod}{algebraic}}: spinor is an element of a minimal left ideal of a Clifford algebra.
    \item[\textcolor{DarkOrchid}{$\triangleright$}] \textbf{\textcolor{DarkOrchid}{geometrical}}: spinor is a section of the associated vector bundle  $\Spin(\M)\cross_{\varrho} S$ of the principal bundle $\Spin(\M)$ with respect to the Spin structure on $(\M,g)$ and the spin representation $\varrho$.
\end{itemize}

\noindent The classical and the algebraic definition were seen to be equivalent since irreducible representation spaces are minimal left ideals. The classic definition of a spinor is also equivalent to the geometrical one. The key to relating these two is understanding that the spinor bundle and the spin representation $\varrho$ of the group  $\Spin(n)$ are connected through the bundle structure. The space $S$ represents the typical fibre of the associated bundle and carries the spin representation, that is, the representation dictates how the elements of $S$ transform under the action of the spin group, hence $S$ encodes the algebraic properties and symmetries that spinors must satisfy. Locally, in each point $x \in U$, a section, namely, a spinor $\psi$ is simply $\psi: U \to S$ with $\psi(x) \in S$. Globally the space of spinors is $\Gamma(\M, \Spin(\M)\cross_{\varrho} S)$ however, locally, it can be simply realised as $S$ as the classical definition establishes.

Nevertheless, the bundle of spinors can be defined in terms of the \textcolor[rgb]{0.08,0.18,0.88}{\textbf{Kähler-Atiyah bundle}}, this encapsulate both the geometry of the manifold and the algebraic properties of spinors with respect to the Clifford algebra, thus establishing a comprehensive framework for studying spinors in relation to both their geometric and algebraic contexts. To be more precise, first the bundle of real pinors is introduced \cite{bab2}.

{\colorlet{shadecolor}{BlueViolet!15}\begin{shaded}
The \textbf{bundle of real pinors} $\mathcal{P}$ over the manifold  $\M$ is defined as being the real bundle whose fibers are spaces that carry the irreducible representation of the fibers $\cl(T^{*}_{x}\M,g_{x}^{*})$ of the Clifford bundle $\cl(T^{*}\M)$, for all $x \in U \subset \M$.
\end{shaded}}

\noindent Where $U$ denotes an open set in $\M$. The bundle of real pinors is equipped with a morphism $\upgamma: (\bigwedge T^{*}\M, \diamond) \to (\text{End}(\mathcal{P}), \circ)$ that maps the Kähler--Atiyah bundle of $(\M,g)$ to the bundle of endomorphisms of $\mathcal{P}$, where here $\circ$ represents the product in the space of  endomorphisms.

{\colorlet{shadecolor}{BlueViolet!15}\begin{shaded}
The induced mapping on global sections, with the same notations, wits \cite{bab2}
\begin{equation}\label{eq_mapagamapin}
    \upgamma : (\Omega (\M), \diamond) \to \Gamma\left(\M,\text{End}(\mathcal{P})\right), \circ).
\end{equation}
\end{shaded}}

\noindent For each point $x \in \M$, the fibre $\upgamma_x$ is an irreducible representation of the Clifford algebra $(\bigwedge T^{*}_x\M, \diamond_x) \simeq \Cl_{p,q} $ in the 
$\mathbb{R}$-vector space $\mathcal{P}_x$, which denotes the fibre of $\mathcal{P}$ at the point $x \in \M$. A section of the bundle of real pinors is called a \textbf{pinor field}. Since we are interested in spinor fields, it is natural to consider the bundle of real spinors.

{\colorlet{shadecolor}{BlueViolet!15}\begin{shaded}
The \textbf{bundle of real spinors}  $\mathcal{S}$ of $(\M,g)$ consists of a bundle of modules over the even Kähler--Atiyah bundle $(\bigwedge T^{*}\M^{\textsc{even}}, \diamond)$. The fibres of the bundle of real spinors comprise objects that carry the irreducible representation spaces of $\cl^{\textsc{even}}(T^{*}_x \M, g_{x}^{*})$ in $\cl^{\textsc{even}}(T^{*}\M)$, for all $x \in U$. Likewise, a \textbf{spinor field} is defined as a section of the bundle of real spinors and each fibre $\mathcal{S}_{x}$ arises from the mapping
\begin{equation}\label{eq_mapaspinn}
    \upgamma^{\textsc{even}}: (\bigwedge T^{*}\M^{\textsc{even}}, \diamond) \to (\text{End}(\mathcal{S}), \circ). 
\end{equation}
\end{shaded}}

The restriction $\upgamma^{\textsc{even}}$ of $\upgamma$ to the subbundle $\bigwedge T^{*}\M^{\textsc{even}} \subset \bigwedge T^{*}\M$ makes any bundle of real pinors $(\mathcal{P}, \upgamma)$ to drop into a bundle of real spinors $(\mathcal{S}, \upgamma^{\textsc{even}})$. Henceforth, for the sake of simplicity, the notation ($\mathcal{S}, \upgamma$) is employed to emphasise the approach to spinors.

\paragraph{ } The bundle of spinors $\mathcal{S}$ defined through the Kähler-Atiyah bundle is the associated bundle $\mathcal{S} = \Spin(\M) \cross_{\upgamma} S$. Note that for the definition of Clifford bundles including the Kähler-Atiyah bundle, the spin structure on $\M$ was not required. However, the spin structure is essential in defining the Clifford multiplication on the bundle of spinors. Since the space of spinors $S$ is a left ideal of Clifford algebra $\cl_{n}$ we want that the Clifford multiplication $\rr^{n} \cross S \to S$ , given by $(\'v,\psi) \mapsto \upgamma(\'v)\psi$ to extend to the bundle of spinors, therefore we want it to extend to a map of sections $\Gamma(T^{*}\M) \cross \mathcal{S} \to \mathcal{S}$ such that $(\uptheta, \psi) \mapsto \upgamma(\uptheta)\psi$ with the morphism $\upgamma$ realised as the spin representation. This mapping defines the Clifford multiplication on the bundle of spinors \cite{Konstantin}. Indeed, a trivialising chart $U_{\alpha}$ for $\SO(\M)$ trivialises both $T^{*}\M$ and $\mathcal{S}$ by construction. Over $U_{\alpha}$ a local section $\uptheta$ of $T^{*}\M$ is  $\uptheta_{\a}: U_{\a} \to \rr^{n}$, and a local section $\psi$ of $\mathcal{S}$ is given by $\psi_{\a}: U_{\a} \to S$. One defines

\begin{equation}
    (\upgamma(\uptheta)\psi)_{\a} =  \upgamma(\uptheta_{\a})\psi_{\a}
\end{equation}

\noindent To show that this is a section of $\mathcal{S}$ it is required to show that it
transforms correctly, namely

\begin{equation}
    (\upgamma(\uptheta)\psi)_{\b} = f_{\a\b}(\upgamma(\uptheta)\psi)_{\a},
\end{equation}

\noindent with respect to the transition functions $f_{\a\b}$. Indeed, the transition functions on the overlap $U_{\a\b}$ for $T^{*}\M$ is $g_{\a\b}$, however, within a spin structure,  $g_{\a\b}$ lifts to $f_{\a\b}$ through $\text{Ad}(f_{\a\b}) = g_{\a\b}$, therefore: 
\begin{align}
    \begin{aligned}
        (\upgamma(\uptheta)\psi)_{\b}  &= \upgamma(\uptheta_{\b}) \psi_{\b}\\
        &= g_{\a\b}\upgamma(\uptheta_{\a})f_{\a\b} \psi_{\a}\\
        &= \text{Ad}(f_{\a\b})\upgamma(\uptheta_{\a})f_{\a\b} \psi_{\a}\\
        &= f_{\a\b}\upgamma(\uptheta_{\a})f_{\a\b}^{-1}f_{\a\b} \psi_{\a}\\
        &= f_{\a\b}\upgamma(\uptheta_{\a})\psi_{\a}\\\
        &=  f_{\a\b}(\upgamma(\uptheta)\psi)_{\a}.
    \end{aligned}
\end{align}

\paragraph{ } The geometric algebra approach to the bundle of spinors $\mathcal{S}$ explicitly ties together the geometric properties of the manifold $\M$ and the algebraic structure captured by the Kahler-Atiyah bundle of algebras. At each point the fibre captures the essence of spinors as irreducible representations of the Clifford algebra $\cl_{p,q}$, thereby unifying the local and global perspectives on spinors, establishing a comprehensive framework for studying spinors in relation to both their geometric and algebraic contexts.

\colorlet{chapter}{Fuchsia!50}
\chapter{Generalised Bilinear Covariants}\label{chap_bil2}

\hypersetup{
  colorlinks = true,
  linkcolor  = Fuchsia,
  citecolor = Violet,
}

\paragraph{ } In this Chapter we demonstrate that \textcolor[rgb]{0.47,0.81,0.1}{bilinear covariants} can be understood and extended to any dimension and signature, along with the spinor classification. The \textcolor[rgb]{0.08,0.18,0.88}{Kähler-Atiyah bundle} described in Chapter \ref{chap3} plays a fundamental role in the essence of the \textcolor[rgb]{0.74,0.06,0.88}{bundle of spinors} defined in Chapter \ref{chap_spingeo}. It provides a robust framework with efficient techniques to analyse the generalised bilinear covariants. Depending on the dimension and metric signature, certain homogeneous differential forms, which act as bilinear covariants, may vanish due to algebraic constraints. Consequently, some spinor bilinear covariants can have null values, this behaviour is governed by generalized geometric Fierz identities \cite{bab1, bab2, Lazaroiu:2012fw, Lazaroiu:2013fg, Babalic:2013iza, Babalic:2013iha, Babalic:2014fua, Babalic:2015xia, Lazaroiu:2012kxa}. Consequently, Lounesto's classification of spinor fields on four-dimensional Lorentzian manifolds can be successfully extended to other dimensions and metric signatures.

\paragraph{ } The main reference of this Section is Ref. \cite{bab2}. The first step is to present the concept of the \textcolor{Fuchsia}{generalised bilinear covariants}, that is, we need to define a type of inner product in the space of spinors, which in this context refers to the sections of the bundle of spinors. This inner product is based on the concept of a bilinear form being deemed \textit{admissible}. From this admissible bilinear form, we construct the generalised bilinear covariants, which are given here by the \textit{Fierz isomorphisms.} Additionally, we recall that bilinear covariants in Minkowski space-time are constrained by the Fierz-Pauli-Kofink identities. We aim to generalise and emulate these identities for the broader case as well. Generalising the Fierz-Pauli-Kofink relations beyond the Minkowski spacetime setup brings unexpectedly prominent panoramas for higher dimensional theories in Physics \cite{daRocha:2017vqh}. 

\paragraph{ } References \cite{bab1, bab2} show that the admissible bilinear form exhibits certain symmetries that give rise to obstructions depending on the dimension and signature of the Clifford algebra to which they are related.  The geometric Fierz identities strongly obstruct the number and the existence of classes of spinor fields and enable the classification of spinors to be carried out as have been done in the context of compactifications underlying supergravity and string theory \cite{Bonora:2014dfa,deBrito:2016qzl,Yanes:2018krn}. 

\paragraph{ } In the end of this Chapter, we present an original result regarding the classification of spinors in the space of an eight-dimensional manifold $\M_8$ which is derived from an eleven-dimensional manifold being the compactification AdS$_3 \times \M_8$.  This eleven-dimensional manifold is considered within the framework of supergravity. Therefore, the space $\M_8$ is connected to supersymmetry conditions in AdS$_3$. We consider Killing spinors that are constrained by supersymmetry conditions in these spaces and analyse the covariant bilinears of these spinors to achieve our classification result. It is important to emphasise that the physical aspects related to supersymmetry and supergravity are beyond the scope of this dissertation. We mention these aspects only as motivation and context for the manifold we are considering for the spinor classification.

\paragraph{ } Let us start by recalling the mapping $\upgamma$ of Eq. \eqref{eq_mapaspinn}. Studying some aspects of this mapping is extremely important as it is related to the symmetries of the admissible bilinear form that we want to define to construct the generalised bilinear covariants. This mapping induced on sections is defined on a local coframe by $\upgamma^{p} = \upgamma(e^{p}) \in \Gamma(U,\text{End}\,\mathcal{S})$ and  satisfies the morphism property   

\begin{align}
    \begin{aligned}
   \color{Fuchsia} \begin{cases}\color{black}
           \upgamma(\upalpha \diamond \upbeta) = \upgamma(\upalpha) \circ \upgamma(\upbeta) \text{ for all } \upalpha, \upbeta \in \Omega(\M),\\\color{black}
        \upgamma(1_{\M}) = 1_{\mathcal{S}}.\color{black}
    \end{cases}\color{black}
    \end{aligned}
\end{align}

\noindent  Moreover, 
\begin{equation}
    \upgamma(e^{m_1 \ldots m_k}) = \upgamma^{m_1 \ldots m_k} = \upgamma^{m_1} \circ \cdots \circ \upgamma^{m_k}. 
\end{equation}

Let $\upalpha$ in $\Omega(\M)$ be an inhomogeneous differential form. One can represent it with respect to a local coframe $\{e^{1},\ldots,e^{n}\}$ as follows
 \begin{equation}\label{app}
    \upalpha = \sum_{k=0}^{n} \upalpha^{k} =  \sum_{k=0}^{n} \frac{1}{k!} \upalpha^{k}_{m_1 \ldots m_k} e^{m_1 \ldots m_k},
 \end{equation}
 \noindent where $\upalpha^{k} \in \Omega^{k}(\M)$ and $\upalpha^{k}_{m_1 \ldots m_k}$ are $\mathcal{C}^{\infty}$-functions on $U \subset \M$. Applying $\upgamma$ on Eq. (\ref{app}) yields
 \begin{equation}
    \upgamma (\upalpha) =\sum_{k=0}^{n} \frac{1}{k!} \upalpha^{k}_{m_1 \ldots m_k} \upgamma^{m_1\ldots m_k}.
\end{equation}

\paragraph{ } The surjectivity or the injectivity properties of the mapping $\upgamma$ are not always verified, being contingent on the classification of $\cl_{p,q} \simeq \bigwedge T_{x}^{*}\M$ with techniques to consider the inverse $\upgamma^{-1}$ in many situations as presented in the Reference \cite{bab2}. 

\begin{table}[H]\centering
\begin{tabular}{
>{\columncolor{Fuchsia!15}}l |cc}
               & \cellcolor{Fuchsia!15}injective                                                                                                              & \cellcolor{Fuchsia!15}non-injective \\ \hline
surjective     & $\mathbf{0}(\rr)$, $\mathbf{2}(\rr)$                                                                         & $\mathbf{1}(\rr)$    \\
non-surjective & $\mathbf{3}(\cc)$, $\mathbf{7}(\cc)$, $\mathbf{4}(\hh)$, $\mathbf{6}(\hh)$ & $\mathbf{5}(\hh)$   
\end{tabular}\caption{Classification of the morphism $\upgamma$ within the real Clifford algebras classification.}\label{tablesa65dfr87}
\end{table}

\noindent In the above Table \ref{tablesa65dfr87}, the number in bold represents the factor $p - q \mod 8$ and the field in parenthesis is the representation space with respect to the Table \ref{table1}. The bundle morphism $\upgamma$ is then fibrewise injective iff. $\cl_{p,q}$ is simple as an associative real algebra, that is, if $p - q \not\equiv_{8} 1,5$. In the case $p - q \equiv_{8} 1,5$ when $\upgamma$ fails to be injective, one has for the volume form $\boldsymbol{\uptau}\in \Omega(\M)$:

\begin{equation}
    \upgamma(\boldsymbol{\uptau}) = \varepsilon_{\upgamma}1_{\mathcal{S}},
\end{equation}

\noindent such that the sign factor $\varepsilon_{\upgamma} \in \{\pm1\}$ is called the signature of $\upgamma$. Recall the operator $P_{\pm}$ defined in Eq. \eqref{eq_asdfgh62525234} and consider $\upgamma \circ P_{\pm}$. For $\upalpha \in \Omega(\M)$, it follows that

\begin{align}
    \begin{aligned}
        \upgamma \circ P_{\pm} (f) &=  \frac{1}{2} \left( \upgamma ( \upalpha  \pm \upalpha  \diamond \boldsymbol{\uptau} )\right)\\
    &=  \frac{1}{2} \left(  \upgamma (\upalpha ) \pm \upgamma(\upalpha  \diamond \boldsymbol{\uptau})\right) \\
    &=  \frac{1}{2} \left( \upgamma(\upalpha ) \pm  \upgamma(\upalpha ) \diamond \upgamma(\boldsymbol{\uptau})\right)\\
    &=  \frac{1}{2} \left( \upgamma(\upalpha ) \pm  \upgamma(\upalpha ) \diamond \varepsilon_\upgamma 1_{\mathcal{S}}\right).
    \end{aligned}
\end{align}

\noindent We thus have four cases:
\begin{itemize}
    \item [\textcolor{Fuchsia}{\textit{\textbf{1.}}}] $P_{+}$ and $ \varepsilon_\upgamma = +1$
\begin{align}
    \begin{aligned}
         \upgamma \circ P_{+} (\upalpha) = \frac{1}{2} \left( \upgamma(\upalpha ) + \upgamma(\upalpha ) \diamond 1_{\mathcal{S}}\right) = \upgamma(\upalpha ).
    \end{aligned}
\end{align}

    \item [\textcolor{Fuchsia}{\textit{\textbf{2.}}}] $P_{+}$ and $ \varepsilon_\upgamma = -1$
\begin{align}
    \begin{aligned}
         \upgamma \circ P_{+} (\upalpha) =   \frac{1}{2} \left( \upgamma(\upalpha ) + \upgamma(\upalpha ) \diamond -1_{\mathcal{S}}\right) = 0.
    \end{aligned}
\end{align}

    \item [\textcolor{Fuchsia}{\textit{\textbf{3.}}}] $P_{-}$ and $ \varepsilon_\upgamma = +1$
\begin{align}
    \begin{aligned}
        \upgamma \circ P_{+} (\upalpha) =   \frac{1}{2} \left( \upgamma(\upalpha ) - \upgamma(\upalpha ) \diamond 1_{\mathcal{S}}\right) = 0.
    \end{aligned}
\end{align}

    \item [\textcolor{Fuchsia}{\textit{\textbf{4.}}}]$P_{-}$ and $ \varepsilon_\upgamma = -1$
\begin{align}
    \begin{aligned}
          \upgamma \circ P_{+} (\upalpha) =   \frac{1}{2} \left( \upgamma(\upalpha ) - \upgamma(\upalpha ) \diamond -1_{\mathcal{S}}\right) = \upgamma(\upalpha ).
    \end{aligned}
\end{align}

\end{itemize}

\noindent In summary,

\begin{align}
    \begin{aligned}\label{eq_4223pnjkhuytrdf}
\color{Fuchsia}\begin{cases}\color{black}
 \upgamma \circ P_{\varepsilon_{\upgamma}}  = \upgamma, \color{black}\\
        \color{black}\upgamma \circ P_{-\varepsilon_{\upgamma}}  = 0.\color{black}   
\end{cases}        \color{black}
    \end{aligned}
\end{align}

\noindent That is, for this case, there is then two choices for this sign factor which lead to two non equivalent choices for $\upgamma$. By considering the direct sum

\begin{equation}
    \bigwedge T^{*}\M = \bigwedge\,\!^{\varepsilon_{\upgamma}} T^{*}\M \oplus \bigwedge\!\,\!^{-\varepsilon_{\upgamma}} T^{*}\M
\end{equation}

\noindent the Eq. \eqref{eq_4223pnjkhuytrdf} shows that $\upgamma$ vanishes when restricted to the sub-bundle $\bigwedge\!\,\!^{-\varepsilon_{\upgamma}} T^{*}\M$. Since $\ker(\upgamma) = \bigwedge\!\,\!^{-\varepsilon_{\upgamma}} T^{*}\M$ the corresponding map
on sections (which has the same notation $\upgamma$) has kernel $\ker(\upgamma) = \Omega^{-\varepsilon_{\upgamma}}(\M)$. On the other hand, $\bigwedge\!\,\!^{\varepsilon_{\upgamma}} T^{*}\M$ is a sub-bundle of algebras of the Kähler-Atiyah bundle of $(\M, g)$ and is called the effective domain for $\upgamma$ \cite{bab2}. The following notation allows treating injective and non-injective cases simultaneously \cite{bab1}:

\begin{align}
    \begin{aligned}
        \Omega^{\upgamma}(\M) &:= \color{Fuchsia} \begin{cases}\color{black}
            \Omega(\M),\;\;\;  \text{if $\upgamma$ is fibrewise injective $(p - q \not\equiv_{8} 1,5)$},\\\color{black}
            \Omega^{\varepsilon_{\upgamma}}(\M),   \;\;\;\text{if $\upgamma$ is not fibrewise injective $(p - q \equiv_{8} 1,5)$},\color{black}
        \end{cases}\color{black}\\
        \color{black}\Omega^{-\upgamma}(\M) &:= \color{Fuchsia}\begin{cases}\color{black}
            0,\;\;\; \text{if $\upgamma$ is fibrewise injective $(p - q \not\equiv_{8} 1,5)$},\\\color{black}
            \Omega^{-\varepsilon_{\upgamma}}(\M),   \;\;\;\text{if $\upgamma$ is not fibrewise injective $(p - q \equiv_{8} 1,5)$}.\color{black}
        \end{cases}\color{black}
    \end{aligned}
\end{align}

\noindent With this notation, one always have $\ker(\upgamma) = \Omega^{-\upgamma}(\M)$ and

\begin{equation}
     \Omega(\M) =  \Omega^{\upgamma}(\M) \oplus \Omega^{-\upgamma}(\M).
\end{equation}

\noindent Therefore, the injection of the restriction of the -- map on sections induced by -- $\upgamma$ to the effective domain $\Omega^{\upgamma}(\M)$ is guaranteed. A partial inverse  $\upgamma^{-1} : \text{End}(\mathcal{S}) \to \Omega^{\upgamma}(\M)$ can be defined, satisfying the following relations\cite{bab2}:

\begin{align}
    \begin{aligned}
\color{Fuchsia}\begin{cases}\color{black}\upgamma \circ \upgamma^{-1}   \color{black}= 1_{\text{End}(\mathcal{S})},\\ \color{black}
        \upgamma^{-1} \circ \upgamma   \color{black}= P_{\varepsilon_{\upgamma}}, \\ \color{black}
 \upgamma \circ P_{\varepsilon_{\upgamma}}  = \upgamma, \color{black}\\
        \color{black}\upgamma \circ P_{-\varepsilon_{\upgamma}}  = 0.\color{black}   
\end{cases}        \color{black}
    \end{aligned}
\end{align}

\noindent The partial inverse is used to define the so-called \textit{‘dequantization’} map\cite{bab1}

\begin{equation}
    \check{T} := \upgamma^{-1}(T) \in \Omega^{\upgamma}(\M).
\end{equation}

\noindent which is a (generally inhomogeneous) differential form defined on $\M$. By employing the partial inverse, statements about operators acting on spinors can be transferred to to statements about differential forms. The fundamental Fierz identities that constrain differential forms constructed as bilinears in sections of $\mathcal{S}$ can be systematically formulated using \textit{Fierz isomorphisms} constructed from the partial inverse. This approach provides algebraic constraints on specific systems of differential forms defined on $(\M,g)$. The resulting algebra of constraints on these differential forms can be expressed succinctly, facilitating the analysis of their structural properties. We proceed to the definition of the generalised Fierz identities in the next section.

\section{Generalised Fierz Identities}

\begin{definrosa}
    A non-degenerate bilinear mapping $B$ defined on the bundle of real spinors $(\mathcal{S},\upgamma)$ is said to be \textbf{admissible} if, for every $\upxi, \upxi' \in \Gamma(\M,\mathcal{S})$ and $\upalpha \in \Omega^{k}(\M)$, the following conditions hold \cite{bab1}:
 \begin{itemize}
     \item [\textcolor{Fuchsia}{\textit{\textbf{1.}}}]
         $B(\upxi, \upxi') = \upsigma(B) B(\upxi', \upxi)$
  such that $\upsigma(B) = \pm 1$ is the symmetry of $B$. The positive [negative] sign mimics a self-adjoint [anti-self-adjoint] bilinear mapping.
      \item [\textcolor{Fuchsia}{\textit{\textbf{2.}}}] $B(\upgamma(\upalpha)\upxi, \upxi') = B(\upxi, (-1)^{\frac{k(k-\upepsilon(B))}{2}}\upgamma({\upalpha}) \upxi')$, such that $\upepsilon(B) = \pm 1$ is the type of $B$. 
      \item [\textcolor{Fuchsia}{\textit{\textbf{3.}}}] Whenever $p - q \equiv_8 0, 4, 6, 7$, the splitting spaces $\mathcal{S}^{\pm}$ of $\mathcal{S}$ with respect to $P_{\pm}$ are either: a) isotropic, where $B(\Gamma\mathcal{(M,S^{\pm})},\Gamma\mathcal{(M,S^{\pm})}) = 0$, or b) orthogonal, for which $B\left(\Gamma\mathcal{(M,S^{\pm})},\Gamma\mathcal{(M,S^{\mp})}\right) = 0$.
 \end{itemize} 
\end{definrosa}

\noindent The above properties of $B$ depend on the dimension $n$ and the metric signature $(p,q)$ of the manifold. These relations can be found classified in Ref. \cite{bab1}. For $\upalpha \in \Omega^{k}({\M})$, the expression

\begin{equation}
    (-1)^{\frac{k(k-\upepsilon(B))}{2}}\upgamma({\upalpha}) 
\end{equation}

\noindent in the item \textcolor{Fuchsia}{\textit{\textbf{2.}}} can be realised as an anti-automorphism $\varsigma_{B}: (\Omega,\diamond) \to (\Omega,\diamond)$ of the Kähler-Atiyah algebra:

\begin{equation}
   (-1)^{\frac{k(k-\upepsilon(B))}{2}}\upgamma({\upalpha}) = \upgamma(\varsigma_{B}({\upalpha})).  
\end{equation}

\noindent With respect to the grade involution, reversion and conjugation operators respectively given by

\begin{equation}\label{involution2345rtyhjgf}
    \yhwidehat{\upalpha} = (-1)^{k} \upalpha, \hspace{0.5cm}\qquad \yhwidetilde{\upalpha} = (-1)^{\frac{k(k-1)}{2}} \upalpha, \hspace{0.5cm} \qquad \overline{\upalpha} = \yhwidehat{\yhwidetilde{\upalpha}} =  (-1)^{\frac{k(k+1)}{2}}\upalpha,
\end{equation}

\noindent the operator $\varsigma_{B}$ can be read as

\begin{equation}
    \varsigma_{B} =\color{Fuchsia}\begin{cases}\color{black}
        \yhwidetilde{\;\cdot \;}, \text{ if } \upepsilon(B) = +1,\\\color{black}
        \yhwidetilde{\;\cdot \;} \circ \yhwidehat{\;\cdot \;} \color{black}= \overline{\;\cdot \;}, \color{black}\text{ if } \upepsilon(B) = -1.      \color{black}
    \end{cases}\color{black}
\end{equation}

\noindent The transpose of $T \in \Gamma(\M, \text{End}(\mathcal{S})$ with respect to $B$ is defined through:

\begin{equation}
    B(T\upxi,\upxi') = B(\upxi,T^{\intercal}\upxi')
\end{equation}

\noindent for all $\upxi,\upxi' \in \Gamma(\M,\mathcal{S})$ satisfying $(T^{\intercal})^{\intercal} = T$ and $(1_{\mathcal{S}})^{\intercal} = 1_{\mathcal{S}}$, as consequence of the signed symmetry property of $B$ listed in the item \textcolor{Fuchsia}{\textit{\textbf{1.}}}. Therefore, the item \textcolor{Fuchsia}{\textit{\textbf{2.}}} leads to 

\begin{equation}
    \upgamma(\varsigma_{B}({\upalpha})) = \upgamma({\upalpha})^{\intercal} \implies ({\;\cdot \;})^{\intercal} \circ \upgamma = \upgamma \circ \varsigma_{B}.
\end{equation}

\noindent Thus, the operation of taking the $B$-transposition $T \to T^{\intercal}$  defines a $\Cinf(\M,\rr)$-linear anti-automorphism of the algebra $(\Gamma(\M, \text{End}(\mathcal{S})), \circ)$.

\paragraph{ } The non-degenerated bilinear form $B$ induces a bundle isomorphism $f: \mathcal{S} \to \mathcal{S}^{*}$, whose action on sections (denoted by the same letter) is given by

\begin{align}\begin{aligned}
 f \colon \Gamma(\M,\mathcal{S}) &\longrightarrow \Gamma(\M,\mathcal{S})^{*} \\
 \upxi &\longmapsto \begin{alignedat}[t]{2} 
& f(\upxi) & \colon \Gamma(\M,\mathcal{S}) & \longrightarrow \mathbb{R} \\
 & & \upxi' & \longmapsto B(\upxi',\upxi).
\end{alignedat}\end{aligned}
\end{align}
\noindent A natural bundle isomorphism is also given by 
\begin{align}\begin{aligned}
 h \colon \Gamma(\M,\mathcal{S}) \otimes \Gamma(\M,\mathcal{S})^{*}  &\longrightarrow \text{End}(\Gamma(\M,\mathcal{S})) \\
 \upxi \otimes T &\longmapsto \begin{alignedat}[t]{2} 
& h(\upxi  \otimes T) & \colon \Gamma(\M,\mathcal{S}) & \longrightarrow \Gamma(\M,\mathcal{S}) \\
 & & \upxi' & \longmapsto T(\upxi')\upxi.
\end{alignedat}\end{aligned}
\end{align}

   {\colorlet{shadecolor}{Violet!15}\begin{shaded}The combination of those isomorphisms can define the following mapping:
\begin{equation}
    E := h \circ (1_{\mathcal{S}} \otimes f) : \Gamma(\M,\mathcal{S}) \otimes \Gamma(\M,\mathcal{S}) \longrightarrow  \text{End}(\Gamma(\M,\mathcal{S})),
\end{equation}\end{shaded}}

\noindent which induces the endomorphism, $ E_{\upxi_{1} , \upxi_{2}} := E (\upxi_{1} \otimes \upxi_{2}): \Gamma(\M,S) \longrightarrow \Gamma(\M,S)$, having the  explicit form as follows
\begin{align}
    \begin{aligned}
        E_{\upxi_{1}, \upxi_{2}} (\uppsi) &= ((g \circ (1 \otimes f))(\upxi_{1} \otimes \upxi_{2}))(\uppsi)\\&= (g(\upxi_{1} \otimes f(\upxi_{2})))(\uppsi)\\
        &= (f(\upxi_{2})\upxi_{1})(\uppsi)\\&= B(\uppsi, \upxi_{2}) \upxi_{1}.
    \end{aligned}
\end{align}
\noindent Moreover, for $\upxi_{1},\upxi_{2},\upxi_{3},\upxi_{4},\uppsi \in \Gamma(\M,S)$ one has

\begin{align}
    \begin{aligned}
        (E_{\upxi_{1},\upxi_{2}} \circ E_{\upxi_{3},\upxi_{4}})(\uppsi) &= E_{\upxi_{1},\upxi_{2}}( E_{\upxi_{3},\upxi_{4}}(\uppsi)) 
        \\&= E_{\upxi_{1},\upxi_{2}} (B(\uppsi, \upxi_{4})\upxi_{3})
        \\
        &= B(\uppsi, \upxi_{4})E_{\upxi_{1},\upxi_{2}}(\upxi_{3})\nonumber\\&= B(\uppsi, \upxi_{4}) B(\upxi_{3}, \upxi_{2}) \upxi_{1} \\
        &=  B(\upxi_{3}, \upxi_{2}) B(\uppsi, \upxi_{4}) \upxi_{1}
        \\&= B(\upxi_{3}, \upxi_{2}) E_{\upxi_{1}, \upxi_{4}} (\uppsi).
    \end{aligned}
\end{align}

  {\colorlet{shadecolor}{Violet!15}\begin{shaded} This defines the generalised Fierz relation, as 
\begin{equation}\label{fie}
     E_{\upxi_{1},\upxi_{2}} \circ E_{\upxi_{3},\upxi_{4}} = B(\upxi_{3},\upxi_{2}) E_{\upxi_{1}, \upxi_{4}}.
\end{equation}\end{shaded}}

\noindent Note that $E$ depends on the choice
of the bilinear form $B$ since $f$ does. Eq. \eqref{fie} encodes the seed for constructing the non-trivial classes of spinor fields according to their bilinear covariants. It is possible to to transfer the fibrewise composition of operators from End$\mathcal{S}$ to an associative and bilinear fibrewise composition $\bullet$ defined on the bundle of bispinors $\mathcal{S} \tensor \mathcal{S}$, whose action on sections takes
the form for all $u,v \in \Gamma(\M,\mathcal{S} \tensor \mathcal{S})$ \cite{bab2}:

\begin{equation}
    u \bullet v = E^{-1}(E(v) \circ E(u)).
\end{equation}

\noindent By Eq. \eqref{fie}, for $\upxi_{1},\upxi_{2},\upxi_{3},\upxi_{4} \in \Gamma(\M,\mathcal{S})$, this operation satisfies:

\begin{align}
    \begin{aligned}
        (\upxi_{1} \tensor \upxi_{2}) \bullet (\upxi_{3} \tensor \upxi_{4}) &= E^{-1}(E(\upxi_{1} \tensor \upxi_{2}) \circ E(\upxi_{3} \tensor \upxi_{4}))\\
        &=B(\upxi_{3},\upxi_{2})(\upxi_{1} \tensor \upxi_{4}).
    \end{aligned}
\end{align}

\noindent The  bundle of bispinors $\mathcal{S} \tensor \mathcal{S}$ equipped with this composition $\bullet$ turns out to be a bundle of unital associative algebras which is isomorphic with the bundle of algebras $(\text{End}(\mathcal{S}), \circ)$ with the unit section $1_{\mathcal{S}}$ of $\text{End}(\mathcal{S})$ being mapped to $E^{-1}(1_{\mathcal{S}})$ of $\mathcal{S} \tensor \mathcal{S}$. Therefore, smooth sections of
the bispinor bundle $\Gamma(\M,\mathcal{S} \tensor \mathcal{S} \simeq \Gamma(\M,\mathcal{S}) \tensor \Gamma(\M, \mathcal{S})$ is said to be the bispinor algebra of $\mathcal{S}$.

\paragraph{ } The partial inverse $\upgamma^{-1}$ can be employed to define the mapping
\begin{equation}
    \check{E} = \upgamma^{-1} \circ E,
\end{equation}

\noindent Let us assume from now on that we are in the the case when $\kk = \cc$ and $\kk = \rr$ with $p - q \equiv_{8} = 0,1,2$. In this case, the mapping $\upgamma$ is fibrewise surjective \cite{bab2}. Therefore, the isomorphism 

\begin{equation}
    \upgamma^{-1} : \text{End}(\mathcal{S}) \to \bigwedge\,\!^{\upgamma} T^{*}\M
\end{equation}

\noindent can be transported to the bundle of bispinors to get an isomorphism of bundles of
algebras:

\begin{equation}
     \check{E} = \upgamma^{-1} \circ E : (\mathcal{S} \tensor \mathcal{S}, \bullet) \to (\bigwedge\,\!^{\upgamma} T^{*}\M, \diamond)
\end{equation}

\noindent which is called \textbf{Fierz isomorphism}. On sections,denoted by the same symbol. this induces a isomorphisms of algebras that identifies the bispinor algebra with the subalgebra $(\Omega^{\upgamma}(\M)$ of the Kähler-Atiyah algebra:

\begin{equation}
     \check{E} = \upgamma^{-1} \circ E : (\Gamma(\M, \mathcal{S} \tensor \mathcal{S}), \bullet) \to (\Omega^{\upgamma}(\M), \diamond).
\end{equation}

\noindent For $\upxi,\upxi' \in \Gamma(\M,\mathcal{S})$ one sets

\begin{equation}
    \check{E}_{\upxi,\upxi'} := \check{E}(\upxi \tensor \upxi') = \upgamma^{-1} \circ E (\upxi \tensor \upxi') = \upgamma^{-1} (E_{\upxi,\upxi'}) \in \Omega^{\upgamma}(\M).
\end{equation}

{\colorlet{shadecolor}{Violet!15}\begin{shaded}\noindent Thus, Eq. \eqref{fie} implies the analogous identity in the subalgebra $\Omega^{\upgamma}(\M)$:

\begin{equation}\label{eq_fvd54gr8des6gfr2345}
   \check{E}_{\upxi_1,\upxi_2} \diamond \check{E}_{\upxi_3,\upxi_4} = B(\upxi_3,\upxi_2) \check{E}_{\upxi_1, \upxi_4}.
\end{equation}
 \end{shaded}}

\noindent The above Eq.\eqref{eq_fvd54gr8des6gfr2345} succinctly encapsulates the Fierz identities involving four spinors. These identities indicate that $\upgamma$ (and consequently $\check{E}$) acts as an isomorphism between bundles of algebras, not just between vector bundles. The construction of $\check{E}$ is summarised in the following commutative diagram of the morphisms on sections:

{\colorlet{shadecolor}{Violet!15}\begin{shaded} \begin{center}
\begin{tikzcd}
\Gamma(\M, \mathcal{S} \otimes \mathcal{S}) \arrow[r, "1_{\mathcal{S}} \otimes f", "\sim"'] \arrow[d, "\check{E}"', "\sim"] \arrow[dr, "\sim"', "E"] & \Gamma(\M, \mathcal{S} \otimes \mathcal{S}^*) \arrow[d, "h", "\sim"'] \\
\Omega^{\upgamma}(\M) \arrow[r, "\upgamma", "\sim"'] & \Gamma(\M, \text{End}(\mathcal{S}))
\end{tikzcd}
\end{center}\end{shaded}}

\noindent which applies for the case when $\kk = \cc$ and $\kk = \rr$ with $p - q \equiv_{8} = 0,1,2$. In this case, the explicit local expansion of $E$ is  \cite{bab2}
\begin{equation}\label{both}
    E_{\upxi, \upxi'} =  \frac{1}{2^{\left [  \frac{n+1}{2} \right]}}\sum^{n}_{k=0} \frac{1}{k!} B(\upgamma_{m_k \ldots m_1}\upxi, \upxi')\upgamma^{m_1 \ldots m_k},
\end{equation}
\noindent  for all $\upxi, \upxi' \in \Gamma(U,S)$. Applying $\upgamma^{-1}$ to both sides  of Eq. \eqref{both}, the local expansion for the Fierz isomorphism $\check{E}$ can be  expressed as
\begin{equation}
    \check{E} = \frac{1}{2^{\left [  \frac{n+1}{2} \right]}} \sum^{n}_{k=0} \check{\mathbf{E}}^{(k)}_{\upxi, \upxi'},
\end{equation}
\noindent such that 

{\colorlet{shadecolor}{Violet!15}\begin{shaded}\begin{align}
    \begin{aligned}\label{eq563}
        \check{\mathbf{E}}^{(k)}_{\upxi, \upxi'} = \frac{1}{k!} B(\upgamma_{m_k \ldots m_1}\upxi, \upxi')e^{m_1 \ldots m_k} = \frac{1}{k!} \upepsilon(B)^{k} B(\upxi, \upgamma_{m_1 \ldots m_k}\upxi')e^{m_1 \ldots m_k},
    \end{aligned}
\end{align}\end{shaded}}

\noindent are the geometric Fierz identities.

\section{Algebraic Constraints as Differential Forms}

\begin{definrosa}
     Let $\mathcal{S}$ be a bundle of spinors over $\M$. A \textbf{constrained generalised Killing spinor} over $\M$ is a section $\upxi \in \Gamma(\M,\mathcal{S})$ which satisfies the constrained generalized Killing spinor equations

     \begin{equation}
         D\upxi = Q_{1}\upxi = \ldots = Q_{\chi}\upxi = 0,
     \end{equation}

\noindent where $D: \Gamma(\M, \mathcal{S}) \to \Gamma(T^{*}\M \tensor \mathcal{S})$ is a linear connection on $\mathcal{S}$ and $Q_{1},\ldots, Q_{\chi} \in \Gamma(\M, \text{End}(\mathcal{S}))$ a finite collection of bundle endomorphisms called algebraic constraints satisfied by $\upxi$.
\end{definrosa}

Generalised Killing spinors are crucial in supergravity and string theory \cite{vintedois,vintetres,vintequatro}. They appear in these theories through the concept of “supersymmetric configuration”, which involves spinors that are parallel under a connection on the bundle of spinors. This leads to the concept of supergravity Killing spinor equations--special cases of (systems of) constrained generalized Killing spinor equations that are tailored to the specific physical theory being studied. Pseudo-Riemannian manifolds that possess parameterising geometric structures allowing for non-trivial solutions to these equations are referred to as supersymmetric configurations. They are termed supersymmetric solutions if they additionally satisfy the equations of motion of the corresponding supergravity theory.

\paragraph{ } Let us consider the case of a single algebraic constraint $Q\upxi = 0$. The space of solutions of this equation forms a $\Cinf(\M,\kk)$-submodule of $\Gamma(\M,\mathcal{S})$ denoted by $\mathcal{K}(Q)$

\begin{tflemma}\label{propos_wer234gfe}
   Consider the bundle of spinors space equipped with an admissible bilinear form $(\mathcal{S},\upgamma, B)$. Let $\upxi \in \mathcal{K}(Q)$ be a constrained generalised Killing spinors over $\M$. It follows that

    \begin{equation}
        Q\upxi = 0 \iff B(\upxi', \upgamma_{a_{1}\ldots a_{k}} \circ Q\upxi) = 0.
    \end{equation}

\noindent for all $\upxi' \in \Gamma(\M, \mathcal{S})$.
\end{tflemma}
\noindent \textcolor{Fuchsia}{\textit{Proof.}} For $ \upxi \in \mathcal{K}(Q)$ the bilinear form yields $B(\upxi', \upgamma_{a_{1}\ldots a_{k}} \circ Q\upxi) = 0$ for all $\upxi' \in \Gamma(\M, \mathcal{S})$ since $Q\upxi = 0$. Conversely, assume that $B(\upxi', \upgamma_{a_{1}\ldots a_{k}} \circ Q\upxi) = 0$ for all $\upxi' \in \Gamma(\M, \mathcal{S})$. Since $B$ is non-degenerate it is possible to take $\uppsi \in  \Gamma(\M, \mathcal{S})$ such that $B(\uppsi, \upxi) \neq 0$. Therefore, $B(\uppsi, \upgamma_{a_{1}\ldots a_{k}} \circ Q\upxi) = 0$ implies that $Q\upxi =0.\;\; \textcolor{Fuchsia}{\Box}$

\begin{tfpropos}
    It holds for two $\upxi, \upxi' \in \mathcal{K}(Q)$  constrained generalised Killing spinors over $\M$:

    \begin{equation}
        Q\upxi = Q\upxi' = 0 \iff B(\upxi', Q^{\intercal}\circ \upgamma_{a_{1}\ldots a_{k}}\upxi) = B(\upxi', \upgamma_{a_{1}\ldots a_{k}} \circ Q\upxi) = 0.
    \end{equation}
\end{tfpropos}
\noindent \textcolor{Fuchsia}{\textit{Proof.}} The statement for $Q$ follows from Lemma \ref{propos_wer234gfe}. The statement for $Q^{\intercal}$ also follows from the same Lemma together with the fact that the $B$-transposition is an anti-automorphism of the algebra $(\Gamma(\M, \text{End}(\mathcal{S})), \circ)$, upon noticing the relation 

\begin{equation}
    B(\upxi', Q^{\intercal}\circ \upgamma_{a_{1}\ldots a_{k}}\upxi) = B(\upxi', (\upgamma_{a_{1}\ldots a_{k}} \circ Q)^{\intercal}\upxi) = B(\upgamma_{a_{1}\ldots a_{k}} \circ Q\upxi',\upxi).\;\;\; \textcolor{Fuchsia}{\Box}
\end{equation}

As a consequence of the above results, with respect to the Fierz isomorphism, constrained generalised Killing spinors $\upxi,\upxi' \in \mathcal{K}(Q)$ are equivalent to \cite{bab2}:

\begin{equation}
     \check{E}_{\upxi,\upxi'} \diamond \check{Q}  = \check{Q} \diamond \check{E}_{\upxi',\upxi} = 0
\end{equation}

\noindent where $ \check{Q} = \upgamma^{-1}(Q)$. This formalism allows to describe supersymmetric configurations as solutions to a differential system of equations for a polyform that belongs to a certain semi-algebraic body in the Kähler-Atiyah bundle of $(\M, g)$. In Appendix \ref{app9} we show where this discussion appears in our context which is related to a Riemannian $8$-manifold, $\M_8$ composing a warped flux compactification AdS$_3\times \M_8$, whose metric and fluxes preserve one supersymmetry in AdS$_3$ in which we identity new classes of spinors in the next section.  However, for further discussion in the topic presented in this section, Ref. \cite{Cortes:2019xmk} presents a framework for studying generalised Killing spinors, demonstrating that any generalised Killing spinor equation, potentially with constraints, can be equivalently expressed as a system of partial differential equations. These equations involve a polyform that adheres to algebraic relations within the Kähler–Atiyah bundle, which is constructed by quantising the exterior algebra bundle of the underlying manifold.

\section{New Classes of Spinor Fields}

\paragraph{ } From the symmetry properties of the admissible pairing $B$, specific constraints may arise on the geometric Fierz identities in the Eq. \eqref{eq563}, forcing some of them to vanish which in turn, enables the classification of spinor fields within this approach based on those quantities: {\colorlet{shadecolor}{Violet!15}\begin{shaded}\begin{itemize}
    \item [\textcolor{Fuchsia}{$\star$}] New spinors on $M_{7,0}$ manifolds: L. Bonora, K. Brito, R. da Rocha. \textit{``Spinor Fields Classification in Arbitrary Dimensions and New Classes of Spinor Fields on 7-Manifolds''} \cite{Bonora:2014dfa}.
\item [\textcolor{Fuchsia}{$\star$}] New spinors on $M_{4,1}$ manifolds: K. Brito, R. da Rocha. \textit{``New fermions in the bulk'' }\cite{deBrito:2016qzl}.
\item [\textcolor{Fuchsia}{$\star$}] New spinors on $M_{1,6}$ manifolds: L. Bonora, R. da Rocha. \textit{``New Spinor Fields on Lorentzian 7-Manifolds''}
\cite{Bonora:2015ppa}. 
\item [\textcolor{Fuchsia}{$\star$}] New spinors on $M_{1,2}$ manifolds: R. Lopes, R. da Rocha. \textit{``New spinor classes on the Graf-Clifford algebra''}\cite{Lopes:2018cvu}.
\item [\textcolor{Fuchsia}{$\star$}] New spinors on $M_{9,0}$ manifolds: R. Lopes, R. da Rocha. \textit{``New spinor classes on the Graf-Clifford algebra''}\cite{Lopes:2018cvu}.
\item [\textcolor{Fuchsia}{$\star$}] New spinors on $M_{8,0}$ manifolds: D. Fabri, R. da Rocha. \textit{``New constrained generalized Killing spinor field classes in warped flux compactifications''}\cite{Goncalves:2023pty}.
\end{itemize}\end{shaded}}

Therefore, Lounesto's classification of spinor fields on four-dimensional Lorentzian manifolds \cite{Lou01} can be effectively extended to other dimensions and metric signatures, which are crucial in the study of fermionic fields in flux compactifications. In Refs. \cite{Bonora:2014dfa,Yanes:2018krn}, Moufang loops on the 7-sphere, part of the AdS$_4\times S^7$ compactification, were examined, leading to the discovery of new spinor classes. These spinor fields were demonstrated to properly transform under the Moufang loop generators on $S^7$. Additionally, Ref. \cite{deBrito:2016qzl} explored and identified new spinor field classes in the AdS$_5\times S^5$ compactification, presenting novel fermionic solutions within the AdS/CFT framework. Further, Ref. \cite{Lopes:2018cvu} introduced new classes of spinor fields in the cone and cylinder formalisms, addressing $M$-theory compactifications with one supersymmetry.

The paper in Reference \cite{Goncalves:2023pty} written by the author of this dissertation and the supervisor Prof. Dr. Roldão da Rocha contains an example of the geometric Fierz identities in the context of the spin geometry in a Riemannian 8-manifold, $\M_8$ composing a warped flux compactification AdS$_3\times \M_8$, whose metric and fluxes preserve one supersymmetry in AdS$_3$, this context is explained in the Appendix \ref{app9}. The Fierz aggregate in $(\M_{8}, \mathtt{g})$ is expressed as the following inhomogeneous differential form
\begin{equation}
    \check{E} = \frac{1}{16} \sum_{k=0}^{8} \check{\mathbf{E}}^{(k)}. 
\end{equation}

\noindent  Considering the admissible pairing $B$ with $\upsigma(B)= \upepsilon(B) = +1$. Eq. \eqref{eq563} gives
\begin{equation}
    \check{\mathbf{E}}^{(k)}_{\upxi, \upxi'} = \frac{1}{k!}  B(\upxi, \upgamma_{m_1 \ldots m_k}\upxi)e^{m_1 \ldots m_k}
\end{equation}

\noindent for all $m_1, \ldots, m_k = 1,\ldots,8$ and  $k = 0,\ldots,8$. The cases $k=2,3,6,7$ yield  $\check{\mathbf{E}}^{(k)} = 0$. Indeed, 
\begin{align}
    \begin{aligned}
        B(\upxi, \upgamma_{m_1 \ldots m_k} \upxi) &=  \upsigma(B) B(\upgamma_{m_1 \ldots m_k} \upxi,  \upxi) =  \upsigma(B) (-1)^{\frac{k(k-\upepsilon(B))}{2}} B( \upxi, \upgamma_{m_1 \ldots m_k}\upxi).       
    \end{aligned}
\end{align}
Therefore, since $\upepsilon(B) = \upsigma(B) = +1$ for  $k = 2,3,6,7$  one has $B(\upxi, \upgamma_{m_1 \ldots m_k} \upxi) = 0$, and consequently $\check{\mathbf{E}}^{(k)} = 0$. On the other hand, for $k = 0,1,4,5,8$, the non-vanishing bilinear covariants are:

\begin{subequations}
\begin{align}
    \label{nonvanishingbilinears}
        &\check{\mathbf{E}}^{(0)} = B(\upxi, \upxi),\\
        &\check{\mathbf{E}}^{(1)} = B(\upxi, \upgamma_{m}\upxi)e^{m},\label{nonvanishingbilinears1}\\
        &\check{\mathbf{E}}^{(4)} = \frac{1}{4!} B(\upxi, \upgamma_{{m_1 \ldots m_4}}\upxi)e^{{m_1 \ldots m_4}},\\
        &\check{\mathbf{E}}^{(5)} = \frac{1}{5!} B(\upxi, \upgamma_{{m_1 \ldots m_5}}\upxi)e^{{m_1 \ldots m_5}},\\
        &\check{\mathbf{E}}^{(8)} = \frac{1}{8!} B(\upxi, \upgamma_{{m_1 \ldots m_8}}\upxi)e^{{m_1 \ldots m_8}}.\label{nonvanishingbilinears2}
\end{align}
\end{subequations}

\noindent The non-vanishing bilinear covariants determine a unique class of spinor fields \cite{bab2}. Specifically, the set where the associated bilinear covariants are matched with the homogeneous $k$-forms is such that $\check{\mathbf{E}}^{(k)} = 0$ for $k=2,3,6,7$, and $\check{\mathbf{E}}^{(k)} \neq 0$ for $k=0,1,4,5,8$. Following the procedures outlined in Refs. \cite{Bonora:2015ppa, bab1}, one can examine a complexification procedure of bilinear covariants within $\cl_{8,0}$. Consider the complex structure $J$ on the real bundle of spinors, where $J^2 = -I$ \cite{bab1, Alekseevsky:2003vw}. Given the splitting $\mathcal{S} = \mathcal{S}^{+} \oplus \mathcal{S}^{-}$, with $\mathcal{S}^{\pm} = P_{\pm}(\mathcal{S})$, we have $\mathcal{S} = \mathcal{S}^{+} \oplus J(\mathcal{S}^{+})$, since $J(\mathcal{S}^{\pm}) = \mathcal{S}^{\mp}$. Thus, for the spinor field $\upxi$ expressed as $\upxi_{R} + J(\upxi_{I})$, the bilinear form yields

\begin{align}\label{bc}
        B(\upxi, \upgamma_{{m_1 \ldots m_k}} \upxi) &= B\left(\upxi_{R} + J(\upxi_{I}), \upgamma_{{m_1 \ldots m_k}}\upxi_{R} + J(\upxi_{I})\right)\nonumber \\\nonumber
&= B(\upxi_{R}, \upgamma_{{m_1 \ldots m_k}} \upxi_{R}) 
+ B(\upxi_{R}, \upgamma_{{m_1 \ldots m_k}} J(\upxi_{I}))\\\nonumber &\qquad+  B(J(\upxi_{I}), \upgamma_{{m_1 \ldots m_k}} \upxi_{R})
+ B(J(\upxi_{I}), \upgamma_{{m_1 \ldots m_k}} J(\upxi_{I}))\\\nonumber
&=  B(\upxi_{R}, \upgamma_{{m_1 \ldots m_k}} \upxi_{R}) 
+ B(\upxi_{R},(J \circ \upgamma_{{m_1 \ldots m_k}} )\upxi_{I})\\\nonumber&\qquad+(-1)^{\frac{n(n+1)}{2}} B(\upxi_{I}, (J\circ\upgamma_{{m_1 \ldots m_k}}) \upxi_{R})
- (-1)^{\frac{n(n+1)}{2}} B(\upxi_{I}, \upgamma_{{m_1 \ldots m_k}} \upxi_{I})\\\nonumber
&=  B(\upxi_{R}, \upgamma_{{m_1 \ldots m_k}} \upxi_{R}) 
-  B(\upxi_{I}, \upgamma_{{m_1 \ldots m_k}} \upxi_{I}) \\ &\qquad +B(\upxi_{R},(J \circ \upgamma_{{m_1 \ldots m_k}} )\upxi_{I})  + B(\upxi_{I}, (J\circ\upgamma_{{m_1 \ldots m_k}}) \upxi_{R}),
\end{align}
\noindent where $J^{\intercal} = (-1)^{\frac{n(n+1)}{2}} J$ \cite{bab1}. As a result, the complexified bilinear form $\mathcal{B}$ can be expressed as

\begin{align}    
\nonumber
       \mathcal{B}(\upxi, \upgamma_{{m_1 \ldots m_k}}\upxi)& =  B(\upxi_{R}, \upgamma_{{m_1 \ldots m_k}} \upxi_{R}) 
-  B(\upxi_{I}, \upgamma_{{m_1 \ldots m_k}} \upxi_{I}) \\ &\qquad +\, i \left( B(\upxi_{R}, \upgamma_{{m_1 \ldots m_k}} \upxi_{I})\right.\left. + B(\upxi_{I}, \upgamma_{{m_1 \ldots m_k}} \upxi_{R}). \right)
\end{align}

\noindent The bilinear covariants can now be generalized from the standard Majorana spinor field $\upxi \in \Gamma(\M,\mathcal{S})$ to sections of $\Gamma(\M,\mathcal{S}_{\mathbb{C}})$, by redefining the bilinear covariants as follows:

\begin{equation}
    \check{\mathcal{E}}^{(k)} = \frac{1}{k!}\mathcal{B}(\upxi, \upgamma_{{m_1 \ldots m_k}}\upxi)e^{m_1 \ldots m_k}. 
\end{equation}
\noindent Thus, since the terms in both the real and imaginary parts of the complex bilinear form can cancel each other out, the generalized bilinear covariants \eqref{nonvanishingbilinears} -- \eqref{nonvanishingbilinears2} can result in either non-zero or zero values. Consequently, 32 new classes of spinor fields can be identified based on the values of their generalized bilinear covariants. We present all the possible cases below.

\begin{itemize}
    \item[1.] Five classes of spinor fields with one non-null bilinear covariant.
\begin{subequations}
\begin{align}
    &&&&\check{\mathcal{E}}^{(0)} \neq 0, &&&&\check{\mathcal{E}}^{(1)} = 0, &&&&\check{\mathcal{E}}^{(4)} = 0, &&&&\check{\mathcal{E}}^{(5)} = 0, &&&&\check{\mathcal{E}}^{(8)} = 0,\label{46a}\\
    &&&&\check{\mathcal{E}}^{(0)} = 0, &&&&\check{\mathcal{E}}^{(1)} \neq 0, &&&&\check{\mathcal{E}}^{(4)} = 0, &&&&\check{\mathcal{E}}^{(5)} = 0, &&&&\check{\mathcal{E}}^{(8)} = 0,\label{46b} \\
    &&&&\check{\mathcal{E}}^{(0)} = 0, &&&&\check{\mathcal{E}}^{(1)} = 0, &&&&\check{\mathcal{E}}^{(4)} \neq 0, &&&&\check{\mathcal{E}}^{(5)} = 0, &&&&\check{\mathcal{E}}^{(8)} = 0, \\
    &&&&\check{\mathcal{E}}^{(0)} = 0, &&&&\check{\mathcal{E}}^{(1)} = 0, &&&&\check{\mathcal{E}}^{(4)} = 0, &&&&\check{\mathcal{E}}^{(5)} \neq 0, &&&&\check{\mathcal{E}}^{(8)} = 0, \\
    &&&&\check{\mathcal{E}}^{(0)} = 0, &&&&\check{\mathcal{E}}^{(1)} = 0, &&&&\check{\mathcal{E}}^{(4)} = 0, &&&&\check{\mathcal{E}}^{(5)} = 0, &&&&\check{\mathcal{E}}^{(8)} \neq 0.
\end{align}
\end{subequations}
\end{itemize}

\begin{itemize}
    \item[2.] Ten classes of spinor fields, each one containing two non-null bilinear covariants:
\begin{subequations}
\begin{align}
    &&&&\check{\mathcal{E}}^{(0)} \neq 0, &&&&\check{\mathcal{E}}^{(1)} \neq 0, &&&&\check{\mathcal{E}}^{(4)} = 0, &&&&\check{\mathcal{E}}^{(5)} = 0, &&&&\check{\mathcal{E}}^{(8)} = 0, \\
    &&&&\check{\mathcal{E}}^{(0)} \neq 0, &&&&\check{\mathcal{E}}^{(1)} = 0, &&&&\check{\mathcal{E}}^{(4)} \neq 0, &&&&\check{\mathcal{E}}^{(5)} = 0, &&&&\check{\mathcal{E}}^{(8)} = 0, \\
    &&&&\check{\mathcal{E}}^{(0)} \neq 0, &&&&\check{\mathcal{E}}^{(1)} = 0, &&&&\check{\mathcal{E}}^{(4)} = 0, &&&&\check{\mathcal{E}}^{(5)} \neq 0, &&&&\check{\mathcal{E}}^{(8)} = 0, \\
    &&&&\check{\mathcal{E}}^{(0)} \neq 0, &&&&\check{\mathcal{E}}^{(1)} = 0, &&&&\check{\mathcal{E}}^{(4)} = 0, &&&&\check{\mathcal{E}}^{(5)} = 0, &&&&\check{\mathcal{E}}^{(8)} \neq 0, \\
    &&&&\check{\mathcal{E}}^{(0)} = 0, &&&&\check{\mathcal{E}}^{(1)} \neq 0, &&&&\check{\mathcal{E}}^{(4)} \neq 0, &&&&\check{\mathcal{E}}^{(5)} = 0, &&&&\check{\mathcal{E}}^{(8)} = 0, \\
    &&&&\check{\mathcal{E}}^{(0)} = 0, &&&&\check{\mathcal{E}}^{(1)} \neq 0, &&&&\check{\mathcal{E}}^{(4)} = 0, &&&&\check{\mathcal{E}}^{(5)} \neq 0, &&&&\check{\mathcal{E}}^{(8)} = 0, \\
    &&&&\check{\mathcal{E}}^{(0)} = 0, &&&&\check{\mathcal{E}}^{(1)} \neq 0, &&&&\check{\mathcal{E}}^{(4)} = 0, &&&&\check{\mathcal{E}}^{(5)} = 0, &&&&\check{\mathcal{E}}^{(8)} \neq 0, \\
    &&&&\check{\mathcal{E}}^{(0)} = 0, &&&&\check{\mathcal{E}}^{(1)} = 0, &&&&\check{\mathcal{E}}^{(4)} \neq 0, &&&&\check{\mathcal{E}}^{(5)} \neq 0, &&&&\check{\mathcal{E}}^{(8)} = 0, \\
    &&&&\check{\mathcal{E}}^{(0)} = 0, &&&&\check{\mathcal{E}}^{(1)} = 0, &&&&\check{\mathcal{E}}^{(4)} = 0, &&&&\check{\mathcal{E}}^{(5)} \neq 0, &&&&\check{\mathcal{E}}^{(8)} \neq 0,\\
    &&&&\check{\mathcal{E}}^{(0)} = 0, &&&&\check{\mathcal{E}}^{(1)} = 0, &&&&\check{\mathcal{E}}^{(4)} \neq 0, &&&&\check{\mathcal{E}}^{(5)} = 0, &&&&\check{\mathcal{E}}^{(8)} \neq 0.
\end{align}
\end{subequations}
\end{itemize}
\begin{itemize}
    \item[3.] Ten classes of spinor fields with three non-vanishing bilinear covariants:
 \begin{subequations}
\begin{align}
    &&&&\check{\mathcal{E}}^{(0)} \neq 0, &&&&\check{\mathcal{E}}^{(1)} \neq 0, &&&&\check{\mathcal{E}}^{(4)} \neq 0, &&&&\check{\mathcal{E}}^{(5)} = 0, &&&&\check{\mathcal{E}}^{(8)} = 0, \\
    &&&&\check{\mathcal{E}}^{(0)} \neq 0, &&&&\check{\mathcal{E}}^{(1)} \neq 0, &&&&\check{\mathcal{E}}^{(4)} = 0, &&&&\check{\mathcal{E}}^{(5)} \neq 0, &&&&\check{\mathcal{E}}^{(8)} = 0, \\
    &&&&\check{\mathcal{E}}^{(0)} \neq 0, &&&&\check{\mathcal{E}}^{(1)} \neq 0, &&&&\check{\mathcal{E}}^{(4)} = 0, &&&&\check{\mathcal{E}}^{(5)} = 0, &&&&\check{\mathcal{E}}^{(8)} \neq 0, \\
    &&&&\check{\mathcal{E}}^{(0)} \neq 0, &&&&\check{\mathcal{E}}^{(1)} = 0, &&&&\check{\mathcal{E}}^{(4)} \neq 0, &&&&\check{\mathcal{E}}^{(5)} \neq 0, &&&&\check{\mathcal{E}}^{(8)} = 0, \\
    &&&&\check{\mathcal{E}}^{(0)} \neq 0, &&&&\check{\mathcal{E}}^{(1)} = 0, &&&&\check{\mathcal{E}}^{(4)} \neq 0, &&&&\check{\mathcal{E}}^{(5)} = 0, &&&&\check{\mathcal{E}}^{(8)} \neq 0, \\
    &&&&\check{\mathcal{E}}^{(0)} \neq 0, &&&&\check{\mathcal{E}}^{(1)} = 0, &&&&\check{\mathcal{E}}^{(4)} = 0, &&&&\check{\mathcal{E}}^{(5)} \neq 0, &&&&\check{\mathcal{E}}^{(8)} \neq 0, \\
    &&&&\check{\mathcal{E}}^{(0)} = 0, &&&&\check{\mathcal{E}}^{(1)} \neq 0, &&&&\check{\mathcal{E}}^{(4)} \neq 0, &&&&\check{\mathcal{E}}^{(5)} \neq 0, &&&&\check{\mathcal{E}}^{(8)} = 0, \\
    &&&&\check{\mathcal{E}}^{(0)} = 0, &&&&\check{\mathcal{E}}^{(1)} \neq 0, &&&&\check{\mathcal{E}}^{(4)} \neq 0, &&&&\check{\mathcal{E}}^{(5)} = 0, &&&&\check{\mathcal{E}}^{(8)} \neq 0, \\
    &&&&\check{\mathcal{E}}^{(0)} = 0, &&&&\check{\mathcal{E}}^{(1)} \neq 0, &&&&\check{\mathcal{E}}^{(4)} = 0, &&&&\check{\mathcal{E}}^{(5)} \neq 0, &&&&\check{\mathcal{E}}^{(8)} \neq 0, \\
    &&&&\check{\mathcal{E}}^{(0)} = 0, &&&&\check{\mathcal{E}}^{(1)} = 0, &&&&\check{\mathcal{E}}^{(4)} \neq 0, &&&&\check{\mathcal{E}}^{(5)} \neq 0, &&&&\check{\mathcal{E}}^{(8)} \neq 0.
\end{align}
\end{subequations}
\end{itemize}
\begin{itemize}
    \item[4.] Five classes of spinor fields with four non-null bilinear covariants:
\begin{subequations}
\begin{align}
    &&&&\check{\mathcal{E}}^{(0)} \neq 0, &&&&\check{\mathcal{E}}^{(1)} \neq 0, &&&&\check{\mathcal{E}}^{(4)} \neq 0, &&&&\check{\mathcal{E}}^{(5)} \neq 0, &&&&\check{\mathcal{E}}^{(8)} = 0, \\
    &&&&\check{\mathcal{E}}^{(0)} \neq 0, &&&&\check{\mathcal{E}}^{(1)} \neq 0, &&&&\check{\mathcal{E}}^{(4)} \neq 0, &&&&\check{\mathcal{E}}^{(5)} = 0, &&&&\check{\mathcal{E}}^{(8)} \neq 0, \\
    &&&&\check{\mathcal{E}}^{(0)} \neq 0, &&&&\check{\mathcal{E}}^{(1)} \neq 0, &&&&\check{\mathcal{E}}^{(4)} = 0, &&&&\check{\mathcal{E}}^{(5)} \neq 0, &&&&\check{\mathcal{E}}^{(8)} \neq 0, \\
    &&&&\check{\mathcal{E}}^{(0)} \neq 0, &&&&\check{\mathcal{E}}^{(1)} = 0, &&&&\check{\mathcal{E}}^{(4)} \neq 0, &&&&\check{\mathcal{E}}^{(5)} \neq 0, &&&&\check{\mathcal{E}}^{(8)} \neq 0, \\
    &&&&\check{\mathcal{E}}^{(0)} = 0, &&&&\check{\mathcal{E}}^{(1)} \neq 0, &&&&\check{\mathcal{E}}^{(4)} \neq 0, &&&&\check{\mathcal{E}}^{(5)} \neq 0, &&&&\check{\mathcal{E}}^{(8)} \neq 0.
\end{align}
\end{subequations}
\end{itemize}

\begin{itemize}
    \item[5.] A single class of spinor fields with all five non-vanishing bilinear covariants. 
    \begin{align}
&&&&\check{\mathcal{E}}^{(0)} \neq 0, &&&&\check{\mathcal{E}}^{(1)} \neq 0, &&&&\check{\mathcal{E}}^{(4)} \neq 0, &&&&\check{\mathcal{E}}^{(5)} \neq 0, &&&&\check{\mathcal{E}}^{(8)} \neq 0.\label{50}
\end{align}
\end{itemize}

\begin{itemize}
    \item[6.] A single class consisting of null bilinear covariants (trivial class):
        \begin{align}
&&&&\check{\mathcal{E}}^{(0)} = 0, &&&&\check{\mathcal{E}}^{(1)} = 0, &&&&\check{\mathcal{E}}^{(4)} = 0, &&&&\check{\mathcal{E}}^{(5)} = 0, &&&&\check{\mathcal{E}}^{(8)} = 0. \label{trivialclass}
\end{align}
\end{itemize}

For the AdS$_3\times \M_8$ compactification considered in this study, Eq. (\ref{50}) was previously reported in Ref. \cite{bab2}. The bilinear form \eqref{bc} allows for the implementation of $32$ additional spinor field classes, expanding beyond the single class identified in Ref. \cite{bab2}. Since the context of Lounesto's classification, identifying new classes of spinors is highly significant, new classes of spinors can be considered for explaining phenomena, solving physical problems, advancing theoretical physics, and enhancing our understanding of the universe. Novel spinor classes can contribute to the study of black holes, quantum field theory, and the development of unified theories in physics as we discussed on Chapter \ref{chap_bil}. Given this, it is natural to consider that the new spinor classes identified here can be relevant within the realms of supersymmetry, supergravity, M-theory and string theory \cite{StrinGeo, supersymmetry}. These fields often require intricate spinor structures to explore possible solutions and symmetries \cite{Lukierski:2002ux, Carrion:2003ve, Kuznetsova:2007rb, Kuznetsova:2005cd, Kuznetsova:2005mv, Gonzales:2009ye, Berkovits:2018gbq, Berkovits:2019szu, Meert:2018qzk}. Finding 32 new classes of spinor fields in the AdS$_3\times \M_8$ compactification extends the previous results and can accommodate new supersymmetric fermionic solutions. The aim of this dissertation was to develop and study the entire geometric and algebraic framework in which spinors are defined and classified. By finding these new classes, we concretely demonstrated how this procedure functions in practical terms. In doing so, we have provided a comprehensive framework that not only aligns with the established mathematical structures but also open doors to future research and discoveries in theoretical physics. 
\addcontentsline{toc}{chapter}
{\;\;\;\;\;\textsc{conclusion}}
\chapter*{\textsc{conclusion}}


\paragraph{ } By bridging geometric and algebraic concepts, this work provided a comprehensive study of Clifford structures on bundles, spinor fields and spinor classification. The exploration of the Kähler-Atiyah bundle, which encapsulates the essence of Clifford algebras, has offered profound insights into the algebraic structures underlying geometric frameworks. The aim of this work was to journey from the fundamental concepts of Clifford algebras to the classification of spinors, taking two distinct approaches. The first, referred to as the "algebraic approach," involved studying Clifford algebras and achieving the classification of spinors in Minkowski space-time. Subsequently, with the goal of defining a concept that allows for the generalisation of spinor classification, we transitioned to the "geometric approach." This approach involved examining Clifford structures in fibre bundles, leading to the desired generalisation and culminating in a concrete example of spinor classification with new results in the context of a compactified eleven-dimensional manifold. In this conclusion, we will recap and summarise what was accomplished in each of these approaches.

\paragraph{ } We began by presenting the foundational aspects of Clifford algebras, demonstrating their definition, classification, and representation in any quadratic space. This set the stage for understanding their periodicity and natural relation to matrix algebra, which are crucial for appreciating their broader applications, particularly in the context of spinors. The second Chapter, that was focused on orthogonal transformations and the groups defined within Clifford algebras, leading to the introduction of the Pin and Spin groups. These groups were essential for defining spinors, which carry representations of the Spin group, a double cover of the special orthogonal group. In the third chapter, we presented the definition and classification of algebraic and classical spinors, as well as a discussion on spin representation and also the presentation of classical spinors from the literature: the Weyl, Pauli and Dirac spinors. Based on the fundamental bases of the second chapter, classical spinors were defined as elements of the irreducible representation space of the spin group and as the irreducible representation space forms a minimal algebraic left ideal of Clifford algebras, algebraic spinors were defined within Clifford algebras. The fourth chapter were focused on bilinear covariants, quantities constructed from spinor fields and their adjoints that transform under symmetry operations and are constrained by the Fierz-Pauli-Kofink identities. These covariants play crucial roles in describing physical phenomena in Minkowski space-time. We emphasised the significant role of bilinear covariants, which, understood within the framework of Clifford algebra, are indispensable for various applications in quantum mechanics, particle physics, and gravitation. We concluded this chapter, and with it the first part of the dissertation, by presenting Lounesto's classification of spinors and engaging in important discussions about this classification. This served as a motivation for the second part of the dissertation, where one of the objectives was to achieve a generalisation along with an original result. Nonetheless, this first part of the work held significant importance on its own. It not only elucidated the entire pathway from Clifford algebras to the definition and classification of spinors but also provided a solid foundation that fully supported the second part of the dissertation and was indispensable for all subsequent definitions and discussions.

\paragraph{ } Having concluded the first part, we advanced to the part where geometry comes to the forefront. The Kähler-Atiyah bundle forms the foundation for this second part that starts in the fifth chapter. Thus, we introduced the "geometric approach" with the necessary preliminaries by defining essential concepts such as tangent and cotangent bundles, vector bundles, and principal bundles. This foundational exploration was crucial for understanding spin structures and the geometric context in which spinors are defined. Based on the fundamental concepts of Clifford algebras set out in the first chapter and the geometric framework established in the fifth chapter, in the sixth chapter, we delved into the Clifford algebra on the exterior bundle from the formalism of the Kähler-Atiyah bundle. This bundle's non-commutative multiplication induces a non-commutative associative multiplication on inhomogeneous differential forms, enriching the study of geometric algebra and spin geometry. We also explored the volume form and key isomorphic subalgebras, emphasising the Kähler-Atiyah bundle's significance in bridging geometric and algebraic concepts. This was followed by the seventh chapter, where spinors were explored within the context of differential geometry. Here, we defined spin structures and their associated bundle of spinors, highlighting the Kähler-Atiyah bundle's role in providing a comprehensive understanding of spin geometry and spinors. The investigation of spin structures and the associated bundles of spinors within the framework of geometric algebra provided a profound understanding of the interplay between geometry and algebra on manifolds. By using the Kähler-Atiyah bundle, the local and global perspectives were unified, encapsulating both the algebraic properties and the geometric structure of spinors. The last chapter, we demonstrated that bilinear covariants, understood within the Kähler-Atiyah framework, can be extended to any dimension and metric signature, guided by generalised geometric Fierz identities, along with the spinor classification. We highlighted the role of the Kähler–Atiyah bundle in analysing generalized bilinear covariants and discussed how algebraic constraints can cause some spinor bilinears to vanish. Additionally, we presented an original result by identifying new classes of spinor fields within an eight-dimensional manifold derived from an eleven-dimensional warped flux compactification preserving one supersymmetry, AdS$_3 \times \M_8$, thereby extending Lounesto's spinor classification to different dimensions and metric signatures.

\paragraph{ } In conclusion, this dissertation has successfully bridged algebraic and geometric perspectives on Clifford bundles of algebras, spinors, and spinor classification. By integrating the algebraic foundations of Clifford algebras with the geometric framework of fibre bundles, particularly through the Kähler-Atiyah bundle, we have extended the spinor classification to new dimensions and signatures.


\backmatter
\appendix
\renewcommand{\thesection}{\Alph{section}}
\setcounter{section}{0}
\setcounter{equation}{0}
\renewcommand{\theequation}{\thesection.\arabic{equation}}
\setcounter{theorem}{0}
\renewcommand{\thetheorem}{\thesection.\arabic{theorem}}

\setcounter{lemma}{0}
\renewcommand{\thelemma}{\thesection.\arabic{lemma}}

\setcounter{defin}{0}
\renewcommand{\thedefin}{\thesection.\arabic{defin}}

\setcounter{propos}{0}
\renewcommand{\thepropos}{\thesection.\arabic{propos}}

\setcounter{ex}{0}
\renewcommand{\theex}{\thesection.\arabic{ex}}

\setcounter{cor}{0}
\renewcommand{\thecor}{\thesection.\arabic{cor}}

\setcounter{rem}{0}
\renewcommand{\therem}{\thesection.\arabic{rem}}
\addcontentsline{toc}{chapter}
{\;\;\;\;\;\textsc{appendix}}
\chapter*{\textsc{appendix}}
\hypersetup{
  colorlinks = true,
  linkcolor  = gray,
  citecolor = gray,
}
\chaptermark{}
\section{Multilinear Algebra}\label{app1}
\hspace{0.5cm} The basis of Clifford algebra is the multilinear algebra. This Appendix \ref{app1} contains a summary of certain topics in multilinear algebra that are required for the sequel. We briefly introduce some concepts on multilinear algebra, set up notation and terminology about multilinear mappings and quadratics spaces. Our aim here is to establish the fundamental ingredients for a Clifford algebra to be defined: quadratic spaces and the musical isomorphisms (for the Clifford product).


\begin{figure}[htp]
\centering
\begin{tikzpicture}
\coordinate (A) at (-3.5,0) {};
\coordinate (B) at ( 3.5,0) {};
\coordinate (C) at (0,5.3) {};
\path[name path=AC,draw=none] (A) -- (C);
\path[name path=BC,draw=none] (B) -- (C);
\filldraw[draw=black, ultra thick,fill=gray!10] (A) -- (B) -- (C) -- cycle;
\foreach \y/\A in {
    0/{Multilinear algebra \small\upshape\mdseries},
    1/{Tensor algebra \small\upshape\mdseries},
    2/{Exterior algebra},
    3/{\parbox{3cm}{\centering Clifford\\ algebra}}} {
    \path[draw=none, very thick, dashed, name path=horiz] (A|-0,\y) -- (B|-0,\y);
    \draw[draw=black, very thick, dashed, 
          name intersections={of=AC and horiz,by=P},
          name intersections={of=BC and horiz,by=Q}] (P) -- (Q)
          node[midway,above,font=\bfseries\scshape,color=gray] {\A};
}
\end{tikzpicture}
\end{figure}

\begin{defincinza}
Let $V_1, V_2, \ldots, V_p$ and $W$ be a finite family of vector spaces over the same field $\mathbb{K}$. The mapping

\begin{equation}
    \varphi : V_1 \cross V_2 \cross \cdots \cross V_p \to W
\end{equation}

\noindent is called a \textbf{multilinear mapping} (in this case p-linear) if it is linear in each argument, when the others are fixed, which means that given arbitrary $\lambda \in \mathbb{K}$, $\mathbf{v}_1,\mathbf{v}'_1 \in V _1$, $\mathbf{v}_2, \mathbf{v}'_2 \in V_2,\ldots,\mathbf{v}_n, \mathbf{v}'_n \in V_n$, then
\end{defincinza}
\begin{align}
    \begin{aligned}
  &\textcolor{gray}{\textbf{(i)}} \;\; \varphi (\mathbf{v}_1,\ldots,\mathbf{v}_i,\ldots,\mathbf{v}_p) + \varphi(\mathbf{v}_1,\ldots,\mathbf{v}'_i,\ldots,\mathbf{v}_p) = \varphi (\mathbf{v}_1,\ldots,\mathbf{v}_i + \mathbf{v}'_i,\ldots,\mathbf{v}_p), \\
  &\textcolor{gray}{\textbf{(ii)}} \;\; \lambda \varphi(\mathbf{v}_1,\mathbf{v}_2,\ldots,\mathbf{v}_i,\ldots,\mathbf{v}_p) = \varphi(\mathbf{v}_1,\mathbf{v}_2,\ldots,\lambda \mathbf{v}_i,\ldots,\mathbf{v}_p)
    \end{aligned}
\end{align}

\noindent \textit{for $i = 1,..,p$. When $p=1$ the mapping is said to be linear and when $p=2$, bilinear.}\\

We denote the vector space of such multilinear mappings by $\mathcal{L}( V_1, V_2, \ldots, V_p;W)$. In our work, it is necessary to consider a symmetric bilinear mapping and its quadratic form. A bilinear mapping $g : V \cross V \to \mathbb{K} $ is said to be \textit{symmetric} if for all $\mathbf{v}_1, \mathbf{v}_2 \in V$ the property $g(\mathbf{v}_1,\mathbf{v}_2) = g(\mathbf{v}_2,\mathbf{v}_1)$ holds. A  bilinear form is said to be \textit{non-degenerate} if

\begin{align}
    \begin{aligned}
    g(\mathbf{v}_1,\mathbf{v}_2) &= 0 \text{ for all } \mathbf{v}_2 \in V \text{ implies }  \mathbf{v}_1 = 0 \text{ and }\\
    g(\mathbf{v}_1,\mathbf{v}_2) &= 0 \text{ for all } \mathbf{v}_1 \in V \text{ implies } \mathbf{v}_2 = 0.
    \end{aligned}
\end{align}

\begin{defincinza}\label{def_5t4r8uemc9dwzasex}
Let $V$ a vector space over the field $\mathbb{K}$. A  \textbf{quadratic form} $Q$ on the space $V$ is a mapping $Q : V \to \mathbb{K}$ for which there exists a bilinear form $g: V \cross V \to \mathbb{K}$ such that for all $\mathbf{v} \in V$

\begin{equation}
    Q(\mathbf{v}) = g(\mathbf{v},\mathbf{v}).
\end{equation}
\end{defincinza}

If the characteristic of the field $\mathbb{K}$ is different of $2$, then for any quadratic form $Q$ there exists a unique symmetric bilinear form $g$ with the property $Q(\mathbf{v}) = g(\mathbf{v},\mathbf{v})$, called the \textit{polarisation} of $Q$ \cite{Kos89}. In terms of its quadratic form $Q: V \to \mathbb{K}$ the bilinear form $g$ can be expressed as

\begin{equation}
    g(\mathbf{v},\mathbf{w}) = \frac{1}{2}(Q(\mathbf{v} + \mathbf{w}) - Q(\mathbf{v}) - Q(\mathbf{w})).
\end{equation}

\noindent for every $\mathbf{v},\mathbf{w} \in V$. A vector space $V$ equipped with a symmetric bilinear mapping $g: V \cross V \to \mathbb{K}$ is said to be a \textbf{quadratic space} which is the main structure on which we define a Clifford algebra.

\begin{ex}\label{example1inner}
{{\textcolor{gray}{{$\blacktriangleright$\;}}}} Inner product. \normalfont In a vector space $V$ an inner product $\langle \;, \;\rangle : V \cross V \to \mathbb{K}$ is a positive definite symmetric bilinear form, that is, for every $\mathbf{v} \in V$, $\langle \mathbf{v} , \mathbf{v} \rangle \geq 0$, whereas the equality holds if and only if $\mathbf{v} = 0$. The standard inner product in $\mathbb{R}^n$ for $\mathbf{v} = (v_1,\ldots,v_n)$ and $\mathbf{w} = (w_1,\ldots,w_n) \in \rr^n$ is given by 
\begin{equation}
     \langle \mathbf{v} , \mathbf{w} \rangle = \sum_{i\,=\,1}^{n} v_i w_i. \;\;\; \textcolor{gray}{{\blacktriangleleft}}
\end{equation}\end{ex}

\begin{ex}
{{\textcolor{gray}{{$\blacktriangleright$\;}}}}Signature of a general bilinear form. \normalfont Let $V$ a vector space over $\mathbb{R}$ with basis $\{\mathbf{e}_1, \ldots, \mathbf{e}_n\}$. For $0 \leq p \leq 0$ and $0 \leq r \leq 0$ and $\alpha_1, \ldots, \alpha_n, \alpha_{p+r+1}, \ldots, \alpha_n \in \mathbb{R}^{*}_{+}$ some bilinear form $f$ can be defined more generally as

\begin{equation}
    f(\mathbf{v},\mathbf{w}) = - \sum_{i \,=\, 1}^{p} \alpha_i v_i w_i + \sum_{i \,=\, p+1}^{p+q} \alpha_i v_i w_i + \sum_{i \,=\, p+q+1}^{n} 0 v_i w_i.
\end{equation}

\noindent For $\mathbf{v} = (v_1,\ldots,v_n),\mathbf{w} = (w_1,\ldots,w_n) \in V$. If $r = n - (p+q) \neq 0$ the bilinear form is degenerate, the \textbf{signature} of the bilinear form is the numbers $(p,q,r)$. In the non-degenerate case $(r = 0)$ the signature of the bilinear form is $(p,q)$.\;\textcolor{gray}{{$\blacktriangleleft$}}\end{ex}

\subsection{Musical Isomorphisms}

\hspace{0.5cm} Let $V$ be a vector space and let us consider its dual space denoted by $V^*$. There does not exist a natural isomorphism between those spaces like there exists for $V$ and its bidual space $(V^{*})^{*}$ \cite{Roc16}. An additional structure is required to define the isomorphism  $V \simeq V^*$ and is called \textit{correlation}. In other words, an isomorphism between $V$ and $V^*$ must to be chosen. One can introduce the correlation as the linear mapping $\tau : V \to V^*$ that naturally defines a bilinear functional $B: V \cross V \to \mathbb{R}$ by the equation

\begin{equation}
    B(\mathbf{v},\mathbf{u}) = \tau(\mathbf{v}) (\mathbf{u}), \;\;\; \text{for all}\; \mathbf{u},\mathbf{v} \in V.
\end{equation}

\noindent Since dim $V = $ dim $V^*$, if $\tau$ is non-degenerated, i.e., ker $\tau = \{0\}$, it follows that $\tau$ is an \textit{isomorphism}.

\begin{defincinza}\label{defin_w23485rte}In the case of quadratic spaces, the correlations $\tau : V \to V^*$ and $\tau^{-1} : V^* \to V$  are denoted by

\begin{align}
    \begin{aligned}
    \flat : V \to V^*, \;\;\;
    \sharp: V^* \to V  \\
    \end{aligned}
\end{align}

\noindent in such way $\flat = \sharp^{-1}$ and $\sharp = \flat^{-1}$. Such isomorphisms are called \textbf{musical isomorphisms} with respect to $g$. \end{defincinza}

Alternatively, it is also used the expression

\begin{equation}
    \mathbf{v}_{\flat} = \flat(\mathbf{v}), \;\;\; \alpha^{\sharp} = \sharp(\alpha).
\end{equation}

\noindent Hence by definition it follows that

\begin{equation}\label{eq_isomusi545}
     \mathbf{v}_{\flat}(\mathbf{u}) = g(\mathbf{v},\mathbf{u}), \;\;\; g(\alpha^{\sharp}, \mathbf{v}) = \alpha(\mathbf{v}).
\end{equation}

\noindent The musical isomorphism is important to characterise the Clifford product on Clifford algebras as in the Corollary \ref{cor_235qwsad} and the geometric product in the Eq. \eqref{Eq_cliffproduct} with respect to the Kähler-Atiyah bundle.

\section{Tensor Algebra}\label{app2}

\hspace{0.5cm} The theory of tensor algebra is fundamental to our work, through tensor algebra we are able to construct the exterior algebra and the Clifford algebra. In addition, the tensor product of spaces is considered oftentimes in theorems concerning the Clifford algebra structure. Hence, this appendix is devoted to establishing some notations, concepts and results about the tensor product together with a motivation and a discussion about it.

\subsection{Tensor Product}

The tensor product is defined in Ref. \cite{Roc17} as follows 

\begin{defincinza}\label{def_tensorr}
Let $U$ and $V$ vector spaces over the same field $\mathbb{K}$. The \textbf{tensor product} between $U$ and $V$ is a vector space $T$ with a bilinear mapping 

\begin{align}
    \begin{aligned}
 \otimes : U \cross V &\to  T \\
    (\mathbf{u},\mathbf{v}) &\mapsto (\mathbf{u} \tensor \mathbf{v} )
    \end{aligned}
\end{align}

\noindent satisfying the following condition:
 if $\{\mathbf{e}_1,\ldots,\mathbf{e}_m\}$ and $\{\mathbf{f}_1,\ldots,\mathbf{f}_n\}$ are basis for $U$ and $V$ respectively, then $ \mathbf{e}_i \tensor \mathbf{f}_j$, $1 \leq i \leq m$, $1 \leq j \leq n$ is a basis for $T$.
\end{defincinza}

\noindent Such condition in Definition \ref{def_tensorr} does not depend on the choice of basis for $U$ and $V$ and can be expressed in the form of table as

\begin{center}
\begin{tabular}{cc|ccccc}
 & $\tensor$ &  $\mathbf{f}_1$  & $\mathbf{f}_2$ & $\cdots$ & $\mathbf{f}_n$ \tabularnewline
\cline{2-6} 
&  $\mathbf{e}_1$ & $\mathbf{e}_1 \tensor \mathbf{f}_1$ & $\mathbf{e}_1 \tensor \mathbf{f}_2$ & $\cdots$ & $\mathbf{e}_1 \tensor \mathbf{f}_n$  \tabularnewline
&  $\mathbf{e}_2$ & $\mathbf{e}_2 \tensor \mathbf{f}_1$ & $\mathbf{e}_2 \tensor \mathbf{f}_2$ & $\cdots$ & $\mathbf{e}_2 \tensor \mathbf{f}_n$  \tabularnewline
&  $\vdots$ & $\vdots$ & $\vdots$ & $\cdots$ & $\vdots$ \tabularnewline
&  $\mathbf{e}_m$ & $\mathbf{e}_m \tensor \mathbf{f}_1$ & $\mathbf{e}_m \tensor \mathbf{f}_2$ & $\cdots$ & $\mathbf{e}_m \tensor \mathbf{f}_n$  \tabularnewline
 \end{tabular}
\par \end{center}

\noindent Therefore, dim $U \tensor V$ = dim $U$ dim $V$. The consequence about this fact is that not every tensor $x \in U \otimes V$ has the form of a pure tensor $x = \mathbf{u} \otimes \mathbf{v}$ for some $\mathbf{u} \in U$ and $\mathbf{v} \in V$, nevertheless, since the dimension of $U \tensor V$ is the product of the dimensions of $U$ and $V$ not the sum of them, it follows that one general element from $U \tensor V$ is not a pure tensor, but rather a finite linear combination of pure tensors. In this way, the vector space $U \tensor V$ is the vector space of linear combinations of elements $\mathbf{u} \tensor \mathbf{v}$, $\mathbf{u} \in U$ and $\mathbf{v} \in V$ satisfying

\begin{align}
    \begin{aligned}
     (\mathbf{u}_1 + \mathbf{u}_2) \tensor \mathbf{v} &=  \mathbf{u}_1  \tensor \mathbf{v}  + \mathbf{u}_2 \tensor \mathbf{v}, \\ 
      \mathbf{u} \tensor (\mathbf{v}_1 + \mathbf{v}_2)  &=  \mathbf{u}  \tensor \mathbf{v}_1  + \mathbf{u} \tensor \mathbf{v}_2, \\ 
      (\lambda\mathbf{u}) \tensor \mathbf{v} &= \mathbf{u} \tensor (\lambda\mathbf{v}) =  \lambda (\mathbf{u} \tensor \mathbf{v}),
      \end{aligned}
\end{align}

\noindent for $\lambda \in \mathbb{K}$, $\mathbf{u}, \mathbf{u}_1,\mathbf{u}_2 \in U$, $\mathbf{v}, \mathbf{v}_1,\mathbf{v}_2 \in V$, since $\otimes$ is a bilinear mapping. Right, it is easier to understand the motivation about the tensor product once it has been defined, so now we are in a position to present a motivation behind the tensor product through multilinear mappings and another definition that follows from it. \\

With respect to linear mappings we know that the composition of two linear mappings is a linear mapping, however, it is not true that the composition of two multilinear mappings is still a multilinear mapping. For instance, let us consider the following two multilinear mappings

\begin{align}
    \begin{aligned}
    &f: V_1 \cross V_2 \to W_1 \cross W_2 \cross W_3, \\
    &g:  W_1 \cross W_2 \cross W_3 \to U.
    \end{aligned}
\end{align}

\noindent We have that $f$ is bilinear and $g$ is trilinear, but what about the composition $g \circ f$?

\begin{equation*}
\begin{tikzcd}
V_1 \cross V_2 \rar{f}\arrow[gray, bend right]{rr}[black,swap]{g \,\circ\, f}  & {W_1 \cross W_2 \cross W_3} \rar{g}  & {U} 
\end{tikzcd} 
\end{equation*}{}

\noindent Let us investigate what happens with respect to that composition, writing $f$ as

\begin{equation}
    f(\mathbf{v}_1,\mathbf{v}_2) = (f_1(\mathbf{v}_1,\mathbf{v}_2), f_2(\mathbf{v}_1,\mathbf{v}_2), f_3(\mathbf{v}_1,\mathbf{v}_2))
\end{equation}

\noindent for $\mathbf{v}_1 \in V_1, \mathbf{v}_2 \in V_2$ we have that since $f$ is bilinear, so is its components $f_1,f_2,f_2$. Now, inspecting $g \circ f (\lambda\mathbf{v}_1, \mathbf{v}_2)$ for $\lambda \in \mathbb{K}$ we have that

\begin{align}
    \begin{aligned}
    g \circ f (\lambda\mathbf{v}_1, \mathbf{v}_2) &=   g (f (\lambda\mathbf{v}_1, \mathbf{v}_2))\\
    &= g(f_1(\lambda\mathbf{v}_1,\mathbf{v}_2), f_2(\lambda\mathbf{v}_1,\mathbf{v}_2), f_3(\lambda\mathbf{v}_1,\mathbf{v}_2))\\
    &= g((\lambda f_1(\mathbf{v}_1,\mathbf{v}_2),\lambda f_2(\mathbf{v}_1,\mathbf{v}_2), \lambda f_3(\mathbf{v}_1,\mathbf{v}_2))\\
    &= \lambda^3\, g(f_1(\mathbf{v}_1,\mathbf{v}_2),f_2(\mathbf{v}_1,\mathbf{v}_2),  f_3(\mathbf{v}_1,\mathbf{v}_2)) \\
    &=  \lambda^3 \,g \circ f (\mathbf{v}_1, \mathbf{v}_2).
    \end{aligned}
\end{align}

\noindent Therefore, linearity or multilinearity does not hold for $g \circ f$. For that reason, we conclude that the multilinearity is not a well-behaved condition, which motivates us to construct a new vector space, starting from the spaces that we are already considering, with the property that for every multilinear mapping on these product spaces there exists a unique linear mapping on this new vector space.

\begin{defincinza}\label{def_tensoruniversal}
Given two vector spaces $U$ and $V$, the \textbf{tensor product} between these spaces is a new vector space $U \otimes V$ and a bilinear mapping $\otimes: U \cross V \to U \otimes V $ such that for any other bilinear mapping $f : U \cross V \to W$, where $W$ is a vector space, there exists a unique linear mapping $\overline{f}: U \otimes V \to W$ such that $f = \overline{f} \circ \otimes.$
\end{defincinza}

\[
  \begin{tikzcd}
    U \cross V  \arrow{r}{\otimes} \arrow[swap]{dr}{f} & [20pt] U \tensor V \arrow[gray, dashed]{d}[black]{\overline{f}} \\[20pt]
     &  W
  \end{tikzcd}
\]

\noindent The aim of the Definition \ref{def_tensoruniversal} is that the new vector space $U \otimes V$ linearises the bilinear mappings out of $U \cross V$ and owns the information of the original spaces as well. In addition, the Definition \ref{def_tensoruniversal} establish $\mathcal{L}(U \tensor V, W) \simeq \mathcal{L}(U,V;W)$ such that $(\overline{f}: U \tensor V \to W) \mapsto (f: U \cross V \to W)$. If $\{\mathbf{e}_1,\ldots,\mathbf{e}_m\}$ and $\{\mathbf{f}_1,\ldots,\mathbf{f}_n\}$ are basis for $U$ and $V$ respectively, then the linear mapping $\overline{f}$ is given by \cite{Roc17}

\begin{equation}
    \overline{f}(\mathbf{e}_i \tensor \mathbf{f}_j) = f(\mathbf{e}_i,\mathbf{f}_j).
\end{equation}

\noindent Furthermore, the condition shown in the Definition \ref{def_tensoruniversal} is called universality, such property is quite common in algebra

\begin{ex}\label{ex_teohomomo}
{{\textcolor{gray}{{$\blacktriangleright$\;}}}}Homomorphism theorem. \normalfont Let $V$ a vector space, $L \subset V$ vector subspace over $\mathbb{K}$ and $\pi : V \to V/L$ the quotient mapping then for any linear mapping $\varphi : V \to W$ satisfying $L \subset \ker\varphi$ there exists a unique linear mapping $\overline{\varphi}: V/L \to W$ such that $\overline{\varphi} \circ \pi = \varphi$ \cite{RomanS}
\[
  \begin{tikzcd}
    V  \arrow{r}{\pi} \arrow[swap]{dr}{\varphi} & [20pt] V/L \arrow[gray, dashed]{d}[black]{\overline{\varphi}} \\[20pt]
     &  W
  \end{tikzcd}
\]
\noindent i.e., the above diagram commutes. \;\textcolor{gray}{{$\blacktriangleleft$}}\end{ex}

\hspace{0.5cm} The universality of the tensor product has several benefits, the universality guarantees the uniqueness of the tensor product and in situations when we desire to prove certain isomorphism between spaces, the universality property provides us one starting point: the existence of a morphism.

\begin{propos}
If the tensor products exists, it is unique up to a isomorphism.
\end{propos}
\noindent \textit{Proof.} Let $U,V$ vector spaces and suppose there are two vector spaces $U \tensor_1 V$ and $U \tensor_2 V$ corresponding to $\tensor_1: U\cross V \to U \tensor_1 V$ and $\tensor_2: U\cross V \to U \tensor_2 V$, respectively,
satisfying the conditions of the Definition \ref{def_tensoruniversal}. Hence, by the universal property there exists a unique linear mapping $\overline{\tensor_2}: U \tensor_1 V \to U \tensor_2 V$ such that the following diagram commutes

\[
  \begin{tikzcd}
    U \cross V  \arrow{r}{\otimes_1} \arrow[swap]{dr}{\otimes_2} & [20pt] U \tensor_1 V \arrow[gray, dashed]{d}[black]{\overline{\otimes_2}} \\[20pt]
     &   U \tensor_2 V
  \end{tikzcd}
\]

\noindent By the same argument,  there exists a unique linear mapping $\overline{\tensor_1}: U \tensor_2 V \to U \tensor_1 V$ such that the following diagram commutes
\[
  \begin{tikzcd}
    U \cross V  \arrow{r}{\otimes_2} \arrow[swap]{dr}{\otimes_1} & [20pt]  U \tensor_2 V \arrow[gray, dashed]{d}[black]{\overline{\otimes_1}} \\[20pt]
     &  U \tensor_1 V
  \end{tikzcd}
\]

\noindent Define

\begin{align}
    \begin{aligned}
    f_1 &= \overline{\tensor_1} \circ \overline{\tensor_2} : U \tensor_1 V \to  U \tensor_1 V \\
    f_2 &= \overline{\tensor_2} \circ \overline{\tensor_1} :  U \tensor_2 V \to  U \tensor_2 V
    \end{aligned}
\end{align}

\noindent Then we have the following commutative diagrams

 \[     \begin{tikzcd}
    U \cross V  \arrow{r}{\otimes_1} \arrow[swap]{dr}{\otimes_1} & [20pt] U \tensor_1 V \arrow[gray, dashed]{d}[black]{f_1} \\[20pt]
     &   U \tensor_1 V
  \end{tikzcd}  \;\;\;\;\;\;\;\; \begin{tikzcd}
    U \cross V  \arrow{r}{\otimes_2} \arrow[swap]{dr}{\otimes_2} & [20pt]  U \tensor_2 V \arrow[gray, dashed]{d}[black]{f_2} \\[20pt]
     &  U \tensor_2 V
  \end{tikzcd} \]

\noindent Conversely, the identity mappings $id_1 : U \tensor_1 V \to  U \tensor_1 V$,   $id_2 :  U \tensor_2 V \to  U \tensor_2 V$ also make the respective diagrams commute. By uniqueness, $f_1 = id_1$ and $f_2 = id_2$, i.e., $\overline{\tensor_1}$ and $\overline{\tensor_2}$ are inverses of each other. Therefore we conclude that $U \tensor_1 V \simeq U \tensor_2 V$ as desired. $\Box$

\begin{propos}
The following isomorphisms between tensor products of vector spaces hold \cite{Roc17, Kos89, RomanS}:
\end{propos}
\begin{itemize}
    \item[\textcolor{gray}{\textbf{(i)}}] $U \tensor V \simeq V \tensor U$,
    \item[\textcolor{gray}{\textbf{(ii)}}] $U \tensor (V \tensor W) \simeq (U \tensor V) \tensor W$,
    \item[\textcolor{gray}{\textbf{(iii)}}] $V \tensor_{\mathbb{K}} \mathbb{K} \simeq V$,
    \item[\textcolor{gray}{\textbf{(iv)}}] $U^{*} \tensor V \simeq \mathcal{L}(U,V)$,
    \item[\textcolor{gray}{\textbf{(v)}}] $U^{*} \tensor V^{*} \simeq (U \tensor V)^{*}$.
\end{itemize}

\noindent The tensor product is not commutative although the above item \textcolor{gray}{\textbf{(i)}} establishes that there is an isomorphism between these spaces. The item \textcolor{gray}{\textbf{(ii)}} ensures that we can take the tensor product of an arbitrary number of vector spaces consistently. The item  \textcolor{gray}{\textbf{(iv)}} and \textcolor{gray}{\textbf{(v)}} holds only for finite dimensions and it is worth to pointing out that for the item \textcolor{gray}{\textbf{(iv)}}, given $\alpha \in U^{*}$ and $v \in V $ one can define the following linear mapping  \cite{Roc17}

\begin{align}
    \begin{aligned}
 \otimes : U^{*} \cross V &\to  \mathcal{L}(U,V) \\
    (\a,\mathbf{v}) &\mapsto (\alpha \tensor \mathbf{v} ) : U \to V \\  
    &\;\;\;\;\;\;\;\;\;\;\;\;\;\;\;\;\;\;\;\;\;\;\;\mathbf{u} \;\mapsto (\alpha \tensor \mathbf{v}) (\mathbf{u}) := \alpha (\mathbf{u}) \mathbf{v}.
    \end{aligned}
\end{align}

The Definition \ref{def_tensoruniversal} can be generalised for $p \in \mathbb{Z}$ vector spaces over the same field $\mathbb{K}$.

\begin{defincinza} \label{def_tensoruniversalgeral}

Let $V_1, \ldots, V_p$ be $\mathbb{K}$-vector spaces. The \textbf{tensor product} (over $\mathbb{K}$) for $V_1, \ldots, V_p$ is the $2$-uple $(\tensor, T)$ formed by a vector space $T$ and a  $p$-linear mapping $\otimes : V_1 \cross  \cdots \cross V_p \to T$ such that for any other $p$-linear mapping $f: V_1 \cross \cdots \cross V_p \to W$, where $W$ is a vector space, there exists a unique linear mapping $\overline{f}: T \to W$ such that $f = \overline{f} \circ \otimes$ i.e., the following diagram commutes

\[
  \begin{tikzcd}
   V_1 \cross \cdots \cross V_p \arrow{r}{\otimes} \arrow[swap]{dr}{f} & [20pt] T \arrow[gray, dashed]{d}[black]{\overline{f}} \\[20pt]
     &  W
  \end{tikzcd}
\]
\end{defincinza}

The next goal is to prove the existence of the tensor product, some important definitions must be stated first. Let $\mathcal{F}= \{V_i \;|\;i \in I \}$  be any family of vector spaces over $\mathbb{K}$ \cite{RomanS}:

\begin{defincinza}
The \textbf{direct product} of $\mathcal{F}$ is the vector space

\begin{equation}
    \prod_{i \,\in \,I} V_i = \left \{ f : I \to \bigcup_{i \, \in I} V_i \;|\; f(i) \in V_i \right\} 
\end{equation}
    
\noindent thought of as a subspace of the vector space of all functions from $I$ to $\bigcup V_i$.
\end{defincinza}

\begin{defincinza}
The \textbf{support} of a function $f: I \to \bigcup V_i$ is the set
\end{defincinza}

\begin{equation}
    \text{supp}(f) = \{i \in I \;| \; f(i) \neq 0 \}.
\end{equation}

\noindent Thus, a function $f$ has finite support if $f(i) = 0$ for all but finitely many index $i \in I$.

\begin{defincinza}
The \textbf{external direct sum} of the family $\mathcal{F}$ is the vector space
\end{defincinza}

\begin{equation}
    \bigoplus_{i \,\in \,I}^{\text{ext}} V_i =\left \{f: I \to \bigcup_{i \, \in \, I} V_i \;|\; \text{supp}(f) < \infty  \right\}
\end{equation}

\noindent \textit{thought of as a subspace of the vector space of all functions from $I$ to $\bigcup V_i$}. \\

The following result is fundamental to our goal.

\begin{lemma}\label{lemma_tensor1}
Given a set $X$ there exists a $\mathbb{K}$-vector space $F(X)$ such that dim $F(X) = |X|$, i.e., there is a bijective function between $X$ and any basis of $F(X)$.
\end{lemma}
\noindent \textit{Proof.} Consider $F(X)$ as
\begin{equation}
    F(X)  = \bigoplus_{x \,\in \,X}^{\text{ext}} \mathbb{K} =\left \{f: X \to \mathbb{K} \;|\; \text{supp}(f) < \infty  \right\}.
\end{equation}

\noindent In this way, given $f \in F(X)$ we can write $f$ in a unique way as $f = \sum_{x \in X} f(x) \delta_x$ such that $\delta_x : X \to \mathbb{K}$ is defined by

\begin{equation}
   \delta_x(y) = \begin{cases}
    1_{\mathbb{K}}, \; \text{if} \; x = y, \\
    0_{\mathbb{K}},\;  \text{if} \;x \neq y.
    \end{cases}
\end{equation}

\noindent Since supp$(\delta_x) < \infty$, $\delta_x \in F(X)$. Also, $f = \sum_{x \in X} f(x) \delta_x < \infty$ since  $f(x) \neq 0$ for all but finite $x \in X$. Therefore, $\{ \delta_x \;|\; x \in X\}$ is a basis for $F(X)$ and one can identify $X$ and $\{ \delta_x \;|\; x \in X\}$ through the bijection $x\mapsto \delta_x$, which gives the desired result. $\Box$

\begin{theorem}
\textbf{(Existence of the Tensor Product)} For any family $V_1,\ldots,V_p$ of $\mathbb{K}$-vector spaces there exists the tensor product $(\tensor,V)$.
\end{theorem}
\noindent \textit{Proof.} Consider the set $V_1 \cross \cdots \cross V_p$, according to the previous Lemma \ref{lemma_tensor1}, there is a $\mathbb{K}$-vector space $F = F(V_1 \cross \cdots \cross V_p)$ such that 
    $\varphi: V_1 \cross \cdots \cross V_p \to \mathcal{B} $ is a bijection and $\mathcal{B}$ is some basis for $F$ since dim $F = |V_1 \cross \cdots \cross V_p|.$ Now, let us consider a vector subspace $U$ of $F$ generated by elements of the form

\begin{align}\label{eq_tensormultiprova}
    \begin{aligned}
   &\textcolor{gray}{\textbf{(i)}}\;\; \varphi (\mathbf{v}_1,\ldots,\mathbf{v}_i + \mathbf{u}_i,\ldots, \mathbf{v}_p) - \varphi (\mathbf{v}_1,\ldots,\mathbf{v}_i,\ldots, \mathbf{v}_p) - \varphi (\mathbf{v}_1,\ldots, \mathbf{u}_i,\ldots, \mathbf{v}_p).\\
   &\textcolor{gray}{\textbf{(ii)}}\;\; \varphi (\mathbf{v}_1,\ldots,\lambda \mathbf{v}_i,\ldots, \mathbf{v}_p) - \lambda \varphi (\mathbf{v}_1,\ldots,\mathbf{v}_i,\ldots, \mathbf{v}_p).
    \end{aligned}
\end{align}

\noindent Right, let $\tensor :  V_1 \cross \cdots \cross V_p \to F/U = V $ given by $\tensor = \pi \circ \iota \circ \varphi$ s.t $\pi : F \to F/U$ is the quotient mapping and $\iota : \mathcal{B} \hookrightarrow F$ is the inclusion.

\begin{equation*}
\begin{tikzcd}
V_1 \cross \cdots \cross V_p \rar{\varphi}\arrow[bend left]{rrr}[black]{\tensor}  & {\mathcal{B}} \rar[hookrightarrow]{\iota}  & {F} \rar{\pi} & {F/U = V}
\end{tikzcd} 
\end{equation*}{}

\noindent We claim that $\tensor$ is $p$-linear. Indeed, because of the item \textcolor{gray}{\textbf{(i)}} written in the expression \eqref{eq_tensormultiprova}, it follows that

\begin{align}
    \begin{aligned}
    \tensor((\mathbf{v}_1,\ldots,\mathbf{v}_i + \mathbf{u}_i,\ldots, \mathbf{v}_p) &= (\pi \circ \iota \circ \varphi)(\mathbf{v}_1,\ldots,\mathbf{v}_i + \mathbf{u}_i,\ldots, \mathbf{v}_p) \\
    &= \pi(\varphi(\mathbf{v}_1,\ldots,\mathbf{v}_i + \mathbf{u}_i,\ldots, \mathbf{v}_p)) \\
    &= [\varphi(\mathbf{v}_1,\ldots,\mathbf{v}_i + \mathbf{u}_i,\ldots, \mathbf{v}_p)] \\
    &= [\varphi(\mathbf{v}_1,\ldots,\mathbf{v}_i,\ldots, \mathbf{v}_p)] + [\varphi (\mathbf{v}_1,\ldots, \mathbf{u}_i,\ldots, \mathbf{v}_p)].
    \end{aligned}
\end{align}

\noindent Continuing in this way,

\begin{align}
    \begin{aligned}
    \tensor(\mathbf{v}_1,\ldots,\mathbf{v}_i + \mathbf{u}_i,\ldots, \mathbf{v}_p) &= [\varphi(\mathbf{v}_1,\ldots,\mathbf{v}_i,\ldots, \mathbf{v}_p)] + [\varphi (\mathbf{v}_1,\ldots, \mathbf{u}_i,\ldots, \mathbf{v}_p)] \\
    &= \pi (\varphi(\mathbf{v}_1,\ldots,\mathbf{v}_i,\ldots, \mathbf{v}_p)) + \pi (\varphi (\mathbf{v}_1,\ldots, \mathbf{u}_i,\ldots, \mathbf{v}_p))\\
    &= (\pi \circ \iota \circ \varphi) (\mathbf{v}_1,\ldots,\mathbf{v}_i,\ldots, \mathbf{v}_p)) + (\pi \circ \iota \circ \varphi) (\mathbf{v}_1,\ldots, \mathbf{u}_i,\ldots, \mathbf{v}_p)) \\
    &= \tensor (\mathbf{v}_1,\ldots,\mathbf{v}_i,\ldots, \mathbf{v}_p)) + \tensor (\mathbf{v}_1,\ldots, \mathbf{u}_i,\ldots, \mathbf{v}_p)). \\
    \end{aligned}
\end{align}

\noindent Analogously, by the item \textcolor{gray}{\textbf{(ii)}} written in the Eq. \eqref{eq_tensormultiprova} it also holds that for $\lambda \in \mathbb{K}$

\begin{equation}
    \tensor(\mathbf{v}_1,\ldots,\lambda \mathbf{v}_i,\ldots, \mathbf{v}_p) = \lambda \tensor(\mathbf{v}_1,\ldots,\mathbf{v}_i,\ldots, \mathbf{v}_p) .
\end{equation}

\noindent Which proves the $p$-linearity of $\tensor$. We claim that universality holds for $(\tensor, V)$. Indeed, let $W$ be an arbitrary vector space over $\mathbb{K}$ and $\psi: V_1 \cross \cdots \cross V_p$ an arbitrary $p$-linear mapping. 

\begin{equation*}
\begin{tikzcd}
V_1 \cross \cdots \cross V_p \arrow[black]{d}[black,swap]{\psi} \rar{\varphi} \arrow[bend left]{rrr}[black]{\tensor}  & {\mathcal{B}} \rar[hookrightarrow]{i}  & {F} \rar{\pi} & {F/U = V} \\ [20pt]W
\end{tikzcd} 
\end{equation*}{}

\noindent Since $\varphi$ is a bijection one can consider $\varphi^{-1}$. In addition, since $\mathcal{B}$ is a basis for $F$ and $\psi \circ \varphi^{-1}: \mathcal{B} \to W$ is a mapping defined by basis, $\psi \circ \varphi^{-1}$ can be extended uniquely to a linear mapping  $\Psi_1 : F \to W$ such that $\Psi_1 \circ \iota = \psi \circ \varphi^{-1}$, therefore, $\psi = \Psi_1 \circ \iota \circ \varphi$

\begin{equation*}
\begin{tikzcd}
V_1 \cross \cdots \cross V_p \arrow[black]{d}[black,swap]{\psi} \rar{\varphi} \arrow[bend left]{rrr}[black]{\tensor}  & {\mathcal{B}} \rar[hookrightarrow]{\iota}  & {F} \arrow[darkgray, dashed, swap]{dll}[black]{\Psi_1} \rar{\pi} & {F/U = V} \\ [20pt]W
\end{tikzcd} 
\end{equation*}{}

\noindent It is of our interest to investigate how $\Psi_1$ acts on the vector space $U \subset F$. We claim that $U \subset \ker \Psi_1$. It is sufficient to ascertain $\Psi_1(x)$ for $x$ as a generator of $U$ written in the Eq. \eqref{eq_tensormultiprova}. For instance, let the generator $x \in U$ be as the item \textcolor{gray}{\textbf{(ii)}}

\begin{equation}
    x = \varphi (\mathbf{v}_1,\ldots,\lambda \mathbf{v}_i,\ldots, \mathbf{v}_p) - \lambda \varphi (\mathbf{v}_1,\ldots,\mathbf{v}_i,\ldots, \mathbf{v}_p).
\end{equation}

\noindent Therefore by the linearity of $\Psi_1$ and the $p$-linearity of $\psi$, we have

\begin{align}
    \begin{aligned}
    \Psi_1(x) &=  \Psi_1 (\varphi (\mathbf{v}_1,\ldots,\lambda \mathbf{v}_i,\ldots, \mathbf{v}_p) - \lambda \varphi (\mathbf{v}_1,\ldots,\mathbf{v}_i,\ldots, \mathbf{v}_p)) \\
    &=  \Psi_1 (\varphi (\mathbf{v}_1,\ldots,\lambda \mathbf{v}_i,\ldots, \mathbf{v}_p) - \lambda \Psi_1 (\varphi (\mathbf{v}_1,\ldots,\mathbf{v}_i,\ldots, \mathbf{v}_p)) \\
     &=  \Psi_1 (\iota \circ \varphi (\mathbf{v}_1,\ldots,\lambda \mathbf{v}_i,\ldots, \mathbf{v}_p) - \lambda \Psi_1 (\iota \circ \varphi (\mathbf{v}_1,\ldots,\mathbf{v}_i,\ldots, \mathbf{v}_p)) \\
     &=  \psi (\mathbf{v}_1,\ldots,\lambda \mathbf{v}_i,\ldots, \mathbf{v}_p) - \lambda \psi (\mathbf{v}_1,\ldots,\mathbf{v}_i,\ldots, \mathbf{v}_p)) \\
     &=  \lambda \psi (\mathbf{v}_1,\ldots, \mathbf{v}_i,\ldots, \mathbf{v}_p) - \lambda \psi (\mathbf{v}_1,\ldots,\mathbf{v}_i,\ldots, \mathbf{v}_p)) = 0. \\
    \end{aligned}
\end{align}

Analogously, for $x \in U$ as the item \textcolor{gray}{\textbf{(i)}} in the Eq. \eqref{eq_tensormultiprova} we have that $\Psi_1(x) = 0.$ Hence, by concluding that for any generator $x$ of $U$, $\Psi(x) = 0$, we have that $U \subset \,$ $\ker(f)$. By the Example \ref{ex_teohomomo}, it follows that there exists a unique linear mapping $\Psi_2 : F/U \to W$ such that $\Psi_2 \circ \pi = \Psi_1 $

\begin{equation*}
\begin{tikzcd}
V_1 \cross \cdots \cross V_p \arrow[black]{d}[black,swap]{\psi} \rar{\varphi} \arrow[bend left]{rrr}[black]{\tensor}  & {\mathcal{B}} \rar[hookrightarrow]{\iota} & {F} \arrow[darkgray, dashed,swap]{dll}[black]{\Psi_1} \rar{\pi} & {F/U = V} \arrow[gray, dashed]{dlll}[black]{\Psi_2} \\ [20pt]W
\end{tikzcd} 
\end{equation*}{}

\noindent Therefore, we have that

\begin{equation} \label{eq_tensorvariousmappings}
    \Psi_2 \circ \tensor = \Psi_2 \circ (\pi \circ \iota \circ \varphi) = (\Psi_2 \circ \pi) \circ \iota \circ \varphi = \Psi_1 \circ \iota \circ \varphi = \psi
\end{equation}

\noindent Right, we claim that $\Psi_2$ is unique. Indeed, suppose that the diagram commutes with respect to another linear mapping $f: F/U \to W$

\begin{equation*}
\begin{tikzcd}
V_1 \cross \cdots \cross V_p \arrow[black]{d}[black,swap]{\psi} \rar{\tensor} & {F/U = V} \arrow[gray, dashed]{dl}[black]{f} \\ [20pt]W
\end{tikzcd} 
\end{equation*}{}

\noindent By Eq. \eqref{eq_tensorvariousmappings}, we have that

\begin{equation}
    f \circ \tensor =  \psi = \Psi_2 \circ \tensor
\end{equation}

\noindent Which means that $f$ and $\Psi_2$ coincides on the elements of the form 
\begin{align}
    \begin{aligned}
    \tensor(\mathbf{v}_1,\ldots,\mathbf{v}_i,\ldots, \mathbf{v}_p) &= \pi \circ \iota \circ \varphi (\mathbf{v}_1,\ldots,\mathbf{v}_i,\ldots, \mathbf{v}_p ) \\
    &= [\varphi (\mathbf{v}_1,\ldots,\mathbf{v}_i,\ldots, \mathbf{v}_p ) ].
    \end{aligned}
\end{align}

\noindent We have that $[\varphi (\mathbf{v}_1,\ldots,\mathbf{v}_i,\ldots, \mathbf{v}_p ) ]$ is the equivalent class of the basis elements of $F$, namely, the generator set of $F/U$. Hence, $f$ and $\Psi_2$ coincides on entire space $F/U$ and the uniqueness holds. We conclude that $(\tensor, V)$ with $V = F/U$, is the tensor product of for $V_1, \ldots, V_p$ by the Definition \ref{def_tensoruniversalgeral}. $\Box$ \\

 It is worth to mention that one can also write the tensor product $(\tensor, V)$ of $V_1, \ldots, V_p$ as $V_1 \tensor \cdots \tensor V_p$. It follows that $\mathcal{L}(V_1, \ldots, V_p;W) \simeq \mathcal{L}(V_1\tensor \cdots \tensor V_p,W)$ by the universality.  Furthermore, if the vector spaces are all equal $V_i = V$ we denote the tensor product as \cite{Roc17}

\begin{equation}
    T_{q} (V) = \color{gray}
      \overbrace{\color{black}\;V \tensor \cdots \tensor V\;}^{\text{\color{black}$q$ factors}}  \color{black}= V^{\tensor^{q}}
\end{equation}

\noindent Analogously, the tensor product of $p$ dual spaces $V^*$ is written as

\begin{equation}
    T^{p} (V) =
    \color{gray}
      \overbrace{\color{black}\;V^* \tensor \cdots \tensor V^*\;}^{\text{\color{black} $p$ factors}}  \color{black}= (V^*)^{\tensor^{p}}
\end{equation}

\noindent In addition, 

\begin{align}\begin{aligned}
  T^{p}\,_{q}(V) &= (V^*)^{\tensor^{p}} \tensor V^{\tensor^{q}} \\ T_{q}\,^{p}(V) &= V^{\tensor^{q}}  \tensor (V^*)^{\tensor^{p}} 
\end{aligned}
\end{align}

\noindent We adopt the convention that $T^{p}_{q}(V)$ refers to the space $T^{p}\,_{q}(V)$. A basis for the vector space  $T^{p}_{q}(V)$ is given by the set of tensor products \cite{Roc16}

\begin{equation}
    \{\mathbf{e}^{\mu_1} \tensor \mathbf{e}^{\mu_2} \tensor \cdots \tensor \mathbf{e}^{\mu_p} \tensor \mathbf{e}_{\nu_1} \tensor \mathbf{e}_{\nu_2} \tensor \cdots \tensor \mathbf{e}_{\nu_q} \},
\end{equation}

\noindent An arbitrary element $T \in T^{p}_{q}(V)$ can be written as

\begin{equation}
    T =  T^{\nu_1 \nu_2 \cdots \nu_q}_{\mu_1 \mu_2 \cdots \mu_p} \mathbf{e}^{\mu_1} \tensor \mathbf{e}^{\mu_2} \tensor \cdots \tensor \mathbf{e}^{\mu_p} \tensor \mathbf{e}_{\nu_1} \tensor \mathbf{e}_{\nu_2} \tensor \cdots \tensor \mathbf{e}_{\nu_q},
\end{equation}

\noindent such that

\begin{equation}
T^{\nu_1 \nu_2 \cdots \nu_q}_{\mu_1 \mu_2 \cdots \mu_p} = T(\mathbf{e}_{\mu_1} , \mathbf{e}_{\mu_2}, \ldots, \mathbf{e}_{\mu_p} , \mathbf{e}^{\nu_1} ,\mathbf{e}^{\nu_2} , \ldots,\mathbf{e}^{\nu_q}).
\end{equation}

\noindent The multilinear functionals $T \in T^{p}_{q}(V)$ are called \textit{tensors of type (p,q)}. The quantities $T^{\nu_1 \nu_2 \cdots \nu_q}_{\mu_1 \mu_2 \cdots \mu_p}$ are the components of the tensor $T$ in the given basis. The tensors of type $(p,0)$ are called \textit{covariant tensors} as well as tensors of type $(0,q)$ are called \textit{contravariant tensors.}

\begin{ex} {{\textcolor{gray}{{$\blacktriangleright$\;}}}} Type of tensors. \normalfont It is assumed that tensors of type $(0,0)$ are scalars, tensors of type $(0,1)$ are vectors and tensors of type $(1,0)$ are covectors, in other words, $T^{0}_{0}(V) = \mathbb{K},\; T^{0}_{1}(V) = V,\; T^{1}_{0}(V) = V^*$ and more generally \cite{Roc17}

\begin{align}
    \begin{aligned}
    T^{q}_{0}(V) &= \mathcal{L}(V, \ldots, V; \mathbb{K}),\\
    T^{q}_{1}(V) &= \mathcal{L}(V, \ldots, V; V).
    \end{aligned}
\end{align}

\noindent In particular tensors of type $(0,2)$ are bilinear mappings and tensors of type $(1,1)$ are linear mappings. \;\textcolor{gray}{{$\blacktriangleleft$}}\end{ex}

\subsection{Tensor Algebra}\label{appb2_tensoralg}
\hspace{0.5cm} We are now in a position to introduce the tensor algebra. Given two tensors $T$ and $S$ of type $(p,q)$ we define their sum as the tensors $T + S$ of type $(p,q)$ in terms of their components by \cite{Roc16}

\begin{equation}
    (T + S)^{\nu_1 \nu_2 \cdots \nu_q}_{\mu_1 \mu_2 \cdots \mu_p} = T^{\nu_1 \nu_2 \cdots \nu_q}_{\mu_1 \mu_2 \cdots \mu_p} + S^{\nu_1 \nu_2 \cdots \nu_q}_{\mu_1 \mu_2 \cdots \mu_p}.
\end{equation}

\noindent If $T$ is a tensor of type $(p,q)$ and $S$ is a tensor of type $(r,s)$, the tensor product $T \tensor S$, which is a tensor of type $(p+r, q + s)$, is defined in terms of their components as

\begin{equation}
    (T \tensor S)^{\nu_1 \nu_2 \cdots \nu_q \rho_1 \rho_2 \cdots \rho_s}_{\mu_1 \mu_2 \cdots \mu_p \sigma_1 \sigma_2 \cdots \sigma_r} = T^{\nu_1 \nu_2 \cdots \nu_q}_{\mu_1 \mu_2 \cdots \mu_p}  S^{\rho_1 \rho_2 \cdots \rho_s}_{\sigma_1 \sigma_2 \cdots \sigma_r} 
\end{equation}

\noindent The product $\tensor$ is distributive with respect to sum and is associative but is not commutative.

\begin{defincinza}
The direct sum of all vector spaces $T^{p}_{q}(V)$ endowed with the operations of sum and the tensor product is called the \textbf{tensor algebra} associated to the vector space $V$
\end{defincinza}

The tensor algebra is a graded algebra. In the general case, the grading is given by $G = \mathbb{Z} \cross \mathbb{Z}$ and it is positive. Two cases are particularly important: the algebra of covariant and contravariant tensors. The algebra of the covariant tensors, of type $(p,0)$, is denoted by

\begin{equation}
    T^{*}(V) = \bigoplus_{p = 0}^{\infty} T^{p}(V)
\end{equation}

\noindent whereas
\begin{equation}
    T(V) = \bigoplus_{p = 0}^{\infty} T^{p}(V)
\end{equation}

\noindent is the covariant tensor algebra of $(0,q)$ tensors.\\

The tensor algebra is fundamental because many other algebras arise as quotient algebras of T(V). For instance, the exterior algebra and the Clifford algebra are introduced in this work as a quotient of tensor algebra. 

\section{Exterior Algebra}\label{app3}
\hspace{0.5cm} In the study of alternating multilinear mappings and their ramifications, exterior algebra stands out. Introduced by Hermann Grassmann in 1844, from a geometric point of view, exterior algebra is an algebraic construction of multidimensional vectors in which they have the meaning of points, oriented line segment, oriented plane fragment, oriented volume fragment, and so on. This algebra has a great richness in its structure and its importance lies not only in its applications in the construction of physical theories, algebraic topology, differential forms but also in algebra, whose consequences is explored since the exterior algebra provides the multivector structure for Clifford algebra and it is a Clifford algebra itself. Therefore in this appendix, we introduce the general concepts about exterior algebra. The starting point is be the alternator operator and from such operator, we define the exterior algebra elements, the underlying vector space, the exterior product, properties and operations that it is used in this entire work. 

\subsection{Permutations and the Alternator}

\hspace{0.5cm} Let $\{1,2,\ldots,p\}$ be a set of $p$ elements. A \textbf{permutation} is a bijective function $\sigma: \{1,2,\ldots,p\} \to \{1,2,\ldots,p\}$  represented by the cycle

\begin{equation}
    \begin{pmatrix}
    1 & 2 & \cdots & p \\
     \sigma(1) & \sigma(2) & \cdots & \sigma(p) 
    \end{pmatrix}
\end{equation}

\noindent A permutation $\sigma$ such that $\sigma(k) = k$ for all $k \neq i, k \neq j$ and $\sigma(i) = j, \sigma(j) = i$, is called a \textit{transposition} and can be denoted by $(ij)$. A permutation of $n$ elements is said to be \textit{even} or \textit{odd} if the permutation can be written respectively as an even or an odd number of transpositions. Therefore the \textit{sign} $\varepsilon(\sigma)$ of a permutation $\sigma$ is defined to be $\varepsilon(\sigma) = +1$ if the permutation is even and $\varepsilon(\sigma) = -1$ if the permutation is odd. The composition of two permutations is also a permutation and the set of all permutations form a group, namely, \textit{symmetric group} $S_p$ and it has $p!$ elements. \\

Let $X_1 \tensor X_2 \tensor \cdots \tensor X_p$ be either a contravariant or covariant tensor such that where $X$ denotes respectively either a vector or a covector, and the indexes enumerate such elements.

 {\colorlet{shadecolor}{gray!15}\begin{shaded} 
The \textbf{operator alternator} denoted by Alt and defined as follows

\begin{equation}\label{eq_assd4123a}
    \text{Alt}(X_1 \tensor X_2 \tensor \cdots \tensor X_p) = \frac{1}{p!} \sum_{\sigma \in S_p} \varepsilon(\sigma)X_{\sigma(1)} \tensor X_{\sigma(2)} \tensor \cdots \tensor X_{\sigma(p)}
\end{equation}\end{shaded}}

\noindent The operator Alt is a projection operator (Alt$^2$ = Alt). The alternator is the starting point in the construction of the exterior algebra.

\begin{ex}
{{\textcolor{gray}{{$\blacktriangleright$\;}}}} $S_3$. \normalfont Consider the symmetric group $S_3$ given by the permutations of $3$ elements. Those permutations are represented by

\begin{align}
    \begin{aligned}
    &\begin{pmatrix}
    1 & 2 & 3 \\
    1 & 2 & 3
    \end{pmatrix} \leftrightarrow (1)(2)(3), &&\begin{pmatrix}
    1 & 2 & 3 \\
    2 & 3 & 1
    \end{pmatrix} \leftrightarrow (12)(13), &&\begin{pmatrix}
    1 & 2 & 3 \\
    3 & 1 & 2
    \end{pmatrix} \leftrightarrow (13)(12),\\
     &\begin{pmatrix}
    1 & 2 & 3 \\
    3 & 2 & 1
    \end{pmatrix} \leftrightarrow (13)(2), &&\begin{pmatrix}
    1 & 2 & 3 \\
    2 & 1 & 3
    \end{pmatrix} \leftrightarrow (12)(3), &&\begin{pmatrix}
    1 & 2 & 3 \\
    1 & 3 & 2
    \end{pmatrix} \leftrightarrow (23)(1).
    \end{aligned}
\end{align}

\noindent The elements in the first row of the above equation are all even permutations whereas the ones in the second row are all odd. For either a covariant or contravariant tensor $X_1 \tensor X_2 \tensor X_3$ the alternator is given by

\begin{align}
    \begin{aligned}
    \text{Alt}(X_1 \tensor X_2 \tensor X_3) = \; &\frac{1}{6}(X_1 \tensor X_2 \tensor X_3 + X_2 \tensor X_3 \tensor X_1 + X_3 \tensor X_1 \tensor X_2 \\
    &- X_3 \tensor X_2 \tensor X_1 - X_2 \tensor X_1 \tensor X_3 - X_1 \tensor X_3 \tensor X_2). \;\textcolor{gray}{{\blacktriangleleft}}   
    \end{aligned}
\end{align}
\end{ex}

\hspace{0.5cm} Another way to represent the action of the alternation is when a covariant tensor is taken into account. Those objects are multilinear functionals acting on vectors. Without loss of generality, let $\alpha_1 \tensor \alpha_2 \tensor \cdots \tensor \alpha_p$ be a covariant tensor, it reads:

\begin{equation}
    \alpha_1 \tensor \alpha_2 \tensor \cdots \tensor \alpha_p  (\'v_1, \'v_2, \ldots, \'v_p) = \alpha_1(\'v_1) \tensor \alpha_2(\'v_2) \tensor \cdots \tensor \alpha_p(\'v_p).
\end{equation}

\noindent The action of Alt on a contravariant tensor is defined by 

\begin{equation}
    \alt (\alpha_1 \tensor \alpha_2 \tensor \cdots \tensor \alpha_p)  (\'v_1, \'v_2, \ldots, \'v_p) = \frac{1}{p!}\begin{vmatrix}
\alpha_1(\'v_1) & \alpha_1(\'v_2) & \cdots & \alpha_1(\'v_p)\\
\alpha_2(\'v_1) & \alpha_2(\'v_2) & \cdots & \alpha_2(\'v_p)\\
\vdots & \vdots &\ddots &\vdots \\
\alpha_p(\'v_1) & \alpha_p(\'v_2) & \cdots & \alpha_p(\'v_p)
\end{vmatrix}.
\end{equation}

\noindent Such that at the right-hand side of the above equation there is the determinant of the
associated matrix. This result follows from the definition of
the determinant, we have that if $A$ is the matrix of order $p$ with entries $A_{ij}$ then
the determinant det $A$ is given by

\begin{equation}
    \text{det} A = \sum_{\sigma \in S_p} \varepsilon(\sigma)A_{1\sigma(1)}A_{2\sigma(2)} \cdots A_{p\sigma(p)}.
\end{equation}

\subsection{$p$-vectors and $p$-covectors}

\hspace{0.5cm} Once the alternator has been presented, we are now in a position to introduce some elements that can be constructed from it.

\begin{defincinza}\label{def_aaalt}
A \textbf{p-vector} is an alternating contravariant tensor of order $p$ denoted by $A_{[p]}$. A \textbf{p-covector} is an alternating covariant tensor of order $p$ denoted by $\Psi^{[p]}$. The $p$-vector and the $p$-covector are characterised by \normalfont

\begin{align}
    \begin{aligned}
     A_{[p]} = \alt (A_{[p]}), \\
      \Psi^{[p]} = \alt (\Psi^{[p]}).
    \end{aligned}
\end{align}
\end{defincinza}

Let $V$ be a real vector space, the space of $p$-vectors and $p$-covectors are denoted respectively by $\bigwedge_p(V)$ and $\bigwedge^p(V)$ in such way that
\begin{align}
    \begin{aligned}
    \bigwedge_0(V) = \bigwedge^0(V)  = \rr,
    &&\bigwedge_1(V) = V, &&&\bigwedge^1(V) = V^*. \\
    \end{aligned}
\end{align}

\noindent Both $0$-vectors and $0$-covectors are scalars, $1$-vector is a synonym of vector, as well as $1$-covector is a synonym of covector. It is also common to call a $2$-vector a bivector, $3$-vector a trivector, and so on. The important point to note here is the same once one construction involving $p$-vectors is accomplished, the same reasoning applies to the construction concerning $p$-covectors. Thus, just the case involving $p$-vectors shall be considered subsequently. Our next concern will be to introduce a product between those elements.

\subsection{Exterior Product}\label{sec_app_extrprod}

\hspace{0.5cm} Let $\Ap$ be a $p$-vector and $\Bq$ be a $q$-vector, the result of $\Ap \tensor \Bq$ is a contravariant tensor of order $p + q$ however it is not alternating. Meanwhile, $\Alt (\Ap \tensor \Bq)$ is an alternating contravariant tensor of order $p + q$, namely, a $(p+q)$-vector. That motivates the following definition.

\begin{defincinza}\label{definprod_ex}
Let $V$ be a vector space, $\Ap \in \bigwedge_p(V)$ a $p$-vector and $\Bq \in \bigwedge_q(V)$ a $q$-vector. The \textbf{exterior product} $\wedge : \bigwedge_p(V) \cross \bigwedge_q(V) \to \bigwedge_{p+q}(V)$ is defined as
\normalfont
\begin{equation}
    \Ap \wedge \Bq = \alt (\Ap \tensor \Bq).
\end{equation}
\end{defincinza}
\noindent Since the tensor product is associative and bilinear, the exterior product inherits the associativity and bilinearity \cite{Roc16}. That is: for $\Ap \in \bigwedge_p(V), \Bq \in \bigwedge_q(V), \Cr \in \bigwedge_r(V), a \Ap \in \bigwedge_0(V)$ it holds

{\colorlet{shadecolor}{gray!15}\begin{shaded} 
\begin{itemize}
    \item []\textcolor{darkgray}{\textbf{(i)}} $(\Ap \wedge \Bq) \wedge \Cr = \Ap \wedge (\Bq \wedge \Cr),$
    \item []\textcolor{darkgray}{\textbf{(ii)}} $\Ap \wedge (\Bq + \Cr) = \Ap \wedge \Bq + \Ap \wedge \Cr,$
    \item   []\textcolor{darkgray}{\textbf{(iii)}}  $a\wedge \Ap = a\Ap, $
   \item[]\textcolor{darkgray}{\textbf{(iv)}} $\Ap \wedge \Bq =   (-1)^{pq}   \Bq \wedge \Ap.$
\end{itemize}
\end{shaded}}

In order to derive the item \textcolor{darkgray}{\textbf{(iv)}}, it is of our interest to study the case involving the exterior product between two vectors. It follows from Definition \ref{definprod_ex} that

\begin{equation}\label{eq_ext232}
    \'v \wedge \'u = \frac{1}{2}(\'v \tensor \'u - \'u \tensor \'v)
\end{equation}

\noindent which gives  \begin{equation}
    \'v \wedge \'u = - \'u \wedge \'v,
\end{equation} 

\noindent that is, the exterior product is anti-commutative. In particular $\'v \wedge \'v = 0$. For the general case, a $p$-vector $\Ap$ and a $q$-vector $\Bq$ can be written in the form

\begin{equation}\label{eq_pex834743783}
    \Ap = \'v_1 \wedge \cdots \wedge \'v_p, \;\;\;\;\; \Bq = \'u_1 \wedge \cdots \wedge \'u_p.
\end{equation}

\noindent That is due to the fact that the exterior product is bilinear and associative. Hence, the exterior product $\Ap \wedge \Bq$ reads

\begin{equation}
     \Ap \wedge \Bq = \'v_1 \wedge \cdots \wedge \'v_p \wedge \'u_1 \wedge \cdots \wedge \'u_p.
\end{equation}

\noindent In order to interchange the vectors involved in the exterior products, we conclude from Eq. \eqref{eq_pex834743783} that
\begin{equation}
     \'v_1 \wedge \cdots \wedge \'v_p \wedge \'u_1 \wedge \cdots \wedge \'u_q =  (-1)^{pq}\'u_1 \wedge \cdots \wedge \'u_p \wedge \'v_1 \wedge \cdots \wedge \'v_p
\end{equation}

\noindent which gives the $\textcolor{darkgray}{\textbf{(iv)}}$ property of the exterior product $     \Ap \wedge \Bq =   (-1)^{pq}   \Bq \wedge \Ap.$

We emphasise that a $p$-vector that can be written as the exterior product of a $p$ number of vectors as in Eq. \eqref{eq_pex834743783}, is called a \textit{simple} $p$-vector. Considering vector spaces $V$ such that $\dim V \leq 3$, every $p$-vector is simple for higher dimensions not all $p$-vectors are simple \cite{Roc16}.

\hspace{0.5cm} Regarding the space of the $p$-vectors, an important point about it is with respect to its basis. Let $V$ be a vector space with basis $\{\'e_1,\ldots,\'e_n \}$, a basis for each one of the spaces $\bigwedge_p(V)$ can be constructed from the basis of $V$. Let us first examine the space $\bigwedge_2(V)$ and the exterior products $\'e_i \wedge \'e_j$. Since the exterior product is anti-commutative, the linearly independent set of bivectors is provided by

\begin{align}
    \begin{aligned}
    \'e_1 \wedge \'e_2, \; \'e_1 \wedge \'e_3, \; \'e_1 \wedge \'e_4, \;  \cdots , \; \'e_1 \wedge \'e_n, \\
    \'e_2 \wedge \'e_3, \; \'e_2 \wedge \'e_4, \; \cdots , \; \'e_2 \wedge \'e_n, \\
    \vdots \;\;\;\;\;\; \\
    \'e_{n-1} \wedge \'e_n.
    \end{aligned}
\end{align}

\noindent Therefore, the dimension of $\bigwedge_2(V)$ is the number of possible combinations of $n = \dim V$ vectors taken $2$ at a time. For the general case,

\begin{equation}
    \dim \bigwedge_p (V) = \binom{n}{p} = \frac{n!}{(n-p)!p!}.
\end{equation}

\noindent An arbitrary element $\Ap \in \bigwedge_p(V)$ can be written as

\begin{align}
    \begin{aligned}
    \Ap &= \frac{1}{p!} \sum_{\mu_1\mu_2\cdots\mu_p}A^{\mu_1\mu_2\cdots\mu_p} \'e_{\mu_1} \wedge \'e_{\mu_2} \wedge \cdots \wedge \'e_{\mu_p}\\
    &= \sum_{\mu_1 < \mu_2< \cdots< \mu_p}A^{\mu_1\mu_2\cdots\mu_p} \'e_{\mu_1} \wedge \'e_{\mu_2} \wedge \cdots \wedge \'e_{\mu_p}
    \end{aligned}
\end{align}

Regarding $V$ a vector space with $\dim V = n$, one may ask now how many spaces $\bigwedge_{p}(V)$ can be constructed from $V$ taking into account the dimension of $V$. We begin with a general result about the exterior product

   {\colorlet{shadecolor}{gray!15}\begin{shaded} \begin{propos}
If $p>n = \dim V$, the exterior product of $p$ vectors is null.
\end{propos} \end{shaded}}

\proof Consider the exterior product of $n +1$ vectors, $\'v_1 \wedge \cdots \wedge \'v_n \wedge \'v_{n+1}$. Since $\dim V = n$, the $n+1$ given vectors are necessarily linearly dependent and we can write one of those vectors as a linear combination of the others. There is no loss of generality in assuming that $\'v_{n+1} = \sum_{i=1}^{n} a^{i}\'v_i$.  On account of the anticommutativity and the fact that $\'v_i \wedge \'v_i = 0$ we have that:

\begin{align}
    \begin{aligned}
    \'v_1 \wedge \'v_2 \wedge \cdots \wedge \'v_n \wedge \'v_{n+1} &= \'v_1 \wedge \'v_2 \wedge \cdots \wedge \'v_n \wedge \left ( a^{1}\'v_1 + a^{2}\'v_2 +\cdots + a^{n}\'v_n \right ) \\
    &= (-1)^{n-1}a^{1}\'v_1 \wedge \'v_1 \wedge \'v_2 \wedge \cdots \wedge \'v_n \\
    &\;\;\;+(-1)^{n-2}a^{2}\'v_1 \wedge \'v_2 \wedge \'v_2 \wedge \cdots \wedge \'v_n \\
    &\;\;\;+ \cdots + a^{n}\'v_1 \wedge \'v_2 \wedge \cdots \wedge \'v_n  \wedge \'v_n\\
    &= 0
    \end{aligned}
\end{align}

\noindent which completes the proof. $\Box$\\

\hspace{0.5cm} More generally, is proved that 

  {\colorlet{shadecolor}{gray!15}\begin{shaded} $$\mathbf{v}_1 \wedge \cdots \wedge \mathbf{v}_p = 0 \Longleftrightarrow \{    \mathbf{v}_1, \ldots, \mathbf{v}_p \}\;\; \text{is linearly dependent}.$$ \end{shaded}}

\noindent That discussion clarifies that it does not exists the vector space $\bigwedge_{p}(V)$ if $p > n$. Therefore the spaces that can be constructed are

\begin{equation}
    \bigwedge_0(V),\bigwedge_1(V), \bigwedge_2(V), \ldots, \bigwedge_{n-1}(V), \bigwedge_p(V).
\end{equation}

\noindent Such that

\begin{equation}
        \dim \bigwedge_p (V) = \binom{n}{p} = \frac{n!}{(n-p)!p!} = \binom{n}{n - p} =  \dim \bigwedge_{n-p} (V)
\end{equation}

\noindent Although the spaces $\bigwedge_p (V)$ and $\bigwedge_{n-p} (V)$ are isomorphic, there is not a natural isomorphism between them. However, by considering additional structures on the vector space $V$ , it is possible to construct an isomorphism 
called \textit{Hodge isomorphism} \cite{Roc16}.

\subsection{Exterior Algebra}\label{sec4_app3}

\hspace{0.5cm} Recall the exterior product $\wedge$ as being defined on $\bigwedge_p(V) \cross \bigwedge_q(V) \to \bigwedge_{p+q}(V)$. Consider the vector space $\bigwedge(V)$ defined by the direct sum of the vector spaces $\bigwedge_{p}(V), \; (p = 0,1,2,\ldots,n):$

\begin{equation}
    \bigwedge(V) = \bigwedge_0(V) \oplus  \bigwedge_1(V) \oplus \bigwedge_2(V) \oplus \cdots \oplus \bigwedge_n(V) = \bigoplus_{p = 0}^{n} \bigwedge_p(V).
\end{equation}

\noindent The space $\bigwedge(V)$ is thereby closed by the exterior product, that is, $\bigwedge(V) \cross \bigwedge(V) \to \bigwedge(V)$.

\begin{defincinza}
The pair $(\bigwedge(V), \wedge)$ is named \textbf{exterior algebra} associated to the vector space $V$.
\end{defincinza}

The elements of the space $\bigwedge(V)$ are called \textit{multivectors}. An arbitrary multivector $A \in \bigwedge(V)$ is written as

\begin{align}
\begin{aligned}
   A &= \color{darkgray}
      \underbrace{\color{black}a}_{\color{black} scalar}  \color{black} + \color{gray}
      \underbrace{\color{black}v^{i}\'e_i}_{\color{black} vector}  \color{black} + \color{gray}
      \underbrace{\color{black} F^{ij} \'e_i \wedge \'e_j}_{\color{black} 2-vector}  \color{black}\\ &\;\;\;\;+ \color{gray}
      \underbrace{\color{black} T^{ijk} \'e_i \wedge \'e_j \wedge \'e_k}_{\color{black} 3-vector}  \color{black} + \cdots + \color{gray}
      \underbrace{\color{black}p\'e_1 \wedge \cdots \wedge \'e_n}_{\color{black} n-vector}  \color{black} \in \bigwedge(V).
\end{aligned}
\end{align}

According to our previous discussion, it follows that the dimension of $\bigwedge(V)$ is given by

\begin{equation}\label{eq_app3dimeext}
  \dim \bigwedge(V) =  \sum_{p=0}^{n} \dim \bigwedge_{p}(V) = \sum_{p=0}^{n}    \binom{n}{p} = 2^n.
\end{equation}

\begin{ex}
{{\textcolor{gray}{{$\blacktriangleright$\;}}}} The exterior algebra $\bigwedge (\mathbb{R}^3)$. \normalfont Consider the $3$-dimensional euclidean vector space $V = \rt$ with orthonormal basis $\{\'{e}_{1}, \'{e}_{2}, \'{e}_{3}\}$. These basis elements satisfies the following relation with respect to the exterior product,

\begin{align}
\begin{aligned}\label{eq_prodexrelbase124}
\'{e}_{i} \wedge \'{e}_{i} &= 0, &&&&&& (\text{for}\;\; i = 1,2,3).\\
\'{e}_{i} \wedge \'{e}_{j} &= -\'{e}_{j} \wedge \'{e}_{i},  &&&&&& (\text{for} \;\; i \neq j).
\end{aligned}
\end{align}

\noindent The exterior algebra is the space
\begin{equation}
    \bigwedge (\mathbb{R}^3) = \bigoplus_{p=0}^{3} \bigwedge ^p (\mathbb{R}^3).
\end{equation}

\noindent equipped with the exterior product. Consider the subspace of the bivectors $\bigwedge ^{2} (\mathbb{R}^3)$ with basis
$\{\'{e}_{1} \wedge \'{e}_{2}, \'{e}_{1} \wedge \'{e}_{3}, \'{e}_{2} \wedge \'{e}_{3}\}$. In a geometric point of view, each element  $\'{e}_{i} \wedge \'{e}_{j}$ represents an oriented plane fragment generated by two vectors as represented in the following scheme showing the geometry of the elements $\{\'{e}_{1} \wedge \'{e}_{2}, \'{e}_{1} \wedge \'{e}_{3}, \'{e}_{2} \wedge \'{e}_{3}\}$:

\begin{center}

\tikzset{every picture/.style={line width=0.75pt}} 

\begin{tikzpicture}[x=0.75pt,y=0.75pt,yscale=-1,xscale=1]

\draw  [fill={rgb, 255:red, 230; green, 230; blue, 230 }  ,fill opacity=0.6 ] (82.5,54.5) -- (177.4,54.5) -- (177.4,149.4) -- (82.5,149.4) -- cycle ;
\draw    (82.5,149.4) -- (175.4,149.4) ;
\draw [shift={(177.4,149.4)}, rotate = 180] [color={rgb, 255:red, 0; green, 0; blue, 0 }  ][line width=0.75]    (10.93,-3.29) .. controls (6.95,-1.4) and (3.31,-0.3) .. (0,0) .. controls (3.31,0.3) and (6.95,1.4) .. (10.93,3.29)   ;
\draw    (82.5,149.4) -- (82.5,56.5) ;
\draw [shift={(82.5,54.5)}, rotate = 90] [color={rgb, 255:red, 0; green, 0; blue, 0 }  ][line width=0.75]    (10.93,-3.29) .. controls (6.95,-1.4) and (3.31,-0.3) .. (0,0) .. controls (3.31,0.3) and (6.95,1.4) .. (10.93,3.29)   ;
\draw    (48,89) -- (48,182.9) ;
\draw    (48,89.5) -- (83,54.5) ;
\draw    (49.41,182.49) -- (82.5,149.4) ;
\draw [shift={(48,183.9)}, rotate = 315] [color={rgb, 255:red, 0; green, 0; blue, 0 }  ][line width=0.75]    (10.93,-3.29) .. controls (6.95,-1.4) and (3.31,-0.3) .. (0,0) .. controls (3.31,0.3) and (6.95,1.4) .. (10.93,3.29)   ;
\draw    (143.9,182.9) -- (177.4,149.4) ;
\draw    (49.5,182.9) -- (143.9,182.9) ;
\draw    (93.9,120.9) .. controls (81.9,125.9) and (88.4,143.4) .. (98.9,143.9) .. controls (109.08,144.38) and (115.98,139.22) .. (109.99,121.57) ;
\draw [shift={(109.4,119.9)}, rotate = 69.78] [color={rgb, 255:red, 0; green, 0; blue, 0 }  ][line width=0.75]    (10.93,-3.29) .. controls (6.95,-1.4) and (3.31,-0.3) .. (0,0) .. controls (3.31,0.3) and (6.95,1.4) .. (10.93,3.29)   ;
\draw    (104.4,167.9) .. controls (107.4,160.4) and (109.9,154.4) .. (89.9,154.9) .. controls (70.7,155.38) and (76.38,160.47) .. (89.25,169.28) ;
\draw [shift={(90.9,170.4)}, rotate = 214.16] [color={rgb, 255:red, 0; green, 0; blue, 0 }  ][line width=0.75]    (10.93,-3.29) .. controls (6.95,-1.4) and (3.31,-0.3) .. (0,0) .. controls (3.31,0.3) and (6.95,1.4) .. (10.93,3.29)   ;

\draw (31.5,183.9) node [anchor=north west][inner sep=0.75pt]    {$\mathbf{e}_{1}$};
\draw (183,138.4) node [anchor=north west][inner sep=0.75pt]    {$\mathbf{e}_{2}$};
\draw (73,27.4) node [anchor=north west][inner sep=0.75pt]    {$\mathbf{e}_{3}$};
\draw (125.5,57.4) node [anchor=north west][inner sep=0.75pt]    {$\mathbf{e}_{2} \land \mathbf{e}_{3}$};
\draw (122.66,151.4) node [anchor=north west][inner sep=0.75pt]  [font=\small,xslant=0.72]  {$\mathbf{e}_{1} \land \mathbf{e}_{2}$};
\draw (47.85,92.71) node [anchor=north west][inner sep=0.75pt]  [font=\footnotesize,rotate=-313.57,xslant=0.62]  {$\mathbf{e}_{1} \land \mathbf{e}_{3}$};

\end{tikzpicture}
\end{center}

\noindent The exterior product between two vectors $\'a,\'b \in \bigwedge ^{1} (\mathbb{R}^3)$ is a bivector and can be computed through the determinant of the following matrix:

\begin{align}
\begin{aligned}\label{eq_aexprodd2ab}
\'{a} \wedge \'{b} &= \left| \begin{array}{rcc}
     \'e_{2} \wedge \'e_{3} & \'e_{3} \wedge \'e_{1}  & \'e_{1} \wedge \'e_{2} \\
     a_1 & a_2 & a_3\\
     b_1 & b_2  & b_3
    \end{array} \right|  \\
    &=(a_2b_3 - a_3b_2)\'e_{2} \wedge \'e_{3} + (a_3b_1 - a_1b_3)\'e_{3} \wedge \'e_{1} + (a_1b_2 - a_2b_1)\'e_{1} \wedge \'e_{2}\\
 &= B_{23}\'e_{2} \wedge \'e_{3} + B_{31}\'e_{3} \wedge \'e_{1} + B_{12}\'e_{1} \wedge \'e_{2} = \'{B}. 
 \end{aligned}
\end{align}

 \noindent Note that the cross product of $\'a$ and $\'b$ leads to

 \begin{equation}
     \'a \cross \'b = (a_2b_3 - a_3b_2)\'e_{1}  + (a_3b_1 - a_1b_3)\'e_{2} + (a_1b_2 - a_2b_1)\'e_{3}
 \end{equation}

\begin{center}

\tikzset{every picture/.style={line width=0.75pt}} 

\begin{tikzpicture}[x=0.75pt,y=0.75pt,yscale=-1,xscale=1]

\draw  [fill={rgb, 255:red, 191; green, 191; blue, 191 }  ,fill opacity=0.28 ] (87.89,126.78) -- (191.92,133.36) -- (116.92,177.49) -- (12.89,170.9) -- cycle ;
\draw    (87.89,126.78) -- (189.93,133.24) ;
\draw [shift={(191.92,133.36)}, rotate = 183.62] [color={rgb, 255:red, 0; green, 0; blue, 0 }  ][line width=0.75]    (10.93,-3.29) .. controls (6.95,-1.4) and (3.31,-0.3) .. (0,0) .. controls (3.31,0.3) and (6.95,1.4) .. (10.93,3.29)   ;
\draw    (87.89,126.78) -- (14.61,169.89) ;
\draw [shift={(12.89,170.9)}, rotate = 329.53] [color={rgb, 255:red, 0; green, 0; blue, 0 }  ][line width=0.75]    (10.93,-3.29) .. controls (6.95,-1.4) and (3.31,-0.3) .. (0,0) .. controls (3.31,0.3) and (6.95,1.4) .. (10.93,3.29)   ;
\draw    (102.41,152.13) .. controls (99.12,178.96) and (121.92,169.36) .. (125.12,165.36) .. controls (128.14,161.58) and (132.05,158.06) .. (119.82,149) ;
\draw [shift={(117.52,147.36)}, rotate = 34.48] [fill={rgb, 255:red, 0; green, 0; blue, 0 }  ][line width=0.08]  [draw opacity=0] (8.93,-4.29) -- (0,0) -- (8.93,4.29) -- cycle    ;
\draw    (87.89,126.78) -- (87.89,16.16) ;
\draw [shift={(87.89,14.16)}, rotate = 90] [color={rgb, 255:red, 0; green, 0; blue, 0 }  ][line width=0.75]    (10.93,-3.29) .. controls (6.95,-1.4) and (3.31,-0.3) .. (0,0) .. controls (3.31,0.3) and (6.95,1.4) .. (10.93,3.29)   ;

\draw (71.33,135.53) node [anchor=north west][inner sep=0.75pt]  [font=\small]  {$\mathbf{a} \land \mathbf{b}$};
\draw (99.33,17.53) node [anchor=north west][inner sep=0.75pt]  [font=\small]  {$\mathbf{a} \times \mathbf{b}$};

\end{tikzpicture}   
\end{center}

 \noindent The direction of $\'a \cross \'b$ is perpendicular the plane $\'{a} \wedge \'{b}$ and the length/norm of the cross product $\'a \cross \'b$ coincide with the area/norm of the exterior product $\'{a} \wedge \'{b}.$ However, the exterior product does not require a metric while the cross product requires or induces a metric, which in turn, gets involved in positioning the vector $\'a \cross \'b$ perpendicular to  $\'{a} \wedge \'{b}$ \cite{Lou01}. 
 $ \;\textcolor{gray}{{\blacktriangleleft}} $
\end{ex}

\subsection{Exterior Algebra as a Quotient of Tensor Algebra}

\hspace{0.5cm} This discussion is devoted to an explicit construction of the exterior algebra as quotient of tensor algebra. Besides being an interesting mathematical construction itself, an analogous approach is used for the Clifford algebras.

\begin{defincinza}
Let $\A$ be an algebra. A set $I_{L} \subset \A$ is said to be a \textbf{left ideal} of $\A$ if $\forall\, a \in \A, \forall \, x \in I_L, ax \in I_L$. Analogously, $I_{R} \subset \A$ is said to be a \textbf{right ideal} of $\A$ if $\forall\, a \in \A, \forall \, x \in I_L, xa \in I_R$. The set $\mathcal{I} \subset \A$ is said to be a \textbf{two-sided ideal} or simply an \textbf{ideal} if  $\forall\, a,b \in \A, \forall \, x \in \mathcal{I}, axb \in \mathcal{I}$
\end{defincinza}

\hspace{0.5cm} Let $\A$ be an algebra and write $A$ as sum of spaces $\A = \mathcal{B} + \mathcal{C}$. Given $a,b \in \A$ following equivalence relation can be defined:

\begin{equation}
    a \sim b \Longleftrightarrow a = b + x, \;\;\;\;\;\; x \in \mathcal{C}.
\end{equation}

\noindent The set of the equivalence classes $A/\sim$ has a natural vector space structure with sum and multiplication by scalar defined by

\begin{align}
    \begin{aligned}
    [a] + [b] &= [a + b],\\
    \lambda[a] &= [\lambda a].
    \end{aligned}
\end{align}

\noindent To $A/\sim$ be an algebra, a natural way to define the product between equivalence classes is

\begin{equation}
    [a][b] = [ab]
\end{equation}

\noindent Since we know by definition that $[a] = [a + x], [b] = [b + y], x,y \in \mathcal{C}$, it follows that

\begin{equation}
    [a][b] = [a + x][b + y] = [ (a + x)(b + y)] = [ab + ay + xb + xy].
\end{equation}

\noindent The last two equations result in

\begin{equation}
    [ab] = [ab + ay + xb + xy].
\end{equation}

\noindent That means that in order to $A/\sim$ be an algebra $ay + xb + xy$ must be an element of $\mathcal{C}$ which only holds if $\mathcal{C}$ is a two-sided ideal. In that case, $A/\sim$ is named the quotient algebra of $\A$ by $\mathcal{C}$, denoted by $\A / \mathcal{C}$.

\hspace{0.5cm} Let $T(V)$ be the algebra of the contravariant tensors. Consider the ideal $I$ of $T(V)$ generated by elements $\'v \tensor \'v, \'v \in V$. The elements of $I$ consists of the sums

\begin{equation}
    \sum_{i} A_i \tensor \'v_i \tensor \'v_i \tensor B_i,
\end{equation}

\noindent where $\'v_i \in V$ and $A_i,B_i \in T(V)$. We can notice that

\begin{equation}
    \'v \tensor \'u + \'u \tensor \'v = (\'v + \'u) \tensor (\'v + \'u) - \'v \tensor \'v - \'u \tensor \'u.
\end{equation}

\noindent That is, we can also consider that the ideal $\mathcal{I}$ is generated by the elements $\'v \tensor \'u + \'u \tensor \'v$ where $\'v, \'u \in V$. One important point to note here lies in the fact that 

\begin{equation}
    \'v \tensor \'u + \'u \tensor \'v \in \ker \alt
\end{equation}

\noindent and $\ker \alt = \mathcal{I}$. Our next claim is that the exterior algebra is isomorphic to the quotient algebra $T(V)/\mathcal{I}.$ The respective equivalence relation is given by
\begin{align}
    \begin{aligned}
     A \sim B \Longleftrightarrow A = B + x, \;\;\;\;\;\; x \in \mathcal{I}.
    \end{aligned}
\end{align}

\noindent Moreover, the product between them is denoted by

\begin{equation}
    [A] \wedge [B] = [A \tensor B].
\end{equation}

\noindent For $\'v, \'u \in V$ we have that

\begin{equation}
    \'v \tensor \'u = \frac{1}{2}(\'v \tensor \'u - \'u \tensor \'v) + \frac{1}{2}(\'v \tensor \'u + \'u \tensor \'v)
\end{equation}

\noindent such that $\frac{1}{2}(\'v \tensor \'u + \'u \tensor \'v) \in \mathcal{I}$. Therefore, 

\begin{equation}
[\'v] \wedge [\'u] = [\'v \tensor \'u] = [\frac{1}{2}(\'v \tensor \'u + \'u \tensor \'v)] = [\Alt (\'v \tensor \'u)].
\end{equation}

\noindent For the general case, the above result is extended as

\begin{equation}
    \'v_1 \tensor \cdots \tensor \'v_p \sim \Alt (\'v_1 \tensor \cdots \tensor \'v_p) = \'v_1 \wedge \cdots \wedge \'v_p
\end{equation}

\noindent which establishes the desired isomorphism

\begin{equation}
    \bigwedge(V) \simeq T(V)/\mathcal{I}.
\end{equation}

\subsection{Contraction}

\hspace{0.5cm} Let $\Ap$ be a $p$-vector and $\alpha$ a covector. The final task in this appendix is to present an operation that action on $\Ap \in \bigwedge_{p}(V)$ and gives an element of $\bigwedge_{p-1}(V)$, i.e., a $(p-1)$-vector. Such operation is very important and it is used to define the Clifford product on Clifford algebras.

\begin{defincinza}\label{def_fd65g415sad}
The \textbf{left contraction} of a $p$-vector $\Ap $ by a covector $\alpha$, denoted by $\alpha \rfloor$, is defined as

\begin{align}
\begin{aligned}\label{eq_contractiondefinition0}
\rfloor :  \bigwedge^{1}(V)&\times \bigwedge_{p}(V)  &\longrightarrow \;\;\;\;\; &\bigwedge_{p-1} (V) \\
&(\alpha, \Ap)  &\longmapsto \;\;\;\;\; &(\alpha \rfloor \Ap)(\alpha_1,\ldots,\alpha_{p-1}) = p\Ap(\alpha,\alpha_1,\ldots,\alpha_{p-1}).
\end{aligned}
\end{align}
\noindent \textit{such that $(\alpha_1,\ldots,\alpha_{p-1})$ stands for arbitrary covectors.}\end{defincinza}

\noindent On the right-hand side of the above expression, for $\Ap = \'v_1 \wedge \cdots \wedge \'v_p$ it follows that

\begin{align}
    \begin{aligned}
    (\'v_1 \wedge \cdots \wedge \'v_p)(\alpha,\alpha_1, \ldots, \alpha_{p-1})  = \frac{1}{p!} \sum_{\sigma \in Sp}\varepsilon(\sigma) \alpha(\'v_{\sigma(1)})\alpha_1(\'v_{\sigma(2)})\cdots \alpha_{p-1}(\'v_{\sigma(p)}).
    \end{aligned}
\end{align}

\noindent It follows immediately that for $p = 1$

\begin{equation}
    \alpha \rfloor \'v = \alpha (\'v).
\end{equation}

\noindent For an element of $\bigwedge_{0}(V)$ it is assumed that $\alpha \rfloor 1 = 0.$ Our next concern will be the contraction of a $2$-vector $\'v \wedge \'u.$ It reads:

\begin{align}
    \begin{aligned}
    (\alpha \rfloor (\'v \wedge \'u)) (\beta) &= 2 (\'v \wedge \'u)(\alpha, \beta) \\
    &= \alpha (\'v)\beta(\'u) - \alpha(\'u)\beta(\'v)\\
    &=(\alpha(\'v)\'u - \alpha(\'u)\'v)(\b)\\
    &= ((\a\rfloor\'v)\'u - (\a\rfloor\'u)\'v)(\b).
    \end{aligned}
\end{align}

\noindent Since $\b$ is arbitrary, we conclude that

\begin{equation}\label{eq_bivectextcontr32}
    \alpha \rfloor (\'v \wedge \'u) = ((\a\rfloor\'v)\'u - (\a\rfloor\'u)\'v) =(\alpha(\'v)\'u - \alpha(\'u)\'v).
\end{equation}

We proceed to develop the contraction for a $3$-vector $\'v \wedge \'u \wedge \'w$ using the above results. It follows that:

\begin{align}
    \begin{aligned}
    (\a \rfloor (\'v \wedge \'u \wedge \'w) (\b,\gamma) &= 3(\'v \wedge \'u \wedge \'w)(\a,\b,\gamma)\\
    &= 3 \frac{1}{3!} [\a(\'v) \b(\'u) \gamma(\'w)) + \b(\'v) \gamma(\'u) \a(\'w)) + \gamma(\'v) \a(\'u) \beta(\'w)) \\
    &\;\;\;\;- \gamma(\'v) \b(\'u) \a(\'w)) - \b(\'v) \a(\'u) \gamma(\'w)) - \a(\'v) \gamma(\'u) \b(\'w))] \\
    &= \a(\'v) \frac{1}{2!}[\b(\'u)\gamma(\'w) - \b(\'w)\gamma(\'u)] \\
    &\;\;\;\;- \a(\'u) \frac{1}{2!}[\b(\'v)\gamma(\'w) - \b(\'w)\gamma(\'v)]\\
    &\;\;\;\;+\a(\'w) \frac{1}{2!}[\b(\'v)\gamma(\'u) - \b(\'u)\gamma(\'v)]\\
    &= \a(\'v)(\'u \wedge \'w)(\b,\gamma) - \a(\'u)(\'v \wedge \'w)(\b,\gamma) \\
    &\;\;\;\;+ \a(\'w)(\'v \wedge \'u)(\b,\gamma).
    \end{aligned}
\end{align}

\noindent Since $\b$ and $\gamma$ are arbitrary,

\begin{align}
    \begin{aligned}
    \a \rfloor (\'v \wedge \'u \wedge \'w) &=(\a\rfloor\'v)(\'u \wedge \'w) - (\a \rfloor\'u)(\'v \wedge \'w) + (\a \rfloor \'w)(\'v \wedge \'u) \\
    &= \a(\'v)(\'u \wedge \'w) - \a(\'u)(\'v \wedge \'w) + \a(\'w)(\'v \wedge \'u).
    \end{aligned}
\end{align}

\noindent Now using Eq. \eqref{eq_bivectextcontr32} for the $2$-vector case and the last one for the $3$-vector case, together with the associativity of the exterior product, we have that

\begin{align}\label{eq_81ext}
    \a \rfloor ((\'v \wedge \'u) \wedge \'w) &= (\a \rfloor (\'v \wedge \'u)) \wedge \'w + \'v \wedge \'u(\alpha \rfloor \'w),   \end{align} \begin{equation}\label{eq_82ext}
        \a \rfloor (\'v \wedge (\'u \wedge \'w)) = (\a \rfloor \'v) \'u \wedge \'w - \'v \wedge (\alpha \rfloor (\'u \wedge \'w)).
    \end{equation}

\noindent Both equations \eqref{eq_81ext} and \eqref{eq_82ext} considers the contraction of the exterior product between a $1$-vector and a $2$-vector. One important point to note here is the fact that the presence of either a positive or a negative sign according to respectively to the presence of a $2$-vector in Eq. \eqref{eq_81ext} and a $1$-vector in Eq. \eqref{eq_82ext}. For the general case, since the grade involution takes into account the different signs, for $A, B \in \bigwedge(V)$ it follows that:

\begin{equation}\label{eq_contracttermsap}
    \alpha \rfloor (A \wedge B) = (\alpha \rfloor A) \wedge B + \yhwidehat{A} \wedge (\alpha \rfloor B).
\end{equation}

The right contraction is defined analogously as the left contraction. For an arbitrary multivector $A$, the left and the right contraction are related by \cite{Roc16} 

\begin{equation}
    \alpha \rfloor A = - \yhwidehat{A} \lfloor \alpha.
\end{equation}

\section{Clifford Algebras Classification}\label{app4}\label{clifclassapp}
\begin{itemize}
    \item[\textcolor{gray}{{$\diamond$}}]\textcolor{gray}{\textbf{ $p + q = 0$}}
    \item[ ] $\cl_{0,0} \simeq \rr$
\end{itemize}

\begin{itemize}
  \item[\textcolor{gray}{{$\diamond$}}]\textcolor{gray}{\textbf{ $p + q = 1$}}
    \item[ ] $\cl_{0,1} \simeq \cc$
    \item[ ] $\cl_{1,0} \simeq \rsr$
\end{itemize}

\begin{itemize}
  \item[\textcolor{gray}{{$\diamond$}}]\textcolor{gray}{\textbf{ $p + q = 2$}}
    \item[ ] $\cl_{0,2} \simeq \hh$
    \item[ ] $\cl_{2,0} \simeq \matreald$
    \item[ ] $\cl_{1,1} \simeq \cl_{2,0} \simeq \matreald$
\end{itemize}

\begin{itemize}
 \item[\textcolor{gray}{{$\diamond$}}]\textcolor{gray}{\textbf{ $p + q = 3$}}
    \item[ ] $\cl_{0,3} \simeq \cl_{0,2} \tensor \cl_{1,0} \simeq \hh \tensor (\rr \oplus \rr) \simeq \hh \oplus \hh$
    \item[ ] $\cl_{3,0} \simeq \cl_{2,0} \tensor \cl_{0,1} \simeq \matreald \tensor \cc \simeq \text{Mat}(2,\cc)$
    \item[ ] $\cl_{1,2} \simeq \cl_{1,1} \tensor \cl_{0,1} \simeq \matreald \tensor \cc \simeq \text{Mat}(2,\cc)$
    \item[ ] $\cl_{2,1} \simeq \cl_{1,1} \tensor \cl_{1,0} \simeq \matreald \tensor \rsr \simeq \text{Mat}(2, \rsr)$
\end{itemize}

\begin{itemize}
  \item[\textcolor{gray}{{$\diamond$}}]\textcolor{gray}{\textbf{ $p + q = 4$}}
    \item[ ] $\cl_{0,4} \simeq \cl_{0,2} \tensor \cl_{2,0} \simeq \hh \tensor \matreald \simeq \text{Mat}(2,\hh)  $ 
    \item[ ] $\cl_{4,0} \simeq \cl_{2,0} \tensor \cl_{0,2} \simeq  \matreald \tensor \hh \simeq \text{Mat}(2,\hh) $ 
    \item[ ] $\cl_{1,3} \simeq \cl_{1,1} \tensor \cl_{0,2} \simeq  \matreald \tensor \hh \simeq \text{Mat}(2,\hh)$
     \item[ ] $\cl_{3,1} \simeq \cl_{1,1} \tensor \cl_{2,0} \simeq  \matreald \tensor \matreald \simeq \text{Mat}(4,\rr)$
     \item[ ] $\cl_{2,2} \simeq \cl_{1,1} \tensor \cl_{1,1} \simeq  \matreald \tensor \matreald \simeq \text{Mat}(4,\rr)$
    \end{itemize}
    
\begin{itemize}
 \item[\textcolor{gray}{{$\diamond$}}]\textcolor{gray}{\textbf{ $p + q = 5$}}
    \item[ ] $\cl_{0,5} \simeq \cl_{0,2} \tensor \cl_{3,0} \simeq \hh \tensor \matcomp \simeq \hh \tensor \cc \tensor \matreald \simeq \matcomp \tensor \matreald \simeq \matcompq$ 
    \item[ ] $\cl_{5,0} \simeq \cl_{2,0} \tensor \cl_{0,3} \simeq \matreald \tensor \hh \oplus \hh \simeq  \text{Mat}(2,\hsh) $
    \item[ ] $\cl_{1,4} \simeq \cl_{1,1} \tensor \cl_{0,3} \simeq \matreald \tensor \hsh \simeq \text{Mat}(2, \hsh)$
    \item[ ] $\cl_{4,1} \simeq \cl_{1,1} \tensor \cl_{3,0} \simeq \matreald \tensor \matcomp \simeq \matcompq$
    \item[ ] $\cl_{2,3} \simeq \cl_{1,1} \tensor \cl_{1,2} \simeq \cl_{1,1} \tensor \cl_{1,1} \tensor \cl_{0,1} \simeq \matrealq \tensor \cc  \simeq \matcompq$
    \item[ ] $\cl_{3,2} \simeq \cl_{1,1} \tensor \cl_{2,1} \simeq \cl_{1,1} \tensor \cl_{1,1} \tensor \cl_{1,0} \simeq \matrealq \tensor \rsr  \simeq \text{Mat}(4,\rsr)$
    \end{itemize}

\begin{itemize}
     \item[\textcolor{gray}{{$\diamond$}}]\textcolor{gray}{\textbf{ $p + q = 6$}}
    \item[ ] $\cl_{0,6} \simeq \cl_{0,2} \tensor \cl_{4,0} \simeq \cl_{0,2} \tensor \cl_{2,0} \tensor \cl_{0,2} \simeq  \hh \tensor \text{Mat}(2, \rr) \tensor \hh \simeq  \text{Mat}(2,\hh) \tensor \hh \simeq \text{Mat}(2, \hh \tensor \hh)\simeq \text{Mat}(2, \cl_{0,2} \tensor \cl_{0,2}) \simeq \text{Mat}(2, \cl_{2,2}) \simeq \text{Mat}(2,\text{Mat}(4,\rr)) \simeq \text{Mat}(8,\rr) $
    \item [ ]$\cl_{6,0} \simeq \cl_{2,0} \tensor \cl_{0,4} \simeq  \cl_{2,0} \tensor \cl_{0,2} \tensor \cl_{2,0} \simeq
    \matreald \tensor \hh \tensor \text{Mat}(2,\rr) \simeq \text{Mat}(4,\hh) $
    \item [ ] $\cl_{1,5} \simeq \cl_{1,1} \tensor \cl_{0,4} \simeq  \cl_{1,1} \tensor \cl_{0,2} \tensor \cl_{2,0} \simeq
    \matreald \tensor \hh \tensor \text{Mat}(2,\rr) \simeq \text{Mat}(4,\hh) $
     \item [ ] $\cl_{5,1} \simeq \cl_{1,1} \tensor \cl_{4,0}  \simeq \cl_{1,1} \tensor \cl_{2,0} \tensor \cl_{0,2}
     \simeq \matreald \tensor \text{Mat}(2,\rr) \tensor \hh \simeq \text{Mat}(4,\hh)$
      \item [ ] $\cl_{2,4} \simeq \cl_{1,1} \tensor \cl_{1,3} \simeq \cl_{1,1} \tensor \cl_{1,1} \tensor \cl_{0,2} 
      \simeq \matreald \tensor \text{Mat}(2,\rr) \tensor \hh \simeq \text{Mat}(4,\hh)$
      \item [ ] $\cl_{4,2} \simeq \cl_{1,1} \tensor \cl_{3,1} \simeq \cl_{1,1} \tensor \cl_{1,1} \tensor \cl_{2,0} \simeq \matreald \tensor \matreald \tensor\matreald  \simeq \text{Mat}(8,\rr)$
      \item [ ] $\cl_{3,3} \simeq \cl_{1,1} \tensor \cl_{1,1} \tensor \cl_{1,1} \simeq \matreald \tensor \matreald \tensor\matreald  \simeq \text{Mat}(8,\rr)$
\end{itemize}

\begin{itemize}
   \item[\textcolor{gray}{{$\diamond$}}]\textcolor{gray}{\textbf{ $p + q = 7$}}
    \item[ ] $\cl_{0,7} \simeq \cl_{0,2} \tensor \cl_{5,0} \simeq \hh \tensor \text{Mat}(2,\hsh) \simeq \text{Mat}(2,(\hh  \tensor \hh ) \oplus (\hh \tensor \hh)) \simeq \text{Mat}(2,\hh  \tensor \hh ) \oplus \text{Mat}(2,\hh  \tensor \hh ) \simeq \text{Mat}(8,\rr) \oplus \text{Mat}(8,\rr) \simeq \text{Mat}(8,\rsr)$
     \item[ ] $\cl_{7,0} \simeq \cl_{2,0} \tensor \cl_{0,5} \simeq \matreald \tensor \matcompq \simeq \text{Mat}(8,\cc) $
      \item[ ] $\cl_{1,6} \simeq \cl_{1,1} \tensor \cl_{0,5} \simeq \matreald \tensor \matcompq \simeq \text{Mat}(8,\cc) $
      \item[ ] $\cl_{6,1} \simeq \cl_{1,1} \tensor \cl_{5,0} \simeq \matreald \tensor \text{Mat}(2, \hsh) \simeq \text{Mat}(4,\hsh) $
        \item[ ] $\cl_{2,5} \simeq  \cl_{1,1} \tensor \cl_{1,4} \simeq \cl_{1,1} \tensor \cl_{1,1} \tensor \cl_{0,3} \simeq \text{Mat}(4,\rr) \tensor \hsh \simeq \text{Mat}(4, \hsh)$
       \item[ ] $\cl_{5,2} \simeq \cl_{1,1} \tensor \cl_{4,1} \simeq \cl_{1,1} \tensor \cl_{1,1} \tensor \cl_{3,0} \simeq \text{Mat}(4,\rr) \tensor \matcomp \simeq \text{Mat}(8, \cc)$
       \item[ ] $\cl_{3,4} \simeq \cl_{2,0} \tensor \cl_{4,1} \simeq \cl_{2,0} \tensor \cl_{1,1} \tensor \cl_{3,0} \simeq \text{Mat}(4,\rr) \tensor \matcomp \simeq \text{Mat}(8, \cc)$
       \item[ ] $\cl_{4,3} \simeq \cl_{2,0} \tensor \cl_{3,2} \simeq \cl_{2,0} \tensor \cl_{1,1} \tensor \cl_{1,1} \tensor \cl_{3,0} \simeq \text{Mat}(8,\rr) \tensor \rsr \simeq \text{Mat}(8, \rsr)$
\end{itemize}

Suppose that $p>q$ and take $p-q = 8k + r$ with $r<8$. One can use the relations shown in the Theorem \ref{teo_casperiod1} and Eq. \eqref{eq_acsclpqppqq} to obtain:

\begin{align}
    \begin{aligned}
    \cl_{p,q} &\simeq \cl_{q,q} \tensor \cl_{p-q,0} \\
    &\simeq \cl_{q,q} \tensor \cl_{8k + r,0} \\
    &\simeq \cl_{q,q} \tensor \cl_{(8(k-1) + r + 6) + 2,0}\\
     &\simeq \cl_{q,q} \tensor \cl_{2,0} \tensor \cl_{0,8(k-1) + r + 6}. 
    \end{aligned}
\end{align}

\noindent Recalling the Eq. \eqref{eq_r4p10}, a similar relation can be obtained as consequence of Corollary \ref{cor_casperiod1}, that is

\begin{align}
    \begin{aligned}
    \cl_{p,q+8k'} \simeq \cl_{0,8k'} \tensor \clpq
    \end{aligned}
\end{align}

\noindent for any $k'>0$, therefore, we have

\begin{equation}
    \cl_{0,8(k-1) + r + 6} \simeq \cl_{0,8(k-1)} \tensor \cl_{0,r + 6}.
\end{equation}

\noindent We also have that
\begin{align}
    \begin{aligned}
     \cl_{0,r + 6} &\simeq \cl_{0,2} \tensor  \cl_{2,0} \tensor \cl_{0,2} \tensor \cl_{r,0} \\
     &\simeq \cl_{2,2} \tensor \cl_{2,0} \tensor \cl_{r,0} 
    \end{aligned}
\end{align}

\noindent Hence,

\begin{align}
    \begin{aligned}
    \cl_{p,q} &\simeq \cl_{q,q} \tensor \cl_{2,0} \tensor \cl_{0,8(k-1) + r + 6}\\
    &\simeq  \cl_{q,q} \tensor \cl_{2,0} \tensor \cl_{0,8(k-1)} \tensor \cl_{2,2} \tensor \cl_{2,0} \tensor \cl_{r,0}\\
    &\simeq \text{Mat}(2^q, \rr) \tensor \matreald \tensor \text{Mat}(16^{k-1}, \rr) \tensor \text{Mat}(4,\rr) \tensor \text{Mat}(2,\rr) \tensor \cl_{r,0}\\
    &\simeq \text{Mat}(2^q, \rr) \tensor \matreald \tensor \text{Mat}(2^{4(k-1)}, \rr) \tensor \text{Mat}(2^2,\rr) \tensor \text{Mat}(2,\rr) \tensor \cl_{r,0}\\
     &\simeq \text{Mat}(2^{q+4k},\rr) \tensor \cl_{r,0}
    \end{aligned}
\end{align}

\noindent Now, if $q > p$ and by considering $q - p = 8k +r$, it yields \cite{Roc16}

\begin{align}
    \begin{aligned}
    \cl_{p,q} &\simeq \cl_{p,p} \tensor \cl_{0,q-p} 
    \simeq \cl_{p,p} \tensor \cl_{0, 8k + r} 
    \simeq \cl_{p,p} \tensor   \cl_{0, 8k} \cl_{0,r} \\
    &\simeq \text{Mat}(2^p, \rr) \tensor \text{Mat}(2^{4k},\rr) \tensor \cl_{0,r} 
    \simeq \text{Mat}(2^{p+4k}, \rr) \tensor \cl_{0,r}
    \end{aligned}
\end{align}

\noindent Therefore, the Clifford algebra is determined by $r = p - q \mod 8$. One may analyse the possibilities as follows. 

\begin{itemize}
  \item[\textcolor{gray}{{$\bullet$}}]\textcolor{gray}{\textbf{ $r = 0$}}
  \begin{itemize}
  \item [$\diamond $] $p \geq q \;\; (p - q \mod 8 = 0):$
\begin{equation*}
    \cl_{p,q} \simeq \text{Mat}(2^{q+4k}, \rr) \tensor \cl_{0,0} \simeq \text{Mat}(2^{q+4k},\rr)
\end{equation*}  
  
    \item [$\diamond $]  $p < q \;\; (p - q \mod 8 = 0):$
    
    \begin{equation*}
        \cl_{p,q} \simeq \text{Mat}(2^{p+4k}, \rr) \tensor \cl_{0,0} \simeq \text{Mat}(2^{p+4k},\rr)
    \end{equation*}
    \end{itemize}
\end{itemize}

\begin{itemize}
  \item[\textcolor{gray}{{$\bullet$}}]\textcolor{gray}{\textbf{ $r = 1$}}
    \begin{itemize}
  \item [$\diamond $] $p > q \;\; (p - q \mod 8 = 1):$
  \begin{align*}
      \begin{aligned}
          \cl_{p,q} &\simeq \text{Mat}(2^{q+4k}, \rr) \tensor \cl_{1,0}\\&\simeq \text{Mat}(2^{q+4k},\rr)  \tensor  (\rsr)  \simeq \text{Mat}(2^{q+4k},\rr)  \oplus \text{Mat}(2^{q+4k},\rr)
      \end{aligned}
  \end{align*}
    \item [$\diamond $]  $p < q \;\; (q - p \mod 8 = 1 \Rightarrow p - q \mod 8 = 7):$
        \begin{equation*}
        \cl_{p,q} \simeq \text{Mat}(2^{p+4k}, \rr) \tensor \cl_{0,1} \simeq \text{Mat}(2^{p+4k},\rr) \tensor \cc  \simeq \text{Mat}(2^{p+4k},\cc) 
    \end{equation*}
      \end{itemize}
\noindent In the above cases, $q + 4k = \left [ \frac{2q+8k + r}{2} \right ] = \left [\frac{p+q}{2} \right ] = \left [ \frac{n}{2} \right ]$, where $\left[s \right]$ denotes the integer part of $s$.

\end{itemize}
\begin{itemize}
\item[\textcolor{gray}{{$\bullet$}}]\textcolor{gray}{\textbf{ $r = 2$}}
      \begin{itemize}
  \item [$\diamond $] $p > q \;\; (p - q \mod 8 = 2):$
   \begin{equation*}
        \cl_{p,q} \simeq \text{Mat}(2^{q+4k}, \rr) \tensor \cl_{2,0} \simeq \text{Mat}(2^{q+4k},\rr) \tensor \matreald \simeq \text{Mat}(2^{q+4k+1},\rr)
    \end{equation*}
  
    \item [$\diamond $]  $p < q \;\; (p - q \mod 8 = 6):$
       \begin{equation*}
        \cl_{p,q} \simeq \text{Mat}(2^{p+4k}, \rr) \tensor \cl_{0,2} \simeq \text{Mat}(2^{p+4k},\rr) \tensor \hh  \simeq \text{Mat}(2^{p+4k},\hh) 
    \end{equation*}
    \end{itemize}
\end{itemize}
\begin{itemize}
 \item[\textcolor{gray}{{$\bullet$}}]\textcolor{gray}{\textbf{ $r = 3$}}
      \begin{itemize}
  \item [$\diamond $] $p > q \;\; (p - q \mod 8 = 3):$
     \begin{equation*}
        \cl_{p,q} \simeq \text{Mat}(2^{q+4k}, \rr) \tensor \cl_{3,0} \simeq \text{Mat}(2^{q+4k},\rr) \tensor \matcomp \simeq \text{Mat}(2^{q+4k+1},\cc)
    \end{equation*}
  
    \item [$\diamond $]  $p < q \;\; (p - q \mod 8 = 5):$
\begin{align*}
    \begin{aligned}
            \cl_{p,q} &\simeq \text{Mat}(2^{p+4k}, \rr) \tensor \cl_{0,3} \\&\simeq \text{Mat}(2^{p+4k},\rr)  \tensor  (\hsh)  \simeq \text{Mat}(2^{p+4k},\hh)  \oplus  \text{Mat}(2^{p+4k},\hh)     
    \end{aligned}
\end{align*}
    
    \end{itemize}\end{itemize}
\noindent In the cases where $r= 2,3$, if $p>q$ then $q+4k+1 = \left [ \frac{2q+8k+r}{2} \right] = \left [ \frac{n}{2} \right]$ and if $p<q$, $p+4k = \left [ \frac{2p+8k+r}{2} \right] - 1 = \left [ \frac{n}{2} \right] - 1$.

\begin{itemize}
  \item[\textcolor{gray}{{$\bullet$}}]\textcolor{gray}{\textbf{ $r = 4$}}
      \begin{itemize}
  \item [$\diamond $] $p > q \;\; (p - q \mod 8 = 4):$
     \begin{equation*}
        \cl_{p,q} \simeq \text{Mat}(2^{q+4k}, \rr) \tensor \cl_{4,0} \simeq \text{Mat}(2^{q+4k},\rr) \tensor \text{Mat}(2,\hh) \simeq \text{Mat}(2^{q+4k+1},\hh)
    \end{equation*}
    \item [$\diamond $]  $p < q \;\; (p - q \mod 8 = 4):$
     \begin{equation*}
        \cl_{p,q} \simeq \text{Mat}(2^{p+4k}, \rr) \tensor \cl_{0,4} \simeq \text{Mat}(2^{p+4k},\rr) \tensor \text{Mat}(2,\hh) \simeq \text{Mat}(2^{p+4k+1},\hh) 
    \end{equation*}
    \end{itemize}
\end{itemize}

\begin{itemize}
  \item[\textcolor{gray}{{$\bullet$}}]\textcolor{gray}{\textbf{ $r = 5$}}
      \begin{itemize}
  \item [$\diamond $] $p > q \;\; (p - q \mod 8 = 5):$

  \begin{align*}
      \begin{aligned}
      \cl_{p,q} &\simeq \text{Mat}(2^{q+4k}, \rr) \tensor \cl_{5,0} \simeq \text{Mat}(2^{q+4k},\rr) \tensor \text{Mat}(2,\rr)  \tensor (\hsh) \\ &\simeq \text{Mat}(2^{q+4k},\rr) \tensor (\text{Mat}(2,\hh) \oplus \text{Mat}(2,\hh)) \\
      &\simeq \text{Mat}(2^{q+4k+1},\hh)  \oplus \text{Mat}(2^{q+4k+1},\hh). 
      \end{aligned}
  \end{align*}

    \item [$\diamond $]  $p < q \;\; (p - q \mod 8 = 3):$
      \begin{equation*}
        \cl_{p,q} \simeq \text{Mat}(2^{p+4k}, \rr) \tensor \cl_{0,5} \simeq \text{Mat}(2^{p+4k},\rr) \tensor \text{Mat}(4,\cc) \simeq \text{Mat}(2^{p+4k+2},\cc) 
    \end{equation*}
    \end{itemize}\end{itemize}

    \noindent In the cases where $r=4,5$, if $p>q$ or $r = 4$ then $q+ 4k+1 = \left[\frac{n}{2} \right] -1$. If $p<q$ and $r=5$, then $p+4k+2 = \left[\frac{n}{2} \right].$

\begin{itemize}
 \item[\textcolor{gray}{{$\bullet$}}]\textcolor{gray}{\textbf{ $r = 6$}}

      \begin{itemize}
  \item [$\diamond $] $p > q \;\; (p - q \mod 8 = 6):$
    \begin{equation*}
        \cl_{p,q} \simeq \text{Mat}(2^{q+4k}, \rr) \tensor \cl_{6,0} \simeq \text{Mat}(2^{q+4k},\rr) \tensor \text{Mat}(4,\hh) \simeq \text{Mat}(2^{q+4k+2},\hh)
    \end{equation*}  
  
    \item [$\diamond $]  $p < q \;\; (p - q \mod 8 = 2):$
      \begin{equation*}
        \cl_{p,q} \simeq \text{Mat}(2^{p+4k}, \rr) \tensor \cl_{0,6} \simeq \text{Mat}(2^{p+4k},\rr) \tensor \text{Mat}(8,\rr) \simeq \text{Mat}(2^{p+4k+3},\cc) 
    \end{equation*}
    \end{itemize}
\end{itemize}

\begin{itemize}
\item[\textcolor{gray}{{$\bullet$}}]\textcolor{gray}{\textbf{ $r = 7$}}
      \begin{itemize}
  \item [$\diamond $] $p > q \;\; (p - q \mod 8 = 7):$
      \begin{equation*}
        \cl_{p,q} \simeq \text{Mat}(2^{q+4k}, \rr) \tensor \cl_{7,0} \simeq \text{Mat}(2^{q+4k},\rr) \tensor \text{Mat}(8,\cc) \simeq \text{Mat}(2^{q+4k+3},\cc)
    \end{equation*}  
    \item [$\diamond $]  $p < q \;\; (p - q \mod 8 = 1):$
\begin{align*}
    \begin{aligned}
      \cl_{p,q} &\simeq \text{Mat}(2^{p+4k}, \rr) \tensor \cl_{0,7} \simeq \text{Mat}(2^{p+4k},\rr) \tensor \text{Mat}(8,\rsr) \\ &\simeq \text{Mat}(2^{p+4k},\rr)  \tensor \text{Mat}(8,\rr) \tensor (\rsr)\\ &\simeq \text{Mat}(2^{p+4k+3},\rr) \oplus \text{Mat}(2^{p+4k+3},\rr) 
    \end{aligned}
\end{align*}
    \end{itemize}
\end{itemize}

\noindent In the cases where $r=6,7$, if $p<q$, therefore $p+4k+3 = \left[\frac{n}{2} \right]$; and if $p>q$ and $r=6$, hence $q + 4k +2 = \left [ \frac{n}{2} \right] -1$.

\section{Orthogonal Transformations}\label{app5}
{\colorlet{shadecolor}{gray!15}\begin{shaded} 
\begin{defin}\label{def_as54dfd3245}
Let $g$ be a symmetric bilinear form endowing the vector space $V$. A linear mapping $T: V \to V$ is said to be an \textbf{isometry} or an \textbf{orthogonal transformation} if for all $\mathbf{v},\mathbf{u} \in V$

\begin{equation}
    g(T(\mathbf{v}), T(\mathbf{u})) = g(\mathbf{v},\mathbf{u})
\end{equation}
\end{defin}\end{shaded}}

Consider a basis $\{\'e_{i}\}$ for the vector space $V$. Therefore, one can write $T(\'e_{i}) = T_{i}^{j} \'e_j$ and $g_{ij} = g(\'e_i, \'e_j)$. By virtue of the properties of bilinearity and linearity,

\begin{equation}
        T_{i}^{k} g_{kl} T_{j}^{l} = g_{ij} \;\;\longleftrightarrow\;\; T^{\intercal}GT = G
\end{equation}

\noindent Such that $G$ is the \textit{Gram matrix} of the basis $\{\'e_{i}\}$ with respect to $g$ \cite{Kos89} and $T$ denotes a matrix with entries $\{T_{i}^{j}\}$ and $T^{\intercal}$ its associated transposed matrix. Since for any matrices $A,B$ it follows that $\det(AB) = \det(A)\,\det(B)$ and $\det(A) = \det(A^{\intercal})$, as a result from the previous equations:

\begin{align}
\begin{aligned}\label{eq_grdet1}
\text{det}(T^{\intercal}GT) &= \text{det}(T)\\
\text{det}(T^{\intercal})\,\text{det}(G)\,\text{det}(T) &= \text{det}(G) \\
\text{det}(T)^{2} &= 1\\
\text{det}(T) &= \pm 1.
\end{aligned}
\end{align}

\noindent The orthogonal transformations such that $\det T = 1$ are called \textit{rotations} and those ones for which $\det T = -1$ are \textit{reflections}. Consider $V = \rr^{p,q}$ with $p+q = n$ and the general linear group $\text{GL}(n,\rr)$. The set of isometries forms a subgroup of $\text{GL}(n,\rr)$ called the orthogonal group and it is denoted by $\O(p,q).$ In particular, the rotations form a subgroup of $\O(p,q)$ called the special orthogonal group denoted by $\SO(p,q)$. 

\subsection{The Components of The Orthogonal Group}\label{sec_sd54fd}

\paragraph{ } Let $G$ be a group, a continuous mapping $\phi : [0,1] \to G$ is called a \textit{path}. A subset $G'$ of $G$ is said to be \textit{connected} if for any elements $g_0,g_1 \in G'$ there exists a path $\phi(t)$ linking these elements, i.e., $\phi(0) = g_0$ and $\phi(1) = g_1$. A \textit{component} of the group $G$ is a connected subset that is not contained in any other connected subset. \cite{Roc16}\\
\indent The orthogonal group $\O(p,q)$ has four components. In fact, let us consider an orthonormal basis $\{\'e_1,\ldots, \'e_{p}, \'e_{p+1}, \ldots, \'e_{p+q}\}$ of $\rr^{p,q}$, one can write the matrix $G$ that represents the symmetric bilinear form $g$ as

\begin{equation}
    G = \begin{pmatrix} 1_p & 0 \\ 0 & -1_q \end{pmatrix},
\end{equation}

\noindent such that $1_p$ and $1_q$ are the identity matrices of order $p$ and $q$, respectively.  The matrix that represents an arbitrary orthogonal transformation $T$ in $\rr^{p,q}$ can be written as

\begin{equation}
    T = \begin{pmatrix} A_p & B_{p,q} \\ C_{q,p} & D_q \end{pmatrix}.
\end{equation}

\noindent In that case, $A_p$ is a $p \times p$ matrix, $B_{p,q}$ a $p \times q$ matrix and so on. The orthogonality condition $T^{\intercal}GT = G$ establish that

\begin{align}
    \begin{aligned}
     \begin{pmatrix} A_p^{\intercal} &  C_{p,q}^{\intercal}  \\ B_{q,p}^{\intercal} & D_q^{\intercal} \end{pmatrix} \begin{pmatrix} 1_p & 0 \\ 0 & -1_q \end{pmatrix} \begin{pmatrix} A_p & B_{p,q} \\ C_{q,p} & D_q \end{pmatrix} = \begin{pmatrix} 1_p & 0 \\ 0 & -1_q \end{pmatrix}
    \end{aligned}
\end{align}

\noindent which implies that

\begin{align}\label{eq_acg007}
    \begin{aligned}
     A_p^{\intercal} A_p &-  C_{p,q}^{\intercal} C_{q,p} = 1_p, \\
     D_q^{\intercal} D_q &- B_{q,p}^{\intercal}B_{p,q} = 1_q,\\
     A_p^{\intercal}  A_p &= C_{p,q}^{\intercal}D_q.
    \end{aligned}
\end{align}

We claim that the matrices $A_p$ and $D_q$ satisfies $\det Ap \neq 0$ and $\det D_q \neq 0$. In fact, let us consider the case involving the matrix $A_p$. It follows from the Eq. \eqref{eq_acg007} that

\begin{equation}
    (\det A_p)^2 = \det (1_p +C_{p,q}^{\intercal} C_{q,p})
\end{equation}

\noindent Suppose, by contradiction, that $\det Ap = 0$. Therefore, since  $\det (1_p +C_{p,q}^{\intercal} C_{q,p})$, the equation $(1_p +C_{p,q}^{\intercal} C_{q,p})X$ has a non-trivial solution $X$ that can be settled as $X = -C_{p,q}^{\intercal} C_{q,p}X$, multiplying the both sides of this equation by $X^{\intercal}$ gives

\begin{equation}
    X^{\intercal}X = -    X^{\intercal}C_{p,q}^{\intercal} C_{q,p}X = -(C_{q,p}X)^{\intercal} (C_{q,p}X).
\end{equation}

\noindent We have that $    X^{\intercal}X = (X_1)^2 + \cdots + (X_n)^2 > 0$. On the other hand,  $-(C_{q,p}X)^{\intercal} (C_{q,p}X) \leq 0$, which is a contradiction. Hence $\det A_p$ must be non-zero. The $D_q$ case is precisely analogous. Moreover, it implies that

\begin{equation}
    (\det A_p)^2 \neq 0, \;\;\;\; (\det D_q)^2 \neq 0.
\end{equation}

   {\colorlet{shadecolor}{gray!15}\begin{shaded}
The four components of the orthogonal group $\O(p,q)$ are established:
\begin{align}\label{eq_lfhjknb654}
    \begin{aligned}
    &&\textcolor{gray}{\textbf{(i)}} \;\;\;\;\;\;\;\O^{\uparrow}_{+}(p,q) =\{T \in \O(p,q) \;:\; \det A_p > 0, \;\; \det D_q >0 \},\\
    &&\textcolor{gray}{\textbf{(ii)}} \;\;\;\;\;\;    \O^{\uparrow}_{-}(p,q) =\{T \in \O(p,q) \;:\; \det A_p > 0, \;\; \det D_q < 0 \},\\
   &&\textcolor{gray}{\textbf{(iii)}}\;\;\;  \;\;  \O^{\downarrow}_{+}(p,q) =\{T \in \O(p,q) \;:\; \det A_p < 0, \;\; \det D_q >0 \},\\
    &&\textcolor{gray}{\textbf{(iv)}}\;\;\;  \;\;  \O^{\downarrow}_{-}(p,q) =\{T \in \O(p,q) \;:\; \det A_p < 0, \;\; \det D_q < 0 \}.
    \end{aligned}
\end{align}\end{shaded}}

\noindent The subgroups of $\O(p,q)$ are the following sets \cite{Roc16}

\begin{align}\label{eq_as564d1f}
    \begin{aligned}
    &&\textcolor{gray}{\textbf{(i)}} \;\;\;\;\;\;\;&\O^{\uparrow}_{+}(p,q),\\
    &&\textcolor{gray}{\textbf{(ii)}} \;\;\;\;\;\;    &\O^{\uparrow}(p,q) =\O^{\uparrow}_{+}(p,q) \cup \O^{\uparrow}_{-}(p,q),\\
   &&\textcolor{gray}{\textbf{(iii)}}\;\;\;  \;\;  &\O_{+}(p,q) =\O^{\uparrow}_{+}(p,q) \cup \O^{\downarrow}_{+}(p,q),\\
    &&\textcolor{gray}{\textbf{(iv)}}\;\;\;  \;\; &\O^{\uparrow}_{+}(p,q) \cup \O^{\downarrow}_{-}(p,q).
    \end{aligned}
\end{align}

In addition, we can realise that

\begin{equation}\label{eqagcr511}
    \SO_{+}(p,q) = \SO^{\uparrow}(p,q)= \SO^{\uparrow}_{+}(p,q) = \O_{+}(p,q) \cup  \O^{\uparrow}(p,q).
\end{equation}

\begin{ex}
{{\textcolor{gray}{{$\blacktriangleright$\;}}}} The four connected components of the orthogonal group $O(1,1)$. \normalfont Consider an arbitrary matrix $T \in \O(1,1)$ given by

\begin{equation}
    T = \begin{pmatrix} a & b \\ c & d \end{pmatrix}
\end{equation}

\noindent The orthogonal condition $T^{\intercal}GT = G$ yields:

\begin{align}
    \begin{aligned}
    \begin{pmatrix} a & c \\ b & d \end{pmatrix}\begin{pmatrix} 1 & 0 \\ 0 & -1 \end{pmatrix} \begin{pmatrix} a & b \\ c & d \end{pmatrix} = \begin{pmatrix} 1 & 0 \\ 0 & -1 \end{pmatrix} \implies \color{gray}\begin{cases}\color{black} a^{2} - c^2 = 1,\\ \color{black} d^{2} - b^2 = 1, \\ \color{black} ab = cd. \end{cases}\color{black} 
    \end{aligned}
\end{align}

\noindent The two first equations hold for $a = \pm \cosh \a, c = \sinh \a, d= \pm \cosh \b, b = \sinh \b$, whereas the last one provides us four cases that are:

\begin{itemize}
    \item[\textcolor{gray}{\textit{ 1.}}] If $a = \cosh \a$ and $d = \cosh \b$ then $\a = \b$;
    \begin{equation}
        T_1 = \begin{pmatrix} \cosh \a & \sinh \a \\ \sinh \a & \cosh \a \end{pmatrix} \in  \O^{\uparrow}_{+}(1,1).
    \end{equation}
    \item[\textcolor{gray}{\textit{ 2.}}] If $a = -\cosh \a$ and $d = \cosh \b$ then $\a = -\b$;
      \begin{equation}
        T_2 = \begin{pmatrix} \cosh \a & -\sinh \a \\ \sinh \a & -\cosh \a \end{pmatrix} \in \O^{\uparrow}_{-}(1,1)
    \end{equation}
       \item[\textcolor{gray}{\textit{ 3.}}] If $a = \cosh \a$ and $d = -\cosh \b$ then $\a = -\b$.
         \begin{equation}
        T_3 = \begin{pmatrix} -\cosh \a & -\sinh \a \\ \sinh \a & \cosh \a \end{pmatrix} \in \O^{\downarrow}_{+}(1,1).
    \end{equation}
         \item[\textcolor{gray}{\textit{ 4.}}] If $a = -\cosh \a$ and $d = -\cosh \b$ then $\a = \b$.
           \begin{equation}
        T_4 = \begin{pmatrix} -\cosh \a & \sinh \a \\ \sinh \a & -\cosh \a \end{pmatrix} \in \O^{\downarrow}_{-}(1,1).
    \end{equation}
\end{itemize}

\noindent That way it is clear that there is no path that links any two matrices $T_i$ and $T_j$ for $i \neq j$.\;$\textcolor{gray}{{\blacktriangleleft}} $ \end{ex}

\section{$SU(2)$ and $SO(3)$}\label{app6}


{\colorlet{shadecolor}{gray!15}\begin{shaded}
\begin{propos}\label{propos_app6_inverse}
    The mapping $\mu: \rr^{3} \to \mathcal{H}$, $x = (x_1,x_2,x_3) \mapsto  x \cdot \sigma = x_1\sigma_{1} + x_2\sigma_{2} + x_3\sigma_{3}$ has inverse $\mu^{-1}: \mathcal{H} \to \rr^{3}$, $H \mapsto \frac{1}{2}\Tr(H\sigma^{i})\mathbf{e}_{i}$.
\end{propos}
\end{shaded}}
\textit{Proof.} \textcolor{gray}{\textbf{i)}} $\mu^{-1} (\mu (x)) = x$. In fact,

\begin{align}
\begin{aligned}
   \mu^{-1} (\mu (x)) &= \frac{1}{2} \Tr(\mu(x) \cdot \sigma^{i})\'e_i\\
&=   \frac{1}{2} \Tr(x^{j}\sigma_{j} \sigma^{i})\'e_i\\
&=   \frac{1}{2} \Tr(x^{j}\sigma_{j} \sigma_{k}g^{ki})\'e_i\\
&= \frac{1}{4} \Tr(x^{j}(\sigma_{j}\sigma_{k} + \sigma_{k}\sigma_{j}))g^{ki}\'e_i\\
&= \frac{1}{4} \Tr(x^{j}2g_{kj}\mathbb{1})g^{ki}\'e_i\\
&= \frac{1}{2} x^{j}\Tr(\mathbb{1})g_{kj}g^{ki}\'e_i\\
&= \delta_{j}^{i}x^{j}\'e_i\\
&= x^{i}\'e_i = x.
\end{aligned}
\end{align}

\noindent Where we use that: $\sigma_{j}\sigma_{k} + \sigma_{k}\sigma_{j} = 2g_{kj}\mathbb{1}$, $\sigma_{i} = g_{ij}\sigma^{j}$ and the cyclicality of the trace. Moreover:  \textcolor{gray}{\textbf{ii)}} $\mu (\mu^{-1} (H)) = H$. Similarly,

\begin{align}
    \begin{aligned}
\mu(\mu^{-1} (H)) &=  \mu \left(\frac{1}{2} \Tr(H\sigma^{i})  \'e_{i}\right)\\
&=\frac{1}{2}\Tr(h^{j}\sigma_{j}\sigma^{i}) \mu(\'e_{i})\\
&=\delta^{i}_{j} h^{j}\mu(\'e_{i})\\
&=h^{i}\sigma_{i} = H. \; \Box
    \end{aligned}
\end{align}

{\colorlet{shadecolor}{gray!15}\begin{shaded}
\begin{propos}\label{proposappp6_innerprodiso}
$(\rr^{3}, \langle \,\cdot \,,\, \cdot\, \rangle_{\rr^{3}}) \simeq (\mathcal{H}, \langle \,\cdot \,,\, \cdot\, \rangle_{\mathcal{H}})$ as inner product spaces.
\end{propos}
\end{shaded}}
\textit{Proof.} We claim that $\langle x ,y \rangle_{\rr^{3}} = \langle \mu(x),\mu(y) \rangle_{\mathcal{H}}$. In fact, using the same computations from Proposition \ref{propos_app6_inverse}, that is,

\begin{equation}
    \frac{1}{2}\Tr(\sigma_{i}\sigma_{j}) = \frac{1}{4}\Tr(\sigma_{i}\sigma_{j} + \sigma_{i}\sigma_{j}) = \frac{1}{4}\Tr(2g_{ij}\mathbb{1}) = g_{ij}\frac{1}{2}\Tr(\mathbb{1}) = g_{ij}.
\end{equation}

\noindent Consequently,
\begin{align}
    \begin{aligned}
         \langle \mu(x),\mu(y) \rangle_{\mathcal{H}} &= \frac{1}{2} \Tr(\mu(x)\mu(y))\\
         &=\frac{1}{2} \Tr(x^{i}\sigma_{i} y^{j}\sigma_{j})\\
         &=\frac{1}{2}x^{i} y^{j} \Tr(\sigma_{i} \sigma_{j})\\
         &= x^{i} y^{j} g_{ij} \\&= \langle x ,y \rangle_{\rr^{3}}. \; \Box
    \end{aligned}
\end{align}

{\colorlet{shadecolor}{gray!15}\begin{shaded}
\begin{propos}\label{propos_ik6gh5f}
The inner product $\langle \,\cdot \,,\, \cdot\, \rangle_{\mathcal{H}}$ on ${\mathcal{H}}$ is \normalfont $\text{SU}(2)$\textit{-invariant.} 
\end{propos}
\end{shaded}}
\textit{Proof.} In fact, take $U \in \text{SU}(2)$ and $X,Y \in \mathcal{H}$.

\begin{align}
    \begin{aligned}
        \langle  \text{Ad}_{U} X,\text{Ad}_{U} Y \rangle_{\mathcal{H}} &= \frac{1}{2}\Tr(\text{Ad}_{U} X\text{Ad}_{U} Y)\\
        &= \frac{1}{2}\Tr(UXU^{\dagger}UYU^{\dagger})\\
        &= \frac{1}{2}\Tr(UXYU^{\dagger})\\
         &= \frac{1}{2}\Tr(XYU^{\dagger}U)\\
           &= \frac{1}{2}\Tr(XY)\\
           &=  \langle X, Y \rangle_{\mathcal{H}}.\; \Box
    \end{aligned}
\end{align}

{\colorlet{shadecolor}{gray!15}\begin{shaded}
\begin{propos} \label{propos_hjgfbdh}
There is a Lie group homomorphism \normalfont $\rho: \text{SU}(2) \to \SO(3)$, $U \mapsto \rho(U): \rr^{3} \to \rr^{3} $ \textit{such that} $\rho(U):= \frac{1}{2}\Tr(U\sigma_{i}U^{\dagger} \sigma^{j})\mathbf{e}_{j}\tensor e^{i}$ \textit{with kernel} $\ker \rho = \zz_2$  
\end{propos}
\end{shaded}}
\textit{Proof.} First note that since  $\rr^{3} \tensor (\rr^{3})^{*} \simeq \text{Hom}(\rr^{3},\rr^{3})$ and since the coefficients of $\rho(U)$ are linear,  $\rho(U) \in \text{GL}(3,\rr)$. We claim that $\rho(U) \in \SO(3)$. Therefore, we need to show that 

\begin{equation}
    \langle x ,y \rangle_{\rr^{3}} = \langle \rho(U)x ,\rho(U)y \rangle_{\rr^{3}}
\end{equation}

\noindent and $\det \rho(U) = +1$. First, it follows from Proposition \ref{proposappp6_innerprodiso} that

\begin{align}
    \begin{aligned}
         \langle \rho(U)x ,\rho(U)y \rangle_{\rr^{3}} =  \langle \mu(\rho(U)x) ,\mu(\rho(U)y) \rangle_{\mathcal{H}}.
    \end{aligned}
\end{align}

\noindent One may notice that

\begin{equation}\label{eq_36f14sdag}
    \rho(U) = \frac{1}{2}\Tr(U\sigma_{i}U^{\dagger} \sigma^{j})\mathbf{e}_{j}\tensor e^{i} = \mu^{-1}(U\sigma_{i}U^{\dagger})\tensor e^{i}.
\end{equation}

\noindent Therefore,

\begin{align}
    \begin{aligned}
        \langle \rho(U)x ,\rho(U)y \rangle_{\rr^{3}} &=  \langle \mu(\rho(U)x) ,\mu(\rho(U)y) \rangle_{\mathcal{H}}\\
        &= \langle \mu(\mu^{-1}(U\sigma_{i}U^{\dagger})\tensor e^{i}(x)),\mu(\mu^{-1}(U\sigma_{i}U^{\dagger})\tensor e^{j}(y)) \rangle_{\mathcal{H}}\\
        &=  \langle U\sigma_{i}U^{\dagger}x^{i},U\sigma_{i}U^{\dagger}y^{j} \rangle_{\mathcal{H}}\\
        &=\frac{1}{2}\Tr(U\sigma_{i}U^{\dagger}x^{i}U\sigma_{i}U^{\dagger}y^{j})\\
        &= \frac{1}{2}x^{i}y^{j}\Tr(U\sigma_{i}U^{\dagger}U\sigma_{i}U^{\dagger})\\
        &=x^{i}y^{j} \frac{1}{2}\Tr(\sigma_{i}\sigma_{j})\\
        &= x^{i}y^{j}g_{ij}\\
        &=\langle x ,y \rangle_{\rr^{3}}.
    \end{aligned}
\end{align}

\noindent Hence, $\rho(U) \in \text{O}(3).$ We proceed to show that $\det \rho(U) = +1$. However, notice that since 
every element $U \in \text{SU}(2)$ can be written as $U = \exp(i\phi\cdot\sigma/2)$, $U$ can be continuously deformed to the identity by making $\phi$ smaller. Since determinant is a continuous function as well as $\rho$ is a continuous function of $U$, $\det \rho(U)$ cannot jump discontinuously for small changes of $U$. Once $\det \rho(U) = \pm 1$ and $\det \mathbb{1}_{3\times3} = +1$, we must have $\det \rho(U) = + 1$ and therefore $\rho(U) \in \SO(3)$. If,

\begin{align}
    \begin{aligned}
        \mu(y) &= \text{Ad}_{U_{1}} \mu(x),\\
        \mu(z) &= \text{Ad}_{U_{2}} \mu(y).
    \end{aligned}
\end{align}

\noindent Then by applying $\mu^{-1}$,

\begin{align}
\begin{aligned}
\mu^{-1}(\text{Ad}_{U_{1}} \mu(x))&= \mu^{-1}(\mu(y))\\
\mu^{-1}(U_{1}x\cdot \sigma U^{\dagger}) &= y\\
\mu^{-1}(U_{1}\sigma_{i} U^{\dagger}) \tensor e^{i}(x) &= y \implies \rho(U_{1})x  = y.    
\end{aligned}
\end{align}

\noindent The implication follows from \eqref{eq_36f14sdag}. The same result holds for $\mu(z)$. Therefore, with $ \rho(U_{1})x  = y$ and $ \rho(U_{2})y  = z$, then $\rho(U_{2})\rho(U_{1})x = z$. However,

\begin{align}
    \begin{aligned}
z &= \rho(U_{2})y \\&= \mu^{-1} (U_{2}\sigma_{i}U_{2}^{\dagger})\tensor e^{i}(y) \\ &= \mu^{-1}(U_{2}y\cdot \sigma U_{2}^{\dagger}) \\
&= \mu^{-1}(U_{2}\mu(y) U_{2}^{\dagger})\\
&= \mu^{-1}(U_{2}\text{Ad}_{U_{1}}\mu(x)U_{2}^{\dagger})\\
&= \mu^{-1}(U_{2} U_{1}\mu(x)U_{1}^{\dagger}U_{2}^{\dagger})\\
&= \mu^{-1}(U_{2} U_{1}x \cdot \sigma (U_{2}U_{1})^{\dagger})\\
&= \mu^{-1}(U_{2} U_{1}\sigma_{i}(U_{2}U_{1})^{\dagger}) \tensor e^{i}(x) \\
&= \rho(U_{2}U_{1})x.
    \end{aligned}
\end{align}

\noindent Consequently, 

\begin{equation}
    \rho(U_{2})\rho(U_{1})x = \rho(U_{2}U_{1})x. 
\end{equation}

Since it holds for arbitrary vectors, we conclude that for any $U_{1},U_{2} \in \text{SU}(2)$:

\begin{equation}
    \rho(U_{2})\rho(U_{1}) = \rho(U_{2}U_{1}). 
\end{equation}

which shows that the group multiplication is preserved, making $\rho: \text{SU}(2) \to \text{SO}(3)$ into a Lie group homomorphism. The kernel

\begin{equation}
    \ker \rho = \{U \in \text{SU}(2) : \frac{1}{2}\Tr(U\sigma_{i}U^{\dagger}\sigma^{j}) = \delta_{i}^{j}\}
\end{equation}

\noindent is such that for $U \in \text{SU}(2)$

\begin{align}
    \begin{aligned}
        \rho(U)x &= x\\
        \mu^{-1}(U\sigma_{i}U^{\dagger}) \tensor e^{i}(x) &= x\\
        \mu^{-1}(Ux^{i}\sigma_{i}U^{\dagger}) &= x\\
        \mu^{-1}(U\mu(x)U^{\dagger}) &= x\\
    \end{aligned}
\end{align}

Therefore, $(U\mu(x)U^{\dagger}) = \mu(x)$ holds for the identity $2 \times 2$ up to a signal. That is, 

\begin{equation}
    \ker \rho = \{\pm \mathbb{1}\} \simeq \mathbb{Z}_{2}. \; \Box
\end{equation}

\section{Lorentz Group and Representations}\label{app62}

\paragraph{ } The Lorentz group determines how fields transform under spacetime rotations and boosts. Representations labelled as '$(j_{1},j_{2})$'
specify the spin properties of these fields, with $j_{1}, j_{2}$ corresponding to the two $\textbf{su}(2)$ components of the Lorentz group's complexified form $\textbf{so}(1,3;\cc) \simeq \textbf{su}(2) \times \textbf{su}(2)$. These labels define the transformation behaviour and spin characteristics of the fields. This was first introduced in Chapter \ref{chap_spinors} and oftentimes mentioned in \ref{chap_bil}. To explain where those labels come from, we start by presenting the Lorentz group composed by the Lorentz transformations.

\subsection{Lorentz Group}

In general, a \textit{Lorentz transformation} is a real linear transformation that occurs in spacetime that preserves the fundamental quadratic form $g_{\mu\nu}x^{\mu}x^{\nu}$ and does not mix past and future \cite{LorentzG}. The behavior of a vector $x \in \rr^{1,3}$ under a Lorentz transformation can be represented by:

\begin{equation}
    x' = \Lambda x  \;\;\;\; \text{or}   \;\;\;\; x^{\mu'} = \Lambda ^{\mu}_{\nu}x^{\nu},
\end{equation}

\noindent where $\Lambda ^{\mu}_{\nu}$ is the real elements of a $4 \times 4$ matrix representing a Lorentz transformation. Assuming that the form $g_{\mu\nu}x^{\mu}x^{\nu}$ is invariant under this transformation, it follows that

\begin{align}
\begin{aligned}\label{eq_lorentzmatt00}
g_{\alpha \beta}x^{\alpha'}x^{\beta'} &= g_{\mu\nu}x^{\mu}x^{\nu} \\
g_{\alpha \beta}\Lambda ^{\alpha}_{\mu}x_{\mu} \Lambda ^{\beta}_{\nu}x_{\nu} &= g_{\mu\nu}x^{\mu}x^{\nu} \\
g_{\alpha \beta}\Lambda ^{\alpha}_{\mu} \Lambda ^{\beta}_{\nu}x_{\mu}x_{\nu} &= g_{\mu\nu}x^{\mu}x^{\nu} \\
g_{\alpha \beta}\Lambda ^{\alpha}_{\mu} \Lambda ^{\beta}_{\nu}&= g_{\mu\nu},\\
\end{aligned}
\end{align}

\noindent In matrix form:

\begin{equation}\label{eq_lorentzmat111sim}
    \Lambda^{\intercal}g\Lambda= g
\end{equation}

\noindent This last equation \eqref{eq_lorentzmat111sim} shows us concisely what a matrix representing a Lorentz transformation must satisfy obeying the imposed symmetry. Some information can be taken from this relation. Starting from the Eq. \eqref{eq_lorentzmat111sim} and the fact that $g^2 = I$ one has to multiply the right and left by $g$

\begin{align}
\begin{aligned}\label{eq_lorentzmat1343}
  g\Lambda^{\intercal}g\Lambda &= gg = I, \\
  \Lambda^{\intercal}g\Lambda g &= gg = I.
\end{aligned}
\end{align}

\noindent Such that $(g\Lambda^{\intercal}) = (g\Lambda)^{-1}$. It follows that

\begin{align}
\begin{aligned}\label{eq_lorentzmat1233}
  \Lambda^{\intercal}g\Lambda  &= \Lambda g\Lambda^{\intercal} = I.
\end{aligned}
\end{align}

\noindent In components,

\begin{equation}\label{eq_hntgbrnvw}
 g^{\alpha \beta}\Lambda _{\alpha}^{\mu} \Lambda _{\beta}^{\nu}= g^{\mu\nu}
\end{equation}

\noindent In addition, again from the equation \eqref{eq_lorentzmat111sim}, one has

\begin{align}
\begin{aligned}\label{eq_lorentzmat2222}
\text{det}(\Lambda) &= \pm 1.
\end{aligned}
\end{align}

\noindent The matrices that represent the Lorentz transformations are unimodular. The set of matrices $4 \times 4$ with real entries, equipped with the usual matrix product satisfying
$\Lambda^{\intercal}g\Lambda = g$ with $g =$ diag$(+1,-1,-1,-1)$ forms a group, such group is precisely the group $\O(1,3)$ called the Lorentz group and describes the isometries on the Minkowski space-time. The most common form of the transformation $(t,x,y,z) \to (t',x',y',z') \in \rr^{1,3}$, in terms of the real constant $v$ representing a velocity confined to the $x$-direction, is expressed as \cite{LorentzG}

\begin{align}
\begin{aligned}\label{eq_grupoincxlinhatlinha}
t' &= \gamma \left ( t - \frac{vx}{c^2} \right ),\\
x' &= \gamma (x - vt), \\
y' &= y,\\
z' &= z.
\end{aligned}
\end{align}

\noindent where $c$ is the speed of light and $\gamma$ is the \textit{Lorentz factor} given by

\begin{equation}
     \gamma = \frac{1}{\sqrt{1 - \frac{v^2}{c^2}}}.
\end{equation}

The four connected components of the Lorentz group are categorised by:

\begin{align}
    \begin{aligned}
        &\O^{\uparrow}_{+}(1,3), \;\;\; \textcolor{gray}{\textbf{(proper orthochronous)}};\\
        &\O^{\uparrow}_{-}(1,3), \;\;\,\textcolor{gray}{\textbf{(non-proper orthochronous)}}; \\
        &\O^{\downarrow}_{+}(1,3), \;\;\;\textcolor{gray}{\textbf{(proper non-orthochronous)}};\\
        &\O^{\downarrow}_{-}(1,3), \;\;\,\textcolor{gray}{\textbf{(non-proper non-orthochronous)}}.
    \end{aligned}
\end{align}

\noindent An important subgroup of the Lorentz group is $\SO_{+}(1,3) \simeq \SO^{\uparrow}(1,3) \simeq \SO^{\uparrow}_{+}(1,3)$. Although the components of the Lorentz group are not subgroups they can be described from $\SO_{+}(1,3)$. In fact, it can be shown that considering discrete transformations:

\begin{itemize}
 \item[\textcolor{gray}{\textbf{(i)}}] \textit{parity inversion}, $x \to -x$: represented by the matrix $P =$diag$(1,-1,-1,-1)$.
 \item[\textcolor{gray}{\textbf{(ii)}}] \textit{time inversion}, $t \to -t$: represented by the matrix $T =$ diag$(-1,1,1,1)$
 \item[\textcolor{gray}{\textbf{(iii)}}] \textit{space-time inversion}: represented by the matrix $PT =$ diag $(-1,-1,-1,-1) = -I$
\end{itemize}

\noindent Consequently,
\begin{equation}
 P\text{O}^{\uparrow}_{+}(1,3) = \O^{\uparrow}_{-}(1,3), \;\;\;\; T\text{O}^{\uparrow}_{+}(1,3) = \O^{\downarrow}_{-}(1,3), \;\;\;\; PT\text{O}^{\uparrow}_{+}(1,3) = \O^{\downarrow}_{-}(1,3).
\end{equation}

\noindent In this way, the transformations $P,T,PT \in \SO_{+}(1,3)$ communicate the various components of the Lorentz Group, that is, the different sectors of $\O(1,3)$ are connected by Lorentz transformations of the subgroup $\SO_{+}(1,3)$. In addition to rotations, the parity inversion and the time inversion, there is another Lorentz transformation called \textit{boost}. Boosts are transformations that relate the coordinates of one inertial frame to another moving at a constant velocity relative to the first. They specifically change the time and spatial coordinates without involving rotations, effectively mixing space and time coordinates in a manner consistent with special relativity. Boosts are a subset of Lorentz transformations that preserve the spacetime interval between events. Boosts and rotations, which are continuously connected to the identity, are proper and orthochronous. Lorentz boosts do not constitute a group on their own because two consecutive boosts in different directions result in a combination of a boost and a rotation. Therefore, boosts and rotations must be combined to form a complete group. In contrast, rotations do form a group independently. As we will see in the next subsection of this Appendix, \textit{boosts} and \textit{rotations} can be formed by consecutive infinitesimal transformations around the identity \textit{I}, that is, they are \textit{‘continuously connected’} to the identity. On the other hand, the parity inversion $P$ and the time inversion $T$, being a discrete transformation, they are \textit{‘disconnected’} from the identity. Any product of boosts, rotation, $T$, and $P$ belongs to the Lorentz group. We write symbolically, \cite{LorentzG}

{\colorlet{shadecolor}{gray!15}\begin{shaded}\centering Lorentz Group = boosts + rotations +  parity inversion +  time inversion.
\end{shaded}}

\subsection{Infinitesimal Transformations and Generators of the Lorentz Group}
\paragraph{ } Let us have a look on infinitesimal transformations and the generators of the Lorentz group. This subsection will give us the insight about the $\textbf{su}(2)$ algebra sitting inside $\textbf{so}(1,3)$. In this sense, we focus on the proper and orthochronous Lorentz group, we recall that any other elements of the Lorentz group can be obtained by multiplying $P$, $T$ and $PT$ to the elements of this group. We start by considering Lorentz transformations infinitesimally close to the identity:

\begin{equation}
    \Lambda _{\nu}^{\mu} =  g^{\mu}_{\nu} + \omega_{\nu}^{\mu}.
\end{equation}

\noindent In this case, $\omega_{\nu}^{\mu}$ is a set of small real numbers. By considering the group condition in the Eq. \eqref{eq_hntgbrnvw} and keeping the terms related to the first order in $\omega$, it follows that is $\omega_{\nu\mu}$ is antisymmetric (when the indices are both subscript or
both superscript). In fact,

\begin{align}
    \begin{aligned}
        g_{\alpha \beta} &= \Lambda _{\nu \alpha} \Lambda^{\mu}_{\beta}\\
        &= (g_{\nu\alpha} + \omega_{\nu \alpha})(g^{\nu}_{\beta} + \omega^{\nu}_{\beta})\\
        &= g_{\nu\alpha}g^{\nu}_{\beta} + \omega_{\nu\alpha}g^{\nu}_{\beta} + g_{\nu \alpha} \omega_{\beta}^{\nu} +\cancel{\omega_{\nu\alpha}\omega^{\nu\beta}}\\
        &= g_{\alpha \beta} + \omega_{\beta \alpha} + \omega_{\alpha \beta}.
    \end{aligned}
\end{align}

\noindent That is,  

\begin{equation}
    \omega_{\beta \alpha} = -\omega_{\alpha \beta}.
\end{equation}

\noindent The following matrix

\begin{equation}
  \begin{pmatrix}
0 & \omega_{01} & \omega_{02}  & \omega_{03} \\ 
-\omega_{01} & 0 & \omega_{12} & \omega_{13} \\ 
-\omega_{02} & -\omega_{12}  & 0 & \omega_{23} \\ 
-\omega_{03} &  -\omega_{13} & -\omega_{23} & 0
\end{pmatrix}
\end{equation}

\noindent can be parametrised by using $6$ antisymmetric matrices \cite{LorentzG}

\begin{align}
    \begin{aligned}
            [\omega_{\alpha \beta}] &= \omega_{01}M^{01}_{\alpha \beta} +  \omega_{02}M^{02}_{\alpha \beta} + \omega_{03}M^{03}_{\alpha \beta} + \omega_{12}M^{12}_{\alpha \beta} + \omega_{13}M^{13}_{\alpha \beta} + \omega_{23}M^{23}_{\alpha \beta}\\
            &= \sum_{\mu < \nu}\omega_{\mu\nu}M^{\mu\nu}_{\alpha \beta}.
    \end{aligned}
\end{align}

\noindent with

\begin{align}
    \begin{aligned}
       M^{01}_{\alpha \beta} =  \begin{pmatrix}
0 & 1 & 0 & 0\\ 
-1 & 0 & 0 & 0\\ 
0 & 0 & 0 & 0\\ 
0 & 0 & 0 & 0
\end{pmatrix}, \;\;\;        M^{02}_{\alpha \beta} = \begin{pmatrix}
0 & 0 & 1 & 0\\ 
0 & 0 & 0 & 0\\ 
-1 & 0 & 0 & 0\\ 
0 & 0 & 0 & 0
\end{pmatrix},\\
       M^{03}_{\alpha \beta} = \begin{pmatrix}
0 & 0 & 0 & 1\\ 
0 & 0 & 0 & 0\\ 
0 & 0 & 0 & 0\\ 
-1 & 0 & 0 & 0
\end{pmatrix}, \;\;\;        M^{23}_{\alpha \beta}= \begin{pmatrix}
0 & 0 & 0 & 0\\ 
0 & 0 & 0 & 0\\ 
0 & 0 & 0 & 1\\ 
0 & 0 & -1 & 0
\end{pmatrix},\\
       M^{13}_{\alpha \beta} = \begin{pmatrix}
0 & 0 & 0 & 0\\ 
0 & 0 & 0 & 1\\ 
0 & 0 & 0 & 0\\ 
0 & -1 & 0 & 0
\end{pmatrix}, \;\;\; 
       M^{12}_{\alpha \beta} = \begin{pmatrix}
0 & 0 & 0 & 0\\ 
0 & 0 & 1 & 0\\ 
0 & -1 & 0 & 0\\ 
0 & 0 & 0 & 0
\end{pmatrix}.
    \end{aligned}
\end{align}

\noindent The elements $M^{\mu\nu}_{\alpha \beta}$ can be written in a coherent manner in terms of the metric as follows

\begin{equation}
    M^{\mu\nu}_{\alpha \beta} = g_{\mu \alpha} g^{\nu}_{\beta} - g^{\mu}_{\beta} g^{\nu}_{\alpha}
\end{equation}

\noindent and satisfy $M^{\mu\nu}_{\alpha \beta} = -M^{\nu\mu}_{\alpha \beta}$. One has, 

\begin{equation}
    \omega_{\alpha\beta} = \sum_{\mu < \nu}\omega_{\mu\nu}M^{\mu\nu}_{\alpha \beta} =  \sum_{\mu > \nu}\omega_{\mu\nu}M^{\mu\nu}_{\alpha \beta} =  \frac{1}{2}\omega_{\mu\nu}M^{\mu\nu}_{\alpha \beta}.
\end{equation}

That is, 

\begin{align}
    \begin{aligned}
        \Lambda^{\alpha}_{\beta} &= g^{\alpha \beta} +  \frac{1}{2}\omega_{\mu\nu}M^{\mu\nu\alpha}_{\beta}.\\
        \Lambda &= I + \frac{1}{2}\omega_{\mu\nu}M^{\mu\nu}
    \end{aligned}
\end{align}

\noindent Therefore, by suppressing the matrix indices there are six independent antisymmetric generators $M^{\mu\nu}$. By writing them alternatively for the ones with two space indices 

\begin{equation}
    J_{i} = \frac{1}{2}\epsilon_{ijk}M_{jk}
\end{equation}

\noindent and

\begin{equation}
    K_{i} = M_{0i}
\end{equation}

\noindent for those with one space and one time index the six different Lorentz transformations generators naturally decompose into three rotations $J_i$ generators and three boosts $K_{i}$ generators as follows

\begin{align}
    \begin{aligned}
        &J_1 = \begin{pmatrix}
0 & 0 & 0 & 0 \\
0 & 0 & 0 & 0 \\
0 & 0 & 0 & 1 \\
0 & 0 & 1 & 0
\end{pmatrix},\;\;\; J_2 = \begin{pmatrix}
0 & 0 & 0 & 0 \\
0 & 0 & 0 & 1 \\
0 & 0 & 0 & 0 \\
0 & 1 & 0 & 0
\end{pmatrix},\;\;\; J_3 = \begin{pmatrix}
0 & 0 & 0 & 0 \\
0 & 0 & 1 & 0 \\
0 & 1 & 0 & 0 \\
0 & 0 & 0 & 0
\end{pmatrix},\\
&K_1 = \begin{pmatrix}
0 & 1 & 0 & 0 \\
1 & 0 & 0 & 0 \\
0 & 0 & 0 & 0 \\
0 & 0 & 0 & 0
\end{pmatrix},\;\;\;K_2 = \begin{pmatrix}
0 & 0 & 1 & 0 \\
0 & 0 & 0 & 0 \\
1 & 0 & 0 & 0 \\
0 & 0 & 0 & 0
\end{pmatrix},\;\;\;K_3 = \begin{pmatrix}
0 & 0 & 0 & 1 \\
0 & 0 & 0 & 0 \\
0 & 0 & 0 & 0 \\
1 & 0 & 0 & 0
\end{pmatrix}.
    \end{aligned}
\end{align}

The rotation matrices are hermitian whereas the boost matrices are anti hermitian. The generators obey commutation relation

\begin{align}
    \begin{aligned}
    [J_i,J_j] &= i\epsilon_{ijk}J_k,\\
    [J_i, K_j] &= -i\epsilon_{ijk}K_{k},\\
        [K_i, K_j] &= -i\epsilon_{ijk}J_{k}.
    \end{aligned}
\end{align}

By taking the exponential of the generators the transformation matrices for the Lorentz transformations can be obtained. For a rotation around the axis \(\hat{n}\) by an angle \(\theta\):
    $R(\theta, \hat{n}) = \exp(i \theta \hat{n} \cdot \vec{\sigma} / 2)$ one take $R_{i}(\theta)= \exp(\theta J_{i})$. For a boost along the axis \(\hat{n}\) with a boost parameter \(\eta\):
  $  B(\eta, \hat{n}) = \exp(\eta \hat{n} \cdot \vec{K}) = \exp(\eta \hat{n} \cdot \frac{i \vec{\sigma}}{2})$ one take $B_{i}(\eta)= \exp(\eta K_{i})$. Therefore, the four-by-four Lorentz transformations representation is given by:

\begin{align}
    \begin{aligned}\label{eq_mjnhtr45ertygd}
        &R_x(\theta) = \begin{pmatrix}
1 & 0 & 0 & 0 \\
0 & \cos\theta & -\sin\theta & 0 \\
0 & \sin\theta & \cos\theta & 0 \\
0 & 0 & 0 & 1
\end{pmatrix}, \quad B_x(\eta) = \begin{pmatrix}
\cosh\eta & \sinh\eta & 0 & 0 \\
\sinh\eta & \cosh\eta & 0 & 0 \\
0 & 0 & 1 & 0 \\
0 & 0 & 0 & 1
\end{pmatrix},\\
        &R_y(\theta) = \begin{pmatrix}
\cos\theta & 0 & \sin\theta & 0 \\
0 & 1 & 0 & 0 \\
-\sin\theta & 0 & \cos\theta & 0 \\
0 & 0 & 0 & 1
\end{pmatrix}, \quad B_y(\eta) = \begin{pmatrix}
\cosh\eta & 0 & \sinh\eta & 0 \\
0 & 1 & 0 & 0 \\
\sinh\eta & 0 & \cosh\eta & 0 \\
0 & 0 & 0 & 1
\end{pmatrix},\\
        &R_z(\theta) = \begin{pmatrix}
\cos\theta & -\sin\theta & 0 & 0 \\
\sin\theta & \cos\theta & 0 & 0 \\
0 & 0 & 1 & 0 \\
0 & 0 & 0 & 1
\end{pmatrix}, \quad B_z(\eta) = \begin{pmatrix}
\cosh\eta & 0 & 0 & \sinh\eta \\
0 & 1 & 0 & 0 \\
0 & 0 & 1 & 0 \\
\sinh\eta & 0 & 0 & \cosh\eta
\end{pmatrix}. 
    \end{aligned}
\end{align}

\noindent Related to the fact that $\SO(3) \simeq \text{SU}(2)/ \mathbb{Z}_{2}$, the rotations form an $\textbf{su}(2)$-subalgebra. However, by combining the generators one can find two mutually commuting $\textbf{su}(2)$ algebras sitting inside $\textbf{so}(1,3)$ by considering the linear combinations \cite{LorentzG}

\begin{align}
    \begin{aligned}
        A_{i} = \frac{1}{2}(J_{i} + iK_{i}) \text{ and } B_{i} = \frac{1}{2}(J_{i} - iK_{i}) 
    \end{aligned}
\end{align}

\noindent Now, both of these are hermitian and obey

\begin{align}
    \begin{aligned}
    [A_i,A_j] &= i\epsilon_{ijk}A_k,\\
    [B_i, B_j] &= i\epsilon_{ijk}B_{k},\\
        [A_i, B_j] &= 0.
    \end{aligned}
\end{align}

These commutation relations indicate that the Lie algebra for the Lorentz group has two commuting subalgebras, the algebra generated by $A_{i}$ or $B_{i}$ is the $3$D rotation algebra, namely, $\textbf{so}(3)\simeq \textbf{su}(2)$. Therefore, this leads to

\begin{equation}
    \textbf{so}(1,3) = \textbf{su}(2) \oplus \textbf{su}(2).
\end{equation}

That is, representations of $\textbf{su}(2) \oplus \textbf{su}(2)$ determine representations of the Lorentz group. Particles in physics are classified by their mass, electric charge and another fundamental property called spin, which is either an integer or a half-integer. For instance, for spin $0$ one has the Higgs boson, for spin $\frac{1}{2}$ one has quarks, eletrons and neutrinos; photons and gluons are examples of spin $1$ particles. Spin $0$ particles are represented by scalars, spin $\frac{1}{2}$ particles are represented by spinors and spin $1$ particles are represented by vectors or realised as '\textit{spinors $\tensor$ spinors}'. On the other hand, each irreducible representation of $\text{SU}(2)$ is labelled by a quantum number $j$ (where $j =0, \frac{1}{2}, 1, \frac{3}{2},..$) acting on a vector space with with dimension $(2j +1)$ \cite{Zee:2003mt}. When considering two copies of $\textbf{su}(2)$ algebra, $\textbf{su}(2) \times \textbf{su}(2)$, the representations are labeled by pairs of half-integers $(j_1,j_2)$, hence, the irreducible representations of $\text{SU}(2) \times \text{SU}(2)$ have dimensions $(2j_{1}+1,2j_{2}+1)$.  Each component of $(j_1,j_2)$ represents a spin value for each $\textbf{su}(2)$ factor. For instance,

{\colorlet{shadecolor}{gray!15}\begin{shaded}\begin{itemize}\centering    \item[] $(0, 0)$ : scalar
\item[] $( \frac{1}{2} , 0)$ : left-handed Weyl spinor\item[] $(0, \frac{1}{2} ) $: right-handed Weyl spinor\item[] $(\frac{1}{2}  , \frac{1}{2} )$ : vector\item[] $(1, 0)$ : self-dual 2-form\item[] $(0, 1)$ : anti-self-dual 2-form\end{itemize}   \end{shaded}}

It is important to know how a particle will behave when subjected to a physical transformation, like a rotation or a boost. In this sense, since the groups $\SO^{+}(1,3)$ and $\SO(3)$ act on $3$D and $4$D vectors, this matrices are part of the Spin-$1$ representation of $3$D rotations and $4$D Lorentz transformations. With respect to another particles, the group $\text{SU}(2) \simeq \Spin(3)$ acts on the Pauli spinor and the group $\text{SL}(2,\cc) \simeq \Spin_{+}(1,3)$ acts on the Weyl spinor and they are are said to be Spin$\frac{1}{2}$-representation of $3$D rotations and $4$D Lorentz transformations. The Spin$\frac{1}{2}$ representation generated by the Pauli matrices are particular interesting and fundamental since any other spin $j$ representation can be constructed from them. The role of  $\text{SL}(2,\cc)$ is introduced in the Example \ref{ex_re2654h56984ytrg} therefore, as we know from Chapter \ref{chap_spinors}, the group $\text{SL}(2,\cc) \simeq \Spin_{+}(1,3)$ is a double cover for $\SO_{+}(1,3)$ thus all finite-dimensional irreducible representations of $\SO_{+}(1,3)$ descend from irreducible finite-dimensional representations of $\text{SL}(2,\cc)$.  By comparing Eq. \eqref{eq_mjnhtr45ertygd} and Eq. \eqref{eq_gnfkjbdn6788} the transformations of the Spin-${\frac{1}{2}}$-representation are given by

\begin{table}[H]\centering
\begin{tabular}{
>{\columncolor{gray!30}}c |cc}
              & \multicolumn{1}{c|}{\cellcolor{gray!30}Two-by-Two} & \cellcolor{gray!30}Four-by-Four \\ \hline
$R_x(\theta)$ & $\begin{pmatrix}
\cos\left(\frac{\theta}{2}\right) & -i \sin\left(\frac{\theta}{2}\right) \\
-i \sin\left(\frac{\theta}{2}\right) & \cos\left(\frac{\theta}{2}\right)
\end{pmatrix}$                                                       & $ \begin{pmatrix}
1 & 0 & 0 & 0 \\
0 & \cos\theta & -\sin\theta & 0 \\
0 & \sin\theta & \cos\theta & 0 \\
0 & 0 & 0 & 1
\end{pmatrix}$                                  \\ \cline{2-3} 
$R_y(\theta)$ & $ \begin{pmatrix}
\cos\left(\frac{\theta}{2}\right) & -\sin\left(\frac{\theta}{2}\right) \\
\sin\left(\frac{\theta}{2}\right) & \cos\left(\frac{\theta}{2}\right)
\end{pmatrix}$                                                       & $\begin{pmatrix}
\cos\theta & 0 & \sin\theta & 0 \\
0 & 1 & 0 & 0 \\
-\sin\theta & 0 & \cos\theta & 0 \\
0 & 0 & 0 & 1
\end{pmatrix}$                                    \\ \cline{2-3} 
$R_z(\theta)$ & $ \begin{pmatrix}
\text{exp}\left(\frac{-i\theta}{2}\right) & 0 \\
0 & \text{exp}\left(\frac{i\theta}{2}\right)  
\end{pmatrix}$                                                       & $\begin{pmatrix}
\cos\theta & -\sin\theta & 0 & 0 \\
\sin\theta & \cos\theta & 0 & 0 \\
0 & 0 & 1 & 0 \\
0 & 0 & 0 & 1
\end{pmatrix}$                                    \\ \cline{2-3} 
$B_x(\eta)$   & $ \begin{pmatrix}
\cosh\left(\frac{\eta}{2}\right) & -\sinh\left(\frac{\eta}{2}\right) \\
-\sinh\left(\frac{\eta}{2}\right) & \cosh\left(\frac{\eta}{2}\right)
\end{pmatrix}$                                                      & $\begin{pmatrix}
\cosh\eta & \sinh\eta & 0 & 0 \\
\sinh\eta & \cosh\eta & 0 & 0 \\
0 & 0 & 1 & 0 \\
0 & 0 & 0 & 1
\end{pmatrix}$                                    \\ \cline{2-3} 
$B_y(\eta)$   & $\begin{pmatrix}
\cosh\left(\frac{\eta}{2}\right) & i \sinh\left(\frac{\eta}{2}\right) \\
-i \sinh\left(\frac{\eta}{2}\right) & \cosh\left(\frac{\eta}{2}\right)
\end{pmatrix}$                                                      & $\begin{pmatrix}
\cosh\eta & 0 & \sinh\eta & 0 \\
0 & 1 & 0 & 0 \\
\sinh\eta & 0 & \cosh\eta & 0 \\
0 & 0 & 0 & 1
\end{pmatrix}$                                    \\ \cline{2-3} 
$B_z(\eta)$   & $\begin{pmatrix}
\text{exp}\left(\frac{-\eta}{2}\right)  & 0 \\
0 & \text{exp}\left(\frac{\eta}{2}\right)
\end{pmatrix}$                                                       & $\begin{pmatrix}
\cosh\eta & 0 & 0 & \sinh\eta \\
0 & 1 & 0 & 0 \\
0 & 0 & 1 & 0 \\
\sinh\eta & 0 & 0 & \cosh\eta
\end{pmatrix}$                                   
\end{tabular}\caption{Two-by-two and four-by-four representations of the Lorentz group.}
\end{table}

\section{Simple Algebras and Representation}\label{app7}
\paragraph{ } In this Appendix, we discuss how to explicitly obtain a representation of a Clifford algebra. First, we present some concepts involving algebra in general that aid us in the discussion of Clifford algebras representations itself \cite{Roc16}.

\paragraph{ } Let $\A$ be an algebra where there is involved naturally an underlying vector space structure. Consider the set of endomorphisms $\endo (\A)$. 

\begin{defincinza}\label{defin_lkjhgcve}
   The \textbf{regular representation} is defined by $L : \A \to \endo (\A)$ as

\begin{equation}
    L(a)b = ab, \;\;\; \forall\, b \in \A.
\end{equation}
\end{defincinza} 
\noindent It is straightforward to see that $L(1) = 1$ and $L(ab) = L(a)L(b)$ thus $L$ is indeed a representation.

{\colorlet{shadecolor}{gray!15}\begin{shaded} 
\begin{propos}
A regular representation is faithful.
\end{propos}\end{shaded}}
\proof Indeed, $L(a)c = L(b)c \implies L(a-b)c = (a-b)c = 0$. By taking $c = 1$ it follows that $a = b$, i.e., $\ker L = {0}$. $\Box$ 

\begin{ex}
{{\textcolor{gray}{{$\blacktriangleright$\;}}}} Regular and a non-regular representation for $\mathbb{C}$. \normalfont Consider the representation $\rho$ and $\bar{\rho}$ shown in the Example \ref{ex_reprec1}. For any $z_1 = a + ib ,z_2 = c + id  \in \cc$ it follows that 

\begin{align}
    \begin{aligned}
    \rho(z_1 z_2) &= \rho ((a + ib)(c + id)) 
    = \rho ((ac - bd) + i(bc + ad)) = (ac - bd) + i(bc + ad) \\
    &= (a + ib)(c + id) = \rho((a + ib)(c +id) = \rho(z_1)z_2;\\
    \\
    \bar{\rho}(z_1 z_2) &= \bar{\rho} ((a + ib)(c + id)) 
    = \bar{\rho} ((ac - bd) + i(bc + ad)) = (ac - bd) - i(ad + bc) \\
    &\neq \bar{\rho}(z_1) z_2 = \bar{\rho} (a + ib)(c + id) = (a - ib)(c + id) = (ac + bd) + i(ad - bc).
    \end{aligned}
\end{align}

\noindent Therefore, we conclude that $\rho$ is regular and $\bar{\rho}$ is not regular. \,$\textcolor{gray}{{\blacktriangleleft}}$  \end{ex}

\paragraph{ } Suppose there exists invariant subspaces $\B_1$ and $\B_2$ of $\A$ with respect to the regular representation. This means that $L(a)(\B_1) \subset \B_1$ and $L(a)(\B_2) \subset \B_2,$ $\forall \, a \in \A$. Hence it is possible to express $L = L_1 \oplus L_2$, where $L_1 \A \to \endo(\B_1)$. Moreover, if there exists other invariant subspaces with respect to $L_1$ and $L_2$ the same procedure is employed up a space $S$ and a representation $\mathcal{L} : \A \to \endo(S)$ are obtained, such that the unique invariant subspaces are $S$ and $\{0\}$. We denominate such representation $\mathcal{L}$ an \textit{irreducible representation}. The subspace $S$ is a \textit{minimal left ideal} with respect to the algebra $\A$ since $\mathcal{L}(a)(S) \subset S, \forall\, a \in  \A$, i.e, $ \forall \, x \in S, a x \in S$ and $S$ does not contain any non-trivial sub-ideal. 

 {\colorlet{shadecolor}{gray!15}\begin{shaded}
To summarise, the representation space associated to an irreducible regular representation is a left minimal ideal of the algebra.
\end{shaded}}

\begin{defincinza}
Let $\A$ be an algebra. An element $f \in \A$ is an \textbf{idempotent} if $f^2 = f$.
\end{defincinza}
\begin{defincinza}
An algebra $\A$ is called \textbf{division algebra} if every element different of zero has
a inverse, equivalently, if $ab = 0 $ or $ba = 0$, $\forall \, b \in \A$, then $a = 0$.
\end{defincinza}

\noindent If the algebra $\A$ is a division algebra, then the unique idempotent is the identity $1$. Indeed, for a division algebra: $f^2 = f \implies f = f^{1}f^2 = f^{-1}f = 1$. However, most of Clifford algebras are not division algebras. Besides, we also have that:

\begin{itemize}
    \item[\textcolor{gray}{{$\diamond$}}] Two idempotents $f_1, f_2$ are said to be \textit{orthogonal} if $f_1 f_2 = f_2 f_1 = 0.$ 
    \item[\textcolor{gray}{{$\diamond$}}] An idempotent $f$ is said to be \textit{primitive} if it can not be written as the sum of other two orthogonal idempotents, i.e., $f \neq f_1 + f_2$ where $(f_1)^2 = f_1, (f_2)^2 = f_2$ and $f_1f_2 = f_2f_1 = 0.$
\end{itemize}

\paragraph{ } Inspecting the Clifford algebra classification table, one can realise that every Clifford algebra can be expressed either by $\mathbb{K} \tensor \text{Mat}(N,\rr)$ or by $[\kk \tensor \text{Mat}(N,\rr)] \oplus [\kk \tensor \text{Mat}(N,\rr)],$ for $\kk = \rr, \cc,\hh$, which are division algebras, and some $N=0,1,2,\ldots$.

\begin{defincinza}
An algebra $\A$ is said to be \textbf{simple} if the unique two-sided ideals of $\A$ are $\A$ and $\{0\}$.
\end{defincinza}

   {\colorlet{shadecolor}{gray!15}\begin{shaded}
\begin{propos}\label{propos_matsimplealgebra}
Every algebra $\A = \kk \tensor \text{Mat}(N,\rr)$ is a simple algebra.
\end{propos}\end{shaded}}
\proof Consider the basis for $\text{Mat}(N,\rr)$ given by the set of the $N^2$ matrices $E_{A B}\; (A,B = 1,\ldots, N)$ defined in the following way: all the matrix entries $E_{A B}$ equal 0 except the entry that corresponds to the $A^{th}$ row and the $B^{th}$ column, which equals 1. In other words, using the Kronecker delta, the matrices $\{E_{A B}\}$ are defined by $(E_{A B})_{C D} = \delta_{A C}\delta_{B D}$ satisfying $E_{A B}E_{C D}= \delta_{BC}E_{AD} \; (A,B,C,D = 1, \ldots, N)$. Consider $I \subset \A$ a two-sided ideal and let $x \in I$, this way, $x$ can be written as $x = \sum_{AB} x_{AB}E_{A B}$, with $x_{AB} \in \kk$. If $x \neq 0$ then at least one of the components $x_{AB}$ is non-null. It follows that

\begin{align}
    \begin{aligned}
    \sum_{C} E_{CA} x E_{BC} &= \sum_{CDE} x_{DE}E_{CA}E_{DE}E_{BC} \\
    &=\sum_{CDE} x_{DE}\delta_{AD}E_{CE}E_{BC} \\
    &=\sum_{CDE} x_{DE}\delta_{AD}\delta_{BE}E_{CC} \\
    &= x_{AB} \sum_{C} E_{CC} = x_{AB}1,
    \end{aligned}
\end{align}

\noindent where $ \sum_{C} E_{CC} = 1$ is the identity matrix. Since $x_{AB} \in \kk$ and $\kk$ is an associative division algebra, there exists $x_{AB}^{-1} \in \kk$, therefore it follows from the above expression that

\begin{equation}
     \sum_{C} x_{AB}^{-1} E_{CA} x E_{BC} = 1.
\end{equation}

\noindent It implies that $1 \in I$. Therefore, $I = \A$. Otherwise, if $x = 0$ then $I = \{0\}$ and the proof is concluded. $\Box$

{\colorlet{shadecolor}{gray!15}\begin{shaded} One can conclude that: every Clifford algebra is either a simple algebra or the direct sum of simple algebras.
\end{shaded}}

\begin{defincinza}
An algebra that is the sum of simple algebras is said to be a \textbf{semisimple} algebra.
\end{defincinza}

\subsubsection{Simple Algebras Representation}

\paragraph{ } In order to develop Clifford algebra representation method, since every Clifford algebra is either a simple algebra or the direct sum of them, we now present the representation of simple algebras. Consider a set of $N$ primitive idempotents $\{f_1, \ldots, f_N \}$ mutually orthogonal in a simple algebra $\A$, that is \cite{Roc16}

\begin{equation}
    f_{A}f_{B} = \delta_{AB} f_{A}.
\end{equation}

\noindent Consider now the set $\A_{AB}$ defined by $\A_{AB} = f_{A} \A f_B$ \; $(A,B = 1, \ldots, N).$ For those sets:

\begin{equation}
    \A_{AB}\A_{CD} = f_A \A f_B f_C \A f_D = \delta_{BC} f_A \A f_B \A f_D. 
\end{equation}

\noindent Meanwhile, the set $\A f_B \A $ is a two-sided ideal of $\A$ and since the algebra is simple and $f_B \neq 0$, then $\A f_B \A = \A$. Therefore,

\begin{equation}\label{eq_algsimpidemp}
    \A_{AB}\A_{CD} = \delta_{BC} f_A \A f_D = \A_{AD}. 
\end{equation}

\noindent On the other hand, for every $A = 1, \ldots, N$ there exists idempotents $f_A \in \A_{AA}$. For instance, take $f_1 \in A_{11}$. Due to Eq. \eqref{eq_algsimpidemp} it follows for any value of $A$ that $f_1 \in \A_{11} = \A_{1A}\A_{A1}$. Hence there exists $\E_{1A} \in \A_{1A}$ and $\E_{A1} \in \A_{A1}$ such that $f_1 = \E_{1A}\E_{A1}$

\begin{equation}
   \color{gray}
      \underbrace{\color{black}\A_{11}}_{\color{black} f_1}  \color{black}  = \color{gray}
      \underbrace{\color{black}\A_{1A}}_{\color{black} \E_{1A}}  \color{black}  \color{gray}
      \underbrace{\color{black}\A_{A1}}_{\color{black} \E_{A1}} \color{black}.
\end{equation}

\noindent Now choose $\E_{1A}$ and $\E_{A1}$ such that

\begin{equation}
    f_B \E_{A1} = \delta_{AB} \E_{A1}, \;\;\;\;\; \E_{1A}f_B  = \delta_{AB}\E_{1A}.
\end{equation}

\noindent Let us define the quantities $\E_{AB}$ as

\begin{equation}\label{eq_1827fhjs}
    \E_{AB} = \E_{A1} \E_{1B},
\end{equation}

\noindent it yields

\begin{equation}\label{eq_qw928983}
    f_C \E_{AB} = \delta_{AC}\E_{AB}, \;\;\;\;\; \E_{AB}f_C = \delta_{BC}\E_{AB}.
\end{equation}

\noindent From the above definition it reads that

\begin{align}\label{eq_as239284555}
    \begin{aligned}
    \E_{AB}\E_{CD} &= \E_{A1}\E_{1B}\E_{C1}\E_{D1}\\
    &= \delta_{BC}\E_{A1}\E_{1B}\E_{B1}\E_{D1}\\
    &= \delta_{BC}\E_{A1}f_1\E_{D1}\\
    &= \delta_{BC}\E_{A1}\E_{D1}\\
      &= \delta_{BC}\E_{AD}
    \end{aligned}
\end{align}

\noindent As the set of the $N^2$ matrices $E_{A B}\; (A,B = 1,\ldots, N)$ presented in the proof of Proposition \ref{propos_matsimplealgebra}, the above Eq. \eqref{eq_as239284555} shows that the quantities $\{\E_{AB}\} \; (A,B = 1,\ldots, N)$ is also a basis for the space of $N \cross N$ matrices. In addition it shows that $\E_{AA}$ is an idempotent. Since

\begin{equation}
    1_{\A}\E_{BC} = \sum_{A} \E_{AA} \E_{BC} = \sum_{A} \delta_{AB}\E_{AC} = \E_{BC}
\end{equation}

\noindent then the identity $1_{\A}$ is given by

\begin{equation}
     1_{\A} = \sum_{A} \E_{AA}
\end{equation}

\paragraph{ } The set $\A_{AA} = f_A \A f_A$ is an ideal of $\A$, for any $x \in \A_{AA}$ it follows that $x f_A = x = f_A x$, this means that the idempotent $f_A$ is the identity of $\A_{AA}$.

 {\colorlet{shadecolor}{gray!15}\begin{shaded} 
\begin{propos}
The idempotent $f_A$ is the unique idempotent of $\mathcal{A}_{AA}$ if and only if $f_A$ is primitive.
\end{propos}\end{shaded}}
\proof Suppose that $f_A$ is not primitive. Then, one can write $f_A = g_A + h_A$ where $g_A, h_A$ are orthogonal idempotents, $g_A h_A = h_A g_A = 0.$ Therefore,

\begin{align}
\begin{aligned}
f_A g_A f_A = (g_A + h_A)g_A (g_A + h_A) = g_A, \\
f_A h_A f_A = (g_A + h_A)h_A(g_A + h_A) = h_A,
\end{aligned}
\end{align}

\noindent which means that $g_A, h_A \in \A_{AA}$. Therefore $f_A$ is not the unique idempotent of $\A_{AA}$. Now, suppose that $f_A$ is not the unique idempotent of $\A_{AA}$, then there exists another idempotent $g_A \in \A_{AA}$. It follows that $f_A - g_A$ is another idempotent. Indeed, since $g_A \in \A_{AA}$ it follows that $f_A g_A = g_A f_A = g_A$. Consequently,

\begin{align}
    \begin{aligned}
    (f_A - g_A)(f_A - g_A) = f_A - f_Ag_A - g_A f_A + g_A = f_A - g_A.
    \end{aligned}
\end{align}

\noindent We also have that the idempotents $f_A - g_A)$ and $g_A$ are orthogonal, 

\begin{align}
    \begin{aligned}
    (f_A - g_A)g_A = f_Ag_A - g_Ag_A = g_A -g_A = 0,
    g_A(f_A - g_A) = g_Af_A - g_Ag_A = g_A -g_A = 0.
    \end{aligned}
\end{align}

\noindent Therefore, the idempotent $f_A$ is not primitive. $\Box$

\paragraph{ } Continuing in this way, the idempotent $\E_{AA} \in \A_{AA} = f_A \A f_A$ and $f_A$ is primitive, as a result:

\begin{equation}
    \E_{AA} = f_A.
\end{equation}

   {\colorlet{shadecolor}{gray!15}\begin{shaded}
\begin{propos}
$\A = f_A \A f_A $ is a division algebra.
\end{propos}\end{shaded}}

\proof Let $I_{A} = \A f_A$ be a left ideal of $\A$. Since $f_A$ is primitive, this ideal is minimal. Let $J_A$ be non-null left ideal of $\A_{AA}$. Therefore:

\begin{equation}
    J_A \subset \A_{AA} \implies \A J_A  \subset \A \A_{AA} = \A f_A \A f_A = \A f_A I_A \subset I_A
\end{equation}

\noindent However, since $I_A$ is minimal and  $J_A$ is non-null, we conclude that  $\A J_A = I_A$. On the other hand,

\begin{equation}
    \A_{AA} = f_A \A f_A = f_A I_A = f_A \A J_A \subset J_A
\end{equation}

\noindent  and one has $\A_{AA} = J_A$. Hence, the unique left ideal of $\A_{AA}$ are either $\A_{AA}$ itself or $\{0\}$. Now, consider a non-null element $z \in \A_{AA}$. The set $\A_{AA}z$ is a left ideal of $\A_{AA}$ but by above result, since $\A_{AA}$ does not contain non-trivial subideals and $z \neq 0$ it follows that $\A_{AA} = \A_{AA}z$. Therefore, it means that there exists a non-null $z' \in \A_{AA}$ such that $z z' = f_A$, such that $f_A$ has been shown to be the identity of $\A_{AA}$. By the same way one can show that $z'z = f_A$, which shows that $\A_{AA}$ is a \textit{division algebra with identity} $f_A$. $\Box$

 {\colorlet{shadecolor}{gray!15}\begin{shaded} 
\begin{propos}
$\A_{AA} \simeq \A_{BB}\; (A,B = 1,\ldots, N). $
\end{propos}\end{shaded}}

\proof Let $x_A \in \A_{AA}, x_B \in \A_{BB}$. We define the linear map $\phi_{AB} : \A_{AA} \to \A_{BB}$ by setting

\begin{equation}\label{eq1aejo83}
   \phi_{AB} (x_A) = \E_{BA} x_A \E_{AB}. 
\end{equation}

\noindent Since $\E_{AA} = f_A$ is the identity, by Eq. \eqref{eq_as239284555} it follows that $\phi_{AB}$ is an algebra homomorphism. Indeed,

\begin{align}
    \begin{aligned}
    \phi_{AB}(x_A y_A) &= \E_{BA} x_A y_A \E_{AB} \\
    &=\E_{BA} x_A f_A y_A \E_{AB}\\
    &=\E_{BA} x_A \E_{AA} y_A \E_{AB}\\
    &=\E_{BA} x_A \E_{AB}\E_{BA} y_A \E_{AB}\\
    &=\phi_{AB} (x_A) \phi_{AB} (y_A).
    \end{aligned}
\end{align}

\noindent Moreover, we also have that $\phi_{AB}^{-1} = \phi_{BA}$. Indeed,

\begin{align}
    \begin{aligned}
    \phi_{BA} (\phi_{AB} (x_A)) &= \phi_{BA} (\E_{BA} x_A \E_{AB})\\
    &=\E_{AB}\E_{BA} x_A \E_{AB}\E_{BA}\\
    &= f_A x_A f_A \\
    &= x_A,
    \end{aligned}
\end{align}

\noindent which proves that $\phi_{AB}$ is an algebra isomorphism and all the division algebras $\A_{AA}\; (A= 1,\ldots,N) $ isomorphic. $\Box$

\paragraph{ } It is possible to write $x \in \mathbb{K}$ in terms of $x_1 \in \A_{11}$ and $x_A = \E_{A1}x_1\E_{1A}$ as \cite{Roc16}

\begin{equation}\label{eq1982894lk}
    x = \sum_{A} x_A = \sum_A \E_{A1}x_1\E_{1A}.
\end{equation}

\noindent Such expression defines a linear map $\varphi : \A_{11} \to \kk$ and its inverse $\varphi^{-1} : \kk \to \A_{11}$

\begin{align}
    \begin{aligned}
    \varphi(x_1) &= \sum_A \E_{A1}x_1\E_{1A} = x \in \kk\\
    \varphi^{-1}(x) &= \E_{11}x\E_{11} = x_1 \in \A_{11}
    \end{aligned}
\end{align}

\noindent Hence it implies that $\kk \simeq \A_{11} \simeq \A_{AA}\; (A = 1,\ldots,N).$ Besides, $x \in \kk$ commutes with any matrix in the set generated by $\{\E_{AB} \}$. Indeed,

\begin{align}
    \begin{aligned}
    x\E_{AB} &= \sum_C \E_{C1}x_1\E_{1C} \E_{AB}\\
    &= \sum_C \delta_{CA} \E_{C1}x_1\E_{1C} \E_{1B}\\
    &= \E_{A1}x_1\E_{1B}\\
    &= \E_{AB}\E{B1}x_1\E_{1B}\\
    &= \sum_C \E_{AB}\E{C1}x_1\E_{1C}\\
    &= \E_{AB}x
    \end{aligned}
\end{align}

Consider now $x \in \A$, since

\begin{equation}
    1_A = \sum_A f_A  = \sum_A \E_{AA},
\end{equation}

\noindent then it is possible to write

\begin{equation}\label{eq_932479387}
    x =  \sum_A \E_{AA} x  \sum_B \E_{BB} =  \sum_A \E_{AA} x \E_{BA}\E_{AB} = \sum_{ABC} \E_{CA} x \E_{BC}\E_{AB}.
\end{equation}

\noindent Right, defining $(x_{C})_{AB}$ as

\begin{equation}
    (x_{C})_{AB} = \E_{CA}x\E_{BC} \in \A_{CC}
\end{equation}

\noindent it follows from Eq. \eqref{eq_932479387} that 

\begin{equation}\label{eq_abc2397}
    x = \sum_{ABC}  (x_{C})_{AB}  \E_{AB}.
\end{equation}

\noindent In addition, one can notice that
\begin{align}
    \begin{aligned}
    (x_D)_{AB}  &= \E_{DC} x \E_{BD} \\
    &= \E_{DC}\E_{CA} x \E_{BC} \E_{CD}\\
    &= \E_{DC}(x_C)_{AB} \E_{CD},
    \end{aligned}
\end{align}

\noindent which agrees with Eq. \eqref{eq1aejo83} as we expected.  Continuing in this way, according to Eq.   \eqref{eq1982894lk} we define

\begin{equation}\label{eqx10239029}
    x_{AB} = \sum_C (x_C)_{AB}
    = \sum_C \E_{CA}x\E_{BC}.
\end{equation}

\noindent Therefore, it follows from Eq. \eqref{eq_abc2397}

\begin{align}\label{eq_8271872f}
    \begin{aligned}
        x &= \sum_{ABC}  (x_{C})_{AB}  \E_{AB} \\
&= \sum_{ABC} \E_{CA}x\E_{BC}\E_{AB} \\
&= \sum_{AB} x_{AB} \E_{AB}
    \end{aligned}
\end{align}

\noindent where $x_{AB} \in \kk$ and $\{ \E_{AB} \}$ is a basis for the space of $N \cross N$ matrices. This result shows that if $\A$ is a simple algebra, it can be written as $\kk \tensor \text{Mat}(N,\rr)$. Moreover, it means that there is a representation $\rho$ given by

\begin{equation}
    \rho (x) =  \begin{pmatrix} 
    x_{11} & x_{12} & \cdots & x_{1N} \\
      x_{21} & x_{22} & \cdots & x_{2N}\\
    \vdots & \vdots & \ddots & \vdots\\
    x_{N1} & x_{N2} & \cdots & x_{NN}
    \end{pmatrix}
\end{equation}

\noindent It is worth mentioning that the expression \eqref{eq_8271872f} is not unique. Indeed, any invertible element $u \in \A$ can be used to define another basis of $ \text{Mat}(N,\rr)$ by
\begin{equation}
    \E'_{AB} = u \E_{AB}u^{-1},
\end{equation}
\noindent which implies

\begin{equation}
    x = \sum_{AB} x'_{AB}\E'_{AB}.
\end{equation}

Due our choice of the idempotent $f_1$ when the quantities $\E_{AB} = \E_{A1} \E_{1B} \; (A,B = 1, \ldots, N)$ were defined via Eq. \eqref{eq_1827fhjs}, the representation space that has been constructed is the minimal left ideal $\A_{f_1}$ with basis $\{\E_{A1}\}$ \cite{Roc16}. Indeed, any $x \in \A$ can be written as

\begin{align}
    \begin{aligned}
    xf_1 &= \sum_{AB}x_{AB} \E_{AB}\E_{11} = \sum_{A} x_{A1}\E_{A1}.
    \end{aligned}
\end{align}

\noindent Since $\A \simeq \kk \tensor \text{Mat}(N,\rr)$ then $\A f_1 \simeq \kk \tensor \rr^{N}$, which such isomorphism shall be explored in the chapter concerning the spinors.

\subsection{Clifford Algebras Representation}

\paragraph{ } As a result of all the previous studies, there is a method to obtain matrix representation of Clifford algebras \cite{Roc16}

{\colorlet{shadecolor}{gray!15}\begin{shaded}
\begin{itemize}
    \item [\textcolor{gray}{\textbf{1)}}] Choose a set of $N$ primitive idempotents $f_A \;(A = 1,\ldots, N)$ of $\clpq$ such that $\sum_A f_A  = 1$ and among them a primitive idempotent, for instance $f_1$.
    \item [\textcolor{gray}{\textbf{2)}}] Choose elements $\{\E_{A1}\}$ and $\{\E_{1A}\} \;(A = 1,\ldots, N)$ such that $f_1 =  \E_{1A}\E_{A1}$ and Eq. \eqref{eq_qw928983} holds, namely, $f_B\E_{A1} = \delta_{AB}\E_{A1}$ and $\E_{1A}f_B = \delta_{AB}\E_{1A}.$
\end{itemize}
\end{shaded}}

The above steps can be interchanged by another equivalent ones:

 {\colorlet{shadecolor}{gray!15}\begin{shaded}
\begin{itemize}
    \item [\textcolor{gray}{\textbf{1)}}] Choose a primitive idempotent $f_1$.
    \item [\textcolor{gray}{\textbf{2)}}] Find a basis of the ideal $\clpq f_1$, denoted by $\{\E_{A1}\}$, and the associated dual basis $\{\E_{1A}\} $, which satisfies $  \E_{1A}\E_{B1}= \delta_{AB}f_1$.
\end{itemize}
\end{shaded}}

Whatever the above possibility, the third step is:

 {\colorlet{shadecolor}{gray!15}\begin{shaded}
\begin{itemize}
    \item [\textcolor{gray}{\textbf{3)}}] Define a basis to $\text{Mat}(N,\rr)$ as $\E_{AB} = \E_{A1}\E_{1B}$. If $\{\gamma_i = \gamma(\textbf{e}_{i}) \} \; (i = 1,\ldots, n)$ are the generators of $\clpq \; (p+q = n)$, its matrix representation is given by Eq. \eqref{eqx10239029}:
    \begin{equation}
        (\gamma_i)_{AB} = \sum_C \E_{CA}\gamma_i \E_{BC}.
    \end{equation}
\end{itemize}
\end{shaded}}

\noindent The scalars are isomorphic to the set $f_1 \clpq f_1$ where $f_1$ is the identity and the representation space is isomorphic to the minimal left ideal $\clpq f_1$

\begin{ex}
{{\textcolor{gray}{{$\blacktriangleright$\;}}}} $\cl_{2,0}$. \normalfont Consider the quadratic space $\rr^{2,0}$ with orthonormal basis $\{\'e_1,\'e_2\}$ and the associated Clifford algebra $\cl_{2,0}$. One has that $(\'e_1)^2 = (\'e_2)^2 = 1$ and $\'e_1\'e_2 + \'e_2\'e_1 = 0$. One also has:

\begin{align}
    \begin{aligned}
     f_{1} = \frac{1}{2}(1 + \'e_1),  &&f_{2} = \frac{1}{2}(1 - \'e_1)
    \end{aligned}
\end{align}

\noindent are primitive idempotents of $\cl_{2,0}$ satisfying

\begin{equation}
    f_{1} + f_{2}  =  \frac{1}{2}(1 + \'e_1) +  \frac{1}{2}(1 - \'e_1) = 1.
\end{equation}

\noindent Other idempotents are 

\begin{align}
    \begin{aligned}
     g_{1} = \frac{1}{2}(1 + \'e_2),
      &&g_{2} = \frac{1}{2}(1 - \'e_2)
    \end{aligned}
\end{align}

\noindent as well as suitable linear combinations of $f_{1}, f_{2}, g_{1}, g_{2}$. Let us follow the presented steps, therefore, let us choose $f_1$ as the first step stated. For the next step, one has to choose elements $\{\E_{A1}\}$ e $\{\E_{1A}\}$ $(A = 1,2)$ such that

\begin{align}
    \begin{aligned}
    f_1 = \E_{1A}\E_{A1}
    \end{aligned}
\end{align}

\noindent and
\begin{align}
    \begin{aligned}
    f_B\E_{A1} = \delta_{AB}\E_{A1}, \;\;&&\E_{1A}f_B = \delta_{AB}\E_{1A} \;\;\text{for} \; A,B  = 1,2.
    \end{aligned}
\end{align}

\noindent Since $\{\E_{A1}\}, \{\E_{1A}\} \in \cl_{2,0}$, these elements are written in the form

\begin{equation}
    \phi = a  + b\'e_1 + c\'e_2 + d\'e_1\'e_2.
\end{equation}

\noindent Now is required to compute $f_1 \phi, \phi f_1,f_2 \phi, \phi f_2$. By noticing that $f_1 \'e_1 = f_1, f_2 \'e_1 = -f_2, \'e_2 f_1 = f_2 \'e_2, \'e_2 f_2 = f_1\'e_2$ the results are

\begin{align}
    \begin{aligned}\label{eq9238results}
    f_1 \phi = (a + b)f_1 + (c+d)f_1\'e_2,
     \;&&&f_2 \phi = (a - b)f_2 + (c - d)f_2\'e_2,\\
       \phi f_1 = (a + b)f_1 + (c - d)f_2\'e_2,
        \;&&&\phi f_2  = (a - b)f_2 + (c+d)f_2\'e_2.
    \end{aligned}
\end{align}

\noindent Since $f_1f_2 = 0$, 

\begin{equation}
    f_1 \phi f_1 = (a+b)f_1 = a' f_1 = a'\begin{pmatrix}
    1 & 0 \\
    0 & 0
    \end{pmatrix} = \begin{pmatrix}
    a' & 0 \\
    0 & 0
    \end{pmatrix} \implies   f_1 \phi f_1 \simeq \rr.
\end{equation}

\noindent The conditions $ f_1\E_{11} = \E_{11}f_1 = \E_{11}$ and    $f_2\E_{11} = \E_{11}f_2 = 0$ are satisfied for $\E_{11} = f_1$, analogously $\E_{22} = f_2$. For $\E_{12}$ and $\E_{21}$:

\begin{align}
    \begin{aligned}
     &\E_{12}: \begin{cases}
   f_2\E_{12} = \E_{21}f_1 = 0,\\
   f_1\E_{12} = \E_{12}f_2 = \E_{12}.
  \end{cases}\\
  &\E_{21}: \begin{cases}
   f_1\E_{21} = \E_{21}f_2 = 0,\\
   f_2\E_{21} = \E_{21}f_1 = \E_{21}.
  \end{cases}
  \end{aligned}
\end{align}

\noindent Therefore, it follows from Eq. \eqref{eq9238results} for $\E_{12}$ that

\begin{align}
    \begin{aligned}
    f_2\E_{12} = \E_{21}f_1 = 0 \implies f_2\phi = \phi f_1 \implies (a-b)f_2 = (a+b)f_1 = 0 \implies b = -a\'e_1
    \end{aligned}
\end{align}

\noindent and also that

\begin{align}
    \begin{aligned}
 f_1\E_{12} = \E_{12}f_2 = \E_{12} \implies \E_{12} &= (a + b)f_1 + (c + d)f_1\'e_2\\
 &=  (a  -a\'e_1)f_1 + (c + d)f_1\'e_2 \\
  &=  (af_1 - a f_1) + (c + d)f_1\'e_2 \\
  &= (c + d)f_1\'e_2 \\
  &= \frac{1}{2}(c + d)(1 + \'e_1)\'e_2.
    \end{aligned}
\end{align}

\noindent Doing the same procedure for $\E_{21}$:

\begin{align}
    \begin{aligned}
    \E_{21} &= (c'-d')f_2\'e_2 = \frac{1}{2}(c' - d')(1 - \'e_1)\'e_2.
    \end{aligned}
\end{align}

\noindent Continuing in this way, by another condition one has:

\begin{align}
    \begin{aligned}
    f_1 &= \E_{12}\E_{21} =  \left ( (c + d)f_1\'e_2 \right) \left( (c' - d')f_2\'e_2\right)\\
    &= (c + d)(c'-d') f_1\'e_2f_2\'e_2  =(c + d)(c'-d') f_1\'e_2\'e_2 f_1\\
    &= (c + d)(c'-d') f_1.
    \end{aligned}
\end{align}

\noindent Hence, $(c + d)(c'-d') =1$. The simplest solution is $(c + d) = (c'-d') = 1$ and then:

\begin{align}\label{eq_clif1289}
    \begin{aligned}
      &\E_{11} = \frac{1}{2}(1 + \'e_1), \;\;\; &&\E_{22} = \frac{1}{2}(1 - \'e_1)\\
        &\E_{21} = \frac{1}{2}(1 - \'e_1)\'e_2,  \;\;\; &&\E_{12} = \frac{1}{2}(1 + \'e_1)\'e_2.
    \end{aligned}
\end{align}

\noindent In order to find a matrix representation of an element $\phi$ the above equations \eqref{eq_clif1289} can be used to find the matrix representation of the basis elements $1, \'e_1,\'e_2, \'e_{1}\'e_{2}$ as follows. 

\begin{align}\label{eq_clif1282319}
    \begin{aligned}
      &1 = \E_{11} + \E_{22}, \;\;\; &&\'e_{1} = \E_{11} - \E_{22},\\
        &\'e_{2} = \E_{12} + \E_{21}  \;\;\; &&\'e_{1}\'e_{2} = \E_{12} - \E_{21}. 
    \end{aligned}
\end{align}

\noindent This method is certainly more straightforward than calculate the matrix component $\phi_{AB} = \sum_C\E_{CA} \phi\E_{BC}$. Therefore, it follows from the equations \eqref{eq_clif1282319} that 

\begin{align}\label{eq_clirep282319}
    \begin{aligned}
\rho(1) &= \begin{pmatrix}
1 & 0 \\
0 & 0 
\end{pmatrix} + \begin{pmatrix}
0 & 0 \\
0 & 1 
\end{pmatrix} = \begin{pmatrix}
1 & 0 \\
0 & 1 
\end{pmatrix}, \\
\rho(\'e_1) &= \begin{pmatrix}
1 & 0 \\
0 & 0 
\end{pmatrix} -  \begin{pmatrix}
0 & 0 \\
0 & 1 
\end{pmatrix} = \begin{pmatrix}
1 & 0 \\
0 & -1 
\end{pmatrix}, \\
\rho(\'e_2) &= \begin{pmatrix}
0 & 1 \\
0 & 0 
\end{pmatrix} + \begin{pmatrix}
0 & 0 \\
1 & 0 
\end{pmatrix} = \begin{pmatrix}
0 & 1 \\
1 & 0 
\end{pmatrix}, \\
\rho(\'e_1\'e_2) &= \begin{pmatrix}
0 & 1 \\
0 & 0 
\end{pmatrix} - \begin{pmatrix}
0 & 0 \\
1 & 0 
\end{pmatrix} = \begin{pmatrix}
0 & 1 \\
-1 & 0 
\end{pmatrix}.
    \end{aligned}
\end{align}

\noindent The above representations is precisely that one shown in the Eq. \eqref{eq_ac_cl2isomorhicmat2r} about the discussion of the isomorphism $\cl_{2,0} \simeq \matreald$ located in the Section \ref{section_acsclass}. \paragraph{ } Now, let us consider the second procedure for the same example. By Eq. \eqref{eq9238results}: $\phi f_1 = (a + b)\frac{1}{2}(1 + \'e_1)+ (c - d)\frac{1}{2}(1 - \'e_1)\'e_2$, then it follows that the ideal $\cl_{2,0}f_1$ is given by

\begin{equation}
    \cl_{2,0} f_1 = \{  a' \frac{1}{2}(1 + \'e_1)+ b' \frac{1}{2}(1 - \'e_1)\'e_2 \; | \; a',b' \in \mathbb{R} \}
\end{equation}
 
\noindent Then $\mathfrak{B} = \{\frac{1}{2}(1 + \'e_1), \frac{1}{2}(1 - \'e_1)\'e_2 \} = \{\E_{11},\E_{21}\}$ is a basis for the ideal $\cl_{2,0} f_1$. The dual basis $\{\E_{1A}\} = \{\E_{11}, \E_{12} \}$ can be straightforwardly obtained, since $\E_{1A}(\E_{B1}) = \delta_{AB}f_1 = \E_{1A}\E_{B1}$, one has:
\begin{equation}
    f_1 = \E_{12}\E_{21} = \E_{12}\'e_{1}f_1 \implies \E_{12} = f_1\'e_{1} = \frac{1}{2}(1 + \'e_1)\'e_2 .
\end{equation}

\noindent Therefore the dual basis is $\mathfrak{B}^{*} = \{\E_{11}, \E_{12} \} =  \{\frac{1}{2}(1 + \'e_1), \frac{1}{2}(1 + \'e_1)\'e_2 \}$ and the objects $\E_{AB} \; (A,B = 1,2)$ are the exactly the ones already discovered and therefore the representation is derived. \,\textcolor{gray}{{$\blacktriangleleft$}}  \end{ex}
\section{Bilinear Covariants Calculations}\label{app8}

 {\colorlet{shadecolor}{gray!15}\begin{shaded}    \begin{propos}\label{propos_weyltodiracapp}
The Dirac matrices are equivalent to the Weyl matrices by a similarity transformation
\end{propos} \end{shaded}}

\noindent \textit{Proof.} Indeed, take $S \in \matcompq$ such that

\begin{align}
\begin{aligned}
&S = \frac{1}{\sqrt{2}}  \begin{pmatrix} 
    1 & 0 & 1 & 0 \\
    0 & 1 & 0 & 1 \\
    1 & 0 & -1 & 0\\
    0 & 1 & 0 & -1
    \end{pmatrix},
\end{aligned}
\end{align}

\noindent the matrix $S$ is invertible  $S^{-1} = S$. Note that,

\begin{equation}
    \gamma_{\mu} = g_{ij}\gamma^{\mu},
\end{equation}

\noindent such that $g_{ij}$ is the metric of the four-dimentional spacetime, in our case, the signature is $(1,3)$. That is, $\gamma_{0} = \gamma^{0}, \gamma_{1} = -\gamma^{1}, \gamma_{2} = -\gamma^{2}, \gamma_{3} = -\gamma^{3}$, valid for the Dirac or Weyl matrices. This way, one has that for  $\mu = 1,2,3$, $S\gamma^{\mu}_{\text{Weyl}}S^{-1} = -S\gamma_{\mu, \text{Weyl}}S^{-1}$. As a result, by taking the Weyl matrices defined in Eq.  \eqref{eq_matweylquatro}, it is possible to verify that 
\begin{align*}
\begin{aligned} S\gamma^{0}_{\text{Weyl}}S^{-1}& \!=\! \frac{1}{2}\!   \begin{pmatrix} 
    \!1 & \!0 & \!1 & \!0 \\
    \!0 & \!1 & \!0 & \!1 \\
    \!1 & \!0 & \!-1 & \!0\\
    \!0 & \!1 & \!0 & \!-1
    \end{pmatrix}\! \begin{pmatrix} 
    \!0 & \!0 & \!1 & \!0 \\
    \!0 & \!0 & \!0 & \!1\\
    \!1 & \!0 & \!0 & \!0\\
    \!0 & \!1 & \!0 & \!0
    \end{pmatrix}\! \begin{pmatrix} 
    \!1 & \!0 & \!1 & \!0 \\
    \!0 & \!1 & \!0 & \!1 \\
    \!1 & \!0 & \!-1 & \!0\\
    \!0 & \!1 & \!0 & \!-1
    \end{pmatrix}  \! =\! \begin{pmatrix} 
    \!1 & \!0 & \!0 & \!0 \\
    \!0 & \!1 & \!0 & \!0\\
    \!0 & \!0 & \!-1 & \!0\\
    \!0 & \!0 & \!0 & \!-1
    \end{pmatrix} \!=\!  \gamma^{0}_{\text{Dirac}},
\end{aligned}
\end{align*}
\begin{align*} 
\begin{aligned} S\gamma^{1}_{\text{Weyl}}S^{-1}& \!= \!-\frac{1}{2}\!\!   \begin{pmatrix} 
    \!1 & \!0 & \!1 & \!0 \\
    \!0 & \!1 & \!0 & \!1 \\
    \!1 & \!0 & \!-1 & \!0\\
    \!0 & \!1 & \!0 & \!-1
    \end{pmatrix}\!\! \begin{pmatrix} 
    \!0 & \!0 & \!0 & \!-1 \\
    \!0 & \!0 & \!-1 & \!0\\
    \!0 & \!1 & \!0 & \!0\\
    \!1 & \!0 & \!0 & \!0
    \end{pmatrix}\!\!\begin{pmatrix} 
    \!1 & \!0 & \!1 & \!0 \\
    \!0 & \!1 & \!0 & \!1 \\
    \!1 & \!0 & \!-1 & \!0\\
    \!0 & \!1 & \!0 & \!-1
    \end{pmatrix}\!\! =\! \! \begin{pmatrix} 
    \!0 & \!0 & \!0 & \!1 \\
    \!0 & \!0 & \!1 & \!0\\
    \!0 & \!-1 & \!0 & \!0\\
    \!-1 & \!0 & \!0 & \!0
    \end{pmatrix} \!=\! \gamma^{1}_{\text{Dirac}},\\
\end{aligned}
\end{align*}
\begin{align*}
\begin{aligned} S\gamma^{2}_{\text{Weyl}}S^{-1}& \!=\!- \frac{1}{2}  \!\! \begin{pmatrix} 
    \!1 & \!0 & \!1 & \!0 \\
    \!0 & \!1 & \!0 & \!1 \\
    \!1 & \!0 & \!-1 & \!0\\
    \!0 & \!1 & \!0 & \!-1
    \end{pmatrix}\!\! \begin{pmatrix} 
    \!0 & \!0 & \!0 & \!i \\
    \!0 & \!0 & \!-i & \!0\\
    \!0 & \!-i & \!0 & \!0\\
    \!i & \!0 & \!0 & \!0
    \end{pmatrix} \!\! \begin{pmatrix} 
    \!1 & \!0 & \!1 & \!0 \\
    \!0 & \!1 & \!0 & \!1 \\
    \!1 & \!0 & \!-1 & \!0\\
    \!0 & \!1 & \!0 & \!-1
    \end{pmatrix}\!\! = \! \!\begin{pmatrix} 
    \!0 & \!0 & \!0 & \!-i \\
    \!0 & \!0 & \!i & \!0\\
    \!0 & \!i & \!0 & \!0\\
    \!-i & \!0 & \!0 & \!0
    \end{pmatrix} \!=\! \gamma^{2}_{\text{Dirac}},\\
\end{aligned}
\end{align*}
\begin{align*}
\begin{aligned} S\gamma^{3}_{\text{Weyl}}S^{-1}& \!=\!- \frac{1}{2}\!   \begin{pmatrix} 
    \!1 & \!0 & \!1 & \!0 \\
    \!0 & \!1 & \!0 & \!1 \\
    \!1 & \!0 & \!-1 & \!0\\
    \!0 & \!1 & \!0 & \!-1
    \end{pmatrix}\!\!\begin{pmatrix} 
    \!0 & \!0 & \!-1 & \!0 \\
    \!0 & \!0 & \!0 & \!1\\
    \!1 & \!0 & \!0 & \!0\\
    \!0 & \!-1 & \!0 & \!0
    \end{pmatrix}\!\!\begin{pmatrix} 
    \!1 & \!0 & \!1 & \!0 \\
    \!0 & \!1 & \!0 & \!1 \\
    \!1 & \!0 & \!-1 & \!0\\
    \!0 & \!1 & \!0 & \!-1
    \end{pmatrix}\!\!=\!\! \begin{pmatrix} 
    \!0 & \!0 & \!1 & \!0 \\
    \!0 & \!0 & \!0 & \!-1\\
    \!-1 & \!0 & \!0 & \!0\\
    \!0 & \!1 & \!0 & \!0
    \end{pmatrix} \!=\! \gamma^{3}_{\text{Dirac}}.\\
\end{aligned}
\end{align*}

\noindent Consequently,

\begin{equation}
    \gamma^{\mu}_{\text{Dirac}} = S\gamma^{\mu}_{\text{Weyl}}S^{-1},\\
\end{equation}

\noindent as claimed. $\Box$ \\

  {\colorlet{shadecolor}{gray!15}\begin{shaded} \begin{propos}\label{propos_asdfg65app}
   The bilinear covariants are independent of the representation.
\end{propos}   \end{shaded}}

\noindent \textit{Proof.} Indeed, consider the transformation matrix:

\begin{equation}
    S = \frac{1}{\sqrt{2}}  \begin{pmatrix} 
    1 & 0 & 1 & 0 \\
    0 & 1 & 0 & 1 \\
    1 & 0 & -1 & 0\\
    0 & 1 & 0 & -1
    \end{pmatrix}.
\end{equation}

\noindent As a result from Proposition \ref{propos_weyltodirac}, one has  $   \gamma^{\mu}_{\text{Dirac}} = S\gamma^{\mu}_{\text{Weyl}}S^{-1}$. In addition the spinor transforms from Dirac to Weyl (and vice-versa) by:  $\psi \mapsto (S\psi) $ and the conjugate by $\bar\psi \mapsto (\bar\psi S^{-1})$. Consequently,

\noindent $\sigma = \bar\psi \psi $ transforms by

\begin{equation}
    \sigma' = (\bar\psi S^{-1}) (S\psi) = \bar\psi (S^{-1} S)\psi =  \bar\psi \psi = \sigma.
\end{equation}

\noindent $J^{\mu} = \bar\psi \gamma^a \psi$ transforms by
 $$J^{{\mu}'} = (\bar\psi S^{-1}) (S\gamma^{\mu} S^{-1}) (S\psi) = \bar\psi (S^{-1} S)\gamma^{\mu} (S^{-1}S)\psi =  \bar\psi \gamma^a \psi = J^{\mu}.$$

\noindent $S^{\mu\nu} = \frac{1}{2}i\bar{\psi}\gamma^{\mu\nu}\psi$ transforms by

\begin{equation}
    S^{\mu\nu'} \!=\! \frac{1}{2}i(\bar{\psi}S^{-1}\!)(S\gamma^{\mu}S^{-1}\!)(S\gamma^{\nu}S^{-1}\!)(S\psi) \!=\! \frac{1}{2}i\bar{\psi}(S^{-1}\!S)\gamma^{\mu}(S^{-1}\!S)\gamma^{\nu}(S^{-1}\!S)\psi \!=\! \frac{1}{2}i\bar{\psi}\gamma^{\mu\nu}\psi \!=\! S^{\mu\nu}.
\end{equation}

\noindent $K^{\mu} = i\bar{\psi}\gamma^{0123}\gamma^{\mu}\psi$ transforms by

\begin{equation}
    K^{\mu'} \!\!=\! i(\bar{\psi}S^{-1}\!)(S\gamma^{0123}\!S^{-1}\!)(S\gamma^{\mu}S^{-1}\!)(S\psi) \!=\! i\bar{\psi}(S^{-1}\!S)\gamma^{0123}\!(S^{-1}\!S)\gamma^{\mu}(S^{-1}\!S) \psi \!=\! i\bar{\psi}\gamma^{0123}\!\gamma^{\mu}\psi \!=\!  K^{\mu}.
\end{equation}

\noindent $\omega =  \bar{\psi}\gamma^{0123}\psi$ transforms by

\begin{equation}
 \omega^{'} = (\bar{\psi}S^{-1})(S\gamma^{0123}S^{-1})(S\psi) = \bar{\psi}(S^{-1}S)\gamma^{0123}(S^{-1}S)\psi  = \bar{\psi}\gamma^{0123}\psi   = \omega.
\end{equation}

\noindent That ends the proof. $\Box$

 {\colorlet{shadecolor}{gray!15}\begin{shaded} \begin{ex}\label{ex_espcalcbi}
Explicit computation of the bilinear covariants of the spinor \begin{equation}
   \psi = \begin{pmatrix} 
    -i\beta^{*} \\
    i\alpha^{*} \\
    \alpha \\
    \beta 
    \end{pmatrix},
\end{equation}
\noindent with $\alpha,\beta : \rr^{1,3} \to \mathbb{C}$. 
\end{ex}
 \end{shaded}}

\noindent The Weyl representation is used. First, define $\gamma_{5} = \pm i\gamma_{0}\gamma_{1}\gamma_{2}\gamma_{3}$ such that
\begin{align}
\begin{aligned}\label{eq_matwyel1}
\gamma_{5} = \begin{pmatrix} 
    \mathbb{I} & \mathbb{O} \\
    \mathbb{O} & -\mathbb{I}
    \end{pmatrix}.
\end{aligned}
\end{align}

\noindent one has $\gamma_{5} = + i\gamma_{0}\gamma_{1}\gamma_{2}\gamma_{3}$.

\noindent For $\sigma = \bar{\psi}\psi$, 
\begin{align}
\begin{aligned}
\sigma &= \bar{\psi}\psi = \psi^{\dagger}\gamma_{0} \psi \\
&= \begin{pmatrix} 
    i\beta & -i\alpha & \alpha^{*} & \beta^{*}
    \end{pmatrix} \begin{pmatrix} 
    \;0 & \;0 & \;1 & \;0 \\
    \;0 & \;0 & \;0 & \;1\\
    \;1 & \;0 & \;0 & \;0\\
    \;0 & \;1 & \;0 & \;0
    \end{pmatrix}\begin{pmatrix} 
    -i\beta^{*} \\
    i\alpha^{*} \\
    \alpha \\
    \beta 
    \end{pmatrix} \\
    &= i\beta \alpha - i \alpha \beta + \alpha^{*} i \beta^{*} - \beta^{*} i \alpha = 0.
\end{aligned}
\end{align}

 \noindent This way, the first result is $\sigma = 0$. For $J_{\mu} = \bar{\psi}\gamma_{\mu}\psi$, it follows that:
 
\noindent For $\mu = 0$:
\begin{align}
\begin{aligned}
J_{0} &= \bar{\psi}\gamma_{0}\psi = \psi^{\dagger}\gamma_{0}\gamma_{0} \psi =  \psi^{\dagger} I \psi = \begin{pmatrix} 
    i\beta & -i\alpha & \alpha^{*} & \beta^{*}
    \end{pmatrix}\begin{pmatrix} 
    -i\beta^{*} \\
    i\alpha^{*} \\
    \alpha \\
    \beta 
    \end{pmatrix} \\&= -i^2 \beta \beta^{*} + i^2 \alpha \alpha^{*} + \alpha^{*}\alpha + \beta^{*}\beta  = ||\beta||^2 - ||\alpha||^2 + ||\alpha||^2 + ||\beta||^2 = 2||\beta||^2.
\end{aligned}
\end{align}

\noindent For $\mu = 1$:

\begin{align}
\begin{aligned}
J_{1} &= \bar{\psi}\gamma_{1}\psi = \psi^{\dagger}\gamma_{0}\gamma_{1} \psi \\&= \begin{pmatrix} 
    i\beta & -i\alpha & \alpha^{*} & \beta^{*}
    \end{pmatrix}\begin{pmatrix} 
    0 & 1 & 0 & 0 \\
    1 & 0 & 0 & 0\\
    0 & 0 & 0 & -1\\
    0 & 0 & -1 & 0
    \end{pmatrix}\begin{pmatrix} 
    -i\beta^{*} \\
    i\alpha^{*} \\
    \alpha \\
    \beta 
    \end{pmatrix} =\begin{pmatrix} 
    i\beta & -i\alpha & \alpha^{*} & \beta^{*}
    \end{pmatrix} \begin{pmatrix} 
     i\alpha^{*}\\
    -i\beta^{*} \\
    -\beta  \\
    -\alpha
    \end{pmatrix} \\
&= i^2\beta \alpha + i^2\alpha\beta^{*} - \alpha^{*}\beta - \beta^{*}\alpha = - (\alpha\beta + 2\alpha\beta^{*} + \alpha^{*}\beta). 
\end{aligned}
\end{align}

\noindent For $\mu = 2$:

\begin{align}
\begin{aligned}
J_{2} &= \bar{\psi}\gamma_{2}\psi = \psi^{\dagger}\gamma_{0}\gamma_{2} \psi \\
&= \begin{pmatrix} 
    i\beta & -i\alpha & \alpha^{*} & \beta^{*}
    \end{pmatrix}\begin{pmatrix} 
    0 & -i & 0 & 0 \\
    i & 0 & 0 & 0\\
    0 & 0 & 0 & i\\
    0 & 0 & -i & 0
    \end{pmatrix}\begin{pmatrix} 
    -i\beta^{*} \\
    i\alpha^{*} \\
    \alpha \\
    \beta 
    \end{pmatrix} =\begin{pmatrix} 
    i\beta & -i\alpha & \alpha^{*} & \beta^{*}
    \end{pmatrix} \begin{pmatrix} 
     \alpha^{*}\\
    \beta^{*} \\
    i\beta  \\
    -i\alpha
    \end{pmatrix} \\
&= i\beta \alpha^{*} - i\alpha\beta^{*} + i\alpha^{*}\beta - i\beta^{*}\alpha = 2i(\alpha^{*}\beta - \alpha\beta^{*}). 
\end{aligned}
\end{align}

\noindent For $\mu = 3$:

\begin{align}
\begin{aligned}
J_{3} &= \bar{\psi}\gamma_{3}\psi = \psi^{\dagger}\gamma_{0}\gamma_{3} \psi \\
&= \begin{pmatrix} 
    i\beta & -i\alpha & \alpha^{*} & \beta^{*}
    \end{pmatrix} \begin{pmatrix} 
    1 & 0 & 0 & 0 \\
    0 & -1 & 0 & 0\\
    0 & 0 & -1 & 0\\
    0 & 0 & 0 & 1
    \end{pmatrix}\begin{pmatrix} 
    -i\beta^{*} \\
    i\alpha^{*} \\
    \alpha \\
    \beta 
    \end{pmatrix} =\begin{pmatrix} 
    i\beta & -i\alpha & \alpha^{*} & \beta^{*}
    \end{pmatrix} \begin{pmatrix} 
    -i\beta^{*} \\
    -i\alpha^{*} \\
    -\alpha \\
    \beta 
    \end{pmatrix} \\
&= -i^2\beta \beta^{*} + i^2\alpha\alpha^{*} - \alpha^{*}\alpha + \beta^{*}\beta = 2( ||\beta||^2 - ||\alpha||^2). 
\end{aligned}
\end{align}

\noindent We conclude that $\mathbf{J} \neq 0$. For $S_{\mu \nu} = \frac{1}{2} i\bar{\psi}\gamma_{\mu}\gamma_{\nu}\psi $, one has \\
\noindent For $\mu = 0$ and $\nu = 0,$
\begin{align}
\begin{aligned}
 S_{00} &= \frac{1}{2} i\bar{\psi}\gamma_{0}\gamma_{0}\psi = \frac{1}{2} i\psi^{\dagger}\gamma_{0}\gamma_{0}\gamma_{0}\psi \\
 &= \frac{1}{2}\begin{pmatrix} 
    -\beta & \alpha & i\alpha^{*} & i\beta^{*}
    \end{pmatrix}\begin{pmatrix} 
    0 & 0 & 1 & 0 \\
    0 & 0 & 0 & 1\\
    1 & 0 & 0 & 0\\
    0 & 1 & 0 & 0
\end{pmatrix}\begin{pmatrix} 
    -i\beta^{*} \\
    i\alpha^{*} \\
    \alpha \\
    \beta 
    \end{pmatrix} = \frac{1}{2}\begin{pmatrix} 
    -\beta & \alpha & i\alpha^{*} & i\beta^{*}
    \end{pmatrix}\begin{pmatrix} 
    \a \\
    \b \\
    -i\b^{*} \\
    i\a^{*}
    \end{pmatrix} \\
&= \frac{1}{2} (-\b\a + \a\b - i^2\a^{*}\b^{*} + i^2 \b^{*}\a^{*}) = 0.
\end{aligned}
\end{align}
\noindent For $\mu = 0$ and $\nu = 1,$
\begin{align}
\begin{aligned}
 S_{01} &= \frac{1}{2} i\bar{\psi}\gamma_{0}\gamma_{1}\psi = \frac{1}{2} i\psi^{\dagger}\gamma_{0}\gamma_{0}\gamma_{1}\psi \\
 &= \frac{1}{2}\begin{pmatrix} 
    -\beta & \alpha & i\alpha^{*} & i\beta^{*}
    \end{pmatrix} \begin{pmatrix} 
    0 & 0 & 0 & -1 \\
    0 & 0 & -1 & 0\\
    0 & 1 & 0 & 0\\
    1 & 0 & 0 & 0
\end{pmatrix}\begin{pmatrix} 
    -i\beta^{*} \\
    i\alpha^{*} \\
    \alpha \\
    \beta 
    \end{pmatrix} = \frac{1}{2}\begin{pmatrix} 
    -\beta & \alpha & i\alpha^{*} & i\beta^{*}
    \end{pmatrix}\begin{pmatrix} 
    -\b \\
    -\a \\
    i\a^{*} \\
    -i\b^{*}
    \end{pmatrix} \\
&= \frac{1}{2} (\b\b - \a\a + i^2\a^{*}\a^{*} - i^2 \b^{*}\b^{*}) = \frac{1}{2} (\b^2 - \a^2 - \a^{*2} + \b^{*2}).
\end{aligned}
\end{align}
\noindent For $\mu = 0$ and $\nu = 2,$

\begin{align}
\begin{aligned}
 S_{02} &= \frac{1}{2} i\bar{\psi}\gamma_{0}\gamma_{2}\psi = \frac{1}{2} i\psi^{\dagger}\gamma_{0}\gamma_{0}\gamma_{2}\psi \\
 &= \frac{1}{2}\begin{pmatrix} 
    -\beta & \alpha & i\alpha^{*} & i\beta^{*}
    \end{pmatrix}\begin{pmatrix} 
    0 & 0 & 0 & i \\
    0 & 0 & -i & 0\\
    0 & -i & 0 & 0\\
    i & 0 & 0 & 0
\end{pmatrix}\begin{pmatrix} 
    -i\beta^{*} \\
    i\alpha^{*} \\
    \alpha \\
    \beta 
    \end{pmatrix} = \frac{1}{2}\begin{pmatrix} 
    -\beta & \alpha & i\alpha^{*} & i\beta^{*}
    \end{pmatrix}\begin{pmatrix} 
    i\b \\
    -i\a \\
     \a^{*}\\
    \b^{*}
    \end{pmatrix}    \\ 
&= \frac{1}{2} (-i\b\b - i\a\a + i\a^{*}\a^{*} + i \b^{*}\b^{*})= \frac{i}{2} (-\b^2 - \a^2 + \a^{*2} + \b^{*2}) 
\end{aligned}
\end{align}

\noindent For $\mu = 0$ and $\nu = 3,$

\begin{align}
\begin{aligned}
 S_{03} &= \frac{1}{2} i\bar{\psi}\gamma_{0}\gamma_{3}\psi = \frac{1}{2} i\psi^{\dagger}\gamma_{0}\gamma_{0}\gamma_{3}\psi \\
 &= \frac{1}{2}\begin{pmatrix} 
    -\beta & \alpha & i\alpha^{*} & i\beta^{*}
    \end{pmatrix} \begin{pmatrix} 
    0 & 0 & -1 & 0 \\
    0 & 0 & 0 & 1\\
    1 & 0 & 0 & 0\\
    0 & -1 & 0 & 0
\end{pmatrix}\begin{pmatrix} 
    -i\beta^{*} \\
    i\alpha^{*} \\
    \alpha \\
    \beta 
    \end{pmatrix} = \frac{1}{2}\begin{pmatrix} 
    -\beta & \alpha & i\alpha^{*} & i\beta^{*}
    \end{pmatrix} \begin{pmatrix} 
    -\a  \\
    \b  \\
     -i\b^{*}\\
    -i\a^{*}
    \end{pmatrix} \\ 
&= \frac{1}{2} (\b\a + \a\b - i^2\a^{*}\b^{*} - i^2\b^{*}\a^{*})= \a\b + \a^{*}\b^{*}.
\end{aligned}
\end{align}
\noindent For $\mu = 1$ and $\nu = 0,$
\begin{align}
\begin{aligned}
 S_{10} &= \frac{1}{2} i\bar{\psi}\gamma_{1}\gamma_{0}\psi = \frac{1}{2} i\psi^{\dagger}\gamma_{0}\gamma_{1}\gamma_{0}\psi = - \frac{1}{2} i\psi^{\dagger}\gamma_{0}\gamma_{0}\gamma_{1}\psi = - S_{01}.
\end{aligned}
\end{align}
\noindent For $\mu = 1$ and $\nu = 1,$
\begin{align}
\begin{aligned}
 S_{11} &= \frac{1}{2} i\bar{\psi}\gamma_{1}\gamma_{1}\psi = \frac{1}{2} i\psi^{\dagger}\gamma_{0}\gamma_{1}\gamma_{1}\psi \\
 &= \frac{1}{2}\begin{pmatrix} 
    -\beta & \alpha & i\alpha^{*} & i\beta^{*}
    \end{pmatrix} \begin{pmatrix} 
    0 & 0 & -1 & 0 \\
    0 & 0 & 0 & -1\\
    0 & -1 & 0 & 0\\
    -1 & 0 & 0 & 0
\end{pmatrix}\begin{pmatrix} 
    -i\beta^{*} \\
    i\alpha^{*} \\
    \alpha \\
    \beta 
    \end{pmatrix} = \frac{1}{2}\begin{pmatrix} 
    -\beta & \alpha & i\alpha^{*} & i\beta^{*}
    \end{pmatrix}\begin{pmatrix} 
    -\a   \\
    -\b \\
     -i\a^{*}\\
    i\b^{*}
    \end{pmatrix} \\
&= \frac{1}{2} (\b\a - \a\b - i^2\a^{*}\a^{*} + i^2\b^{*}\b^{*}) = 0.
\end{aligned}
\end{align}
\noindent For $\mu = 1$ and $\nu = 2,$
\begin{align}
\begin{aligned}
 S_{12} &= \frac{1}{2} i\bar{\psi}\gamma_{1}\gamma_{2}\psi = \frac{1}{2} i\psi^{\dagger}\gamma_{0}\gamma_{1}\gamma_{2}\psi \\
 &= \frac{1}{2}\begin{pmatrix} 
    -\beta & \alpha & i\alpha^{*} & i\beta^{*}
    \end{pmatrix} \begin{pmatrix} 
    0 & 0 & -i & 0 \\
    0 & 0 & 0 & i\\
    -i & 0 & 0 & 0\\
    0 & i & 0 & 0
\end{pmatrix}\begin{pmatrix} 
    -i\beta^{*} \\
    i\alpha^{*} \\
    \alpha \\
    \beta 
    \end{pmatrix} = \frac{1}{2}\begin{pmatrix} 
    -\beta & \alpha & i\alpha^{*} & i\beta^{*}
    \end{pmatrix} \begin{pmatrix} 
    -i\a   \\
    i\b \\
     -\b^{*}\\
    -\a^{*}
    \end{pmatrix} \\   
&= \frac{1}{2} (i\b\a + i\a\b - i\a^{*}\b^{*} - i\b^{*}\a^{*})= i(\a\b - \a^{*}\b^{*}).
\end{aligned}
\end{align}
\noindent For $\mu = 1$ and $\nu = 3,$
\begin{align}
\begin{aligned}
 S_{13} &= \frac{1}{2} i\bar{\psi}\gamma_{1}\gamma_{3}\psi = \frac{1}{2} i\psi^{\dagger}\gamma_{0}\gamma_{1}\gamma_{3}\psi \\
 &= \frac{1}{2}\begin{pmatrix} 
    -\beta & \alpha & i\alpha^{*} & i\beta^{*}
    \end{pmatrix}\begin{pmatrix} 
    0 & 0 & 0 & 1\\
    0 & 0 & -1 & 0\\
    0 & 1 & 0 & 0\\
    -1 & 0 & 0 & 0
\end{pmatrix}\begin{pmatrix} 
    -i\beta^{*} \\
    i\alpha^{*} \\
    \alpha \\
    \beta 
    \end{pmatrix}= \frac{1}{2}\begin{pmatrix} 
    -\beta & \alpha & i\alpha^{*} & i\beta^{*}
    \end{pmatrix}\begin{pmatrix} 
    \b   \\
    -\a \\
     i\a^{*}\\
    i\b^{*}
    \end{pmatrix} \\
&= \frac{1}{2} (-\b\b - \a\a + i^2\a^{*}\a^{*} + i^2\b^{*}\b^{*})= \frac{1}{2} (-\b^2 - \a^2 -\a^{*2} - \b^{*2}).
\end{aligned}
\end{align}
\noindent For $\mu = 2$ and $\nu = 0,$
\begin{align}
\begin{aligned}
 S_{20} &= \frac{1}{2} i\bar{\psi}\gamma_{2}\gamma_{0}\psi = \frac{1}{2} i\psi^{\dagger}\gamma_{0}\gamma_{2}\gamma_{0}\psi = -  \frac{1}{2} i\psi^{\dagger}\gamma_{0}\gamma_{0}\gamma_{2}\psi = - S_{02}.
\end{aligned}
\end{align}
\noindent For $\mu = 2$ and $\nu = 1,$
\begin{align}
\begin{aligned}
 S_{21} &= \frac{1}{2} i\bar{\psi}\gamma_{2}\gamma_{1}\psi = \frac{1}{2} i\psi^{\dagger}\gamma_{0}\gamma_{2}\gamma_{1}\psi= -\frac{1}{2} i\psi^{\dagger}\gamma_{0}\gamma_{1}\gamma_{2}\psi = - S_{12}.
\end{aligned}
\end{align}
\noindent For $\mu = 2$ and $\nu = 2,$
\begin{align}
\begin{aligned}
 S_{22} &= \frac{1}{2} i\bar{\psi}\gamma_{2}\gamma_{2}\psi = \frac{1}{2} i\psi^{\dagger}\gamma_{0}\gamma_{2}\gamma_{2}\psi \\
 &= \frac{1}{2}\begin{pmatrix} 
    -\beta & \alpha & i\alpha^{*} & i\beta^{*}
    \end{pmatrix} \begin{pmatrix} 
    0 & 0 & -1 & 0 \\
    0 & 0 & 0 & -1\\
    -1 & 0 & 0 & 0\\
    0 & -1 & 0 & 0
\end{pmatrix}\begin{pmatrix} 
    -i\beta^{*} \\
    i\alpha^{*} \\
    \alpha \\
    \beta 
    \end{pmatrix} = \frac{1}{2}\begin{pmatrix} 
    -\beta & \alpha & i\alpha^{*} & i\beta^{*}
    \end{pmatrix}\begin{pmatrix} 
    -\a   \\
    -\b \\
     i\b^{*}\\
    -i\a^{*}
    \end{pmatrix} \\  
&= \frac{1}{2} (\b\a - \a\b + i^2\b^{*}\a^{*} - i^2\b^{*}\a^{*})= 0.
\end{aligned}
\end{align}
\noindent For $\mu = 2$ and $\nu = 3,$
\begin{align}
\begin{aligned}
 S_{23} &= \frac{1}{2} i\bar{\psi}\gamma_{2}\gamma_{3}\psi = \frac{1}{2} i\psi^{\dagger}\gamma_{0}\gamma_{2}\gamma_{3}\psi \\
 &= \frac{1}{2}\begin{pmatrix} 
    -\beta & \alpha & i\alpha^{*} & i\beta^{*}
    \end{pmatrix} \begin{pmatrix} 
    0 & 0 & 0 & -i \\
    0 & 0 & -i & 0\\
    0 & -i & 0 & 0\\
    -i & 0 & 0 & 0
\end{pmatrix}\begin{pmatrix} 
    -i\beta^{*} \\
    i\alpha^{*} \\
    \alpha \\
    \beta 
    \end{pmatrix} = \frac{1}{2}\begin{pmatrix} 
    -\beta & \alpha & i\alpha^{*} & i\beta^{*}
    \end{pmatrix}\begin{pmatrix} 
    -i\b   \\
    -i\a \\
     \a^{*}\\
    -\b^{*}
    \end{pmatrix} \\  
&= \frac{1}{2} (i\b\b - i\a\a + i\a^{*}\a^{*} - i\b^{*}\b^{*})= \frac{i}{2} (\b^2 - \a^2 + \a^{*2} - \b^{*2}).
\end{aligned}
\end{align}
\noindent For $\mu = 3$ and $\nu = 0,$
\begin{align}
\begin{aligned}
 S_{30} &= \frac{1}{2} i\bar{\psi}\gamma_{3}\gamma_{0}\psi = \frac{1}{2} i\psi^{\dagger}\gamma_{0}\gamma_{3}\gamma_{0}\psi = -\frac{1}{2} i\psi^{\dagger}\gamma_{0}\gamma_{0}\gamma_{3}\psi = - S_{03}.
\end{aligned}
\end{align}
\noindent For $\mu = 3$ and $\nu = 1,$
\begin{align}
\begin{aligned}
 S_{31} &= \frac{1}{2} i\bar{\psi}\gamma_{3}\gamma_{1}\psi = \frac{1}{2} i\psi^{\dagger}\gamma_{0}\gamma_{3}\gamma_{1}\psi = -\frac{1}{2} i\psi^{\dagger}\gamma_{0}\gamma_{1}\gamma_{3}\psi = - S_{13}.
\end{aligned}
\end{align}
\noindent For $\mu = 3$ and $\nu = 2,$
\begin{align}
\begin{aligned}
 S_{32} &= \frac{1}{2} i\bar{\psi}\gamma_{3}\gamma_{2}\psi = \frac{1}{2} i\psi^{\dagger}\gamma_{0}\gamma_{3}\gamma_{2}\psi = -\frac{1}{2} i\psi^{\dagger}\gamma_{0}\gamma_{2}\gamma_{3}\psi = - S_{23}.
\end{aligned}
\end{align}
\noindent For $\mu = 3$ and $\nu = 3,$
\begin{align}
\begin{aligned}
 S_{33} &= \frac{1}{2} i\bar{\psi}\gamma_{3}\gamma_{3}\psi = \frac{1}{2} i\psi^{\dagger}\gamma_{0}\gamma_{3}\gamma_{3}\psi \\
&= \frac{1}{2}\begin{pmatrix} 
    -\beta & \alpha & i\alpha^{*} & i\beta^{*}
    \end{pmatrix}\begin{pmatrix} 
    0 & 0 & -1 & 0 \\
    0 & 0 & 0 & -1\\
    -1 & 0 & 0 & 0\\
    0 & -1 & 0 & 0
\end{pmatrix}\begin{pmatrix} 
    -i\beta^{*} \\
    i\alpha^{*} \\
    \alpha \\
    \beta 
    \end{pmatrix} = \frac{1}{2}\begin{pmatrix} 
    -\beta & \alpha & i\alpha^{*} & i\beta^{*}
    \end{pmatrix}\begin{pmatrix} 
    -\a   \\
    -\b \\
     i\b^{*}\\
    -\a^{*}
    \end{pmatrix} \\
&= \frac{1}{2} (\b\a - \a\b + i\a^{*}\b^{*} - i\b^{*}\a^{*})= 0.
\end{aligned}
\end{align}

\noindent We conclude that $\mathbf{S} \neq 0$. Moreover, for $K_{\mu} = i\bar{\psi}\gamma_{5}\gamma_{\mu}\psi $, one has: \\

\noindent For $\mu = 0$
\begin{align}
\begin{aligned}
K_{0} &= i\bar{\psi}\gamma_{5}\gamma_{0}\psi = i\psi^{\dagger}\gamma_{0}\gamma_{5}\gamma_{0}\psi \\
&= \begin{pmatrix} 
    -\beta & \alpha & i\alpha^{*} & i\beta^{*}
    \end{pmatrix} \begin{pmatrix} 
    -1 & 0 & 0 & 0 \\
    0 & -1 & 0 & 0\\
    0 & 0 & 1 & 0\\
    0 & 0 & 0 & 1
    \end{pmatrix}\begin{pmatrix} 
    -i\beta^{*} \\
    i\alpha^{*} \\
    \alpha \\
    \beta 
    \end{pmatrix} = \begin{pmatrix} 
    -\beta & \alpha & i\alpha^{*} & i\beta^{*}
    \end{pmatrix}\begin{pmatrix} 
    i\beta^{*} \\
    -i\alpha^{*} \\
    \alpha \\
    \beta 
    \end{pmatrix} \\
&= -i\b \b^{*} -i\a \a^{*} +i\a^{*} \a +i\b^{*} \b= 0.
\end{aligned}
\end{align}

\noindent For $\mu = 1$
\begin{align}
\begin{aligned}
K_{1} &= i\bar{\psi}\gamma_{5}\gamma_{1}\psi = i\psi^{\dagger}\gamma_{0}\gamma_{5}\gamma_{1}\psi \\
&= \begin{pmatrix} 
    -\beta & \alpha & i\alpha^{*} & i\beta^{*}
    \end{pmatrix} \begin{pmatrix} 
    0 & -1 & 0 & 0 \\
    -1& 0 & 0 & 0\\
    0 & 0 & 0 & -1\\
    0 & 0 & -1 & 0
    \end{pmatrix}\begin{pmatrix} 
    -i\beta^{*} \\
    i\alpha^{*} \\
    \alpha \\
    \beta 
    \end{pmatrix} = \begin{pmatrix} 
    -\beta & \alpha & i\alpha^{*} & i\beta^{*}
    \end{pmatrix}\begin{pmatrix} 
    -i\a^{*} \\
    i\b^{*} \\
    -\b \\
    -\a 
    \end{pmatrix} \\
&= i\b \a^{*} +  i\a \b^{*} -  i\a^{*} \b - i \b^{*} \a =  0.
\end{aligned}
\end{align}

\noindent For $\mu = 2$
\begin{align}
\begin{aligned}
K_{2} &= i\bar{\psi}\gamma_{5}\gamma_{2}\psi = i\psi^{\dagger}\gamma_{0}\gamma_{5}\gamma_{2}\psi \\
&= \begin{pmatrix} 
    -\beta & \alpha & i\alpha^{*} & i\beta^{*}
    \end{pmatrix} \begin{pmatrix} 
    0 & i & 0 & 0 \\
    -i & 0 & 0 & 0\\
    0 & 0 & 0 & i\\
    0 & 0 & -i & 0
    \end{pmatrix}\begin{pmatrix} 
    -i\beta^{*} \\
    i\alpha^{*} \\
    \alpha \\
    \beta 
    \end{pmatrix} = \begin{pmatrix} 
    -\beta & \alpha & i\alpha^{*} & i\beta^{*}
    \end{pmatrix}\begin{pmatrix} 
    -\a^{*} \\
    -\b^{*} \\
    i\b \\
    -i\a 
    \end{pmatrix} \\
&= \b\a^{*} - \a\b^{*} -\a^{*}\b + \b^{*}\a = 0.
\end{aligned}
\end{align}

\noindent For $\mu = 3$
\begin{align}
\begin{aligned}
K_{3} &= i\bar{\psi}\gamma_{5}\gamma_{3}\psi = i\psi^{\dagger}\gamma_{0}\gamma_{5}\gamma_{3}\psi \\
&= \begin{pmatrix} 
    -\beta & \alpha & i\alpha^{*} & i\beta^{*}
    \end{pmatrix} \begin{pmatrix} 
    -1 & 0 & 0 & 0 \\
    0 & 1 & 0 & 0\\
    0 & 0 & -1 & 0\\
    0 & 0 & 0 & 1
    \end{pmatrix}\begin{pmatrix} 
    -i\beta^{*} \\
    i\alpha^{*} \\
    \alpha \\
    \beta 
    \end{pmatrix} = \begin{pmatrix} 
    -\beta & \alpha & i\alpha^{*} & i\beta^{*}
    \end{pmatrix}\begin{pmatrix} 
    i\b^{*} \\
    i\a^{*} \\
    -\a \\
    \b 
    \end{pmatrix} \\
&= -i\b\b^{*} + i\a\a^{*} - i\a^{*}\a + i\b^{*} \b = 0.
\end{aligned}
\end{align}

\noindent As a result, $\mathbf{K} = 0$. Finally, for $ \omega =  \bar{\psi} \gamma_{5} \psi $, it follows that:
\begin{align}
\begin{aligned}
\omega &= \bar{\psi} \gamma_{5} \psi = \psi^{\dagger}\gamma_{0}\gamma_{5} \psi \\
&= \begin{pmatrix} 
    i\beta & -i\alpha & \alpha^{*} & \beta^{*}
    \end{pmatrix} \begin{pmatrix} 
    0 & 0 & -1 & 0 \\
    0 & 0 & 0 & -1\\
    1 & 0 & 0 & 0\\
    0 & 1 & 0 & 0
    \end{pmatrix}\begin{pmatrix} 
    -i\beta^{*} \\
    i\alpha^{*} \\
    \alpha \\
    \beta 
    \end{pmatrix} =\begin{pmatrix} 
    i\beta & -i\alpha & \alpha^{*} & \beta^{*}
    \end{pmatrix} \begin{pmatrix} 
    -\alpha \\
    -\beta \\
    -i\beta^{*}\\
    i\alpha^{*}
    \end{pmatrix} \\
&=  -i\b\a + i\a\b - i\a^{*}\b^{*} + i\b^{*}\a^{*} = 0.
\end{aligned}
\end{align}

\noindent That way, we conclude that for the spinor $\psi =  (-i\beta^{*} \;\;\;
    i\alpha^{*} \;\;\;
    \alpha \;\;\;
    \beta )^{\intercal}$ we have $\sigma = \omega = \mathbf{K} = 0$ and $\mathbf{J} \neq 0,\; \mathbf{S} \neq 0.$

\subsection{FPK Identities}\label{fpkidapp8}

The example \ref{ex_644t4e} is very useful for demonstrating the Fierz-Pauli-Kofink identities.

 {\colorlet{shadecolor}{gray!15}\begin{shaded} \begin{equation}
     \mathbf{J}^2 = \sigma^{2} + \omega^{2},
 \end{equation}  \end{shaded}}

\noindent Computing $\mathbf{J}^2$ yields

\begin{align}
    \begin{aligned}
J_0^2 &=     |a|^4 \!+\! 2 |a|^2 |b|^2 \!+\! |b|^4 \!+\! 2 |a|^2 |c|^2 \!+\! 2 |b|^2 |c|^2 \!+\! |c|^4 \!+\! 2 |a|^2 |d|^2 \!+\! 2 |b|^2 |d|^2 \!+\! 2 |c|^2 |d|^2 \!+\! |d|^4\\
J_1^2 &=  a^2 b^{*2} \!+\! b^2 a^{*2} \!-\! 2 b d a^* c^* \!-\! 2 b c a^* d^* \!-\! 2 a d b^* c^* \!-\! 2 a c b^* d^* \!+\! 2 |a|^2|b|^2 \!+\! c^2 d^{*2} \!+\! d^2 c^{*2} \!+\! 2 |c|^2|d|^2\\
J_2^2 &=  \!-\!a^2 b^{*2} \!-\! b^2 a^{*2} \!+\! 2 b d a^* c^* \!-\! 2 b c a^* d^* \!-\! 2 a d b^* c^* \!+\! 2 a c b^* d^* \!+\! 2 |a|^2|b|^2 \!-\! c^2 d^{*2} \!-\! d^2 c^{*2} \!-\! 2|c|^2|d|^2\\
J_3^2 &=  |a|^4 \!-\! 2 |a|^2 |b|^2 \!+\! |b|^4 \!-\! 2 |a|^2 |c|^2 \!+\! 2 |b|^2 |c|^2 \!+\! |c|^4 \!+\! 2 |a|^2 |d|^2 \!-\! 2 |b|^2 |d|^2 \!-\! 2 |c|^2 |d|^2 \!+\! |d|^4 \\
&\implies \mathbf{J}^2 = J_0^2 - J_1^2 - J_2^2 - J_3^2 = 4|a|^2|c|^2  + 4|b|^2 |d|^2 + 4b c a^* d^* + 4 a d b^* c^*.\\
    \end{aligned}
\end{align}

\noindent On the other hand for  $\sigma^{2} + \omega^{2}$, it follows that

\begin{align}\label{eq_jotaquadrado}
    \begin{aligned}
    \sigma^{2} &=  a^2 c^{*2} \!+\! 2 c d a^* b^* \!+\! 2 b c a^* d^* \!+\! 2 a d b^* c^* \!+\! 2 a b c^* d^* \!+\! c^2 a^{*2} \!+\! 2 |a|^2|c|^2 \!+\! b^2 d^{*2} \!+\! d^2 b^{*2} \!+\! 2 |b|^2|d|^2 \\
    \omega^{2} &=  \!-\! a^2 c^{*2} \!-\! 2 c d a^* b^* \!+\! 2 b c a^* d^* \!+\! 2 a d b^* c^* \!-\! 2 a b c^* d^* \!-\! c^2 a^{*2} \!+\! 2|a|^2|c|^2 \!-\! b^2 d^{*2} \!-\! d^2 b^{*2} \!+\! 2 |b|^2|d|^2  \\
    &\implies \sigma^{2} + \omega^{2} =  4|a|^2|c|^2 + 4|b|^2|d|^2 + 4 b c a^* d^* + 4 a d b^* c^* 
    \end{aligned}
\end{align}

\noindent Therefore, when comparing the quantities we conclude that $\mathbf{J}^2 = \sigma^{2} + \omega^{2}$.

   {\colorlet{shadecolor}{gray!15}\begin{shaded} \begin{equation}
   \mathbf{K}^{2} = - \mathbf{J}^{2} 
\end{equation} \end{shaded}}

\noindent Computing $\mathbf{K}^2$ result as:

\begin{align}\label{eq_kaquadrado}
    \begin{aligned}
K_0^2 &= |a|^4 \!+\! 2 |a|^2 |b|^2 \!+\! |b|^4 \!-\! 2 |a|^2 |c|^2 \!-\! 2 |b|^2 |c|^2 \!+\! |c|^4 \!-\! 2 |a|^2 |d|^2 \!-\! 2 |b|^2 |d|^2 \!+\! 2 |c|^2 |d|^2 \!+\! |d|^4\\
K_1^2 &=  a^2 b^{*2} \!+\! b^2 a^{*2} \!+\! 2 b d a^* c^* \!+\! 2 b c a^* d^* \!+\! 2 a d b^* c^* \!+\! 2 a c b^* d^* \!+\! 2 |a|^2|b|^2 \!+\! c^2 d^{*2} \!+\! d^2 c^{*2} \!+\! 2 |c|^2|d|^2\\
K_2^2 &=  \!-\!a^2 b^{*2} \!-\! b^2 a^{*2} \!-\! 2 b d a^* c^* \!+\! 2 b c a^* d^* \!+\! 2 a d b^* c^* \!-\! 2 a c b^* d^* \!+\! 2  |a|^2|b|^2 \!-\! c^2 d^{*2} \!-\! d^2 c^{*2} \!+\! 2 |c|^2|d|^2\\
K_3^2 &=  |a|^4 \!-\! 2 |a|^2 |b|^2 \!+\! |b|^4 \!+\! 2 |a|^2 |c|^2 \!-\! 2 |b|^2 |c|^2 \!+\! |c|^4 \!-\! 2 |a|^2 |d|^2 \!+\! 2 |b|^2 |d|^2 \!-\! 2 |c|^2 |d|^2 \!+\! |d|^4\\
&\implies \mathbf{K}^2  = K_0^2 - K_1^2 - K_2^2 - K_3^2 =  - 4|a|^2|c|^2  - 4|b|^2 |d|^2 - 4b c a^* d^* - 4 a d b^* c^*.\\
    \end{aligned}
\end{align}

\noindent Therefore, with the result $\mathbf{K}^2$ of Eq. \eqref{eq_kaquadrado} with $\mathbf{J}^2$ in the Eq. \eqref{eq_jotaquadrado}, we conclude that in fact $\mathbf{K}^{2} = - \mathbf{J}^{2}$.

   {\colorlet{shadecolor}{gray!15}\begin{shaded} \begin{equation}
   \mathbf{J} \cdot \mathbf{K} = 0
\end{equation} \end{shaded}}
\noindent Computing $J_{\mu}K_{\mu} $, yields:

\begin{align}
    \begin{aligned}
     J_{0}K_0 &= (|a|^2 + |b|^2 + |c|^2 + |d|^2)( -|a|^2 - |b|^2 + |c|^2 + |d|^2) \\
     &=-|a|^4 - 2 |a|^2 |b|^2 - |b|^4 + |c|^4 + 2 |c|^2 |d|^2 + |d|^4 , \\
    J_{1}K_1 &=  (b a^* + a b^* - d c^* - c d^*)(-b a^* - a b^* - d c^* - c d^*)\\
    &= -a^2 (b^*)^2 - b^2 (a^*)^2 - 2|a|^2|b|^2 + c^2 (d^*)^2 + d^2 (c^*)^2 +  2|c|^2|d|^2 , \\
     J_{2}K_2 &=  i(- b a^* +  a b^* +  d c^* -  c d^*)i( b a^* -  a b^* +  d c^* -  c d^*)\\
     &= a^2 (b^*)^2 + b^2 (a^*)^2 - 2 |a|^2|b|^2  - c^2 (d^*)^2 - d^2 (c^*)^2 +  2 |c|^2|d|^2, \\
     J_{3}K_3 &=   (|a|^2 - |b|^2 - |c|^2 + |d|^2)(-|a|^2 + |b|^2 - |c|^2 + |d|^2)\\
     &= -|a|^4 + 2 |a|^2 |b|^2 - |b|^4 + |c|^4 - 2 |c|^2 |d|^2 + |d|^4.
    \end{aligned}
\end{align}

\noindent Consequently,

\begin{align}
    \begin{aligned}
      \mathbf{J} \cdot \mathbf{K} &= J^{0}K^{0} - J^{1}K^{1} - J^{2}K^{2} - J^{3}K^{3}
      \\&= - J_{0}K_{0} + J_{1}K_{1} + J_{2}K_{2} + J_{3}K_{3} \\
      &= -|a|^4 - 2 |a|^2 |b|^2 - |b|^4 + |c|^4 + 2 |c|^2 |d|^2 + |d|^4 
     \\&+ a^2 (b^*)^2 + b^2 (a^*)^2 + 2|a|^2|b|^2  - c^2 (d^*)^2 - d^2 (c^*)^2 - 2|c|^2|d|^2
     \\&- a^2 (b^*)^2 - b^2 (a^*)^2 + 2|a|^2|b|^2  + c^2 (d^*)^2 + d^2 (c^*)^2 - 2 |c|^2|d|^2
     \\&+ |a|^4 - 2 |a|^2 |b|^2 + |b|^4 - |c|^4 + 2 |c|^2 |d|^2 - |d|^4\\
     &= 0.\\
    \end{aligned}
\end{align}

\subsection{Recovering Spinors from their Bilinear Covariants}\label{recov_spinapp}
\paragraph{ } Let $\psi$ be a spinor and $\sigma,\, \mathbf{J} , \, \mathbf{S} , \,\mathbf{K} , \, \omega$ its bilinear covariants. Take any spinor $\eta$ such that $\yhwidetilde{\eta}\,^*\psi \neq 0$ in $\cc \otimes \clm$ or equivalently $\eta^{\dagger} \gamma_0 \psi \neq 0$ in $\matcompq$. Then, the spinor $\psi$ is proportional to \cite{Lou01}

\begin{equation}
\psi \simeq Z\eta \;\;\;\;\; \text{with}\;\;\;\;\; Z = \sigma + \mathbf{J} + \mathbf{S} + i\mathbf{K}\gamma_{0123} + \omega \gamma_{0123},
\end{equation}

\noindent where $\psi$ and $Z \eta$ differ only by a phase factor. The spinor $\psi$ can be retrieved by an algorithm, taking \cite{Lou01, Crawford:1985qg, Takahashi:1982ib}.

\begin{align}
    \begin{aligned}
    N &= \sqrt {\langle \yhwidetilde{\eta}^* Z \eta \rangle_{0}} = \frac{1}{2}\sqrt{\eta^{\dagger}\gamma_0 Z \eta},\\
    e^{-i\alpha} &= \frac{4}{N}{\langle \yhwidetilde{\eta}\,^*\psi \rangle_{0}} = \frac{1}{N} \eta^{\dagger}\gamma_0\psi,\\
    \psi &= \frac{1}{4N} e^{-i\alpha} Z \eta.
    \end{aligned}
\end{align}

\noindent Furthermore, with the choice $\eta = f = \frac{1}{2}(1 + \gamma_0)\frac{1}{2}(1 + \gamma_{12})$, we get

\begin{align}
    \begin{aligned}
    N &= \sqrt {\langle Zf \rangle_{0}} = \frac{1}{2}\sqrt{\sigma + \mathbf{J} \cdot \gamma_0 - \mathbf{S} \rfloor \gamma_{12} - \mathbf{K} \cdot \gamma_{3}} \, ,\\
    e^{-i\alpha} &= \frac{\psi_{1}}{|\psi_{1}|}.
    \end{aligned}
\end{align}

\noindent Once the spinor $\psi$ has been recovered, we can also write

\begin{align}
    \begin{aligned}
    N &=  4|{\langle \yhwidetilde{\eta}^* \psi \rangle_{0}}| = |\eta^{\dagger}\gamma_0  \psi|,\\
    e^{-i\alpha} &= \frac{\langle \yhwidetilde{\eta}^* \psi \rangle_{0}}{|\langle \yhwidetilde{\eta}^* \psi \rangle_{0}|} = \frac{\eta^{\dagger} \gamma_0  \psi}{|\eta^{\dagger} \gamma_0 \psi|}.
    \end{aligned}
\end{align}

\noindent Thus, the spinor $\psi$ is determined by its bilinear covariants $\sigma,\, \mathbf{J} , \, \mathbf{S} , \,\mathbf{K} , \, \omega$, by a phase factor $e^{-i\alpha}$, and from the aggregate $Z = \sigma + \mathbf{J} + \mathbf{S} + i\mathbf{K}\gamma_{0123} + \omega \gamma_{0123}$ we can extract from $\eta$ the relevant parts parallel to $\psi$. Moreover, in the case where $\Omega = \sigma + \omega\gamma_{0123} \neq 0$ we can say that if $\sigma,\, \mathbf{J} , \, \mathbf{S} , \,\mathbf{K} , \, \omega$ are bilinear covariants satisfying the Fierz identities, then the aggregate $Z$ can be factored as follows \cite{Crawford:1985qg, Lou01}

\begin{equation}
    Z = (\Omega + \mathbf{J})(1 + i\Omega^{-1} \mathbf{K}\gamma_{0123}).
\end{equation}

\noindent From this factorisation, Crawford \cite{Crawford:1985qg} proved that if the multivectors $\sigma,\, \mathbf{J} , \, \mathbf{S} , \,\mathbf{K} , \, \omega$ satisfy the Fierz identities and also $J^0 > 0$ with $4\langle \yhwidetilde{\eta}^* Z \eta \rangle_{0} = \eta^{\dagger}\gamma_0 Z \eta > 0 \;\;\; \forall \eta \neq 0$, then $\sigma,\, \mathbf{J} , \, \mathbf{S} , \,\mathbf{K} , \, \omega$ are bilinear covariants for a spinor $\psi$ such that

\begin{align}
    \begin{aligned}
\psi &= \frac{1}{4N} Z \eta, \\
N &= \sqrt {\langle \yhwidetilde{\eta}^* Z \eta \rangle_{0}} = \frac{1}{2} \sqrt{\eta^{\dagger} \gamma_0 Z \eta},
    \end{aligned}
\end{align}

\noindent such that $\psi$ is obtained by different choices of $\eta$ which differ only in their phases.
\section{Flux Compactifications in Warped Geometries}\label{app9}
\paragraph{ } In this appendix, we provide the context for our classification of spinors on the eight-dimensional manifold $\M_8$, derived from the compactification of AdS$_3 \times \M_8$ within the framework of eleven-dimensional supergravity. We outline the formulation of the supercovariant derivative and related equations in the local orthonormal frame and discuss the constraints imposed by supersymmetry conditions in AdS$_3$. We also address the conditions for Killing spinors and their associated bilinear covariants, which underpin our classification results. We emphasise that the physical aspects of supersymmetry and supergravity are not on the focus of this dissertation, however, some aspects are introduced here just to provide a background and motivation for the spinor classification on $\M_8$. All the contents included here come from the Refs. \cite{Goncalves:2023pty, bab2}

\paragraph{ } Consider eleven-dimensional supergravity on a connected, oriented eleven-manifold $\M_{11}$  endowed with a pseudo-Riemannian metric $\mathring{g}$ admitting spin structure. The action governing supergravity accommodates a $3$-form potential associated with $4$-form field strength $\mathring{G} \in \Omega^{4}(\M_{11})$ and the gravitino $\Psi_{M}$ that is described by a spin-$3/2$ real spinor field. 
In supersymmetric theories, particularly when examining the bosonic sector of supersymmetric backgrounds, the behaviour of the gravitino and its interactions with supersymmetry are critical. The spin bundle $\mathring{\mathcal{S}}$ over $\M_{11}$ can be conceptualised as a bundle of simple modules over the Clifford bundle of $T^*\M_{11}$. Within this framework, the supersymmetry generator $\eta^{\circ} \in \Gamma(\M_{11}, \mathring{\mathcal{S}})$ is a Majorana spinor realised as a section of $\mathring{\mathcal{S}}$.  For supersymmetric backgrounds, both the vacuum expectation value of the gravitino and its supersymmetry variation must vanish. This requirement translates into the condition that the supersymmetry variation of the gravitino is zero. Additionally, this implies the existence of a non-trivial solution $\mathring{\eta}$ to the first-order equation of motion:

\begin{equation}\label{eq_wedhusj}
    \mathring{\mathcal{D}}_M \mathring{\eta}= 0.
\end{equation}

\noindent Here, $\mathring{\eta}$ can also be interpreted as a supersymmetry generator represented by a Majorana spinor field, which carries an irreducible representation of the $\Spin(1,10)$ group and is viewed as a smooth section of the  bundle of spinors $\mathring{\mathcal{S}}$.

In a local orthonormal frame $(e_M)_{M=0,\ldots,10}$ of $T\M$, the supercovariant connection $\mathring{\mathcal{D}}_M$ is given by:

\begin{equation}
    \mathring{\mathcal{D}}_M = \nabla^{\mathring{\mathcal{S}}}_M - \frac{1}{288} \left(\mathring{G}_{PNRQ} \mathring{\upgamma}^{PNRQ}_M - 8 \mathring{G}_{MNRQ} \mathring{\upgamma}^{NRQ}\right),
\end{equation}

\noindent where 

\begin{equation}
    \nabla^{\mathring{\mathcal{S}}}_M = \partial_M + \frac{1}{4} \mathring{\Omega}_{MNP} \mathring{\upgamma}^{NP}
\end{equation}

\noindent is the spin connection on $\mathring{\mathcal{S}}$ induced by the Levi-Civita connection of $(\M_{11}, \mathring{g})$. The generators $\mathring{\upgamma}^M$ are the gamma matrices of the Clifford algebra $\cl_{1,10}$, within the $32$-dimensional real Majorana irreducible representation. Here, $\mathring{\upgamma}^{0\ldots10} = \mathring{\upgamma}^0 \circ \cdots \circ \mathring{\upgamma}^{10}$ acts as the volume element in $\M_{11}$. A significant application of these theories involves compactification on an $11$-dimensional manifold $\M_{11}$ to an $\text{AdS}_3$ space. In this case, Ref. \cite{Martelli:2003ki} considered a compactification to an AdS$_3$ space with cosmological constant $\Lambda = -8 \kappa^2$, where $\kappa$ is a positive real parameter. Therefore, $\M_{11}$ can be decomposed as $\M_{11} = A \cross B$, where $A$ is an oriented 3-manifold diffeomorphic with $\rr^{3}$ and carrying the $\text{AdS}_3$ metric while $B$ is an $8$-dimensional Riemannian manifold equipped with a metric $\mathtt{g}$. We say,

\begin{equation}
    \M_{11} = \text{AdS}_3 \times \M_8.
\end{equation}

\noindent The warped metric on $\M_{11}$ is a warped product given by

\begin{equation}
     \mathring{g} = \mathring{g}_{MN}dx^M dx^N = e^{2\Updelta} \left(ds^2_3 + \mathtt{g}_{mn} dx^m dx^n\right),
\end{equation}

\noindent such that the warp factor $\Updelta$ is a smooth function on $\M_8$, and $ds^2_3$ denotes the AdS$_3$ metric, the so-called squared length
element. The ansatz used for the 4-form field strength $\mathring{G}$ is:

\begin{equation}
    \mathring{G} = e^{3\Delta} \left(\boldsymbol{\uptau_{3}} \wedge f_1 + F_4\right),
\end{equation}

\noindent where $f_1 = f_m e^m \in \Omega^1(\M_8)$, $F_4=\frac{1}{4!}F_{mnrs} e^{mnrs}\in \Omega^4(\M_8)$, and $\boldsymbol{\uptau_{3}}$ is the volume $3$-form of AdS$_3$. Lowercase Latin indexes label run in the range $1$ to $8$ correspond to a choice of frame on $\M_8$.  The equations of motion and Bianchi identities for $\mathring{G}$ are:

\begin{align}
    \begin{aligned}
        d\left(e^{3\Delta} F_4\right) &= 0,\\
        e^{-6\Delta} d\left(e^{6\Delta} \star_8 f_1\right) - \frac{1}{2} F_4 \wedge F_4 &= 0,\\
        e^{-6\Delta} d\left(e^{6\Delta} \star_8 F_4\right) - f_1 \wedge F_4 &= 0,
    \end{aligned}
\end{align}

\noindent such that $\star_8$ denotes the Hodge star operator related to the $8$-manifold $\M_8$ metric. The supersymmetry conditions in the context of AdS$_3$ compactification can be analysed by considering the ansatz for the Majorana spinor field ${\mathring \upeta}$:

\begin{equation}
    \mathring{\eta} = e^{\frac{\Updelta}{2}} \eta,
\end{equation}

\noindent where $\eta = \psi \otimes \upxi$, with $\upxi$ being a real Majorana–Weyl spinor on $(\M_{8}, \mathtt{g})$ carrying the irreducible representation of $\text{Spin}(8,0)$ and $\psi$ being a Majorana spinor on AdS$_3$ carrying the irreducible representation of $\text{Spin}(1,2)$. Geometrically, $\upxi$ is a section of the bundle of real spinors on $\M_{8}$, a real vector bundle of rank $16$ on $\M_{8}$. It also carries a representation of the Clifford algebra $\cl_{8,0}$. Since $p-q \equiv_8 0$ for $p = 8$ and $q = 0$, the normal simple case is considered, and the structure $\upgamma: (\bigwedge T^\ast \M_{8}, \diamond) \rightarrow (\text{End}(\mathcal{S}), \circ)$, underlying the Kähler–Atiyah bundle, is an isomorphism. In the Euclidean $(8,0)$ signature, there is a Spin$(8)$-invariant admissible bilinear pairing $B$ on the bundle of real spinors $\mathcal{S}$ with with $\upsigma(B) = 1$ and $\epsilon(B) = 1.$ Now, when assuming that $\psi$ from $\eta = \psi \otimes \upxi$ is a Killing spinor on the AdS$_3$ space, the supersymmetry condition in Eq. \eqref{eq_wedhusj} splits into a system - \textit{Generalised Killing conditions} - consisting of the following constraints on $\upxi$:

\begin{align}\label{eq_rt4gf65q}
    \begin{aligned}
        \mathcal{D}_m \upxi &= 0,\\
\mathcal{Q} \upxi &= 0,
    \end{aligned}
\end{align}

\noindent  where $\mathcal{D}_m$ is a linear connection on $\mathcal{S}$ and $\mathcal{Q} \in \Gamma(\M_{8}, \text{End}(\mathcal{S}))$ is an endomorphism of the bundle of real spinors. Analogously to Refs. \cite{Martelli:2003ki,Tsimpis,Ashmore:2022ydf}, the Majorana spinor field $\upxi$ is not assumed to have definite
chirality. The space of solutions of Eq.\eqref{eq_rt4gf65q} is a finite-dimensional real subespace $\mathcal{K}(\mathcal{D},\mathcal{Q} )$ of the space $\Gamma(\M_{8},\mathcal{S})$ of smooth sections of $\mathcal{S}$. Refs. \cite{bab2,Babalic:2014fua} focused on obtaining a set of metrics and fluxes on $\M_8$ preserving a fixed number of supersymmetries in AdS$_3$. Equivalently, the set of metrics and fluxes on $\M_8$ is consistent with the $s$-dimensional subspace $\mathcal{K}(\mathcal{D},\mathcal{Q} )$, for a given  $s\in\mathbb{N}$. Therefore, the amount $\mathcal{N}$ of supersymmetry that is preserved in three dimensions correspond to the $\dim \mathcal{K}(\mathcal{D},\mathcal{Q} ) = s$.  The case of supergravity regarding $\mathcal{N}=1$ supersymmetry on $3$-dimensional manifolds  was considered in Refs. 
\cite{Martelli:2003ki,Tsimpis,bab2,Babalic:2014fua,Becker:2003wb}, which reported  the
explicit expressions for $\mathcal{D}_m$ and $\mathcal{Q}$ in Eqs. \eqref{eq_rt4gf65q} as 
\begin{align}
    \begin{aligned}
        \mathcal{D}_m &= {\mathbf{\nabla}}^S_m + \frac{1}{4} f_p \upgamma_m (\star_8 \upgamma^p) + \frac{1}{24} F_{mnpq} \upgamma^{npq} + \upkappa (\star_8 \upgamma_m),\\
        \mathcal{Q} &= \frac{1}{2} \upgamma^m \partial_m \Updelta - \frac{1}{288} F_{mpqr} \upgamma^{mpqr} - \frac{1}{6} f_p (\star_8 \upgamma^p) - \upkappa \upgamma^{1 \ldots 8}.
    \end{aligned}
\end{align}

\noindent These expressions reflect the dependence on the AdS$_3$ cosmological parameter, and solutions can be explored by considering bilinear covariants ${\boldsymbol{\check{E}}}^{(k)}_{\upxi,\upxi'}$, which satisfy Fierz identities. Within this framework, the number of supersymmetries preserved in AdS$3$ is reflected in the dimension $s$ of $\mathcal{K}(\mathcal{D},\mathcal{Q} )$. The solution space of Eqs. \eqref{eq_rt4gf65q} can be reinterpreted using equations involving bilinear covariants
${\boldsymbol{\check{E}}}^{(k)}_{\upxi,\upxi'}=\frac{1}{k!}B(\upxi,\upgamma_{m_1\ldots m_k}\upxi') e^{m_1\ldots m_k}$, provided that the spinor fields $\upxi,\upxi'$ satisfy Eqs. \eqref{eq_rt4gf65q}.  The equations governing the bilinear covariants are derived when the algebraic constraints $\mathcal{Q}\upxi=\mathcal{Q}\upxi'=0$ are expressed as

\begin{equation}
B\left(\upxi,\left(\mathcal{Q}^\intercal\circ\upgamma_{m_1\ldots m_k}\pm \upgamma_{m_1\ldots m_k}\circ \mathcal{Q}\right) \upxi'\right) =0
\end{equation}

Additionally, the remaining constraints $D_m \upxi=D_m\upxi'=0$ can be resolved using an algorithm detailed in Ref. \cite{Martelli:2003ki} and re-obtained in Ref. \cite{bab2} within the Fierz isomorphism groundwork. For the simplest case of $s=1$, corresponding to $\mathcal{N}=1$ supersymmetry in AdS$_3$, Eqs. \eqref{eq_rt4gf65q} should admit at least one non-trivial solution $\upxi$ \cite{bab2,Babalic:2014fua}. This allows for considering CGK spinor equations \eqref{eq_rt4gf65q} on $(\M_8, \mathtt{g})$ under the assumption of a $1$-dimensional solution space, which corresponds to $s=1$. In this scenario, the $\mod 8$ equivalence $p-q\equiv_8 0$ results in the standard simple case in which we can consider the Fierz isomorphism. These spinor bilinear covariants are subject to generalised Fierz identities, similar to Lounesto's classification in the context of AdS$_3\times \M_8$ compactification. By appropriately combining the bilinear forms, one can derive $32$ new classes of spinor fields.

\paragraph{ } The (off-shell) underlying structure of $\mathcal{N} = 1$ supergravity in three dimensions has been long comprehended \cite{Howe:1977us,Brown:1979ma,Siegel:1979fr}.  Significant advancements in minimal $3$-dimensional supergravity, including the $\mathcal{N} = 1$ massive case, have been documented in Refs. \cite{vanNieuwenhuizen:1985cx,Uematsu:1984zy,Howe:1995zm,Kuzenko:2011xg,Kuzenko:2016qwo}.  An approach to the gravitino in supergravity, utilising the quadratic spinor Lagrangian and the classification of spinor fields, was explored in Ref. \cite{daRocha:2009gb}. For further information on the context of eight-dimensional manifolds in M-theory compactification, see \cite{Babalic:2014fua}, which explores the characterisation of eight-manifolds arising in $\mathcal{N} = 1$ flux compactifications and provides detailed analyses of foliations and their G2 structures. The geometric framework for studying supersymmetric backgrounds using Clifford bundles is developed in \cite{Lazaroiu:2012kxa}, offering a method for translating supersymmetry conditions into geometric terms. The study of G-structures and their role in eight-manifold compactifications is elaborated in \cite{Babalic:2015xia}, where the authors describe the stratifications of manifolds induced by Majorana spinors and their implications for supersymmetry. Lastly, \cite{Babalic:2013iha} investigates warped compactifications preserving $\mathcal{N} = 2$ supersymmetry using geometric algebra techniques, shedding light on the supersymmetry constraints and the role of vector fields derived from pinor bilinears on the compactification manifold. 
\clearpage
\phantomsection
\addcontentsline{toc}{chapter}{Índice Remissivo}
\printindex
\end{document}